\documentclass[a4paper,reqno]{amsbook} 
\usepackage{cite}
\usepackage{amsmath, amsxtra, mathrsfs, amsfonts, amssymb,hyperref}
\usepackage{amsthm}
\usepackage{graphicx}
\usepackage{xspace, varioref}
\usepackage[margin=4cm]{geometry}
\usepackage{fancyhdr}

\newcounter{mycounter}

\DeclareMathOperator{\acl}{acl}
\DeclareMathOperator{\fo}{fo}

\newcommand{\disj}{\mathrel{||}}

\newcommand{\nocontentsline}[3]{}
\newcommand{\tocless}[2]{\bgroup\let\addcontentsline=\nocontentsline#1{#2}\egroup}
\DeclareMathOperator{\Sym}{Sym}
\DeclareMathOperator{\Id}{Id}
\DeclareMathOperator{\Forb}{Forb}
\DeclareMathOperator{\Exprfo}{Expr-fo}
\DeclareMathOperator{\Exprep}{Expr-ep}
\DeclareMathOperator{\Exprex}{Expr-ex}
\DeclareMathOperator{\Exprpp}{Expr-pp}
\DeclareMathOperator{\id}{id}
\newcommand{\majority}{\textit{majority}}
\newcommand{\minority}{\textit{minority}}
\DeclareMathOperator{\Csp}{CSP}
\DeclareMathOperator{\CD}{D} % canonical Database
\DeclareMathOperator{\CQ}{Q} % canonical Query
\DeclareMathOperator{\Th}{Th}
\DeclareMathOperator{\Qcsp}{QCSP}
\newcommand{\suc}{{\textit{succ}}}

\newcommand{\To}{\rightarrow}

\newcommand{\C}{{\mathcal C}}
\newcommand{\F}{{\mathcal F}}

\newcommand{\B}{{\mathcal B}}
\newcommand{\R}{{\mathcal R}}
\renewcommand{\P}{{\mathcal P}}

\newcommand{\N}{{\mathcal N}}
\newcommand{\V}{{\mathcal V}}
\newcommand{\cl}[1]{\langle #1 \rangle}
\DeclareMathOperator{\Pol}{Pol} 
\DeclareMathOperator{\Clo}{Clo} 
 
\DeclareMathOperator{\Alg}{Alg} 
\DeclareMathOperator{\Inv}{Inv} 
\DeclareMathOperator{\Aut}{Aut}
\DeclareMathOperator{\End}{End}

\DeclareMathOperator{\pol}{Pol} 
\DeclareMathOperator{\inv}{Inv} 
\DeclareMathOperator{\aut}{Aut}

\DeclareMathOperator{\fin}{fin}
\DeclareMathOperator{\HS}{HS}
\DeclareMathOperator{\HSP}{HSP}
\DeclareMathOperator{\HHH}{H}
\DeclareMathOperator{\PPP}{P}
\DeclareMathOperator{\PPPfin}{P^{\fin}}
\DeclareMathOperator{\SSS}{S}
\DeclareMathOperator{\HSPfin}{HSP^{\fin}}
\DeclareMathOperator{\HSSP}{HSSP}

\newcommand{\ignore}[1]{}

\newcommand{\sw}{{\it sw}}
\DeclareMathOperator{\Bool}{Boole}
\DeclareMathOperator{\inj}{inj}
\DeclareMathOperator{\Gsat}{Graph-SAT}
\DeclareMathOperator{\tp}{tp}
\DeclareMathOperator{\maxi}{max}
\DeclareMathOperator{\mini}{min}
\DeclareMathOperator{\xor}{xor}
\DeclareMathOperator{\xnor}{xnor}
\newcommand{\nin}{\notin}
\DeclareMathOperator{\OIT}{1IN3}
\DeclareMathOperator{\NAE}{NAE}
\DeclareMathOperator{\Cycl}{Cycl}
\DeclareMathOperator{\Betw}{Betw}
\DeclareMathOperator{\btw}{Betw}
\DeclareMathOperator{\Sep}{Sep}

\newcommand{\NNE}{{N}{N}{E}}
\newcommand{\NEN}{{N}{E}{N}}
\newcommand{\ENN}{{E}{N}{N}}
\newcommand{\NNN}{{N}{N}{N}}
\newcommand{\EEE}{{E}{E}{E}}
\newcommand{\EEN}{{E}{E}{N}}
\newcommand{\ENE}{{E}{N}{E}}
\newcommand{\NEE}{{N}{E}{E}}
\newcommand{\EE}{{E}{E}}
\newcommand{\NN}{{N}{N}}
\newcommand{\NE}{{N}{E}}
\newcommand{\EN}{{E}{N}}
\newcommand{\EEQ}{{E}{=}}
\newcommand{\NEQ}{{N}{=}}
\newcommand{\NEQEQ}{{\neq}{=}}
\newcommand{\NEQNEQ}{{\neq\neq}}
\DeclareMathOperator{\NEQNEQNEQ}{\neq\neq\neq}

\newtheorem{theorem}{Theorem}[section]
\newtheorem{observation}[theorem]{Observation}
\newtheorem{proposition}[theorem]{Proposition}
\newtheorem{definition}[theorem]{Definition}
\newtheorem{corollary}[theorem]{Corollary}
\newtheorem{lemma}[theorem]{Lemma}

\newtheorem{question}{Question}[chapter]
\newtheorem{conjecture}{Conjecture}[chapter]
\theoremstyle{remark}
\newtheorem{example}[theorem]{Example}

\newcommand{\cal}[1]{\mathcal{#1}}

\newcommand{\mult}{\times}

\newcommand{\DR}{{\rm DR}}
\newcommand{\PO}{{\rm PO}}
\newcommand{\PP}{{\rm PP}}
\newcommand{\EQ}{{\rm EQ}}
\newcommand{\PPI}{{\rm PPI}}

\newcommand{\yca}{{\it yca}}

\newcommand{\cproblem}[3]{
\vspace{.2cm}
\noindent {\bf #1} \\
INSTANCE: #2 \\
QUESTION: #3 \\}
\newcommand{\Nesetril}{Ne\v{s}et\v{r}il}
\newcommand{\Fresse}{Fra\"{i}ss\'{e}}

\newcommand{\true}{\textit{true}}
\newcommand{\false}{\textit{false}}

\newcommand{\Neg}{{\leftrightarrow}}
\newcommand{\Cyc}{{\circlearrowright}}
\newcommand{\cyc}{{\circlearrowright}}
\newcommand{\AQ}{\ensuremath{\Aut((\mathbb Q;<))}}
\newcommand{\mi}{{\it mi}}
\renewcommand{\min}{{\it min}}
\newcommand{\mx}{{\it mx}}

\renewcommand{\max}{{\it max}}

\newcommand{\lex}{\text{\it lex}}
\newcommand{\lele}{\text{\it ll}}

\newcommand{\pp}{\text{\it pp}}
\newcommand{\dpp}{\text{\it dual{-}pp}}
\newcommand{\dll}{\text{\it dual{-}ll}}

\newcommand{\bF}{\mathfrak{F}}

\newcommand{\bA}{\ensuremath{\mathfrak{A}}}
\newcommand{\bB}{\ensuremath{\mathfrak{B}}}
\newcommand{\bC}{\ensuremath{\mathfrak{C}}}
\newcommand{\bD}{\ensuremath{\mathfrak{D}}}
\newcommand{\bE}{\ensuremath{\mathfrak{E}}}
\newcommand{\bG}{\ensuremath{\mathfrak{G}}}
\newcommand{\bH}{\ensuremath{\mathfrak{H}}}
\newcommand{\bN}{\ensuremath{\mathfrak{N}}}
\newcommand{\bM}{\ensuremath{\mathfrak{M}}}
\newcommand{\bP}{\ensuremath{\mathfrak{P}}}
\newcommand{\bS}{\ensuremath{\mathfrak{S}}}

\newcommand{\bT}{\mathfrak{T}}
\newcommand{\mA}{{\mathbb A}}
\newcommand{\mB}{{\mathbb B}}

\newcommand{\mV}{{\mathbb V}}
\newcommand{\mN}{{\mathbb N}}
\newcommand{\mQ}{{\mathbb Q}}
\newcommand{\mR}{{\mathbb R}}
\newcommand{\mZ}{{\mathbb Z}}

\newcommand{\cB}{{\mathscr B}}
\newcommand{\cC}{{\mathscr C}}
\newcommand{\cD}{{\mathscr D}}
\newcommand{\cE}{{\mathscr E}}
\newcommand{\cF}{{\mathscr F}}
\newcommand{\cG}{{\mathscr G}}

\newcommand{\cM}{{\mathscr M}}
\newcommand{\cO}{{\mathscr O}}
\newcommand{\cOn}{{\mathscr O}^{(n)}}

\newcommand{\cP}{{\mathscr P}}

\newcommand{\fA}{{\bf A}}
\newcommand{\fB}{{\bf B}}
\newcommand{\fC}{{\bf C}}

\newcommand{\fG}{{\bf G}}

\newcommand{\fN}{{\bf N}}

\newcommand{\fS}{{\bf S}}

\begin{document}
\thispagestyle{empty}
\begin{center}
\Huge{Complexity Classification in Infinite-Domain Constraint Satisfaction} \\
\end{center}
\vspace{2cm}
\begin{center}
\Large
M\'emoire pour l'obtention d'une habilitation \`a diriger des recherches \\
Universit\'e Paris Diderot -- Paris 7 \\
Sp\'ecialit\'e Informatique \\
\end{center}
\vspace{1cm}
\begin{center}
\Large
pr\'esent\'e par \\
Manuel Bodirsky
\end{center}
\vspace{1cm}

\begin{center}
soutenue publiquement le 19 janvier 2012, devant le jury compos\'e de : \\
\vspace{.5cm}
\begin{tabular}{lll}
Arnaud & DURAND & Universit\'e Paris 7, IMJ, \'Equipe Logique Math\'ematique \\
Christoph & D\"URR & CNRS / LIP6, Universit\'e Paris 6 \\
Markus & JUNKER &  Mathematisches Institut, Albert-Ludwigs-Universit\"at Freiburg \\
Pascal & KOIRAN & ENS Lyon \\
Luc & SEGOUFIN & INRIA / LSV, CNRS+ENS Cachan 
\end{tabular} \\
\vspace{1cm}
Les rapporteurs sont : \\
\vspace{.5cm}
\begin{tabular}{lll}
V\'ictor & DALMAU & Universit\'e Pompeu Fabra, Barcelone, Espagne \\
Arnaud & DURAND & Universit\'e Paris 7, IMJ, \'Equipe Logique Math\'ematique \\
Peter & JONSSON & Universit\'e de Link\"oping, Suede \\
\end{tabular}
\end{center}

\begin{abstract}
A \emph{constraint satisfaction problem (CSP)} is a computational problem 
where the input consists of a finite set of variables and a finite
set of constraints, and where the task is to determine whether there
exists an assignment of values to the variables that satisfies
all the given constraints. Depending on the type of constraints that
we allow in the input, a constraint satisfaction problem might
be tractable, or computationally hard.
In recent years, very general criteria have
been discovered that imply that a constraint satisfaction problem is
polynomial-time tractable, or that it is NP-hard (and most likely not polynomial-time tractable). Those results usually
focus on the situation when the set of values that a variable can attain is \emph{finite}. The complexity of CSPs has become a major common research focus of graph theory, artificial intelligence, and finite model theory. It turned out that the key questions for complexity classification
of finite domain CSPs are closely
linked to central questions in universal algebra.

This thesis extends the powerful techniques for constraint satisfaction to CSPs with \emph{infinite} domains. 
The generalization to infinite domains enhances dramatically the range of computational problems that can be modeled as a CSP. 
Many problems from areas that have so far seen no interaction with constraint satisfaction theory can be formulated using infinite domains (and not with finite domains), for instance problems from temporal and spatial reasoning, phylogenetic reconstruction, computer algebra, and operations research. 

It turns out that the universal-algebraic approach can also
be applied to study large classes of infinite-domain CSPs,
and that it can be used to obtain elegant complexity classification results. The universal-algebra approach can also be applied to analyze and generalize existing algorithmic methods, and to discover new polynomial-time tractable classes of CSPs. 
%To employ the universal algebraic approach for infinite domain constraint satisfaction 
%we need fundamental concepts from model theory. 
A new tool in this thesis and that becomes relevant particularly 
for infinite domains is Ramsey theory.
We demonstrate the feasibility of our approach with two complete complexity classification results: one on CSPs in temporal reasoning,
the other on a generalization of Schaefer's theorem for propositional logic to propositional logic over graphs. 
We also present results about the limits of complexity classification,
and show that certain rich classes of computational problems provably do not exhibit a complexity dichotomy into hard and easy problems. 
% BOOK VERSION: The thesis closes with a list of proposals for future research 
%directions and open problems. 
\end{abstract}

\newpage 

{\bf Acknowledgements.}
I want to thank my institution, the CNRS, for the great freedom in research that allowed me to write this text. 
The research leading to the results presented here has also received funding from the European Research Council under the European Community's Seventh Framework Program (FP7/2007-2013 Grant Agreement no. 257039).
I also want to thank all of my constraint satisfaction co-authors for the good time we had with our joint work. 

My first address at the \'Equipe de Logique Math\'ematique has always
been Arnaud Durand, and I am indebted to him for his great help with
many things over the years, including the process of the habilitation. Many thanks also to Micha\l\ Wrona, Fran\c{c}ois Bossi\`{e}re, Trung van Pham, Antoine Mottet, Johannes Greiner, Michael Kompatscher, Christian Pech, and Florian Starke for reporting mistakes in earlier versions of this text. 

Special thanks to Christoph D\"urr for the permission to 
include his wonderful pictures at the beginning of many chapters. 
More of his artwork can be found under
\begin{center}
\emph{http://picasaweb.google.com/xtof.durr/LaVieEstDurr}.
\end{center}
The pictures for Chapters~\ref{chap:ecsp}, \ref{chap:schaefer}, and~\ref{chap:tcsp} are
drawings that I kept from discussions with Jan K\'ara, Martin Kutz, and Jaroslav \Nesetril. 
The picture for Chapter~\ref{chap:examples} is by Lewin Bodirsky, Summer 2010, and the picture for Chapter~\ref{chap:topology} is by Otto Bodirsky, Summer 2011.

\tableofcontents
\thispagestyle{empty}
\setcounter{page}{1}  
\chapter{Introduction}
\label{chap:intro}
\begin{center}
\includegraphics[scale=.6]{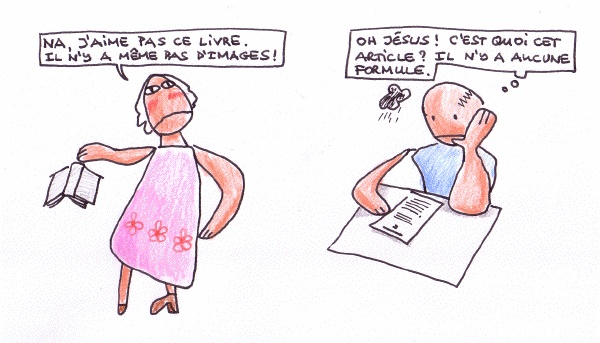}
\end{center}
\thispagestyle{empty}

Constraint satisfaction problems (CSPs) 
appear in almost every area of theoretical computer science, for instance in artificial intelligence, scheduling, computational linguistics,
computational biology, verification, and algebraic computation. 
Many computational problems studied in those areas can be modeled by appropriately
choosing a set of constraint types, the \emph{constraint language},
that are allowed in the input instance of a CSP.
In the last decade, huge progress
was made to find general criteria for constraint languages that imply that the corresponding CSP can be solved efficiently~\cite{FederVardi,JBK,Bulatov,Conservative,IMMVW,Maltsev,BoundedWidth}. 

Lately, the complexity of the CSP became a topic that vitalizes the
 field of universal algebra, since it turned out that questions about the
computational complexity of CSPs translate to important universal-algebraic questions about algebras that can be associated to CSPs. This approach
is now known as the \emph{algebraic approach} to constraint
satisfaction complexity. The algebraic approach has raised questions
that are of central importance in universal algebra.

Another reason why the complexity of CSPs attracts attention 
is an exciting conjecture due to Feder and Vardi~\cite{FederVardi}, which is still unresolved, and which is known as the \emph{dichotomy
  conjecture}. This conjecture says that every CSP \emph{with a finite
domain} is either polynomial-time tractable (i.e., in P) or NP-complete. According to
a well-known result by Ladner, it is known that there are
NP-\emph{intermediate} computational problems, i.e., problems in NP that are
neither tractable nor NP-complete (unless P=NP). But the known
NP-intermediate problems are extremely artificial.  
It would be interesting from a complexity theoretic perspective
to discover more natural candidates for NP-intermediate problems. 
Unlike many questions in computational complexity that are wide open,
the dichotomy conjecture allows many promising partial results and different approaches (see the collection of survey articles in~\cite{CSPSurveys}), and therefore is an attractive research topic.

Any outcome of the dichotomy conjecture is significant: a negative answer might provide
relatively natural NP-intermediate problems, which would be 
interesting for complexity theorists. 
A positive answer probably comes
with a criterion which describes the NP-hard CSPs (and it would
probably even provide algorithms for the polynomial-time tractable CSPs). But then we
would have a fascinatingly rich catalogue of computational problems
where the computational complexity is known. Such a catalogue would be
a valuable tool for deciding the complexity of computational problems: 
since CSPs are abundant, one might
derive algorithmic results by reducing the problem of interest to a
known tractable CSP, and one might derive hardness results by reducing
a known NP-hard CSP to the problem of interest.

Even though very powerful partial results on the dichotomy conjecture have
been obtained in recent years, 
the impact of constraint satisfaction complexity theory
on other fields in theoretical computer science has so far been 
modest. A reason might be that the range of problems in the literature 
that can be described by specifying a constraint language over a finite
domain, and that have been studied independently from the CSP framework, is quite limited,
and mostly focussed on specialized graph theoretic problems or Boolean satisfiability problems.

If we consider the class of all problems that can be formulated 
by specifying a constraint language over an \emph{infinite} domain,
the situation changes drastically. Many problems that have been studied independently 
in temporal reasoning,
%(see Section~\ref{sssect:temporal}) 
spatial reasoning, %(Section~\ref{sssect:spatial}, 
phylogenetic reconstruction, %~\ref{sssect:phylo}, 
and computational linguistics %~\ref{sssect:ling}
 can be directly formulated as CSPs.
Also feasibility problems in linear (and also non-linear) programming (over the rationals, the integers, or other domains) can be cast as CSPs. 
%(see Subsection~\ref{sssect:lp}).

The goal of this thesis is to generalize the universal-algebraic approach to infinite domains. It turns out that this is possible when the constraint language, viewed as a relational
structure $\bB$ with an infinite domain, is \emph{$\omega$-categorical}. 
Many of the CSPs in the mentioned application areas can be formulated with $\omega$-categorical constraint languages --- in particular, problems coming from
so-called \emph{qualitative calculi} in artificial intelligence tend to have
formulations with $\omega$-categorical constraint languages.
While $\omega$-categoricity is a quite strong assumption from
a model-theoretic point of view (and, for example, constraint languages for linear programming cannot be $\omega$-categorical), the class of computational
problems that can be formulated with $\omega$-categorical 
constraint languages is still a very large generalization of the class of CSPs that can be formulated with a constraint language over a finite domain.
This will be amply demonstrated by examples of $\omega$-categorical constraint languages from many different areas in computer science in Chapter~\ref{chap:examples}.

There are several general results for
$\omega$-categorical structures that are relevant when studying the computational complexity of the respective CSPs.
Every $\omega$-categorical structure is homomorphically equivalent to an $\omega$-categorical structure which is \emph{model-complete} and a \emph{core}.
Model-complete cores have many good properties: for example, those structures have quantifier elimination once expanded by all primitive positive definable relations; this is treated in Chapter~\ref{chap:mt}. Since homomorphically equivalent structures have the same CSP, we can therefore focus on constraint languages that have 
those properties.

Moreover, 
it can be shown that the so-called \emph{polymorphism clone} of an $\omega$-categorical structure $\bB$
fully captures the computational complexity of the corresponding CSP (Chapter~\ref{chap:algebra}). 
By this observation, universal-algebraic
techniques can be used to analyze the computational complexity of
the CSP for $\bB$. Indeed, the study of CSPs has triggered questions that are
of central interest in universal algebra, and that have led to considerable new activity 
(see e.g.~\cite{BartoKozikLICS10,MarotiMcKenzie,Siggers,BIMMVW-TAMS}). 

Another tool that becomes useful specifically for polymorphisms over
infinite domains is \emph{Ramsey theory} (Chapter~\ref{chap:ramsey}). 
 The basic idea here is to apply Ramsey theory to show that
polymorphisms must act \emph{canonically} on large parts of their domain.
Typically there are only finitely many possibilities for canonical behavior,
and so this technique allows to perform combinatorial analysis when proving classification results. 
With this approach we can also show that, 
under further assumptions on $\bB$,
many questions about the expressive power of $\bB$ become decidable, 
such as the question whether a given quantifier-free first-order formula is in
$\bB$ equivalent to a primitive positive formula. 

An important feature of the universal-algebraic approach
is that tractability of a CSP can be linked to the existence of
polymorphisms of the constraint language.
This link can be exploited in several directions: first, when we already know that a constraint language of interest has a polymorphism satisfying good properties, then this polymorphism can guide the search for an efficient algorithm for the corresponding CSP. Another direction is that we already have an algorithm (or an algorithmic technique), and that we want to
 know for which CSPs the algorithm is a correct decision procedure: again, polymorphisms are the key tool for this task.
 Finally, we might use the absence of polymorphisms with good properties to prove that a CSP is NP-hard. There are several instances
 where these three directions of the algebraic approach have been used very successfully for CSPs with finite domain constraint languages~\cite{Maltsev,SLAAC,IMMVW,BoundedWidth} or $\omega$-categorical constraint languages~\cite{tcsps-journal,BodPin-Schaefer-Both}.
 
In Chapter~\ref{chap:schaefer} and Chapter~\ref{chap:tcsp} we use polymorphisms to classify the computational complexity in some large families of constraint satisfaction problems. In Chapter~\ref{chap:schaefer}, we study constraint languages definable over
the random graph, and in Chapter~\ref{chap:tcsp} constraint languages definable over $(\mQ;<)$. Even though the two underlying structures are very different from a model-theoretic point of view, and even though the classification proofs are very different
in both cases, we can give a common formulation of the two classification results 
that delineates also the border between polynomial-time solvable and NP-complete
CSPs. 
%The resulting classification results provide dichotomy statements that are 
%meaningful even when P=NP, since they turn out to be dichotomy statements
%about the logical expressive power of constraint languages.

\subsection*{Chapter outline}
Constraint satisfaction problems can appear in several different forms,
because there are several ways how CSPs can be formalized. 
%Those forms can be viewed as different ways of formalizing the same class of computational problems.
The differences in formalizing constraint satisfaction problems
are related to the way how instances are coded and to how the
problem itself is described.
In the next sections
we present four formalisms; each of those formalisms is 
attached to a different line of research.
In later sections some arguments are more natural from one perspective than from the other, so it will be convenient to have them all discussed here. 
See Figure~\ref{fig:perspectives} for an illustration
how the four perspectives we discuss can be put into relationship to each other.

\begin{figure}[h]
\begin{center}
\begin{tabular}{|l||l|l|}
\hline
Perspective & Instance & Problem Description \\
\hline \hline
Homomorphism & Structure & Structure \\
Sentence Evaluation & Sentence & Structure \\
Satisfiability & Sentence & Sentences \\
Existential Second-Order & Structure & Sentence \\
\hline
\end{tabular}
\end{center}
\caption{The four perspectives on the definition of CSPs.}
\label{fig:perspectives}
\end{figure}

%!TEX root = 0.tex

\section{The Homomorphism Perspective}
\label{sect:homo}
%Here, put the perspective that a CSP is just a class of structures with signature $\tau$ that is closed under disjoint union and whose complement is closed under homomorphisms.
A \emph{relational signature} $\tau$ is a set of relation symbols
$R_i$, each of which has an associated finite arity $k_i$. 
A relational {\em structure} $\bA$ over the signature $\tau$ (also called $\tau$-\emph{structure}) consists of a set $A$ (the {\em domain} or \emph{base set}) together with a relation $R^\bA \subseteq A^k$ for each relation symbol $R$ of arity $k$ from $\tau$. It causes no harm to allow structures whose domain is empty.

A \emph{homomorphism} $h$ from a structure $\bA$ with domain $A$ 
to a structure $\bB$ with domain $B$ and the same signature $\tau$ is a mapping from
$A$ to $B$ that \emph{preserves} each relation
for the symbols in $\tau$; that is, 
if $(a_1,\dots,a_k)$ is in
$R^\bA$, then $(h(a_1),\dots,h(a_k))$ must
be in $R^\bB$.
An \emph{isomorphism} is a bijective homomorphism $h$ such that the
inverse mapping $h^{-1} \colon B \rightarrow A$ that sends $h(x)$ to $x$
is a homomorphism, too.

In this thesis, a \emph{(non-uniform) constraint satisfaction problem (CSP)} is a computational problem that is specified by a single structure with a finite relational signature, called the \emph{template} (or the \emph{constraint language}; the name `constraint language' is typically used in the context of the second perspective on CSPs that we present in Section~\ref{sect:csp-logical}). 
%Relational structures
%that denote templates for CSPs will be denoted by capital fraktur letters $\bA$, $\mB$ (and their domain by $A$, $B$, respectively).

\begin{definition}[$\Csp(\bB)$]\label{def:csp-hom}
Let $\bB$ be a (possible infinite) structure with a finite relational signature $\tau$.
Then \emph{$\Csp(\bB)$} is the computational problem to decide
whether a given finite $\tau$-structure $\bA$ homomorphically maps to $\bB$.
\end{definition}
$\Csp(\bB)$ can be considered to be a class ---
the class of all finite $\tau$-structures that homomorphically map to $\bB$.

A homomorphism from a given $\tau$-structure $\bA$ to $\bB$ is called a \emph{solution} of $\bA$ for $\Csp(\bB)$.
%We sometimes also write $\Csp(D, R_1,\dots,R_l)$ instead
%of $\Csp((D,R_1,\dots,R_l))$.
It is in general not clear
how to represent solutions for $\Csp(\bB)$ on a computer; however, for 
the definition of the problem $\Csp(\bB)$ we do not need to represent solutions,
since we only have to decide the \emph{existence} of solutions.
To represent an input structure $\bA$ of $\Csp(\bB)$
we can fix any representation of the relation symbols in the signature $\tau$, 
due to the assumption that $\tau$ is \emph{finite}.
Thus, $\Csp(\bB)$ is a well-defined computational 
problem for \emph{any}
infinite structure $\bB$ with finite relational signature. 

\begin{example}[Digraph acyclicity]\label{expl:acycl}
Next, consider the problem $\Csp(({\mathbb Z}; <))$. 
Here, the relation $<$ denotes the strict linear order of the integers 
$\mathbb Z$. An instance $\bA$ of this problem can be viewed as a directed graph (also called \emph{digraph}), potentially with loops.
It is easy to see that $\bA$ homomorphically maps to $({\mathbb Z}; <)$ if and only if there is no directed cycle 
in $\bA$ (loops are considered to be directed cycles, too). 
It is easy to see and well-known that this can be tested in linear time, for example 
by performing a depth-first search on the digraph $\bA$. 
\qed
\end{example}

\begin{example}[Betweenness]\label{expl:betw}
The so-called \emph{betweenness problem}~\cite{Opatrny} can be modeled as
$\Csp(({\mathbb Z}; \Betw))$
where $\Betw$ is the ternary relation $$\{(x,y,z) \in {\mathbb Z}^3 \; | \; (x < y < z) \vee (z < y < x)\} \; .$$ 
This problem is one of the NP-complete problems listed in the book of Garey and Johnson~\cite{GareyJohnson}.  \qed
\end{example}

\begin{example}[Cyclic-Ordering]\label{expl:cycl}
The \emph{Cyclic-order problem}~\cite{GalilMegiddo} 
can be modeled as
$\Csp(({\mathbb Z}; \Cycl))$
where $\Cycl$ is the ternary relation 
$$\{(x,y,z) \in {\mathbb Z}^3 \; | \; (x < y < z) \vee (y < z < x) \vee (z < x < y) \} \; .$$
This problem is again NP-complete and can be found in~\cite{GareyJohnson}.  \qed
\end{example}

\begin{example}[$\bH$-coloring problems]
Let $\bH$ be an (undirected) graph. We view undirected graphs
as $\tau$-structures where $\tau$ contains
a single binary relation symbol $E$, which denotes a symmetric and anti-reflexive relation.
Then the \emph{$\bH$-coloring problem} is the computational problem to decide for a given finite graph $\bG$ whether there exists a homomorphism from $\bG$ to $\bH$.
For instance, if $\bH$ is the graph $K_3$ (the complete graph on three vertices), then the $\bH$-coloring problem is the famous \emph{3-colorability problem} (see e.g.~\cite{GareyJohnson}). Similarly, for every fixed $k$, the $k$-colorability problem
can be modeled as $\Csp(\bH)$, for an appropriate graph $\bH$.
\qed \end{example}

The next lemma (Lemma~\ref{lem:fund})
is a useful test to determine
whether a computational problem can be formulated as
$\Csp(\bB)$ for an infinite relational structure $\bB$.
An \emph{(induced) substructure} of a $\tau$-structure
$\bA$ is a $\tau$-structure $\bB$ with $B \subseteq A$
and $R^{\bB}=R^\bA \cap B^n$ for each $n$-ary $R\in\tau$; we also
say that $\bB$ is \emph{induced by $B$ in $\bA$}, and write $\bA[B]$ for $\bB$.
The
\emph{union} of two $\tau$-structures $\bA,\bB$ is the $\tau$-structure
$\bA\cup\bB$ with domain $A \cup B$ and relations
$R^{\bA\cup\bB}=R^{\bA}\cup R^{\bB}$ for all $R\in\tau$. 
The \emph{intersection} $\bA \cap \bB$ of $\bA$ and $\bB$ is defined analogously. 
A \emph{disjoint
  union} of $\bA$ and $\bB$ is the union of isomorphic copies of
$\bA$ and $\bB$ with disjoint domains. 
As disjoint unions are unique up
to isomorphism, we usually speak of \emph{the} disjoint union of $\bA$ and
$\bB$, and denote it by $\bA \uplus \bB$.
The disjoint union of a set of $\tau$-structures $\cal C$ is defined
analogously (and the disjoint union of an empty set of structures
is the $\tau$-structure with empty domain).
A structure is called \emph{connected} if it is not the disjoint
union of two non-empty structures. 
A maximal connected substructure of $\bB$ is called a \emph{connected component}
of $\bB$.

\begin{definition}\label{def:inv-hom-disj-union}
We say that a class $\cal C$ of relational structures is 
\begin{itemize}
\item \emph{closed under homomorphisms} iff whenever $\bA \in \cal C$ and $\bA$ homomorphically maps to $\bB$ then $\bB \in \cal C$;
\item \emph{closed under inverse homomorphisms} iff whenever $\bB \in \cal C$ and $\bA$ homomorphically maps to $\bB$ then $\bA \in \cal C$;
\item \emph{closed under (finite) disjoint unions} iff whenever $\bA,\bB \in \cal C$
then the disjoint union of $\bA$ and $\bB$ is also in $\cal C$. 
\end{itemize}
\end{definition}

Note that a class $\cal C$ of $\tau$-structures is closed under inverse homomorphisms 
if and only if its complement in the class of all $\tau$-structures is closed under homomorphisms. When a class is closed under inverse homomorphisms, or closed under
homomorphisms, 
it is in particular closed under isomorphisms. The following is a simple, but 
fundamental lemma for CSPs. 
When $\cal N$ is a class of $\tau$-structures,
we say that a structure $\bA$ is \emph{$\N$-free} if no $\bB \in \N$ homomorphically maps to $\bA$.
The class of all finite $\N$-free structures we denote by $\Forb(\N)$.

% DOES IT WORK FOR INFINITE SIGNATURES?! YES!

\begin{lemma}\label{lem:fund}
Let $\tau$ be a finite relational signature, and $\cal C$ a class of finite $\tau$-structures.
Then the following are equivalent.
\begin{enumerate}
\item ${\cal C} = \Csp(\bB)$ for some $\tau$-structure
$\bB$.
\item ${\cal C} = \Forb(\N)$ for a class of finite connected $\tau$-structures $\N$. 
\item $\cal C$ is closed under disjoint unions and inverse homomorphisms.
\item ${\cal C} = \Csp(\bB)$ 
for a countably infinite $\tau$-structure $\bB$.
\end{enumerate}
\end{lemma}
\begin{proof}
It suffices to prove the implications $(1) \Rightarrow (2) \Rightarrow (3) \Rightarrow (4)$. 
% (1) implies (2), (2) implies (3), and (3) implies (4). 
For the implication from (1) to (2), let $\N$ be the class of
all finite connected $\tau$-structures that do not homomorphically
map to $\bB$. Then by transitivity of the homomorphism relation, 
a $\tau$-structure $\bA$ homomorphically maps to $\bB$ if and only if
no $\bC \in \N$ homomorphically maps to $\bA$.
%if a structure $\bA$ homomorphically maps to $\bB$ then no $\bC \in \N$ homomorphically maps to $\bA$. Conversely, if a structure $\bC \in \N$ homomorphically maps to a $\tau$-structure $\bA$, 
%then $\bA$ does not homomorphically map to $\bB$. 

(2) implies (3). Suppose (2), and let $\bA_1$ and $\bA_2$
be two structures from $\Forb(\N)$. If there were a homomorphism
from one of the structures $\bC \in \N$ into $\bA_1 \uplus \bA_2$,
then because $\bC$ is connected, it must already be a homomorphism into $\bA_1$ or $\bA_2$, which is impossible. Hence, $\Forb(\N)$ is closed under disjoint unions. Closure under inverse homomorphisms
follows straightforwardly from transitivity of the homomorphism relation. 

(3) implies (4). Suppose that $\cal C$ is a class of
finite relational structures that is closed under disjoint unions 
and inverse homomorphisms. Let $\cal C'$ be a subclass of $\cal C$
where we select one structure from each isomorphism class of structures in $\cal C$. Let $\bB$ be the (countably infinite) 
disjoint union over all structures in $\cal C'$ (if $\C$ is empty then $\bB$ is by definition
the empty structure\footnote{Structures with an empty domain are often forbidden in model theory. Lemma~\ref{lem:fund} is one of the places that motivates our decision to allow them in this text.}). 
Clearly, every structure in $\cal C$ homomorphically maps to $\bB$.
Now, let $\bA$ be a finite structure with a homomorphism $h$ to $\bB$.
By construction of $\bB$, the set $h(A)$ is contained 
in the disjoint union $\bC$ of a finite
set of structures from $\cal C$. Since $\cal C$ is closed under 
disjoint unions, $\bC$ is in $\cal C$. Clearly, 
$\bA$ homomorphically maps to $\bC$, and because
$\cal C$ is closed under inverse homomorphisms, 
$\bA$ is in $\cal C$ as well.
%The implication from (4) to (1) is trivial.
\end{proof}

\begin{figure}
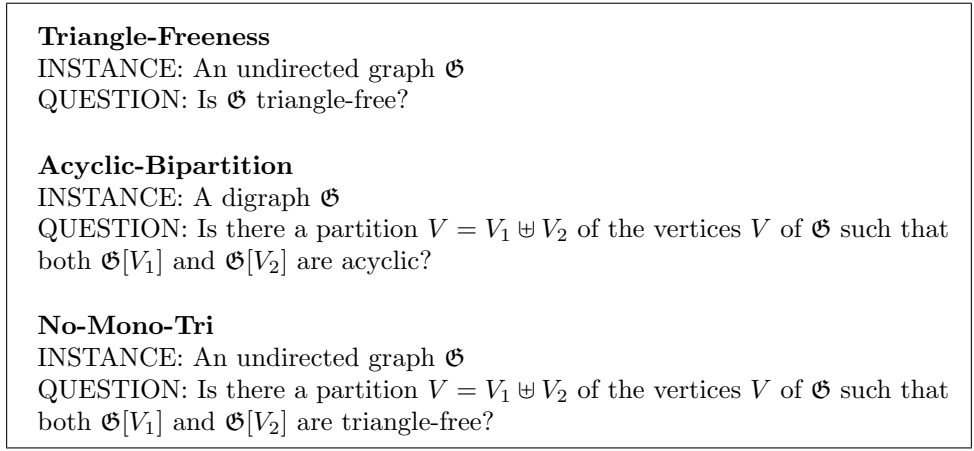

\fbox{
\begin{tabular}{p{12cm}}
\vspace{-0.3cm}
\cproblem{Triangle-Freeness}
{An undirected graph $\bG$}
{Is $\bG$ triangle-free?}
\cproblem{Acyclic-Bipartition}
{A digraph $\bG$}
{Is there a partition $V = V_1 \uplus V_2$ of the vertices $V$ of $\bG$
such that both $\bG[V_1]$ and $\bG[V_2]$ are acyclic?}
\cproblem{No-Mono-Tri}
{An undirected graph $\bG$}
{Is there a partition $V = V_1 \uplus V_2$ of the vertices $V$ of $\bG$
such that both $\bG[V_1]$ and $\bG[V_2]$ are triangle-free?}
\end{tabular}}
\caption{Three computational problems that are closed under disjoint unions and inverse homomorphisms.}
\label{fig:cproblems2}
\end{figure}

\begin{example}\label{expl:other-problems}
The computational problems in Figure~\ref{fig:cproblems2} 
are closed under disjoint unions
and inverse homomorphisms. Hence, Lemma~\ref{lem:fund}
shows that they can be formulated as $\Csp(\bB)$
for some relational structure $\bB$.
It is easy to see that none of those three problems 
can be formulated as $\Csp(\bB)$ for
a \emph{finite} structure
$\bB$. 

We verify this for the problem of Triangle-freeness. 
For a fixed $n$, consider the graph that contains vertices $x_1,\dots,x_n$,
and that contains for every pair $i,j$ with $1 \leq i < j \leq n$ 
two additional vertices $u_{i,j},v_{i,j}$ and the edges $(x_i,u_{i,j})$, $(u_{i,j},v_{i,j})$,
$(v_{i,j},x_j)$. The resulting graph is clearly triangle-free. But note
that every homomorphism $f$ from this graph to a graph $\bH$ 
with strictly less than $n$ vertices must identify at least two of the vertices $x_1,\dots,x_n$. So suppose that $f(x_i)=f(x_j)$. Because $f$ is a homomorphism, we have that $(f(x_i),f(u_{i,j})),(f(u_{i,j},f(v_{i,j})),(f(v_{i,j}),f(x_j))$
are edges in $\bH$. Hence, $\bH$ either contains a triangle or a loop. In both cases,
$\bH$ cannot be the template for Triangle-Freeness. Hence we have ruled out all 
templates of size $n-1$. This concludes the proof since $n$ was chosen arbitrarily.
\qed \end{example}

We close with an important concept for finite structures
$\bB$, the notion of \emph{core structures}; generalizations to infinite structures $\bB$ are presented
in Section~\ref{sect:cores}. 
Two structures $\bA$ and $\bB$ are called \emph{homomorphically equivalent} if there exists a homomorphism from $\bA$ to $\bB$
and vice versa. 
An \emph{embedding} of $\bA$ into $\bB$ is an injective map $f \colon A \rightarrow B$ such that $(a_1,\dots,a_k)$ is in
$R^\bA$ if and only if $(f(a_1),\dots,f(a_k))$ is in $R^\bB$.
An \emph{endomorphism} of a structure $\bB$
is a homomorphism from $\bB$ to $\bB$. 

\begin{definition}\label{def:cores}
A structure $\bB$ is a \emph{core} if all its endomorphisms are embeddings\footnote{For finite structures $\bB$, injective self-maps must be bijective, and
in fact every injective homomorphism of a structure $\bB$ must be an isomorphism.
For infinite structures, however, this need not be true, and for
reasons that become clear in Chapter~\ref{chap:mt} we chose
the present definition.}.
For structures $\bA,\bB$ of the same signature, the structure $\bB$ is called a \emph{core of $\bA$}
if $\bB$ is a core and homomorphically equivalent to $\bA$.
\end{definition}

In fact, we speak of \emph{the core} of a finite structure 
$\bA$, due to the following fact, whose proof is easy and left to the reader.
 
\begin{proposition}\label{prop:finite-cores}
Every finite structure $\bA$ has a core. All cores of $\bA$ are isomorphic.   
\end{proposition}

Core structures $\bB$ have many pleasant properties when it comes to studying the computational complexity of $\Csp(\bB)$
(see for instance Proposition~\ref{prop:orbits-in-finite-cores} below). 
Clearly, when $\bA$ and $\bB$ are homomorphically equivalent, then $\Csp(\bA) = \Csp(\bB)$.
Therefore, and because of  Proposition~\ref{prop:finite-cores}, 
we can assume without loss of generality that a finite structure $\bB$ is a core when studying $\Csp(\bB)$. We finally remark that structures with a
one-element core have a trivial CSP.

\begin{proposition}\label{prop:const-core}
Let $\bB$ be a relational structure with a finite relational signature and a one-element core. 
Then $\Csp(\bB)$ is in P.
\end{proposition}
\begin{proof}
Let $\bC$ be the core of $\bB$, and let $c$ be the unique element of $\bC$. 
The problem $\Csp(\bB)$ can be solved
as follows. Let $\bA$ be an input structure of $\Csp(\bB)$. If there is $(t_1,\dots,t_n) \in R^\bA$ such that $(c,\dots,c) \notin R^\bC$, then reject. Otherwise accept. 
\end{proof}

\section{The Sentence Evaluation Perspective}
\label{sect:csp-logical}
Let $\tau$ be a relational signature. 
A first-order $\tau$-formula $\phi(x_1,\dots,x_n)$ is called 
\emph{primitive positive} if it is of the form
$$ \exists x_{n+1},\dots,x_m (\psi_1 \wedge \dots \wedge \psi_l)$$
where $\psi_1, \dots, \psi_l$ are \emph{atomic $\tau$-formulas},
i.e., formulas of the form $R(y_1,\dots,y_k)$ with $R \in \tau$ 
and $y_i \in \{x_1,\dots,x_m\}$, of the form $y = y'$ for $y,y' \in \{x_1,\dots,x_m\}$, of the form $\bot$ or $\top$ for \emph{false} and \emph{true}, respectively. Note that if the domain is non-empty
then we do not need a symbol $\top$ for \emph{true}, 
since we can use the primitive positive sentence $\exists x. \, x=x$ to express it. 
As usual, formulas without free variables are called \emph{sentences}.

From a model-checking perspective, CSPs are defined as follows.
We will see (in Propositions~\ref{prop:hom-to-logic} and~\ref{prop:logic-to-hom}) that this definition is essentially the same
definition as Definition~\ref{def:csp-hom}, and that the differences
are a matter of formalization\footnote{A small difference between the homomorphism perspective and the sentence evaluation problem results from the fact that we \emph{do} allow equality in primitive positive formulas; as we will see, adding equality to the constraint language does not affect the complexity of the CSP up to log-space reductions. There are articles, though, that study the complexity of CSPs at an even finer level than logspace-reducibility, and in those papers equality is not automatically allowed in the input to a constraint satisfaction problem.}.

\begin{definition}
Let $\bB$ be a (possibly infinite) structure with a finite relational signature $\tau$. Then \emph{$\text{CSP}(\bB)$} is the computational
problem to decide whether a given primitive positive $\tau$-sentence $\phi$
is true in $\bB$.
\end{definition}

%It is straightforward to verify that $\Csp(\Gamma)$ is polynomial-time equivalent to the following computational problem
%(in fact, the two problems can be considered to be the same computational problem,
%up to formalization). The input consists of a primitive positive $\tau$-sentence $\phi$
%(i.e., a primitive positive $\tau$-formula without free variables),
%and the question is whether $\phi$ is holds true in $\Gamma$.
The given primitive positive $\tau$-sentence $\phi$ is also called an
\emph{instance} of $\Csp(\bB)$. 
The conjuncts of an instance $\phi$ are 
called the \emph{constraints} of $\phi$.
A mapping from the variables of $\phi$ to the elements of $B$ that is a satisfying 
assignment for the quantifier-free part of $\phi$ is also called a \emph{solution}
to $\phi$.

Some authors omit the (existential) quantifier-prefix in instances $\phi$
of $\Csp(\bB)$, and the question is then whether $\phi$ is
\emph{satisfiable} over $\bB$. Clearly, this is just re-phrasing the problem above, but it explains the terminology of \emph{satisfiable} and \emph{unsatisfiable} (rather than \emph{true} and \emph{false}) instances of $\Csp(\bB)$.

%Remark that it therefore makes perfect sense to talk about
%satisfiability of instances of the CSP. Also remark that 
%we sometimes also view quantifier-free conjunctions of atomic formulas 
%as instances of the CSP.

% THIS WILL BE COVERED BY THE THIRD FORMULATION
%From this formulation of the constraint satisfaction problem
%it is obvious that $\Csp(\Gamma)$ is fully determined by the
%\emph{first-order theory of $\Gamma$}
%(i.e., the set of first-order sentences that
%are valid in $\Gamma$). 
%We would like to remark that the logic perspective on $\Csp(\Gamma)$ is closely related to the evaluation problem for conjunctive queries studied in database theory. 

%\qed
%An instance of the betweenness problem 
%consists of a finite
%set $A$ and a collection $C$ of ordered triples $(a,b,c)$ from $A$. 
%The question is whether there is a function $f: A \rightarrow \mathbb Q$
%such that, for each $(a,b,c) \in C$, we either have $f(a)<f(b)<f(c)$
%or $f(c) < f(b) < f(a)$. \\

\begin{figure}
\fbox{
\begin{tabular}{p{12cm}}
\vspace{-.3cm}
\cproblem{3SAT}
{A propositional formula in conjunctive normal form (CNF) with at most three literals per clause}
{Is there a Boolean assignment for the variables such that in each clause at least one literal is true?} 
\cproblem{Positive 1-in-3-3SAT}
{A propositional 3SAT formula with only positive literals}
{Is there a Boolean assignment for the variables such that in each clause exactly one literal is true?} 
\cproblem{Positive Not-All-Equal-3SAT}
{A propositional 3SAT formula with only positive literals}
{Is there a Boolean assignment for the variables such that in each clause neither all three literals are true nor all three are false?}
\end{tabular}}
\caption{Three Boolean satisfiability problems from the list of NP-complete problems of~\cite{GareyJohnson} that can be formulated as $\Csp(\bB)$ for appropriate $\bB$.}
\label{fig:cproblems}
\end{figure}

\begin{example}[Boolean satisfiability problems]
\label{expl:sat}
There are many Boolean satisfiability problems that
can be cast as CSPs. Well-known examples are 3SAT (see Figure~\ref{fig:cproblems}), and the restricted
versions of 3SAT called 1-in-3-3SAT and NOT-ALL-EQUAL-3SAT~\cite{GareyJohnson}.
These three problems are NP-complete.
An interesting feature of the last two problems is that they remain NP-complete
even when all clauses in the input only contain positive literals.
With this additional restriction, the problems are called positive 1-in-3-3SAT and positive NOT-ALL-EQUAL-3SAT, and their definition can be found in
Figure~\ref{fig:cproblems}.

All of these problems can be formulated as $\Csp(\bB)$, for an 
appropriate 2-element structure $\bB$. Positive 1-in-3-3SAT can be formulated as $\Csp(\bB)$ for
the template 
$$\bB=(\{0,1\}; \OIT) \quad
\text{where } \OIT = \{ (0,0,1), (0,1,0),(1,0,0) \} \; ,$$
and Positive-Not-All-Equal-3SAT as $\Csp(\bB)$ for 
the template
$$\bB=(\{0,1\}, \NAE) \quad 
\text{where } \NAE = \{0,1\}^3 \setminus \{(0,0,0),(1,1,1)\} \; .$$

These problems can also be formulated as CSPs if we do \emph{not} 
impose the restriction that all literals are positive; the corresponding problems are then called \emph{1-in-3-3SAT} and
\emph{Not-All-Equal-3SAT}, respectively. 
The idea is to use a different ternary relation for each of the eight ways
how three distinct variables in a clause with three literals might be negated. In this way, we can also model the classical problem of 3SAT (again, see Figure~\ref{fig:cproblems})
as a CSP. Clauses of the type $x \vee y \vee \neg z$
in the 3SAT problem will then be viewed as constraints $R^{++-}(x,y,z)$, where $R^{++-} =  \{0,1\}^3 \setminus \{(0,0,1)\}$ (here, $x,y,z$ are not necessarily distinct variables).
Similarly, the well-known 2SAT problem can be viewed as 
$\Csp((\{0,1\}; R^{++},R^{+-},R^{-+},R^{--}))$ where
\begin{align*}
R^{++} & = \{(0,1),(1,0),(1,1)\}, \\
R^{+-} & = \{(0,0),(1,1),(1,0)\}, \\
R^{-+} & = \{(1,1),(0,0),(0,1)\}, \text{ and } \\
R^{--} & = \{(1,0),(0,1),(0,0)\} \; . 
\end{align*}
\qed \end{example}

\begin{example}[Disequality constraints]\label{expl:basic-ecsp}
Consider the problem $\Csp(({\mathbb N};=,\neq))$.
An instance of this problem can be viewed as an (existentially quantified) set of variables, 
some linked by equality, some by disequality\footnote{We deliberately use the word \emph{disequality} instead of \emph{inequality}, since we reserve the word \emph{inequality} for the relation $x \leq y$.} constraints.
Such an instance is false in $({\mathbb N};=,\neq)$ if and only if there
is a path $x_1,\dots,x_n$
from a variable $x_1$ to a variable $x_n$ that uses only equality edges, i.e., 
`$x_i = x_{i+1}$' is a constraint in the instance for each $1 \leq i \leq n-1$,
and additionally `$x_1 \neq x_n$' is a constraint in the instance.
Clearly, it can be tested in linear time in the size of the input instance
whether the instance contains such a path. 
\qed
\end{example}

\subsection{Canonical conjunctive queries}
\label{ssect:can-query}
To every finite relational $\tau$-structure $\bA$ 
we can associate a $\tau$-sentence,
called the 
\emph{canonical conjunctive query} of $\bA$, and denoted
by \emph{$\CQ(\bA)$}. The variables of this sentence are the elements of $\bA$,
all of which are existentially quantified in the quantifier prefix of the formula, which is
followed by 
the conjunction of all formulas of the form
$R(a_1, \ldots, a_k)$ 
for $R \in \tau$
and tuples $(a_1, \ldots, a_k) \in R^\bA$.

For example, the canonical conjunctive query $\CQ(K_3)$
of the complete graph on three vertices $K_3$ is
the formula
$$\exists u \exists v \exists w \;
 \big(E(u, v) \wedge E(v,u) \wedge E(v, w) \wedge E(w,v) \wedge E(u, w) \wedge E(w,u)\big) \; .$$

The proof of the following proposition is straightforward.
\begin{proposition}\label{prop:hom-to-logic}
Let $\bB$ be a structure with finite relational signature $\tau$, 
and let $\bA$ be a finite $\tau$-structure.
Then there is a homomorphism from $\bA$ to $\bB$ if and only if
$\CQ(\bA)$ is true in $\bB$.
\end{proposition}

\subsection{Canonical databases}
To present a converse of Proposition~\ref{prop:hom-to-logic},  we
define the \emph{canonical database $\CD(\phi)$} 
of a primitive positive $\tau$-formula, which
is a relational $\tau$-structure defined as follows. 
We require that $\phi$ 
does not contain $\bot$. 
If $\phi$ contains an atomic formula of the form
$x=y$, we remove it from $\phi$, and replace all occurrences of $x$ in $\phi$ by $y$. 
Repeating this step if necessary, we may assume that
$\phi$ does not contain atomic formulas of the form $x=y$. 

%\footnote{Defining the canonical database for a formula without variables that is always false would lead to formal problems.}).
% the problem is:
% the naive solution would be to allow empty structures, and to put in a relation
% with the empty tuple that denotes the empty relation.
% but then every structure would have to map to this structure, according
% to the properties of the canonical database?  WHY?
% certainly there can't be a map from a non-empty set to an empty set, so a fortiori
% no homomorphism. 

Then the domain of $\CD(\phi)$ 
is the set of variables (both the free variables and the existentially quantified variables) that occur in $\phi$. There is a tuple $(v_1,\dots,v_k)$ in a relation $R$ of $\CD(\phi)$
iff $\phi$ contains the conjunct $R(v_1,\dots,v_k)$. 
 The following is similarly straightforward as Proposition~\ref{prop:hom-to-logic}.

\begin{proposition}\label{prop:logic-to-hom}
Let $\bB$ be a structure with signature $\tau$, and let $\phi$
be a primitive positive $\tau$-sentence other than $\bot$. 
Then $\phi$ is true in $\bB$ if and only
if $\CD(\phi)$ homomorphically maps to $\bB$.
\end{proposition}

%\begin{lemma}\label{lem:cm}
%Let $\Gamma$ be a $\tau$-structure. 
%For any conjunctive query $Q$ with free variables $x_1,\dots,x_k$, 
%and any sequence $v_1,\dots,v_k$ of elements from $\Gamma$
%the following are equivalent.
%\begin{itemize}
%\item $\Gamma$ satisfies $\CQ(v_1,\dots,v_k)$.
%\item There is a homomorphism from $A(Q)$ to $\Gamma$ that maps $x_1,\dots,x_k$ to $v_1,\dots,v_k$.
%\end{itemize}
%\end{lemma}

Due to Proposition~\ref{prop:logic-to-hom} and Proposition~\ref{prop:hom-to-logic},  we may freely switch between the  homomorphism and the logic perspective whenever this is convenient. 
In particular, instances of $\Csp(\bB)$ can from now on be either
finite structures $\bA$ or primitive positive sentences $\phi$.

\subsection{Expansions}
\label{ssect:expansion}
% The following observations are most natural from the model-checking perspective. % We need the following standard terminology in model theory. 
Let $\bA$ be a $\tau$-structure, and let $\bA'$ be a
$\tau'$-structure with $\tau \subseteq \tau'$. If
$\bA$ and $\bA'$ have the same domain and 
$R^\bA = R^{\bA'}$ for all $R \in \tau$, then $\bA$
is called the \emph{$\tau$-reduct} (or simply \emph{reduct}) of $\bA'$,
and $\bA'$ is called a \emph{$\tau'$-expansion} (or simply \emph{expansion}) of $\bA$. When $\bA$ is a structure, and $R$ is a relation over the domain of $\bA$, then we denote the expansion of $\bA$ by $R$ by 
$(\bA,R)$. 

The following lemma says that we can expand structures  by primitive positive
definable relations without changing the complexity of the corresponding
CSP. Hence, primitive positive definitions are an important tool to prove NP-hardness: 
to show that $\Csp(\bB)$
is NP-hard, it suffices to show that there is a primitive positive 
definition of a relation $R$ such that $\Csp((\bB,R))$ is already known to be NP-hard. Stronger tools to prove NP-hardness of CSPs will be introduced 
in Section~\ref{sect:pseudo-var}.

\begin{lemma}\label{lem:pp-reduce}
Let $\bB$ be a structure with finite relational signature, 
and let $R$ be a relation that has a primitive positive definition in $\bB$. Then $\Csp(\bB)$ and 
$\Csp((\bB,R))$ are linear-time equivalent.
They are also equivalent under deterministic log-space reductions.
\end{lemma}
\begin{proof}
It is clear that $\Csp(\bB)$ reduces to the new problem.
So suppose that $\phi$ is an instance of $\Csp((\bB,R))$.   
Replace each conjunct $R(x_1,\dots,x_l)$ of $\phi$ by its primitive positive
definition $\psi(x_1,\dots,x_l)$. Move all quantifiers
to the front, such that the resulting formula is in \emph{prenex normal form} and hence primitive positive. 
Finally, equalities can be eliminated one by one: for equality $x=y$, remove $y$ from the quantifier prefix, and replace all remaining
occurrences of $y$ by $x$. Let $\psi$ be the formula obtained in this way.

It is straightforward to verify that $\phi$ is true in $(\bB,R)$ if and only
if $\psi$ is true in $\bB$, and it
is also clear that $\psi$ can be constructed in linear time in the representation size of $\phi$. For the observation that
the reduction is deterministic log-space, we need the recent result that
undirected reachability can be decided in deterministic log-space~\cite{Reingold}. 
\end{proof}

\begin{example}\label{expl:oit-defines-nai}
The relation $\NAE(x_1,x_2,x_3)$ has the following primitive positive definition
in $(\{0,1\};\OIT)$.
\begin{align*} 
\exists u_1,u_2,u_3,v_1,v_2,v_3,z_1,z_2,z_3 & \big( \OIT(x_1,u_1,v_1) \wedge \OIT(x_2,u_2,v_2) \wedge \OIT(x_3,u_3,v_3)  \\
\wedge \, \OIT(v_1,u_2,z_1) \, \wedge 
& \, \OIT(v_2,u_3,z_2) \wedge \OIT(v_3,u_1,z_3) \wedge \OIT(z_1,z_2,z_3) \big)
\end{align*}

To see that this works, note that when $x_1=x_2=x_3=1$, then the first three conjuncts imply that $u_1=v_1=u_2=v_2=u_3=v_3=0$, and the next three conjuncts imply that $z_1=z_2 = z_3=1$, and hence the last conjunct is violated. When $x_1=x_2=x_3=0$, then the first conjunct implies that 
$u_1=0$ and $v_1=1$, or $u_1=1$ and $v_1=0$. In both cases, the fourth conjunct
implies that $z_1=0$. Similarly, we can infer that $z_2 = z_3=0$. Whence, the last conjunct is violated. 

Now consider the case when exactly one out of $x_1,x_2,x_3$ is $0$. Since the formula is symmetric with respect to $x_1,x_2,x_3$,
we assume without loss of generality that $x_1=0, x_2=1,x_3=1$. Then we can set $u_1=z_1=z_2=1$, and $v_1=u_2=v_2=u_3=v_3=z_3=0$ and satisfy all conjuncts. Similarly, when exactly two out of $x_1,x_2,x_3$ are $0$,
we assume without loss of generality that $x_1=1$, $x_2=x_3=0$. Then we can set $u_1=v_1=u_2=u_3=z_2=z_3=0$ and $z_1=v_2=v_3=1$ 
and satisfy all conjuncts. \qed
\end{example}

An \emph{automorphism} of a structure $\bB$ with domain $B$ is an isomorphism between $\bB$ and itself. When applying an automorphism $\alpha$ to an element $b$ from $B$ we omit brackets, that is, we write $\alpha b$ instead of $\alpha(b)$.  
The set of all automorphisms $\alpha$ of $\bB$ is denoted by $\text{Aut}(\bB)$,
and $\alpha^{-1}$ denotes the inverse map of $\alpha$.
Let $(b_1,\dots,b_k)$ be a $k$-tuple of elements of $\bB$. 
A set of the form $S = \{ (\alpha b_1,\dots,\alpha b_k) \; | \; \alpha \in \text{Aut}(\bB)\}$ is called an \emph{orbit of $k$-tuples} (the \emph{orbit of $(b_1,\dots,b_k)$}). 
%When $k=1$, we also say that $S$ is an \emph{orbit} of $\bB$.

\begin{lemma}\label{lem:constant-expansion}
Let $\bB$ be a structure with a finite relational signature and domain $B$, and 
let $R = \{(b_1,\dots,b_k)\}$ be a $k$-ary relation that only contains one tuple $(b_1,\dots,b_k) \in B^k$. 
If the orbit of $(b_1,\dots,b_k)$ 
in $\bB$ is primitive positive definable, then
there is a polynomial-time reduction from $\Csp((\bB,R))$ to $\Csp(\bB)$.
\end{lemma}
\begin{proof}
Let $\phi$ be an instance of $\Csp((\bB,R))$ with variable set $V$. 
If $\phi$ contains two 
constraints $R(x_1,\dots,x_k)$ and $R(y_1,\dots,y_k)$, 
then replace each occurrence
of $y_1$ by $x_1$, then each occurrence of $y_2$ by $x_2$, and so on,
and finally each occurrence of $y_k$ by $x_k$.
We repeat this step until all constrains that involve
$R$ are imposed on the same tuple of variables $(x_1,\dots,x_k)$.
Replace $R(x_1,\dots,x_k)$ by the primitive positive definition $\theta$ of its orbits in $\bB$. 
Finally, move all quantifiers
to the front, such that the resulting formula $\psi$ is in prenex normal form and thus an instance of $\Csp(\bB)$. Clearly, $\psi$ can be
computed from $\phi$ in polynomial time. We claim that $\phi$
is true in $(\bB,R)$ if and only if $\psi$ is true in $\bB$. 

Suppose $\phi$ has a solution $s \colon V \rightarrow B$. Let $s'$ be the restriction of $s$ to the variables of $V$ that also appear in $\phi$. Since $(b_1,\dots,b_n)$ satisfies $\theta$, we can extend $s'$ to the existentially quantified variables of $\theta$ to obtain
a solution for $\psi$. 
In the opposite direction, suppose that $s'$ is a solution
to $\psi$ over $\bB$. Let $s$ be the restriction of $s'$ to $V$.
Because $(s(x_1),\dots,s(x_k))$ satisfies $\theta$ it
lies in the same orbit as $(b_1,\dots,b_k)$. Thus, there exists an automorphism $\alpha$ of
$\bB$ that maps $(s(x_1),\dots,s(x_k))$ to $(b_1,\dots,b_k)$.
Then the extension of the map 
$x \mapsto \alpha s(x)$ that maps variables $y_i$ of $\phi$ 
that have been replaced by $x_i$ in $\psi$ to the value $b_i$ is a solution
to $\phi$ over $(\bB,R)$.
\end{proof}

Recall from Section~\ref{sect:homo} that every
finite structure $\bC$ is homomorphically equivalent to
a core structure $\bB$, which is unique up to isomorphism.
For core structures, all orbits are primitive positive definable. 
This fact has a simple proof for finite structures $\bB$;
however, the same fact is true for a large class of infinite
structures, and presented in Chapter~\ref{chap:mt}, Theorem~\ref{thm:mc-core}. 
Since Theorem~\ref{thm:mc-core} implies the following proposition, 
we omit the proof at this point.

\begin{proposition}\label{prop:orbits-in-finite-cores}
Let $\bB$ be a finite core structure. Then orbits of $k$-tuples
of $\bB$ are primitive positive definable.
\end{proposition}

Proposition~\ref{prop:orbits-in-finite-cores} and Lemma~\ref{lem:constant-expansion} have the following well-known consequence.

\begin{corollary}
Let $\bB$ be a finite core structure with elements $b_1,\dots,b_n$ and finite signature.
Then $\Csp(\bB)$ and $\Csp((\bB,\{b_1\},\dots,\{b_n\}))$ are 
polynomial time equivalent. 
%Let $\bB$ be a finite core structure, and let $\bB'$ be the expansion
%of $\bB$ by the relation $\{b\}$ for each element $b$ of $B$. 
%Then $\Csp(\bB)$ and $\Csp(\bB')$ are polynomial-time equivalent.
\end{corollary}

\section{The Satisfiability Perspective}
\label{sect:sat}
% LATERTD: in the sat perspective, clarify that for every T we
% can find a universal-negative T' such that CSP(T) and CSP(T') have the same CSP. For such T', we actually have finite model
% property!
Yet another perspective on the constraint satisfaction problem 
translates not only the instances, but also the
template of the CSP into logic. This leads to a natural perspective for
various model-theoretic considerations in Chapter~\ref{chap:mt}.
Moreover, this perspective is 
convenient when discussing the literature that uses 
 \emph{relation algebras} in the context of constraint satisfaction\cite{LadkinMaddux,Duentsch}; 
 the connection will be described 
 in Section~\ref{ssect:relation-algebras} 
 and Section~\ref{ssect:network-satisfaction}. 
 
We use the opportunity to introduce some inevitable
terminology from logic. We assume that the reader is already
familiar with basic terminology of first-order logic; a highly recommendable text-book is Hodges~\cite{Hodges}.

\subsection{Theories}
A (first-order) \emph{theory} is a set of first-order sentences. 
When the first-order sentences are over the signature $\tau$,
we also say that $T$ is a \emph{$\tau$-theory}.
A \emph{model} of a $\tau$-theory $T$ is a $\tau$-structure $\bB$ such that $\bB$ satisfies all sentences in $T$. Theories that have a model are called \emph{satisfiable}.

\begin{definition}
Let $\tau$ be a finite relational signature, and let $T$ be a 
$\tau$-theory. Then \emph{$\Csp(T)$} is the computational problem to decide 
for a given primitive positive $\tau$-sentence $\phi$ whether $T \cup \{\phi\}$ is satisfiable.
\end{definition}

The satisfiability perspective on CSPs stresses the fact that the problem $\Csp(\bB)$ is fully determined by the first-order theory of $\bB$, that is, by the theory that contains exactly those sentences that are true in $\bB$. 
In fact, it is already determined by the primitive positive sentences that are false in $\bB$.

\begin{example}
Let $T$ be the theory that consists of the following sentences. 
%\begin{itemize}
%\item 
\begin{align*}
& \forall x,y,z \; ((x < y \wedge y < z) \rightarrow x < z) &&
\text{(transitivity)} \\
%\item
& \forall x,y \; \neg (x<x) && \text{(irreflexivity)} \\
%\item 
& \forall x,y,z \; ((x < y) \vee (y < x) \vee (x=y)) && 
\text{(totality)}
\end{align*}
%\end{itemize}
It is straightforward to verify that $\Csp(T)$ equals $\Csp(({\mathbb Z}; <))$ (Example~\ref{expl:acycl}).
\qed \end{example}

When $T$ is a theory and $\phi$ a sentence, 
we say that $T$ \emph{entails} $\phi$, in symbols $T \models \phi$, if every model of $T$ satisfies $\phi$. 
The following is clear from the definitions.

\begin{proposition}
Let $\tau$ be a finite relational signature, and let $T$ be a
$\tau$-theory. 
Suppose that $T$ entails exactly those negations of primitive positive sentences $\phi$ such that $\bB \models \phi$.
Then $\Csp(T)$ and $\Csp(\bB)$ are the same problem. 
\end{proposition}

We have already seen that two structures that are homomorphically equivalent have the same CSP;
the following provides a necessary and sufficient condition that describes when two \emph{theories} have the same CSP.
Its proof is simple once the relevant notions from logic are introduced, and will be given in Section~\ref{sect:diagrams}. 

\begin{proposition}\label{prop:csp-equiv}
Let $T$ and $T'$ be two first-order theories. Then the following are equivalent. 
\begin{itemize}
\item $\Csp(T)$ equals $\Csp(T')$.
\item Every model of $T'$ has a homomorphism to some model of $T$,
and every model of $T$ has a homomorphism to some model of $T'$. 
\item $T$ and $T'$ entail the same negations of primitive positive sentences. 
\end{itemize}
\end{proposition}

We now present a couple of basic observations relating the
definition of $\Csp(T)$ for a theory $T$ with the definition of
$\Csp(\bB)$ for a relational structure $\bB$.
We start with the observation that there are theories $T$ such that $\Csp(T)$ cannot be formulated as $\Csp(\bB)$.

\begin{example}
Let $\tau$ be the signature $\{R,G\}$, where $R$ and $G$ are unary relation symbols, and let $T$ be the $\tau$-theory 
$\{ \forall x,y \; \neg(R(x) \wedge G(y))\}$.
There is no structure $\bB$ such that $\Csp(\bB)$ equals
$\Csp(T)$. To see this, observe that $T \cup \{\exists x. R(x)\}$ is satisfiable,
and $T \cup \{\exists x. G(x)\}$ is satisfiable. But any structure $\bB$ that satisfies
both $\exists x. R(x)$ and $\exists x. G(x)$ also satisfies $\exists x,y (R(x) \wedge R(y))$,
which shows that $\Csp(\bB)$ and $\Csp(T)$ are different. 
\qed \end{example}

We next characterize those satisfiable theories $T$ that have a model $\bB$ such that $\Csp(\bB)$ and $\Csp(T)$ are the same problem. 

\begin{proposition}\label{prop:sat-csp}
Let $\tau$ be a finite relational signature, and
let $T$ be a satisfiable first-order $\tau$-theory. The following are equivalent.
\begin{enumerate}
\item There is a structure $\bB$ such that $\Csp(\bB)$
and $\Csp(T)$ are the same problem.
\item There is a model $\bB$ of $T$ such that $\Csp(\bB)$
and $\Csp(T)$ are the same problem.
%\item For all primitive positive $\tau$-sentences $\phi_1$ and 
%$\phi_2$, if $T \models \neg (\phi_1 \wedge \phi_2)$, then 
%$T \models \neg \phi_1$
%or $T \models \neg \phi_2$.
\item For all primitive positive $\tau$-sentences $\phi_1$ and 
$\phi_2$, if $T \cup \{\phi_1\}$ is satisfiable and $T \cup \{\phi_2\}$ is satisfiable
then $T \cup \{\phi_1, \phi_2\}$ is satisfiable as well. 
\item $T$ has the \emph{Joint Homomorphism Property (JHP)}, that is, 
when $T$ has models $\bA$ and $\bB$, then it also has a model $\bC$ such that
both $\bA$ and $\bB$ homomorphically map to $\bC$.  
\end{enumerate}
\end{proposition}
We defer the proof of this fact to Section~\ref{sect:diagrams} when we have
some more concepts from logic available. 

\subsection{Relation Algebras}
\label{ssect:relation-algebras}
Many interesting infinite-domain CSPs,  in particular in spatial and temporal reasoning, have been studied in the context of 
\emph{relation algebras} (many examples will be given in Section~\ref{sect:csp-examples} and Chapter~\ref{chap:examples}).
%Sections~\ref{ssect:allen}, \ref{ssect:branching-time}, \ref{ssect:cornell}, \ref{ssect:spatial}). 
In Artificial Intelligence, relation algebras are used as a framework to formalize and study qualitative reasoning problems~\cite{LadkinMaddux,Duentsch,HirschAlgebraicLogic}.
From the perspective of this thesis, the relation algebra approach does not
bring substantially new tools, and Section~\ref{ssect:relation-algebras} and Section~\ref{ssect:network-satisfaction} can be safely skipped.
Here we nonetheless give a quick introduction in order to link the relation algebra terminology with the satisfiability perspective on the CSP (Section~\ref{ssect:network-satisfaction}). 

Relation algebras are designed to handle binary relations in an algebraic way; we follow the presentation in~\cite{HirschAlgebraicLogic}. 

\begin{definition}\label{def:proper-rel-alg}
A \emph{proper relation algebra} is a domain $D$ together with
a set $\mathcal B$ of binary relations over $D$ such that 
\begin{enumerate}
\item $\Id := \{(x,x) \; | \; x \in D\} \in \mathcal B$;
\item If $B_1$ and $B_2$ are from $\mathcal B$, then $B_1 \vee B_2 := B_1 \cup B_2 \in \mathcal B$;
\item $1 := \bigcup_{B \in \mathcal B} B \in \mathcal B$;
\item $0 := \emptyset \in \mathcal B$;
\item If $B \in \mathcal B$, then $-B := 1 \setminus B \in \mathcal B$;
\item If $B \in \mathcal B$, then $B^{\smallsmile} := \{(x,y) \; | \; (y,x) \in B\} \in \mathcal B$;
\item If $B_1$ and $B_2$ are from $\mathcal B$, then $B_1 \circ B_2 \in \mathcal B$; where
$$ B_1 \circ B_2 := \{(x,z) \; | \; \exists y ((x,y) \in B_1 \wedge (y,z) \in B_2)\} \; .$$
\end{enumerate}
\end{definition}

We want to point out that in this standard definition of proper relation algebras it is \emph{not} required that $1$ denotes $D^2$ (and this will be used for instance in the proof of Proposition~\ref{prop:disj-union-networks}).
However, in most examples that we encounter, 
$1$ indeed denotes $D^2$.
The minimal non-empty elements of $\mathcal B$ with respect to
set-wise inclusion are called the \emph{basic relations} of the relation algebra. 

\begin{example}[The Point Algebra]\label{expl:point-algebra}
Let $D={\mathbb Q}$ be the set of rational numbers,
and consider %set $\mathcal B$ containing the eight binary relations 
$$\mathcal B = \{\emptyset,=,<,>,\leq,\geq,\neq,{\mathbb Q}^2\} \; .$$
Those relations form a proper relation algebra (with atoms $<,>,=$, and where $1$ denotes ${\mathbb Q}^2$) which is one
of the most fundamental relation algebras and known under the name \emph{point algebra}.
\qed \end{example}

When $\mathcal B$ is finite,
every relation in $\mathcal B$ 
can be written as a finite union of basic relations, and we abuse
notation and sometimes write $R = \{B_1,\dots,B_k\}$ when
$B_1, \dots,B_k$ are basic relations, $R \in \mathcal B$, and 
$R = B_1 \cup \dots \cup B_k$. 
Note that composition of basic 
relations determines the composition of all relations in the relation algebra, since $$R_1 \circ R_2 = \bigcup_{B_1 \in R_1, B_2 \in R_2}
B_1 \circ B_2 \;.$$

An \emph{abstract relation algebra} (Definition~\ref{def:rel-algebra} below) is an algebra with signature
$\{\vee,-,0,1,\circ,^{\smallsmile},\Id\}$
%$\Id,0,1,-,^{\smallsmile},\vee,\circ$ 
that satisfies laws that we expect from
those operators in a proper relation algebra.

\begin{definition}[Compare~\cite{HirschAlgebraicLogic,Duentsch,LadkinMaddux}]\label{def:rel-algebra} An (abstract) relation algebra $\bf A$ is an algebra
with domain $A$ and signature $\{\vee,-,0,1,\circ,^{\smallsmile},\Id\}$ such that
\begin{itemize}
\item the structure $(A; \vee,\wedge,-,0,1)$ is a Boolean algebra where 
  $\wedge$ is defined by $(x,y) \mapsto -(-x \vee -y)$ from $-$ and $\vee$;
\item $\circ$ is an associative binary operation on $A$;
\item $(a^{\smallsmile})^{\smallsmile} = a$ for all $a \in A$; 
%\item $\Id \circ~a = a \circ \Id = a$ for all $a \in A$; 
\item $a \circ (b \vee c) =  a \circ b \vee a \circ c$;
\item $(a \vee b)^{\smallsmile} = a^{\smallsmile} \vee b^{\smallsmile}$;
\item $(-a)^{\smallsmile} = -(a^{\smallsmile})$;
\item $(a \circ b)^{\smallsmile} = b^{\smallsmile} \circ a^{\smallsmile}$;
\item $(a \circ b) \wedge c^{\smallsmile} = 0 \; \Leftrightarrow \; (b \circ c) \wedge a^{\smallsmile} = 0$.
\end{itemize}
\end{definition}
We define $x \leq y$ by $x \wedge y = x$.
A \emph{subalgebra} $\bf B$ of a relation algebra $\bf A$ with domain $A$ is a relation algebra with domain $B \subset A$ such that
for every function $f$ of $\bf A$, the element obtained by applying
$f$ to elements from $B$ is again in $B$. 

A \emph{representation} $(D,i)$ of $\bf A$ consists of a set $D$
and a mapping $i$ from the domain $A$ of $\bf A$ to binary relations
over $D$ such that the image of $i$ induces a proper relation algebra $\B$, and $i$ is an isomorphism with respect to the functions (and constants)
$\{\vee,-,0,1,\circ,^{\smallsmile},\Id\}$.
In this case, we also say that $\bf A$ is the \emph{abstract relation algebra of $\B$}. 

There are finite relation algebras that do not have a representation~\cite{LyndonRelationAlgebras}. 
Note that when $(D,i)$ is a representation of $\bf A$, then
$i(a)$ is a basic relation of the induced proper relation algebra
if and only if $a \neq 0$, and for every $b \leq a$ we have $b=a$ or $b=0$; we call $a$ an \emph{atom} of $\bf A$. Using the axioms of relation algebras, it can be shown 
that the composition operator is uniquely determined by the
composition operator on the atoms. Similarly, the inverse of an
element $a \in A$ is the disjunction of the inverses of all the atoms below $a$.

%Abstract relation algebras can be used to give a finite description
%of an infinite relational structure with binary constraints. 
%Usually, but not always, some information is lost when we go from 
%a proper relation algebra to an abstract relation algebra.

\begin{example}\label{expl:a-pa}
The (abstract) point algebra
is a relation algebra with 8 elements and 3 atoms, $=$, $<$, and $>$, and can be described as follows. 
The composition operator of the basic relations of the point algebra is shown in the table of Figure~\ref{fig:point-algebra}.
By the observation we just made, this table determines the full composition table.
The inverse of $<$ is $>$, and $\Id$ denotes $=$ which is its own inverse. 
This fully determines the relation algebra.

\begin{figure}
\begin{tabular}{|l||l|l|l|}
\hline
$\circ$ & $=$ & $<$ & $>$ \\
\hline \hline
$=$ & $=$ & $<$ & $>$ \\
\hline
$<$ & $<$ & $<$ & $1$ \\
\hline
$>$ & $>$ & $1$ & $>$ \\
\hline
\end{tabular}
\caption{The composition table for the basic relations in the point algebra.}
\label{fig:point-algebra}
\end{figure}

We can obtain a representation of the abstract point algebra
from the point algebra with domain $\mQ$ presented in Example~\ref{expl:point-algebra} in the obvious way. \qed
\end{example}

\subsection{Network Satisfaction Problems}
\label{ssect:network-satisfaction}
The central computational problems that have been studied for relation algebras are \emph{network satisfaction problems}~\cite{LadkinMaddux,Duentsch,HirschAlgebraicLogic}.
Let $\bf A$ be a finite relation algebra with domain $A$. 
An \emph{(${\bf A}$-) network}~$N = (V;f)$ consists of a finite set of nodes $V$ and a partial function $f \colon V^2 \rightarrow A$. Here, we slightly deviate from the 
definition given in the papers listed above 
in that we allow $f$ to be undefined on some pairs
of nodes. 

Two types of network satisfaction problems have been studied for ${\bf A}$-networks.
The first is the \emph{network satisfaction problem for a (fixed) representation of $\fA$}, 
defined as follows. 

\begin{definition}
Let $(D,i)$ be a representation of a finite relation algebra $\fA$.
Then the \emph{network satisfaction problem for $(D,i)$}
is the computational problem to decide whether a given
$\fA$-network $N=(V;f)$ is \emph{satisfiable with respect to $(D,i)$}, that is, whether there exists a mapping $s \colon V \rightarrow D$ such that $(s(u),s(v)) \in i(f(u,v))$ for all $u,v \in V$ where $f$ is defined. 
\end{definition}

The second problem is the \emph{(general) network satisfaction problem for ${\bf A}$}.

\begin{definition}
Let $\fA$ be a finite relation algebra. Then the \emph{network satisfaction problem for ${\bf A}$} is the computational
problem to decide whether a given $\fA$-network $N$ is \emph{satisfiable}, i.e., whether 
there \emph{exists} a representation $(D,i)$ of ${\bf A}$ such that $N$ is satisfiable with respect to $(D,i)$.
\end{definition}

%In many situations we can find a single representation $(D,i)$ of a relation algebra ${\bf A}$
%so that the general network consistency problem for ${\bf A}$ and the
%network consistency problem for this representation are the same computational problem.
% NO, ALWAYS!
%This is for instance the case for the (abstract) point algebra from Example~\ref{expl:a-pa}:
%it is easy to see that the general network satisfaction problem for the point algebra is 
%the same computational problem as the network satisfaction problem for the
%representation of the point algebra that is described in Example~\ref{expl:point-algebra}. 

It is not surprising that every network satisfaction problem
for a fixed representation is closely related to a corresponding constraint satisfaction problem; this correspondence will be described in the following. 
It is maybe less obvious that the same
 also applies to the \emph{general} network satisfaction problem:
every finite relation algebra $\fA$ that has a representation also has a representation $(D,i)$ such that the general network satisfaction problem for $\fA$ and the network satisfaction problem for $(D,i)$ are one and the same problem (Proposition~\ref{prop:disj-union-networks}).

To present the link between network satisfaction problems and CSPs as defined earlier we need the following notation. Let $\tau_{\bf A}$ be a signature
consisting of a binary relation symbol $R_a$ 
for each element $a \in A$. 
When $(D,i)$ is a representation of $\tau_\fA$, then
this gives rise to a $\tau_\fA$-structure $\bB_{D,i}$ in a natural way:
the domain of the structure is $D$, and the relation symbol $R_a$ is interpreted 
by $i(a)$. 
We can associate to each $\bf A$-network $N = (V;f)$ a 
primitive positive $\tau_{\bf A}$-sentence $\phi_N$,
in the following straightforward way: 
the variables of $\phi_N$ are $V$, and $\phi$ contains the conjunct $R_a(u,v)$ iff $f(u,v)=a$. 
Conversely, we can associate to each primitive positive $\tau_{\bf A}$-sentence $\phi$ with variables $V$ a network $N_\phi$ as follows.
The domain of $N_\phi$ is $V$.
Let $u,v \in V$, and list by $a_1,\dots,a_k$ all those elements $a$ of $A$ such that $\phi$ contains the conjunct $R_a(u,v)$.
Then define $f(u,v)=a$ for $a = (a_1 \wedge a_2 \wedge \dots \wedge a_k)$; if $k=0$, then $f(u,v)$ is undefined.

The following link between the network satisfaction problem for a fixed representation
$(D,i)$ of ${\bf A}$, and the constraint satisfaction problem for $\bB_{D,i}$ is straightforward from the definitions.
 
\begin{proposition}\label{prop:netw-sat-wrt-repr}
Let $\fA$ be a finite relation algebra with 
representation $(D,i)$. 
Then an ${\bf A}$-network $N$ is satisfiable with respect to $(D,i)$ if and only if $\bB_{D,i} \models \phi_N$. 
Conversely, $\bB_{D,i}$ satisfies 
a primitive positive $\tau_{\bf A}$-sentence $\phi$ 
if and only if $N_\phi$ is satisfiable with respect to $(D,i)$.
\end{proposition}

% The following would be the structure variant: 
%\begin{proposition}\label{prop:netw-sat-wrt-repr}
%Let $\fA$ be a finite relation algebra with 
%representation $(D,i)$. 
%Then an ${\bf A}$-network $N$ is satisfiable with respect to $(D,i)$ if and only if
%$\bS_N$ homomorphically maps to $\bB_{D,i}$. Moreover, 
%a finite $\tau_{\bf A}$-structure $\bS$ homomorphically maps to $\bB_{D,i}$
%if and only if $N_S$ is satisfiable with respect to $(D,i)$.
%\end{proposition}

Proposition~\ref{prop:netw-sat-wrt-repr} shows that
network satisfaction problems for fixed representations essentially \emph{are} constraint satisfaction problems, and that the differences are only a matter of formalization. 
To also relate the \emph{general} network satisfaction problem for a finite relation algebra $\fA$ to a constraint satisfaction problem,
%For every finite relation algebra $\bf A$, the 
%network satisfaction problem for $\bf A$ 
%can be formulated as $\Csp(T_{\bf A})$, for a finite first-order theory $T_{\bf A}$ 
%that can be obtained from $\bf A$ as follows 
we define in Figure~\ref{fig:ta} the first-order $\tau_{\bf A}$-theory $T_{\bf A}$ (as in~\cite{HirschAlgebraicLogic}, Section 2.3).
The models
of $T_{\bf A}$ correspond to the representations of $\bf A$, as described in the following.

\begin{figure}
\begin{align}
T_{\bf A} := \; & \big\{ \forall x,y (\neg 0(x,y) \wedge (\Id(x,y) \Leftrightarrow x=y)) \big \} \\
\cup & \big \{ \forall x,y (1(x,y) \Leftrightarrow \bigvee_{a \in A} R_a(x,y)) \big \} \\
\cup & \bigcup_{a \in A} \big \{\forall x,y (R_{a^{\smallsmile}}(x,y) \Leftrightarrow R_a(y,x) \wedge (R_{-a}(x,y) \Leftrightarrow \neg R_{a}(x,y) ) \big \} \\
\cup & \bigcup_{a,b \in A} \big \{\forall x,y (R_{a \vee b}(x,y) \Leftrightarrow (R_a(x,y) \vee R_b(x,y))) \big \} \\
\cup & \bigcup_{a,b \in A} \big \{\forall x,z (R_{a \circ b}(x,z) \Leftrightarrow \exists y (R_a(x,y) \wedge R_b(y,z))) \big \}
\end{align}
\caption{The definition of the $\tau_{\bf A}$-theory $T_{\bf A}$.}
\label{fig:ta}
\end{figure}

\begin{proposition}\label{prop:representation-theory}
Let $\bf A$ be a finite relation algebra. 
When $\bB$ models $T_{\bf A}$, then $(B,i)$ 
where $B$ is the domain of $\bB$ 
and $i$ is given by $i(a)=R^\bB_a$ is a representation of $\bf A$.
Conversely, for every representation $(D,i)$ of $\fA$ the $\tau_{\fA}$-structure $\bB_{D,i}$ is a model of $T_{\fA}$.
\end{proposition}
\begin{proof}
The proof is straightforward by matching the sentences
in $T_{\fA}$ with the items of Definition~\ref{def:proper-rel-alg}.
% Every representation satisfies the theory:
%The verification that $\bB_{D,i} \models T_{\fA}$ for
%every representation $(D,i)$ of $\fA$ is straightforward, and it suffices to match
%the sentences in the definition of $T_{\fA}$ with the corresponding items in Definition~\ref{def:proper-rel-alg}.
% The sentence $(1)$ 
%of the definition of $T_{\fA}$ holds 
%since $R_0$ denotes $\emptyset$ in $\bB_{D,i}$, 
%and since $R_{\Id} = \{(x,x) \; | \; x \in D\}$, by item (1) and (3) of Definition~\ref{def:proper-rel-alg}. Sentence (2) of the definition of $T_{\bf A}$ holds by item (3) of Definition~\ref%{def:proper-rel-alg}.  ETC ALL TRIVIAL AND CHECKED.
%
% Every model of the theory is a representation: 
% have to verify the isomorphism
%Let $\bB$ be a model of $T_A$, and $(D,i)$ be as described in the statement. We have to show that $(D,i)$ is a representation of ${\bf A}$.
\end{proof}

\begin{corollary}\label{cor:network-sat-csp}
Let $\bf A$ be a finite relation algebra. Then an $\bf A$-network $N$
is satisfiable if and only if $\phi_N \cup T_{\bf A}$ 
is satisfiable. Conversely, when $\phi$ is a primitive positive
$\tau_{\bf A}$-sentence, then the $\bf A$-network 
$N_\phi$ is satisfiable if and only if $\phi \cup T_{\bf A}$ is satisfiable.
\end{corollary}
% Das ist genauso trivial, Beweis weglassen.
%\begin{proof}
%Let $N$ be a $\bf A$-network.
%Suppose first that $N$ is satisfiable; that is, there is an interpretation $(D,i)$ of ${\bf A}$
%and a mapping $s \colon V \rightarrow D$ such that for all $u,v \in V$ $$(s(u),s(v)) \in i(f(u,v)) \; .$$
%Then the structure $\bB(D,i)$ from Proposition~\ref{prop:representation-theory} is
%a model of $T_{\bf A} \cup \{\phi_N(s(v_1),\dots,s(v_n))\}$,  where $V=\{v_1,\dots,v_n\}$.
%Conversely, when $T_{\bf A}$ 
%\end{proof}

It is easy to see that $T_{\bf A}$ has the Joint Homomorphism Property (JHP, introduced in 
Proposition~\ref{prop:jhp}); in fact,
the disjoint union of two models of $T_{\bf A}$ is again
a model of $T_{\bf A}$. 

\begin{proposition}\label{prop:disj-union-networks}
Every finite relation algebra $\bf A$ that has a representation also has a representation $(D;i)$ whose network satisfaction problem is the same problem as the general
 network satisfaction problem for $\fA$.
\end{proposition}
\begin{proof}
Since $\fA$ has a representation, 
and by Proposition~\ref{prop:representation-theory}, 
the theory $T_\fA$ is satisfiable. Since $T_{\bf A}$ also has the JHP, we can apply 
Proposition~\ref{prop:jhp} to obtain
a model $\bB$ of $T_{\fA}$ with domain $B$ be such that $\Csp(\bB)$ and $\Csp(T_\fA)$ are the same problem.
Then by Proposition~\ref{prop:representation-theory},
for $i$ given by $i(a)=R^\bB_a$, the relation algebra $\fA$ has the representation $(B,i)$.

We then have for all $\fA$-networks $N$ the following equivalences. 
\begin{align*}
N \text{ is satisfiable} \;
\Leftrightarrow & \; \phi_N \cup T_{\bf A} \text{ is satisfiable } && \text{(Corollary~\ref{cor:network-sat-csp})}\\
\Leftrightarrow & \; \bB \models \phi_N && \text{(by the properties of $\bB$)} \\
\Leftrightarrow & \; \phi_N \text{ is satisfiable wrt. $(B,i)$} && \text{(Proposition~\ref{prop:netw-sat-wrt-repr})}
\end{align*}
This concludes the proof that the representation $(B,i)$ of $\fA$ has
a network satisfaction problem that equals the general network satisfaction problem for $\fA$.
%By Corollary~\ref{cor:network-sat-csp}, 
%an $\bf A$-network $N$
%is satisfiable if and only if $\phi_N \cup T_{\bf A}$ 
%is satisfiable, and by the property of $\bB$ 
%this is the case if and only if $\bB$ satisfies $\phi_N$.
%By Proposition~\ref{prop:netw-sat-wrt-repr}, this is
%the case if and only if $\phi_N$ is satisfiable with 
%respect to the representation $(B,i)$ of $\fA$. 
\end{proof}

% Das folgende finde ich zu schwerfaellig:
%When $\bf A$ is a finite relation algebra, and $\bB$ is a 
%$\tau_{\bf A}$-structure, 
%we say that $\Csp(\bB)$ and the network satisfaction problem for $\bf A$ 
%are \emph{essentially the same problem} if
%, similarly as in Corollary~\ref{cor:network-sat-csp}, we have that 
%\begin{itemize}
%\item an $\bf A$-network $N$
%is satisfiable if and only if $\bB \models \phi_N$, and
%\item a primitive positive
%$\tau_{\bf A}$-sentence $\phi$ holds in $\bB$ if and only if
%the $\bf A$-network $N_\phi$ is satisfiable.
%\end{itemize}
%Note that when $\Csp(\bB)$ and the network satisfaction problem
%for $\bf A$ are essentially the same problem, then they are in particular polynomial-time equivalent.

In combination with Proposition~\ref{prop:netw-sat-wrt-repr},
this implies that also every general network satisfiability problem is essentially the same problem as a CSP for an infinite template.

%\begin{theorem}\label{thm:disj-union-networks}
%Let $\bf A$ be an arbitrary finite relation algebra. Then
%there exists a $\tau_{\bf A}$-structure
%$\bB$ such that $\Csp(\bB)$ and the network satisfaction
%problem for $\bf A$ are essentially the same problem.
%\end{theorem}
%\begin{proof}
%We first show that when
%$\phi_1$ and $\phi_2$ are primitive positive $\tau_{\bf A}$-sentences such that $T_{\bf A} \cup \psi_1$ has a model $\bA_1$ and $T_{\bf A} \cup \psi_2$ has a model $\bA_2$, then $T_{\bf A} \cup \{\psi_1,\psi_2\}$ has a model $\bA$.
%The domain of $\bA$ is the disjoint union of the domains $A_1$ and $A_2$ of $\bA_1$ and $\bA_2$, respectively. 
%For every $a \in A$, the relation $R^{\bA}_a$ is
%$R^{{\bA}_1}_a \cup R^{{\bA}_2}_a$. 
%(Note that we exploit that $R_1$, the relation for the element of $A$ denoted by `1', need not be $A \times A$ in 
%$\bA$.) 
%It follows from Proposition~\ref{prop:sat-csp} that there exists an infinite structure $\bB$ such that $\Csp(\bB)$ and $\Csp(T_{\bf A})$ are (exactly) the same problem. Then 
%Corollary~\ref{cor:network-sat-csp} shows that $\Csp(\bB)$
%and the network consistency problem are essentially the same problem. 
%\end{proof}

%In fact, we see that the network satisfaction problem for $\bf A$ and
%$\Csp(\bB)$ are not only polynomial-time equivalent, but essentially
%one and the same problem, up to formalization of how
%the input is given (networks versus primitive positive formulas).

We close this section by discussing the weaknesses of the relation
algebra approach to constraint satisfaction. 
First of all, the class
of problems that can be formulated as a network satisfiability
problem for finite relation algebra $\bf A$ is \emph{severely} restricted.
The relations that we allow in the input network
are closed under unions; this introduces a sort of restricted disjunction 
that quickly leads to NP-hardness, and indeed only a few exceptional situations have a
polynomial-time tractable network satisfiability problem~\cite{HirschAlgebraicLogic}. 
The typical work-around here is to introduce another parameter,
which is a subset $B$ of the domain of $\bf A$,
and to study the network satisfaction problem for networks
$N=(V;f)$ where the image of $f$ is contained in $B$.
Such subsets $B$ are often called a \emph{fragment} of $\fA$. Note that such an additional parameter is not necessary
for CSPs as studied in this thesis: with the techniques of this section, we can also formulate the network satisfaction problems for fragments of $\fA$ as CSPs.

Also note that the network satisfaction problem is restricted to \emph{binary} relations, whereas
many important CSPs can only be formulated in a natural way with
relations of higher arity (see e.g.\ Section~\ref{ssect:phylo} or Section~\ref{ssect:and-or}). As we have seen in Proposition~\ref{prop:disj-union-networks}, every network satisfaction problem can be formulated
as $\Csp(\bB)$ for an appropriate infinite structure $\bB$;
but as the above remarks show, only a very small fraction of CSPs
can be formulated as a network satisfaction problem.
 Even though only very specific CSPs can be formulated
as the network satisfaction problem 
for a finite relation algebra $\bf A$, 
there are hardly any additional techniques available for
studying network satisfaction problems. The tools we
have for network satisfaction usually also apply to
constraint satisfaction problems.

%For instance, the main computational technique that has been studied for the network satisfaction problem is local consistency (such as path consistency); however, this technique is also applicable
%to infinite-domain CSPs (see Section~\ref{ssect:datalog}). This technique is particularly powerful
%for problems of the form $\Csp(\bB)$ where $\bB$ is $\omega$-categorical (see Chapter~\ref{chap:mt}). When the network satisfiability problem under consideration cannot be formulated as $\Csp(\bB)$ for an $\omega$-categorical structure $\bB$, then not much is known about
%the power of consistency techniques for the network satisfiability problem, either. 

%Note that the composition of two relations in a structure is in particular primitive positive definable in this structure. The converse does not hold: there might be binary primitive positive definable relations in a binary structure such that this relation cannot be obtained by closure under the functions in the corresponding relation algebra.
%This demonstrates another drawback of relation algebras in the context of constraint satisfaction: 

The study of composition of
relations in the context of the network satisfiability problem is usually justified by the fact that a network with constraints over the relation $R \circ S$ 
can be simulated by networks that only
have constraints over the relation $R$ and over the relation $S$. To study the computational complexity of the network satisfaction problem for a fragment $B$ of a relation algebra $\fA$,
one therefore typically computes the closure of $B$ under 
the operations of the relation algebra. 
But note that every binary relation in the closure of $B$
is also primitive positive definable in any representation of $\fA$, and that the converse of this statement is false. 
% TODO: add example
Since the computational complexity is preserved also for expansions by primitive positive definable relations (see Lemma~\ref{lem:pp-reduce}), primitive positive definitions therefore appear to be the more appropriate tool for studying network satisfaction problems. 
Apart from being more powerful, primitive positive definability
has another advantage in comparison to closure in relation algebras:
while the latter is intricate and not well-understood,
we can offer a powerful Galois theory to study primitive positive definability of relations (see Chapter~\ref{chap:algebra}). 

\section{The Existential Second-Order Perspective}
\label{sect:snp}
By a famous result of Fagin, which will be reviewed below, 
the complexity class NP corresponds exactly to those problems that
can be formulated in existential second-order logic (ESO). 
An important fragment of ESO that is particularly natural when it comes to the formulation
of CSPs is the logic called \emph{SNP} (for \emph{strict NP}; see \cite{PapaYanna} and~\cite{FederVardi}), introduced by Kolaitis and Vardi under the name \emph{strict $\Sigma_1^1$}~\cite{KolaitisVardi87}. 
An existential second-order sentence is in SNP if
its first-order part is universal.
There are many links
between constraint satisfaction and the complexity class SNP;
many of those go back to~\cite{FederVardi} and~\cite{FederVardiNegation}, some others that we present here are new. 

SNP is often a convenient way to specify CSPs. 
However, not every problem in SNP is a CSP.
In this section we present a syntactic condition that implies that an SNP sentence describes a problem of the form $\Csp(\bB)$ for an infinite structure $\bB$.
Conversely, if an SNP sentence describes a CSP, then there is an equivalent
SNP sentence that satisfies the syntactic condition. 

The special case in which all existentially quantified relations are unary, known as \emph{monadic SNP}, deserves special attention, and will be discussed at the end of this section.

\subsection{Fagin's theorem}
\label{ssect:fagin}
We start by reviewing Fagin's theorem 
(see e.g.~\cite{EbbinghausFlum}).
Fix a finite relational signature $\tau$.
Let $\mathcal C$ be a class of finite $\tau$-structures 
that is closed under isomorphisms (that is, if $\bB \in \mathcal C$,
and $\bA$ is isomorphic to $\bB$, then $\bA \in \mathcal C$).
We also fix some standard way to code relational structures as finite
strings so that they can be given as an input to a Turing machine, see again~\cite{EbbinghausFlum}. We say that $\mathcal C$ is in NP when there exists a non-deterministic polynomial time algorithm that accepts exactly the structures from $\mathcal C$ under this representation. 

A sentence of the form $\exists R_1,\dots,R_m. \, \phi$
where $\phi$ is a first-order sentence with signature $\tau \cup \{R_1,\dots,R_m\}$ is called an \emph{existential second-order sentence}.
When a structure $\bA$ satisfies $\Phi$ (and this is defined in the obvious way, see e.g.~\cite{EbbinghausFlum}), we write
$\bA \models \Phi$.
\begin{theorem}[Fagin's Theorem, see e.g.~\cite{EbbinghausFlum}]
\label{thm:fagin}
An isomorphism-closed class of finite $\tau$-structures 
is in NP if and only if there exists an existential second-order sentence $\Phi$ 
that \emph{describes} $\mathcal C$ in the sense that
 $$\bA \in \mathcal C \text{ if and only if } \bA \models \Phi \; .$$
\end{theorem}

\subsection{SNP}
\label{ssect:snp}
An \emph{SNP sentence} is an existential second-order sentence with a universal first-order part, i.e., a sentence of the form $$\exists R_1,\dots,R_k. \; \forall x_1,\dots,x_n. \; \phi $$
where $\phi$ is quantifier-free
and over the signature $\tau \cup \{R_1,\dots,R_k\}$.
%We deviate from our convention that equality is always part of first-order logic, and do \emph{not} allow equality or inequality in $\phi$.
%Hence, we always consider what has been called \emph{SNP without inequality} in~\cite{FederVardi}; this will avoid some technical complications. 
The class of problems that can
be described by SNP sentences is called SNP, too.

\begin{example}\label{expl:acycl-snp}
The problem $\Csp(({\mathbb Z};<))$ can be described by the following SNP sentence.
\begin{align*}
\exists T \, \forall x,y,z \big ( & (x<y \Rightarrow T(x,y)) \\
\wedge & \big ( (T(x,y) \wedge T(y,z)) \Rightarrow T(x,z) \big) \wedge \neg T(x,x) \big )
\end{align*}
\qed
\end{example}

\begin{example}\label{expl:betw-snp}
The \emph{Betweenness problem} $\Csp(({\mathbb Z};\Betw))$ (Example~\ref{expl:betw})
can be described by the following SNP sentence.
\begin{align*}
\exists T \, \forall x,y,z \big ( & \neg T(x,x) \wedge \big((T(x,y) \wedge T(y,z)) \Rightarrow T(x,z) \big) \\
\wedge & \; \big (\Betw(x,y,z) \Rightarrow \big( (T(x,y) \wedge T(y,z)) \vee (T(z,y) \wedge T(y,x)) \big )\big)
\end{align*}
\qed
\end{example}

\begin{example}\label{expl:triangle-bipart-snp}
The problem whether a given undirected graph 
can be partitioned into two triangle-free graphs (this problem has been called No-Mono-Tri in Example~\ref{expl:other-problems})
can be described by
the SNP sentence.
\begin{align*}
\exists M \, \forall x,y,z \; & \big ( \neg \big (M(x) \wedge M(y) \wedge M(z) \wedge E(x,y) \wedge E(y,z) \wedge E(z,x) \big) \\
\wedge &
\neg \big (\neg M(x) \wedge \neg M(y) \wedge \neg M(z) \wedge E(x,y) \wedge E(y,z) \wedge E(z,x) \big ) \big ) 
\end{align*}
\end{example}

The following fundamental lemma for SNP sentences is due to Feder and Vardi~\cite{FederVardiNegation},
and can be shown by a simple compactness argument (Theorem~\ref{thm:compactness}).

\begin{lemma}[from~\cite{FederVardiNegation}]\label{lem:snp-compactness}
Let $\bA$ be an infinite structure, and $\Phi$ an SNP sentence. Then $\bA \models \Phi$ if and only if
$\bA' \models \Phi$ for all finite induced substructures $\bA'$ of $\bA$.
\end{lemma}

Since every finite induced substructure of $\bB$ homomorphically maps to $\bB$, and therefore satisfies $\Phi$,
we have the following consequence.

\begin{corollary}\label{cor:snp-on-template}
Let $\Phi$ be an SNP sentence that describes $\Csp(\bB)$ for a structure $\bB$. Then $\bB$ itself satisfies $\Phi$.
\end{corollary}

\subsection{SNP and CSPs} 
\label{ssect:snp-csp}
We say that two SNP sentences $\Phi$ and $\Psi$ are \emph{equivalent} if for all structures (equivalently: all finite structures) 
$\bA$ we have 
$\bA \models \Phi$ if and only if
$\bA \models \Psi$.
We assume in the following
that the first-order part $\phi$ of $\Phi$ is written in conjunctive normal form. 

\begin{definition}
Let $\Phi$ be an SNP sentence whose unquantified relation symbols
are from the signature $\tau$.
Then $\Phi$ is called \emph{monotone} if each literal
of $\Phi$ with a symbol from $\tau \cup \{=\}$ is negative, that is,
of the form $\neg R(\bar x)$, for $R \in (\tau \cup \{=\})$.
\end{definition}

In particular, monotone SNP sentences do not contain literals of the form $x=y$ 
(hence, in the terminology of Feder and Vardi~\cite{FederVardi}, we work here with
\emph{monotone SNP without inequality}; the reason why Feder and Vardi add the attribute \emph{without inequalities}
is that for them, SNP sentences are written in negation normal form, so forbidding literals of the form $x=y$ amounts to
forbidding inequalities in negation normal form).
 
We also assume that monotone SNP sentences 
do not contain literals of the form $x \neq y$.
This is without loss of generality, since every monotone SNP sentence is equivalent
to one which does not contain literals of the form $x \neq y$. To obtain the equivalent sentence,
we remove literals of the form $x \neq y$ and replace all occurrences of $y$ in the same clause by $x$. Note that the SNP sentences given in 
Example~\ref{expl:acycl-snp}, \ref{expl:betw-snp}, and~\ref{expl:triangle-bipart-snp}
can be easily re-written into equivalent monotone SNP sentences.

The class of structures that satisfy a given monotone SNP sentences is clearly closed under
inverse homomorphisms. The converse is a result by 
Feder and Vardi~\cite{FederVardiNegation};
it shows that for SNP, the semantic restriction of closure under inverse homomorphisms and the syntactic restriction
of monotonicity match.

% Isn't it true that we could add a third item to the following: 
% namely that the class of all models of the first-order part of \Phi has the joint embedding property?
% Problem: this doesn't follow from closure under disjoint unions, since we might have to change
% the expansions to combine them. And, as Cherlin explicitly says somewhere, JEP might be undecidable
% for a given universal fo sentence. 

\begin{theorem}[from~\cite{FederVardiNegation}]
\label{thm:monotone-snp}
Let $\Phi$ be an SNP sentence. Then the class of structures that satisfy $\Phi$ is closed
under inverse homomorphisms if and only if $\Phi$ is equivalent to a monotone SNP sentence.
\end{theorem}

\begin{definition}[Connected SNP]
When $\psi$ is a clause of a first-order $\sigma$-formula $\phi$ in conjunctive normal form, let 
%$\neg \psi_1,\dots,\neg \psi_k$
%be the negative literals in $\psi$, i.e., for all $i$
%the formula $\psi_i$ is of the form $R(\bar x)$ for 
%$R \in \tau \cup \{R_1,\dots,R_k\}$. 
%Then the \emph{canonical database of $\psi$} is 
%the canonical database (see Section~\ref{sect:csp-logical}) of $\psi_1 \wedge \dots \wedge \psi_k$ (that is, in this definition the positive literals in $\psi$ are ignored).
$\bC$ be the $\sigma$-structure 
whose vertices are
the variables of $\psi$, and where $(x_1,\dots,x_n) \in R^{\bC}$ if and only if $\psi$ contains a
negative literal of the form 
$\neg R(x_1,\dots,x_n)$. 
We say that $\psi$ is \emph{connected} 
if $\bC$ is connected.
We say that an SNP sentence $\Phi$ is \emph{connected} if all clauses of the first-order part $\phi$ of $\Phi$ are connected.
\end{definition}

% THE FOLLOWING IS BUGGY (24/2/2014)
%When $\psi$ is a clause of a first-order formula $\phi$ in conjunctive normal form, let $\neg \psi_1,\dots,\neg \psi_k$
%be the negative literals in $\psi$, i.e., for all $i$
%the formula $\psi_i$ is of the form $R(\bar x)$ for 
%$R \in \tau \cup \{R_1,\dots,R_k\}$. 
%Then the \emph{canonical database of $\psi$} is 
%the canonical database (see Section~\ref{sect:csp-logical}) of $\psi_1 \wedge \dots \wedge \psi_k$
%(that is, in this definition the positive literals in $\psi$ are ignored).
%We say that $\psi$ is \emph{connected}
%if the canonical database 
%of $\psi$ is connected (see Section~\ref{sect:homo}).
%We say that an SNP sentence $\Phi$ is \emph{connected} if all clauses of the first-order part $\phi$ of $\Phi$ are connected.

\begin{theorem}\label{thm:connected-snp}
Let $\Phi$ be an SNP sentence. Then the class of structures that satisfy $\Phi$ is closed
under disjoint unions if and only if $\Phi$ is equivalent to a connected SNP sentence.
\end{theorem}

\begin{proof}
Let $\Phi$ be of the form $\exists R_1,\dots,R_k \, \forall x_1,\dots,x_l. \,  \phi$ where $\phi$ is a quantifier-free first-order formula 
over the signature $\sigma = (\tau \cup \{R_1,\dots,R_k\})$.

Suppose first that $\Phi$ is connected, and that $\bA_1$ and $\bA_2$ both satisfy $\Phi$.
In other words, there is a $\sigma$-expansion $\bA_1^*$ of $\bA_1$ 
and a $\sigma$-expansion $\bA_2^*$ of $\bA_2$
such that those expansions satisfy $\forall \bar x. \phi$.
We claim that the disjoint union $\bA^*$ of $\bA_1^*$ and $\bA_2^*$ also satisfies $\forall \bar x. \phi$;
otherwise, there would be a clause $\psi$ in $\phi$ and elements $a_1,\dots,a_q$
of $A_1 \cup A_2$ such that $\psi(a_1,\dots,a_q)$ is false in $\bA^*$.
Since $\bA_1^*$ and $\bA_2^*$ satisfy $\forall \bar x. \psi$,
there must be $i,j$ such that $a_i \in A_1$ and $a_j \in A_2$. But then the canonical database for $\psi$ is disconnected,
a contradiction. 

For the opposite direction of the statement, assume that the class of structures that satisfy $\Phi$ is closed
under disjoint unions.
Consider the SNP sentence $\Psi = \exists R_1,\dots,R_k,E. \; \forall x_1,\dots,x_l. \; \psi$ where $\psi$ is the conjunction 
of the following clauses (we assume without loss of generality that $l \geq 3$). 
\begin{itemize}
\item For each relation symbol $R \in \tau$, say of arity $p$, 
and each $i<j\leq p$, add the conjunct $\neg R(x_1,\dots,x_p) \vee E(x_i,x_j)$ to $\psi$.
%\item For each relation symbol $R \in \tau$, say of arity $p$, add to $\psi$ the conjunct $$R(x_1,\dots,x_p) \rightarrow \bigwedge_{i<j \leq p} E(x_i,x_j) \; .$$ 
%Note that this clause can be re-written into an equivalent conjunction of connected clauses. 
\item Add the conjunct $\neg E(x_1,x_2) \vee \neg E(x_2,x_3) \vee E(x_1,x_3)$ 
to $\psi$.
\item Add the conjunct $\neg E(x_1,x_2) \vee E(x_2,x_1)$ to $\psi$.
\item For each clause %$\phi_1 \vee \dots \vee \phi_k$ 
$\phi'$ of $\phi$ with variables $y_1,\dots,y_q \subseteq \{x_1,\dots,x_l\}$, add to $\psi$ the conjunct $$\phi' \vee \bigvee_{i<j \leq q} \neg E(y_i,y_j)\; .$$
\end{itemize}
We claim that the connected monotone SNP sentence $\Psi$ is equivalent to $\Phi$. 
Suppose first that $\bA$ is a finite structure that satisfies $\Phi$.
Then there is a $\sigma$-expansion $\bA'$ of $\bA$ that satisfies $\forall \bar x. \phi$.
The expansion of $\bA'$ by the relation $E = A^2$ shows that $\bA$ also satisfies $\forall \bar x. \psi$.

Now suppose that $\bA$ is a finite structure with domain $A$ that satisfies $\Psi$.
Then there is a $(\sigma \cup \{E\})$-expansion $\bA'$ of $\bA$ that satisfies $\forall \bar x. \psi$.
Write $\bA' = \bA'_1 \uplus \dots \uplus \bA_l'$ for connected $\sigma$-structures $\bA'_1, \dots, \bA_l'$.
Note that the clauses of $\psi$ force that the relation $E$ denotes $A_i^2$ in the structure $\bA'_i$, for each $i \leq l$.
Let $\bA_i$ be the $\sigma$-reduct of $\bA'_i$. Then $\bA_i$ satisfies $\forall \bar x. \phi$, because if there was a clause
$\phi'$ from $\phi$ violated in $\bA_i$ then the corresponding clause in $\psi$ would be violated in $\bA_i'$.
Hence, $\bA_i \models \Phi$ for all $i \leq l$, and since $\Phi$ is closed under disjoint unions,
we also have that $\bA \models \Phi$. 
\end{proof}

Theorem~\ref{thm:monotone-snp} combined with the previous result shows the following.

\begin{corollary}\label{cor:snp-csp}
An SNP sentence $\Phi$ describes a problem of the form $\Csp(\bB)$
for an infinite structure $\bB$ 
if and only if $\Phi$ is equivalent to a monotone and connected SNP sentence $\Psi$.
\end{corollary}
\begin{proof}
Suppose first that $\Phi$ is a monotone SNP sentence with connected clauses.
To show that $\Phi$ describes a problem of the form $\Csp(\bB)$ we
can use Lemma~\ref{lem:fund}.
It thus suffices to show that the class of structures that satisfy $\Phi$ is closed under disjoint unions and inverse homomorphisms.
But this has already been observed in Theorem~\ref{thm:monotone-snp} and Theorem~\ref{thm:connected-snp}.

For the implication in the opposite direction, suppose that $\Phi$ describes a problem
of the form $\Csp(\bB)$ for some infinite structure $\bB$.
In particular, the class of structures that satisfy $\Phi$ is closed
under inverse homomorphisms. By Theorem~\ref{thm:monotone-snp}, $\Phi$ is equivalent to a monotone
SNP sentence. Moreover, the class of structures that satisfy $\Phi$ is closed under disjoint unions,
and hence $\Phi$ is also equivalent to a connected SNP sentence. 
By inspection of the proof of  Theorem~\ref{thm:connected-snp}, we see that 
when $\Phi$ is already monotone, then the connected SNP sentence in the proof of  Theorem~\ref{thm:connected-snp}
will also be monotone.
It follows that $\Phi$ is also equivalent to a connected monotone SNP sentence.
\end{proof}

\subsection{Monadic SNP}
\label{ssect:monadic-snp}
When we further restrict monotone SNP by only allowing \emph{unary}
existentially quantified relations, the corresponding class of problems,
called \emph{montone monadic SNP} (or, short, \emph{MMSNP}),
gets very close to finite domain constraint satisfaction problems.
Indeed, Feder and Vardi showed that the class MMSNP
exhibits a complexity dichotomy if and only if
the class of all finite domain CSPs exhibits a complexity dichotomy
(that is, if the \emph{dichotomy conjecture} 
mentioned in the introduction is true).
In one direction, this is obvious since MMSNP obviously contains
$\Csp(\bB)$ for all finite structures $\bB$ (we may use a unary
relation symbol for each element of $\bB$).
In the other direction, Feder and Vardi showed that
every problem in MMSNP is equivalent under randomized
Turing-reductions to a finite domain constraint satisfaction problem. 
The reduction has subsequently been derandomized by Kun~\cite{Kun}. 

\begin{theorem}[of~\cite{FederVardi} and~\cite{Kun}; see~\cite{MadelaineStewartSicomp} for a formalization]
Every problem in monotone monadic SNP is polynomial-time Turing 
equivalent to $\Csp(\bB)$ for a finite structure $\bB$.
\end{theorem}

Similarly as in the previous section, we might ask for a syntactic
characterization of those monadic SNP sentences that describe a CSP. Note that this does not directly follow from Corollary~\ref{cor:snp-csp}, since the reductions used there introduce additional existentially
quantified relations that are not monadic. 
However, we have the following monadic version of Theorem~\ref{thm:monotone-snp}.

\begin{theorem}[Theorem 3 in~\cite{FederVardiNegation}]
\label{thm:mmsnp}
Let $\Phi$ be a monadic SNP sentence. Then the class of structures that satisfy $\Phi$ is closed
under inverse homomorphisms if and only if $\Phi$ is equivalent to a monotone monadic SNP sentence.
\end{theorem}

Moreover, one can show the following monadic version of 
Proposition~\ref{thm:connected-snp}. 

\begin{proposition}\label{prop:connected-monotone-msnp}
Let $\Phi$ be a monadic SNP sentence. Then the class of structures that satisfy $\Phi$ is closed under disjoint unions if and only if $\Phi$ is equivalent to a connected monadic SNP sentence.
\end{proposition}

\begin{proof}
Let $V$ be the set of variables of the first-order part $\phi$ of $\Phi$, 
let $P_1, \dots, P_k$ be the existential monadic predicates in $\Phi$,
and let $\tau$ be the input signature so that $\phi$ has signature
$\{P_1,\dots,P_k\} \cup \tau$. 
If $\Phi$ is connected, then it describes a problem that is closed
under disjoint unions; this follows from Theorem~\ref{thm:connected-snp}. 

For the opposite direction, suppose that $\Phi$ describes a problem
that is closed under disjoint unions. 
We can assume without loss of generality that $\Phi$ is minimal in
the sense that if we remove literals from some of the clauses the resulting SNP sentence is inequivalent. 
We shall show that then $\Phi$ must be connected. 
Let us suppose that this is
not the case, and that there is a clause $\psi$ in $\phi$ that is not connected. The clause $\psi$ can be written
as $\psi_1 \vee \psi_2$ where the set of variables $X \subset V$ of $\psi_1$ and
the set of variables $Y \subset V$ of $\psi_2$ are non-empty and disjoint.
Consider the formulas
$\Phi_X$ and $\Phi_Y$ obtained from $\Phi$ by replacing $\psi$ by $\psi_1$ and $\psi$ by $\psi_2$, respectively. By minimality of $\Phi$ there is a $\tau$-structure $\bA_1$ that satisfies $\Phi$ but not $\Phi_X$, and
similarly there exists a $\tau$-structure $\bA_2$ that satisfies $\Phi$ but
not $\Phi_Y$. By assumption, 
the disjoint union $\bA$ of $\bA_1$ and $\bA_2$ satisfies
$\Phi$. So there exists a $\tau \cup \{P_1,\dots,P_k\}$-expansion $\bA'$ of $\bA = \bA_1 \uplus \bA_2$ that satisfies the first-order
part of $\Phi$. 
Consider the substructures $\bA'_1$ and $\bA'_2$ of
$\bA'$ induced by $A_1$ and $A_2$, respectively. 
We have that $\bA'_1$ does not satisfy $\psi_1$ 
(otherwise $\bA_1$ would satisfy
$\Phi_X$). Consequently, there is an assignment 
$s_1 \colon V \rightarrow A_1$ of the
universal variables that falsifies $\psi_1$. 
By similar reasoning we can infer that
there is an assignment $s_2 \colon V \rightarrow A_2$ 
that falsifies $\psi_2$. Finally, fix any
assignment $s \colon V \rightarrow A_1 \cup A_2$ that coincides with $s_1$ over $X$ and with $s_2$
over $Y$ (such an assignment exists because $X$ and $Y$ are
disjoint). Clearly, $s$ falsifies $\psi$ and $\bA$ does not satisfy
$\Phi$, a contradiction. 
\end{proof}

Similarly as in Corollary~\ref{cor:snp-csp} for SNP, 
we can combine the conditions of closure under inverse homomorphisms and closure under disjoint unions, 
and arrive at the following.

\begin{corollary}\label{cor:msnp-csp}
A monadic SNP sentence $\Phi$ describes a problem of the form $\Csp(\bB)$
for an infinite structure $\bB$ 
if and only if $\Phi$ is equivalent to a connected monotone monadic SNP sentence.
\end{corollary}

We want to remark that the problems that can be described by 
connected monotone monadic SNP sentences are exactly the problems called \emph{forbidden patterns problems}
in the sense of Madelaine~\cite{Madelaine}. Clearly, for every finite $\bB$ the problem $\Csp(\bB)$ is a forbidden patterns problem. In~\cite{MadelaineStewart} is has been shown that the problems
in MMSNP are exactly finite unions of forbidden patterns problems (going back to ideas from~\cite{FederVardi}).

\ignore{ Too similar to the one above
\begin{proof}
Suppose first that $\Phi$ is a monotone SNP sentence with connected clauses.
To show that $\Phi$ describes a problem of the form $\Csp(\bB)$ we
can use Lemma~\ref{lem:fund}.
It thus suffices to show that the class of structures that satisfy $\Phi$ is closed under disjoint unions and inverse homomorphisms.
But this has already been observed in Theorem~\ref{thm:monotone-snp} and Theorem~\ref{thm:connected-snp}.

For the implication in the opposite direction, suppose that $\Phi$ describes a problem
of the form $\Csp(\bB)$ for some infinite structure $\bB$.
In particular, the class of structures that satisfy $\Phi$ is closed
under inverse homomorphisms. By theorem Theorem~\ref{thm:monotone-snp}, $\Phi$ is equivalent to a monotone
SNP sentence. Moreover, the class of structures that satisfy $\Phi$ is closed under disjoint unions,
and hence $\Phi$ is also equivalent to a connected SNP sentence. 
By inspection of the proof of  Theorem~\ref{thm:connected-snp}, we see that 
when $\Phi$ is already monotone, then the connected SNP sentence in the proof of  Theorem~\ref{thm:connected-snp}
will also be monotone. It follows that $\Phi$ is also equivalent to a connected monotone SNP sentence.
\end{proof}}

\begin{figure}[h]
\begin{center}
\includegraphics[scale=.5]{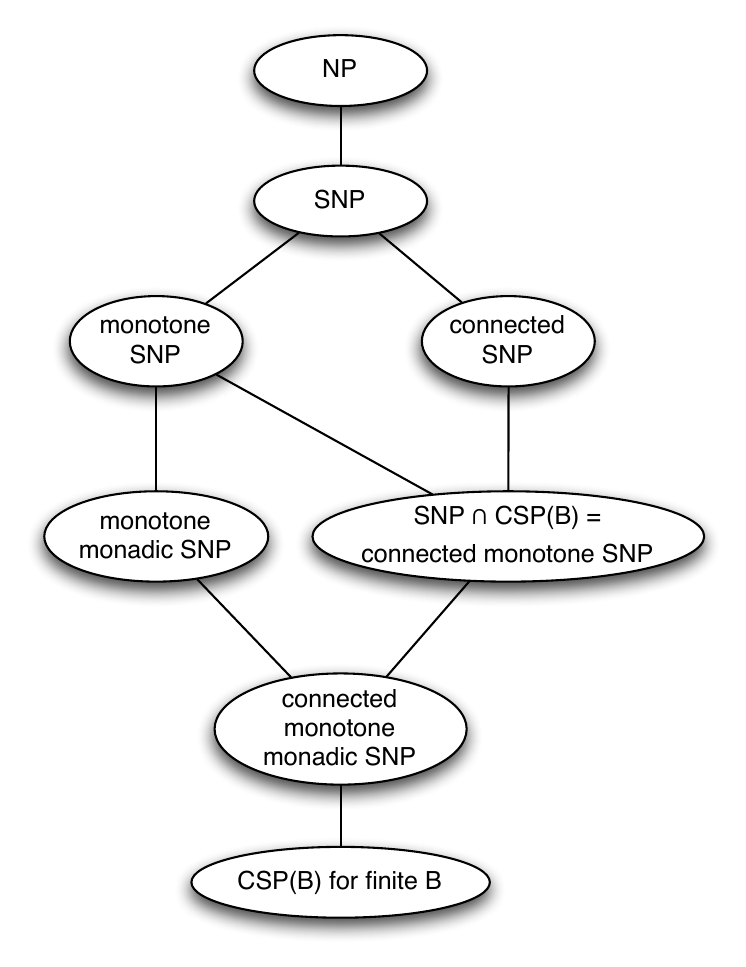} 
\end{center}
\caption{Fragments of SNP.}
\label{fig:snp-pre}
\end{figure}

We summarize the landscape of classes of computational
problems from this section in Figure~\ref{fig:snp-pre}.

% !TEX root = 0.tex

\section{Examples}
\label{sect:csp-examples}
We present computational problems that have been studied
in various areas of theoretical computer science, and 
that can be formulated as constraint satisfaction problems
in the sense of Section~\ref{sect:homo}, \ref{sect:csp-logical}, \ref{sect:sat}, or~\ref{sect:snp}. We describe each problem from the 
perspective in which the computational
problem has appeared first in the literature.

Our list is by far not exhaustive;
computational problems that can be 
exactly formulated as $\Csp(\bB)$ for an infinite
structure $\bB$ are abundant in almost every area of theoretical computer science.

\subsection{Allen's interval algebra}\label{ssect:allen}
Allen's interval algebra~\cite{Allen} is a formalism that is famous 
in artificial intelligence, and which has been introduced to 
reason about intervals %(e.g., to model events), 
and about the relationships between intervals.

Formally, Allen's interval algebra is a proper relation algebra
(see Section~\ref{ssect:relation-algebras}); we can also
view it as a structure with a binary relational signature.
The domain is the set $\mathbb I$ 
of all closed intervals $[a,b]$ of rational numbers,
where $a,b \in \mathbb Q, a<b$.
When $x = [a,b]$ is an interval, then $-x$ denotes the interval $[-b,-a]$. For $R \subseteq {\mathbb Q}^2$,
$R^-$ denotes the relation $\{(-x,-y) \; | \; (x,y) \in R\}$. 
Recall that in proper relation algebras, 
$R^{\smallsmile}$ denotes the relation $\{(y,x) \; | \; (x,y) \in R\}$.

\begin{figure}[h]
\begin{center}
\begin{tabular}{|l|l|l|}
\hline
Relation Symbol & Definition & Explanation \\
\hline
$P$ & $\{([a,b],[c,d]) \; | \; b < c \}$ & $[a,b]$ preceeds $[c,d]$ \\
$M$ & $\{([a,b],[c,d]) \; | \; b = c \}$ & $[a,b]$ meets $[c,d]$ \\
$O$ & $\{([a,b],[c,d]) \; | \; a < c < b < d \}$ & $[a,b]$ overlaps with $[c,d]$ \\
$S$ & $\{([a,b],[c,d]) \; | \; a=c \text{ and } b < d \}$ & $[a,b]$ starts $[c,d]$ \\
$D$ & $\{([a,b],[c,d]) \; | \; c < a < b < d \}$ & $[a,b]$ is during $[c,d]$  \\
$E$ & $\{([a,b],[c,d]) \; | \; a=c, b=d \}$ & $[a,b]$ equals $[c,d]$ \\
 \hline
\end{tabular}
\end{center}
\caption{The definitions for the basic relations of Allen's interval algebra.} \label{fig:allen-basic-defs}
\end{figure}

The basic relations of Allen's interval algebra 
%(again, see Section~\ref{ssect:relation-algebras}) 
are the 13 relations $P,M,O,S,D,E$ (defined in Figure~\ref{fig:allen-basic-defs}), $P^-,M^-,O^-,S^-$,
and the inverse of $S$, $D$, and $S^-$,
denoted by $S^\smallsmile$, $D^\smallsmile$, and $(S^-)^\smallsmile$, respectively.
Note that those 13 relations are pairwise disjoint, and that
their union equals ${\mathbb I}^2$.
Recall our convention that when $\mathcal R$ is a subset of the basic relations, we write $x {\mathcal R} y$ if $(x,y) \in \bigcup_{R \in \mathcal R} R$.
For example, $x \{P,P^-\} y$ signifies that the intervals $x$ and $y$
are disjoint. The $2^{13}$ relations that arise in this way will
be called the \emph{relations of Allen's interval algebra}.

An important computational problem for Allen's interval algebra is the 
network satisfaction problem % (Section~\ref{ssect:relation-algebras})
for Allen's interval algebra. 
%Theorem~\ref{prop:disj-union-networks},
This problem can be viewed as $\Csp(\bA)$
where $\bA$ is a structure with domain $\mathbb I$ 
and a signature containing $2^{13}$ binary relation symbols
(see Section~\ref{ssect:relation-algebras}). 
More on this structure
can be found in Chapter~\ref{chap:mt}, Example~\ref{example:allen}. 
We are sometimes sloppy and write \emph{Allen's interval algebra} 
when we mean $\bA$ (rather than $\bf A$).

%\cproblem{Network satisfaction problem for Allen's interval algebra}
%{a set of variables $V$, and constraints of the form
%$x \mathcal R y$ where $\mathcal R$ stands for one of the
%relations from Allen's interval algebra.}
%{is there an assignment from $V \rightarrow \mathbb I$ that
%satisfies all the constraints?}

The problem $\Csp(\bA)$ is NP-complete~\cite{Allen}.
The complexity of the CSP for (binary) reducts of Allen's interval algebra has been completely classified in~\cite{KrokhinAllen}.

\subsection{Phylogenetic reconstruction problems}
\label{ssect:phylo}
In modern biology it is believed that 
the species in the evolution of life on earth 
developed in a mostly tree-like fashion:
at certain time periods, species separated into sub-species.
The goal of phylogenetic reconstruction is to determine
the evolutionary tree from given partial information about the tree.
This motivates the computational problem of 
\emph{rooted triple satisfiability} (also called \emph{rooted triple consistency}), defined below.
In 1981, Aho, Sagiv, Szymanski, and Ullman~\cite{ASSU}
presented a quadratic time algorithm to this problem, motivated independently from computational biology by questions in database theory.

Let $\bT$ be a tree with vertex set $T$ and 
with a distinguished vertex $r$, 
the \emph{root} of $\bT$. 
For $u,v \in T$, we say that $u$ \emph{lies below} $v$
if the path from $u$ to $r$ passes through $v$. 
We say that $u$ \emph{lies strictly below} $v$ if
$u$ lies below $v$ and $u \neq v$.
The \emph{youngest common ancestor (yca)} of two vertices $u,v \in T$ is the node $w$
such that both $u$ and $v$ lies below $w$ and $w$ has 
maximal distance from $r$. 
%Note that the yca, viewed as a binary
%operation, is commutative and associative, and hence there is a canonical definition of the yca of a set of elements $u_1,\dots,u_k$.
%The \emph{rooted triple consistency problem} is a 
%special case of the subtree consistency problem,
%in which each tree in the input has three leaves only.
%However, instances of the subtree consistency problem have a straightforward translation into rooted triple consistency problems
%(see ~\ref{exe:trees-to-triples}).
%We therefore focus on the rooted triple consistency problem here.
 % HOW MUCH COMMON SUPERTREE PROBLEM
% and infinite constraint language discussion should be here?

\cproblem{Rooted-Triple Satisfiability}
{A finite set of variables $V$, and a set of triples $xy|z$ for $x,y,z \in V$.}
{Is there a rooted tree $\bT$ with leaves $L$ and a mapping $s \colon V \rightarrow L$ such that for every triple $xy|z$ the yca of $s(x)$ and $s(y)$ lies strictly below the yca of $s(x)$ and $s(z)$ in $\bT$?}

%The \emph{rooted triple consistency problem} is a 
%special case of the subtree consistency problem,
%in which each tree in the input has three leaves only.
%However, instances of the subtree consistency problem have a straightforward translation into rooted triple consistency problems
%(see ~\ref{exe:trees-to-triples}).
%We therefore focus on the rooted triple consistency problem here.
 
Another famous problem that has been studied in this context
is the quartet satisfiability problem, which is NP-complete~\cite{Steel}.
 
\cproblem{Quartet Satisfiability}
{A finite set of variables $V$, and a set of quartets $xy{:}uv$ with $x,y,u,v \in V$.}
{Is there a tree $\bT$ with leaves $L$ and a mapping $s \colon V \rightarrow L$ such that for every quartet $xy{:}uv \in R^{\yca}$ the shortest path from $x$ to $y$ is disjoint to the shortest path from $u$ to $v$?}

It is straightforward to check that the class of positive
instances (viewed as relational structures) of each of those two computational problems
is closed under disjoint unions and
inverse homomorphisms. By Lemma~\ref{lem:fund}, both the rooted triple satisfaction problem and the quartet satisfaction problem can be formulated as
$\Csp(\bB)$ for an infinite structure $\bB$. We come back to those 
CSPs in Chapter~\ref{chap:examples}.

\subsection{Branching-time constraints}
\label{ssect:branching-time}
An important model in temporal reasoning 
is \emph{branching time}, where 
for every time point 
the past is linearly ordered, but the future is only partially ordered.

This motivates the so-called \emph{left-linear point algebra}~\cite{HirschAlgebraicLogic,Duentsch}, 
which is a relation algebra with four basic relations, denoted by 
$=$, $<$, $>$, and $|$. Here we imagine that `$x < y$' signifies that
$x$ is \emph{earlier in time than $y$}, and \emph{to the left of $y$} when we draw points in the plane, and this motivates the name \emph{left linear point algebra}.
The composition operator on those four basic relations is given in Figure~\ref{fig:left-linear}. The inverse of $<$ is $>$, $\Id$ denotes $=$,
and $|$ is its own inverse, and the relation algebra is uniquely
given by this data. 

\begin{figure}
\begin{tabular}{|c||c|c|c|c|}
\hline
$\circ$ & $=$ & $<$ & $>$ & $|$ \\
\hline \hline
$=$ & $=$ & $<$ & $>$ & $|$ \\
\hline
$<$ & $<$ & $<$ & $\{<,=,>\}$ & $\{<,|\}$ \\
\hline
$>$ & $>$ & $1$ & $>$ & $|$ \\
\hline
$|$ & $|$ & $|$ & $\{>,|\}$ & 1 \\ 
\hline
\end{tabular}
\caption{The composition table for the basic relations in the left-linear point algebra.}
\label{fig:left-linear}
\end{figure}

 As explained in Section~\ref{ssect:relation-algebras}, the network consistency problem for the left-linear point algebra can be viewed as $\Csp(\bB)$ for
an appropriate infinite structure with $16=2^4$ binary relations (one
for each subset of $\{=,<,>,|\}$). In this structure, for
every $x$ the set $\{y \; | \; y < x\}$ is linearly ordered by by $<$.
An explicit example of such a structure is given in Section~\ref{ssect:mt:branching-time}.
The network consistency problem of the left-linear point algebra is polynomial-time equivalent to the following problem, which we call \emph{branching-time satisfiability problem}. 

\cproblem{Branching-Time Satisfiability}
{A finite relational structure $\bA = (A;\leq,\parallel,\neq)$ where 
$\leq$, $\parallel$, and $\neq$ are binary relations.}
{Is there a rooted tree $\bT$ and a mapping $s \colon A \rightarrow T$ such that in $\bT$ the following is satisfied: 
%\begin{itemize}
a) 
%\item 
If $(x,y) \in \; \leq^\bA$, then $s(x)$ lies above $s(y)$; 
b) 
%\item 
If $(x,y) \in \; ||^{\bA}$, then neither $s(x)$ 
lies strictly above $s(y)$,
 nor $s(y)$ strictly above $s(x)$; 
%\item
c) 
If $(x,y) \in \; \neq^\bA$, then $s(x) \neq s(y)$.
%\end{itemize}
}

The idea why this problem is polynomial-time equivalent to the network satisfaction problem of the left-linear point algebra is the observation 
that in any representation $\mathfrak B$ of the left-linear point algebra,
the relation $x \{<,>,=\} y$  has the primitive positive 
definition $$\exists z \; (x \{<,=\} z \wedge y \{<,=\} z) \; ,$$ 
and the relation
$x \{<,|,=\} y$ has the primitive positive definition 
$$\exists z \; ( x \{<,=\} z \wedge z \{|,=\} y) \; ;$$ 
we can then use Theorem~\ref{lem:pp-reduce} (for details, see~\cite{BodirskyKutzAI}).

The branching-time satisfiability problem can be formulated as $\Csp(\bC)$ for
the structure with domain $C := \{0,1\}^*$ and relations $\leq$, $\parallel$, and $\neq$, where $\leq$ denotes the relation
$$\{(u,v) \in C^2 \; | \; u \text{ is a prefix of } v\} \; .$$
The relation $\neq$ is the disequality relation, and $u \parallel v$ holds
if $u$ and $v$ are equal or incomparable with respect to $\leq$. 
Let $<$ denote the intersection of $\leq$ and $\neq$.
Note that the structure $\mathfrak C$ can \emph{not} be used to obtain 
a representation of the left-linear point algebra, since $(<) \circ (<)$
does not equal $<$.

The first polynomial-time algorithm for the branching-time consistency problem (and therefore also for the network satisfaction problem of the left-linear point algebra) is 
due to Hirsch~\cite{HirschAlgebraicLogic}, and has a worst-case running time in $O(n^5)$.
This has been improved by Broxvall and Jonsson~\cite{BroxvallJonsson}, who
presented an algorithm running in $O(n^{3.376})$
(this algorithm uses an $O(n^{2.376})$ algorithm 
for fast integer matrix multiplication).
A simpler algorithm which does not use fast matrix multiplication and
runs in $O(nm)$ has been found in~\cite{BodirskyKutz}.

\subsection{Cornell's tree description constraints}
\label{ssect:cornell}
Motivated by problems in computational linguistics, Cornell~\cite{Cornell} introduced
the following computational problem\footnote{I feel personally committed to Cornell's problem since it was the first CSP with an $\omega$-categorical template I met.}. It is a strictly more expressive
problem than the branching time satisfaction problem from the previous section, but has been introduced independently from~\cite{HirschAlgebraicLogic} and~\cite{BroxvallJonsson}.
There are many equivalent formulations of this problem. One is as the general network satisfaction problem
for the relation algebra $\fC$ with atoms $=$, $<$, $>$, $\prec$, and $\succ$ which is given by the composition table in Figure~\ref{fig:cornell}. 
The idea is that $<$ denotes a dense semilinear order (see Section~\ref{ssect:branching-time}),
and $\prec \cup <$ denotes a linear order. 
The idea how to use this in natural language grammar formalisms like dependency grammars is that $<$ represents the syntactic structure of a natural language sentence whereas $\prec \cup <$ stands for the word order. 

\begin{figure}
\begin{tabular}{|c||c|c|c|c|c|}
\hline
$\circ$ & $=$ & $<$ & $>$ & $\prec$ & $\succ$ \\
\hline \hline
$=$ & $=$ & $<$ & $>$ & $\prec$ & $\succ$ \\
\hline
$<$ & $<$ & $<$ & $\{<,>\}$ & $\{<,\prec\}$ & $\{<,\succ\}$ \\
\hline
$>$ & $>$ & $1$ & $>$ & $\prec$ & $\succ$ \\
\hline
$\prec$ & $\prec$ & $\prec$ & $\{>,\prec\}$ & $\prec$ & 1 \\ 
\hline
$\succ$ & $\succ$ & $\succ$ & $\{>,\succ\}$ & 1 & $\succ$ \\ 
\hline
\end{tabular}
\caption{The composition table for the basic relations of Cornell's tree algebra $\fC$.}
\label{fig:cornell}
\end{figure}

Similarly as in Section~\ref{ssect:branching-time}, 
%it is possible to reduce
%this network satisfaction problem to Cornell's Tree Description Constraints by expressing 
all $2^5$ relations of $\fC$ can be obtained by repeated compositions and intersections of the four relations $\{<,=\}$, $\{\prec,=\}$, 
$\{\prec,\succ,=\}$, and $\{<,>,\prec,\succ\}$; for details, see~\cite{BodirskyKutzAI}.
% BOOKTD: when there is a chapter on datalog, make hints to consistency here
The algorithm presented for the general network satisfaction problem for $\fC$ in~\cite{Cornell} is not complete. 
A polynomial-time algorithm has been found in~\cite{BodirskyKutzAI}. 

\subsection{Set constraints}
\label{ssect:setcsps}
Many fundamental problems in artificial intelligence, knowledge representation, and verification involve reasoning about sets
and relations between sets and can be modeled as constraint satisfaction problems. 
One of the most fundamental problems of this type is the following.
We denote the set of all subsets of $\mN$ by $\P(\mN)$.

\cproblem{Basic Set Constraint Satisfiability}
{A finite set of variables $V$, and a set $\phi$ of constraints of the form $x \subseteq y$, $x \disj y$, or $x \neq y$, for $x,y \in V$.}
{Is there a mapping $s \colon V \rightarrow {\P}(\mN)$
such that \\
a) If $x \subseteq y$ is in $\phi$, then $s(x)$ is contained in $s(y)$; \\
b) If $x \disj y$ is in $\phi$, then $s(x)$ and $s(y)$ are disjoint sets; \\
c) If $x \neq y$ is in $\phi$, then $s(x)$ and $s(y)$ are distinct sets.}

This problem has the shorter description 
$\text{CSP}(({\mathcal P}({\mathbb N}); \subseteq, \disj, \neq))$ where $\subseteq$, $\disj$, $\neq$ are binary relations over ${\P}({\mathbb N})$, standing for the binary relations containment, disjointness, and inequality between sets.
Drakengren and Jonsson~\cite{DrakengrenJonssonSets} showed that basic set constraint satisfiability can be decided in polynomial time. 
They also showed that the generalization of the problem can be solved in polynomial time 
where each constraint has the form
$$ x_1 \neq y_1 \vee \dots \vee x_k \neq y_k \vee x_0 R y_0$$
where $R$ is either $\subseteq$, $\disj$, or $\neq$, and where
$x_0,\dots,x_k,y_0,\dots,y_k$ are not necessarily distinct variables.

\subsection{Spatial reasoning}
\label{ssect:general-spatial}
Qualitative spatial reasoning (QSR) is concerned with representation
formalisms that are considered close to conceptual schemata used by
humans for reasoning about their physical environment---in particular,
about processes or events and about the spatial environment in which
they are situated. %
The approach in qualitative reasoning is to develop relational schemas
that abstract from concrete metrical data of entities (for example
time points, coordinate positions, or distances) by subsuming similar
metric or topological configurations of entities into one
qualitative representation.

There are many formalisms for qualitative spatial reasoning. 
In particular, several relation algebras 
(see Section~\ref{ssect:relation-algebras})
have been studied in this context.
A basic example is the RCC5 relation algebra (with 5 atoms; the RCC5 relation algebra is also known under the name \emph{containment algebra}~\cite{Bennett,Duentsch}), and the RCC8 relation algebra (with 8 atoms).
In both formalisms, the variables denote `non-empty regions'.
In RCC5, the five atoms are denoted by \DR, \PO, \PP, \PPI, \EQ, and they
stand for \emph{disjointness}, \emph{proper overlap}, \emph{proper containment (proper-part-or)}, its inverse, and equality, respectively.
In RCC8, we further distinguish how the `boundaries' of two regions
relate to each other. We do not further discuss RCC8, for details, see~\cite{Duentsch,BodirskyWoelfl}.

There are many equivalent ways to formally define RCC5.
Often, this is done by specifying the composition table for atomic relations, but we find this tedious. Here, we rather define RCC5 as 
%as subalgebra of Allen's interval algebra, namely the one
%induced by the elements $\{P, P^\smallsmile\}, \{M,O,M^\smallsmile,O^\smallsmile\}$, $\{D,S,S^-\}$, $\{D^\smallsmile,S^\smallsmile,(S^-)^\smallsmile\}$, and $E$ 
%in Allen's interval algebra.
%Those elements become the atoms of RCC5,
%where they are called $\DR, \PO, \PP, \PPI, \EQ$, in this order.
% BOOKTD: reference for this fact?! Because of missing
% reference, I had to take this out again.
%Another way of defining RCC5 is via 
the proper
relation algebra whose domain are all open (or all closed) disks 
in $\mR^2$, and where the basic relations are
disjointness (empty intersection), proper overlap, containment, the inverse of containment, and equality of disks. Then RCC5 is the abstract relation algebra of the proper relation algebra of closed disks (see Section 2.1.5 in~\cite{Duentsch}). 

The network satisfaction problem for RCC5 is NP-complete;
the computational complexity of the CSP for the (binary) reducts of $\bB$
has been classified in~\cite{NebelRenz,RCC5JD}.
A polynomial-time tractable case of particular interest is the \emph{network satisfaction problem for the basic relations of RCC5}~\cite{NebelRenz}, i.e., the
network satisfaction problem for RCC5 when the input is restricted to networks $N = (V;f)$
where $f$ maps to $1$ or the atoms in RCC5 only.

In any representation of RCC5, the atomic relations 
satisfy the following set of axioms $T$.
We use $P(x,y)$ as a shortcut for $\PP(x,y) \vee \EQ(x,y)$.

\begin{align*}
 T \; := \; \big \{ \;
\forall x,y,z \;  (DC(x,y) \wedge P(z,y) & \rightarrow DC(x,z)) \\ 
\forall x,y,z \;  (PO(x,y) \wedge P(y,z) & \rightarrow (PO(x,z) \vee PP(x,z))) \\ 
\forall x,y,z \;   (PP(x,y) \wedge PP(y,z) & \rightarrow PP(x,z))  \\ 
\forall x,y,z \;   (P^{-1}(x,y) \wedge P(y,z) & \rightarrow \neg DC(x,y)) \; \big \}
\end{align*}

It is easy to see that the network satisfaction problem for the basic
relations of RCC5 is 
 essentially the same problem as
$\Csp(T)$, where $T$ is the first-order theory defined as above.
It can be checked easily that $T$ satisfies item (2) in the statement of Proposition~\ref{prop:sat-csp}, and hence there exists an infinite structure
$\bB$ such that $\Csp(\bB)$ equals the satisfiability problem for RCC5. We will give more explicit descriptions of such an 
infinite structure $\bB$
in Chapter~\ref{chap:examples}
(and it turns out that there are close links with the problem
from Section~\ref{ssect:setcsps}).

\subsection{Horn-SAT}\label{ssect:horn-sat}
The following problem is an important P-complete problem~\cite{KleineBueningLettmann}. 
It can be solved in linear time in the size of the input~\cite{horn-linear}.

\cproblem{Horn-SAT}
{A propositional formula in conjunctive normal form (CNF) with at most one positive literal per clause.}
{Is there a Boolean assignment for the variables such that in each clause at least one literal is true?} 

We cannot model this problem as $\Csp(\bB)$ for a finite signature structure;
however, note that a clause $\neg x_1 \vee \dots \vee \neg x_k \vee x_0$
is equivalent to 
$$ \exists y_1,\dots,y_{k-1} \;\big( (\neg x_1 \vee \neg x_2 \vee y_1) \wedge (\neg y_1 \vee \neg x_3 \vee y_2) \wedge \dots \wedge (\neg y_{k-1} \vee \neg x_{k} \vee x_0) \big )\; .$$

Hence, by introducing new variables, 
there is a straightforward reduction of Horn-SAT to the restriction 
of Horn-SAT where every clause has at most three literals. 
This restricted problem, which we call Horn-3SAT, can be formulated as $\Csp(\bB)$ for \begin{align*}
\bB =  \big (\{0,1\}; & \{(x,y,z) \mid (x \wedge y) \Rightarrow z\},
\{(x,y,z) \mid \neg x \vee \neg y \vee \neg z\}, \\
 & \{((x,y) \mid x \Rightarrow y\},
\{((x,y) \mid \neg x \vee \neg y\},
 \{0\},\{1\} \big).
 \end{align*}

\subsection{Precedence constraints in scheduling}
\label{ssect:and-or}
The following problem has been studied in~\cite{and-or-scheduling} in scheduling:
given is a finite set of variables $V$, and a finite set of \emph{and/or precedence constraints}, i.e., constraints
of the form 
\begin{align}
x_0 > x_1 \vee \dots \vee x_0 > x_k 
\label{eq:and-or}
\end{align}
for $x_0,x_1,\dots,x_k \in V$. 
The question is whether there exists an assignment 
$V \rightarrow {\mQ}$ (equivalently, we can replace $\mQ$ by $\mZ$, or any other 
infinite linearly ordered set).
%$\mN$, or $\mR$) such that all those constraints are satisfied. 

As in the case of Horn-SAT, we cannot directly model this problem as $\Csp(\bB)$ for
a finite signature structure $\bB$. However, note that
Formula~(\ref{eq:and-or}) is equivalent to 
\begin{align*}
\exists y_1,\dots,y_{k-1} \big( \; & (x_0 > x_1 \vee x_0 > y_1) \wedge \\
& (y_1 > x_2 \vee y_1 > y_2) \wedge \dots \wedge (y_{k-1} > x_{k-1} \vee y_{k-1} > x_k) \big ) \; .
\end{align*}

This shows that and/or precedence constraints can be translated into
conjunctions of constraints of the form $x_0 > x_1 \; \vee \; x_0 > x_2$
by introducing new existentially quantified variables. 
Hence, the problem whether 
a given set of and/or precedence constraints is satisfiable
reduces naturally to $\Csp(({\mathbb Q}; R^{\text{min}}))$
where $R^{\text{min}}$ is the ternary relation 
$\{(a,b,c) \; | \; a > b \vee a > c\}$. Note that $R^{\text{min}}$ holds on exactly
those triples $(a,b,c)$ where $a$ is larger than the minimum of $b$ and $c$.
The problem $\Csp(({\mathbb Q}; R^{\text{min}}))$ can be solved in polynomial time;
this is essentially due to~\cite{and-or-scheduling}.
For more expressive
constraint languages over ${\mathbb Q}$ that contain the
relation $R^{\min}$ and whose CSP can still be solved in polynomial
time, see Section~\ref{ssect:min} or Section~\ref{ssect:ll}.

\subsection{Ord-Horn constraints}\label{ssect:ord-horn}
In this section we work with first-order formulas over the signature $\{<\}$. 
We write $x \leq y$ as a shortcut for $(x < y) \vee (x=y)$ (recall our convention that equality is part of first-order logic).
A formula over the signature $\{<\}$ and with variables $V$ 
is called \emph{Ord-Horn} if
it is a conjunction of disjunctions of the form
$$(x_1 = y_1) \vee \dots \vee (x_k = y_k) \vee (x_0 R y_0)$$
where $x_0,x_1,\dots,x_k,y_0,y_1,\dots,y_k \in V$,
and $R$ is either $\leq$, $<$, $\neq$, or $=$.

\cproblem{Ord-Horn Satisfiability}
{A finite set of variables $V$, and a finite set of Ord-Horn formulas with variables from
$V$.}
{Is there an assignment $V \rightarrow {\mathbb Q}$ that satisfies all 
the given formulas over $(\mathbb Q;<)$?}%

%Again, it does not play a role whether we ask for satisfiability over $({\mathbb N}, <)$ instead of $({\mathbb Q};<)$.
Nebel and B\"urckert~\cite{Nebel} 
showed that Ord-Horn Satisfiability can be solved in polynomial time.
A relation $R \subseteq {\mathbb Q}^k$ 
is called \emph{Ord-Horn} if it is definable by an Ord-Horn formula over $({\mathbb Q}; <)$.
As in the case of Horn-SAT and of and/or precedence constraints, 
there are structures $\bB$ with finitely many Ord-Horn relations
such that all Ord-Horn relations have a primitive positive definition in $\bB$. It can be shown that the following structure has this property (see Chapter~\ref{chap:ecsp}).

\begin{align*}
\big ({\mathbb Q}; \leq, \neq, \{(x,y,u,v) \; | \; (x=y) \Rightarrow (u=v)\} \big)
\end{align*}

In Section~\ref{ssect:ll} 
we see that constraint languages that contain and/or precedence constraints
\emph{and} Ord-Horn constraints can still be solved in polynomial time.

\subsection{Ord-Horn interval constraints}
For some (binary) reducts $\bB$ of Allen's interval algebra 
 the problem $\Csp(\bB)$ can be solved
in polynomial time. The most important of these reducts 
is the class of \emph{Ord-Horn interval constraints}, which 
has been introduced by Nebel and B\"urkert~\cite{Nebel}.
It consists of all the relations $R$ of Allen's interval algebra
such that the relation 
$\big \{ (x,y,u,v) \; | \; ([x,y],[u,v]) \in R \big \}$ is Ord-Horn (see Section~\ref{ssect:ord-horn}).
Now it is not hard to see that satisfiability for Ord-Horn interval constraints has
a polynomial-time reduction to Ord-Horn satisfiability. This type of reduction will be 
studied in Section~\ref{sect:pseudo-var}.

\subsection{Linear program feasibility}\label{ssect:lp}
Linear Programming is a computational problem of outstanding theoretical and practical importance (see e.g.~\cite{Schrijver}).
It is known to be computationally
equivalent to the problem to decide whether a given set
of linear (non-strict) inequalities is \emph{feasible}, i.e., defines a non-empty set. 

\cproblem{Linear Program Feasibility}
{A finite set of variables $V$; a finite set of linear inequalities of the form $a_1x_1+ \dots+a_k x_k \leq a_0$ where $x_1,\dots,x_k \in V$ and $a_0,\dots,a_k$ are
rational numbers where numerator and denominator are represented in binary.}
{Does there exist an $x \in {\mathbb R}^{|V|}$ that satisfies all inequalities?}

Kachyian showed in~\cite{Khachiyan} that 
Linear Program Feasibility can be solved in polynomial time.
It is clearly not possible to formulate this problem as $\Csp(\bB)$ 
for a structure $\bB$ with a \emph{finite} relational signature.
However, we show below that it is polynomial-time equivalent
to $\Csp\big((\mR; \{(x,y,z) \; | \; x+y=z\},\{1\},\leq)\big)$.
For this, we need the following lemma.

\begin{lemma}[from~\cite{Essentially-convex}]\label{lem:pp-rational}
Let $n_0,\ldots,n_l \in {\mathbb Q}$ be arbitrary rational numbers.
Then the relation $\{(x_1,\ldots,x_l) \; | \; n_1x_1 + \ldots + n_l x_l = n_0\}$ is primitive positive definable 
in $(\mathbb R; \{(x,y,z) \; | \; x+y=z\},\{1\})$. Furthermore, the primitive positive formula that
defines the relation can be computed in polynomial time.
\end{lemma}

The idea to prove this is to use iterated doubling to define large
numbers with small primitive positive formulas.
By extending the previous result to inequalities, one can prove
the following.
%that $\Csp(\lingamma)$ and linear program feasibility are
%polynomial-time equivalent problems.

\begin{proposition}[from~\cite{Essentially-convex}]\label{prop:lp-pp}
The linear program feasibility problem for linear programs
is polynomial-time equivalent to $\Csp((\mathbb R; \{(x,y,z) \; | \; x+y=z\},\{1\},\leq))$.
\end{proposition}

%\subsection{Integer Program Feasibility}
%General problem in finitary formulation. 
%Linear equalities subcase (in P).
%Two variable fragment subcase. 

\subsection{The max-atoms problem}
In our list of problems from the literature that can be formulated
as $\Csp(\bB)$, we also want to include one problem in NP
where it is not known whether $\Csp(\bB)$ is in P or NP-hard.
Our problem is closely related to the following problem, 
which has been introduced in~\cite{Max-atoms} and, independently, in~\cite{and-or-scheduling}. 

\cproblem{The Max-Atoms Problem}
{A finite set of variables $V$; a finite set of constraints of the form $x_0 \leq \max(a_1x_1, \dots, a_k x_k)$ where $x_1,\dots,x_k \in V$ and $a_0,\dots,a_k$ (the \emph{coefficients}) are integers represented in binary.}
{Does there exist an $x \in {\mathbb Q}^{|V|}$ that satisfies all inequalities?}

It is known that the Max-atoms problem is computationally equivalent to mean-payoff games~\cite{and-or-scheduling}, and therefore
it is contained in NP $\cap$ coNP. It also follows that deciding the winner in Parity games and satisfiability of the propositional $\mu$-calculus can be reduced to the
max-atoms problem. Bezem, Nieuwenhuis and Rodr\'{\i}guez-Carbonell~\cite{Max-atoms} give an alternative proof that the problem is in NP $\cap$ coNP.
In the same paper, they also shown that a certain hypergraph reachability problem, and an intensively studied problem in max/+ algebra are equivalent to the max-atoms problem.
Moreover, they show that the problem is in P when the coefficients in the input are represented in unary. 

Similarly as for linear program feasibility, the Max-atoms problem
cannot be formulated as $\Csp(\bB)$ for a structure $\bB$ with
a finite relational signature. The problem we introduce instead is
\begin{align*}
\text{CSP}(({\mathbb Q}; \{(x,y) \; | \; y=x+1\}, \{(x,y) \; | \; y=2x\}, R_\leq^{\min}))
\end{align*}
where $R^{\min}_\leq = \{ (x,y,z) \; | \; x \geq y \vee x \geq z\}$ is a variant of 
the relation $R^{\min}$ from Section~\ref{ssect:and-or}.
The max-atoms problem can be reduced to this problem:
we replace
expressions of the form $x_i + a_i$ by a new variable $y_i$,
and add a primitive positive formula $\phi(x,y)$ that defines
$y_i=x_i+a_i$ and can be computed in polynomial time in the
input size of the max-atoms problem.
We do not know how to prove hardness for the CSP above, and rather think that the problem might well be in P.

\subsection{Unification}
Unification (and unification modulo equational theories) is an proper field 
in computational logic, and the complexity of the
unification problem has been studied in numerous variants~\cite{BaaderNipkow}.
Many unification problems can be viewed as $\Csp(\bB)$,
for an appropriate infinite structure $\bB$, as we will see in the following. We start with the most fundamental unification problem.

Let $\sigma:=\{f_1,\dots,f_k\}$ be a finite set function symbols,
and let $x$ be a variable symbol.
Then ${\mathcal F}(x)$ denotes the set of all terms that can
be constructed from $\tau$ and the variable $x$.
The \emph{unnested unification problem over $\tau$} is the following problem\footnote{This problem is
known to be equivalent to the standard unification problem
%~\cite{BaaderNipkow} 
where the input is a single equation $t_1 \approx t_2$ for `nested' terms $t_1,t_2 \in {\mathcal F}(x)$.}.

\cproblem{Unnested Unification Problem over $\tau$} 
{a finite set of variables $V$, and a finite set of `un-nested' term equations, i.e., expressions of the form $y_0 \approx f(y_1,\dots,y_k)$ for 
$y_0,y_1,\dots,y_k \in V$ and $f \in \tau$.}
{is there an assignment $s \colon V \rightarrow {\mathcal F}(x)$ 
such that for every expression $y_0=f(y_1,\dots,y_k)$ in the input we have $s(y_0) = f(s(y_1),\dots,s(y_k))$?}

For fixed $\tau$ as above,  
let ${\mathfrak T} = ({\mathcal F}(x); F_1,\dots, F_k)$ be the structure
where $F_i$ is the relation 
$\{(t_0,t_1,\dots,t_r) \in ({\mathcal F}(x))^{r+1} \; | \; t_0 = f_i(t_1,\dots,t_r)\}$ (here, $r$ is the arity of $f_i$). 
It is clear that the unnested unification problem over $\tau$ 
can be described as $\Csp({\mathfrak T})$. In a similar way, equational unification problems (see~\cite{BaaderNipkow}) can be viewed as CSPs.
 
\section{Overview}
This text develops the universal-algebraic approach 
for complexity analysis of constraint satisfaction problems with countably infinite \emph{$\omega$-categorical} templates. Parts of the corresponding 
theory follow or generalize the universal-algebraic approach 
for CSPs with finite
templates, whereas other parts are specific to infinite domains, 
such as the way in which we apply Ramsey theory.
We then present two 
complexity classification results that have been obtained using this approach: the classification of temporal CSPs in Chapter~\ref{chap:tcsp}, and Schaefer's theorem for graphs in Chapter~\ref{chap:schaefer}. 
We close with Chapter~\ref{chap:nodich} on classes of 
computational problems that probably do \emph{not} exhibit a complexity dichotomy. 

\subsection*{Publication Note}
My PhD-thesis~\cite{Bodirsky} also treated constraint satisfaction with $\omega$-categorical templates, and already hinted at the relevance of polymorphisms and universal algebra. But it is only here that we fully present the universal-algebraic approach and its applications for classification projects of large classes of infinite-domain constraint satisfaction problems. The self-contained presentation in this thesis is
collecting and re-combining results that have been fragmented over
various publications, and often come with new proofs.

Parts of the content of this thesis have been published by co-authors and myself in conferences or journals.
The example sections in Chapter~\ref{chap:intro} and Chapter~\ref{chap:examples} present CSPs studied in~\cite{phylo-long,BodirskyKutzAI,
Cores-journal,BodDalJournal,HornOrFull,BodHilsKrim,BodirskyWoelfl}.
Chapter~\ref{chap:mt} contains a new proof, 
to be published in the journal version of~\cite{BodHilsMartin-Journal}, of the main result in~\cite{Cores-journal}. 
Chapter~\ref{chap:algebra} covers original results from~\cite{BodirskyNesetrilJLC}, \cite{OligoClone},
and the survey~\cite{BodirskySurvey}. 
Many results in Chapter~\ref{chap:ramsey} are from~\cite{BPT-decidability-of-definability} and the survey~\cite{BP-reductsRamsey}. 

The classification for equality constraint satisfaction problems
has first been obtained in~\cite{ecsps}, but the proof presented
here is new and borrows results and ideas from~\cite{Maximal} 
and~\cite{RandomMinOps}. 
The classification for temporal constraint satisfaction problems 
in Chapter~\ref{chap:tcsp} is based on~\cite{tcsps-journal} and~\cite{ll},
with some additions from the survey~\cite{BP-reductsRamsey}.
Schaefer's theorem for graphs is based on~\cite{BodPin-Schaefer-Both}
and~\cite{RandomMinOps}, again with additions from~\cite{BP-reductsRamsey}
Finally, Chapter~\ref{chap:nodich} also contains some results from~\cite{BodirskyGrohe}.

\section{Uncovered Topics}
When choosing the material to be included in this thesis,
certain restrictive choices had to be made. We comment
on related lines of research or facets of the area that we had
to skip. 

\subsection{Infinite Signatures}
Several natural computational problems could be formulated in the form $\Csp(\bB)$ when we would allow that the structure $\bB$ has
a countably infinite signature. For example, we might want to view 
the feasibility problem
for linear programs (Section~\ref{ssect:lp}) as $\Csp(\bB)$ where
$\bB$ contains all relations of the form $\{(x_1,\dots,x_k) \; | \; a_1x_1 + \dots + a_kx_k \leq a_0\}$, for all rational numbers $a_0,a_1,\dots,a_k$. Indeed, several general results for constraint
satisfaction that we present in this thesis would carry over 
to infinite signatures with no problems.

If we wanted to extend the present definition of $\Csp(\bB)$ to
structures $\bB$ with an infinite signature, we are faced with the
difficulty to specify how the constraints in input instances
of $\Csp(\bB)$ are \emph{represented}. When $\bB$ has a finite
signature, this causes no problems, since we can fix any representation for the finite number of relation symbols; since $\bB$
is considered to be fixed, the precise choice of the representation
is irrelevant. When $\bB$ has an infinite signature, 
a good choice how to code the constraints in the input 
very much depends on the structure $\bB$. In the example
of linear programming feasibility, for instance, we 
might want to represent the constraint $a_1x_1 + \dots + a_kx_k \leq a_0$ by specifying the coefficients $a_0,a_1,\dots,a_k$ in binary.
%Note that by a pure cardinality argument, there must
%exist structures $\bB$ with an infinite domain and an infinite signature such that there is no way to code the relations 
%such that

Note that the issue of finite versus infinite constraint languages
is not specific to infinite domains, but becomes relevant already
for finite domains. Typically, for infinite constraint languages
over a finite domain each constraint in the input is represented
by listing all tuples of the corresponding relation in the constraint language. But this is not the only, and sometimes not even the most
natural way to represent the constraints. 
For instance for the Horn-SAT problem (see Section~\ref{ssect:horn-sat}), the most natural way to present the constraint is by writing 
them as conjunctions of Horn-clauses. 
In the general setting, several representations have been proposed, some of which are more concise than listing 
all tuples~\cite{ChenGrohe}, and some of which are
less concise~\cite{TruthTablesMarx}. 

It turns out that typically when a constraint satisfaction problem with an infinite
constraint language is computationally hard, 
then there is a finite set of relations in this language
such that the CSP for this sub-language is already NP-hard.
For infinite constraint languages over a finite domain, 
and when the constraints are represented by explicitly listing
all satisfying assignments to the variables of the constraint,
it has even been conjectured~\cite{BulatovJeavons} that this might be true in general; that is, when $\Csp(\bB)$ is NP-hard
under this representation, then $\bB$ has a finite signature reduct with a hard CSP. This conjecture is still open.
We also want to mention a conditional non-dichotomy result for infinite constraint
languages from~\cite{BodirskyGrohe}.

%BOOKTD: We would like to remark that there are infinite-domain structures $\bB$ with infinite signature such that even with the very natural representations of the relation symbols of $\bB$, an analogous conjecture is not true, unless P=NP. 
%The following example is due to Peter Jonsson ....

We have decided to keep in this thesis the focus on CSPs for \emph{finite} constraint languages, for the following reasons.
This allows to work with one and the same definition
of the computational problem $\Csp(\bB)$ for \emph{all}
infinite structures $\bB$ with a finite signature.
Moreover, for all of the algorithms presented in this thesis
it will be immediately clear under which input assumption they might
be generalized to deal with an infinite constraint language.
This does not prevent us from stating relevant mathematical facts in full
generality when they also hold for structures with an infinite signature;
only when it comes to statements about CSP$(\bB)$ we insist that $\bB$ has finite relational signature.

\subsection{Complexity classes below P}
Besides the mentioned progress on the dichotomy conjecture
for finite domain CSPs, there has been considerable research
activity to localize the exact complexity of CSPs inside the 
complexity class P, or with respect to definability in certain logics.
By \emph{definability} of $\Csp(\bB)$ we mean that
there exists a sentence $\Phi$ is some logic (typically extensions of first-order logic and restrictions of least fixed point logics) such that
$\bA \models \Phi$ if and only if $\bA$ homomorphically
maps to $\bB$ (that is, in this case it is most natural to consider 
the definitions of the CSP presented in Section~\ref{sect:homo}
and in Section~\ref{sect:snp}).

One motivation for studying computational complexity within P is the question
whether it is possible to solve problems faster in parallel models of
computation. Another motivation, in particular for definability
of CSPs in certain logics, is the goal to better understand
the scope of existing algorithmic techniques to solve CSPs 
(such as Datalog, or restrictions of Datalog). %, see Section~\ref{ssect:datalog}).

In this line of research, the computational complexity of $\Csp(\bB)$ 
has been completely classified when $\bB$
is a two-element structure~\cite{AllenderSchaefer}.
Each problem in this class is complete for one of the
complexity classes NP, P, $\oplus L$, NL, L, and AC$^0$ under
AC$^0$ isomorphisms.
For general finite domains, several universal-algebraic conditions are known that
imply hardness for various complexity classes~\cite{AtseriasBulatovDawar,LaroseTesson}.
Concerning definability of CSPs, there are precise characterizations
 of those CSPs that are definable by a first-order sentence~\cite{LLT,Atserias,Rossman08}. 
Moreover, if $\Csp(\bB)$ is not first-order definable, 
then it is L-complete under AC$^0$-reductions~\cite{LaroseTesson} (also see~\cite{LinearDatalog,EgriLaroseTessonLogspace}). %, and definable in symmetric Datalog, a fragment of linear Datalog~\cite{EgriLaroseTessonLogspace}.
For infinite domain constraint satisfaction, it appears that 
there are no general results about localizing the complexity of CSPs within the complexity class P yet. 
However, we would like to remark that already in some concrete and model-theoretically well-behaved structures $\bB$
the precise complexity of $\Csp(\bB)$ within P is open. We give one example. 

\begin{example}
Consider the problem $$\Csp(({\mathbb Q}; \neq, \{(x,y,z) \; | \: (x=y \Rightarrow y \leq z) \wedge x \leq y)) \; .$$
This problem is NL-hard since there is an easy reduction from directed reachability to this problem, and directed
reachability is an NL-complete problem. By careful inspection of Tarjan's linear-time algorithm for strongly connected
components~\cite{TarjanSCCs} we see that the problem can be solved in linear time. However, the precise complexity
of this problem is not known; it might be that the problem is contained in NL, but it might also be P-hard. 
\end{example}

% BOOKTD \subsection{Left-hand Side Restrictions}
% This becomes even more relevant when we have a Datalog Section
% (in which case the thesis also covers a left-hand-side restriction
% result.)

\subsection{Quantified CSPs}
Let $\bB$ be a structure with a finite relational signature. 
Then the \emph{quantified constraint satisfaction problem for $\bB$}, denoted $\Qcsp(\bB)$, is the computational problem to decide for a given first-order sentence $\phi$
in prenex normal form and
without disjunction and negation symbols whether $\phi$ is true in $\bB$. 
The difference of $\Qcsp(\bB)$ from $\Csp(\bB)$ as we have presented it in Section~\ref{sect:homo} is that \emph{universal quantification} is permitted in the input sentences $\phi$. 

The additional expressiveness often comes at the prize of higher computational complexity;
whereas for finite structures $\bB$, the CSP for $\bB$ is always in NP,
there are finite structures $\bB$ where $\Qcsp(\bB)$ is PSPACE-complete.
But quite surprisingly, several constraint languages with a polynomial-time tractable
CSP also have a polynomial-time QCSP. This is for instance the case for 2SAT~\cite{AspvallPlassTarjan} (see Example~\ref{expl:sat}), or for Horn-3SAT~\cite{QHorn} 
(see Section~\ref{ssect:horn-sat}). Similarly, it can be shown that the temporal constraint languages presented in Section~\ref{ssect:min} and Section~\ref{ssect:mx} are not only tractable for the CSP, but also for the QCSP.
These are attractive results, since they assert that we can solve 
an even more expressive computational problem than the CSP for the same constraint
language without loosing polynomial-time tractability.
From a methodological point of view, we remark that the universal-algebraic approach
can also be applied to study the complexity of the QCSP~\cite{qcsp}; 
as in the case of the CSP,
the computational complexity of QCSP$(\bB)$ is captured by the \emph{polymorphisms} of $\bB$ (see Chapter~\ref{chap:algebra}).
Classifications of the QCSP typically rely on the corresponding classification for the CSP.
In particular, any hardness result for the CSP immediately translates into a hardness result 
for the QCSP. Moreover, in the cases where the $\Csp(\bB)$ is tractable, the algorithmic
insight is often the starting point for further investigations of $\Qcsp(\bB)$.

However, complexity classifications for QCSPs are typically harder to obtain than the corresponding complexity classifications for CSPs. 
One of the reasons is that several relevant universal-algebraic facts require the assumption that the algebra be 
\emph{idempotent} (see Section~\ref{ssect:tractability}). The complexity of the QCSP, however, is not preserved by homomorphic 
equivalence, and when we study $\Qcsp(\bB)$ we can thus not pass to the core of $\bB$. 
Hence, we can in general not make the assumption that the polymorphism clone of
$\bB$ is idempotent. 

For $\Csp(\bB)$, a powerful way of proving NP-hardness is to give a primitive
positive interpretation (see Section~\ref{sect:pseudo-var})
of a Boolean template with a hard CSP. This is no longer possible for the QCSP. 
There are for example
3-element templates $\bB$ that are preserved by a semi-lattice operation (and hence
no hard Boolean CSP can be interpreted in $\bB$) where $\Qcsp(\bB)$ is PSPACE-complete~\cite{qcsp}.
Finally, we would like to mention that PSPACE-hardness proofs for the QCSP 
are often much harder than NP-hardness proofs for the CSP~\cite{qcsp,qecsps}.

From the above is not surprising that a full classification of the QCSP complexity 
for three-element structures is still open. Similarly, there is no classification of
the QCSP for the class of temporal constraint languages presented in Chapter~\ref{chap:tcsp}.
There are concrete temporal constraint languages where the QCSP is of unknown
computational complexity, for instance the QCSP for
$$\big({\mathbb Q}; \{(x,y,z) \; | \; x=y \Rightarrow y \geq z\}\big) \; .$$
For this problem, we do not know hardness for any complexity class above P, and do not know containment in any complexity class below PSPACE.

\subsection{Non $\omega$-categorical templates}
Most methods presented in this thesis crucially rely on the assumption 
that the constraint languages are $\omega$-categorical.
%, and various techniques are exactly
%applicable over the class of all $\omega$-categorical templates
%(characterization of first-order definability with automorphisms,
%existence of canonical Datalog program in appropriate template, etc
However, systematic complexity classification is also possible
for large classes of constraint languages that are not
$\omega$-categorical.

\vspace{.2cm}
\subsubsection{Distance CSPs}
\label{ssect:succ}
The structure $({\mathbb Z}; \suc)$ 
of the integers with the sucessor relation $\suc = \{(x,y) \; | \; x=y+1\}$
 constitutes one of the simplest
infinite structures with a finite signature that is not $\omega$-categorical.
Structures with a first-order definition in $({\mathbb Z}; \suc)$ are particularly well-behaved from a model-theoretic perspective: 
they are strongly minimal~\cite{Marker,Hodges}, and therefore uncountably categorical (but usually not $\omega$-categorical). 
In~\cite{BodDalMarPin}, the complexity of $\Csp(\bB)$ has been studied when $\bB$ is first-order
definable in $({\mathbb Z}; \suc)$. 
 As an example, consider the directed graph with vertex set
${\mathbb Z}$ which has an edge between $x$ and $y$
if the difference between $x$ and $y$ is either $1$ or $3$.
This graph can be viewed as the structure $({\mathbb Z}; R_{\{1,3\}})$
where $R_{\{1,3\}} = \{(x,y) \; | \; x-y \in \{1,3\} \}$,  which has a first-order definition over $({\mathbb Z}; \suc)$ since $R_{\{1,3\}}(x,y)$ iff
$$\suc(x,y) \vee \exists u,v \, (\suc(x,u) \wedge \suc(u,v) \wedge \suc(v,y))\; .$$
Another example is the undirected graph with vertex set ${\mathbb Z}$
where two integers $x,y$ are linked if the \emph{distance} between $x$ and $y$ is one or two.

The corresponding class of CSPs contains many natural
combinatorial problems. For instance, the CSP for the structure
$({\mathbb Z}; R_{\{1,3\}})$ is the computational problem to
label the vertices of a given directed graph $G$ such that if $(x,y)$
is an arc in $G$, then the difference between the label for $x$
and the label for $y$ is one or three. 
It follows from the results in~\cite{BodDalMarPin} that this problem is in P. 
The CSP for the undirected graph $\big({\mathbb Z}; \big \{(x,y) \; \big | \; |x-y| \in \{1,2\} \big\} \big)$ mentioned above is exactly the 3-coloring problem, and therefore NP-complete.
In general, those problems
have the flavor of assignment problems where the task is to
map the variables to integers such that various given constraints on differences and distances (and Boolean combinations thereof) are satisfied.
Therefore, CSPs whose template is definable over $({\mathbb Z}; \suc)$ are called
\emph{distance CSPs}.

The complexity classification for distance CSPs presented in~\cite{BodDalMarPin} is incomplete in two respects. 
First, it might be the case that a structure $\bB$ with a first-order definition in $({\mathbb Z}; \suc)$ 
has a finite core $\bA$. Those cores have the property that they have a single orbit (see Section~\ref{ssect:expansion}); this follows
easily from the fact that the structure $({\mathbb Z}; \suc)$ also has only one orbit.
But the dichotomy conjecture for finite domain CSPs has not yet been established for the particular case of 
finite templates that have only one orbit. 
The second point in which the complexity classification for distance CSPs given in~\cite{BodDalMarPin} is incomplete is that
it only studies templates $\bB$
that are \emph{locally finite}, an additional finiteness condition, defined as follows.
A graph is called \emph{locally finite} if every vertex is contained in a finite number of edges; a relational structure is called locally finite if its Gaifman graph (definition given in Section~\ref{sect:structs}) is locally finite.
The method that is applied in~\cite{BodDalMarPin} is relational, and not universal-algebraic.

\vspace{.2cm}
\subsubsection{Tractable Expansions of Linear Program Feasibility}
Linear Programming is a computational problem of outstanding theoretical and practical importance. As we have seen, it is computationally
equivalent to the CSP we have presented in Section~\ref{ssect:lp}.
It seems to be interesting to investigate how far the constraint language of all
linear inequalities can be expanded such that the corresponding
constraint satisfaction problem remains polynomial-time solvable.
It has been shown in~\cite{Essentially-convex} that every \emph{first-order expansion} of linear programming 
is contained in a class called Horn-DLR from~\cite{JonssonBaeckstroem} and polynomial-time tractable, 
or otherwise NP-hard.

An important class of relations over the real numbers that generalizes the class
of relations defined by linear inequalities is the class of all
\emph{semi-algebraic} relations, i.e., relations with a first-order definition in $(\mathbb R; +,*)$. By the fundamental theorem of Tarski and Seidenberg it is known that a relation $S \subseteq {\mathbb R}^n$
is semi-algebraic if and only if it has a quantifier-free first-order definition in $(\mathbb R; +,*,0,1,\leq)$. Geometrically, we 
can view semi-algebraic sets as unions of intersections of the solution
sets of strict and non-strict polynomial inequalities.
The classification of the computational complexity of CSPs for real-valued semi-algebraic constraint languages is an ambitious research project, and has important
links to semidefinite programming: every semidefinite representable set 
is semi-algebraic and \emph{convex}.
Surprisingly, there are many fundamental questions in this area that
are wide open, for instance the complexity of semi-linear programming feasibility (see e.g.\ Section 6.4.4 
in~\cite{SDP-Handbook}, or~\cite{Ramana}), 
or the conjecture that all convex semi-algebraic
set are \emph{semidefinite representable}~\cite{HeltonNie}, i.e., 
primitive positive definable over the structure that has as its relations
all the solution spaces of semi-definite programs.
These important questions from real algebraic geometry
are out of the scope of this thesis.

% other prelims: o-notation -- omit
\chapter{Preliminaries in Logic}
\label{chap:logic}
\begin{center}
\includegraphics[scale=.7]{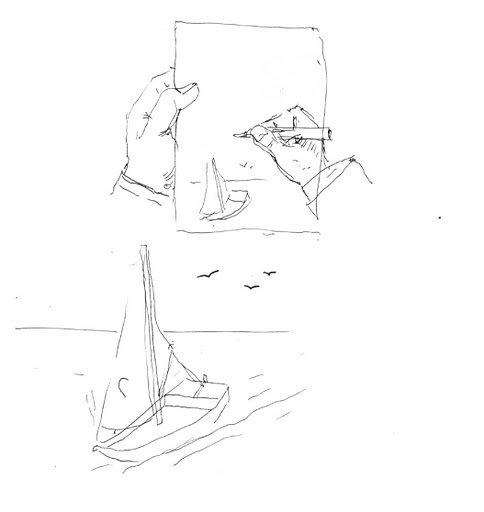}
\end{center}

This chapter collects some basic terminology and facts from logic.
The notation mostly follows Hodges' text book~\cite{Hodges}, so many readers may safely skip this chapter; they can consult it later, if needed, for particular concepts that we introduce here. 
% Later-TD: selection of $k$-th element from a tuple by []

\tocless\section{Structures}
\label{sect:structs}
In Section~\ref{sect:homo} we have already
defined \emph{relational structures}; we now give the general
definition of structures that might also contain functions, since
we need those later.
One occasion where we need functions rather than relations
is in Chapter~\ref{chap:algebra} when we consider algebras 
(by which we mean structures with a purely functional signature) 
that arise from the set of polymorphisms of a structure.
Moreover, several templates are most naturally defined over a structure having a functional signature, see e.g.~Section~\ref{ssect:mt-set-constraints}.
Most definitions go parallel for functional and relational signatures, so we give them together in this section.

A \emph{signature} $\tau$ is a set of relation and function symbols, each
equipped with an arity.
A \emph{$\tau$-structure} \bA\ is a set $A$ 
(the \emph{domain} of \bA) together with 
a relation $R^\bA \subseteq A^k$ for each $k$-ary 
relation symbol in $\tau$ and a function $f^\bA\colon A^k \rightarrow A$ for each $k$-ary function symbol in $\tau$; here we allow the case $k=0$ to model constant symbols.  
Unless stated otherwise, $A,B,C,\dots$ denote the domains of the structures
\bA, \bB, \bC, \dots, respectively. 
We sometimes write $(A; R^\bA_1,R^\bA_2,\dots,f^\bA_1,f^\bA_2,\dots)$ for the relational
structure $\bA$ with relations $R_1^\bA,R_2^\bA,\dots$ and
functions $f_1^\bA,f_2^\bA,\dots$
When there is no danger of confusion, we use the same symbol for a
function and its function symbol, 
and for a relation and its relation symbol. 
We say that a structure is 
infinite if its domain is infinite. 
The most important special cases of structures that appear in this
thesis are \emph{relational structures}, that is, structures with a purely
relational signature, and \emph{algebras}, 
that is, structures with a purely functional signature. Algebras with
domain $A,B,C,\dots$ are typically denoted by $\bf A$, $\bf B$, $\bf C$, \dots \\

\begin{example} \label{expl:groups}
A \emph{group} is an algebra ${\bf G}$ with a binary function symbol $\cdot$ for composition, a unary function symbol 
$^{-1}$ for the inverse, and a constant $e$ for the identity element of ${\bf G}$,
satisfying the sentences $\forall x,y,z. \, x \cdot (y \cdot z) = (x \cdot y) \cdot z$,
$\forall x. \, x \cdot x^{-1} = e$, $\forall x. \, e \cdot x = x$, and $\forall x. \, x \cdot e = x$.
In this signature,
the subgroups of ${\bf G}$ are precisely the subalgebras of ${\bf G}$ as defined below.
We typically omit the function symbol $\cdot$ and write $fg$ for the product of elements $f,g$ of $\bf G$. 
Such groups will also be called  \emph{abstract groups} to distinguish them from \emph{permutation groups}; 
a permutation group (over a set $X$) 
is a set of permutations of $X$ closed under composition and inverse, and containing the identity. 
\end{example}

When working with function symbols, it is sometimes convenient to work with
\emph{multi-sorted} structures, where we have distinguished unary predicates,
called \emph{sorts}, that define a partition of the domain,
and where function symbols might only be defined
on some of the sorts (that is, the function symbols might not be defined on
some of the elements). We are sloppy with the formal details since
they can always be worked out easily. In all our applications,
the multi-sorted structures will in fact be two-sorted, in which case
we denote them by $(\bA,\bB)$ --- here one sort induces the
structure $\bA$,
and the other sort induces the structure $\bB$. 

\vspace{.2cm}
\subsection{Expansions and reducts}
Let $\sigma,\tau$ be signatures with $\sigma \subseteq \tau$.
When $\bA$ is a $\sigma$-structure and $\bB$ is a $\tau$-structure,
both with the same domain, such that $R^\bA = R^\bB$ for all relations $R \in \sigma$ and
$f^\bA = f^\bB$ for all functions and constants $f \in \sigma$,
then $\bA$ is called a \emph{reduct} of $\bB$, and $\bB$ is called an
\emph{expansion} of $\bA$. An expansion $\bB$ of $\bA$ is called \emph{first-order}
if all new relations in $\bB$ are first-order definable over $\bA$.
A structure $\bA$ is called a \emph{finite reduct} of $\bB$
if $\bA$ is a reduct of $\bB$ with a finite signature.
We also write $(\bA,R)$ (and, similarly, $(\bA,f)$) for the expansion of $\bA$ by a new relation $R$ (a new function or constant $f$, respectively).

If $\bA$ is a $\tau$-structure and $(a_i)_{i \in I}$ a sequence of elements of $A$ indexed by $I$, then $(\bA; (a_i)_{i \in I})$ is the natural $(\tau \cup \{c_i | i \in I\})$-expansion of $\bA$ with $|I|$ new constants, where $c_i$ is interpreted by $a_i$ for all $i \in I$.

\subsection{Extensions and substructures}
A $\tau$-structure $\bA$ is a \emph{substructure} of a $\tau$-structure $\bB$ iff 
\begin{itemize}
\item $A \subseteq B$,
\item for each $R \in \tau$ and for all tuples $\bar a$ from $A$, $\bar a \in R^\bA$ iff $\bar a \in R^\bB$, and
\item for each $f \in \tau$ we have that $f^\bA(\bar a) = f^\bB(\bar a)$.
\end{itemize} 
In this case, we also say that $\bB$ is an \emph{extension} of
$\bA$. %, and write $\bA \subseteq \bB$. I DON'T THINK THIS IS EVER USED!
Substructures $\bA$ of $\bB$ and 
extensions $\bB$ of $\bA$ are called \emph{proper} if the domains of $\bA$ and $\bB$ are distinct. 
%Let $\langle b_\alpha \rangle_{\alpha <|B|}$ 
%well-order the elements of $\bB$. $\bA$ is an \emph{elementary extension} of $\bB$, denoted $\bB \preceq \bA$, if it is an extension and, for each first-order (fo) $\tau \cup \{c_\alpha : \alpha <|B|\}$-sentence $\phi$, $(\bB,\langle b_\alpha \rangle_{\alpha <|B|}) \models \phi$ iff $(\bA,\langle b_\alpha \rangle_{\alpha <|B|}) \models \phi$.  
%If $\bA$ and $\bB$ are structures, $A \subseteq B$,
%and the inclusion map from $A$ to $B$ is an embedding, then
%we say that $\bA$ is a \emph{substructure} of $\bB$.
Note that for every subset $S$ of the domain of $\bB$ 
there is a unique smallest substructure of $\bB$
whose domain contains $S$, which is called the \emph{substructure of 
$\bB$ generated by $S$}, and which is denoted by $\bB[S]$.
We say that $\bB$ is \emph{finitely generated} if $\bB = \bB[S]$ for a finite set $S$ of elements. 
% Need this for the definition of homogeneity
% in the presence of function symbols. Later also
% in the algebraic parts.
A \emph{subalgebra} of an algebra ${\bf B}$ (induced by $S$) is 
simply a substructure of ${\bf B}$ (generated by $S$) -- recall that
we have defined algebras as functions with a purely functional 
signature.

% ALL THIS HAS ALREADY BEEN INTRODUCED (and it is only for
% relational structures)
%The \emph{union} of two structures $\bA,\bB$ with relational signature $\tau$ is the $\tau$-structure
%$\bC = \bA\cup\bB$ with 
%domain $A \cup B$, and
%relations
%$R^{\bC}=R^{\bA}\cup R^{\bB}$ for all $R\in\tau$.
%The intersection $\bA\cap\bB$ is defined similarly.  A \emph{disjoint
%  union} of $\bA$ and $\bB$ is the union of isomorphic copies of
%$\bA$ and $\bB$ with disjoint domains. 
%As disjoint unions are unique up
%to isomorphism, we usually speak of \emph{the} disjoint union of $\bA$ and
%$\bB$. 
%The disjoint union of a set of $\tau$-structures $\cal C$ is defined
%analogously (and the disjoint union of an empty set of structures
%is the $\tau$-structure with empty domain).
%A relational structure is called \emph{connected} if it is not the disjoint
%union of two nonempty structures. 

The following is a concept that we only define for relational structures.

\begin{definition}\label{def:gaifman}
The Gaifman-graph of a relational structure $\bB$
with domain $B$ is the following undirected graph: the vertex set is
$B$, and there is an edge between distinct elements $x,y \in B$ when
there is a tuple in one of the relations of $\bB$ that has both $x$
and $y$ as entries. 
\end{definition}

A relational structure $\bB$ is readily seen to be connected 
(in the sense of Section~\ref{sect:homo}) if and only if its Gaifman
graph is connected (in the usual graph-theoretic sense).

\subsection{Products}\label{ssect:product}
Let $\bA$ and $\bB$ be two structures with domain $A$ and $B$, and the same signature $\tau$.
Then the (\emph{direct}, or \emph{categorical})  \emph{product} $\bC = \bA \times \bB$ 
is the $\tau$-structure with domain $A \times B$,
which has for each $k$-ary $R \in \tau$ 
the relation that contains a tuple $((a_1,b_1),\dots,(a_k,b_k))$
if and only if $R(a_1,\dots,a_k)$ holds in $\bA$ and $R(b_1,\dots,b_k)$
holds in $\bB$. For each $k$-ary $f \in \tau$ the structure $\bC$ has
the operation that maps $((a_1,b_1),\dots,(a_k,b_k))$ to 
$(f(a_1,\dots,a_k), f(b_1,\dots,b_k))$. 
The direct product $\bA \times \bA$ is also denoted by $\bA^2$,
and the $k$-fold product $\bA \times \cdots \times \bA$, defined analogously, by $\bA^k$.

% Need the following for varieties, decomposition of non-omega-cat permutation groups into transitive constituents, 
 
We generalize the definition of products in the obvious way to infinite products. 
For a sequence of $\tau$-structures $(\bA_i)_{i \in I}$, the direct product $\bB  = \prod_{i \in I} \bA_i$ is the $\tau$-structure on the domain $\prod_{i \in I} A_i$ such that 
for $R \in \tau$ of arity $k$
$$((a^1_i)_{i \in I},\ldots,(a^k_i)_{i \in I}) \in R^\bB \text{ iff } 
(a^1_i,\ldots,a^k_i) \in R^{\bA_i} \text{ for each } i \in I \; , $$
and for $f \in \tau$ of arity $k$, we have
$$f^\bB((a^1_i)_{i \in I},\dots,(a^k_i)_{i \in I}) 
= (f^{\bA_i}(a^1_i,\dots,a^k_i))_{i \in I} \; .$$

\tocless\section{Mappings}
Throughout the text, we use the following conventions.
When $f \colon A \rightarrow B$ is a function, and $S$ is a subset of $A$, then
$f(S)$ denotes the set $\{f(s) \; | \; s \in S \} \subseteq B$. 
When $t=(t_1,\dots,t_k)$ is a $k$-tuple of elements of $B$,
then $f(t)$ denotes the tuple $(f(t_1),\dots,f(t_k))$. 
Moreover, we use the same convention for higher-ary functions
$f \colon B^m \rightarrow B$: when $t^1,\dots,t^m$ are $k$-tuples of elements of $B$, then
$f(t^1,\dots,t^m)$ denotes the $k$-tuple $(f(t^1_1,\dots,t^m_1),\dots,f(t^1_k,\dots,t^m_k))$
(that is, the $k$-tuple is computed componentwise). 

In the following, let $\bA$ be a $\tau$-structure with domain $A$ and
$\bB$ a $\tau$-structure with domain $B$.
A \emph{homomorphism} $h$ from $\bA$
to $\bB$ is a mapping from
$A$ to $B$ that \emph{preserves} each function and each relation
for the symbols in $\tau$; that is, 
\begin{itemize}
\item if $(a_1,\dots,a_k)$ is in
$R^\bA$, then $(h(a_1),\dots,h(a_k))$ must
be in $R^\bB$;
\item 
$f^\bB(h(a_1),\dots,h(a_k))=h(f^\bA(a_1,\dots,a_k))$.
\end{itemize}
When $\fA,\fB$ are algebras with the same signature and domain $A,B$, respectively,
and $f$ is a homomorphism from $\fA$ to $\fB$, then $f(A)$ induces a subalgebra of $\fB$, and this subalgebra is called a
\emph{homomorphic image} of $\fA$. 

When a mapping $h$ preserves a relation $R$, we also say that $R$ is \emph{invariant}
under $h$. If $h$ does not preserve $R$, we also say that $h$ \emph{violates}
$R$. A homomorphism from $\bA$ to $\bB$
is called a \emph{strong homomorphism} 
if it also preserves the complements of the relations from $\bA$.
Injective strong homomorphisms are called \emph{embeddings}. 
Surjective embeddings are called isomorphisms.
A homomorphism from a substructure of $\bA$ to $\bB$ is called
a \emph{partial homomorphism} from $\bA$ to $\bB$.
An embedding from a substructure of $\bA$ into $\bB$ is called
a \emph{partial isomorphism} between $\bA$ and $\bB$.

Homomorphisms and isomorphisms from $\bB$ to itself are called \emph{endomorphisms} and \emph{automorphisms}, respectively.
Structures where the identity is the only automorphism are called
\emph{rigid}.  When $f \colon A \rightarrow B$ and $g \colon B \rightarrow C$, then
$g \circ f$ denotes the composed function $x \mapsto g(f(x))$. 
Clearly, the composition of two homomorphisms (embeddings, automorphisms) is again a homomorphism (embedding, automorphism). Let $\Aut(\bA)$ and $\End(\bA)$ be the sets of automorphisms and endomorphisms, respectively, of $\bA$. The set $\Aut(\bA)$ can be viewed as a group, and $\End(\bA)$ as a monoid with respect to composition;
more on that can be found in Section~\ref{sect:galois} and Section~\ref{ssect:endos}.

\vspace{.2cm}
\tocless\section{Formulas and Theories}
\label{sect:formulas}
We assume familiarity with basic concepts of classical first-order logic; 
see for example~\cite{EbbinghausFlumThomas}. 
In particular, we will use the concepts of
\emph{conjunctive normal form (CNF)}, 
%\emph{quantifier prefix}, \emph{prenex normal form}, % I have checked:
% after this section, this is not used!
%\emph{quantifier-block}, also not used
\emph{free and bound variables}, \emph{terms} and \emph{subterms}, 
\emph{clauses}, \emph{(positive and negative) literals}, 
and \emph{atomic formulas}.

We always allow the first-order formula $x=y$ (for equality) and $\bot$ (for `false'), independently of the signature, 
unless stated otherwise.
A formula without free variables will be called a \emph{sentence}.
A \emph{(first-order) theory} is a set of (first-order) sentences.
A structure $\bB$ is a \emph{model} of a sentence $\phi$ (or a theory $T$) if $\phi$ (all sentences in $T$, respectively) holds true in $\bB$;
in this case we write $\bB \models \phi$ ($\bB \models T$).
The set of all first-order sentences that are true in a given structure
$\bB$ is called the \emph{first-order theory of $\bB$}, and denoted $\Th(\bB)$.
If a sentence or a theory has a model, we call it \emph{satisfiable}.
We state two basic facts that will be used later.

\vspace{.2cm}
\begin{theorem}[Compactness; see Theorem 5.1.1 in~\cite{Hodges}]
\label{thm:compactness}
Let $T$ be a first-order theory. If every finite subset of $T$ is
satisfiable then $T$ is satisfiable.
\end{theorem}

% Book-TD: sorts. Need this for application of Loewenheim-Skolem in Preservation Theorem Section (2 times there!),
% but also in omega-cat.tex to get the compactness generalization of lem:infinst

When $T$ is a theory and $\phi$ a sentence, 
we say that $T$ \emph{entails} $\phi$, in symbols $T \models \phi$,
if every model of $T$ satisfies $\phi$. 
Two theories $T_1$, $T_2$ are said to be \emph{equivalent} if $T_1 \models T_2$ and $T_2 \models T_1$. 

\begin{lemma}[see Lemma 2.3.2 in~\cite{Hodges}]\label{lem:constants}
Let $T$ be a first-order $\tau$-theory, and $\phi$ a first-order $\tau$-formula
with free variables $x_1,\dots,x_n$. Let $c_1,\dots,c_n$ be distinct constants
that are not in $\tau$. Then $ T \models \phi(c_1,\dots,c_k)$ if and only if
$T \models \forall x_1,\dots,x_n. \phi$. 
\end{lemma}

%When $\phi$ has free variables $x_1,\dots,x_n$, we say that $\phi$
%is \emph{satisfiable} if there exists a structure $\bB$ and elements $b_1,\dots,b_n$ from $\bB$ such that $\bB \models \phi(b_1,\dots,b_n)$.
%When $\bB$ is fixed, we then also say that $(b_1,\dots,b_n)$ satisfies $\phi$.

Let $\bB$ be a $\tau$-structure.
When $\phi$ is a first-order $\tau$-formula, 
and when $x_1,\dots,x_n$ is an ordered
list that enumerates all the free variables, then $\phi(x_1,\dots,x_n)$
\emph{defines} over $\bB$ the relation 
$\{(b_1,\dots,b_n) \; | \; \bB \models \phi(b_1,\dots,b_n) \}$.
When $\phi$ is a $\tau$-formula with free variables $x_1,\dots,x_n$,
and $h$ is a $k$-ary function
then $h$ \emph{preserves $\phi$}
if $h$ preserves the $n$-ary relation that is defined by $\phi(x_1,\dots,x_n)$.
We say that a structure $\bA$ is \emph{(first-order) definable} in $\bB$
if $\bA$ and $\bB$ have the same domain, and 
every relation from $\bA$ has a first-order definition in $\bB$.
Two structures $\bA, \bB$ are \emph{(first-order) interdefinable} if $\bA$
is definable in $\bB$ and vice versa. 

A first-order $\tau$-formula $\phi$ is said to be 
\begin{itemize}
\item \emph{quantifier-free} if it does not contain any quantifiers; that is, it is built from the logical connectives $\wedge,\vee,\neg$, the binary relation $=$, the (free) variables, and the symbols from $\tau$ only (also see Section~\ref{sect:qe});
\item \emph{in prenex normal form} if it is of the form 
$Q_1 x_1 \dots Q_n x_n. \psi$ where $Q_i \in \{\forall,\exists\}$ and $\psi$ is quantifier-free;
\item \emph{Horn} if it is written in conjunctive normal form
and every clause has at most one positive literal (those formulas appear e.g.~in Section~\ref{sect:horn});
\item \emph{positive quantifier-free} if $\phi$ is quantifier-free, and if in addition $\phi$ does not contain negation symbols $\neg$.
\item \emph{existential} if it is of the form
$\exists x_1,\dots,x_n. \; \psi$ where $\psi$ is quantifier-free (those formulas appear e.g.~in Section~\ref{sect:mc});
\item \emph{universal} if is of the form
$\forall x_1,\dots,x_n. \; \psi$ where $\psi$ is quantifier-free;
\item $\exists^+$ (\emph{existential positive}) if it existential and if the quantifier-free part of $\phi$ does not contain any negation symbols (those formulas appear e.g.~in Section~\ref{ssect:core-theories});
\item $\forall^-$ (\emph{universal negative}) if it is of the form
$\forall x_1,\dots,x_n. \; \neg \psi$ where $\psi$ is positive quantifier-free;
\item \emph{universal conjunctive} if it is universal and if the quantifier-free part of $\phi$ does not contain any negation or disjunction symbols (those formulas appear e.g.~in Section~\ref{sect:varieties});
\item $\forall \exists$ (\emph{forall-exists}) if it is of the form 
$\forall y_1,\dots,y_m. \; \psi$ where $\psi$ is existential (those formulas appear e.g.~in Section~\ref{sect:mc});
\item $\forall\exists^+$ (\emph{positively restricted forall-exists}) if it is of the form $\forall \bar y. \phi(\bar y)$, where $\phi(\bar y)$ is a positive boolean
combination of quantifier-free formulas and existential positive formulas (those formulas appear throughout Section~\ref{sect:mccore})
\item \emph{primitive positive} if it is of the form
$\exists x_1,\dots,x_n. \; \psi_1 \wedge \dots \wedge \psi_m$, where $\psi_1,\dots,\psi_m$ are atomic (they are of central importance in this thesis).
%$\forall\exists^+$ if it is a universally quantified positive boolean combination of existential positive formulas and negated atomic formulas.
\end{itemize}

We could have equivalently defined positively restricted forall-exists formulas as conjunctions of universally quantified 
disjunctions of primitive positive formulas and negated atomic formulas.
It is easy to see that every $\forall\exists^+$-formula can be re-written into such a formula.

Note that homomorphisms preserve all existential positive formulas.
An important property of primitive positive sentences $\phi$ is that $\bA \times \bB \models \phi$ iff $\bA \models \phi$ and $\bB \models \phi$.
Also note that 
partial isomorphisms preserve quantifier-free formulas,
embeddings preserve existential formulas, 
and isomorphisms preserve first-order formulas.
Embeddings that preserve all first-order formulas are called \emph{elementary}.
When $\bB$ is an extension of $\bA$ such that the identity map from 
$\bB$ to $\bA$ is an elementary embedding, we say that $\bB$ is an \emph{elementary extension} of $\bA$, and that $\bA$ is an \emph{elementary substructure} of $\bB$.

\begin{theorem}[L\"owenheim-Skolem; see Corollary 3.1.4 in~\cite{Hodges}]
\label{thm:LS}
% (so allgemein brauchen wir das gar nicht)
Let $\bA$ be a $\tau$-structure, $X$ a set of elements of $\bA$, 
and $\lambda$ a cardinal such that $|\tau|+|X| \leq \lambda \leq |A|$. 
Then $\bA$ has an elementary substructure $\bB$ of cardinality $\lambda$ with
$X \subseteq B$.
\end{theorem}

A first-order theory $T$ is said to be \emph{existential}
if all sentences
in $T$ are existential, and the set of all existential $\tau$-sentences that is true in a $\tau$-structure $\bB$ is called the \emph{existential theory of $\bB$}. Analogously, 
we define $\forall\exists^+$, $\forall\exists$, universal, existential positive, and universal negative theories. 

\tocless \section{Diagrams}
\label{sect:diagrams}
We need the concept of a \emph{diagram} of a structure,
in various variants. The idea is to transform a structure into a formula,
similarly as in the definition of the \emph{canonical query} given in Section~\ref{sect:csp-logical}. \\

\begin{definition}\label{def:diag}
Let $\bA$ be a $\tau$-structure 
so that in $\bA$ every element is named by a constant. Then 
\begin{itemize}
\item the set of all positive quantifier-free sentences
that hold on $\bA$ 
is denoted by $\text{diag}_+(\bA)$, 
\item the set of all quantifier-free sentences that hold on $\bA$ 
is denoted by $\text{diag}(\bA)$, 
\item the set of all universal negative sentences that hold on $\bA$, 
is denoted by $\text{diag}_{\forall^-}(\bA)$, 
\item the \emph{elementary diagram of $\bA$} is the set of all first-order sentences true in $\bA$,
and is denoted by $\text{diag}_{\text{fo}}(\bA)$.
%\item for \emph{finite} $\bA$, the formula whose variables are the elements of $\bA$, and which consists of a conjunction
%of the atomic formulas $R(a_1,\dots,a_k)$ for all $(a_1,\dots,a_k) \in R^\bA$, is denoted by $\text{diag}_=^0(\bA)$.
% Not so robust: the problem is that it is not so easy to name the free variables in this formula in a canonical way, when we want to use it. 
\end{itemize}
\end{definition}

The following is straightforward from the definitions.
% BOOKTD:  version with function symbols?

\begin{lemma}[Diagram lemma; Lemma 1.4.2.~in~\cite{Hodges}]\label{lem:diagrams}
Let $\bA$ and $\bB$ be relational $\tau$-structures, and let $\bA'$ be $(\tau \cup \rho)$-expansion of $\bA$ by
constant symbols $\rho$ such that every element of $\bA'$ is named by a constant.
Then the following are equivalent.
\begin{enumerate}
\item There is a $(\tau \cup \sigma)$-expansion $\bB'$ of $\bB$ such that 
$\bB' \models \text{diag}_+(\bA')$;
\item There is a homomorphism from $\bA$ to $\bB$.
\end{enumerate}
\end{lemma}

Diagrams are useful in the proof of the following elementary, but important lemma, which  has been called the \emph{existential amalgamation theorem} in~\cite{Hodges}.  
The proof is analogous to the proof of Lemma~\ref{lem:existential-positive-amalgamation} given in full detail below, and we therefore omit it. 
%(it can be found in~\cite{Hodges}).

\begin{proposition}[Theorem 5.4.1 in~\cite{Hodges}]\label{prop:existential-amalgamation}
Let $\bA$ and $\bB$ be $\tau$-structures with domain $A$ and $B$, respectively. 
Suppose that $\bar a$ lists a subset $S$ of $A$, and let $h \colon S \rightarrow B$ be a partial 
homomorphism from $\bA$ to $\bB$
such that every existential sentence true in $(\bB,h(\bar a))$ 
is also true in $(\bA,\bar a)$. Then there exists an elementary 
extension $\bC$ of $\bA$ and an embedding $g \colon \bB \rightarrow \bC$
such that $g(h(\bar a))=\bar a$. 
\end{proposition}

We will need several times a positive variant of Proposition~\ref{prop:existential-amalgamation}, which is explicitly given in~\cite{Hodges}, but without proof.

\begin{lemma}[Theorem 5.4.7 in~\cite{Hodges}]
\label{lem:existential-positive-amalgamation}
Let $\bA$ and $\bB$ be $\tau$-structures with domain $A$ and $B$, respectively.
Suppose that $\bar a$ lists a subset $S$ of $A$, and let $h \colon  S \rightarrow B$ be a partial homomorphism from $\bA$ to $\bB$ 
such that every existential positive sentence true in $(\bB,h(\bar a))$ 
is also true in $(\bA,\bar a)$. Then there exists an elementary 
extension $\bC$ of $\bA$ and a homomorphism 
$g \colon \bB \rightarrow \bC$
such that $g(h(\bar a))=\bar a$. 
\end{lemma}
\begin{proof}
Similar to the proof of 5.4.1 in~\cite{Hodges}. 
%Since $h$ is an embedding, we can replace $\bB$ 
%by an isomorphic copy and assume that $h$ is the identity on $\bar a$. 
Let $\sigma$ be a set of constant symbols 
and $\bA'$ be a $(\tau \cup \sigma)$-expansion of $\bA$
such that every element of $\bA'$ is denoted by some constant symbol from $\sigma$.
Let $\bB'$ be an expansion of $\bB$ by constant symbols such that
\begin{itemize}
\item every element of $\bB'$ is denoted by some constant, and 
\item if $c \in \sigma$ is such that $c^{\bA'} \in S$ in $\bA'$, then $c^{\bB'} = h(c^{\bA'})$. 
\end{itemize}
It suffices to show that the theory 
$T := \text{diag}_{\text{fo}}(\bA') \cup \text{diag}_{+}(\bB')$ has a model $\bC'$, since Lemma~\ref{lem:diagrams} then asserts the existence of a homomorphism from $\bB$ to the $\tau$-reduct $\bC$ of $\bC'$, which will be an elementary extension of $\bA$. Moreover, such a homomorphism must map $h(\bar a)$ to $\bar a$.
 
If $T$ has no model, then by the compactness theorem there is a $\tau$-formula $\phi$
such that $\phi(\bar c) \in \text{diag}_{+}(\bB')$ and $\bA \models \neg \exists \bar y. \phi(\bar y)$. Since $\exists \bar y. \phi(\bar y)$ is existential positive, 
the assumptions imply that 
$\bB \models \neg \exists \bar y. \phi(\bar a,\bar y)$.
This contradicts that
$\bB \models \phi(\bar c)$.
\end{proof}

We can now prove a generalization of the condition given in Proposition~\ref{prop:csp-equiv} from Section~\ref{sect:sat}
% in Chapter~\ref{chap:intro} 
that characterizes when two theories have the same CSP.

\begin{proposition}\label{prop:companions}
Let $T$ and $T'$ be $\tau$-theories. The following are equivalent.
\begin{enumerate}
\item Every model of $T$ has a homomorphism to a model of $T'$, and every model of $T'$ has a homomorphism to a model of $T$.
\item $T$ and $T'$ imply the same universal negative sentences.
\end{enumerate}
\end{proposition}
\begin{proof}
To prove the implication from $(1)$ to $(2)$, 
assume $(1)$, and let $\phi$ be a universal negative sentence implied by $T'$, and let $\bC$ be a model of $T$.
By $(1)$, there is a homomorphism from $\bC$ to a model $\bB$ of $T'$. 
Since $\neg \phi$ is equivalent to an existential positive sentence, it is preserved by homomorphisms,
and hence we have a contradiction to the assumption that $T' \models \phi$. 

For the implication from $(2)$ to $(1)$, assume $(2)$, and let $\bB$ be a model of $T$. 
Let $S$ be the existential positive theory of $\bB$. We claim that $S \cup T'$ is satisfiable.
If not, then by compactness (Theorem~\ref{thm:compactness}) there is some finite subset 
$\{\phi_1,
\dots,\phi_k\}$ of $S$ such that $T' \vdash (\neg \phi_1\vee \cdots \vee \neg \phi_k)$. 
The formula $\neg \phi_1 \vee \cdots \vee \neg \phi_k$ is equivalent to a universal negative sentence $\psi$,
and $T' \vdash \psi$, so by $(2)$ we have that $T \vdash \psi$, and hence $\bB \models \psi$.
We have reached a contradiction, since $\bB \models \phi_i$ for all $i \leq k$. 
So there indeed exists a model $\bA$ of $S \cup T'$. 
Lemma~\ref{lem:existential-positive-amalgamation}
applied to $\bA$ and $\bB$ for the empty sequence $\bar a$ gives a model $\bC$ of $T' \cup S$ and a homomorphism 
from $\bB$ to $\bC$.
\end{proof}

This indeed proves Proposition~\ref{prop:csp-equiv},
since theories that imply the same universal negative sentences have obviously the same CSP. 
It is now also easy to prove Proposition~\ref{prop:sat-csp} from Section~\ref{sect:sat}, characterizing those theories $T$ for which there exists a structure $\bB$ such that
$\Csp(T) = \Csp(\bB)$. We first show the following.

\begin{proposition}\label{prop:jhp}
For any satisfiable theory $T$, the following are equivalent. 
\begin{enumerate}
\item there exists a structure $\bB$ that satisfies an existential positive sentence $\phi$ if and only if
$T \cup \{\phi\}$ is satisfiable. 
\item $T$ has a model $\bB$ that satisfies every existential positive sentence $\phi$ where
$T \cup \{\phi\}$ is satisfiable. 
\item For all existential positive sentences $\phi_1$ and 
$\phi_2$, if $T \cup \{\phi_1\}$ is satisfiable and $T \cup \{\phi_2\}$ is satisfiable,
then $T \cup \{\phi_1, \phi_2\}$ is satisfiable as well. 
\item $T$ has the Joint Homomorphism Property (JHP -- confer Proposition~\ref{prop:sat-csp}).
%}, that is, 
%when $T$ has models $\bA$ and $\bB$, then it also has a model $\bC$ such that
%both $\bA$ and $\bB$ homomorphically map to $\bC$.  
\end{enumerate}
\end{proposition}
\begin{proof}
We prove $(1) \Leftrightarrow (2)$, $(2) \Leftrightarrow (3)$,
$(3) \Leftrightarrow (4)$. 
For the implication from $(1)$ to $(2)$, let $T'$ be the
universal negative theory of $\bB$. By assumption,
$T'$ and $T$ imply the same universal negative sentences,
and hence by Proposition~\ref{prop:companions} 
there is a homomorphism from $\bB$ to a model $\bC$ of $T$.
This model $\bC$ has the desired property: if $\phi$ is existential positive such that $T \cup \{\phi\}$ is satisfiable,
then $\bB$ satisfies $\phi$ and since homomorphisms preserve existential positive formulas, $\bC$ satisfies $\phi$ as well. 

The implication from $(2)$ to $(1)$ and the implication from $(2)$ to $(3)$ are obvious. 
To show that $(3)$ implies $(2)$, assume $(3)$. 
Let $P$ be the set of all existential positive sentences $\phi$
such that $T \cup \phi$ is satisfiable. By the assumption that $T$ is satisfiable, and by $(3)$, all finite subsets of $T \cup P$ are satisfiable, so by compactness
of first-order logic (Theorem~\ref{thm:compactness}) we have that $T \cup P$ has a model $\bB$.

Now we show that $(3)$ implies $(4)$. 
Assume $(3)$, and let $\tau$ be the signature of $T$. 
Let $\bA_1$ and $\bA_2$ be models of $T$. We have to show that there exists a model $\bB$ of $T$ that admits homomorphisms from $\bA_1$ and $\bA_2$. 
Let $\bA'_1$ and $\bA'_2$ be expansions of $\bA_1$ and $\bA_2$, respectively, where every element is denoted by a distinct constant symbol. Consider
the theory $T' := T \cup \text{diag}_+(\bA'_1) \cup \text{diag}_+(\bA'_2)$; we claim that $T'$ is satisfiable. 
By compactness (Theorem~\ref{thm:compactness}), it suffices to show that every finite subset $S$ of $T'$
is satisfiable. Let $S_1 := S \cap \text{diag}_+(\bA)$ and $S_2 := S \cap \text{diag}_+(\bA_2)$. By forming a finite conjunction, we see that $S_1$ and $S_2$ are logically
equivalent to single sentences $\phi_1$ and $\phi_2$, respectively. 
Certainly $T \cup \{\phi_1\}$ and $T \cup \{\phi_2\}$ are satisfiable since $\bA'_1$ and $\bA'_2$ are expansions of models of $T$ and therefore satisfy all sentences from $T$.
By $(3)$,  the theory $T \cup \{\phi_1, \phi_2\}$ is satisfiable as well. 
Therefore the claim is true, and there exists a model $\bB'$ of $T'$. 
Let $\bB$ be the $\tau$-reduct of $\bB'$.
Finally, Lemma~\ref{lem:diagrams} asserts the existence of a homomorphism from $\bA_1$ to $\bB$ and from $\bA_2$ to $\bB$, which proves $(4)$. 

For the implication from $(4)$ to $(3)$, suppose that $T$ has the JHP, and that
$\phi_1$ and $\phi_2$ are existential positive sentences such that $T \cup \{\phi_1\}$ has a model $\bA_1$ 
and $T \cup \{\phi_2\}$ has a model $\bA_2$. By $(4)$, there exists a model $\bB$ of $T$ such that $\bA_1$ and $\bA_2$ homomorphically
map to $\bB$. Then $\bB$ clearly satisfies $T \cup \{\phi_1,\phi_2\}$ since homomorphisms preserve existential positive 
sentences.
\end{proof}

Note that in the statement and the proof above, the phrase \emph{existential positive} can be used interchangeably with the phrase \emph{primitive positive}. 
With the additional assumption that $T$ has a finite relational signature, item $(1)$ in Proposition~\ref{prop:jhp}
becomes the statement that there exists a structure $\bB$ such that $\Csp(\bB)$
and $\Csp(T)$ are the same problem, so we indeed proved in particular Proposition~\ref{prop:sat-csp}.

\tocless\section{Chains and Direct Limits}
\label{sect:limits}
Chains and direct limits of sequences of $\tau$-structures are an important method to construct models of first-order theories, and will be used for instance in Section~\ref{sect:pres} and Section~\ref{sect:completion}. 

Let $({\mathfrak A}_i)_{i < \kappa}$ be a sequence 
of $\tau$-structures for a relational signature $\tau$.
Then $({\mathfrak A}_i)_{i < \kappa}$ is called a \emph{chain} if ${\mathfrak A}_i \subseteq {\mathfrak A}_j$ for all $i < j < \kappa$.
A chain is called an \emph{elementary chain} when for all $i,j < \kappa$, 
the extension $\bA_j$ of $\bA_i$ is elementary.

\begin{definition}
The \emph{union} of the chain 
$({\mathfrak A}_i)_{i < \gamma}$ is a $\tau$-structure $\mathfrak B$ 
defined as follows. The domain of $\mathfrak B$ is $\bigcup_{i \leq \gamma} A_i$, and for each relation symbol $R \in \tau$ we
put $\bar a \in R^{\mathfrak B}$ if $\bar a \in R^{{\mathfrak A}_i}$
for some (or all) ${\mathfrak A}_i$ containing $\bar a$.
\end{definition}

\begin{theorem}[Tarski-Vaught; Theorem~2.5.2 in~\cite{Hodges}]\label{thm:tarski-vaught}
Let $(\bA_i)_{i < \kappa}$ be an elementary chain. 
Then $\bigcup_{i < \kappa} \bA_i$ is an elementary extension of $\bA_i$ for each $i<\kappa$. 
\end{theorem}

We say that a formula $\phi$ is \emph{preserved in chains (of models of $T$)}
if for all chains $(\bA_i)_{0 \leq i<\kappa}$ of $\tau$-structures 
(where all the $\bA_i$ and $\bA := \bigcup_{i<\kappa} \bA_i$ are models of $T$), and every $\bar a \in A^n$ we have $\bA \models \phi(\bar a)$ whenever $\bA_i \models \phi(\bar a)$ for every $i < \kappa$.

\begin{proposition}[Theorem 2.4.4 in~\cite{Hodges}]\label{prop:inductive}
Every $\forall\exists$-formula is preserved in unions of chains.
\end{proposition}

Direct limits can be seen as a positive variant 
of the notion of a union of chains; we essentially replace embeddings in chains by homomorphisms. 
Let $\tau$ be a relational signature, 
and let $\bA_0,\bA_1, \dots$ be a sequence of  
$\tau$-structures such that there are homomorphisms $f_{ij} \colon \bA_i \rightarrow \bA_j$.
Those homomorphisms are called \emph{coherent} if $f_{jk} \circ f_{ij} = f_{ik}$ for every $i \leq j \leq k$. 

\begin{definition}\label{def:limit}
Let $\bA_0,\bA_1, \dots$ be a sequence of 
$\tau$-structures with coherent homomorphisms $f_{ij} \colon \bA_i \rightarrow \bA_j$. 
Then the \emph{direct limit} $\lim_{i<\omega} \bA_i$ is the
$\tau$-structure $\bA$ defined as follows. 
The domain $A$ of $\bA$ 
comprises the equivalence classes of the equivalence relation
$\sim$ defined on $\bigcup_{i < \omega} A_i$ by setting
$x_i \sim x_j$ for $x_i \in A_i, x_j \in A_j$ iff there is a $k$ such that
$f_{ik}(x_i) = f_{jk}(x_j)$.
Let $g_i \colon A_i \rightarrow A$ be the function that maps
$a \in A_i$ to the equivalence class of $a$ in $A$.
For $R \in \tau$ and a tuple $\bar a$ over $A$, define $\bA \models R(\bar a)$ iff there
is a $k$ and a tuple $\bar b$ over $A_k$ such that $\bA_k \models R(\bar b)$
and $\bar a = g_k(\bar b)$.
\end{definition}
Note that the definition of $\lim_{i<\omega} \bA_i$ also depends on the coherent family
$f_{ij}$, but this is left implicit and will be clear from the context.
Also note that $g_i$ defines a homomorphism from $\bA_i$ to $\bA$;  this function is called the
\emph{limit homomorphism} from $\bA_i$ to the direct limit $\bA$. Note that $g_i = g_j \circ f_{ij}$ for all $i<j<\omega$. 

We have seen that unions of chains preserve 
$\forall\exists$-formulas; the analogous
statement for direct limits is as follows.
We say that a first-order formula $\phi(x_1,\dots,x_n)$ is \emph{preserved in direct limits (of models of $T$)} if for all sequences  $(\bA_i)_{0 \leq i<\kappa}$ (where all the $\bA_i$ and $\bA := \lim_{i<\kappa} \bA_i$ are models of $T$), and every  $\bar a \in A^n$ we have $\bA \models \phi(\bar a)$ whenever
there is for every $i < \kappa$ an $n$-tuples $\bar a^i$ where the $j$-th
entry is a representative of the $j$-th entry of $\bar a$, and
$\bA_i \models \phi(\bar a^i)$. The following is Theorem 2.4.6 in~\cite{HodgesLong}; since it is given there without proof, 
we present it here for completeness.

% NOT STRONG ENOUGH;
%\begin{proposition}[see Theorem 2.4.6 in~\cite{HodgesLong}]
%\label{prop:direct-product-preservation}
%Let the sequence $\bA_1,\bA_2,\dots$ be as in Definition~\ref{def:limit}, and let $\bA$ be its direct limit. 
%If $\phi$
%is positively restricted $\forall\exists$
%such that $\bA_i \models \phi$ for all $i$,
%then $\bA \models \phi$.
%\end{proposition}

\begin{proposition}[Theorem 2.4.6 in~\cite{HodgesLong}]
\label{prop:direct-product-preservation}
Every $\forall\exists^+$-formula is preserved in direct limits of models of $T$.
\end{proposition}
\begin{proof}
Let $(\bA_i)_{i<\kappa}$ be a sequence of models of $T$
with coherent homomorphisms $h_{ij} \colon \bA_i \rightarrow \bA_j$, for $i,j < \kappa$, such that $\bA := \lim_{i<\kappa} \bA_i$ is a model of $T$.
Let $g_i$ be the limit homomorphism from $\bA_i$ to $\bA$.
We have to show that if $\bar a \in A^n$ is such that for all $i < \kappa$ there is $\bar a_i$ with 
$g_i(\bar a_i) = \bar a$ and 
$\bA_i \models \phi(\bar a_i)$, then $\bA \models \phi(\bar a)$.
Since $\phi$ is $\forall\exists^+$, we can assume that
$\phi(\bar x)$ is of the form $\forall \bar y. \, \phi'(\bar x,\bar y)$ where $\phi'$
is a disjunction of negated atomic formulas and existential positive
formulas. 
Suppose that $\phi'(\bar a,\bar b)$ is false in $\bA$ for 
some tuple $\bar b$ of elements of $\bA$.
Every disjunct $\psi$ of $\phi'(\bar a,\bar b)$ is
false in $\bA$. 
Then there exists an $i < \kappa$ such that all entries of $\bar b$ have representatives in $\bA_i$, and the negated atomic disjuncts of $\phi'$
are already false in $\bA_i$, by definition of direct limits.
Let $\bar b_i$ be a tuple of elements of $\bA_i$ 
where the $j$-th entry is a representative
of the $j$-th entry in $\bar b$. Since $\bA_i \models \phi(\bar a_i)$, there must exist a disjunct $\psi$ of $\phi'$
such that $\psi(\bar a_i,\bar b_i)$ holds in $\bA_i$.
The limit homomorphism $g_i$ maps $(\bar a_i,\bar b_i)$ to
$(\bar a,\bar b)$ and is a homomorphism from $\bA_i$ to $\bA$, and therefore preserves existential positive formulas, contradicting the assumption that $\phi'(\bar a,\bar b)$ is false in $\bA$. 
\end{proof}

\tocless \section{Types}
\label{sect:types}
% types
A set $\Phi$ of formulas with free variables $x_1,\dots,x_n$ 
is called \emph{satisfiable} over a structure $\bB$
if there are elements $b_1,\dots,b_n$ of $\bB$ 
such that for all sentences $\phi \in \Phi$ we have $\bB \models \phi(b_1,\dots,b_n)$.
We say that $\Phi$ is \emph{satisfiable} if there exists a structure $\bB$ such that
$\Phi$ is satisfiable over $\bB$.
For $n \geq 0$, an \emph{$n$-type} of a theory $T$ is a set $p$ 
of formulas with free variables $x_1,\dots,x_n$ such that $p \cup T$ 
is satisfiable.  An $n$-type $p$ of $T$ is \emph{maximal} if $T \cup p \cup \{\phi(x_1,\ldots,x_n)\}$ is unsatisfiable for any formula $\phi \notin p$.  The set of all complete $n$-types of $\Th(\bA)$
is denoted by $S^{\bA}_n$.  An $n$-type of a structure $\bB$ is an $n$-type of the first-order theory of $\bB$. 

% everything with pp and atomic types
In a similar manner, an \emph{existential positive $n$-type} (ep-$n$-type) of a theory $T$ is a set of existential positive formulas $p$ with free variables $x_1,\dots,x_n$ such that $p \cup T$ is satisfiable. A ep-$n$-type $p$ of $T$ is \emph{maximal} if $T \cup p \cup \{\phi(x_1,\ldots,x_n)\}$ is unsatisfiable for any existential positive formula $\phi \notin p$.  
A \emph{(ep-) $n$-type of a structure $\bA$} is a (ep-) $n$-type of the theory $\Th(\bA)$. 
When $S$ is the universal negative theory of $\bA$, then note that $p \cup \Th(\bA)$ is satisfiable if and only if $p \cup S$ is satisfiable; thus we could equivalently have defined ep-$n$-type with respect to the latter theory.

%\begin{definition}
When $p$ is an $n$-type, and $I \subseteq \{1,\dots,n\}$ with $|I|=k$, then the \emph{subtype of $p$ induced by $I$} is the 
$k$-type obtained from $p$ by existentially quantifying in all formulas in $p$ the variables $x_i$ for $i \in \{1,\dots,n\} \setminus I$,
and then renaming the variables in the resulting set of formulas to 
$x_1,\dots,x_k$ in a way that preserves the order on the indices of the variables. 
%\end{definition}

% realization
An $n$-type $p$ of $\bA$ is \emph{realized} in $\bA$ if there exist $a_1,\ldots,a_n \in A$ such that $\bA \models \phi(a_1,\ldots,a_n)$ for each $\phi \in p$. The set of all first-order formulas with free variables $x_1,\dots,x_n$ satisfied by
an $n$-tuple $\bar a = (a_1,\dots,a_n)$ in $\bA$ is a maximal type of $\bA$, and called the \emph{type of $\bar a$}, and denoted 
by $\tp^\bA(\bar a)$. 

%By Theorem~\ref{thm:LS}, the $n$-types of a structure $\bA$ are exactly the subsets of $\tp^{\bA'}(\bar a)$ for some elementary extension $\bA'$ of $\bA$. % do we need it?

% BRAUCHEN WIR fuer pp-homo extension
For an infinite cardinal $\kappa$, a structure $\bA$ is \emph{$\kappa$-saturated} if, 
for all $\beta < \kappa$ and expansions $\bA'$ of $\bA$ by at most $\beta$ constants, every $1$-type of $\bA'$ is realized 
in $\bA'$. We say that an infinite $\bA$ is \emph{saturated} when it is $|A|$-saturated.
Realization of pp-types and pp-$(\kappa)$-saturation are defined analogously. 

%Suppose $\bA$ is a finite structure over a finite signature with domain $A:=\{a_1,\ldots,a_s\}$. Let $\theta(x_1,\ldots,x_s)$ be the conjunction of the positive facts of $\bA$, where the variables $x_1,\ldots,x_s$ correspond to the elements $a_1,\ldots,a_s$. That is, $R(x_{\lambda_1},\ldots,x_{\lambda_k})$ appears as an atom in $\theta$ iff $(a_{\lambda_1},\ldots,a_{\lambda_k}) \in R^\bA$. Define the \emph{canonical query} $\Phi[\bA]$ of $\bA$ to be the pp-formula $\exists x_1 \ldots x_s. \theta(x_1,\ldots,x_s)$.
%Note that we arrive at an equivalent theory if we substitute existential positive for primitive positive in this definition. Further, a \emph{complete pp-$\tau$-theory} $T$ is a satisfiable set of pp- and negated pp-$\tau$-sentences such that, for all pp-$\tau$-sentences $\phi$, either $\phi$ or $\neg \phi$ is in $T$. 

\begin{theorem}[Corollary 8.2.2 in~\cite{Hodges}]
\label{thm:saturation}
Let $\tau$ be a signature and $\lambda \geq |\tau|$. 
Then every $\tau$-structure $\bB$ has an $\lambda^+$-saturated
elementary extension of cardinality $\leq |B|^\lambda$.
\end{theorem}

%Note that a structure that is $\kappa$-saturated is a fortiori pp-$\kappa$-saturated. 

\chapter{Model Theory}
\label{chap:mt}

\begin{center}
\includegraphics[scale=.5]{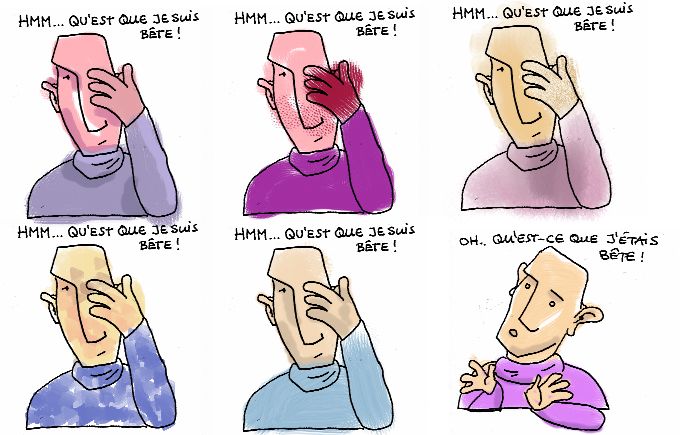}
\end{center}
\vspace{.4cm}

Hodges~\cite{Hodges} writes that \emph{``model theory is about the classification of mathematical structures, maps and sets by means of logical formulas''}. This text is about the computational complexity of constraint satisfaction problems for
infinite structures $\bB$ --- and since the constraint satisfaction problem of $\bB$ (and in particular its complexity)
is fully determined by the first-order theory of
$\bB$, it is not surprising that model theory has a great deal to
say about constraint satisfaction problems. 

Many important infinite-domain constraint satisfaction 
problems can be formulated with templates that 
are \emph{$\omega$-categorical}. 
The concept
of $\omega$-categoricity is of central importance in model theory,
and for reasons that will become clear in Section~\ref{sect:omegacat}
of this chapter, also in permutation group theory. 
From a model-theoretic perspective,
$\omega$-categoricity is a very strong assumption -- but still
 many problems that have been studied in the literature,
in particular constraint satisfaction problems for \emph{qualitative} reasoning formalisms in artificial intelligence\footnote{The question which reasoning formalisms in Artificial Intelligence should be called \emph{qualitative} has been the topic of scientific discussion~\cite{LigozatRenz04}. My own response to this question is: it is qualitative if and only if it can be formulated with an $\omega$-categorical template.}, can be formulated
as CSPs with $\omega$-categorical templates. We will also see that 
every 
connected monotone monadic SNP sentence 
(the corresponding problems have been called~\emph{forbidden patterns problems} 
and studied in~\cite{MadelaineStewartSicomp,Madelaine,MadelaineCP10}) describes a
constraint satisfaction problem of an $\omega$-categorical structure (Section~\ref{ssect:logics}). % of Chapter~\ref{chap:mt}).

In this section we present general results about $\omega$-categorical structures: for example how to construct them (Sections~\ref{sect:omegacat} and~\ref{sect:fraisse}, and~\ref{sect:completion}), and how to algebraically 
characterize syntactically restricted definability of relations over $\omega$-categorical structures 
%by first-order formulas (Section~\ref{sect:galois}) and 
%by formulas from various syntactic fragments of first-order logic 
(Sections~\ref{sect:galois}, \ref{sect:pres}, and~\ref{sect:mccore}).
%A powerful tool to construct $\omega$-categorical structures
%is \Fresse-amalgamation (Section~\ref{sect:fraisse}).
%It is important to note that the concept of $\omega$-categoricity
%is robust with respect to change of signature, since
% all structures that are first definable in an $\omega$-categorical
%structure $\mathfrak A$ are $\omega$-categorical as well; in fact,
%this also holds for structures that are interpretable in $\mathfrak A$  using a finite number of parameters from $\mathfrak A$ (Section~\ref{sect:omegacat}).
Section~\ref{sect:epc} gives an exact characterization
of those constraint satisfaction problems that can be formulated
with $\omega$-categorical templates.

There is already an excellent literature on $\omega$-categoricty:
most notably, the book by Cameron~\cite{Oligo}, the recent survey by Macpherson~\cite{MacphersonSurvey}, and the collection~\cite{Autos}.
Moreover, classical text-books on model theory, such as~\cite{Hodges,Marker,Keisler}, always treat $\omega$-categority, and use $\omega$-categorical structures as a rich source of examples.
The present chapter is different from those in that it focusses on 
techniques and facts that will be relevant for complexity classification
for the corresponding constraint satisfaction problems. It contains
many results that are not contained in any of the sources mentioned above (and which have been published in~\cite{BodHilsMartin-Journal,BodirskyNesetrilJLC,cores,tcsps-journal,ecsps,
BodPin-Schaefer-Both,
RandomMinOps,BodJunker}). 
Some parts can be derived from proofs in Hodges' book~\cite{Hodges} and
its original longer version~\cite{HodgesLong}; our
citation policy is to give the reference to the shorter version, 
whenever this is possible, since it is more widely accessible.
When this is not possible, we quote~\cite{HodgesLong}; 
for what is relevant in this text, the long version subsumes the short version.

\section{$\omega$-categorical Structures}
\label{sect:omegacat}

\begin{flushright}
\emph{``Every statement about all $\omega$-categorical structures is either trivial, or false.''} \\
(Martin Ziegler, 2005)
\end{flushright}
\vspace{.5cm}

A satisfiable first-order theory $T$ is called 
\emph{$\omega$-categorical} if all countable
models of $T$ are isomorphic.
A structure is called \emph{$\omega$-categorical} if its first-order
theory is $\omega$-categorical. Since almost all $\omega$-categorical structures that appear in this text will be countably infinite, we make
the convention that $\omega$-categorical structures are countably infinite. 
One of the first structures that were found to be
$\omega$-categorical (by Cantor~\cite{CantorDicht}) 
is the linear order of the rational numbers $({\mathbb Q};<)$, which we will use as a running example in this section. 
We will see many more examples of $\omega$-categorical structures in this section, in Section~\ref{sect:fraisse}, and in 
Chapter~\ref{chap:examples}. 
One of the standard approaches to verify that a structure is $\omega$-categorical is via a so-called \emph{back-and-forth argument}. 
To illustrate, we give the back-and-forth argument that shows that 
$({\mathbb Q};<)$ is $\omega$-categorical; much more 
about this important concept in model theory can be found in~\cite{Hodges,Pouzat}. 

\begin{proposition}\label{prop:cantor}
The structure $({\mathbb Q};<)$ is $\omega$-categorical.
\end{proposition}
\begin{proof}
Let $\bA$ be a countable model
of the first-order theory $T$ of $({\mathbb Q};<)$. 
It is easy to verify that $T$ contains (and, as this argument will show, 
is uniquely given by)
\begin{itemize}
\item $\exists x. \; x=x$ (no empty model)
\item $\forall x,y,z \; ((x < y \wedge y < z) \Rightarrow x < z)$ (transitivity)
\item $\forall x,y. \; \neg (x<x)$ (irreflexivity)
\item $\forall x,y \; (x < y \vee y < x \vee x=y)$ (totality)
\item $\forall x \, \exists y. \; x < y$ (no largest element)
\item $\forall x \, \exists y. \; y < x$ (no smallest element)
\item $\forall x,z \, \exists y \; (x < y \wedge y < z)$ (density).
\end{itemize}

An isomorphism between $\bA$ and $({\mathbb Q};<)$ can be defined inductively
as follows. 
Suppose that we have already defined $f$ on a finite subset
$S$ of ${\mathbb Q}$ and that $f$ is an embedding of the structure 
induced by $S$ in
$({\mathbb Q};<)$ into $\bA$. Since $<^{\bA}$ is dense and unbounded, we can extend $f$ to any other element of $\mathbb Q$ such that the extension is still an embedding from a substructure of $\mathbb Q$
into $\bA$ (\emph{going forth}). Symmetrically, for every element $v$ of $\bA$ we can find
an element $u \in {\mathbb Q}$ such that the extension of $f$
that maps $u$ to $v$ is also an embedding (\emph{going back}).
We now alternate between going forth and going back; when
going forth, we extend the domain of $f$ by the \emph{next} element
of $\mathbb Q$, according to some fixed enumeration of the elements
in $\mathbb Q$. When going back, we extend $f$ such that the
image of $A$ contains the \emph{next} element of $\bA$, according
to some fixed enumeration of the elements of $\bA$.
If we continue in this way, we have defined the value of $f$
on all elements of $\mathbb Q$. Moreover, $f$ will be surjective,
and an embedding, and hence an isomorphism between $\bA$ and $({\mathbb Q};<)$.
\end{proof}

A second important running example of this section is the \emph{random graph} 
$({\mathbb V}; E)$, 
which is a (simple and undirected) graph with a countably infinite set 
of vertices $\mathbb V$ that 
is defined uniquely up to isomorphism 
by the following \emph{extension property}: for all finite disjoint subsets $U,U'$ of $\mathbb V$ 
there exists a vertex $v \in {\mathbb V} \setminus (U \cup U')$ 
such that $v$ is adjacent to all vertices in $U$ and to no vertex in $U'$.
We will see in Section~\ref{sect:fraisse} that 
such a graph indeed exists (Theorem~\ref{thm:fraisse}).

\begin{proposition}\label{prop:random-graph-cat}
The random graph $({\mathbb V}; E)$ is $\omega$-categorical.
\end{proposition}
\begin{proof}
Note that the defining property of $({\mathbb V};E)$ given above is a first-order property;
a simple back-and-forth argument shows that every countably infinite graph with
this property is isomorphic to $({\mathbb V};E)$.
\end{proof}

\subsection*{The theorem of Ryll-Nardzewski}
There are many equivalent characterizations of $\omega$-categoricity; 
the most important one is in terms of the automorphism group of $\bB$.
In the following, let $\cG$ be a set of permutations of a set $X$. 
We say that $\cG$ is a \emph{permutation group}
if $\cG$ contains the identity $\id_X$, and for $g,f \in \cG$ the functions $x \mapsto g(f(x))$ and $x \mapsto g^{-1}(x)$ are also in $\cG$.
For $n \geq 1$ 
the \emph{orbit} of $(t_1,\dots,t_n) \in X^n$ under $\cG$ is the set $\{(\alpha(t_1),\dots,\alpha(t_n)) \; | \; \alpha \in \cG \}$.

\begin{definition}
A permutation group $\cG$ over a countably infinite set $X$ 
is \emph{oligomorphic} if
$\cG$ has only finitely many orbits 
of $n$-tuples for each $n\geq 1$.
\end{definition}

An accessible proof of the following 
theorem can be found in Hodges' book (Theorem 6.3.1 in~\cite{Hodges}).

\begin{theorem}[Engeler, Ryll-Nardzewski, Svenonius]\label{thm:ryll}
For a countably infinite structure $\bB$ with countable signature, 
the following are equivalent:
\begin{enumerate}
\item $\bB$ is $\omega$-categorical;
\item the automorphism group $\Aut(\bB)$ of $\bB$ is oligomorphic;
\item for each $n\geq 1$, there are finitely many inequivalent formulas with free variables $x_1,\dots,x_n$ over $\bB$;
\item for all $n \geq 1$, every set of $n$-tuples that is preserved
by all automorphisms of $\bB$ is first-order definable in $\bB$.
\end{enumerate}
\end{theorem}

The fourth of those conditions is missing in Theorem 6.3.1 of~\cite{Hodges}; but the implication from $(3)$ to $(4)$ follows from the
proof given there. 
Conversely, suppose that $\Aut(\bB)$ are infinitely many orbits of $n$-tuples, for some $n$. Then the union of any subset of the set of all orbits of $n$-tuples is preserved by all automorphisms of $\bB$; 
but there are only countably many first-order 
formulas over a countable language, so not all the invariant sets of $n$-tuples
can be first-order definable in $\bB$.

The second condition in Theorem~\ref{thm:ryll} provides another possibility to verify that a structure is $\omega$-categorical.
We again illustrate this with the structure $({\mathbb Q};<)$.
It is not difficult but a good exercise to verify that the orbit of an $n$-tuple $(t_1,\dots,t_n)$ from ${\mathbb Q}^n$ with respect to the automorphism group of $({\mathbb Q};<)$ is determined by the weak linear order induced by $(t_1,\dots,t_n)$ in $({\mathbb Q};<)$. We write \emph{weak linear order}, and not \emph{linear order}, because some of the elements $t_1,\dots,t_n$ might be equal. That is, a weak linear order is a total quasiorder.
There are less than $n^n$ such orders, and hence the automorphism group of $({\mathbb Q};<)$ has a finite number of orbits of $n$-tuples, for all $n \geq 1$.

Lemma~\ref{lem:infinst} below states a useful property that 
$\omega$-categorical structures have in common with finite structures, and is an easy consequence of K\"onigs tree lemma. 
%Instead of presenting the proof, we will
%derive it later from a more general theorem (Theorem~\ref{lem:omega-cat-compactness}) for $\omega$-categorical structures.

\begin{lemma}\label{lem:infinst}
Let $\bB$ be a finite or an infinite $\omega$-categorical 
structure with relational
signature $\tau$, and let $\bA$ be a countable $\tau$-structure. 
If there is no homomorphism (embedding)
%, strong homomorphism, injective homomorphism) 
from $\bA$ to $\bB$, then there is a finite substructure of $\bA$
that does not homomorphically map (embed) 
%, strongly homomorphically map, injectively homomorphically map) 
to $\bB$.
\end{lemma}

\begin{proof}
We present the proof for homomorphisms; the proof for embeddings is analogous.
Suppose every finite substructure of $\bA$ homomorphically maps
to $\bB$. We show the contraposition of the lemma,
and prove the existence of a homomorphism from $\bA$ to $\bB$.
Let $a_1, a_2, \dots$ be an enumeration of $\bA$.
We construct a rooted tree with finite out-degree, 
where each node lies on some level $n \geq 0$.
The nodes on level $n$
are equivalence classes of homomorphisms from the substructure of $\bA$
induced by $a_1, \dots, a_n$ to
$\bB$. Hence, there is only one vertex on level $0$, which will be the root
of the tree. Two such homomorphisms $f$ and $g$ 
are equivalent if there is an automorphism
$\alpha$ of $\bB$ such that $\alpha f = g$. Two equivalence classes
of homomorphisms on level $n$ and $n+1$ are adjacent, if there are 
representatives of the classes such that one is a restriction of the other.
Theorem~\ref{thm:ryll} asserts that $\bA$ has
only finitely many orbits of $k$-tuples, for all $k\geq 0$
(clearly, this also holds if $\bB$ is finite).
Hence, the constructed tree 
has finite out-degree. By assumption, there is
a homomorphism from 
the structure induced by $a_1, a_2, \dots, a_n$ to $\bB$ for all $n\geq 0$,
and hence the tree has vertices on all levels. 
K\"onig's lemma asserts the existence of an infinite
path in the tree, which can be used to inductively to
define a homomorphism $h$ from $\bA$ to $\bB$ as follows.

The restriction of $h$ to $\{a_1,\dots,a_n\}$
will be an element from the $n$-th node of the infinite path. Initially,
this is trivially true if $h$ is restricted to the empty set.
Suppose $h$ is already defined on $a_1,\dots,a_n$, for $n\geq 0$. By
construction of the infinite path, we find representatives $h_n$ and
$h_{n+1}$ of the $n$-th and the $(n+1)$-st element on the path such
that $h_n$ is a restriction of $h_{n+1}$. The inductive assumption
gives us an automorphism $\alpha$ of $\bA$ such that $\alpha(h_n(x))=h(x)$
for all $x \in \{a_1,\dots,a_n\}$. We set $h(a_{n+1})$ to be
$\alpha(h_{n+1}(a_{n+1}))$. The restriction of $h$ to $a_1,\dots,a_{n+1}$
will therefore be a member of the $(n+1)$-st element of the infinite
path. The operation $h$ defined in this way is indeed a 
homomorphism from $\bA$ to $\bB$. 
\end{proof}

The assumption that $\bA$ is countable is necessary in 
Lemma~\ref{lem:infinst}; consider for example $\bA := ({\mathbb R};<)$, which does not admit a homomorphism
to $\bB := ({\mathbb Q};<)$ for cardinality reasons, even though any finite substructure of $\bA$ does. 

% A more powerful lemma would be useful in:
% 0) proof of inv-pol
% 1) construction of binary injective polymorphism in chap:ecsp
% 2) construction of terms satisfying equations in Section Variety, several times.
% 3) construction of operations in temporal paper

\begin{corollary}
For any structure $\bC$, there is a finite structure $\bB$ with the same $\Csp$
as $\bC$ if and only if $\bC$ has a finite core.
\end{corollary}
\begin{proof}
If there exists a finite structure $\bB$ with the same $\Csp$ as $\bC$,
then every finite substructure of $\bC$ homomorphically maps to the core
$\bB'$ of $\bB$, and by Lemma~\ref{lem:infinst} there exists
a homomorphism from $\bC$ to the finite core structure $\bB'$ (which is unique up to isomorphism); since $\bB'$ also maps to $\bC$, it is a core of $\bC$. 
The converse is trivial. 
\end{proof}

\begin{corollary}\label{cor:homo-equiv}
Two countable $\omega$-categorical relational $\tau$-structures $\bA$ and $\bB$
have the same $\Csp$ if and only if there is a homomorphism from $\bA$ 
to $\bB$ and a homomorphism from $\bB$ to $\bA$.
\end{corollary}

%One example is clearly $(\mathbb R,<)$ and 
%$(\mathbb Q,<)$, which have the same CSP but clearly there
%is no homomorphism from $(\mathbb R,<)$ to $(\mathbb Q,<)$.
%An example of two countable structures where 
%Lemma~\ref{lem:homo-equiv} fails can be found in~\cite{Bauslaugh}.

Corollary~\ref{cor:homo-equiv} is false for general countable relational structures. Consider for example the structure $(\mathbb Z; \{(x,y) \; | \; y=x+1 \})$ --- the `infinite line', and the structure $(\mathbb N; \{(x,y) \; | \; y=x+1 \})$ --- the `infinite ray'. Clearly, these two structures give rise to the same CSP, but there is no homomorphism from the line to the ray.

%\ignore{
Several times we need variants of Lemma~\ref{lem:infinst} that can be proved in the same way. For instance, we can replace \emph{homomorphism} in the statement and the proof %by \emph{embedding},
by \emph{strong homomorphism}, or 
\emph{injective homomorphism}, or mappings satisfying universal identities such as
$\forall x,y. \; f(x,y) = f(y,x)$.
What is common for all those statements is that the respective property of the function can be expressed by \emph{universal first-order sentences}. 
We make this more precise and derive the following
generalization of Lemma~\ref{lem:infinst} based on 
the compactness theorem. 

%ACHTUNG: die folgende Variante ist FALSCH, selbst fuer omega-categorische abzaehlbare B (siehe z.b. nicht-existenz von majority polymorphism fuer Cliquen). 
%\begin{lemma}\label{lem:omega-cat-compactness}
%Let $\bB$ be a
%structure with relational
%signature $\tau$,
%and let $\sigma$ be a set of function symbols.
%Then for any universal $(\tau \cup \sigma)$-theory $T$ the following are equivalent.
%\begin{enumerate}
%\item $\bB$ can be expanded to a model of $T$;
%\item For every finite subset $F$ of the elements of $\bB$
%there is a $(\tau \cup \sigma)$-expansion $\bB'$ of $\bB$ 
%such that for every sentence 
%$\forall x_1,\dots,x_n. \, \phi$ from $T$ where $\phi$ is quantifier-free
% we have $\bB' \models \phi(a_1,\dots,a_n)$ for all $a_1,\dots,a_n \in F$.
%\end{enumerate}
%\end{lemma}

\begin{lemma}\label{lem:omega-cat-compactness}
Let $\bB$ be a countable $\omega$-categorical or finite
structure with countable relational
signature $\tau$, let $\bA$ be a countably infinite $\tau$-structure, 
and let $\sigma$ be a countable set of function symbols.
Then for any universal $(\tau \cup \sigma)$-theory $T$ the following are equivalent.
\begin{enumerate}
\item The two-sorted
structure $(\bA,\bB)$ has a $(\tau \cup \sigma)$-expansion that satisfies $T$
such that every $f \in \sigma$ denotes a function from $\bA$ to $\bB$.
\item For every finite induced substructure $\bC$ of $\bA$ 
the two-sorted structure $(\bC,\bB)$ has a $(\tau \cup \sigma)$-expansion 
that satisfies $T$ such that every $f \in \sigma$ denotes a function from $\bC$ to $\bB$.
\end{enumerate}
\end{lemma}
\begin{proof}
Any substructure of a model of a universal theory is again a model of the theory,
so (1) implies (2). Conversely, we prove the existence of a homomorphism from 
$\bA$ to $\bB$ by a compactness argument as follows.
Let $P$ be a unary relation symbol not contained in $\tau$. Let $\bA'$ be an expansion of $\bA$
by countably many constants such that every element of $\bA$ is named in $\bA'$ by 
a constant symbol; let $\tau'$ be the (countable) 
signature of $\bA'$. 
Let $D$ be the diagram of $\bA'$,
and let $S$ be a set of universal first-order sentences that 
\begin{itemize}
\item forces that all function symbols from $\sigma$ denote functions from 
the elements in $P$ to the elements that are not in $P$, and
\item expresses that the $\tau$-reduct of the structure induced by the elements not in $P$
has the same first-order theory as $\bB$.
\end{itemize}
We first prove that $D \cup S \cup T$ is satisfiable.
By compactness, it suffices to prove satisfiability of $D' \cup S \cup T$ for all finite subsets $D'$ of $D$.
Let $c_1,\dots,c_n$ be the constant symbols mentioned in $D'$.
Let $\bC'$ be the structure induced by $\{c_1,\dots,c_n\}$ in $\bA'$. 
Clearly, $\bC' \models D'$. Let $\bC$ be the $\tau$-reduct of $\bC'$. By assumption,
the two-sorted structure $(\bC,\bB)$ 
%with the two sorts $P$ and its complement 
can be expanded to a two-sorted $(\tau \cup \sigma)$-structure $\bD$
that satisfies $T$; this structure also satisfies $S$.
%BAUSTELLE HIER.
%where the first sort is $P$ and the second sort its complement; then $\bD$ 
%satisfies the sentences in $S$. 
When we additionally denote the constants
$c_1,\dots,c_n$ as in $\bA'$, 
then the expansion satisfies also
$D'$, and so we have found a model of $D' \cup S \cup T$.

By compactness, there exists an (infinite) model of $D \cup S \cup T$, 
and by Theorem~\ref{thm:LS} and since $\tau' \cup \sigma$ is countable there is also a countably infinite
model $\bM$ of $D \cup S \cup T$. Consider the substructure
of $\bM$ generated by the constants from $\tau'$,
and all the elements in $P_B$.
It can be checked that the resulting structure $\bM'$ still satisfies $D$ and $S$, and since
universal sentences are preserved by taking substructures, $\bM'$ also satisfies $T$.
Note that in $\bM'$, the elements from $P$ 
induce a copy of $\bA$, and the complement induces a structure that is isomorphic to $\bB$, since $\bB$
is finite or $\omega$-categorical. So the functions
of $\bM'$ denoted by the
function symbols from $\sigma$ provide the required $(\tau \cup \sigma)$-expansion
of $(\bA,\bB)$. 
\end{proof}

Lemma~\ref{lem:omega-cat-compactness} is indeed a generalization of Lemma~\ref{lem:infinst}: to make sure that $f$ is a homomorphism, $T$ contains for every relation symbol $R \in \tau$ the sentence $\forall \bar x. (R(\bar x) \Rightarrow R(f(\bar x))$.

\subsection*{First-Order Interpretations}
\label{ssect:oldnew}
%\paragraph{Interpretations.}
%A very flexible way to derive $\omega$-categorical structures
%are first-order interpretations. 
Many $\omega$-categorical structures can be derived from
other $\omega$-categorical structures via first-order interpretations
(our definition follows~\cite{Hodges}).

If $\delta(x_1,\dots,x_k)$ is a first-order $\tau$-formula 
with $k$ free variables $x_1,\dots,x_k$, 
and $\bA$ is a $\tau$-structure, we write $\delta(\bA^k)$ 
for the $k$-ary relation that is defined by $\delta$ on $\bA$.

\begin{definition}
A relational $\sigma$-structure $\bB$ has a \emph{(first-order) interpretation $I$} in a $\tau$-structure $\bA$ if there exists a natural number $d$, called the \emph{dimension} of $I$, and
\begin{itemize}
\item a $\tau$-formula $\delta_I(x_1, \dots, x_d)$ -- called \emph{domain formula},
\item for each atomic $\sigma$-formula $\phi(y_1,\dots,y_k)$ a $\tau$-formula $\phi_I(\overline x_1, \dots, \overline x_k)$ where the $\overline x_i$ denote disjoint $d$-tuples of distinct variables -- called the \emph{defining formulas},
\item a surjective map $h \colon \delta_I(\bA^d) \rightarrow B$ -- called \emph{coordinate map},
\end{itemize}
such that for all atomic $\sigma$-formulas $\phi$ and all tuples $\overline a_i \in \delta_I(\bA^d)$ 
\begin{align*}
\bB \models \phi(h(\overline a_1), \dots, h(\overline a_k)) \; 
& \Leftrightarrow \; 
\bA \models \phi_I(\overline a_1, \dots, \overline a_k) \; .
\end{align*}
\end{definition}
If the formulas $\delta_I$ and $\phi_I$ are all primitive positive, 
we say that the interpretation $I$ is \emph{primitive positive}.
Note that the dimension $d$, the set $S := \delta(\bA^k)$, and the coordinate map $h$ determine the defining formulas up to logical equivalence; hence, we sometimes denote an interpretation by $I = (d,S,h)$. Note that the kernel 
of $h$ coincides with the relation defined by $(x=y)_I$, for which we also write $=_I$, the defining formula for equality. 

\ignore{
We later use the following.
\begin{lemma}[Theorem 5.3.2 in~\cite{HodgesLong}]
Let $I = (d,S,h)$ be an interpretation of the $\tau_2$-structure
$\bB_2$ in the $\tau_1$-structure $\bB_1$.
Then for every first-order $\tau_2$-formula $\phi(x_1,\dots,x_k)$ 
there is a $\tau_1$-formula $\phi_I(x^1_1,\dots,x^{d}_1,\dots,x^1_k,\dots,x_k^{d})$
such that for all $a_1,\dots,a_k \in S)$
$$ \bB_2 \models \phi(h(a_1),\dots,h(a_k)) \;
\Leftrightarrow \; \bB_1 \models \phi_I(a_1,\dots,a_k) \; .$$
\end{lemma}
}

We say that $\bB$ is \emph{interpretable in} $\bA$ \emph{with finitely many parameters} 
% in interpretationen von strukturen mit unendlicher 
% signatur koennte man das auch tatsaechlich brauchen!
if there are $c_1,\dots,c_n \in A$ 
such that $\bB$ is interpretable in the expansion of $\bA$ by the singleton relations $\{c_i\}$ for all $1 \leq i \leq n$. 
A first-order \emph{definition} of one structure in another is in model
theory often the
special case of an interpretation $I$ where $=_I$ is simply the equality relation;
in this text, however,
we say that a structure $\bB$ is \emph{(first-order) definable} in $\bA$
if $\bB$ has a 1-dimensional interpretation $I$ in $\bB$ where $=_I$ is the equality relation and the domain formula $\delta_I$ is logically equivalent to true.

\begin{lemma}[see Theorem 7.3.8 in~\cite{HodgesLong}] %TODO what is it in the short version? e.g.~\cite{Hodges}]
\label{lem:interpret}
Let $\bA$ be an $\omega$-categorical structure. Then every
structure $\bB$ that is first-order interpretable in $\bA$ 
with finitely many parameters is $\omega$-categorical or finite.
\end{lemma}

Note that in particular all reducts of an $\omega$-categorical structure 
and all expansions of an $\omega$-categorical structure by finitely
many constants are again $\omega$-categorical.
%Also note that for finite $k$ the power structure $\mathfrak A^k$ (see Chapter~\ref{chap:logic}) is $\omega$-categorical when $\mathfrak A$ is $\omega$-categorical. % proof? Kommt eh spaeter in Galois section

\begin{example}\label{example:allen}
In Section~\ref{sect:csp-examples} %of Chapter~\ref{chap:intro}
we have described 
\emph{Allen's Interval Algebra} for temporal reasoning in Artificial Intelligence~\cite{Allen}, and the corresponding CSP.
Formally, it is easiest to describe the template $\mathfrak A$ 
for this $\Csp$ by a first-order
interpretation $I$ in $({\mathbb Q};<)$. The dimension of the
interpretation is two, and the domain formula $\delta_I(x,y)$ is
$x<y$. Hence, the elements of $\mathfrak A$ can indeed be viewed as non-empty closed intervals $[x,y]$ over $\mathbb Q$. The template $\mathfrak A$ contains 
for each inequivalent $\{<\}$-formula $\phi$ with four variables 
a binary relation $R$ such that
$(a_1,a_2,a_3,a_4)$ satisfies $\phi$ if and only if
$((a_1,a_2),(a_3,a_4)) \in R$. In particular, $\mathfrak A$ has relations
for equality of intervals, containment of intervals, and so forth.
By Lemma~\ref{lem:interpret}, $\mathfrak A$ is
$\omega$-categorical. \qed
\end{example}

\section{Fra\"{i}ss\'{e} Amalgamation}
\label{sect:fraisse}
% WITH ISOMORPHISM-CLOSED CLASSES 
% (simpler when working with examples, more intuitive)
% WITHOUT FUNCTION SYMBOLS
% (motivation for why no function symbols:
% not didactic to merge it with algebra stuff 
% do not really need it for examples
% The only important examples that come to my mind are the vector space, the Boolean algebra, and the join function for semi-linear orders. For for the atomless Boolean algebra
% we have other ways to construct it; moreover, we can conveniently
% quote Hodges!!
% would complicate the issue 

% 2/2012:
% But for a book, having functions here would be better. 
% Function symbols are the systematic approach to those omega-cat structs that are NOT reducts of homogeneous structures. Function symbols come into the game
% in a very natural way when looking at bi-interpretations and homogeneity.
% Atomless Boolean algebra is also very important. 

A versatile tool to construct $\omega$-categorical structures
is \Fresse-amalgamation. We present it here for the special
case of \emph{relational structures}; this is all that is needed in
the examples we are going to present. For a stronger version
of \Fresse-amalgamation for classes of structures that might involve
function symbols, see~\cite{Hodges}.

In the following, let $\tau$ be a countable relational signature.
The \emph{age} of a $\tau$-structure $\mathfrak A$ is the class of all finite $\tau$-structures 
that embed into $\mathfrak A$.
Let $\bB_1,\bB_2$ be $\tau$-structures such
that $\bA$ is an induced substructure of both $\bB_1$ and $\bB_2$ and all common elements of $\bB_1$ and $\bB_2$ are elements of $\bA$; note that in this case
$\bA = \bB_1 \cap \bB_2$.  Then
we call $\bB_1\cup \bB_2$ the \emph{free amalgam} of $\bB_1,\bB_2$ over $\bA$. More
generally, a $\tau$-structure $\bC$ is an \emph{amalgam of $\bB_1$ and $\bB_2$ over
  $\bA$} if for $i=1,2$ there are embeddings $f_i$ of $\bB_i$ to $\bC$ such that
$f_1(a)=f_2(a)$ for all $a \in \bA$.  
A \emph{strong amalgam} of $\bB_1,\bB_2$ over $\bA$ is an amalgam
of $\bB_1,\bB_2$ over $\bA$ where $f_1(B_1) \cap f_2(B_2) = f_1(A) (=f_2(A))$.

\begin{definition}\label{def:amalgamation-prop}
An isomorphism-closed class $\mathcal C$ of
$\tau$-structures has the \emph{amalgamation property} if for all
$\bA,\bB_1,\bB_2\in\mathcal C$ with $\bA=\bB_1\cap \bB_2$ 
there is a $\bC\in\mathcal C$ that is an amalgam of $\bB_1$ and $\bB_2$ over $A$. 
A class of finite
$\tau$-structures that contains at most countably many non-isomorphic structures,
has the amalgamation property, and is closed under
taking induced substructures and isomorphisms
is called an \emph{amalgamation class}.
\end{definition}

Analogously, an isomorphism-closed class $\mathcal C$ of
$\tau$-structures has the \emph{free amalgamation property} if for all
$\bA,\bB_1,\bB_2\in\mathcal C$ with $\bA=\bB_1\cap \bB_2$ the free amalgam of
$\bB_1,\bB_2$ over $\bA$ is in $\mathcal C$.
The \emph{strong} amalgamation property is defined analogously. 
Note that since we only look at relational structures here (and since we allow structures
to have an empty domain),
the amalgamation property of $\mathcal C$ implies the \emph{joint embedding property (JEP)} for $\mathcal C$, which says that for any two structures $\bB_1,\bB_2 \in \mathcal C$ there exists a structure $\bC \in \mathcal C$ that embeds both $\bB_1$ and $\bB_2$.

%Let $\cal C$ be a
%class of finitely generated structures that is closed under isomorphisms.
%We say that $\cal C$ has the 
%\begin{itemize}
%\item[HP] \emph{Hereditary property}
%if whenever $\bA \in \cal C$ and $\bB$ is a finitely generated substructure of 
%$\bA$ then  $\bB \in \cal C$.
%\item[JEP] \emph{Joint embedding property} if whenever $\bA,\bB \in \cal C$ then there is $\bC \in \cal C$ such that both $\bA \uplus \bB \in \bC$.
%\item[AP] \emph{Amalgamation property} if whenever $\bA,\bB_1,\bB_2 \in \cal C$ and $e_1: \bA \rightarrow \bB_1$ and $e_2: \bA \rightarrow \bB_2$
%are embeddings 
%there exists $\bC \in {\mathcal C}$ and embeddings $f_1: \bB_1 \rightarrow \bC$ and $f_2: \bB_2 \rightarrow \bC$ such that $f_1e_1=f_2e_2$. 
%\end{itemize}
%We say that a
%class of finitely generated structures $\cal C$ is an \emph{amalgamation class} if it is closed under isomorphisms, and has HP, JEP, and AP.

A structure $\bA$ is \emph{homogeneous} 
(sometimes also called \emph{ultra-homogeneous}~\cite{Hodges})
if every isomorphism between
finitely generated 
substructures of $\bA$ can be extended to an automorphism of $\bA$.

\begin{theorem}[\Fresse~\cite{OriginalFraisse,Fraisse}; see~\cite{Hodges}]\label{thm:fraisse}
Let $\tau$ be a countable relational signature and let 
$\cal C$ be an amalgamation class of $\tau$-structures. 
Then there is a homogeneous and at 
most countable $\tau$-structure $\bC$ whose age equals $\cal C$.
The structure $\bC$ is unique up to isomorphism, and called
the \emph{\Fresse-limit} of $\cal C$.
\end{theorem}

When $\cal C$ is a strong amalgamation class, then
the \Fresse-limit of $\cal C$ has a remarkable property. 
Let $\Gamma$ be a structure, and $A$
a finite set of elements of $\Gamma$. Then $\acl_\Gamma(A)$ denotes the model-theoretic algebraic closure of $A$ in $\Gamma$, i.e., the elements of $\Gamma$ that lie in finite sets that are first-order definable over $\Gamma$ 
with parameters from $A$.
In $\omega$-categorical structures, this is 
precisely the set of elements of $\Gamma$ that
lie in finite orbits in $\Aut(\Gamma)_{(A)}$. 
We say that a structure $\Gamma$ has 
\emph{no algebraicity} if $\acl_\Gamma(A)=A$ for all finite sets of parameters $A$.

\begin{theorem}[See (2.15) in~\cite{Oligo}]\label{thm:no-algebraicity}
A homogeneous $\omega$-categorical 
structure $\Gamma$ 
has no algebraicity if and only if the
age of $\Gamma$ has strong amalgamation.
\end{theorem}

\ignore{ % For book?!
Theorem~\ref{thm:fraisse} has an inverse.
For \emph{finite} structures $\bA$, $\bB_1$, $\bB_2$, this is Theorem~7.1.7 in~\cite{Hodges};
we need the statement also for countable $\bA$, $\bB_1$, and $\bB_2$ later, so we present the
simple proof. 

\begin{proposition}\label{prop:fraisse-inverse}
Let $\bC$ be a countable homogeneous $\tau$-structure. Then the age of $\bC$ is an amalgamation class. 
%In fact, the amalgam for any two
%countable substructures of $\bC$ over a finite structure embeds into $\bC$.
%In fact, when $\bB_1$
%and $\bB_2$ embed into $\bC$, and $\bA$ is a finite structure of both $\bB_1$ and $\bB_2$,
%then an amalgam of $\bB_1$ and $\bB_2$
%over $\bA$ also embeds into $\bC$.
% MORE PRACTICAL TO USE:
In fact, for arbitrary structures $\bA$, $\bB_1$
and $\bB_2$ that embed into $\bC$, 
if $\bA$ has embeddings $e_1,e_1$  
into $\bB_1$ and $\bB_2$, then there are
embeddings $f_1,f_2$ of $\bB_1$ and $\bB_2$ into $\bC$ such that $f_1(e_1(a)) = f_2(e_2(a))$ for all elements $a$ of $\bA$.
% OH, only see how to get this in the 
% omega-categorical context. 
\end{proposition}
\begin{proof}
Let $g_1$ and $g_2$ be the embeddings of $\bB_1$ and $\bB_2$ into $\bC$.
For $i \in \{1,2\}$, let $\bA_i$ be the structure induced by the image of $g_i \circ e_i$ in $\bC$. Then $g_2 \circ e_2 \circ e_1^{-1} \circ g_1^{-1}$ is an
isomorphism between the finite substructures $\bA_1$ and $\bA_2$ of $\bC$, and hence can be extended to
an automorphism $\alpha$ of $\bC$.
Then $f_1 = g_1$ and $f_2 = \alpha g_2$
have the required property.
\end{proof}
}

\begin{example}
Let $\cal C$ be the class of all linear orders.
Then $\cal C$ is clearly closed under isomorphisms and induced substructures, and has countably many isomorphism types. To show that it also has the amalgamation property,
let $\bB_1, \bB_2 \in \cal C$, and let $\bA$
be an induced substructure of both $\bB_1$ and $\bB_2$. 
Let $\bC$ be the free amalgam of $\bB_1$ and $\bB_2$ over $\bA$.
Then $\bC$ is an acyclic finite graph;
therefore, any depth-first traversal of $\bC$ leads to a linear ordering of the elements
that is an amalgam (but not a free amalgam) in $\mathcal C$ of $\bB_1$ and $\bB_2$
over $\bA$. It follows that $\mathcal C$ is an amalgamation class.
By homogeneity, the \Fresse-limit of $\mathcal C$ is unbounded and dense, and hence
isomorphic to $({\mathbb Q}; <)$ by Proposition~\ref{prop:cantor}. \qed
\end{example}

\begin{example}
Let $\cal C$ be the class of all finite partially ordered sets. Amalgamation can be shown by computing the transitive closure: when $\bC$ is the free amalgam of $\bB_1$ and $\bB_2$ over $\bA$, then the transitive
closure of $\bC$ gives an amalgam in $\mathcal C$. The \Fresse-limit of $\mathcal C$ is called the \emph{homogeneous universal partial order}.
\qed
\end{example}

\begin{example}\label{expl:random}
Let $\cal C$ be the class of all finite graphs.
It is even easier than in the previous examples to verify that $\cal C$ is an amalgamation class,
since here the free amalgam itself shows the amalgamation property. 
We can use homogeneity to verify that the \Fresse-limit of $\cal C$ satisfies
the defining property of the random graph $(\mV;E)$ (the existence of the random graph was left open in Section~\ref{sect:omegacat}). \qed
\end{example}

\begin{example}\label{expl:Henson}
Henson~\cite{Henson} used \Fresse\ limits to construct $2^\omega$ many
$\omega$-categorical directed graphs. A
\emph{tournament} is a directed graph without self-loops such that for
all pairs $x,y$ of distinct vertices exactly one of the pairs $(x,y)$, $(y,x)$
is an arc in the graph.  Note that for all classes $\cal N$ of finite
tournaments, $\Forb(\cal N)$ is an amalgamation class, because if $\bA_1$
and $\bA_2$ are directed graphs in $\Forb(\cal N)$ such that $\bA=\bA_1\cap
\bA_2$ is an induced substructure of both $\bA_1$ and $\bA_2$, then the free amalgam
$\bA_1\cup \bA_2$ is also in $\Forb(\cal N)$.  

Henson in his proof specified an infinite set $\cal T$ of 
tournaments $\bT_1,\bT_2,\dots$ with the property that
$\bT_i$ does not embed into $\bT_j$ if $i \neq j$; the set $\cal T$ will be described in Section~\ref{sect:coNP}.
Note that this property implies that for two distinct subsets ${\cal N}_1$ 
and ${\cal N}_2$ of $\cal T$ the two sets $\Forb({\cal N}_1)$
and $\Forb({\cal N}_2)$ are distinct as well. Since there
are $2^\omega$ many subsets of the infinite set $\cal T$,
there are also that many distinct homogeneous (and therefore $\omega$-categorical)
directed graphs; they are often referred to as \emph{Henson digraphs}. \qed
\end{example}

The structures from Example~\ref{expl:Henson} can be used to prove various negative results about homogeneous structures with finite signature,
for instance in Section~\ref{sect:decidability} % of Chapter~\ref{chap:ramsey}, 
and in Section~\ref{sect:coNP}. %of Chapter~\ref{chap:nodich}. 
A better behaved class of structures are homogeneous structures whose age is
\emph{finitely bounded} (this is the same terminology as in~\cite{MacphersonSurvey}).

\begin{definition}\label{def:finitely-bounded}
We say that a class $\C$ of finite 
relational $\tau$-structures (or a structure with age $\C$) is \emph{finitely bounded}
if $\tau$ is finite and there exists a finite set of finite $\tau$-structures $\N$ such
that $\C = \Forb(\N)$.
%for all finite $\tau$-structures $\bA$ we have $\bA \in \C$ 
%if and only if no structure from $\N$ embeds into $\bA$.
\end{definition}

\begin{proposition}
When $\bB$ is finitely bounded, then $\Csp(\bB)$ is in NP.
\end{proposition}
\begin{proof}
The problem $\Csp(\bB)$ is in monotone SNP (Section~\ref{sect:snp}).
\end{proof}

\Fresse's theorem can be used to construct 
$\omega$-categorical structures, because homogeneous structures with finite
relational signature are $\omega$-categorical. More generally, we have the following.

\begin{lemma}\label{lem:cat-via-homogen}
Let $\bC$ be a countably infinite homogeneous 
structure such that for each $k$ only a finite number of distinct $k$-ary relations
can be defined by atomic formulas. Then $\bC$ is $\omega$-categorical.
\end{lemma}
\begin{proof}
By homogeneity of $\bC$, the atomic formulas that hold on the elements of 
a tuple $t$ in $\bC$ determine the orbit of $t$ in $\Aut(\bC)$. Since there are
only finitely many inequivalent such atomic formulas, there are finitely many orbits of $k$-tuples in $\Aut(\bC)$. 
The claim follows by Theorem~\ref{thm:ryll}.
\end{proof}

%Note that Lemma~\ref{lem:cat-via-homogen}
%becomes false when we only require that $\bC$
%has for each $k$ only a finite number of distinct
%$k$-ary relations. % There should be a nice
% example using universal metric spaces, but
% it takes too long to describe it here.

It is sometimes convenient to define 
an $\omega$-categorical $\tau$-structure $\bB$ 
by specifying an amalgamation class $\cal C$ with a signature that
is larger than $\tau$ such that $\bB$ is a reduct of the
\Fresse-limit of $\cal C$. If the \Fresse-limit of $\cal C$ 
satisfies the condition of Lemma~\ref{lem:cat-via-homogen},
it will be $\omega$-categorical, and therefore
also all its reducts are $\omega$-categorical (Lemma~\ref{lem:interpret}). 
This method is for instance used in Section~\ref{ssect:c-relation}. 

This technique has also been used in~\cite{HubNes:Homomorphism} to give another proof
of a theorem due to Cherlin, Shelah, and Shi (Theorem~\ref{thm:universal}). The result appears in~\cite{CherlinShelahShi} for the special case where $\tau$ has a single binary relation denoting the edge relationship of undirected graphs. 
The statement for general relational signatures $\tau$ also follows from a result of~\cite{Covington}. 
The theorem of Cherlin, Shelah, and Shi will be useful in
Section~\ref{ssect:logics}.

Let $\cal N$ be a finite set of
finite structures with a finite relational signature $\tau$.
Recall that a $\tau$-structure $\bB$ is called \emph{$\cal N$-free} if there is no homomorphism from any structure in $\cal N$ to $\bB$.  A structure $\bA$ in a class of structures $\cal C$ is
called \emph{universal} for $\cal C$ if it contains all structures
in $\cal C$ as an induced substructure. Recall that a structure is \emph{connected} if it cannot be given as the disjoint union of non-empty structures. 

%%\begin{lemma}\label{lem:connected}
%Let $\cN$ be a set of structures with relational signature $\tau$.
%Then $\Forb(\cN)$ is closed under
%disjoint unions if and only if there is a set of \emph{connected}
%$\tau$-structures $\cN'$ such that $\Forb(\cN)=\Forb(\cN')$.
%\end{lemma}
%\begin{proof}
%Suppose that $\Forb(\cN)$ is closed under disjoint unions. 
%Then $\Forb(\cN) = \Forb(\cN')$ where $\cN'$ is the
%set of all connected components of structures from $\cN$.
%For the other direction, suppose that $\cN$ only contains connected structures, and let $\bA_1, \bA_2$ be structures from $\Forb(\cN)$.
%If there is a structure $\bC$ from $\cN$ that homomorphically maps
%to $\bA_1 \uplus$ 
%suppose that then it
%is clear that when there is no homomorphism from $\bA_1 \in \cN$
%into $\bB \in \Forb(\cN)$ and no homomorphism from $\bA_2$.
%(...)
%\end{proof}

% UNIQUENESS IN THE FOLLOWING THEOREM CAN ONLY BE STATED WHEN WE HAVE MODEL-COMPLETENESS.

\begin{theorem}[of ~\cite{CherlinShelahShi}; also see~\cite{HubNes:Homomorphism}]
\label{thm:universal}
Let $\cal N$ be a finite set of finite connected $\tau$-structures.
Then there is an $\omega$-categorical $\cal N$-free $\tau$-structure $\bB$
that is universal for the class of all countable $\cal N$-free structures. 
The structure $\bB$ can be expanded by finitely many primitive positive definable relations whose complement is existential positive definable so that the expanded structure is homogeneous. 
\end{theorem}

We want to remark that the structure $\bB$ 
from Theorem~\ref{thm:universal} is \emph{uniquely} (up to isomorphism)
given by the fact that it is $\mathcal N$-free, universal for the class of all finite $\mathcal N$-free graphs, and \emph{model-complete} (Theorem~\ref{thm:css-strong}); 
model-completeness will be treated in 
Section~\ref{sect:mc}.

\section{Oligomorphic Permutation Groups}
\label{sect:galois}
We have seen in Section~\ref{sect:omegacat} that
a structure is $\omega$-categorical if and only if its automorphism group is \emph{oligomorphic}, i.e., has
for each $n\geq 1$ only finitely many orbits of $n$-tuples.
This section describes this connection between logic
and permutation groups in more detail. 

\subsection{Closure}
\label{ssect:lc}
Automorphism groups of relational structures $\bB$
have the property that they are \emph{closed},
in the following sense. We write $S(B)$
for the set of all permutations of the set $B$.

\begin{definition}
\label{def:lc}
A set of permutations $\cP$ of a set $B$ is called \emph{closed (in $S(B)$)}
if $\cP$ contains all $\alpha \in S(B)$ with the property that 
for every finite subset $A$ of $B$ there
exists $\beta \in \cP$ such that $
\alpha x=\beta x$ for all $x \in A$. 
\end{definition}

%The following fact is well-known (see for instance \cite{Oligo}, Section 2.3 and 2.4).
%The proof of the following is straightforward, and can be found in~
% HMH cannot find the homogeneity part in Oligo book!

\begin{proposition}\label{prop:loc-clos-group}
Let $\cP$ be a set of permutations of some base set $B$. 
Then the following are equivalent.
\begin{enumerate}
\item $\cP$ is the automorphism group of a relational
structure;
\item $\cP$ is a closed permutation group; 
\item $\cP$ is the automorphism group of a homogeneous relational structure. 
\end{enumerate}
\end{proposition}

In the proof of this proposition, the following concept is useful.
When $\cF$ is a subset of $B \rightarrow B$, then $\Inv(\cF)$
denotes the set of all relations over $B$ that are preserved by all functions from $\cF$. A relational structure over the base set
$B$ whose relations are exactly the relations from $\Inv(\cF)$
is called a \emph{canonical structure}\footnote{Here, we slightly deviate from the definition given in~\cite{Oligo}, which only 
includes a $k$-ary relation for each orbit of $k$-tuples, for all $k$. The difference does not matter here, but becomes important in later sections.} 
for $\cF$.

\begin{proof}[Proof of Proposition~\ref{prop:loc-clos-group}]
For the implication from (1) to (2), let $\cP$ be the automorphism group of a relational structure $\bB$ with domain $B$, 
and let $\alpha \in \overline{\cP}$.
Then $\alpha$ must preserve all relations from $\bB$, because
if $\alpha$ violates a relation from $\bB$, then this
can be seen from the restriction of $\alpha$ to
a finite subset of the domain. Hence, $\alpha \in \cP$.

For the implication from $(2)$ to $(3)$, first note that canonical structures $\bB$ for ${\cP}$ are homogeneous:
when $i$ is an isomorphism between finite substructures of $\bB$, say $i$ has domain $\{a_1,\dots,a_n\}$, consider the relation $\{ (\alpha a_1,\dots, \alpha a_n) \; | \; \alpha \in \cP\}$. This relation is preserved by all operations in $\cP$ and hence belongs to the relations of $\bB$.
Thus, $i$ preserves this relation, and $(i(a_1),\dots,i(a_n))=(\alpha a_1,\dots,\alpha a_n)$ for some $\alpha \in \cP$. This shows that there is an automorphism of $\bB$ that extends $i$. In fact, since $\cP$ is closed,
this also shows that every automorphism of $\bB$ is from $\cP$.

The implication from (3) to (1) is trivial.
\end{proof}

\subsection{The Inv-Aut Galois Connection}
\label{ssect:inv-aut}
When $\bB$ is a relational structure, we denote by
$\langle \bB \rangle_{\fo}$ the set of all
relations that are first-order definable in $\bB$.
We will see in this section that the set
\begin{align*}
Ê\{ \langle \bB \rangle_{\fo} \, | 
\; \bB \text{ first-order definable in } \bCÊ\} \; ,
\end{align*}
partially ordered by inclusion, 
is closely connected to the set of all
closed permutation groups that contain the automorphisms of 
$\bC$, again partially ordered by inclusion; 
the connection is one-to-one when $\bC$ is $\omega$-categorical. 

Recall that the automorphism group of a relational structure $\bB$, i.e., the set of all automorphisms of
$\bB$, is denoted by $\Aut(\bB)$.
In the following it will be convenient to define the operator $\Aut$ 
also on sets $\R$ of relations over the same domain $B$,
in which case $\Aut({\R})$ 
denotes the set of all permutations $\alpha$ of $B$ such that $\alpha$ and
its inverse $\alpha^{-1}$ preserve all relations form $\mathcal R$. 

\begin{definition}\label{def:galois}
An \emph{(anti-tone) Galois connection} is a pair of
functions $F \colon U \rightarrow V$ and $G \colon V \rightarrow U$ between two posets $U$ and $V$, such that $v \leq F(u)$ if and only if $u \leq G(v)$
for all $u \in U, v \in V$. 
\end{definition}

It follows immediately from $F(u) \leq F(u)$ and Definition~\ref{def:galois}
that $u \leq G(F(u))$ for all $u \in U$, 
and similarly that $F(G(v)) \geq v$ for all $v \in V$.
Moreover, $F(u)=F(G(F(u)))$ and $G(v)=G(F(G(v)))$ 
for all $u \in U$, $v \in V$. 

% internal proof:
% Clearly we find that F(G(F(x))) ³ F(x)
%because FG is inflationary as shown above. 
%On the other hand, since GF is deflationary, 
%while F is monotonic, one finds that
%F(G(F(x))) ² F(x). This shows the desired equality.
%The following is obvious.

\begin{proposition}\label{prop:galois}
The operators $\Inv$ and $\Aut$ form a Galois connection between
sets of relations over the base set $B$ and permutation groups of the set $B$, both partially ordered by inclusion.
\end{proposition}
\begin{proof}
Let $\mathcal R$ be a set of relations over the set $B$, and let $\cG$ be a permutation group on the set $B$.
%Then $\cG \subseteq \Aut(\mathcal R)$
%if and only if ${\mathcal R} \subseteq \Inv(\cG)$. 
First suppose that $\cG \subseteq \Aut(\mathcal R)$,
and let $R \in {\mathcal R}$ and $g \in \cG$. 
Then $g \in \Aut(\mathcal R)$ and hence $g$ preserves $\mathcal R$. Thus, ${\mathcal R} \subseteq \Inv(\cG)$.

Conversely, suppose that ${\mathcal R} \subseteq \Inv(\cG)$,
and again let $g \in \cG$ and $R \in \mathcal R$. 
Then $R \in \Inv(\cG)$, and hence $g$ preserves $R$.
Since $g^{-1} \in \cG$, and $g^{-1}$ also preserves $R$,
we have that $g \in \Aut({\mathcal R})$. Thus, 
$\cG \subseteq \Aut({\mathcal R})$. 
\end{proof}

We now present descriptions of the closure operators 
$\cG \mapsto \Aut(\Inv(\cG))$ and 
${\mathcal R} \mapsto \Inv(\Aut({\mathcal R}))$. 

\begin{definition}
Let $\cG$ is a permutation group over a set $B$. 
Then 
%\begin{itemize}
%\item $\cl{\cP}$, the \emph{permutation group generated by $\cP$}, is the smallest permutation group that contains $\cP$. 
%\item 
$\overline{\cG}$, the \emph{closure of $\cG$ in $S(B)$}, is the smallest subset of $S(B)$ that is closed in $S(B)$ and contains $\cG$. 
%\end{itemize}
\end{definition}

The following is a special case of Corollary~1.9 in~\cite{Szendrei}
(which will be presented in full generality %, including a proof, 
in Proposition~\ref{prop:pol-inv} of Chapter~\ref{chap:algebra}).

\begin{proposition}\label{prop:aut-inv}
Let $\cG$ be a permutation group. Then %for every permutation $\alpha$ of $B$, the following are equivalent.
%\begin{enumerate}
%\item $\alpha$ is contained in the smallest permutation group that is closed in $S(B)$ and contains $\cP$;
%\item $\alpha \in \overline{\cl{\cP}}$;
%\item $\alpha \in \Aut(\Inv({\cP}))$.
%\end{enumerate}
%In particular, 
$\Aut(\Inv({\cG})) = \overline{\cG}$.
\end{proposition}
\begin{proof}
To show that $\Aut(\Inv({\cG})) \supseteq \bar \cG$, let $\alpha \in \overline{\cG}$ be arbitrary, and let $R$ be
from $\Inv({\cG})$. We have to show that 
$\alpha$ and $\alpha^{-1}$ preserve $R$. 
Let $t \in R$; since $\alpha \in \overline{\cG}$, 
we have that $\alpha t = \beta t$ for some $\beta \in \cG$. Since $\beta$ preserves $R$, we have that $\alpha t \in R$. The argument for $\alpha^{-1}$ is analogous. 

To show that $\Aut(\Inv({\cG})) \subseteq \bar \cG$, let $\alpha$ be from $\aut(\Inv({\cG}))$. 
It suffices to show that for every finite subset $\{a_1,\dots,a_n\}$ of $B$ there is a $\beta \in \cG$ such that $\alpha a_i=\beta a_i$ for all $i \in \{1,\dots,n\}$. 
Consider the relation $R := \{(\beta a_1,\dots,\beta a_n) \; | \;  \beta \in \cG \}$. 
Note that $R$ is preserved by all permutations in $\cG$. Therefore, 
$\alpha$ preserves $R$. Since $\cG$ contains the identity, $R$ contains $(a_1,\dots,a_n)$, and hence 
$(\alpha(a_1),\dots,\alpha(a_n)) \in R$. Thus, $(\alpha(a_1),\dots,\alpha(a_n)) = (\beta(a_1),\dots,\beta(a_n))$ for some $\beta \in \cG$  as required.
\end{proof}

We now turn to characterisations of the hull operator $\bB \mapsto \Inv(\Aut(\bB))$. First observe the following, 
which is straightforward to prove.

\begin{proposition}\label{prop:fo-preserved}
Let $\mathfrak B$ be any structure. Then $\inv(\aut(\bB))$ contains
$\cl{\bB}_{\fo}$, 
the set of all relations that are first-order definable in $\bB$.
\end{proposition}

An exact characterisation 
of $\bB \mapsto \Inv(\Aut(\bB))$ can be given 
when $\bB$ is $\omega$-categorical.
The analogous statement of
Proposition~\ref{prop:inv-aut-omega-cat} below has been observed 
for finite structures $\bB$ by Krasner~\cite{Krasner}.
The fact that it extends to $\omega$-categorical structures is a direct consequence of the Theorem of Ryll-Nardzewski (Theorem~\ref{thm:ryll}).

\begin{proposition}\label{prop:inv-aut-omega-cat}
Let $\bB$ be a countable $\omega$-categorical structure with base set $B$, and let $R \subseteq B^k$ be a relation. 
Then $R$ is first-order definable in $\bB$
if and only if $R$ is preserved by the automorphisms of $\bB$,
in symbols, $$\Inv(\Aut(\bB))=\langle\bB\rangle_{\fo}\; .$$
\end{proposition}

\ignore{
\begin{proof}
By Proposition~\ref{prop:fo-preserved}, 
first-order formulas are preserved by
automorphisms. The converse follows from Theorem~\ref{thm:ryll},
since every $k$-ary relation that is preserved by the automorphisms of 
$\bB$ is a union of orbits of $k$-tuples under $\Aut(\bB)$.
There is only a finite number of such orbits, by Theorem~\ref{thm:ryll},
say $O_1, \dots, O_m$. 
Moreover, by the same theorem, every orbit $O_i$ has a first-order definition $\phi_i$. Thus, $R$ has the first-order definition
$\phi_1 \vee \dots \vee \phi_m$.
\end{proof}
}

As we have seen in the proof of Proposition~\ref{prop:loc-clos-group}, it follows in particular 
that the expansion of
every $\omega$-categorical structure by all \emph{first-order} definable
relations is homogeneous. 
Recall that Theorem~\ref{thm:ryll} even states that
for countable structures the conclusion in Proposition~\ref{prop:inv-aut-omega-cat} holds \emph{if and only if} $\mathfrak A$ is $\omega$-categorical. 

\ignore{
\begin{proposition}\label{prop:inv-aut-iff-omegacat}
Let $\bB$ be a countable structure with a countable signature. Then 
$\Inv(\Aut(\bB))=\langle\bB\rangle_{\fo}$ if and only if $\bB$ is $\omega$-categorical.
\end{proposition}
\begin{proof}
One direction has been shown in Proposition~\ref{prop:inv-aut-omega-cat}. Now suppose
that $\bB$ is not $\omega$-categorical. 
By Theorem~\ref{thm:ryll} there exists an $n$ such that there are infinitely
many orbits of $n$-tuples of $\bB$. Every union of orbits is preserved by all automorphisms of $\bB$. Hence, there are uncountably
many relations in $\Inv(\Aut(\bB))$, while there are only countably 
many primitive positive formulas over a countable signature. Hence,
there must be relations in $\Inv(\Aut(\bB))$ that are not in 
$\langle\bB\rangle_{\fo}$.
\end{proof}
}

We have the following consequence of Proposition~\ref{prop:aut-inv} 
and Proposition~\ref{prop:inv-aut-omega-cat}. An \emph{anti-isomorphism}
between two posets $U$ and $V$ is a bijection $f$ from the elements of
$U$ to the elements of $V$ such that $u \leq v$ in $U$ if and only if $f(u) \geq f(v)$ in $V$.

\begin{corollary}\label{cor:galois}
    Let $\bC$ be a countable $\omega$-categorical structure. %Then we have the following.
\begin{itemize}     
    \item The set of sets of the form $\langle \bB \rangle_{\fo}$, where
$\bB$ is first-order definable in $\bC$, ordered by inclusion,
forms a lattice. 
    \item The set of closed permutation groups that contain
      $\Aut(\bC)$, ordered by inclusion, forms a lattice.
      \item 
    The operator $\Inv$ is an anti-isomorphism between those two lattices,
    and $\Aut$ is its inverse.
\end{itemize}
\end{corollary}

We explicitly state another consequence. 
%\begin{definition}
    Recall that two structures $\bB, \bC$ on the same domain 
    are said to be \emph{first-order interdefinable} iff all relations of $\bB$ have a first-order definition in $\bC$ and vice-versa.
%\end{definition}
  Then it follows from the above 
  that two $\omega$-categorical structures are first-order interdefinable if and only if they have the same
  automorphisms. 

%We illustrate the connection between $\omega$-categorical structures and oligomorphic permutation groups 
%by translating several concepts from permutation group theory into 
%model-theoretic terminology, and vice versa. 

\subsection{Transitivity and Primitivity}
\label{ssect:trans}
We define concepts from permutation group theory that will be needed later.
A permutation group $\cG$ on a set $B$ is
\begin{itemize}
\item \emph{$k$-transitive}
if for any two $k$-tuples $s,t$ of distinct elements from $B$
there is an $\alpha \in \cG$ such that $\alpha s=t$, where the
action of $\alpha$ on tuples is componentwise, i.e., $\alpha (s_1,\dots,s_k)=(\alpha s_1,\dots,\alpha s_k)$.
We say that $\cG$ is \emph{transitive} if it is $1$-transitive.
\item \emph{$k$-set transitive}
if for any two sets $S,T \subseteq B$ of cardinality $k$
there is an $\alpha \in \cG$ such that $\alpha S = \{\alpha s \; | \; s \in S\} = T$. %Here,  the action of $\alpha$ on sets is element-wise, i.e.,  $\alpha S=...$.
\end{itemize}

It is easy to see that a 2-set transitive permutation group $\cG$ on an infinite set is also transitive.
We prove the contraposition: assume that 
$\cG$ has more than one orbit. There must be an orbit $O$ with
two distinct elements $c_1,c_2$. Let $c_3$ be an element not from $O$.
Then there is no automorphism that maps $\{c_1,c_2\}$ to $\{c_1,c_3\}$,
and hence $\cG$ is not 2-set transitive.
More generally, it holds that the number of
orbits of $n$-subsets is a non-decreasing sequence~\cite{Oligo}.

A \emph{congruence} of $\cG$ is an equivalence relation
on $B$ that is preserved by all permutations in $\cG$.
The equivalence classes of a 
congruence are also called \emph{blocks}.
A congruence is \emph{trivial} if each block contains only one element (and \emph{non-trivial} otherwise),
and it is called \emph{proper} if it is distinct from the equivalence relation that has only one block. 
When $\bB$ is an $\omega$-categorical structure and $\cG$
its oligomorphic automorphism group, 
then the congruences of $\cG$
are exactly the first-order definable equivalence relations in $\bB$
(and so we apply the terminology that we have for congruences also to those equivalence relations).
A permutation group $\cG$ is called \emph{primitive} if $\cG$ is transitive and every proper congruence of $\cG$ is trivial,
and \emph{imprimitive} otherwise.  Clearly, 2-transitive structures are always primitive.

% PRIMITIVITY

An \emph{orbital} is an orbit of pairs, that is, a set of the form 
$\{(\alpha a,\alpha b) \; | \; \alpha \in \cG\}$ for $a,b \in B$.
The \emph{trivial} orbital is the orbital $\{(a,a) \; | \; a \in B\}$. 
When $O$ is an orbital, the
\emph{orbital graph} is the directed 
graph with vertex set $B$ and edges $O$.
The \emph{rank} $r(\cG)$ of $\cG$ is 
the number of distinct orbitals of $\cG$.

% The following I find very nice, but it is not really needed in this text
%\begin{theorem}[Higman's Theorem; see e.g.~\cite{CameronPermutationGroups}]
%A transitive permutation group $\mathcal G$ is primitive
%if and only if the graph of all non-trivial orbitals
%is connected.
%\end{theorem}

For a sequence $\bar a$ of elements of $B$,
the \emph{pointwise stabilizer} 
$\cG_{\bar a}$ of $\cG$ is the set of all elements of $\cG$ that fix $\bar a$.
For a subset $A$ of $B$,
the \emph{setwise stabilizer} $\cG_{A}$ of $\cG$ is the set of 
all elements $\alpha$ of $\cG$ that fix $A$ set-wise, that is, satisfy $\alpha A=A$. 

% BookTD: include the following
%\subsection{Algebraic Closure}
%All the amalgamation classes we have seen in Section~\ref{sect:fraisse} are also \emph{strong} amalgamation classes. \Fresse-limits of strong amalgamation classes can be naturally characterized in terms of their automorphism group.  
%Let $\cG$ be a permutation group 
%on a set $B$.
%Then the \emph{algebraic closure} $\acl(A)$ of a finite subset $A$ of $B$ is the set of all those points of $B$ which lie in finite orbits of 
%$\cG_{\bar a}$. 
%When $\cG = \Aut(\bC)$, then we
%say that a structure $\bC$
%has \emph{no algebraicity} if for any finite subset $A$ of the domain of $\bC$, 
%we have $\acl(A)=A$.
%\begin{proposition}[See (2.15) in~\cite{Oligo}]
%\label{prop:no-algebraicity}
%Let $\bC$ be a homogeneous structure. 
%Then $\Age(\bC)$ has the strong amalgamation property if and only if 
%$\bC$ has no algebraicity.
%\end{proposition}

\subsection{Products}
\label{ssect:products}
% BOOKTO version: the signature product (Hadamard product) 
% for free amalgamation classes should also be introduced.
In this section we review the classical theory how to describe a permutation group in terms of transitive ones.
% BOOKTD: and how to describe transitive permutation groups in terms of primitive ones.
The same idea can be used to construct new oligomorphic permutation groups from known ones. 

\vspace{.2cm}
\subsubsection{Group actions}
It will be convenient to take a more general perspective on permutation groups (and this will again be used in Chapter~\ref{chap:topology}).
We now consider \emph{abstract groups}, that is, algebraic structures $\bf G$ over a set $G$ of group elements, with a function symbol for multiplication of group elements, 
a function for the inverse of a group element, and the constant for the identity. The link to permutation groups is
given by the concept of an \emph{action} of such a group on a set, which is described below.

Let $\Sym(X)$ be the abstract group whose domain is the set of all permutations of $X$, 
and where composition is defined as composition of permutations, the inverse of an element $g$ of $\Sym(X)$ is
the inverse of $g$ as a permutation of $X$, and the identity is the identity permutation. 

\begin{definition}
A \emph{(left) group action} of an (abstract) group $\bf G$ on a set $X$ is a binary function $\cdot \colon G \times X \rightarrow X$
%$(g,x) \mapsto g \cdot x$ x
which satisfies that $(gh) \cdot x = g \cdot (h \cdot x)$
for all $g,h \in G$ and $x \in S$, and $e \cdot x = x$ for every $x \in X$.
The action is \emph{faithful} if for any two distinct $g, h \in G$ there exists an $x \in X$ such that $g \cdot x \neq h \cdot x$.
\end{definition}

Equivalently, a group action of $\bf G$ on a set $X$ can
be viewed as a homomorphism from $\bf G$ into $\Sym(X)$, 
and a faithful group action as an isomorphism between $\bf G$ and a subgroup of $\Sym(X)$. 
Clearly, to every action of $\bf G$ on $X$ we can associate a permutation group as considered before, namely the image
of the action in $\Sym(X)$. 
An action is called \emph{oligomorphic} if the associated permutation group is oligomorphic. 
Conversely, to every permutation group $G$ on a set $X$ we can associate an abstract group
$\bf G$ whose domain is $G$, where composition and inverse are defined in the obvious way, and which acts on $X$ faithfully by $g \cdot x = g(x)$. When $\bB$ is a structure, we call $\bf G$ the \emph{abstract automorphism group of $\bB$} if
there is an action of $\bf G$ on $B$ such that the image of $\bf G$ under this action is the automorphism group of $\bB$. %(as a permutation group).

When $x \in X$, the \emph{orbit} of $x$ with respect to an action of $\bf G$ on $X$ is the set $\{ g \cdot x \; | \; g \in G \}$.
Hence, an \emph{orbit of $k$-tuples} in the corresponding permutation group on $X$ is an orbit of the action
of $\bf G$ on $X^k$ that is defined componentwise, that is, $g$ maps $(x_1,\dots,x_k)$ to $(g x_1,\dots,g x_k)$. 
In this way we can also use other terminology introduced for permutation groups (such a transitivity, congruences, primitivity, etc.) for group actions. 

The \emph{product} of a sequence of groups $({\bf G}_i)_{i \in I}$ is the product of this sequence as defined in general 
in Chapter~\ref{chap:logic}; note that the product is again a group. 
Products appear in several ways when studying permutation groups; the first is 
when we want to describe the relation between a permutation group and its `transitive constituents', described in the following.

\vspace{.2cm}
\subsubsection{The intransitive action of a group product}
When $\bf G$ acts on a set $X$ and $S \subset X$ is an orbit with respect to this action, then $\bf G$ naturally
acts transitively on $S$ by restriction; we call the corresponding group $\bf H$ the \emph{group induced by $S$},
or a \emph{transitive constituent}. 

\begin{proposition}[see~\cite{Oligo}]
Let $\bf G$ be a group acting on a set $X$, and let $({\bf G}_i)_{i \in I}$ be the groups induced by the orbits of $\bf G$ on $X$.
Then $\bf G$ is isomorphic to a subgroup of $\prod_{i \in I} {\bf G}_i$, and there are surjective homomorphisms from $\bf G$
to ${\bf G}_i$, for each $i$. 
\end{proposition}

We can use the same idea to construct new oligomorphic permutation groups from known ones.
\begin{definition}
Let $\fG_1$ and $\fG_2$ be groups acting on disjoint countable sets $X$ and $Y$, respectively. 
Then the action of ${\fG}_1 \times {\fG}_2$ on $X \cup Y$ defined by $(g_1,g_2) \cdot z = g_1 z$ if
$z \in X$ and $g_2 z$ if $y \in Y$ is called the
\emph{natural intransitive action} of ${\fG}_1 \times {\fG}_2$
on $X \cup Y$.
 \end{definition}

Note that when $\fG_1$ and $\fG_2$ act 
oligomorphically on $X$ and $Y$, respectively, then the natural intransitive action of $\fG_1 \times {\bf G}_2$ is
also oligomorphic: when $F_1(n)$ is the number of orbits of the componentwise action
of $\fG_1$ on $X^n$, and $F_2(n)$ is the number of orbits of the componentwise action of $\fG_2$ on $Y$,
then the number of orbits of the componentwise of ${\fG}_1 \times {\fG}_2$ on $X \cup Y$ is $\sum_{0 \leq i \leq n} F_1(i) F_2(n-i)$, and hence finite for all $n$. 

When $\fG_1$ and $\fG_2$ are the automorphism groups of $\omega$-categorical relational structures $\bA$ and $\bB$ with disjoint domains $A$ and $B$, respectively,
then the image of the natural intransitive action on $A \cup B$ 
(as a homomorphism from ${\fG}_1 \times {\fG}_2$ to $\Sym(A \cup B)$) can also be described as the automorphism
group of a relational structure $\bC$: we can take for $\bC$ the disjoint union of $\bA$ and $\bB$ (defined in Section~\ref{sect:homo}), % of Chapter~\ref{chap:intro}), 
expanded by a unary predicate that contains
exactly the elements of $A$. 
Since reducts of $\omega$-categorical structures are again $\omega$-categorical, 
this shows in particular that the disjoint union of two $\omega$-categorical structures is again $\omega$-categorical.

%Define subdirect product of those permutations groups (by defining it just for two we avoid
%the difference between cartesion and (restricted) direct product. This will become
%relevant for the topologies. There it will also be simpler because we have group actions
%and abstract groups.

\vspace{.2cm}
\subsubsection{The product action}
\label{sssect:product-action}
When $\fG_1$ is a group acting on a set $X$, and $\fG_2$ a group acting on a set $Y$, 
there is another important natural action of $\bf G := \fG_1 \times \fG_2$ besides
the intransitive natural action of $\bf G$, which is called
the \emph{product action} of $\bf G$. In this action, $\bf G$ acts on $X \times Y$ by $(g_1,g_2) \cdot (x,y) = (g_1 x, g_1 y)$. 
If the actions of ${\bf G}_1$ and ${\bf G}_2$ are transitive, then the product action is clearly transitive, too.
We claim that when the actions of ${\bf G}_1$ and ${\bf G}_2$ are oligomorphic, then the product action 
is also oligomorphic. 
Let $F_1(n)$ and $F_2(n)$ then the number of orbits of the componentwise action of $G_1$ on $X^n$ and $Y^n$, respectively.
Then the number of orbits of the componentwise action of $\bf G$ on $X \times Y$ is $F_1(n) F_2(n)$, and in particular finite, which proves the claim. 

When ${\bf G}_1$ and ${\bf G}_2$ are the automorphism groups of $\omega$-categorical relational structures $\bA$ and $\bB$, then
the image of the product action of $\bf G$ in $\Sym(A \times B)$ is the automorphism group of the following structure, 
which we call the \emph{full product structure} of $\bA$ and $\bB$, and
denote by $\bA \boxtimes \bB$. 
Let $\sigma$ be the signature of $\bA$, and $\tau$ be the signature of $\bB$; we assume that $\sigma$ and $\tau$ are disjoint,
otherwise we rename the relations so that the assumption is satisfied. 
For each $k$-ary $R \in \sigma$, the structure $\bA \boxtimes \bB$ contains 
the relation 
$\{((a_1,b_1),\dots,(a_k,b_k)) \; | \; (a_1,\dots,a_k) \in R^\bA, b_1,\dots,b_k \in B \}$, and for each $k$-ary $R \in \tau$, it contains the relation $\{((a_1,b_1),\dots,(a_k,b_k)) \; | \; (b_1,\dots,b_k) \in R^\bB, a_1,\dots,a_k \in A \}$.
Finally, we also add the relations $P_1 = \{((a_1,b_1),(a_2,b_2)) \; | \; a_1 = a_2 \}$ and $P_2 = \{((a_1,b_1),(a_2,b_2)) \; | \; b_1 = b_2 \}$ to $\bA \boxtimes \bB$. 

\begin{proposition}\label{prop:full-product} 
The automorphism group of $\bC := \bA \boxtimes \bB$ is $\Aut(\bA) \times \Aut(\bB)$ in its product action on $A \times B$. 
\end{proposition}
\begin{proof}
Let $h$ be the product action of 
 ${\bf G} := \Aut(\bA) \times \Aut(\bB)$
on $A \times B$, viewed as a homomorphism from $\bf G$ to $\Sym(A \times B)$. 
Let $(g_1,g_2)$ be an element of $\bf G$. Then $h((g_1,g_2))$ is the permutation $(x,y) \mapsto (g_1 x,g_2 y)$ of $A \times B$,
and this map preserves $\bC$: when $((a_1,b_1),\dots,(a_k,b_k)) \in R^\bC$, for $R \in \sigma$, then $(a_1,\dots,a_k) \in R^\bA$,
and so $(g_1 a_1,\dots,g_1 a_k) \in R^\bA$. Therefore, $((g_1 a_1,g_2 b_1),\dots,(g_1 a_k,g_2 b_k)) \in R^\bC$. The proof for the relation symbols $R \in \tau$ is analogous. 

We now show that conversely, every automorphism $g$ of $\bC$ is in the image of $h$. 
Note that $P_1$ and $P_2$ are congruences of the automorphism group of $\bC$. %Note that ${\bf H}/{P_1}$  
Fix elements $a_0 \in A, b_0 \in B$. 
Let $g_1$ be the permutation of $A$ that maps $a \in A$ to $a'$ such that $g((a,b_0))=(a',b')$.
Similarly, let $g_2$ be the permutation of $B$ that maps $b \in B$ to $b'$ such that $g((a_0,b))=(a',b')$. 
Since $g$ preserves $P_1, P_2$, the definition of $g_1$ and $g_2$ does not depend on the choice of $a_0$ and $b_0$. 
Moreover, $g_1 \in \Aut(\bA)$, since $g$ preserves the relations for the symbols from $\sigma$.
Similarly, $g_2 \in \Aut(\bB)$. Then $g' := h((g_1,g_2))$ equals $g$, since $g'((a,b))=(g_1 a, g_2 b) = g(a,b)$.
Hence, $g$ is a permutation of $A \times B$ that lies in the image of $h$.
\end{proof}

Note that Proposition~\ref{prop:full-product} becomes false in general when we omit the relations
 $P_1$ and $P_2$ in $\bA \boxtimes \bB$. Consider for example the structure without structure $\bB$ (that is, $\bB$ has
 empty signature). Then the automorphism group of $\bB \boxtimes \bB$ is imprimitive, but
 without the relations $P_1$ and $P_2$, the structure is isomorphic to $\bB$ and hence primitive.
Also note that when $\bA$ and $\bB$ are \emph{ordered} structures (and this will be a typical assumption in Chapter~\ref{chap:ramsey}), we could omit $P_1$ and $P_2$ in the definition of the full product without sacrificing Proposition~\ref{prop:full-product}, since $P_1(x,y)$ is definable from the order $<$ of $\bA$ by the formula $\neg (x < y) \wedge \neg (y < x)$, and similarly $P_2$ is definable from the order of $\bB$.  
 
 Finally we remark that $(\bA \boxtimes \bB) \boxtimes \bC$ and $\bA \boxtimes (\bB \boxtimes \bC)$ have the same automorphism
group (on the domain $A \times B \times C$). We explicitly define the $d$-fold full product as follows.

\begin{definition}[Full product of $d$ structures]\label{def:full-product-structs}
Let $\bB_1,\dots,\bB_d$ be structures with disjoint relational signatures $\tau_1,\dots,\tau_d$. 
We denote by $\bB_1 \boxtimes \cdots \boxtimes \bB_d$
the structure with domain $B := B_1 \times \dots \times B_d$ that contains for every $i \leq d$, and every $m$-ary $R \in (\tau_i \cup \{=\})$ an $m$-ary relation defined by $$\{((x^1_1,\dots,x^d_1),\dots,(x^1_m,\dots,x^d_m)) \in B^m \; | \; (x^i_1,\dots,x^i_m) \in R^{\bB_i} \} \; .$$ 
If $\bB := \bB_1 = \cdots = \bB_k$, then we first rename $R \in \tau_i$ into $R_i$ so that the factors have pairwise disjoint signatures, and then write $\bB^{[d]}$ for $\bB_1 \boxtimes \cdots \boxtimes \bB_d$.
\end{definition}

When $\bA$ and $\bB$ have the same signature $\tau$, then 
the automorphism group of the $\tau$-structure $\bA \times \bB$ (see Definition~\ref{ssect:products})
\emph{contains} the automorphism group of $\bA \times \bB$, and hence $\bA \times \bB$ is 
$\omega$-categorical, by Theorem~\ref{thm:ryll}. As a consequence, the class of all $\omega$-categorical structures forms a \emph{lattice} with respect to the homomorphism order (where disjoint union is the join, and product the meet of two $\omega$-categorical structures).

\section{Preservation Theorems}
\label{sect:pres}
Model-theoretic preservation theorems typically link
definability in (a syntactically restricted fragment of) a given
logic with certain `semantic' closure properties.
For the syntactic restrictions on first-order formulas
that we have introduced in Chapter~\ref{chap:logic}
we have already made remarks about various types of mappings
that automatically preserve the respective formulas.
Surprisingly, very often these 
maps can be used to obtain an exact characterisation of definability
in the corresponding fragment of first-order logic.

In this text, preservation theorems
become relevant in two contexts.
The first is that they can be used to give exact characterizations of existential, existential positive, and quantifier-free definability of relations over an $\omega$-categorical structure,
in a similar way as we characterized first-order 
definability in Section~\ref{sect:galois}. 
These characterizations require that we pass from 
automorphism groups to endomorphism monoids,
and they turn out to be useful
for the complexity analysis of CSPs.
%In this way, we also obtain a different characterization
%of first-order definability in $\omega$-categorical structures. 
%In each case, we will point out how these facts become useful
%in later sections and ultimately the complexity analysis of constraint satisfaction problems.
The various relevant connections
are displayed in Figure~\ref{fig:pres}.

\begin{figure}[h]
\begin{tabular}{|l|l|}
\hline
first-order definitions & automorphisms \\
existential definitions & self-embeddings \\
positive definitions & surjective endomorphisms \\
existential positive definitions & endomorphisms \\
quantifier-free definitions & partial automorphisms \\
%positive quantifier-free definitions & partial endomorphisms \\
\hline
\end{tabular}
\caption{Syntactically restricted definabilities and the corresponding
preservation properties.}
\label{fig:pres}
\end{figure}

The second context where we encounter model-theoretic preservation theorems is when giving syntactic descriptions
of $\omega$-categorical theories themselves
(rather than relations in $\omega$-categorical structures). 
For instance, we will see that for every $\omega$-categorical
structure $\bA$ there exists a homomorphically equivalent
$\omega$-categorical structure $\bB$ whose
first-order theory is $\forall\exists^+$. 
% (this syntactic restriction
%has also been introduced in Chapter~\ref{chap:logic}).

%While preservation under automorphisms corresponds to first-order definability, we have that
%\begin{itemize}
%\item preservation under self-embeddings corresponds to existential definability,
%\item preservation under surjective homomorphisms corresponds to positive definability, 
%\item preservation under endomorphisms corresponds to existential positive definability,
%\item preservation under partial isomorphisms corresponds to quantifier-free definability, and
%\item preservation under partial endomorphisms corresponds to
%positive quantifier-free definability.
%\end{itemize}

%In this section we present characterizations of 
%existential and existential positive definability
%in $\omega$-categorical structures $\mathfrak A$
%in terms of the endomorphism monoid of $\mathfrak A$.

\subsection{Model-theoretic preservation theorems}
%We derive our characterizations of existential, positive, and
%existential positive definability in $\omega$-categorical structures
%as combinations of the theorem of Ryll-Nardzewski (in the form of Proposition~\ref{prop:inv-aut-omega-cat}) and
%classical model-theoretic preservation theorems. 
When $T$ is a first-order theory, we say that $\phi$ and $\psi$
are \emph{equivalent modulo $T$} if $T \models (\phi \Leftrightarrow \psi)$ (see Section~\ref{sect:sat}). % in Chapter~\ref{chap:intro}).
The following theorems are well-known and can be found
in most model theory books.

\begin{theorem}[{\L}os-Tarski; see e.g.~Corollary in~5.4.5 of~\cite{Hodges}]\label{thm:los-tarski}
Let $T$ be a first-order theory.
A first-order formula $\phi$ is equivalent to an existential 
formula modulo $T$ if and only if $\phi$ is preserved
by all embeddings between models of $T$.
\end{theorem}

\begin{theorem}[Lyndon; see e.g.~Corollary in~8.3.5 of~\cite{Hodges}]
\label{thm:lyndon}
Let $T$ be a first-order theory.
A first-order formula $\phi$ is equivalent to a positive 
formula modulo $T$ if and only if $\phi$ is preserved
by all surjective homomorphisms between models of $T$.
\end{theorem}

Note that here the assumption that $\bot$ 
is always part of
first-order logic becomes relevant: the first-order 
formula $\exists x. \, x \neq x$ is preserved by all homomorphisms
between models of $T$, but without $\bot$ it might not
be equivalent to a positive formula modulo $T$ (for instance when $T$ is the empty theory). 

\begin{theorem}[Homomorphism Preservation Theorem; see e.g.~Exercise~2 in Section 5.5 of~\cite{Hodges}]
\label{thm:homo-pres}
Let $T$ be a first-order theory.
A first-order formula $\phi$ is equivalent to an existential positive 
formula modulo $T$ if and only if $\phi$ is preserved
by all homomorphisms between models of $T$.
\end{theorem}

\begin{theorem}[Chang-{\L}o\'s-Suszko Theorem; Theorem~5.4.9 in~\cite{Hodges} and remarks after the proof]
\label{thm:chains}
Let $T$ be a first-order $\tau$-theory. 
\begin{itemize}
\item A set of first-order $\tau$-formulas $\Phi$ is equivalent
to a set of $\forall\exists$-formulas $\Psi$ 
modulo $T$ if and only if $\Phi$ is preserved in unions of chains of models of $T$.
%$(A_i)$ whenever
%$\bigcup A_i$ and all the $A_i$ are models of $T$.
\item A first-order $\tau$-formula $\phi$ is equivalent
to a $\forall\exists$-formula $\psi$ 
modulo $T$ if and only if $\phi$ is preserved in unions of chains of models of $T$.
%$(A_i)$ whenever
%$\bigcup A_i$ and all the $A_i$ are models of $T$.
\end{itemize}
\end{theorem}

Our next preservation theorem, Theorem~\ref{thm:pos-chains}, 
is a positive variant of the 
Chang-{\L}o\'s-Suszko preservation theorem, which we
could not find in explicit form in the literature. Its proof can be
derived from the proof of the Chang-{\L}o\'s-Suszko theorem
given in~\cite{Hodges} by modification of a sequence of lemmata
given there; since some of them require some care, we will 
present those modifications in full detail here. 
Besides the existential positive amalgamation theorem (Lemma~\ref{lem:existential-positive-amalgamation}), 
we need the following lemma.

\begin{lemma}\label{lem:pos-theory-change}
Let $T$ be a first-order theory, 
and let $\bA$ be a model of the
$\forall\exists^+$-consequences of $T$. 
Then $\bA$ can be extended to a model $\bB$ of $T$ such that every 
existential positive formula that holds on a tuple $\bar a$ 
in $\bB$ also holds on $\bar a$ in $\bA$.
\end{lemma}
\begin{proof}
Let $\bA'$ be an expansion of $\bA$ by constants such that all elements
of $\bA'$ are denoted by a constant symbol. 
It suffices to prove that 
$T \cup \text{diag}(\bA') \cup \text{diag}_{\forall^-}(\bA')$ 
has a model $\bB$. Suppose for contradiction that it were
inconsistent; then by compactness, there exists a finite
subset $U$ of $\text{diag}_{\forall^-}(\bA') \cup  \text{diag}(\bA')$ 
such that $T \cup U$ is inconsistent. 
Let $\phi$ be the conjunction over $U$ where all new constant symbols 
are existentially quantified. Then $T \cup \{\phi\}$
is inconsistent as well.
But $\neg \phi$ is equivalent to a $\forall\exists^+$ formula, and a consequence
of $T$. Hence, $\bA \models \neg \phi$, a contradiction.
\end{proof}

The following is a positive version of the Chang-{\L}o\'s-Suszko theorem (Theorem~\ref{thm:chains}). 

\begin{theorem}
\label{thm:pos-chains}
Let $T$ be a first-order $\tau$-theory, and $\Phi$ a set of $\tau$-formulas. Then the following are equivalent.
\begin{enumerate}
\item $\Phi$ is modulo $T$ equivalent
to a set of $\forall\exists^+$-formulas $\Psi$. 
%(If $\Phi$ is finite, then $\Psi$ can be chosen to contain a single formula.)
\item $\Phi$ is preserved in direct limits of sequences of models of $T$;
\item $\Phi$ is preserved in direct limits of countable sequences
of models of $T$.
\end{enumerate}
\end{theorem}
\begin{proof}
The implication from (1) to (2) is Proposition~\ref{prop:direct-product-preservation}.
The implication from (2) to (3) is trivial.
For the 
implication from (3) to (1), assume that $\phi$ is preserved by direct limits
of sequences $(\bA_i)$ as in the statement of the proposition.
We can assume that $\Phi$ is a set of sentences (by adding constants, Lemma~\ref{lem:constants}).
Let $\Psi$ be the set of all $\forall\exists^+$-sentences
that are consequences of $T \cup \Phi$. We first show that 
$T \cup \Psi$ implies $\phi$. It suffices to show that 
every model of $T \cup \Psi$ is elementary equivalent to a direct limit
of a sequence $(\bB_i)_{i < \omega}$ of models of $T \cup \Phi$ where there are coherent homomorphisms $f_{ij} \colon {\mathfrak B}_i \rightarrow {\mathfrak B}_j$ with $f_{jk} \circ f_{ij} = f_{ik}$ 
for all $i \leq j \leq k$. 

To construct this sequence, we define an elementary chain of
models $(\bA_i)_{i < \omega}$ of $T \cup \Psi$ such that there are 
\begin{itemize}
\item homomorphisms $f_i \colon \bA_i \rightarrow \bB_i$, 
with $\bB_i \models T \cup \Phi$, such that for 
every tuple $\bar a_i$ of elements from $\bA_i$ and
every existential positive
formula $\theta$, if $\bB_i \models \theta(f_i(\bar a_i))$,
then $\bA_i \models \theta(\bar a_i)$, and 
\item homomorphisms $g_i \colon \bB_i \rightarrow \bA_{i+1}$, such that $g_i \circ f_i$ is the identity on $\bA_i$.
\end{itemize}

% that is true in 
%$(\bB_i,f_i(a_i))$ is also true in $(\bA_i,a_i)$, where $a_i$ eine
%AufzŠhlung von A_i ist
%\item for each $i < \omega$, 
%every $\forall\exists^+$-formula $\chi$ that holds
%in $(\bA_i,\bar a_i)$ also holds in $(\bB_i,\bar a_i)$.

%2) Konstruktion von g_i\colon folgt unmittelbar aus 6.5.7 (d.h. pos.
%Variante von 6.5.1)
%3) die eigentliche Arbeit: Konstruktion von f_i\colon hier ist aber
%praktisch die ganze Idee im Beweis der pos. Variante von 6.5.8
%bereits enthalten, die ich dir geschrieben habe. Man zeigt das
%folgende

Let $\bA_0$ be a countable model of $T \cup \Psi$. To  construct the rest of the sequence, suppose that $\bA_i$ has been chosen. 
Since $\bA_0$ is an elementary substructure of $\bA_i$, in particular
all the $\forall\exists^+$-consequences of $T \cup \Phi$ hold in $\bA_i$. 
By Lemma~\ref{lem:pos-theory-change},
the structure $\bA_i$ 
can be extended to a model $\bB_i$ of $T \cup \Phi$ such that every ep-sentence that holds
 in $(\bB_i,\bar a_i)$ also holds in $(\bA_i,\bar a_i)$.
By Lemma~\ref{lem:existential-positive-amalgamation}
there are an elementary extension $\bA_{i+1}$ of $\bA_i$ and a homomorphism $g_i \colon \bB_i \rightarrow \bA_{i+1}$ such that $g_i \circ f_i$ is the identity on $\bA_i$. 
Then $\bC := \bigcup_{i < \omega} \bA_i$ equals $\lim_{i < \omega} \bB_i$,
and by the Tarski-Vaught elementary chain theorem (Theorem~\ref{thm:tarski-vaught}) $\bA_0$ is an elementary substructure of $\bC$.
So $\bC$ is a model of $T$, and the direct limit of 
models $\bB_i$ of $T \cup \Phi$, and hence $\bC \models \phi$.
This shows that $T \cup \Psi$ implies $\Phi$.
\end{proof}

By compactness one can show that when $\Phi$
is finite, then the formula $\Psi$ from item (1) in Theorem~\ref{thm:pos-chains} above can be chosen to be finite as well. 
%when $\Phi$ is finite then
%there is a finite subset $\Psi'$ of $\Psi$ such that $T \cup \Psi'$ implies $\phi$. Then $\bigwedge \Psi'$ is a $\forall\exists^+$ sentence, and equivalent to $\phi$ modulo $T$.

\subsection{Endomorphisms and self-embeddings}
\label{ssect:endos}
We apply the model-theoretic preservation theorems from the previous section 
to characterize existential, positive, and existential positive definability 
of relations in $\omega$-categorical structures.

For \emph{finite} structures and existential positive definability 
the corresponding preservation theorem
has already been noted by Krasner~\cite{Krasner}
(for finite structures, self-embeddings 
are necessarily automorphisms, and existential definability is the same as first-order definability). 

\begin{theorem}[from~\cite{Bodirsky} and~\cite{RandomMinOps,BodJunker}]\label{thm:ep}
Let $\bB$ be an $\omega$-categorical structure with base set $B$, 
and $R \subseteq B^k$ be a relation.
\begin{enumerate}
\item 
  $R$ has an existential positive definition in $\bB$ if and only if $R$
  is preserved by all endomorphisms of $\bB$.
\item 
  $R$ has an existential definition in $\bB$ if and only if $R$
  is preserved by all self-embeddings of $\bB$.
\item 
  $R$ has a positive definition
   in $\bB$ if and only if $R$ is preserved by all
   surjective endomorphisms of $\bB$. 
\end{enumerate}
\end{theorem} 

\begin{proof}
We have already remarked 
in Chapter~\ref{chap:logic}
that existential positive formulas are
preserved by endomorphisms, and existential formulas are preserved
by self-embeddings of $\bB$.

For the other direction, note that the endomorphisms and
self-embeddings of $\bB$ contain the automorphisms of $\bB$,
and hence Theorem~\ref{thm:ryll} shows that $R$ has a
first-order definition $\phi$ in $\bB$. 
Suppose for contradiction that $R$ were preserved by all
endomorphisms of $\bB$ but has no existential positive definition
in $\bB$. We use the homomorphism preservation theorem (Theorem~\ref{thm:homo-pres}). Since by
assumption $\phi$ is not equivalent to an existential positive
formula in $\bB$, there are models $\bB_1$ and $\bB_2$ of
the first-order theory of $\bB$ and a homomorphism $h$ from
$\bB_1$ to $\bB_2$ that violates $\phi$. By the theorem of
L\"owenheim-Skolem (Theorem~\ref{thm:LS}) the first-order theory
of the two-sorted structure $(\bB_1,\bB_2;h)$ has a countable
model $(\bB_1',\bB_2';h')$. Since both $\bB_1'$ and
$\bB_2'$ must be countably infinite, and because $\bB$ is
$\omega$-categorical, we have that $\bB_1'$ and $\bB_2'$ are isomorphic to $\bB$, and $h'$ can be seen as an endomorphism of $\bB$ that violates $\phi$; a contradiction.

The argument for existential definitions and positive definitions
is similar, but instead of the homomorphism preservation theorem we use the theorem of {\L}os-Tarski (Theorem~\ref{thm:los-tarski}) and Lyndon's theorem (Theorem~\ref{thm:lyndon}).
\end{proof}

We now present a Galois connection for existential positive definability
and transformation monoids, similar to the Galois connection
for first-order definability and permutation groups.
%between transformation monoids
%containing an oligomorphic permutation group
%and sets of relations with an existential positive definition 
%over an $\omega$-categorical structure.
For a structure $\bB$, we denote the set of relations with 
%an existential definition in $\mathfrak A$ by $\langle {\mathfrak A} \rangle_{\text{ex}$, 
%and the set of relations with 
an existential positive definition in $\bB$ 
by $\langle \bB \rangle_{\operatorname{ep}}$.
%Similarly as in Section~\ref{sect:galois}, we say that an operation $f\colon B\rightarrow B$ is 
%in the \emph{closure} of a set of operations $\cFÊ\subseteq (B \rightarrow B)$ if for every finite subset $A$ of $B$ there exists a $g \in \cF$ such that $f(a) = g(a)$ for all $a \in A$. We say that $\cF$ is \emph{locally closed} if $\cF$ contains all operations that are in the closure of $\cF$. 
Similarly as in Section~\ref{sect:galois}, we say that a set of operations
$\cFÊ\subseteq (B \rightarrow B)$ is \emph{(locally) closed} if it contains every operation $f \colon B \rightarrow B$ such that for every finite subset $A$ of $B$ there exists a $g \in \cF$ such that $f(a) = g(a)$ for all $a \in A$. 
The \emph{closure} of $\cF$ is the smallest locally closed set of operations that contains $\cF$. 

When $\cF$ is a transformation monoid, then $\cl{\cF}$ 
denotes the smallest locally closed transformation monoid
that contains $\cF$. The set of endomorphisms of a relational structure $\bB$ (or the set of operations from $B \rightarrow B$ that preserve a set of relations $\mathcal R$ over the domain $B$) is denoted by $\End(\bB)$ (or by $\End({\mathcal R})$, respectively). 
The following can be shown  in a similarly straightforward way as
Proposition~\ref{prop:loc-clos-group}.

\begin{proposition}\label{prop:loc-clos-monoid}
For every $\cF \subseteq (B \rightarrow B)$, the following are equivalent.
\begin{enumerate}
\item $\cF$ is the transformation monoid of a relational
structure;
\item $\cF$ is a locally closed monoid. 
%\item $\cG$ is the transformation monoid group of a homogeneous relational structure. 
\end{enumerate}
\end{proposition}

The proof of the following statement is similar to the proof of Proposition~\ref{prop:aut-inv}.
%in Section~\ref{sect:galois}.

\begin{proposition}\label{prop:end-inv}
Let $\cF \subseteq (B \rightarrow B)$ 
be a transformation monoid. Then $g \colon B \rightarrow B$ is in the closure of $\cF$ if and only if $g$ preserves all relations in $\Inv(\cF)$. In symbols, $$\End(\Inv(\cF)) = \langle \cF \rangle \; .$$ 
\end{proposition}

Theorem~\ref{thm:ep} now implies the following analog to Corollary~\ref{cor:galois}.

\begin{corollary}\label{cor:end-galois}
Let $\bC$ be an $\omega$-categorical structure.
Then the lattice of locally closed transformation monoids that contain
$\Aut(\bC)$ is anti-isomorphic to the lattice of sets of the form
$\langle \bB \rangle_{\operatorname{ep}}$ where 
$\bB$ is first-order definable in $\bC$.
\end{corollary}

To illustrate the use of this Galois connection, 
we present a simple and 
typical application.

\begin{lemma}\label{lem:constant}
Let $\bB$ be such that $\aut(\bB)$ is $2$-set transitive. 
If $\bB$ has a non-injective endomorphism $f$, 
then $\bB$ also has a constant endomorphism.
\end{lemma}
\begin{proof}
Let $f$ be an endomorphism of $\bB$ such that $f(b)=f(b')$ for two distinct
values $b,b' \in B$. Let $b_1,b_2, \dots$ be an enumeration of $B$.  We
construct an infinite sequence of endomorphisms $e_1, e_2,\ldots$, where $e_i$
is an endomorphism that maps all of the values $b_1, \dots, b_i$ to $b_1$.  This
suffices, since then by local closure the mapping defined by $e(x):=b_1$ for all $x$ is an endomorphism of $\bB$.

For $e_1$, we take the identity map, which clearly is an endomorphism with the
desired properties. To define $e_i$ for $i \geq 2$, let $\alpha$ be an
automorphism of $\bB$ that maps $\{b_1,e_{i-1}(b_i)\}$ to $\{b,b'\}$; such an
automorphism exists because $\aut(\bB)$ is 2-set transitive. 
Then the
endomorphism $f(\alpha e_{i-1}(x))$ is constant on $b_1, \ldots, b_i$; recall
that $b_1=e_{i-1}(b_1)= \cdots = e_{i-1}(b_{i-1})$. 
Since $\bB$ is 2-transitive, it is in particular transitive, and there is an
automorphism $\beta$ that maps $f(b)$ to $b_1$. Then
$e_i \colon x \mapsto \beta f(\alpha e_{i-1}(x))$ is an endomorphism of $\bB$ with the desired properties.
\end{proof}

%22.12.2010: ein self-embedding ist locally
%invertible genau dann wenn es local von den automorphismen erzeugt wird.  Aber in manchen argumenten, zum beispiel bei temporalen CSPs, kriegt man halt manchmal nur das umkehrende embedding, und muss dann zeigen dass es local auch autos gibt, die umkehren.

\subsection{Locally invertible self-embeddings}
\label{ssect:loc-invert}
Let $\mathfrak A$ and $\mathfrak B$ be $\tau$-structures,
let $e$ be an embedding of $\mathfrak A$ into $\mathfrak B$,
and let $f$ be an embedding of $\mathfrak B$ into $\mathfrak A$.
We say that $e$ and $f$ \emph{locally invert each other}
 if
\begin{itemize}
\item 
for every tuple $\bar a$ of elements of $\mathfrak A$
%and every $\alpha_1 \in \text{Aut}(\mathfrak B,e(\bar a))$ 
there are $\beta \in \text{Aut}(\mathfrak B)$
and $\alpha \in \text{Aut}(\mathfrak A)$ such that $\alpha f (\beta e(\bar a)) = \bar a$, and
\item 
for every tuple $\bar b$ of elements of $\mathfrak B$
%and every $\alpha_1 \in \text{Aut}(\mathfrak A,f(\bar b))$ 
there are $\alpha \in \text{Aut}(\mathfrak A)$ and $\beta \in \text{Aut}(\mathfrak B)$ such that $\beta e (\alpha f(\bar b))  = \bar b$.
\end{itemize}
We say that $e$ is \emph{locally invertible} if there exists
a self-embedding $f$ such that $e$ and $f$ locally invert each other.

We will show that locally invertible self-embeddings preserve
first-order formulas. To do so, we need the following concept.
Let $\mathfrak A$ and $\mathfrak B$ be $\tau$-structures.
A \emph{back-and-forth system} from $\mathfrak A$ to $\mathfrak B$ (our definition is taken from~\cite{Hodges})
is a non-empty set $I$ of pairs $(\bar a, \bar b)$ of tuples, with $\bar a$ from $\mathfrak A$ and $\bar b$ from $\mathfrak B$, such that the following hold.
\begin{enumerate}
\item If $(\bar a,\bar b) \in I$ then $\bar a$ and $\bar b$ have the same
length and $(\bA,\bar a)$ satisfies the same atomic formulas
as $(\bB, \bar b)$. 
\item (Going Forth.) For every pair $(\bar a,\bar b) \in I$ and every element $c$
of $\mathfrak A$ there is an element $d$ of $\mathfrak B$ such that the pair
$(\bar ac,\bar bd) \in I$.
\item (Going Back.) For every pair $(\bar a,\bar b) \in I$ and every element $d$
of $\mathfrak B$ there is an element $c$ of $\mathfrak A$ such that the pair
$(\bar ac,\bar bd) \in I$.
\end{enumerate}
There is a back-and-forth system from
$\mathfrak A$ to $\mathfrak B$ if and only if $\mathfrak A$ and $\mathfrak B$ 
are isomorphic (combination of Lemma 3.2.2 and Theorem 3.2.3 (b) in~\cite{Hodges}). 

\begin{theorem}\label{thm:loc-invertible}
A relation $R$ has a first-order definition in an
$\omega$-categorical structure $\bB$ if and only if
$R$ is preserved by all locally invertible self-embeddings
of $\bB$.
\end{theorem}
\begin{proof}
We are in the remarkable situation (in comparison to the other preservation theorems discussed here) that the ``if" direction 
of the statement is easy (it follows directly from the theorem of Ryll-Nardzewski, since automorphisms are locally inverted by their inverse),
and that we only have to show the `only if' direction.

Let $e$ and $f$ be self-embeddings of $\bB$ that locally invert each other, and suppose
that $\bar a$ is a tuple from $\bB$ that satisfies
a first-order formula $\phi$. We claim that $e(\bar a)$ satisfies
$\phi$ as well. It clearly suffices to show that the structures
$(\bB,\bar a)$ and $(\bB,e(\bar a))$ are isomorphic.
We claim that the set 
\begin{align*}
I := \{(\bar u,\bar v) \; | \; & \text{ there are } \gamma,\delta \in \text{Aut}(\bB) \\
& \text{ so that } \delta e \gamma (\bar u) = \bar v \}
\end{align*}
is a back-and-forth system from 
$(\bB,\bar a)$ to $(\bB,e(\bar a))$.

The set $I$ is non-empty, since $(\bar a,e(\bar a)) \in I$ (we have $\gamma=\delta=id$ in the definition of $I$).
It is obvious that $I$ satisfies item (1) in the definition 
of back-and-forth systems since all involved operations are embeddings. Now, let $(\bar u,\bar v)$ be from $I$.
By definition
of $I$, there are $\gamma \in \text{Aut}(\bB)$
and $\delta \in \text{Aut}(\bB)$ so that $\delta e (\gamma \bar u) = \bar v$. 
For going forth, let $c$ be an arbitrary element of $\bB$.
Let $d$ be $\delta e (\gamma c)$. The clearly
$(\bar u c, \bar v d) \in I$.

For going back, let $d$ be an arbitrary element of $\bB$.
Since $e$ is locally inverted by $f$, there exist 
$\alpha,\beta \in \text{Aut}(\bB)$ such that $\alpha f (\beta e (\gamma \bar u)) = \gamma\bar u$. 
Since $e (\gamma \bar u) = \delta^{-1} \bar v$,
this is the same as saying that 
$\alpha f ( \beta \delta^{-1} \bar v) = \gamma \bar u$,
and by multiplication with $\alpha^{-1}$ we note
\begin{align}
f ( \beta \delta^{-1} \bar v) = \alpha^{-1} \gamma \bar u \; .
\label{eq:f-of-p}
\end{align}
We now set $c$ to $\gamma^{-1} \alpha f (\beta \delta^{-1} d)$,
claiming that $(\bar uc, \bar vd) \in I$, which completes the proof.

To show the claim, 
we have to find $\gamma',\delta' \in \text{Aut}(\bB)$ 
such that $\delta' e (\gamma' (\bar uc)) = \bar vd$.
Let $\bar p$ be the tuple $\beta \delta^{-1} (\bar v d)$. 
By the second item in the definition of local inversion,
there
are $\alpha',\beta' \in \text{Aut}(\bB)$ such that 
$\beta' e (\alpha' f(\bar p))  = \bar p$. 

Choose $\gamma' = \alpha' \alpha^{-1} \gamma$ 
and $\delta' = \delta \beta^{-1} \beta'$.
Then 
\begin{align*}
\delta' e (\gamma' (\bar uc)) = & \; \delta \beta^{-1} \beta' e (\alpha' \alpha^{-1} \gamma (\bar uc)) \\
= & \; \delta \beta^{-1} \beta' e (\alpha' (\alpha^{-1} \gamma \bar u,\alpha^{-1} \gamma \bar c)) \\
= & \; \delta \beta^{-1} \beta' e (\alpha' (f (\beta \delta^{-1} \bar v),f (\beta \delta^{-1} d))) & \text{(see (\ref{eq:f-of-p}))} \\
= & \; \delta \beta^{-1} \beta' e (\alpha' f (\beta \delta^{-1} (\bar v d))) \\
= & \; \delta \beta^{-1} \beta \delta^{-1} (\bar v d) \\
= & \; \bar vd \; ,
\end{align*}
and so $(\bar uc, \bar vd) \in I$.
\end{proof}

Theorem~\ref{thm:loc-invertible} will be used in Section~\ref{sect:mccore},
and later also in Chapter~\ref{chap:tcsp},
as a tool for proving that all automorphisms of certain structures
$\bB$ are locally generated by the self-embeddings
of $\bB$. We note the following consequence of Theorem~\ref{thm:loc-invertible}.

\begin{corollary}
An endomorphism $e$ of an $\omega$-categorical structure
is locally invertible if and only if $e$ is locally generated by the automorphisms of $\Gamma$.
\end{corollary}
\begin{proof}
If $e$ is locally generated by the automorphisms, then $e$ is clearly locally invertible. 
The converse follows from Theorem~\ref{thm:loc-invertible} in combination with
Proposition~\ref{prop:end-inv}. 
\end{proof}

\subsection{Partial Automorphisms}
Recall that a formula is called \emph{quantifier-free} if it can
be constructed from atomic formulas
by usage of Boolean connectives only.
Also quantifier-free definability can be characterized by a model-theoretic preservation theorem; in this case, this is very easy to prove.

\begin{proposition}\label{prop:qe}
Let $T$ be a first-order theory over a relational signature. A first-order formula
$\phi$ is equivalent to a quantifier-free formula 
modulo $T$ if and only if $\phi$ is preserved by partial
isomorphisms between models of $T$.
\end{proposition}
\begin{proof}
It is clear that quantifier-free formulas are preserved by partial isomorphisms between models of $T$. For the converse, 
let $\phi$ be preserved by all partial isomorphisms between models
of $T$. 
Let $\Psi$ be the set of all quantifier-free formulas $\psi$
such that $T \models \forall \bar x (\phi(\bar x) \Rightarrow \psi(\bar x))$. It suffices to prove
that $\Psi$ implies $\phi$. Let $\bA$ be a model of 
$T$ and $\bar a$ a tuple from $\bA$ such that $\bar a$ satisfies $\Psi$
in $\bA$.
Let $\bB$ be a model of $T$ and $\bar b$ a tuple from $\bB$ such that $\bar b$ satisfies $\phi$
in $\bB$ (if no such $\bB$ exists, the
statement of the proposition is trivial).
Since $\bar a$ and $\bar b$ satisfy the same atomic formulas, 
the mapping that sends $\bar b$ to $\bar a$ is a partial isomorphism.
Since $\phi$ is by assumption 
preserved by partial isomorphisms, 
$\bar a$ satisfies $\phi$ in $\bA$. 
The set $\Psi$ modulo $T$ equivalent to a
single quantifier-free formula by compactness 
of first-order logic and this concludes the proof. 
\end{proof}

The following can be derived from the previous proposition.
  
\begin{proposition}\label{prop:qe-omegacat}
Let $\bB$ be an $\omega$-categorical structure.
Then a relation $R$ has a quantifier-free definition in $\bB$ if
and only if $R$ is preserved by all partial automorphisms of $\bB$, i.e., preserved by isomorphisms between induced
substructures of $\bB$.
\end{proposition}
\begin{proof}
If a relation $R$ is preserved by all partial endomorphisms
of $\bB$, then
it is in particular preserved by all endomorphisms of $\bB$. 
By Theorem~\ref{thm:ep}, $R$ has an existential positive definition
in $\bB$. 
We can therefore use Proposition~\ref{prop:qe} in the same way as we used model-theoretic preservation theorems to prove Theorem~\ref{thm:ep} to conclude the argument. 
\end{proof}

\section{Existential Positive Completion}
\label{sect:completion}
It might be that the same CSP
can be formulated with different templates.
Consider for example the random graph $({\mV}; E)$, which has exactly the same CSP 
as the template $({\mN}; \{(x,y) \; | \; x \neq y\})$.
Recall that two structures $\bB$, $\bC$  have the same CSP if and only if they have the same existential positive theory $T$,
or equivalently, if they have the same universal negative theory $S$.
Moreover, any model of $T \cup S$ has the same CSP as $\bB$ and $\bC$.
The topic of this section is how to produce a model $\bB$
of $T \cup S$ such that $\bB$ has many good properties
for studying $\Csp(\bB)$.
Good candidates for such models $\bB$ 
are existential-positively
closed models, which will be introduced here.
Much of the material presented here is analogous to the classical facts about 
existential completion, which we briefly review in Section~\ref{sect:ec}.
Existential positive completion is discussed in Section~\ref{sect:epc}. 
The results in this section have been published in~\cite{BodHilsMartin-Journal}.

\subsection{Existential Completion}
\label{sect:ec}
Let $T$ be a first-order theory.
A model $\mathfrak A$ of $T$ is 
\emph{existentially closed for $T$} (we also say that $\bA$ is an \emph{existentially closed model of $T$})
if ${\mathfrak A} \models \phi(\bar a)$ for any embedding $e$ from $\mathfrak A$ into another model
$\mathfrak B$ of $T$, any tuple $\bar a$ from $A$,
and any primitive formula $\phi$ with ${\mathfrak B} \models \phi(e(\bar a))$.

To construct existentially closed models of $T$, we use unions of elementary chains (see Section~\ref{sect:limits}).
A first-order theory $T$ is \emph{inductive} if 
the union of every chain of models of $T$ is also a model of $T$.
Note that by Proposition~\ref{prop:inductive},
when $T$ is a $\forall\exists$-theory, then $T$ is inductive.
The following lemma implies that if $T$ is inductive,
then it has an existentially closed model. For a proof, see~\cite{Hodges}, or the proof of
Lemma~\ref{lem:existence-epc} below, which is very similar.

\begin{lemma}[Corollary 7.2.2 in~\cite{Hodges}]
\label{lem:existence-ec}
Let $T$ be an inductive theory 
and let $\kappa$ be an infinite cardinal.
 Then any model $\mathfrak A$ of $T$ of cardinality at most 
 $\kappa$ embeds into an
 existentially closed model $\mathfrak B$ of $T$ of cardinality
 at most $\kappa$.
\end{lemma}

%The following is one of the crucial properties of existentially closed %models.
%\begin{proposition}[Follows ]\label{prop:ec-types}
%Let $\mathfrak A$ be a countable ec model of a theory $T$. Each of the complete existential types of tuples of $\mathfrak A$ is a maximal existential type of $T$.
%\end{proposition}

\subsection{Existential positive completion}
\label{sect:epc}
Again, let $T$ be a first-order theory.

\begin{definition}
A model $\mathfrak A$ of $T$ is 
\emph{existential-positively closed for $T$} (or short \emph{an epc model of $T$})
if ${\mathfrak A} \models \phi(\bar a)$ for any homomorphism $h$ from $\mathfrak A$ into another model
$\mathfrak B$ of $T$, any tuple $\bar a$ from $A$,
and any existential positive formula $\phi$ with ${\mathfrak B} \models \phi(h(\bar a))$.
\end{definition}

Note that we can equivalently replace `existential positive' by `primitive positive' in the previous definition. 

To show the existence of epc models we apply the \emph{direct limit} construction from Section~\ref{sect:limits}.

\begin{lemma}[from~\cite{BodHilsMartin-Journal}; also see~\cite{BenYaacov}]
\label{lem:existence-epc}
Let $T$ be a $\forall\exists^+$ $\tau$-theory and let $\kappa = \max(\omega,|\tau|)$. 
Then any model $\bA$ of $T$ of cardinality at most $\kappa$ admits a homomorphism to an epc model $\mathfrak B$ of $T$ of cardinality at most $\kappa$.
\end{lemma}
\begin{proof}
Set ${\mathfrak B}_0: = \mathfrak A$. 
Having constructed ${\mathfrak B}_i$ of cardinality at most $\kappa$, for $i<\omega$, let $\{(\phi_\alpha,\bar a_\alpha) \; |Ê\; \alpha < \kappa\}$ be an enumeration
of all pairs $(\phi,\bar a)$ where $\phi$ is existential positive with
free variables $x_1,\dots,x_n$, and $\bar a$ is an $n$-tuple 
from $B_i$. 
We construct a sequence 
$({\mathfrak B}_i^{\alpha})_{0 \leq \alpha < \kappa}$ of models of $T$ of cardinality at most $\kappa$ 
and a coherent sequence
$(f_i^{\mu,\alpha})_{0 \leq \mu < \alpha < \kappa}$
where $f_i^{\mu,\alpha}$ is a homomorphism
from ${\mathfrak B}_i^\mu$ to ${\mathfrak B}_i^{\alpha}$, as follows.

Set ${\mathfrak B}_i^0 = {\mathfrak B}_{i-1}$.
Now let $\alpha = \beta +1 < \kappa$ be a successor ordinal.
Let $\bar b_\beta$
be the image of $\bar a_\beta$ in ${\mathfrak B}_i^\beta$ under $f_i^{0,\beta}$. If there is a model
$\bC$ of $T$ and a homomorphism $h \colon {\mathfrak B}_i^\beta \rightarrow \mathfrak C$ such that $\bC \models \phi_\beta(h(\bar b_\beta))$, then by the theorem of L\"owenheim-Skolem there is
also a model $\bC'$ of cardinality at most 
$\kappa$ of $T$ and a 
homomorphism $h' \colon {\mathfrak B}_i^\beta \rightarrow \bC'$ such that $\bC' \models \phi_\beta(h'(\bar b_\beta))$.
Set ${\mathfrak B}^{\alpha}_i := \bC'$ and $f_i^{\mu,\alpha} := h' \circ f_i^{\mu,\beta}$ for all $\mu < \alpha$. 
Otherwise, if there is no such model $\bC$, we set ${\mathfrak B}^{\alpha}_i := {\mathfrak B}^{\beta}_i$ and $f_i^{\beta,\alpha} := \text{id}$ (the identity mapping) and $f_i^{\mu,\alpha} := f_i^{\mu,\beta}$. 
Finally, for limit ordinals $\alpha<\kappa$, set $\bB^\alpha_i :=
\lim_{\mu<\alpha}\bB^\mu_i$ 
and let $f_i^{\mu,\alpha}$ be the corresponding limit homomorphism from $\bB^\mu_i$ to $\bB^\alpha_i$.

Let $\bB_i$ be 
$\lim_{\alpha<\kappa} \bB_i^\alpha$ and let $g_i \colon \bB_{i-1} \rightarrow \bB_i$ be the limit homomorphism mapping each element of $\bB_{i-1}=\bB_i^0$ 
to its equivalence class in $\bB_i$. In the natural way, 
the $g_i$ give raise to a coherent sequence of homomorphisms, and
by Proposition~\ref{prop:direct-product-preservation}, $\bB := \lim_{i < \omega} \bB_i$ is a model of $T$; let $h_i \colon \bB_i \rightarrow \bB$ for $i<\omega$ be the corresponding limit homomorphisms. 

The structure $\bB$ is epc in $T$.
To verify this, let $g$ be a homomorphism from 
$\bB$ to a model $\bC$ of $T$, and suppose that there is a tuple
$\bar b$ over $B$ and an existential positive formula $\phi$
such that $\bC \models \phi(g(\bar b))$. Then
there is an $i < \omega$ and an $\bar a \in B_i$ such that
$h_i(\bar a)=\bar b$. 
%and all atomic formulas that hold on $\bar b$ hold on $\bar a$ in $\bB_i$. 
Then $g \circ h_i$ is a homomorphism from $\bB_i$ 
to $\bC$, and by construction we have that $\bB_{i+1} \models \phi(g_{i+1}(\bar a))$. 
Note that $h_{i+1} \circ g_{i+1} = h_i$. 
Thus, since $h_{i+1}$ preserves existential positive formulas, we also have that 
$\bB \models \phi(\bar b)$, which is what
we had to show.
\end{proof}

For an equivalent characterization of existentially closed models
in terms of maximal pp-types (Proposition~\ref{prop:epc-types}), 
we need the following lemma, a close relative of Theorem~10.3.1 in \cite{HodgesLong}.

\begin{lemma}\label{lem:sat-hom}
Let $\bA$ and $\bB$ be $\tau$-structures, where $\bB$ is pp-$|A|$-saturated. Suppose that $\mu < |A|$ and that $f$ is a mapping from $\{a_\alpha \, | \, \alpha < \mu\} \subseteq A$ to $B$ such that
all pp-$(\tau \cup \{c_\alpha \, | \, \alpha < \mu\})$-sentences true on $(\bA, (a_\alpha)_{\alpha<\mu})$ are true on $(\bB,(f(a_\alpha))_{\alpha<\mu})$. Then $f$ can be extended to a homomorphism from $\bA$ to $\bB$.
\end{lemma}
\begin{proof}
%If $\bB$ is finite, then $\bB$ is pp-$|A|$-saturated no matter what cardinality $\bA$ has. Suppose $\mu < |A|$. 
Let $(a'_\alpha)_{\alpha < |A|}$ well-order $A$ 
such that $\{a'_\alpha \, | \, \alpha<\mu\} = \{ a_\alpha \, | \, \alpha<\mu\}$ (there is the
implicit and harmless assumption that $(a_\alpha)_{\alpha<\mu}$ contains no repetitions).
Set $(b_\alpha)_{\alpha<\mu}:=(f(a_\alpha))_{\alpha<\mu}$. 

We will construct by transfinite induction on $\beta$ (up to $|A|$) a sequence $(b_\alpha)_{\alpha<\beta}$ such that we maintain the inductive hypothesis
\[ (*) \ \mbox{ all pp-$(\tau \cup \{c_\alpha \, | \, \alpha<\beta\})$-sentences true on 
$(\bA;(a'_\alpha)_{\alpha<\beta})$ are true on $(\bB;(b_\alpha)_{\alpha<\beta})$}.\]
\begin{itemize}
\item (Base Case.) $\beta = \mu$. Follows from the hypothesis of the lemma.
\item (Inductive Step. Limit ordinals.) $\beta = \lambda$. Property $(*)$ holds, since a sentence can only mention a finite collection of constants, whose indices must all be less than some $\gamma < \lambda$. 
\item (Inductive Step. Successor ordinals.) $\beta = \gamma^+ < |A|$. Set
\begin{align*} 
\Sigma := \big \{ \phi(x) \; | \; & \phi \text{ is a pp-}(\tau \cup \{c_\alpha \, | \, \alpha<\gamma\})\text{-formula such that} \\
& (\bA;(a'_\alpha)_{\alpha<\gamma}) \models \phi(a'_\gamma) \big \} \; . 
\end{align*}
By $(*)$, for every $\phi \in \Sigma$, $(\bB; (b_\alpha)_{\alpha<\gamma}) \models \exists x. \phi(x)$. By compactness, since $\Sigma$ is closed under conjunction, we have that $\Sigma$ is a pp-$1$-type of $(\bB;(b_\alpha)_{\alpha<\gamma})$. 
Then $\Sigma$ is realized by some element $b_\gamma \in B$
because $\bB$ is pp-$|A|$-saturated. By construction we maintain that all pp-($\tau \cup \{c_\alpha \, | \, \alpha<\gamma^+\}$)-sentences true on $(\bA;(a'_\alpha)_{\alpha<\gamma^+})$ are true on $(\bB;(b_\alpha)_{\alpha<\gamma^+})$.
\end{itemize}
The result follows by reading $f$ as the function
that maps $a'_\alpha$ to $b_\alpha$ for all $\alpha< |A|$.
\end{proof}

\begin{proposition}\label{prop:epc-types}
Let $T$ be a theory, and let $\bA$ be a model of $T$.
Then $\bA$ is epc for $T$ if and only if every complete pp-$n$-type 
of $\bA$ is a maximal pp-type of $T$.
\end{proposition}
\begin{proof}
(Forwards.) Suppose $p(x_1,\dots,x_n)$ is an pp-$n$-type, realized in $\bA$ 
by the tuple $(a_1,\dots,a_n)$. Let $c_1,\dots,c_n$ be new constant symbols
that denote $a_1,\dots,a_n$ in $\bA$. 
Let $\phi(x_1,\dots,x_n)$ be a primitive positive formula such that 
$T \cup p(c_1,\dots,c_n) \cup \{\phi(c_1,\dots,c_n)\}$ has a model $(\bC;c_1,\dots,c_n)$. 
Now, let $(\bC_{\operatorname{sat}};c_1,\dots,c_n)$ 
be an $|A|$-saturated model of $\operatorname{Th}(\bC;c_1,\dots,c_n)$;
such a model always exists by Theorem~\ref{thm:saturation}. Clearly $(\bC_{\operatorname{sat}};c_1,\dots,c_n)$ is pp-$|A|$-saturated, and all primitive positive formulas true on $(\bA,c_1,\dots,c_n)$ are true on $(\bC_{\operatorname{sat}};c_1,\dots,c_n)$. By Lemma~\ref{lem:sat-hom}, there is a homomorphism $h$ from $(\bA,c_1,\dots,c_n)$ to $(\bC_{\operatorname{sat}})$. Now, since $\phi(c_1,\dots,c_n)$ holds on $\bC_{\operatorname{sat}}$ and $\bA$ is epc for $T$, we find that $(\bA,c_1,\dots,c_n) \models \phi(c_1,\dots,c_n)$,
and conclude that $p$ is a maximal pp-type of $T$.

(Backwards.) Take $\bB \models T$, $h \colon \bA \rightarrow \bB$ a homomorphism, $\bar a$
a tuple of elements of $\bA$, 
and $\phi(x_1,\dots,x_n)$ a primitive positive formula such that 
$\bB \models \phi(h(\bar a))$. Let $p$ be the pp-type of $\bar a$ in $\bA$.
Since $\bB$ is a model of $T$ and $h$ preserves all primitive positive formulas,
it follows that $T \cup p \cup \{\phi\}$ is satisfiable. 
By maximality of $p$, we have that $\phi \in p$, and therefore 
$\bA \models \phi(\bar a)$.
\end{proof}

We close this section with an observation that will be needed later on.
\begin{lemma}\label{lem:epc-limits}
The class of all epc~models of a theory~$T$ is closed under
direct limits.
\end{lemma}
\begin{proof}
Suppose that $\bA = \lim_{\lambda<\kappa} \bA_\lambda$ for a sequence $(\bA_\lambda)_{\lambda < \kappa}$ of epc~models of $T$, $\bar a$ is a tuple from $\bA$, $\phi$ an existential positive
formula, and $h$ is a homomorphism from $\bA$ into another model of $T$ such that $\bB \models \phi(h(\bar a))$. 
Then there exists a $\lambda < \kappa$ such that $\bar a=g_\lambda(\bar a')$ for $\bar a'$ from $\bA_\lambda$ (where $g_\lambda$ is as in the definition of direct limits).
Note that $h \circ g_\lambda$ is a homomorphism from $\bA_\lambda$ to $\bB$, and
since $\bA_\lambda$ is an epc~model of $T$, 
$\bA_\lambda \models \phi(\bar a')$. 
Since $g_\lambda$ preserves existential positive formulas, we
thus also have that $\bA \models \phi(\bar a)$.
\end{proof}
 
\section{Quantifier-elimination, Model-completeness, Cores}
\label{sect:mccore}
\begin{figure*}
\begin{center}
\includegraphics[scale=0.6]{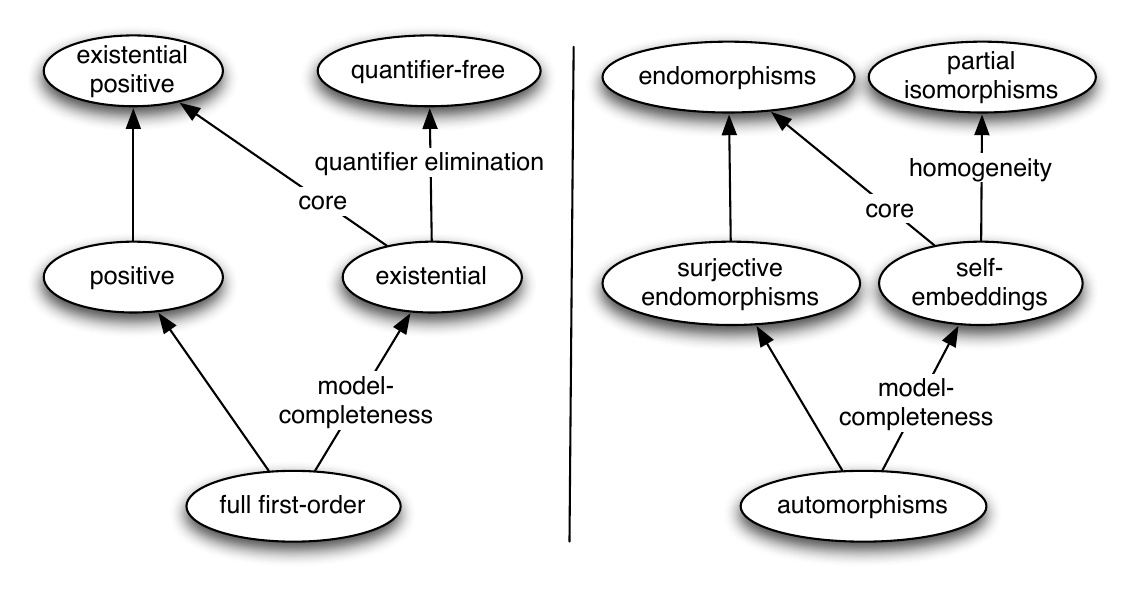} 
\end{center}
\caption{Various forms of definability (left side), ordered by relative strength, and the corresponding class of operations (right side). The labels on the arrows indicate the condition on the structure when the corresponding two forms of definability coincide (left side), and correspondingly when one set of operations locally generates the other (right side).}\label{fig:collaps}
\end{figure*}

This section is concerned with structures
$\bB$ where various forms of syntactic restrictions of first-order logic have equal expressive power. In particular, we consider the situation that in $\bB$ every first-order formula is equivalent to
\begin{itemize}
\item a quantifier-free formula (Section~\ref{sect:qe}),
\item an existential formula (Section~\ref{sect:mc}),
\item an existential positive formula (Sections~\ref{sect:cores} and~\ref{ssect:core-theories}).
%\item every first-order formula is equivalent to an existential positive formula (Section~\ref{sect:mc-cores}), or
\end{itemize}
For $\omega$-categorical structures, such a definability collaps 
translates nicely into a property of the operations that preserve $\bB$,
using the preservation theorems from Section~\ref{sect:pres}. A survey picture is given in 
Figure~\ref{fig:collaps}.
All these collapse results will be useful when studying the complexity of CSPs. 
For example, these results clarify when 
the so-called \emph{constraint entailment problem for $\bA$} can be reduced to the constraint satisfaction problem for $\bA$ 
(see Section~\ref{ssect:core-theories}).

We also develop a theory that can be viewed
as a positive variant of the classical theory of model-completeness and model companions (Section~\ref{ssect:core-theories}
and~\ref{ssect:core-companions}). This allows us to clarify the
question which CSPs can be formulated with an $\omega$-categorical template (Section~\ref{ssect:omega-cat-mc-cores}).

\subsection{Quantifier-elimination}
\label{sect:qe}
%BOOKTD: consider a discussion of QE outside of omega-categoricity.
We say that a $\tau$-structure $\bA$ admits \emph{quantifier elimination} 
if for every first-order $\tau$-formula 
%with at least one free variable
there exists an equivalent quantifier-free $\tau$-formula.

In this context, our assumption
that we allow $\bot$ as a first-order formula (denoting the empty $0$-ary relation) 
becomes relevant; Hodges~\cite{HodgesLong} does not make this assumption, and therefore has to distinguish between \emph{quantifier-elimination} and what he calls \emph{quantifier-elimination for non-sentences}. 
We will later often make use of the following fact.

\begin{lemma}[Statement 2.22 in~\cite{Oligo}] 
\label{lem:qe}
An $\omega$-categorical structure $\bB$ admits quantifier elimination if
and only if it is homogeneous. 
\end{lemma}
\begin{proof}
By Theorem~\ref{thm:ryll}, the orbit
of a $k$-tuple 
of elements of $\bB$ is first-order definable.
Suppose that $\bB$ has quantifier-elimination.
%Every first-order formula is equivalent over $\Gamma$
%to a quantifier-free formula, and therefore
Then two $k$-tuples $\bar a = (a_1,\dots,a_k)$
and $\bar b = (b_1,\dots,b_k)$ are in the same orbit if and only if
the mapping that sends $a_i$ to $b_i$, for $1 \leq i \leq n$,
is an isomorphism between the structures induced
by $\{a_1,\dots,a_k\}$ and by $\{b_1,\dots,b_k\}$.
This proves homogeneity. 

Now suppose that $\bB$ is homogeneous,
and let $\phi(x_1,\dots,x_k)$ be a first-order formula. 
By the theorem of Ryll-Nardzewski (Theorem~\ref{thm:ryll}), there are finitely many orbits $O_1,\dots,O_m$ of
orbits of $k$-typles that satisfy $\phi$. 
Clearly, it suffices to show that each of those
orbits can be defined by a quantifier-free formula. 
Let $a \in B^k$ be such that $\bB \models \phi(a)$. We claim that the set of
quantifier-free formulas that hold on $(a_1,\dots,a_k)$ defines the orbit of $a$ over $\bB$. To see this, let $(b_1,\dots,b_k)$ be another $k$-tuple that satisfies the same quantifier-free formulas as $(a_1,\dots,a_k)$.
Then the mapping that sends $a_i$ to $b_i$
is a partial isomorphism, and  by homogeneity can be extended to an automorphism of $\bB$. Since automorphisms preserve first-order formulas, $(b_1,\dots,b_k)$ also satisfies $\phi$, which proves the claim. 
\end{proof}

% THE FOLLOWING IS OBSOLETE since we have "false"
% and therefore also \not false
%We made the assumption that we restrict to formulas with at least one free variable, since first-order \emph{sentences}, i.e., first-order formulas $\phi$ without free variables, are preserved by all partial isomorphisms of $\bB$; but when $\bB$ contains no constant symbols there is no quantifier-free formula that is equivalent
%to $\phi$. Because of this restriction, Hodges~\cite{Hodges} distinguishes between
%\emph{quantifier-elimination} and \emph{quantifier-elimination for non-sentences}. 
%The difference between the two definitions is not relevant in the following.

\subsection{Model-Completeness}
\label{sect:mc}
The purpose of this section is to recall classical results about
model-completeness; they inspired the new results of the next section about model-complete cores. A theory $T$ is \emph{model-complete} if every embedding 
between models of $T$ is elementary, i.e., preserves
all first-order formulas. There are several 
equivalent characterizations of model-completeness,  
stated in Theorem~\ref{thm:mc} below. 

\begin{theorem}[Theorem~7.3.1 in~\cite{Hodges}]\label{thm:mc}
Let $T$ be a theory. Then the following are equivalent.
\begin{enumerate}
\item $T$ is model-complete.
\item Every model of $T$ is an existentially closed model of $T$.
\item Every first-order formula is equivalent to an existential formula 
modulo $T$.
\end{enumerate}
\end{theorem}
For the proof, we refer to~\cite{Hodges}; but note that the theorem has a positive variant (Theorem~\ref{thm:mc-core} below)
with an analogous proof that will be presented in full length. 

\begin{example}
The structure $({\mathbb Q}^+_0; <)$, where ${\mathbb Q}^+_0$ denotes the non-negative rational numbers, is not
model-complete, because the self-embedding $x \mapsto x+1$ of $({\mathbb Q}^+_0; <)$ does not preserve the formula 
$\phi(x) = \forall y \, (y > x \Rightarrow \exists z \, (x < z \wedge z < y))$ (which is
satisfied only by $0$). 
\end{example}

When $\mathfrak A$ is not model-complete,
we can sometimes find a model-complete structure $\mathfrak B$ that
satisfies the same universal first-order sentences as $\mathfrak A$.

\begin{definition}
A theory $U$ is a \emph{model companion} of a theory $T$ if 
\begin{itemize}
\item $U$ is model-complete;
\item Every model of $U$ embeds into a model of $T$; and 
\item every model of $T$ embeds into a model of $U$.
\end{itemize}
\end{definition}

Note that the last two conditions in this definition are equivalent to saying that 
$U$ and $T$ imply exactly the same existential 
sentences (equivalently, the same universal sentences);
the proof is analogous to the one of Proposition~\ref{prop:companions}. 

If $T$ has a model-companion $U$, then $U$ is unique 
up to equivalence of theories.

\begin{theorem}[Theorem 7.3.6.\ in~\cite{Hodges}]\label{thm:mc-unique}
For any two model-companions $U_1,U_2$ of a theory $T$ 
we have that $U_1 \vdash U_2$ and $U_2 \vdash U_1$.
\end{theorem}

The following theorem by Simmons~\cite{Simmons} 
%BOOK VERSION: (with an extension by Cherlin, Shelah, and Shi~\cite{CherlinShelahShi})
will not be used in this thesis; however, it has an existential positive version, Theorem~\ref{thm:omega-cat-core-companion} below, 
which has important consequences for the study of the CSP. 
Recall the \emph{joint embedding property}, which has been defined for classes 
of structures in Section~\ref{sect:fraisse}; a \emph{theory} $T$ has the joint embedding property (JEP) if for any 
two models $\bB_1,\bB_2$ of $T$ there exists a model $\bC$ of $T$ that embeds both $\bB_1$ and $\bB_2$.

\begin{theorem}[from~\cite{Simmons}]% and~\cite{CherlinShelahShi}
\label{thm:simmons}
Let $T$ be a theory with the JEP. Then the following are equivalent.
\begin{itemize}
\item $T$ has an $\omega$-categorical model companion.
\item For every $n$, $T$ has finitely many maximal existential $n$-types.
%\item In any existentially closed model of $T$ the algebraic closure of finite sets is empty~\cite{CherlinShelahShi}.
\end{itemize}
In particular, every $\omega$-categorical theory has an $\omega$-categorical model companion~\cite{Saracino}.
\end{theorem}

The consequence stated for $\omega$-categorical theories $T$ at the end of Theorem~\ref{thm:simmons} is an earlier result by Saracino~\cite{Saracino}, and clearly follows from the first part.

We say that a structure $\bA$ is model-complete if and only if 
the first-order theory $\Th(\bA)$ of $\bA$ is model-complete.
As we see below, for $\omega$-categorical structures $\bA$ model-completeness of $\bA$ can be translated into a property of the self-embedding monoid of $\bA$, 
and into a property concerning the axiomatization of Th$(\bA)$.
In the following theorem, the equivalence of (1) and (4) can be found in~\cite{RandomMinOps}.
The implication from (5) to (1) has been observed in~\cite{tcsps-journal}.

\begin{theorem}\label{thm:mc-omegacat}
Let $\bB$ be $\omega$-categorical. Then the following are equivalent.
\begin{enumerate}
\item The structure $\bB$ is model-complete.
\item Th$(\bB)$ is equivalent to a $\forall\exists$-theory.
\item $\bB$ has a homogeneous expansion by
relations $R_1,R_2,\dots$ such that both the $R_i$ and their complements have
 existential definitions in $\bB$.
\item Every self-embedding of $\bB$ is locally generated by
the automorphisms of $\bB$.
\item Every self-embedding of $\bB$ is locally invertible (see Section~\ref{ssect:loc-invert}).
\end{enumerate}
\end{theorem}
\begin{proof}
The implication from (1) to (2) holds for all structures
$\bB$ (we do not need $\omega$-categoricity; see e.g.~Theorem 7.3.3 in~\cite{Hodges}). The reverse direction is a direct consequence of a result known 
as \emph{Lindstr\"om's test} (Theorem~7.3.4. in~\cite{Hodges}).

We now prove (1) $\Rightarrow$ (3) 
$\Rightarrow$ (4) $\Rightarrow$ (5) $\Rightarrow$ (1).

% This is (1) -> (4). 
%Suppose that (1) holds. 
%Let $e$ be a self-embedding of $\bB$,
%and let $t$ be a finite tuple of elements of $B$. By Theorem~\ref{thm:mc}, every first-order definable
%relation has an existential definition in $\bB$. 
%Since $e$ preserves existential formulas,
%$e(t)$ satisfies the same first-order sentences as $t$, and by Theorem~\ref{thm:ryll} there
%is also an automorphism that sends $t$ to $e(t)$. Hence, $e$ is in the closure of $\Aut(\bB)$.

Suppose that (1) holds. 
By Theorem~\ref{thm:mc}, every first-order definable relation has an existential definition in $\bB$. Hence, when we expand $\bB$ by all 
existentially definable relations, every first-order formula has a quantifier-free definition.
So Lemma~\ref{lem:qe} shows
that the expansion is homogeneous.

% WRONG WITHOUT THE ASSUMPTION ABOUT THE COMPLEMENTS: (think of 
% semi-linear order, which is not mc)
Now suppose that (3) holds. We claim that every self-embedding
$e$ of $\bB$ is in the closure of the automorphisms
of $\bB$. The restriction $e'$ of $e$ to a finite subset $S$ of the domain of $\bB$
is an isomorphism between finite induced substructures of $\bB$, 
and also an isomorphism between the expansion of $\bB$ by all existentially definable relations. 
%(HERE WAS THE MISTAKE WITHOUT THE STRONGER ASSUMPTION)
Homogeneity of this expansion implies that $e'$ can be extended to an automorphism of $\bB$, which proves the claim. 

The implication from (4) to (5) is trivial. Finally, the implication from (5) to (1) 
it is a direct consequence of Theorem~\ref{thm:loc-invertible}.
\end{proof}

Note that all finite structures are model-complete: self-embeddings of $\bB$
are automorphisms, and hence they are elementary. Every relation that is first-order definable in a finite structure also has an existential definition. 

Using the concept of model completeness, we can restate Theorem~\ref{thm:universal}, 
clarifying in which sense the structure $\bB$ constructed in Theorem~\ref{thm:universal} is unique. In the following, $\tau$ is a finite relational signature. 

\begin{theorem}[a variant of Theorem~\ref{thm:universal}]\label{thm:css-strong}
Let $\cal N$ be a finite set of finite connected $\tau$-structures.
Then there is a model-complete $\tau$-structure $\bB$ whose age is the class of all finite $\cal N$-free structures. 
The structure $\bB$ is unique up to isomorphism.
\end{theorem}
\begin{proof}
Theorem~\ref{thm:universal} states the existence of an $\omega$-categorical $\cal N$-free structure 
which is universal for the class of all countable $\cal N$-free structures which has a homogeneous
expansion by primitive positive definable relations. 
By (3) $\Rightarrow$ (1) in Theorem~\ref{thm:mc-omegacat}, 
the structure $\bB$ is indeed model-complete. 
We have to show that every model-complete structure $\bC$ 
with the same age as $\bB$ is isomorphic to $\bB$. 
Let $T$ be the first-order theory of $\bB$, and $S$ be the first-order theory of $\bC$. Since $\bB$ and $\bC$ have
the same age, $S$ and $T$ imply the same existential sentences. 
By Theorem~\ref{thm:mc-unique}, $S$ and $T$ are equivalent theories. 
By $\omega$-categoricity of $T$,  $\bB$ and $\bC$ are isomorphic.
\end{proof}

\subsection{Cores}
\label{sect:cores}
We have already encountered the concept of a \emph{core} of a finite structure
in Section~\ref{sect:homo}. To recall, a \emph{core} is a structure $\bB$
such that all endomorphisms of $\bB$ are embeddings, and a structure $\bB$ is
\emph{a core} of $\bA$ if $\bB$ is a core and homomorphically equivalent to $\bA$. The concept of the core of a finite relational structure plays an important role in the classification program for finite-domain CSPs.
Three crucial properties of finite cores are:
\begin{itemize}
\item every finite structure $\bA$ has a core $\bB$ (Proposition~\ref{prop:finite-cores});
\item all core structures $\bB$ of $\bA$ are isomorphic (Proposition~\ref{prop:finite-cores});
\item orbits of $k$-tuples in finite cores $\bB$ are primitive positive definable (Proposition~\ref{prop:orbits-in-finite-cores}).
\end{itemize}
Also for every \emph{infinite} structure $\bA$ there is 
a core $\bB$ such that $\Csp(\bA) = \Csp(\bB)$. This follows from Lemma~\ref{lem:existence-epc}
and the following proposition.

\begin{proposition}
\label{prop:epc-cores}
If $\bB$ is an epc model for its universal negative theory, then $\bB$ is a core. 
\end{proposition}
\begin{proof}
Suppose $\bB$ is epc for its universal negative theory $T$, and let $h$ be an endomorphism of $\bB$. By epc, for $b_1,\ldots,b_k$ in $B$, if $\bB \models R(h(b_1),\ldots,h(b_k))$ or $\bB \models (h(b_1)=h(b_2))$, then $\bB \models R(b_1,\ldots,b_k)$ or $\bB \models (b_1= b_2)$, respectively. It follows that $h$ is an embedding.
\end{proof}

There are many equivalent definitions of when
a \emph{finite} structure is a core: for example, 
a finite structure is a core if and only if
all endomorphisms are surjective, or injective, or bijective, or all endomorphisms are automorphisms. 
For infinite structures (even when they are $\omega$-categorical), these definitions are in general not equivalent, 
see~\cite{InfCores,InfDigraphCores,Cores-journal}. 
As we will see in this section, our definition of
cores is the most appropriate definition in many contexts, in particular in the context of constraint satisfaction for  $\omega$-categorical templates.

\begin{example}
The structure $(\mathbb Q;<)$ is easily seen to
be a core: every endomorphism of $(\mathbb Q;<)$ must be injective, and must be strong.
In contrast, the random graph $({\mathbb V};E)$ is not a core. By the defining
property of the random graph, $({\mathbb V};E)$ contains arbitrarily large finite cliques.
By Lemma~\ref{lem:infinst}, it even has an infinite clique as a subgraph.
Therefore, $({\mathbb V};E)$ has endomorphisms with the property that they map pairs
of non-adjacent vertices to pairs of adjacent vertices, and thus is not a core.
\end{example}

Before we prove general results about existence and uniqueness of cores, we state important properties
of cores and model-complete cores 
for $\omega$-categorical structures that follow in a straightforward way from previous facts\footnote{Yet another characterization of when an $\omega$-categorical structure is a model-complete core can be found in Proposition~\ref{prop:pos-restr}.}.

\begin{theorem}\label{thm:mc-core}
Let $\bB$ be $\omega$-categorical. 
Then $\bB$ is a core if and only if every existential formula is equivalent to an existential positive formula over $\bB$. Moreover, the following are equivalent.
\begin{enumerate}
\item $\bB$ is a model-complete core; % every mapping preserves formula
\item $\bB$ has a homogeneous expansion by relations $R_1,R_2,\dots$ such that the relations $R_i$ and their complements have existential positive definitions; 
% (need the complements: ow, take the random graph: it IS already homogeneous, but is not a core.)
\item Every first-order formula is equivalent to an existential positive one over $\bB$; % formula-formula
\item The orbits of $n$-tuples in $\bB$ are primitive positive definable in $\bB$; % every preserved thing definable
\item The automorphisms locally generate the endomorphisms of $\bB$. % mapping-mapping
\end{enumerate}
\end{theorem}
\begin{proof}
The first statement is straightforward from Theorem~\ref{thm:ep}.
To prove the equivalence of (1)-(5), we show implications
in cyclic order. 

For the implication from (1) to (2), 
consider the expansion of $\bB$ 
by all relations with an existential positive definition in $\bB$. 
Since all endomorphisms of $\bB$ also
preserve the complements of those relations,
the complements also have an existential positive definition by Theorem~\ref{thm:ep},
and hence the expansion is of the desired type. The orbits of $n$-tuples of $\bB$ (and its expansion) are by assumption preserved by all endomorphisms of $\bB$, and therefore have an existential positive definition in $\bB$, and thus a quantifier-free definition in the expansion. It follows that the expansion is homogeneous.

For the implication from (2) to (3), 
let $\phi$ be a first-order formula. Then $\phi$
has in the homogeneous expansion of $\bB$
a quantifier-free definition $\psi$; assume without loss of generality that $\psi$ is written
in conjunctive normal form. If we replace all positive literals that involve relations $R_i$ by their existential positive definition in $\bB$,
and all negative literals that involve relations
$R_i$ by the existential positive definition of the complement of $R_i$ in $\bB$, we arrrive
at an equivalent formula which is existential positive in the signature of $\bB$. 

%For the implication from (1) to (2), assume (1) and let $\phi$
%be a first-order formula. To show that $\phi$ is equivalent in $\bB$ to
%an existential positive formula, it suffices to 
%show that every endomorphism of $\bB$ preserves $\phi$, 
%by Theorem~\ref{thm:ep}. Since $\bB$ is a model-complete core,
%every endomorphism of $\bB$ preserves all first-order formulas. 

%For the implication from $(2)$ to $(3)$,
%expand $\bB$ by all relations with an existential positive definition in $\bB$; 
%since every first-order definable relation also has an existential positive definition, 
%the expanded is of the required type.
%Moreover, since in $\omega$-categorical structures orbits of $n$-tuples are first-order definable (Theorem~\ref{thm:ryll}),
%the expanded structure is homogeneous.

For the implication from (3) to (4), let 
$O$ be an orbit of $n$-tuples in $\bB$. By Theorem~\ref{prop:inv-aut-omega-cat}, $O$ has a first-order definition. Assuming (2), $O$ even has an existential positive
definition. Note that every existential positive formula can be written as a disjunction of primitive positive formulas, so let $\phi$ be
such a definition of $O$. 
We can also assume without loss of generality that none of the 
disjuncts in $\phi$ implies another (otherwise, we simply omit it). 
Since $O$ is a minimal first-order definable relation, $\phi$ 
can only contain a single disjunct, and therefore is primitive positive. 

(4) implies (5). Assume (4), and let $e$ be an endomorphism of $\bB$. To show that $e$ is locally generated by the automorphism
of $\bB$, let $t$ be a finite tuple of elements of $\bB$. We have to show that there is an automorphism $\alpha$ of $\bB$ such that $e(t)=\alpha(t)$. The orbit of $t$ is primitive positive definable, 
and hence preserved by $e$. So $e(t)$ is in the same orbit as $t$,
and we are done. 

(5) implies (1). 
Suppose (5), that is, suppose that all endomorphisms are generated
by the automorphisms of $\bB$. Since the automorphisms
preserve all first-order formulas in $\bB$, the same is true for the endomorphisms of $\bB$, by Proposition~\ref{prop:loc-clos-monoid}.
\end{proof}

The fact that in $\omega$-categorical model-complete cores the orbits of $n$-tuples are primitive positive definable
is one of the three key facts for finite cores $\bB$ that we have mentioned above. 
The other two facts, existence and uniqueness of model-complete cores for $\omega$-categorical
structures, follow from more general theorems that apply not only
to $\omega$-categorical structures, as we will see in Section~\ref{ssect:omega-cat-mc-cores}.

\ignore{%with assumption on complement, 
%can perfectly merge this into the previous theorem
We finally note the following consequence of Theorem~\ref{thm:mc-core} for future use. 
% I didn't integrate this into the above theorem, since it is not an iff characterization of mc cores. 
\begin{corollary}
Every $\omega$-categorical model-complete core $\bB$ has a homogeneous expansion by primitive positive definable relations.
\end{corollary}
\begin{proof}
When $\bB$ is an $\omega$-categorical model-complete core, then by Item (3) in Theorem~\ref{thm:mc-core}, all orbits of $n$-tuples are primitive positive definable. So it suffices to show that the expansion $\bC$ by \emph{all}
orbits of $n$-tuples in $\bB$ is homogeneous. Let $i$ be an isomorphism between two finite induced substructures of $\bC$,
and let $\{s_1,\dots,s_n\}$ be the domain of $i$. Since $i$ preserves the orbits of $n$-tuples of $\bB$, 
$(s_1,\dots,s_n)$ lies in the same orbit as $(t_1,\dots,t_n)$. Hence, there exists an automorphism of $\bB$ and therefore also of $\bC$ that extends $i$. \end{proof}
}

%\vspace{.3cm}
\subsection{Core Theories}
\label{ssect:core-theories}
A theory $T$ is called a \emph{core theory} if every homomorphism between models of $T$ is an embedding. Note that
a finite or $\omega$-categorical structure $\bB$ is a core if and only if it has a core theory
(for $\omega$-categorical $\bB$, this is an easy consequence of the L\"owenheim-Skolem theorem -- Theorem~\ref{thm:LS}).

\begin{proposition}\label{prop:core-theories}
Let $T$ be a first-order $\tau$-theory. Then $T$ is a core theory
if and only if every existential formula is equivalent to an
existential positive formula.
\end{proposition}
\begin{proof}
First assume that $T$ is a core theory, and let $\phi$ be an existential formula. Then $\phi$ is preserved by all embeddings between models
of $T$. Since all homomorphism between models of $T$ are embeddings, $\phi$ is also preserved by all homomorphisms between
models of $T$. Hence, Theorem~\ref{thm:homo-pres} implies
that $\phi$ is equivalent modulo $T$ to an existential positive formula.
The converse implication is trivial.
\end{proof}

We would like to point out an interesting corollary 
for CSPs. Let $\bB$ be a structure with finite relational signature $\tau$.
The \emph{constraint entailment problem for $\bB$}
is the following computational problem. The input consists of a
primitive positive $\tau$-formula $\phi$, and a single atomic
$\tau$-formula $\psi$, both $\phi$ and $\psi$
 with free variables $x_1,\dots,x_n$. The question is whether
 $\phi$ implies (\emph{entails}) $\psi$ in $\bB$, 
 i.e., whether $$\bB \models \forall x_1,\dots,x_n \, (\phi \Rightarrow \psi) \; .$$

\begin{corollary}
Let $\tau$ be a finite relational signature, and 
let $\bB$ be a $\tau$-structure whose first-order theory $T$ is
a core theory. Then there is a polynomial-time Turing reduction
from the
constraint entailment problem for $\bB$ 
to $\Csp(\bB)$.
\end{corollary}

\begin{proof}
Let $\phi,\psi$ be an input to the constraint entailment problem for 
$\bB$. Since $T$ is a core theory, $\neg \psi$
is by Proposition~\ref{prop:core-theories} equivalent to
an existential positive $\tau$-formula, and hence equivalent to 
a disjunction $\psi_1 \vee \dots \vee \psi_m$
of primitive positive formulas. Since the signature
$\tau$ is finite, we can consider the size of this disjunction is bounded by a constant, for all possible inputs. 
Then $\phi$ entails $\psi$ if and only if
for all $1 \leq i \leq m$, we have that $\exists x_1,\dots,x_k \, 
(\phi \wedge \psi_i)$ is false in $\bB$.
Like this we have reduced the entailment problem
to solving a constant number of constraint satisfaction problems
for the structure $\bB$.
\end{proof}

%OLD PROOF \begin{proof}
%All orbits of $k$-tuples of Aut$(\mathfrak A)$ 
%are primitive positive definable, by Theorem~\ref{thm:mc-core}. 
%Let $O_1,\dots,O_m$ be the orbits of $k$-tuples
%that do not satisfy $\phi$, and let $\alpha_i$, for $1 \leq i \leq m$,
%be a primitive positive definition of $O_i$ in $\mathfrak A$.
%Then $\psi$ entails $\phi$ if and only if
%for all $1 \leq i \leq m$, we have that $\exists x_1,\dots,x_k. \; 
%\alpha_i \wedge \psi$ is false in $\mathfrak A$.
%Since $k$ is bounded by the maximal arity of relations from $\tau$,
%the number $m$ is for fixed $\mathfrak A$ a constant,
%and therefore we have reduced the entailment problem
%to solving a constant number of constraint satisfaction problems
%for the structure $\mathfrak A$.
%\end{proof}

If we combine the assumption that $T$ is a core theory with the
assumption that it is model-complete, we arrive at Theorem~\ref{thm:mccoretheory}.
Its proof closely follows the proof of Theorem~7.3.1 in~\cite{Hodges}.

\begin{theorem}\label{thm:mccoretheory}
Let $T$ be a first-order theory over signature $\tau$. 
Then the following are equivalent.
\begin{enumerate}
\item $T$ is a model-complete core theory.
\item Every model of $T$ is an existential positive complete model of $T$.
\item If $\bA,\bB$ are models of $T$ and $h$
is a homomorphism from $\bA$ to $\bB$ then there are an elementary extension $\mathfrak C$ of $\bA$ and an embedding $g$ of $\mathfrak B$ into $\mathfrak C$ such that $gh$ is the identity on $\mathfrak A$. 
\item Every first-order formula is equivalent to an existential positive 
formula modulo $T$.
\end{enumerate}
\end{theorem}
\begin{proof}
(1) implies (2) is immediate from the definition of epc~models:
if $\bA$ and $\bB$ are models of $T$ and $h \colon A \rightarrow B$ is 
a homomorphism from $\bA$ to $\bB$, then $h$ must
be an embedding since $T$ is a core theory, and in fact must be
elementary since $T$ is model-complete. Hence, for every tuple
$\bar a$ from $A$ and any existential positive formula $\phi$
such that $h(\bar a)$ satisfies $\phi$ we have that $\bar a$ also 
satisfies $\phi$.

(2) implies (3). Assume (2).
Let $\bA$ and $\bB$ be models of $T$, and let
$h$ be a homomorphism from $\bA$ to $\bB$. 
Choose $\bar a$ to
be a vector that enumerates the elements of $\bA$. Since 
$\bA$ is an epc model of $T$, $h$ is an embedding.
Hence, every existential sentence that holds in $(\bB,h(\bar a))$ also holds in $(\bA,\bar a)$. 
Proposition~\ref{prop:existential-amalgamation} 
now directly implies (3).

(3) implies (4).
We first claim that if (3) holds, then every homomorphism
between models of $T$ preserves all universal $\tau$-formulas.
For if $h$ is a homomorphism of $\bA$ into $\bB$, 
$\bar a$ a tuple from $A$ and $\phi(\bar x)$ 
a universal $\tau$-formula such that 
$\bA \models \phi(\bar a)$, then taking $\bC$ and $g$ as in (3)
we have $\bC \models \phi(g(h(\bar a)))$ and so $\bB \models \phi(h(\bar a))$ since $\phi$ is a universal formula. This proves the claim.
It follows from Theorem~\ref{thm:homo-pres} that all universal 
$\tau$-formulas are equivalent to existential positive $\tau$-formulas.

To finally prove (4), let $\phi(\bar x)$ be any first-order $\tau$-formula, wlog. in prenex normal form. 
By a simple induction on the number of quantifier-blocks we can transform $\phi$ to an existential formula, using the fact that the innermost quantifier block
is either existential or universal, and can therefore be transformed
into an existential formula (see Theorem~7.3.1 in~\cite{Hodges}).
Finally, existential $\tau$-formulas are preserved by homomorphisms between models of $T$, since such homomorphisms must be embeddings. Hence, the entire formula is even equivalent
to an existential positive formula by Theorem~\ref{thm:homo-pres}.

(4) implies (1).
Existential positive formulas are preserved by homomorphisms between models
of $T$. %Hence, when every first-order formula is equivalent to an existential positive
%one, then the homos between models preserve all first-order formulas
\end{proof}

From this we obtain a positive version of a fact known as \emph{Lindstr\"om's test} (Theorem 7.3.4 in~\cite{Hodges}).

\begin{proposition}\label{prop:pos-lindstroem}
Let $T$ be a $\lambda$-categorical $\tau$-theory, for $\lambda \geq |\tau|$, which has no finite models, and whose unique model of cardinality $\lambda$ is epc for $T$.
Then $T$ is a model-complete core theory.
\end{proposition}
\begin{proof} 
We prove that every model of $T$ is an epc model of $T$ and use
Theorem~\ref{thm:mccoretheory}.
So let $\mathfrak A$ and $\mathfrak B$ be two models of $T$ and
let $h$ be a homomorphism from $\mathfrak A$ to $\mathfrak B$.
Let $\bar a$ be a tuple such that $\bB \models \phi(h(\bar a))$ and
suppose for contradiction that $\bA \not\models \phi(\bar a)$.
Then we can put those two structures into a new 2-sorted structure (comprising $\bA$, $\bB$, and the homomorphism
$h$ between them) with first-order theory $T'$, and 
apply the L\"owenheim-Skolem theorem (Theorem~\ref{thm:LS}; here we use the assumption that $\lambda \geq |\tau|$) to produce a countable model of $T'$ (where both sorts have the same cardinality since $T$ has no finite models).
By applying L\"owenheim-Skolem again, this time to $T'$ augmented by sentences expressing a bijection between
the two sorts over a signature expanded by a new function symbol,
we obtain a two-sorted model of $T'$
where each sort has cardinality $\lambda$, inducing structures $\mathfrak C$ and $\mathfrak D$, respectively. 
By assumption there exists an epc model of
cardinality $\lambda$,
and by $\lambda$-categoricity $\mathfrak C$ is an epc model of $T$. This contradicts the fact that we can express in $T'$ that
$\mathfrak A$ is not an epc model of $T$.
\end{proof} 

\begin{proposition}\label{prop:mccoretheory-is-fep}
Let $T$ be a model-complete core theory. Then $T$ is equivalent to a $\forall\exists^+$-theory.
\end{proposition}
\begin{proof}
This is an immediate consequence 
of Theorem~\ref{thm:pos-chains} (where the theory denoted by $T$ in Theorem~\ref{thm:pos-chains} is empty and $\Phi$ from Theorem~\ref{thm:pos-chains} equals the theory $T$ from the statement here) because for any sequence
$(\bB_i)_{i < \kappa}$ of models of $T$ with homomorphisms $g_{ij} \colon \bB_i \rightarrow \bB_j$, the $g_{ij}$ are elementary. By
the Tarski-Vaught theorem (Theorem~\ref{thm:tarski-vaught}), 
we have that $(\lim_{i<\kappa} \bB_i) \models T$.
\end{proof}

\subsection{Core Companions}
\label{ssect:core-companions}
In this section we study when we can pass from a theory $T$
to a model-complete core theory $T'$ that has the same CSP. 

\begin{definition}\label{def:core-companion}
Let $T$ be a first-order $\tau$-theory. Then a $\tau$-theory  $U$
is called a \emph{core companion} of $T$ if
\begin{itemize}
\item  $U$ is a model-complete core theory;
\item every model of  $U$ homomorphically maps to a model of $T$;
\item every model of $T$ homomorphically maps to a model of  $U$.
\end{itemize}
\end{definition}

Recall from Proposition~\ref{prop:companions} that 
the last two items in Definition~\ref{def:core-companion} are equivalent
to requiring that $T$ and $U$ imply the same universal
negative sentences. 

% MOVED THE FOLLOWING INTO THE LOGIC CHAPTER
%This follows from the following.
% I WOULD LIKE TO HAVE BOTH DIRECTIONS IN ONE STATEMENT
% SINCE IT SEEMS THAT WE ONLY NEED THE PROP IN THIS FORM
%\begin{proposition}
%Let $T$ be a first-order theory, and let $T_{\forall^-}$ be the universal negative
%consequences of $T$. Then the models of $T_{\forall^-}$ are precisely the structures
%$\bB$ such that there is a homomorphism from a model $\bC$ of $T$ to $\bB$.
%\end{proposition}

% be a first-order theory, and let $T_{\forall^-}$ be the universal negative
%consequences of $T$. Then the models of $T_{\forall^-}$ are precisely the structures
%$\bB$ such that there is a homomorphism from a model $\bC$ of $T$ to $\bB$.
%\end{proposition}

\begin{proposition} \label{prop:uniqueness-core-companion}
Let $T$ be a $\forall\exists^+$-theory 
with signature $\tau$.
If $T$ has a core companion $U$, then $U$ is up to equivalence of theories unique, and is the theory of the class of all epc~models
of $T$. 
\end{proposition}
\begin{proof}
It suffices to show that the epc models of $T$ are precisely the models of  $U$.
We first show that every model $\bB$ of  $U$ is an epc model for $T$;
that is, we have to show that $\bB$ is a model of $T$, 
and that $\bB$ is epc for $T$. 

Since $U$ is a core companion of $T$, 
there is a homomorphism $e$ from $\bB$ to a model $\bA$ of $T$.
The assumption that $U$ is a core companion of $T$
also implies that there 
exists a homomorphism $f$ from $\bA$ into a model $\bC$ of $U$.
Then $f \circ e$ is a homomorphism between two models of  $U$,
and since $U$ is a model-complete core theory it must be an
elementary embedding. 
This shows in particular that $e$ is an embedding. 

%If $\phi(\bar x)$ is an existential positive $\tau$-formula and $\bar b$ is a 
%tuple of elements of $\bB$ such that $\bA \models \phi(e(\bar b))$, 
%then $\bC \models \phi(f(e(\bar b)))$
%since homomorphisms preserve existential positive formulas.
%Therefore, $\bB \models \phi(f(e(\bar b)))$ since $fe$ is elementary. 

We claim that $\bB$ is a model of the $\forall\exists^+$-theory $T$.
Let $\phi = \forall \bar y. \psi$ be a sentence from $T$ where
$\psi$ is a disjunction of existential positive and negated atomic $\tau$-formulas, and let $\bar b$ be a tuple from $\bB$.
Since $\bA$ is a model of $T$ and therefore satisfies $\forall y. \psi$,
in particular the tuple $e(\bar b)$ satisfies $\psi$. 
If $e(\bar b)$ satisfies a negated atom in the disjunction $\psi$
then $\bar b$ also satisfies $\psi$ as $e$ is an embedding.
Otherwise, $e(\bar b)$ satisfies an existential positive formula
in the disjunction $\psi$, and
$f(e(\bar b))$ satisfies $\phi$ in $\bC$ as well since $f$ is a homomorphism. But this shows that
$\bar b$ satisfies $\psi$ in $\bB$ since $f \circ e$ is elementary. Since this
holds for all $\bar b$, we have proven that $\bB$ satisfies $\phi$. 

The verification that $\bB$ is an epc\ model for $T$ is similar, and as follows.
Let $g$ be a homomorphism from $\bB$ into another model $\bA$ of $T$, $\bar b$ a tuple from $\bB$, and $\phi$ an existential positive formula with $\bA \models \phi(g(\bar b))$. We have to show that $\bB \models \phi(\bar b)$.
Again, since $U$ is a core companion of $T$
there exists a homomorphism $h$ from $\bA$ into a model $\bC$ of
$U$. Since  $U$ is a model-complete core theory, the mapping
$h \circ g$ is elementary. 
Since $h$ preserves existential positive formulas, 
$\bC \models \phi(h(g(\bar b)))$. Since $h \circ g$ is elementary,
$\bB \models \phi(\bar b)$.

Conversely, we show 
that every epc model $\bB$ of $T$ satisfies  $U$.
By Proposition~\ref{prop:mccoretheory-is-fep}, $U$ is equivalent
to a $\forall\exists^+$-theory, 
and thus it suffices to show that $\bB$ satisfies all 
$\forall\exists^+$-consequences $\forall \bar y. \psi(\bar y)$
of  $U$, where $\psi$ is a disjunction of existential positive
and negative atomic $\tau$-formulas.
Let $\bar b$ be a tuple of elements of $\bB$. 
We have to show that
$\bB \models \psi(\bar b)$.
Since $U$ is a core companion, there is a homomorphism $h$
from $\bB$ to a model $\bA$ of  $U$. 
Since $\bA \models \forall \bar y. \psi(\bar y)$, 
at least one disjunct $\theta(h(\bar b))$ of $\psi$
is true in $\bA$. 
If $\theta$ is a negative atomic formula, then 
$\theta(\bar b)$ is also true in $\bB$ since $h$ is a homomorphism.
Now suppose that $\theta$ is an existential positive formula.
Since  $U$ is a core companion of $T$, there is a homomorphism $g$ from $\bA$ to 
a model $\bC$ of $T$. Since $g$ preserves $\theta$ we have that $\bC \models \theta(g(h(\bar b)))$.
Now $\bB \models \theta(\bar b)$, since $\bB$ is an epc model of $T$.
In both cases we can conclude that $\bB \models \psi(\bar b)$.
\end{proof}

\begin{proposition} \label{prop:existence-core-companion}
Let $T$ be a $\forall\exists^+$-theory with signature $\tau$.
Then $T$ has a core companion if and only if the class of epc models of $T$ is axiomatizable by a $\tau$-theory.
\end{proposition}
\begin{proof}
If $T$ has a core companion  $U$, 
then Proposition~\ref{prop:uniqueness-core-companion} above implies that  $U$ 
axiomatizes the epc models of $T$. 

For the converse, 
suppose that the class of epc models of $T$ is the class of all 
models of a $\tau$-theory  $U$. Then every model of  $U$
is in particular a model of $T$, and every model of $T$
homomorphically maps to a model of  $U$ by Lemma~\ref{lem:existence-epc}. So we only have to verify that  $U$ is a model-complete core theory to show that  $U$ is the core companion of $T$.
%Since the class of all epc models of $T$ is closed under direct limits by Lemma~\ref{lem:epc-limits}, 
%we can assume by Theorem~\ref{thm:pos-chains}
%that  $U$ is a $\forall\exists^+$-theory. 
Every model $\bA$ of  $U$ is an epc model of $T$, and in fact an epc model of $U$.
It follows by the equivalence of (1) and (2) in Theorem~\ref{thm:mccoretheory} that $U$ is a model-complete core theory.
\end{proof}

%\vspace{.3cm}
\subsection{$\omega$-categorical model-complete cores}
\label{ssect:omega-cat-mc-cores}
We have already seen in Theorem~\ref{thm:mc-core} that whether an $\omega$-categorical structure is
a model-complete core can be characterized in many different ways.
The results in Section~\ref{ssect:core-theories} 
provide a further characterization
of $\omega$-categorical model-complete cores in terms of their axiomatization, as we see in the following.

\begin{proposition}\label{prop:pos-restr}
Let $\bB$ be a countable $\omega$-categorical structure.
Then $\bB$ is a model-complete core if and only if its theory is equivalent to 
a $\forall\exists^+$-theory.
\end{proposition}
\begin{proof} 
One direction is Proposition~\ref{prop:mccoretheory-is-fep}.
For the other direction, 
assume that the theory of $\bB$ is equivalent to a $\forall\exists^+$-theory $T$. 
Then $T$ has no finite models,
and hence $T$ has a countably infinite model that is epc for $T$, by Lemma~\ref{lem:existence-epc}.
Hence, by Proposition~\ref{prop:pos-lindstroem}, $T$ is a model-complete core theory (since $T$ is $\omega$-categorical, it is easy to see that we can assume for this application that the signature of $T$ is countable), and
so $\bB$ is a model-complete core. 
%Since all countable models of $T$ are isomorphic, 
%the statement follows from the implication $(2) \Rightarrow (1)$ in Theorem~\ref{thm:mccoretheory}.
\end{proof}

We now present an existential positive version of Simmons' theorem (Theorem~\ref{thm:simmons}), which answers the question which CSPs can be formulated
with an $\omega$-categorical template.
For a satisfiable theory $T$, let $\sim_n^T$ be the equivalence relation defined on existential 
positive formulas with $n$ free variables $x_1,\dots,x_n$ (we could have equivalently used primitive positive formulas here) as follows.
For two such formulas $\phi_1$ and $\phi_2$, 
let $\phi_1 \sim^T_n \phi_2$
if for all existential positive formulas $\psi$ with free variables $x_1,\dots,x_n$
we have that $\{\phi_1, \psi\} \cup T$ is satisfiable if
and only if $\{\phi_2,\psi \} \cup T$ is satisfiable. 

%Let $T$ be a theory with the property that when $T \cup \{\phi\}$ is satisfiable and $T \cup \{\psi\}$ is satisfiable,
%then $T \cup \{\phi,\psi\}$ is satisfiable (confer Proposition~\ref{prop:sat-csp}).

\begin{theorem}
\label{thm:omega-cat}
Let $T$ be a theory with the joint homomorphism property (JHP; confer Proposition~\ref{prop:jhp}). 
Then the following are equivalent.
\begin{itemize}
\item[$(i)$] $T$ has a core companion that is either $\omega$-categorical or the theory of a finite structure.  
\item[$(ii)$] $\sim^T_n$ has finitely many equivalence classes for each $n$.
\item[$(iii)$] $T$ has finitely many maximal existential positive $n$-types for each $n$.
\item[$(iv)$] There is a finite or $\omega$-categorical model-complete core $\bB$ that satisfies an existential positive 
sentence $\phi$ if and only if $T \cup \{\phi\}$ is satisfiable. 
\end{itemize}
\end{theorem}

\begin{proof}
We show $(i) \Rightarrow (ii) \Rightarrow (iii) \Rightarrow (iv) \Rightarrow (i)$. 

$(i) \Rightarrow (ii)$.
Let $U$ be the core companion of $T$. 
Since $U$ and $T$ entail the same universal negative sentences, we can deduce that for every existential positive formula $\psi$
the theory $U \cup \{\psi\}$ is satisfiable if and only if $T \cup \{\psi\}$ is satisfiable; from which it follows that the indices of
$\sim_n^U$ and $\sim_n^T$ coincide. 

For a proof by contraposition, assume that $\sim_n^U$
has infinite index for some $n$. Let $\psi_1$ and $\psi_2$ be
two existential positive formulas from different equivalence classes of $\sim^U_n$. Hence, there is an existential positive formula $\psi_3$ such that exactly one of $\{\psi_1,\psi_3\} \cup U$ and $\{\psi_2,\psi_3\} \cup U$ is satisfiable. This shows that
$\psi_1$ and $\psi_2$ are inequivalent modulo $U$.
Therefore there are infinitely many first-order formulas with
$n$ variables that are inequivalent modulo $U$, 
and $U$ can neither be $\omega$-categorical by Theorem~\ref{thm:ryll} nor the theory of a finite structure.

$(ii) \Rightarrow (iii)$.
We show that every maximal ep-$n$-type $p$ is determined completely by the $\sim^T_n$ equivalence classes of the existential positive formulas contained in $p$. Since there are finitely many such classes, the result follows. Let $p$ and $q$ be maximal ep-$n$-types such that for every $\phi_1 \in p$ there exists $\phi'_1 \in q$ such that $\phi_1\sim^T_n \phi'_1$ and for every $\phi_2 \in q$, there exists $\phi'_2 \in p$ such that $\phi_2\sim^T_n \phi'_2$. We aim to prove that $p=q$. If not then there exists, without loss of generality, $\psi \in p$ such that $\psi \notin q$. Since $q$ is maximal, $T \cup q \cup \{\psi\}$ is not satisfiable. By compactness, $T \cup \{\theta, \psi\}$ is not satisfiable for some finite conjunction $\theta$ of formulas from $q$. Now, $\theta \in q$ by maximality and there exists by assumption $\theta' \in p$ such that  $\theta \sim^T_n \theta'$. By definition of $\sim^T_n$ we deduce $T \cup \{ \theta', \psi\}$ satisfiable iff $T \cup \{ \theta, \psi\}$ satisfiable. Since the latter is not satisfiable, we deduce that neither is the former, which yields the contradiction that $T \cup p \cup \{\psi\}$ is not satisfiable.

$(iii) \Rightarrow (iv)$.
An existential positive formula $\phi(\bar{x})$ is said to \emph{isolate} a maximal ep-$n$-type $p(\bar{x})$ of $T$, if $p$ is the only maximal ep-$n$-type of $T$ of which $\phi$ is a member. If there is only a finite number of maximal ep-$n$-types of $T$, then it follows that each has an isolating formula. 
% LATER-TD: expand previous sentence.
Assume $(iii)$, 
and let $S$ be the set of all existential positive sentences $\phi$
such that $T \cup \{\phi\}$ is satisfiable, together with the set of all universal negative consequences of $T$.
By Proposition~\ref{prop:jhp}, $S$ has a model $\bC$, and by Theorem~\ref{thm:LS}
we can assume that $\bC$ is either finite or countable.
Lemma~\ref{lem:existence-epc} gives a homomorphism from $\bC$ to a 
finite or countable epc $\tau$-model $\bB$ of $S$. Note that also $\bB$ satisfies exactly those existential positive
sentences that are satisfiable together with $T$.
We consider the signature $\tau'$, which is the expansion of $\tau$ by $\mu_n$ relations of each arity $n$, 
corresponding to the maximal pp-$n$-types of $T$. 
Any model of $T$ has a canonical (unique) expansion to a $\tau'$-structure, 
by the new relation symbols labeling tuples that attain their type. 
Consider this canonical $\tau'$-expansion $\bB'$ of $\bB$.
We will shortly prove that $\bB'$ is homogeneous. 
From this it will follow that $\bB'$ and $\bB$ are finite, or $\omega$-categorical 
by Lemma~\ref{lem:cat-via-homogen} (since variable identifications are primitive positive, there is only a finite number of inequivalent atomic formulas of each arity $n$), 
whereupon $\omega$-categoricity is inherited by its $\tau$-reduct $\bB$.

To prove that $\bB$ is a model-complete core, we 
use Theorem~\ref{thm:mc-core}, and show that
every first-order formula $\phi$ is equivalent to an existential
positive formula over $\bB$. Since $\bB$ has a homogeneous expansion $\bB'$ by primitive positive definable relations, $\phi$ is
equivalent to a Boolean combination of primitive positive
formulas. 
Because we have added a relation symbol for 
 each maximal pp-$n$-type of $T$, and there are finitely
many of those for each $n$,
 the negation of a primitive positive formula is equivalent to a finite disjunction of maximal pp-$n$-types. This proves 
  that $\phi$ is in $\bB$ equivalent 
 to an existential positive formula. 

It remains to be shown that $\bB'$ is homogeneous. 
Let $f \colon (a_1,\ldots,a_m)\mapsto (b_1,\ldots,b_m)$ be a partial automorphism of $\bB'$ (in the signature $\tau'$). Let $a'$ be an arbitrary element of $B'$. Consider the ep-$n$-types $p(x_1,\ldots,x_m)$ of $(a_1,\ldots,a_m)$ and $q(x_1,\ldots,x_m,y)$ of $(a_1,\ldots,a_m,a')$ in $\bB$. 

By Proposition~\ref{prop:epc-types}, each of these types is maximal, and is isolated by the ep-formulas $\theta_p(x_1,\ldots,x_m)$ and $\theta_q(x_1,\ldots,x_m,y)$, respectively. Furthermore, the type of $(b_1,\ldots,b_m)$ in $\bB$ is $p$, because the partial automorphism of $\bB'$ respects the signature $\tau'$. But now, since $\exists y. \, \theta_q(x_1,\ldots,x_m,y)$ is in $p$ (by maximality), we may deduce a $b'$ such that  $\bB' \models \theta_q(b_1,\ldots,b_m,b')$ and consequently $\bB' \models q(b_1,\ldots,b_m,b')$. It follows that $f' \colon (a_1,\ldots,a_m,a') \mapsto (b_1,\ldots,b_m,b')$ is a partial automorphism of $\bB'$ (in the signature $\tau'$). A simple back-and-forth argument shows that we may extend to an automorphism of $\bB'$, and the result follows.

For the implication $(iv) \Rightarrow (i)$, observe that
a finite or $\omega$-categorical structure $\bB$ is a model-complete core if and only if it has a model-complete core theory -- this is an easy consequence of the L\"owenheim-Skolem theorem (Theorem~\ref{thm:LS}). 
So it suffices to show that the first-order theory of $\bB$ and $T$ have the same universal negative consequences, by Proposition~\ref{prop:companions}.
A universal negative sentence $\phi$ is implied by $T$ if and only if $T \cup \{\neg \phi\}$ is unsatisfiable, which is
the case if and only if $\bB$ does not satisfy $\neg \phi$ (and hence satisfies $\phi$). 
\end{proof}

Theorem~\ref{thm:omega-cat} implies a necessary and sufficient condition 
when an CSP can be formulated with an
$\omega$-categorical template.

\begin{corollary}\label{cor:omega-cat}
Let $\bA$ be a structure with a finite relational signature. Then the following are equivalent. 
\begin{enumerate}
\item There is an $\omega$-categorical template $\bB$ 
such that $\Csp(\bB)=\Csp(\bA)$;
\item $\sim_n^{\Th(\bA)}$ has finite index for all $n$;
\item There exists a structure $\bB$ with
$\Csp(\bB)=\Csp(\bA)$ which has for all $n \geq 1$ finitely many primitive positive definable relations of arity $n$. 
\end{enumerate}
\end{corollary}
\begin{proof}
The implications from $(1)$ to $(3)$ and
from $(3)$ to $(2)$ are easy. 
%Write $T$ for $\Th(\bA)$. 
The implication from $(2)$ to $(1)$ follows from $(ii) \Rightarrow (iv)$ in Theorem~\ref{thm:omega-cat}. 
\end{proof}

We present a simple new proof of the following result from~\cite{Cores-journal}.

\begin{theorem}[of~\cite{Cores-journal}]
\label{thm:omega-cat-core-companion}
Every $\omega$-categorical structure $\bA$ is homomorphically equivalent 
to an $\omega$-categorical model-complete core $\bB$. All model-complete cores of $\bA$
are isomorphic to $\bB$. 
\end{theorem}
\begin{proof}
Let $T$ be the first-order theory of $\bA$; clearly, $T$ has the JHP. Since $T$ is $\omega$-categorical,
$\sim^T_n$ has finite index for each $n$ (Theorem~\ref{thm:ryll}), 
and Theorem~\ref{thm:omega-cat} implies that $T$ has a core companion 
$S$ which is either $\omega$-categorical or the theory of a finite structure. 
By Proposition~\ref{prop:uniqueness-core-companion}, 
the core companion of Th$(\bA)$ is unique up to equivalence of first-order theories.
Since Th$(\bB)$ is $\omega$-categorical or the theory of a finite structure, it follows that
$\bB$ is unique up to isomorphism. 
\end{proof}

Since the model-complete core $\bB$ of $\bA$ from the previous theorem is unique up to isomorphism, 
we call it \emph{the} model-complete core of $\bA$.  
The following gives an indication that the model-complete core of an $\omega$-categorical structure $\bA$
is typically `simpler' than $\bA$. 

\begin{proposition}
Let $\bA$ be an $\omega$-categorical structure, and let $\bB$ its model-complete core. Then 
\begin{itemize} 
\item 
for every $n$,
the number of orbits of $n$-tuples in $\bB$ is at most the number of orbits of $n$-tuples in $\bA$;
\item if $\bA$ is homogeneous, then $\bB$ is homogeneous as well. 
\end{itemize}
\end{proposition}
\begin{proof}
Let $f$ be a homomorphism from $\bA$ to $\bB$, and $g$ be a homomorphism from $\bB$ to $\bA$.
It suffices to show that when two $n$-tuples $t_1,t_2$ from $\bB$ are mapped by $g$ to tuples $s_1,s_2$ 
in the same orbit in $\bA$, then $t_1$ and $t_2$ lie in the same orbit in $\bB$. 
Let $\alpha$ be an automorphism of $\bA$ that maps $s_1$ to $s_2$. Since $\bB$ is an $\omega$-categorical 
model-complete core, there are primitive positive definitions $\phi_1$ and $\phi_2$ of the orbits of $t_1$ and $t_2$. 
Since $g$, $\alpha$, and $f$ preserve primitive positive formulas, the tuple $t_3 := f(\alpha g(t_1))$ satisfies $\phi_1$.
As $\alpha g(t_1) =s_2=g(t_2)$, the tuple $t_3$ can also be written as $f(g(t_2))$, and hence also satisfies $\phi_2$.
Thus, $\phi_1$ and $\phi_2$ define the same orbit, and $t_1$ and $t_2$ are in the same orbit. 

For the second part of the statement, suppose that $h$ is an isomorphism between two finite substructures 
$\bC$ and $\bC'$ of $\bB$. Then $g(\bC)$ induces in $\bA$ a structure that is isomorphic to $\bC$, since
otherwise the endomorphism $e \colon x \mapsto f(g(x))$ of $\bB$ would not preserve all first-order formulas,
contradicting the assumption that $\bB$ is a model-complete core. 
Similarly, $g(\bC')$ induces in $\bA$ a structure that is isomorphic to $\bC'$ and $\bC$,
and by homogeneity of $\bA$ there exists an automorphism $\alpha$ 
of $\bA$ that maps $g(\bC)$ to $g(\bC')$. The mapping $e' \colon x \mapsto f(\alpha g(x))$ is an endomorphism
of $\bB$. By Theorem~\ref{thm:mc-core}, this mapping is locally generated
by the automorphisms of $\bB$, and in particular there exists an automorphism $\beta$ of $\bB$ such that 
$\beta(x)=e'(x)$ for all elements $x$ of $\bC$. 
Since $e$ is locally generated by the automorphisms 
of $\bB$, too,  
there exists $\gamma \in \Aut(\bB)$  
such that $\gamma(x) = e(x)$ for all 
elements $x$ of $\bC'$. 
Then $\gamma^{-1} \circ \beta \in \Aut(\bB)$
maps $\bC$ to $\bC'$. 
This proves the homogeneity of $\bB$.
\end{proof}

When the template of a CSP is a model-complete core, then this can be exploited
in the study of the CSP in many ways. For instance, we have
the following consequence of 
Theorem~\ref{thm:mc-core} and Lemma~\ref{lem:constant-expansion}, which is essentially from~\cite{Cores-journal}.

\begin{corollary}\label{cor:mccore-constants}
Let $\bB$ be an $\omega$-categorical model-complete core,
and let $\bC$ be the expansion of $\bB$ by finitely many unary singleton relations,
that is, relations of the form $\{c\}$ for some element $c$ 
of $\bB$. Then for every finite signature reduct $\bC'$
of $\bC$ there exists a finite signature $\bB'$ of $\bB$
such that $\Csp(\bC')$ has a polynomial-time
reduction to $\Csp(\bB')$. 
%If $\bC$ has a finite signature reduct with a hard CSP,
%then so has $\bB$.
\end{corollary}
\begin{proof}
Let $\{c_1\},\dots,\{c_k\}$ be the relations of $\bC$ that have been added to $\bB$, and let $\bC'$ be a finite signature reduct
of $\bC$.
By Theorem~\ref{thm:mc-core}, the orbit of $(c_1,\dots,c_k)$ 
in $\bB$ has a primitive positive definition $\phi$ in $\bB$. 
Let $\bB'$ be the reduct of $\bB$ whose signature contains 
all relation symbols
mentioned in $\phi$, and the relation symbols of $\bC'$ that are also relation symbols in $\bB$;
observe that the signature of $\bB'$ is finite. 
Lemma~\ref{lem:constant-expansion} implies
that there is a polynomial-time reduction from
$\Csp(\bC')$ to $\Csp(\bB')$.
\end{proof}

% !TEX root = 0.tex
\chapter{Examples}
\label{chap:examples}

\begin{center}
\includegraphics[angle=-90]{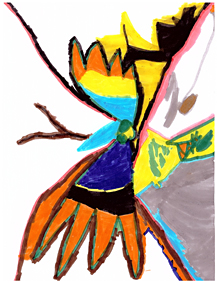}
\end{center}

% LATERTD: discuss finite boundedness for all examples
% also discuss strong amalgamation for examples?
% Give an example of a structure without finite orbits where
% strong amalgamation fails?
The running examples in the previous chapter were the linear order
$(\mathbb Q; <)$ and the random graph $(\mathbb V; E)$.
Structures definable in those structures provide further examples of $\omega$-categorical structures, and they will
be studied in great detail in Chapter~\ref{chap:ecsp}, \ref{chap:schaefer}, and \ref{chap:tcsp}. 
In this chapter, we present other $\omega$-categorical structures
$\bA$ that will not be treated at the same level of detail as it will be done for $(\mathbb Q; <)$ and for $(\mathbb V; E)$ in Chapter~\ref{chap:schaefer} and Chapter~\ref{chap:tcsp}. 
For example, we treat 
homogeneous C-relations, dense semi-linear orders, and  the atomless Boolean algebra.
In each case, we give a brief discussion 
on what is known about CSPs with templates
that can be defined in those $\omega$-categorical structures. Thereby, we revisit many problems from Section~\ref{sect:csp-examples}.
We also discuss $\omega$-categorical structures that serve as templates for problems in connected monotone monadic SNP.

The $\omega$-categorical structures presented in this chapter are chosen so that
they illuminate the diversity of the class of all $\omega$-categorical structures, and so that 
many computational problems and classes of computational problems from the literature can be formulated as CSPs for those structures. 

\section{Phylogeny Constraints and Homogeneous C-relations}
\label{ssect:c-relation}
The rooted-triple satisfiability problem from Section~\ref{ssect:phylo} 
can be formulated as $\Csp(\bB)$ for an $\omega$-categorical template $\bB$ (an observation from~\cite{Cores-journal}). 
There are various different ways how to define such a structure $\bB$; the most convenient for us is via amalgamation.
%An axiomatic approach can be found in~\cite{AdelekeNeumann} and~\cite{phylo-long}.

Let $\cal T$ be the class of all finite rooted binary
trees $\bT$. The \emph{leaf structure}
$\bC$ of a tree $\bT \in \cal T$ with leaves $L$ 
is the relational structure $(L; |)$ 
where $|$ is a ternary relation symbol, and $ab|c$ holds in $\bC$ 
iff $\yca(a,b)$ lies below $\yca(b,c)$ in $\bT$ (recall that $\yca(a,b)$ denotes the youngest common ancestor
of $a$ and $b$ in a rooted tree $\bT$; see Section~\ref{ssect:phylo}). 
We also call $\bT$ the \emph{underlying tree} 
of $\bC$. 
Let $\cal C$ be the class of all leaf structures for trees from $\cal T$. 

\begin{proposition}\label{prop:phylo-amalgamation}
The class $\cal C$ is an amalgamation class.
\end{proposition}

\begin{proof}
Closure under isomorphisms and induced substructures is by definition.
For the amalgamation property, let $\bB_1,\bB_2 \in \cal C$ be such that $\bA=\bB_1 \cap \bB_2$ is an induced substructure of both $\bB_1$ and $\bB_2$. We want
to show that there is an amalgam of $\bB_1$ and $\bB_2$ over $\bA$ in $\C$.
We inductively assume that the statement has been shown for all triples $(\bA,\bB_1',\bB_2')$ where $B_1' \cup B_2'$ is a proper subset of $B_1 \cup B_2$.

Let $\bT_1$ be the rooted binary tree underlying $\bB_1$, and 
$\bT_2$ the rooted binary tree underlying $\bB_2$.
Let $\bB_1^1 \in \cal C$ be the substructure of $\bB_1$ induced by the vertices below the left child of $\bT_1$,
and $\bB_1^2 \in \cal C$ be the substructure of $\bB_1$ induced by the vertices below the right child of $\bT_1$. The structures $\bB_2^1$ and $\bB_2^2$ are defined
analogously for $\bB_2$ instead of $\bB_1$.

\begin{figure}[h]
\begin{center}
\includegraphics[scale=.3]{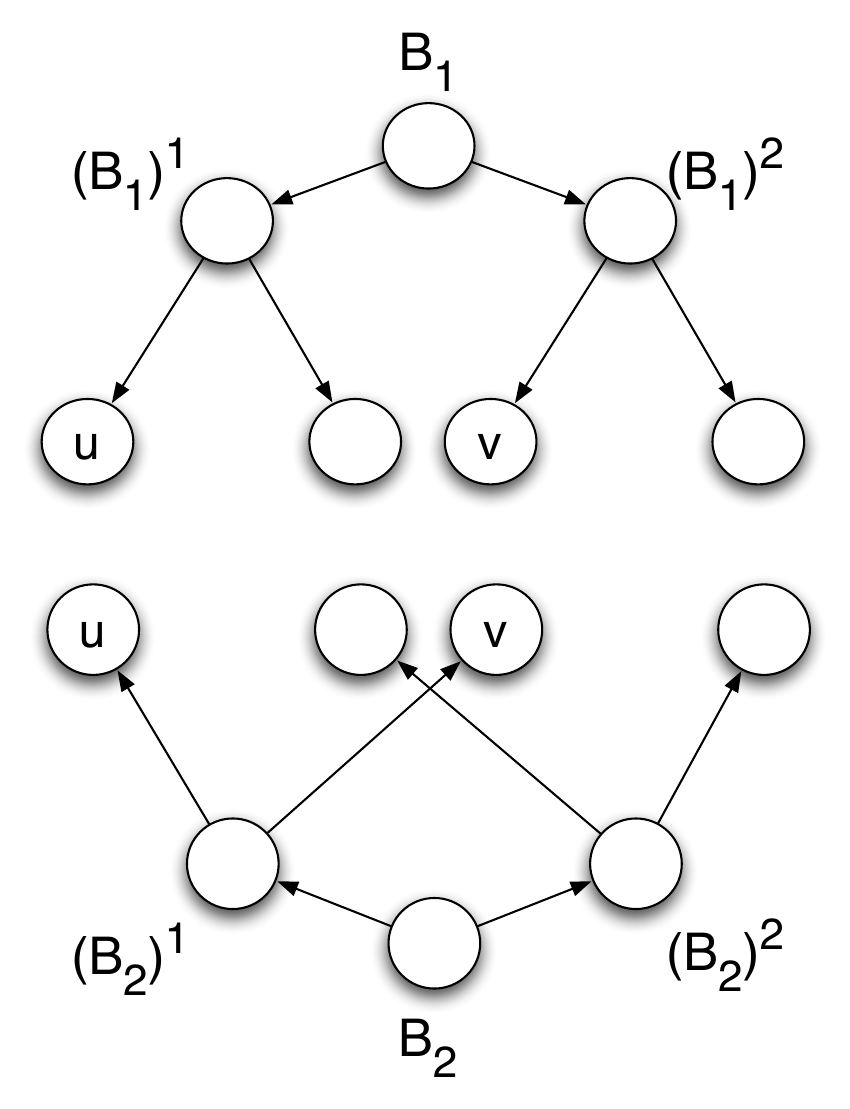} 
\end{center}
%\vspace{1cm}
\caption{Illustration for the proof of Proposition~\ref{prop:phylo-amalgamation}.}
\label{fig:phylo-amalgamation}
\end{figure}

First consider the case that there is a vertex $u$ that lies in both $\bB_1^1$ and $\bB_2^1$, and a vertex $v$ that lies in both $\bB_2^1$ and $\bB_1^2$ (see Figure~\ref{fig:phylo-amalgamation} for an illustration). We claim that in this case no vertex $w$ from $\bB_2^2$ can lie inside $\bB_1$:
for otherwise, $w$ is either in $\bB_1^1$, in which case we have $uw|v$ in $\bB_1$,
or in $\bB_1^2$, in which case we have $vw|u$ in $\bB_1$.
But since $u,v,w$ are in $A$, this is in contradiction to the fact that $uv|w$ holds
in $\bB_2$.  Let $\bC' \in \cal C$ be the amalgam of $\bB_1$ and $\bB_2^1$ over $\bA$, which exists by inductive assumption, and let $\bT' \in \cal T$ be its underlying tree.
Now let $\bT$ be the tree with root $r$ and $\bT'$ as a left subtree, and the
underlying tree of $\bB_2^2$ as a right subtree.
It is straightforward to verify that the leaf structure of $\bT$ 
is in $\cal C$,
and that it is an amalgam of $\bB_1$ and $\bB_2$ over $\bA$ 
(via the identity embeddings).

Up to symmetry, the only remaining essentially different 
case we have to consider is
that $B_1^1 \cup B_2^1$ and  $B_1^2 \cup B_2^2$ are disjoint.
In this case it is similarly straightforward to first amalgamate $\bB_1^1$ with $\bB_2^1$
and $\bB_1^2$ with $\bB_2^2$ to obtain the amalgam of $\bB_1$ and $\bB_2$;
the details are left to the reader.
\end{proof}

Let $\bB$ denote the \Fresse-limit of $\mathcal C$. 
The structure $\bB$ is homogeneous, so it is a fortiori model-complete. It is straightforward to verify that $\bB$ is a core.  
Since the age of $\bB$ is the class of all leaf structures for structures from
$\mathcal T$, it is obvious
that $\Csp(\bB)$ is the rooted triple consistency problem.
The relation $|$ in $\bB$ is closely related to so-called \emph{C-relations}, 
following the terminology of~\cite{AdelekeNeumann}. $C$-relations became 
an important concept in model theory, see e.g.~\cite{C-minimal}. 
They are given axiomatically; the presentation here follows~\cite{AdelekeNeumann,phylo-long}. 

A ternary relation $C$ is said to be a \emph{C-relation} on a set $L$ if for all $a,b,c,d \in L$ the following conditions hold:
\begin{enumerate}
\item [C1] $C(a; b,c) \rightarrow C(a; c,b)$; 
\item [C2] $C(a; b,c) \rightarrow \neg C(b;a,c)$; 
\item [C3] $C(a; b,c) \rightarrow C(a; d,c) \vee C(d; b,c)$; 
\item [C4] $a \neq b \rightarrow C(a;b,b)$.
\end{enumerate}
\noindent A C-relation is called \emph{dense} if it satisfies
\begin{enumerate}
\item [C7] $C(a; b,c) \rightarrow \exists e \, (C(e;b,c) \wedge C(a; b,e))$.
\end{enumerate}
The structure $(L;C)$ is also called a \emph{C-set}. 
A structure $\Gamma$
is said to be \emph{relatively $k$-transitive} if for every partial isomorphism $f$
between induced substructures of $\Gamma$ of size $k$ 
there exists an automorphism
of $\Gamma$ that extends $f$. 
Note that a relatively 3-transitive C-set 
is necessarily 2-transitive (i.e., has a 2-transitive automorphism group, as defined in Section~\ref{ssect:trans}).

\begin{theorem}[Theorem 14.7 in~\cite{AdelekeNeumann}]
Let $(L;C)$ be a relatively 3-transitive C-set. Then $(L;C)$ is $\omega$-categorical.
\end{theorem}

% DON'T NEED THE FOLLOWING, SINCE WE HAVE THE AMALGAMATION CONSTRUCTION
% Theorem 11.2 and 11.3 in~\cite{AdelekeNeumann} show how to construct 
%such a C-relation from a \emph{semi-linear order} that is \emph{dense, normal, and branches everywhere}
%(all these concepts are defined in~\cite{AdelekeNeumann}). 
%Such a semi-linear order is explicitly constructed in Section 5 of~\cite{AdelekeNeumann}. 

In fact, there is, up to isomorphism, a unique 
relatively 3-transitive countable C-set which is dense 
and 
%\begin{itemize} \item is 
\emph{uniform with branching number $2$}, that is, satisfies $\forall x,y,z \, \big ((x \neq y \vee x \neq z \vee y \neq z) \Rightarrow (C(x;y,z) \vee C(y;x,z) \vee C(z;x,y))\big)$ 
%\item is dense, and
%\item satisfies $\neg C(a;a,a)$ for all $a \in L$.
%\end{itemize}
(see the comments in~\cite{AdelekeNeumann} after the statement of Theorem 14.7).
%; the condition that $\neg C(a;a,a)$ for all (equivalently, for some)
%$a \in L$ has been forgotten there, but is necessary to obtain uniqueness.)
% Stimmt doch gar nicht: folgt aus C2!!

It is straightforward to verify that the \Fresse-limit $\bB$ of the amalgamation class from Proposition~\ref{prop:phylo-amalgamation} has the same automorphism group (equivalently, is first-order interdefinable; see Section~\ref{sect:galois})
as the relatively 3-transitive dense countable C-set that is uniform with branching number 2.

%the relation $xy|z$ defined from the $C$-relation 
%by $$xy|z \; \Leftrightarrow \; C(z;x,y) \wedge x \neq y \wedge y \neq z \wedge x \neq z \; .$$
%It is clear that $\Lambda$ has the same automorphism group than
%the dense C-set from which it is defined, and in particular, it is also
%$\omega$-categorical (since by the Theorem of Ryll-Nardzewski, $\omega$-categoricity of a structure only depends on the automorphism group of $\Gamma$, see~\cite{Hodges}).

A similar approach is possible to define a homogeneous template for
the quartet satisfiability problem from Section~\ref{ssect:phylo}.
Alternatively, an $\omega$-categorical template $(B; Q)$ for the quartet satisfiability problem can be given via a first-order definition in the structure $\bB = (B; |)$ constructed above for the rooted triple consistency problem.

The following definition will be useful now, and later. 
Since $\yca$ is associative as a binary operation, it makes
sense to write $\yca(v_1,\dots,v_l)$ for $$\yca(v_1,\dots, \yca(v_{l-1},v_l)\dots) \; .$$

\begin{definition}\label{def:bar}
When $u_1,\dots,u_k$ and $v_1,\dots,v_l$ are leaves
in a rooted tree $\bT$, then we write $u_1\dots u_k| v_1\dots,v_l$
if $u := \yca(u_1,\dots,u_k)$ and $v:=\yca(v_1,\dots,v_l)$
are \emph{disjoint} in $\bT$, i.e., neither $u$ lies above $v$ nor
$v$ lies above $u$ in $\bT$.
\end{definition}

The first-order definition of $Q(x,y,u,v)$ is
\begin{align*}
(xy|uv)  \; \vee \; (uv|x \wedge vx|y) \; \vee \; (xy|u \wedge yu|v)  \; .
\end{align*}
Indeed, when $u,v,x,y \in B$, and $\bT$ is the tree underlying the substructure
of $(B;|)$ induced by $\{u,v,x,y\}$, then
the given formula describes the situation that 
the shortest path from $x$ to $y$ in $\bT$ does not intersect the shortest
path from $u$ to $v$ in $\bT$. Note that
whether this is true is in fact independent
from the position of the root of $\bT$. We leave the verification to the
reader that $\Csp((B;Q))$ indeed describes the quartet satisfiability problem
studied in comptuational biololgy.  
Lemma~\ref{lem:interpret} implies $\omega$-categoricity of $(B;Q)$.
Similarly as for the C-relation given above, an axiomatic treatment of $(B;Q)$ has been given in~\cite{AdelekeNeumann};
there, the relation $Q$ has been called a \emph{D-relation}, and this became standard terminology in model theory. 
As we have mentioned above, the structure $(B;Q)$ can 
also be defined as a \Fresse-limit of finite $D$-structures (see Cameron~\cite{Oligo}).

\section{Branching-Time Constraints}
\label{ssect:mt:branching-time}
The branching-time satisfaction problem from Section~\ref{ssect:branching-time}
% in Chapter~\ref{chap:intro}
 can be formulated as $\Csp(\bB)$ for an
$\omega$-categorical structure $\bB = (D;\leq,\parallel,\neq)$; this has already been observed in~\cite{BodirskyNesetrilJLC}. This time, $\bB$ has an
explicit description.
The domain $B$ consists of the set of all
non-empty finite sequences
of rational numbers. For 
$a = (q_1, q_1, \dots, q_{n}), b = (q'_1, q'_1, \dots, q'_{m})$, $n \leq m$, we write $a < b$ if one of the following conditions holds:
\begin{itemize}
\item $a$ is a proper initial subsequence of $b$, i.e., $n<m$ and
$q_i=q_i'$ for $1 \leq i \leq n$;
\item $q_i=q_i'$ for $1 \leq i < n$, and $q_n < q_n'$.
\end{itemize}
We use $a \leq b$ to denote $(a <  b) \vee (a=b)$, 
and $\parallel$ denotes the binary relation that contains all pairs of elements 
that are incomparable with respect to $<$; in particular, we have $a \parallel a$ for all $a$.
A proof that $\bB=(B;\leq,\parallel,\neq)$ is $1$-transitive and $\omega$-categorical can be found in~\cite{AdelekeNeumann} (Section 5).
% Clear that this is binary branching: starting from a, 
% either prolong, 
% or increase the last element. 
% Contains ramification points? Yes, the element a above
% is the ramification point for the two contains that I described. 

The reduct $(B;\leq)$ of this structure is a semi-linear order, 
i.e., for all $x \in D$, the set $\{y \; | \; y \leq x \}$
is linearly ordered by $<$; such structures have been studied systematically in the context of infinite permutation groups; see~\cite{Droste,Oligo}.
The structure $\bB$ is \emph{not}
homogeneous, and therefore cannot be described as the 
\Fresse-limit of a class of finite structures. 
However, it has an expansion by primitive positive definable relations which is homogeneous. The following is well-known and straightforward to verify.  

\begin{proposition}\label{prop:make-semilinear-homo}
The expansion of $\bB$ by
the ternary relation with the primitive positive definition $$\exists u \, \big((u \leq x) \wedge (u \leq y) \wedge (u \parallel z)\big)$$
and the ternary relation with the primitive positive definition 
$$\exists u \, \big(x || y \wedge x \neq y \wedge (u \leq x) \wedge (u \leq y) \wedge (z \leq u) \wedge (z \neq u) \big)$$
is homogeneous.
\end{proposition}

Therefore it is possible to describe the structure $(D;\leq,\parallel,\neq)$ as the reduct of the \Fresse-limit
of an amalgamation class; the respective proof
is similar to the proof of Proposition~\ref{prop:phylo-amalgamation}. 
Proposition~\ref{prop:make-semilinear-homo} 
 in combination with Theorem~\ref{thm:mc-omegacat} also
shows that $\bB$ is model-complete.
The structure $\bB$ has four orbitals, with the primitive positive definitions $x \leq y \wedge x \neq y$, $y \leq x \wedge x \neq y$,
$x \parallel y \wedge x \neq y$, and $x = y$.
Since all relations of $\bB$ are binary, this
implies that every endomorphism of $\bB$
must be an embedding, and hence $\bB$
is a core. 

The expansion $\bB'$ of $\bB$ by the first of the two ternary relations
given in Proposition~\ref{prop:make-semilinear-homo}
also has an age that has amalgamation; however, 
its \Fresse-limit $\bB''$ is not isomorphic to $\bB'$ because
it \emph{lacks joins}, that is, the second of the ternary relations in Proposition~\ref{prop:make-semilinear-homo} holds for \emph{all} triples $x,y,z$ with $x || y$, $x \neq y$, $z<y$, and $z <y$.

\section{Set Constraints}
\label{ssect:mt-set-constraints}
Here we discuss how the set constraint satisfaction problems
discussed in Section~\ref{ssect:setcsps}, % of Chapter~\ref{chap:intro}, 
and many other set constraint satisfaction problems, can be formulated with $\omega$-categorical templates, following~\cite{BodHilsKrim}.

Here, a \emph{set constraint satisfaction problem} is a CSP for
a template with a first-order definition in the structure
${\mathfrak S}$ with domain ${\cal P}({\mathbb N})$, the set of all subsets of natural numbers, 
and with signature $\{\cap,\cup,c,{\bf 0},{\bf 1}\}$, 
where 
\begin{itemize}
\item $\cap$ is a binary function symbol that denotes intersection, i.e., 
$\cap^{\mathfrak S}=\cap$;
\item $\cup$ is a binary function symbol for union, i.e., 
$\cup^{\mathfrak S}=\cup$;
\item $c$ is a unary function symbol for complementation, i.e., $c^{\mathfrak S}$ is the function
that maps $S \subseteq {\mathbb N}$ to ${\mathbb N} \setminus S$;
\item ${\bf 0}$ and ${\bf 1}$ are constants (treated as $0$-ary function symbols) denoting the empty set $\emptyset$ and the full set $\mathbb N$, respectively. 
\end{itemize}

A \emph{set constraint language} is a relational structure
with a set of relations with a quantifier-free 
first-order definition in $\mathfrak S$; we always allow equality in first-order formulas.
For example, the relation $\{(x,y,z)  \in {\cal P}({\mathbb N})^3 \; | \; x \cap y \subseteq z \}$ 
has the quantifier-free first-order definition $z \cap (x \cap y) = x \cap y$ over $\mathfrak S$.

\begin{proposition}[follows from Proposition 5.8 in~\cite{MarriottOdersky}]
\label{prop:setcsps-in-np}
Let $\bB$ be a set constraint language with a finite signature. Then $\Csp(\bB)$ is in NP.
\end{proposition}

The first-order theory of the structure $\mathfrak S$ is certainly not
$\omega$-categorical -- it is easy to verify that there are infinitely
many pairwise inequivalent first-order formulas with one free variable. 
However, all set constraint satisfaction problems can be formulated
with an $\omega$-categorical template. 
To see this, first note that the structure $({\cal P}({\mathbb N}); \cup,\cap,c,{{\bf 0}},{{\bf 1}})$ is a Boolean algebra, with 
\begin{itemize}
\item ${\bf 0}$ playing the role of false, and ${\bf 1}$ playing the role of true; 
\item $c$ playing the role of $\neg$;
\item $\cap$ and $\cup$ playing the role of $\wedge$ and $\vee$, respectively.
\end{itemize}
%To not confuse logical connectives with the connectives
%of Boolean algebras, we always use the symbols $\cap$, $\cup$, and $c$ instead of the usual function symbols $\wedge$, $\vee$, and $\neg$ in Boolean algebras. 
% GEAENDERT, WEIL HIER UNPROBLEMATISCH
To facilitate the notation, we write
$\bar x$ instead of $c(x)$, and $x \neq y$ instead of $\neg (x = y)$.

An \emph{atom} in a Boolean algebra is an element $x \neq {\bf 0}$ such that for all
$y$ with $x \cap y = y$ and $x \neq y$ we have $y=\bf 0$.
If a Boolean algebra does not contains atoms, it is called \emph{atomless}.
It is well-known that there exist countable atomless Boolean algebras, and that all countable atomless 
Boolean algebras are isomorphic (Corollary 5.16 in~\cite{KoppelbergBoolenAlgebras}; also see Example 4 on 
page 100 in~\cite{HodgesLong}). 
Let $\mathfrak A$ denote such a countable atomless Boolean algebra;
the domain $\mathfrak A$ is denoted by $\mA$. 
Since the axioms of Boolean algebras and the property of not having atoms can all be written as first-order sentences, it follows that $\mathfrak A$ is $\omega$-categorical. 
We also remark that the structure $\mathfrak A$ has quantifier elimination (see Exercise 17 on Page 391 in~\cite{HodgesLong}).
The link between the set constraint satisfaction problems over $2^{\mathbb N}$ mentioned in Section~\ref{sect:csp-examples} %of Chapter~\ref{chap:intro} 
and the atomless Boolean algebra is the following. 

\begin{proposition}\label{prop:omega-cat-set-templates}
Let $\bC$ be a set constraint language. 
Then there exists an $\omega$-categorical structure $\bB$
such that $\bB$ and $\bC$ have the same existential theory. 
In particular, when $\bC$ has finite signature, then $\bB$ and $\bC$ 
have the same $\Csp$.
\end{proposition}

\begin{proof}
Let $\phi_1,\phi_2,\dots$ be quantifier-free first-order formulas
that define the relations $R^{\bC}_1,R^{\bC}_2,\dots$ of $\bC$ over $\mathfrak S = ({\cal P}({\mathbb N}); \cup,\cap,c,{{\bf 0}},{{\bf 1}})$. 
Let $R^{\mathfrak A}_1,R^{\mathfrak A}_2,\dots$ be the
relations defined by $\phi_1,\phi_2,\dots$ over the atomless Boolean algebra $\mathfrak A$.
The structure $\bB = (\mathbb A; R^{\mathfrak A}_1,R^{\mathfrak A}_2,\dots)$ is $\omega$-categorical (see the comment after Lemma~\ref{lem:interpret}). 
To verify that $\bB$ and $\bC$ have the same existential theory, 
let $\Phi$ be a conjunction of atomic formulas 
over the signature $\{R_1,R_2,\dots\}$. Replace each
atomic formula of the form $R_i(x_1,\dots,x_k)$ in $\Phi$ 
by the formula $\phi_i(x_1,\dots,x_k)$. The resulting formula
is a quantifier-free first-order formula in the language of Boolean
algebras, $\{\cup,\cap,c,{{\bf 0}},{{\bf 1}}\}$.
We claim that $\Phi$ is satisfiable in $\mathfrak S$ if and only if it is satisfiable in $\mathfrak A$. This follows from Corollary 5.7 in~\cite{MarriottOdersky}: $\Phi$ is satisfiable in some infinite Boolean algebra if and only if $\Phi$ is satisfiable in all infinite Boolean
algebras.
\end{proof}
A  large class of tractable set constraint languages
has been described in~\cite{BodHilsKrim}; the class given there 
is \emph{maximal tractable} in the sense
that every strictly larger class of set constraint languages contains a finite subset
with an NP-hard CSP.

\section{Spatial Reasoning}
\label{ssect:spatial}
The essential reasons why 
the network satisfaction problem for RCC5 (introduced in Section~\ref{ssect:general-spatial}) is \emph{not} a set constraint satisfaction problem as introduced in the previous section is
that in RCC5 we exclude the empty set as a possible value for the variables. 
To formulate the CSP for the network satisfaction problem of 
RCC5 and its fragments with $\omega$-categorical templates,
we again use structures with a first-order definition
in the atomless Boolean algebra, but restrict those structures
to non-zero elements 
(this observation has already been made in~\cite{Duentsch-rcc}, Proposition 4.4).

Formally, let $\mathfrak A$ be the atomless Boolean algebra, 
and let $\DR$, $\PO$, $\PP$, $\PPI$, $\EQ$ be the binary
relations with the following first-order definition in $\mathfrak A$
(and their intuitive meaning in quotes).

\begin{align*}
  \DR(x,y) \; \text{ iff } \; & (x \cap y = {\bf 0}) \wedge x \neq y \wedge x,y \nin \{{\bf 0},{\bf 1}\} \\ & \text{`$x$ and $y$ are disjoint'} \\
  \PP(x,y) \; \text{ iff } \; & (x \cap y = x) \wedge x \neq y  \wedge x,y \nin \{{\bf 0},{\bf 1}\} \\ & \text{`$y$ properly contains $x$'} \\
  \PPI(x,y) \; \text{ iff } \; & (x \cap y = y) \wedge x \neq y  \wedge x,y \nin \{{\bf 0},{\bf 1}\} \\ & \text{`$x$ properly contains $y$'}  \\
  \EQ(x,y) \; \text{ iff } \; & x = y  \wedge x,y \nin \{{\bf 0},{\bf 1}\} \\ & \text{`$x$ equals $y$'} \\
  \PO(x,y) \; \text{ iff } \; & \neg \DR(x,y) \wedge \neg \PP(x,y) \wedge \neg \PPI(x,y) \wedge x \neq y  \wedge x,y \nin \{{\bf 0},{\bf 1}\} \\
  & \text{`$x$ and $y$ properly overlap'}
\end{align*}
When $\bD$ is the structure that contains
all binary relations that are first-order definable in $(A \setminus \{{\bf 0},{\bf 1}\}; \DR, \PO, \PP, \PPI, \EQ)$ (so that we can associate a binary relation from $\bD$ to every element of RCC5 in the natural way), then $\Csp(\bD)$ 
and the network satisfaction problem for RCC5 are essentially the same problem (in the sense of Section~\ref{ssect:relation-algebras}).
The structure $\bD$ has a (1-dimensional) first-order interpretation
in $\bA$, and hence is $\omega$-categorical by Lemma~\ref{lem:interpret}. It can be shown that the model companion of $\bD$ gives a representation of RCC5. 

\section{CSPs and Fragments of SNP}
\label{ssect:logics}
Recall that a constraint satisfaction problem can be viewed as a class of finite structures 
with finite relational signature (as described 
in Section~\ref{sect:homo} and Section~\ref{sect:snp}), namely the class of all satisfiable
instances of the CSP. 
In this section we study the question when this class
can be \emph{described} by a $\tau$-sentence $\Phi$ from a fixed logic $\mathcal L$ 
in the sense that
for all finite structures $\bA$, we have that $\bA \models \Phi$ 
if and only if $\bA \in \mathcal C$. If there is such a sentence then we say that
$\Csp(\bB)$ \emph{is in $\mathcal L$}.

The first two logics considered here will be first-order logic, and monadic SNP. It turns out that CSPs that can be described by a sentence from those logics
can be formulated with $\omega$-categorical templates. 
Finally, we present a new and more expressive logic,
called Amalgamation SNP. Again, every problem in amalgamation
SNP describes a problem in NP that can be formulated
as $\Csp(\bB)$ for an $\omega$-categorical template $\bB$,
and the universal-algebraic techniques presented in later sections can be applied 
to study the computational complexity of the problems in this logic. 

\vspace{0.2cm}
\subsection{First-order Definable CSPs}
\label{ssect:fo}
We first consider the situation where $\Csp(\bB)$ is in \emph{FO},
i.e.,  can be described by a first-order sentence
 (in the sense just described). The class of CSPs in FO is quite restricted. It is not hard to see that when $\Csp(\bB)$ is in FO, then in particular it can be solved in deterministic
 logarithmic space. We will see that when $\Csp(\bA)$ is
in FO, then there exists an $\omega$-categorical structure $\bB$ 
that has the same CSP\footnote{An exact characterization of those $\omega$-categorical structures that are in FO can be found in~\cite{BodHilsMartin-Journal}, obtained by a slight modification
of a proof for finite domain CSPs in~\cite{LLT}.}.
%This is actually equivalent to a theorem of Cherlin-Shelah-Shi
%that appears in~\cite{CherlinShelahShi} for the special
%case where $\tau$ has a single binary relation denoting
%the edge relationship of undirected graphs. Our proof of this
%fact is new, and only uses Theorem~\ref{thm:omega-cat} from Section~\ref{sect:epc} in this chapter.
%Moreover, the CSP of an $\omega$-categorical $\tau$-structure $\bB$ is in FO if and only if $\bB$ has
%an $n$-ary \emph{one-tolerant polymorphism} for $n \geq 3$, i.e., a polymorphism $f$ that satisfies
%$$(f(a_1^1,\dots,a^n_1),\dots,f(a_k^1,\dots,a_k^n)) \in R^\bB
%\text{ if } |\{j \; | \; (a_1^j,\dots,a_k^j) \in R^\bB\}| \geq n-1Ê\; ,$$ 
%for all elements $R \in \tau$ and all $a_1^1,\dots,a^n_1,\dots,a_k^1,\dots,a_k^n$ of $\bB$. 
%This has been shown in~\cite{BodHilsMartin-Journal}, by slight modification
%of a proof for finite domain CSPs in~\cite{LLT}. 
We use the following famous result.

\begin{theorem}[Homomorphism Preservation in the Finite~\cite{Rossman08}]\label{thm:rossman}
Let $\tau$ be a finite relational signature, and let $\Phi$ be a first-order
$\tau$-sentence. Then $\Phi$ is equivalent to an existential positive first-order sentence on all finite $\tau$-structures if and only if
the class of all finite $\tau$-models of $\Phi$ is closed
under homomorphisms (Definition~\ref{def:inv-hom-disj-union}). 
\end{theorem}

In the rest of this section, $\tau$ is a finite relational signature, and
$\bB$ be a $\tau$-structure. Recall that $\Csp(\bB)$, viewed as a class of finite $\tau$-structures, is closed under inverse homomorphisms and disjoint unions. In particular, the class of all finite $\tau$-structures 
that do \emph{not} homomorphically map to $\bB$ is closed
under homomorphisms, and by Theorem~\ref{thm:rossman}
describable by an existential positive $\tau$-sentence $\Psi$. 
This leads us to the following.

\begin{theorem}
If $\Csp(\bB)$ is in FO, then there exists an $\omega$-categorical
structure $\bB'$ that has the same $\Csp$. 
\end{theorem}
\begin{proof}
From the remarks that precede the statement of the theorem,
Theorem~\ref{thm:rossman} shows that there is an existential positive $\tau$-sentence $\Psi$ such that $\bA$ homomorphically maps to $\bB$ 
 if and only if $\bA$ satisfies $\neg \Psi$. 
 The sentence $\neg \Psi$
 can be re-written as a universal negative sentence in conjunctive normal form; let $\Phi$ be such a universal negative sentence of minimal size. 
 We claim that the canonical database $\bC$ for each conjunct in $\Phi$ is connected. To see this, suppose that $\bC$ has several connected components. If one of them does not homomorphically map to $\bB$, then $\Phi$ was not of minimal size, since the corresponding conjunct could have
 been replaced by the (smaller) conjunctive query for the component. If all components homomorphically map to $\bB$, then so does $\bC$, a contradiction
 to the fact that $\bC$ is the canonical database of a conjunct of $\Phi$.
 
Therefore each conjunct in $\Phi$ is connected, and we can apply Theorem~\ref{thm:universal} to the finite set $\mathcal N$ of canonical databases for all the conjuncts in $\Phi$. From Theorem~\ref{thm:universal} we obtain an $\omega$-categorical $\tau$-structure which is universal for $\Csp(\bB)$, which is what we had to show.
\end{proof}

\subsection{CSPs in Monadic SNP}
Also every CSP in monadic SNP can
be formulated with an $\omega$-categorical template.

\begin{theorem}[from~\cite{BodDalJournal}]
\label{thm:bod-dal}
Let $\bC$ be a structure with a finite relational signature.
If $\Csp(\bC)$ can be described by a monadic SNP sentence $\Phi$,
then there is an $\omega$-categorical $\bB$ such that $\Csp(\bB)=\Csp(\bC)$.
\end{theorem}
\begin{proof}
By Corollary~\ref{cor:msnp-csp}, we can assume without
loss of generality that $\Phi$ is a \emph{connected} and \emph{monotone} monadic SNP sentence. 

We assume without loss of generality that $\Phi$ is written in negation normal form. Let $P_1, \dots, P_k$ be the 
existential monadic predicates in $\Phi$. 
Let $\tau'$ be the signature containing the input relations
from $\tau$, the existential monadic relations $P_i$, and new
symbols $P_i'$ for the negative occurrences of the existential
relations.

By monotonicity, all such literals with
input relations are positive. For each existential monadic relation
$P_i$ we introduce an existentially quantified monadic 
relation symbol $P'_i$, and replace negative
literals of the form $\neg P_i(x)$ in $\Phi$ by $P_i'(x)$.
We shall denote the $\tau'$-formula obtained from $\Phi$ 
after this transformation by $\Phi'$. 
 We define ${\cal N}$ to be the set of $\tau'$-structures
containing for each clause $\psi$ in
$\Phi'$ its canonical database (as defined in Section~\ref{ssect:snp-csp}).
 We shall use the fact that a
$\tau'$-structure $\bA$ satisfies a clause $\psi$ if and only if the  canonical database of $\psi$ is not homomorphic to $\bA$.
Since $\Phi$ is connected, all structures in $\cal N$ are connected.

Then Theorem~\ref{thm:universal} asserts the existence of a $\cal
N$-free $\omega$-categorical $\tau'$-structure $\bB'$ that is
universal for all $\cal N$-free structures. We use $\bB'$ to
define the template $\bB$ with the properties required in the statement of the theorem we are about to prove.
The structure $\bB$ is the $\tau$-reduct of the
 restriction of $\bB'$ to the points
with the property that  for all existential monadic predicates $P_i$, $1 \leq i \leq k$, either $P_i$ or $P_i'$ holds (but not
both $P_i$ and $P_i'$).
%The structure $\cB$ is non-empty, since the problem
%defined by $\Phi$ is non-empty. 
It follows from Theorem~\ref{lem:interpret} that reducts  of $\omega$-categorical structures, and restrictions to first-order
definable subsets  of $\omega$-categorical structures are again
$\omega$-categorical.
Hence, the resulting $\tau$-structure $\bB$ is
$\omega$-categorical.

We claim that a $\tau$-structure $\bA$ satisfies $\Phi$ if and only
if $\bA$ homomorphically maps to $\bB$. 
First, let $\bA$ be a structure that has a homomorphism
$h$ to $\bB$. Let $\bA'$ be the $\tau'$-expansion of
$\bA$ such that for all $i \leq k$ and $a \in A$ the relation 
$P_i(a)$ holds in $\bA'$ if and
only if $P_i(h(a))$ holds in $\bB'$, and $P'_i(a)$ holds in $\bA'$
if and only if $P'_i(h(a))$ holds in $\bB'$. Clearly, $h$ defines
a homomorphism from $\bA'$ to $\bB'$. In
consequence, none of the structures from $\cal N$ maps to $\bA'$.
Hence, the $\tau$-reduct $\bA$ of $\bA'$ satisfies
$\Phi$.

Conversely, let $\bA$ be a $\tau$-structure satisfying $\Phi$. Consequently,
there exists a $\tau'$-expansion $\bA'$ of $\bA$ that satisfies the first-order part of $\Phi'$, and where for every $a \in A$ exactly one of $P_i(a)$ or $P'_i(a)$ holds.
Clearly, no structure in $\cal N$ is homomorphic to $\bA'$,
and by universality of $\bB'$ the $\tau'$-structure
$\bA'$ is an induced substructure of $\bB'$. Since for every $a \in A$ exactly one of $P_i(a)$ and $P_i'(a)$ holds, $\bA'$ is also an
induced substructure of the restriction of $\bB'$ to $B$. 
Consequently, $\bA$ is homomorphic to the $\tau$-reduct of this restriction. 
This completes the proof. 
\end{proof}

\subsection{Amalgamation SNP}
\label{ssect:asnp}
In this section we introduce a logic that describes only CSPs
with $\omega$-categorical templates, and which we call 
\emph{Amalgamation SNP}, or short \emph{ASNP}. 

\begin{definition}\label{def:amalgamation-snp}
\emph{Amalgamation SNP} is the logic that consists of all
%connected: don't need this, is implied by amalgamation via the Joint Embedding Property
monotone SNP sentences $\Phi$ 
where the class of all finite models of the first-order part of $\Phi$
has the amalgamation property.
\end{definition}

It can be verified that the examples of SNP sentences given for CSP$(({\mathbb Z};<))$ in Example~\ref{expl:acycl-snp}  
and for CSP$(({\mathbb Z};\Betw))$ in Example~\ref{expl:betw-snp} are in fact Amalgamation SNP sentences. 
%This will become clear from proof of the following proposition.
%There is a close bi-directional link between Amalgamation SNP and constraint satisfaction problems for certain $\omega$-categorical
%templates; the two directions are presented in Proposition~\ref{prop:asnp-finitely-bounded} and Proposition~\ref{prop:finitely-bounded-asnp}, respectively. 
Recall that a structure is called \emph{finitely bounded} if its age can be described by finitely many
forbidden induced substructures (Definition~\ref{def:finitely-bounded}). 

\begin{proposition}\label{prop:asnp-finitely-bounded}
Every sentence in Amalgamation SNP describes
a problem of the form $\Csp(\bB)$ for an $\omega$-categorical structure $\bB$ that can be expanded to a 
homogeneous finitely bounded structure. 
\end{proposition}
\begin{proof}
To prove the first statement, 
let $\Phi$ be a sentence in ASNP, let $\tau$ be
the input signature of $\Phi$ (i.e., the free relation symbols in $\Phi$), and let $\phi$ be the first-order part of $\Phi$ 
(which is a $\sigma$-formula, for $\sigma \supseteq \tau$). Since the class of all finite models of $\phi$ has the
amalgamation property,  Theorem~\ref{thm:fraisse} asserts the existence of a countable homogeneous
$\sigma$-structure $\bC$ whose age is exactly the class of all finite
$\sigma$-structures that satisfy $\phi$. 
The structure $\bC$ is finitely bounded; this is witnessed by the set $\mathcal N$ of all $\sigma$-structures $\bN$
with a minimal number of vertices that do not satisfy $\phi$. None of those structures $\bN$ can 
have larger size than the number of variables of $\phi$, and hence $\mathcal N$ is finite. 
To see that $\mathcal N$ indeed bounds the age of $\bC$, let $\bA$
be a finite induced substructure of $\bC$. 
Then $\bA$ cannot have a substructure from $\mathcal N$, since any such substructure would falsify $\phi$,
and hence also $\bA$ would not satisfy $\phi$. Now suppose that $\bC$ does \emph{not} contain $\bA$ as a finite
induced substructure. This means that $\bA$ falsifies $\phi$. 
Then there is a minimal number of elements witnessing that the universal
formula $\phi$ is false in $\bA$, and those elements induce a structure from $\mathcal N$, which proves the claim. 
By Lemma~\ref{lem:cat-via-homogen} the structure $\bC$ is $\omega$-categorical.  
The $\tau$-reduct $\bB$ of $\bC$ is also $\omega$-categorical (Lemma~\ref{lem:interpret}).

We finally show that every finite % (in fact, every countable) 
$\tau$-structure $\bA$ satisfies $\Phi$ if and only if 
it homomorphically maps (in fact, embeds) into $\bB$. 
When $\bA$ is a $\tau$-structure that satisfies $\Phi$, then there exists a $\sigma$-expansion $\bA'$ of $\bA$
that satisfies $\phi$. By universality of $\bC$ the same map is an embedding of $\bA'$ into $\bC$, and this
gives an embedding of $\bA$ into $\bB$. 

Conversely, suppose that there is a homomorphism $f$ from
a $\tau$-structure $\bA$ to $\bB$. Then we construct the $\sigma$-expansion $\bA'$ of $\bA$ by putting for
all $S \in \sigma \setminus \tau$ the tuple
$(t_1,\dots,t_n)$ into $S^{\bA'}$ if and only if $(f(t_1),\dots,f(t_n)) \in S^{\bC}$. 
Suppose for contradiction that there were a tuple $t=(t_1,\dots,t_k)$ 
of elements that violates
a clause of $\phi$ in this expansion $\bA'$. 
% The following is not necessary, since we don't have equality in the formulas.
%First consider the case that $f(t_i)=f(t_j)$ for some $i \neq j$.
%Let $\bA_1$ be the structure induced by $\{f(t_1),\dots,f(t_k)\}$ in $\bC$,
%and define $\bA_1'$ from $\bA_1$ by renaming the element $f(t_i)$ of $\bA_1$ to
%$v$. 
%Finally, let $\bA_0$ be the structure induced by $\{f(t_1),\dots,f(t_k)\}\setminus \{t_i,t_j\}$ 
%in $\bC$.
%By strong amalgamation of the age of $\bC$, 
%the strong amalgam of $\bA_1$ with $\bA_2$ over $\bA_0$ embeds into $\bC$
%via an embedding $e$. 
%Clearly, the mapping $h$ from $\bA$ to $\bB$ 
%defined by $h(t_i) = e(f(t_i))$ for $i \neq j$, and $h(t_j)=e(v)$ otherwise,
%is a homomorphism from $\bA$ to $\bB$. Repeating this step, we may assume in the following that $f$ is injective. 
Then the image of $f$ induces a substructure in $\bC$ that also violates
$\phi$, since all relation symbols from $\tau$ appear negatively in $\phi$. This is a contradiction, and hence 
$\bA'$ satisfies $\phi$. We conclude that $\bA$ satisfies $\Phi$.
\end{proof}

% BOOKTD: would be interesting to see whether the following example even works in SNP, not just in NP.
We want to give an example of an $\omega$-categorical
structure $\bB$ such that $\Csp(\bB)$ is in NP, but not in Amalgamation SNP. To show that $\Csp(\bB)$ is not in ASNP, 
the sequence $(m_n)_{n \geq 1}$ of numbers of maximal pp-$n$-types of $\Th(\bB)$ turns out to
be useful (see Section~\ref{ssect:omega-cat-mc-cores}). Note that when two structures have the same CSP, then they have the same number of maximal pp-$n$-types. Hence, when $\Phi$ is a sentence from ASNP, there is a unique sequence $(m_n)_{n \geq 1}$ such that $m_n$ equals the of maximal pp-$n$-types of $\Th(\bB)$ for any $\bB$
such that $\Phi$ describes $\Csp(\bB)$. We call $(m_n)_{n \geq 1}$
the \emph{characteristic sequence} of $\Phi$.

\begin{proposition}\label{prop:asnp-growth}
Let $\Phi$ be a sentence from Amalgamation SNP. Then the characteristic sequence of $\Phi$ is in $O(2^{P(n)})$, for some polynomial $P$.
\end{proposition}
\begin{proof}
Proposition~\ref{prop:asnp-finitely-bounded} shows that there exists an $\omega$-categorical structure $\bB$ such that $\Phi$ describes $\Csp(\bB)$. The proof 
of Proposition~\ref{prop:asnp-finitely-bounded} shows that there exists an expansion $\bC$ of $\bB$ by finitely many relation symbols which is homogeneous. Hence, an orbit of $n$-tuples
in $\bC$ is uniquely described by the atomic formulas that hold on a (equivalently, all)
tuples from this orbit, and since the signature of $\bC$ is finite, this gives
a bound of $2^{P(n)}$ on the number of orbits of $n$-tuples in $\bC$, for some polynomial whose degree equals the maximal arity of the relations in $\bC$.
The number of orbits of $n$-tuples is an upper bound on the number of
maximal pp-$n$-tpyes (since two tuples with the same orbit clearly
have the same pp-type). 
\end{proof}

We can now present the example of an $\omega$-categorical
structure $\bB$ such that $\Csp(\bB)$ is in NP, but not in ASNP.
We use a CSP for a set constraint languages, as introduced
in Section~\ref{ssect:setcsps}. 

\begin{example}
Let $\bB$ be the structure that
contains all relations of arity at most three
with a quantifier-free first-order definition in the atomless Boolean algebra $\bA$.
Since $\bA$ is $\omega$-categorical, the signature of 
$\bB$ is finite. 
By Proposition~\ref{prop:setcsps-in-np},
$\Csp(\bB)$ is in NP.  
To prove that $\Csp(\bB)$ is not in ASNP it suffices
to show that the characteristic sequence of $\Phi$ grows faster than 
$O(2^{P(n)})$, for any polynomial $P$.
We first show that $\bB$ is a model-complete core. 
Trivially, $\bB$ is a core, since with each relation 
also the complement of the relation is a relation of $\bB$.
To see that $\bB$ is model-complete, let $\phi$ be a first-order formula  
that defines a first-order relation $R$ over $\bB$; we have to show that $R$ also has an existential definition over $\bB$. By quantifier-elimination
of $\bA$ (recall that $\bA$ has function symbols $\cup,\cap,c,{\bf 0},{\bf 1}$), there is a quantifier-free first-order formula $\psi$ that defines $R$ over
$\bA$. By un-nesting terms in $\psi$ with the help of new existentially quantified variables, and replacing occurrences of atomic formulas by the corresponding formulas in the language of $\bB$
(for instance replacing formulas of the form $x \cap y = z$ by $S(x,y,z)$ where $S$ is the relation of $\bB$ defined by $x \cap y = z$), we find the required existential definition 
of $R$ in $\bB$.
By Theorem~\ref{thm:mc-core}, the orbits of $n$-tuples 
in $\bB$ are primitive positive
definable, and so $m_n$ equals the number of orbits of $n$-tuples of $\bB$.
 
We show that $m_n \geq 2^{2^{\lfloor n/2 \rfloor}}$.
To see this, let $l = {\lfloor n/2 \rfloor}$ and $X := \{x_1,\dots,x_l\}$ be elements such that for any two distinct subsets $S_1,S_2$ of $X$ the
elements $\cap S_1$ and $\cap S_2$ of $\bA$ are distinct.
Hence, there are $2^l$ many elements $a_1,\dots,a_{2^l}$ 
that can by formed
from $x_1,\dots,x_l$ by applying $\cap$, and there are
$2^{2^l}$ many ways of selecting a tuple $(y_1,\dots,y_l)$ of elements from $\{a_1,\dots,a_{2^l}\}$. For any such selection, and since
the relations $\{(x,y,z) \; | \; x \cap y = z\}$ and $\{(x,y,z) \; | \; x \cup y = z\}$ are in $\bB$, the
tuple $(x_1,\dots,x_l,y_1,\dots,y_l)$ will lie in a distinct orbit, which shows the claim.
\end{example}

From this example and the proof of Proposition~\ref{prop:asnp-growth} we can also see the following, which we note 
for later use. 
\begin{corollary}\label{cor:not-finitely-homogeneous}
There is no relational structure $\bB$ that can define the atomless Boolean algebra and
is homogeneous in a finite relational signature.
\end{corollary}

\subsection{CSPs in SNP without an $\omega$-categorical template}
We have seen in Section~\ref{ssect:snp-csp} %of Chapter~\ref{chap:intro} 
that every CSP in SNP
is also in connected monotone SNP. 
We have also seen a characterization of those CSPs
that can be formulated with an $\omega$-categorical template (Theorem~\ref{thm:omega-cat}).
So it is natural to ask for a concrete example of a connected monotone
SNP sentence that cannot be formulated with an 
$\omega$-categorical template. 

\begin{example}\label{expl:succ-snp}
Let $\Phi$ be the following connected monotone SNP sentence.
\begin{align*}
\exists E,T \; \forall x,y,z \, \big( & \text{`$E$ is equivalence relation'} \\
\wedge & \text{`$T$ is transitive irreflexive and extends $\suc$'} \\
\wedge & ((\suc(x,y) \wedge E(x,z)) \Rightarrow \suc(z,y)) \\
\wedge & ((\suc(x,y) \wedge E(y,z)) \Rightarrow E(x,z)) \\
\wedge & ((\suc(x,y) \wedge \suc(x,z)) \Rightarrow E(y,z)) \\
\wedge & ((\suc(x,y) \wedge \suc(z,y)) \Rightarrow E(y,z)) \\
\wedge & (\neg E(x,y) \vee \neg \suc(x,y)) \big )
\end{align*}
The sentence $\Phi$ describes $\Csp(({\mathbb Z}; \suc))$ where
$\suc = \{(x,y) \in {\mathbb Z}^2 \; | \; y=x+1\}$ 
(as in Section~\ref{ssect:succ}). The idea is that an $\{\suc,E,T\}$-structure
satisfies the quantifier-free part of $\Phi$ if
\begin{itemize}
\item $E(x,y)$ holds if for all homomorphisms from the $\{\suc\}$-reduct of the
structure to $({\mathbb Z}; \suc)$ we have $h(x) = h(y)$, and 
\item $T(x,y)$ holds for all homomorphisms $h$
from the $\{\suc\}$-reduct of the structure to $({\mathbb Z}; \suc)$ we have $h(x)<h(y)$.
\end{itemize}
\qed
\end{example}

\begin{proposition}
$\Csp(({\mathbb Z};\suc))$ cannot be formulated with an
$\omega$-categorical template. 
\end{proposition}
\begin{proof}
The number of maximal pp-$n$-types is the same in any structure $\bB$
where $\Csp(\bB)$ is described by $\Phi$, so by Corollary~\ref{cor:omega-cat} it suffices
to check that $({\mathbb Z}; \suc)$ has an infinite
number of maximal pp-2-types. But this is clear
since the formula $\phi_n(x_0,x_n)$ defined by $\exists x_1,\dots,x_{n-1} \; \bigwedge_{i=1}^{n} \suc(x_{i-1},x_i)$ is for each $n$ 
in a different pp-2-type.
\end{proof}

\ignore{
\begin{example}\label{expl:succ-snp}
Let $\Phi$ be the following connected monotone SNP sentence.
\begin{align*}
\exists E,T \; \forall x,y,z \, \big( & E(x,x) \wedge \neg T(x,x) \\
\wedge & (\suc(x,y) \Rightarrow T(x,y)) \\
\wedge & (E(x,y) \Rightarrow E(y,x)) \\
\wedge & ((E(x,y) \vee E(y,z)) \Rightarrow E(x,z)) \\
\wedge & ((T(x,y) \vee T(y,z)) \Rightarrow T(x,z)) \\
\wedge & ((\suc(x,u) \wedge E(x,y) \wedge \suc(y,v)) \Rightarrow E(u,v)) \\
\wedge & ((\suc(x,u) \wedge E(u,v) \wedge \suc(y,v)) \Rightarrow E(x,y)) \\
\wedge & (\neg E(x,y) \vee \neg \suc(x,y)) \big )
\end{align*}
The sentence $\Phi$ describes $\Csp(({\mathbb Z}; \suc))$ where
$\suc = \{(x,y) \in {\mathbb Z}^2 \; | \; y=x+1\}$ 
(as in Section~\ref{ssect:succ}). The idea is that an $\{E,\suc\}$-structure
satisfies the quantifier-free part of $\Phi$ if
the binary relation $E$ denotes a (symmetric and transitive) relation that holds between
two vertices when all homomorphisms from the $\{\suc\}$-reduct of the
structure to $({\mathbb Z}; \suc)$ map those vertices to the same integer.
The number of maximal pp-$n$-types is the same in any structure $\bB$
where $\Csp(\bB)$ is described by $\Phi$, so by Theorem~\ref{thm:omega-cat} it suffices
to check that $({\mathbb Z}; \suc)$ has an infinite
number of maximal pp-2-types. But this is clear
since the formula $\phi_n(x_0,x_n)$ defined by $\exists x_1,\dots,x_{n-1} \; \bigwedge_{i=1}^{n} \suc(x_{i-1},x_i)$ is for each $n$ 
in a different pp-2-type.
\end{example}}

% !TEX root = ../Book.tex
\chapter{Universal Algebra}
\label{chap:algebra}
\begin{center}
\includegraphics[scale=.8]{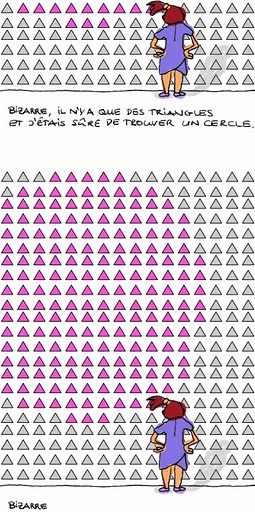}
\end{center}

\vspace{.4cm}

One of the central concerns of universal algebra, similarly as in model-theory, is \emph{classification} of mathematical structures. 
Often, model-theory is considered to be an extension of universal
algebra, as formulated by Chang and Keisler in 
\begin{center} model-theory = univeral algebra + logic.
\end{center} 
Universal algebra leads to classification results with finer distinctions: 
while model theory often considers two relational
structures to be equal when they are first-order interdefinable, universal algebra provides methods that allow to distinguish relational structures up to primitive positive definability.
To do so, we study higher-dimensional generalizations of endomorphisms monoids, called \emph{polymorphism clones}; 
from the perspective
of this text, we therefore have
\begin{center}
 model-theory = one-dimensional universal algebra.
\end{center}

The strongest universal-algebraic classification results are available
on \emph{finite} domains~\cite{HobbyMcKenzie}. 
In recent years, strong links between
deep and central questions in universal algebra
and the Feder-Vardi conjecture have led to renewed activity.
In fact, several important new and purely algebraic results, for example from~\cite{Siggers,BartoKozikLICS10,IMMVW,MarotiMcKenzie}, were originally motivated by questions about CSPs. 

There has been less work on algebras over infinite domains.
However, a considerable amount of universal-algebraic techniques also applies when the algebra under consideration contains as operations all the permutations from an oligomorphic permutation group; we will call such algebras \emph{oligomorphic}. The assumption that the algebra be oligomorphic
seems to provide the necessary amount of `finiteness'
that we need for applying universal-algebraic methods.

We have decided to give a self-contained presentation of this
section, even though that this implies 
a certain redundancy for the reader who has already followed 
Chapter~\ref{chap:mt}. 
The step from algebras with only unary functions to algebras that contain higher-ary
functions is the step where universal algebra becomes interesting. 
At the same time, the step from studying automorphisms and embeddings to studying polymorphisms is the step that is new to
model-theorists, so we found it natural to
divide the material into a chapter on model theoretic and a chapter 
on universal algebraic background.
%Model-theorists might
%read Chapter~\ref{chap:mt} very quickly, whereas 
%for universal algebraists most of Chapter~\ref{chap:algebra} might
%look familiar. 

This chapter contains original material from~\cite{BodirskyNesetrilJLC,BodChenPinsker,OligoClone,
Maximal,BodirskySurvey}.

\section{Oligomorphic Clones}
\label{sect:ua-prelims}
Fix a countably infinite base set $B$, also called the \emph{domain} or \emph{base set}.
For $n\geq 1$, denote by $\cOn$ the
set $B^{B^n}=(B^n\To B)$ of $n$-ary operations on
$B$. 
Then $\cO := \bigcup_{n\geq 1}\cOn$ is the set of all 
operations on $B$ of finite arity. A \emph{clone} $\cC$ is a subset of $\cO$
satisfying the following two properties:
\begin{itemize}
    \item $\cC$ contains all projections, i.e., for all $1\leq
    k\leq n$ the operation $\pi^n_k\in\cOn$ defined by
    $\pi^n_k(x_1,\ldots,x_n)=x_k$, and

    \item $\cC$ is \emph{closed under composition}, that is, for all
    $f\in\cC \cap \cOn$ and $g_1,\ldots,g_n\in\cC \cap \cO^{(m)}$ the operation $f(g_1,\ldots,g_n)\in\cO^{(m)}$ defined by
    $$
        (x_1,\ldots,x_m)\mapsto f(g_1(x_1,\ldots,x_m),\ldots,g_n(x_1,\ldots,x_m))
    $$
    is an element of $\cC$.
\end{itemize}

For our applications of universal algebra in logic,
we are interested in clones that satisfy an additional
topological closure property\footnote{The corresponding topology is defined as follows. Equip $B$ with the discrete topology, and $\cOn=B^{B^n}$ with the corresponding product
(Tychonoff) topology, for every $n\geq 1$. (For background in topology, see Chapter~\ref{chap:topology}.) %We do not make any deep use of topology in this chapter.
A clone $\cC$ is \emph{closed} with respect to this topology iff 
each of its $n$-ary fragments $\cC\cap\cOn$ is a closed subset of $\cOn$.}.
A clone $\cC$ is
called \emph{locally closed}\footnote{In universal algebra, 
locally closed clones are often just called \emph{local clones}. 
Topologists would call 
such objects just \emph{closed clones} since the reference to the specific topology under consideration is clear. Our choice to call those objects \emph{locally closed} is a compromise, and standard~\cite{Szendrei}.} if and only if it satisfies the following
interpolation property:
%Equivalently, a clone $\cC$ is locally closed if 
\begin{quote}
        for all $n\geq 1$ and all $g\in\cOn$, if for all finite $A \subseteq B^n$
        there exists an $n$-ary $f\in\cC$ which agrees with
        $g$ on $A$, then $g\in\cC$.
\end{quote}

The following proposition is folklore in universal algebra, 
see e.g.~\cite{Szendrei}. Its proof is very similar to the proof
of Proposition~\ref{prop:loc-clos-group}, and we leave it to the reader.

\begin{proposition}\label{prop:loc-clos-clone}
Let $\cF \subseteq \cO$ be a set of operations. 
Then the following are equivalent.
\begin{enumerate}
\item $\cF$ is the polymorphism clone of a relational structure;
\item $\cF$ is a locally closed clone.
%\item $\cF$ is the polymorphism clone of a homogeneous relational
%structure.
\end{enumerate}
\end{proposition}

Arbitrary
intersections of clones are clones, and arbitrary
intersections of locally closed sets are locally closed. In fact, 
the set of all locally closed clones on $B$, partially ordered by 
inclusion, forms a complete lattice.
The join of a family
$(\cC_i)_{i\in I}$ can be obtained as follows. First, take the set
 of all operations on $B$ which can be obtained by composing
operations from $\bigcup_{i\in I}\cC_i$; this set is a clone, but
might not be locally closed. For this reason, we have to additionally form the topological
closure of this set in order to arrive
at the join in this lattice. For a set of operations
$\cF\subseteq\cO$, we write $\cl{\cF}$ for the clone \emph{locally generated by $\cF$}, i.e., for the smallest locally closed clone
containing $\cF$. 

%Kommt spaeter mit Beweis. Observe that $\cl{\cF}$ is
%just the intersection of all locally closed clones containing $\cF$, or
%equivalently the topological closure of the set of term operations
%that can be built from $\cF$.

\begin{definition}
Let $\cC \subseteq \cO$ be a clone. A unary operation $e \in \cC$ 
is called \emph{invertible in $\cC$} if there exists a unary $i \in \cC$
such that $i(e(x))=e(i(x))=x$ for all $x \in B$. 
\end{definition}

Suppose that $\cC$ is the polymorphism clone of a structure $\bB$.
Then obviously any invertible operation of $\cC$ is an automorphism
of $\bB$.

\begin{definition}
A clone $\cC \subseteq \cO$ is \emph{oligomorphic} if
the set of operations in $\cC$ that are invertible in $\cC$ forms an oligomorphic permutation group. 
\end{definition}

It is immediate from Theorem~\ref{thm:ryll} that a locally closed clone is oligomorphic if and only if it is the polymorphism clone of an $\omega$-categorical structure.

\section{The Inv-Pol Galois Connection}
\label{sect:inv-pol}
The exposition in this section parallels that of Section~\ref{sect:galois}. 
% in Chapter~\ref{chap:mt}.
Let $f$ be from $\cOn$, and let $R \subseteq B^m$ be a relation. Then we say
that $f$ \emph{preserves} $R$ (and that $R$ is \emph{invariant under $f$}) iff $f(r_1,\ldots,r_n)\in\rho$
whenever $r_1,\ldots ,r_n\in R$, where $f(r_1,\ldots,r_n)$ is
calculated componentwise. 
%This notion of preservation links finitary
%relations on $B$ to finitary operations and is the prime tool for
%studying $\omega$-categorical structures up to primitive positive
%interdefinability.
For a relational structure $\bB$ (or for a set of relations $\R$) with domain $B$, 
we write  $\pol(\bB)$ (or $\pol(\R)$, respectively) for the set of
those operations in $\cO$ which preserve 
all relations from $\bB$ (all relations in $\R$). 
The operations in $\pol(\mathfrak B)$ are called 
\emph{polymorphisms} of $\bB$. Note that the polymorphisms
of $\bB$ are exactly the homomorphism from finite powers of $\bB$ to $\bB$. 

We have seen how to assign sets of operations to sets of relations;
likewise, we can go the other way. Given a set of
operations $\cF\subseteq\cO$, we write $\Inv(\cF)$ for the set of all
relations which are invariant under all $f\in\cF$. 
Using the Galois connection defined by the operators $\pol$ and
$\Inv$, we obtain the following
well-known description of the hull operator $\cF \mapsto \pol(\inv(\cF))$ (confer~\cite{Szendrei}, in particular Corollary 1.9).

\begin{proposition}\label{prop:pol-inv}
    Let $\cF\subseteq\cO$, and $g \in \cO$. 
    Then the following are equivalent.
\begin{enumerate}
\item $g \in \cl{\cF}$;
\item $g$ is in the local closure of the operations of the clone generated by $\cF$;
\item For all $n \geq 1$ and all $a_1,\dots,a_k \in B^n$ there is a $f$ in the clone
generated by $\cF$ such
that $g(a_1,\dots,a_k)=f(a_1,\dots,a_k)$.
\item $g \in \pol(\Inv(\cF))$;
\end{enumerate}
In particular, $\cl{\cF}=\pol(\Inv(\cF))$.
\end{proposition}
\begin{proof}
Note that the set $\cF'$ of all operations that are in the local closure of the clone generated by $\cF$ is a clone and locally closed. Therefore, the clone $\cl{\cF}$ is contained in $\cF'$, and (1) implies (2).

% 2 implies 1 
% Let $\cF'$ be the clone interpolated by the operations of the clone generated by $\cF$.
%We have that $\cl{\cF}$ contains $\cF$, is a clone, and closed under interpolation,
%therefore it contains $\cF'$. 

%For the implication from (2) to (4), let $g$ be a $k$-ary operation
%from $\cl{\cF}$, and let $R$ be from $\inv(\cF)$. We have to show that $g$ preserves $R$. Let $t_1,\dots,t_k$ be from $R$. Since $f$ is interpolated by the clone
%generated by $\cF$, we have that $f(t_1,\dots,t_k) = g(t_1,\dots,t_k)$ for some operation
%$g$ generated from operations in $\cF$ and projections. Since
%all those operations preserve $R$, we have that 
%$f(t_1,\dots,t_k) \in R$. 

For the implication from (2) to (3), let $g$ be a $k$-ary operation
that is in the local closure of the clone $\cF'$ generated by $\cF$. 
Let $a_1,\dots,a_k$  be from $B^n$ for some $n \geq 1$.
Suppose $a_i=(a_i^1,\dots,a_i^n)$ for $i \leq k$, and let $a^j = (a^j_1,\dots,a^j_k)$ for
$j \leq n$. Since $g$ is in the closure of $\cF'$, there exists an $f \in \cF'$ 
that agrees with $g$ on $\{a^1,\dots,a^n\} \subseteq B^k$. 
In particular, $g(a_1,\dots,a_k)=f(a_1,\dots,a_k)$. 

For the implication from (3) to (4), assume (3), and let
$R$ be from $\inv(\cF)$. We have to show that $g$ preserves $R$. 
Let $t_1,\dots,t_k$ be from $R$. By assumption $f(t_1,\dots,t_k) = g(t_1,\dots,t_k)$ 
for some operation $g$ generated from operations in $\cF$ and projections. Since
all those operations preserve $R$, we have that 
$f(t_1,\dots,t_k) \in R$. 

To show that (4) implies (1), let $f$ be a $k$-ary operation from $\pol(\inv(\cF))$. 
Let $\cC$ be the clone generated by $\cF$. It suffices to show that for every finite subset $A$ of $B$ there is
an operation $g \in \cC$ such that $f(\bar a) = g(\bar a)$
for all $\bar a \in A^k$. List all elements of $A^k$ by $a^1,\dots,a^n$, and consider the relation 
$$R:=\{ (g(a^1),\dots,g(a^n)) \; | \; g \in \cC\}.$$ 
Note that $R$ is preserved by all operations in $\cF$ 
and so by assumption $f$ preserves $R$.
Also note that $(a^1_i,\dots,a^n_i) \in R$ since $\cC$
contains the projections. 
Therefore, $(f(a^1),\dots,f(a^n)) \in R$,
and hence $(f(a^1),\dots,f(a^n)) = (g(a^1),\dots,g(a^n))$
for a $g \in \cC$, as required.
\end{proof}

The following is straightforward. %, but we provide a proof for completeness.

\begin{proposition}\label{prop:pp-preserved}
	Let $\mathfrak B$ be any structure. Then $\Inv(\Pol(\bB))$
	contains $\langle \bB \rangle_{\pp}$, the set of all relations that are primitive positive definable in $\bB$.
\end{proposition}
\begin{proof}
Suppose that $R$ is $k$-ary, 
has a primitive positive definition $\psi$, 
and let $f$ be an $l$-ary polymorphism
of $\bB$. To show that $f$ preserves $R$, let 
$t_1,\dots,t_l$ be tuples from $R$. Then there must be witnesses
for the existentially quantified variables $x_{l+1},\dots,x_{n}$ of
$\psi$ that show that $\psi(t_i)$ holds in $\bB$, for all $1 \leq i \leq n$. Write
$s_i$ for the extension of $t_i$ such that $s_i$ satisfies the quantifier-free part $\psi'(x_1,\dots,x_l,x_{l+1},\dots,x_n)$ of $\psi$ (we assume that $\psi$ is written in prenex normal form).
Then the tuple $$(f(s_1[1], \dots, s_l[1]),\dots,f(s_1[n],\dots,s_l[n]))$$
satisfies $\psi'$ as well. 
This shows that $(f(s_1[1], \dots, s_l[1]),\dots,f(s_1[k],\dots,s_l[k]))$
satisfies $\psi$ in $\bB$, which is what we had to show.
\end{proof}

A relation $R$ has a primitive positive definition in a \emph{finite}
structure 
if and only if
$R$ is preserved by all polymorphisms of this structure.
This was discovered by Geiger~\cite{Geiger} and independently by Bodnarcuk et al.~\cite{BoKaKoRo},
and is of central importance in universal algebra.
We have the following generalization of this theorem to $\omega$-categorical structures~\cite{BodirskyNesetril}. 

\begin{theorem}[from~\cite{BodirskyNesetril}]\label{thm:inv-pol}
Let $\bB$ be an $\omega$-categorical or a finite structure.
A relation $R$ has a primitive positive definition in $\bB$
if and only if $R$ is preserved by all polymorphisms of $\bB$; in symbols, $\Inv(\pol(\bB))=\langle \bB \rangle_{\pp}$.
\end{theorem}

\begin{proof}
One direction has been shown in Proposition~\ref{prop:pp-preserved}.
For the other direction, let $R$ be a $k$-ary relation that is preserved
by all polymorphisms of $\bB$. 
In particular, $R$ is preserved by all automorphisms of $\bB$,
and hence $R$ is a union of orbits of $k$-tuples in the automorphism group of $\bB$. 
By item (2) of Theorem~\ref{thm:ryll}, there
is a finite number of such orbits, $O_1,\dots,O_w$.
%, and by
%item (4) of Theorem~\ref{thm:ryll} 
%each
%of these orbits has a first-order definition, $\phi_1(x_1,\dots,x_k),\dots,\phi_w(x_1,\dots,x_k)$. 
If $R$ is empty, there is nothing to show (but we again use the
assumption that $\bot$ is allowed as a primitive positive formula), 
so let us assume that $w \geq 1$.
Fix for each $1 \leq j \leq w$ a $k$-tuple $a_j$ from $O_j$.
Let $B$ be the domain of $\bB$. 
Let $b_1, b_2, \dots$ be an enumeration of all $w$-tuples in $B^w$ with the additional property that for 
$1 \leq i \leq k$ we have $b_i = (a_1[i], \dots, a_w[i])$.
%Let us call a partial mapping from $B^w \rightarrow B$ \emph{bad} 
%if it maps $(b_1, \dots, b_k)$ to a tuple satisfying $ \neg \phi$. 
%Since $R$ is preserved by all polymorphisms, no homomorphism from $\bB^w$ to $\bB$ is bad.

By Lemma~\ref{lem:omega-cat-compactness}, 
if for every finite substructure $\bA$ of $\bB^w$ that contains $b_1,\dots,b_k$ 
there is a homomorphism from $\bA$ to $\bB$ 
that maps $(b_1,\dots,b_k)$ to a tuple that is not in $R$, then
there is also a homomorphism from $\bB^k$ to $\bB$ that maps
$(b_1,\dots,b_k)$ to a tuple that is not in $R$, and this would be a polymorphism of $\bB$ violating $R$ (the properties of a mapping to be a polymorphism and to violate $R$ have universal axiomatizations). 
So there must be a finite substructure $\bA$ containing the vertices $b_1, \dots, b_k$ of $B^w$ such that all homomorphisms from $\bA$ to $\bB$ map $b_1, \dots, b_k$ to a tuple in $R$. 

Let $q_1,\dots,q_l$ be the vertices of $\bA$ without $b_1,\dots,b_k$.
Write $\psi$ for the quantifier-free part of the canonical query of $\bA$
(see Section~\ref{ssect:can-query}).
We claim that the formula $\exists q_1,\dots,q_l. \, \psi$ 
is a primitive positive definition of $R$.
 The above argument shows that $\exists q_1,\dots,q_l. \, \psi$
 implies $R(b_1,\dots,b_k)$. To show that every tuple in $R$ satisfies 
 $\exists q_1,\dots,q_l. \, \psi$,
 let $f \colon \bA \rightarrow \bB$ be a homomorphism such that the tuple $(f(b_1), \dots, f(b_k))$ is from $R$. Then $(f(b_1), \dots, f(b_k)) \in O_j$ for some $1 \leq j \leq w$. 
 There is an automorphism $\alpha$ of $\bB$ sending $a_j$ to $(f(b_1), \dots, f(b_k))$. 
So we can extend $f$ to a homomorphism from $\bB^w$ to $\bB$ by setting $f(x_1,\dots,x_n) := \alpha(x_j)$. This shows in particular that $f(b_1),\dots,f(b_k))$ satisfies $\exists q_1,\dots,q_l. \, \psi$. 
\end{proof}

% After discussions with Hubie, Martin, Barny
%Note that our convention that we also consider $0$-ary operations as polymorphisms
%is important here: if $\bB$ is a structure such that every relation contains the tuple
%$(a,\dots,a)$ for some element $a$ of $\bB$, then the empty relation is not
%primitive positive definable in $\bB$, but the empty relation is preserved by all polymorphisms of arity $n \geq 1$. If we allow $0$-ary operations, then $\bB$
%has the $0$-ary polymorphism $a$, and this polymorphism violates the empty relation.

% DIFFERENTLY:
%Again, note that our convention that $\bot$ is a valid primitive positive formula
%is important here: if $\bB$ is a structure such that every relation contains the tuple
%$(a,\dots,a)$ for some element $a$ of $\bB$, then the empty relation might otherwise
%not be primitive positive definable in $\bB$, but the empty relation is preserved by all polymorphisms of arity $n \geq 1$. 
%If we allow $0$-ary operations, then $\bB$
%has the $0$-ary polymorphism $a$, and this polymorphism violates the empty relation.

Analogously to Corollary~\ref{cor:galois}, we obtain a Galois connection between structures with a first-order definition in 
an $\omega$-categorical structure $\bC$, considered up to primitive positive interdefinability, and
locally closed clones containing $\Aut(\bC)$.

\begin{theorem}\label{thm:pp-galois}
Let $\bC$ be a countable $\omega$-categorical structure. 
Then we have the following. 
    \begin{enumerate}
        \item The set of sets of the form $\langle \bB \rangle_{\pp}$,
        where $\bB$ is first-order definable in $\bC$, ordered by inclusion, forms a lattice.
        \item The set of locally closed clones that contain
        $\aut(\bC)$, ordered by inclusion, forms a lattice.
        \item The operator $\inv$ is an anti-isomorphism between those
        two lattices, and $\pol$ is its inverse.
    \end{enumerate}
\end{theorem}

The above theorem tells us that classifying the reducts of
an $\omega$-categorical structure $\bC$ up to primitive positive
interdefinability really amounts to understanding the lattice of
locally closed clones containing the automorphisms of $\bC$.

%In summary, we have seen that studying $\omega$-categorical structures up to primitive positive definability
%amounts to studying oligomorphic clones.

\section{Essential Arity}
\label{sect:arity}
This section investigates the \emph{bottom} of the lattice
of locally closed clones $\cC$ that contain a fixed oligomorphic permutation group.

\subsection{Essentially unary operations}\label{ssect:unary}
We say that $f \in \cO^{(k)}$ \emph{depends} on the argument
$i \in \{1,\dots,k\}$ if there is no $(k{-}1)$-ary operation $f'$ such that $f(x_1,
\dots, x_k) =$ $f'(x_1, \dots, x_{i-1}, x_{i+1},$ $\dots, x_k)$ for all $x_1,\dots,x_k \in B$. 
We
can equivalently characterize $k$-ary operations that depend on the
$i$-th argument by requiring that there are $x_1, \dots,
x_k \in B$ and $x_i' \in B$ such that $$f(x_1, \dots, x_k) \neq f(x_1, \dots,
x_{i-1}, x'_i, x_{i+1}, \dots, x_k) \; .$$
We say that an operation $f$ is \emph{essentially unary} iff there is an $i \in \{1, \dots, k\}$
and a unary operation $f_0$ such that $f(x_1, \dots, x_k) =
f_0(x_i)$. Operations that are not essentially unary are called
\emph{essential}.\footnote{This is standard in clone theory, and it
makes sense also for us, since the essential operations are those that are
essential for complexity classification.}

\begin{definition}
For any set $B$, the relations $P^B_3$ and $P^B_4$ over $B$ are defined as follows.
\begin{align*}
P^B_3 \; = \; & \big \{ (a,b,c) \in B^3 \; | \; a=b \text{ or } b=c \big \} \\
P^B_4 \; = \; & \big \{ (a,b,c,d) \in B^4 \; | \; a=b \text{ or } c=d \big \} %\\
%E_3 \; = \; & \{ (a,b,c) \in B^3 \; | \; (a=b \neq c) \text{ or } (a \neq b=c)\} 
\end{align*}
\end{definition}

\begin{lemma}\label{lem:unary}
Let $f \in \cO$ be an operation on a set $B$.
% of cardinality at least three. 
% NOT NECESSARY?!
Then
the following are equivalent.
\begin{enumerate}
\item $f$ is essentially unary.
\item $f$ preserves $P^B_3$.
\item $f$ preserves $P^B_4$.
\item $f$ depends on at most one argument.
\end{enumerate}
\end{lemma}
\begin{proof}
Let $k$ be the arity of $f$.
The implication from (1) to (2) is obvious, since unary operations
clearly preserve $P^B_3$.

To show the implication from (2) to (3), we show the contrapositive,
and assume that $f$ violates $P^B_4$. By permuting arguments
of $f$, we can assume that there are an $l \leq k$ and $4$-tuples
$a^1,\dots,a^k \in P^B_4$ with $f(a^1,\dots,a^k) \notin P^B_4$ such that
in $a^1,\dots,a^l$ the first two coordinates are equal,
and in $a^{l+1},\dots,a^k$ the last two coordinates are equal. 
Let $c$ be the tuple $(a^1_1,\dots,a^l_1,a^{l+1}_4,\dots,a^{k}_4)$.
Since $f(a^1,\dots,a^k) \notin P^B_4$ we have $f(a^1_1,\dots,a^k_1) \neq f(a^1_2,\dots,f^k_2)$, and therefore $f(c) \neq f(a^1_1,\dots,a^k_1)$ or $f(c) \neq f(a^1_2,\dots,a^k_2)$.
Let $d = (a^1_1,\dots,a^k_1)$ in the first case, and 
 $d = (a^1_2,\dots,a^k_2)$ in the second case.
Likewise, we have $f(c) \neq f(a^1_3,\dots,a^k_3)$ or $f(c) \neq f(a^1_4,\dots,a^k_4)$, and let $e = (a^1_3,\dots,a^k_3)$ in the first,
and $e = (a^1_4,\dots,a^k_4)$ in the second case.
Then for each $i \leq k$, the tuple $(d_i,c_i,e_i)$ is from $P^B_3$,
but $(f(d),f(c),f(e)) \notin P^B_3$.

The proof of the implication from (3) to (4) is again by contraposition.
Suppose $f$ depends
on the $i$-th and $j$-th argument, $1 \leq i \neq j \leq k$. Hence
there exist tuples $a_1,b_1,a_2,b_2 \in B^k $ such that $a_1,b_1$ and
$a_2,b_2$ only differ at the entries $i$ and $j$, respectively, and such
that $f(a_1) \neq f(b_1)$ and $f(a_2) \neq f(b_2)$. Then $(a_1(l),
b_1(l), a_2(l), b_2(l)) \in P^B_4$ for all $l \leq k$, but $(f(a_1),
f(b_1), f(a_2), f(b_2)) \notin P^B_4$, which shows that $f$ violates $P^B_4$. 

For the implication from (4) to (1), suppose that $f$ depends only on the first argument. 
Let $i \leq k$ be maximal such that there is an operation $g$
with
$f(x_1,\dots,x_k)=g(x_1,\dots,x_i)$. If $i=1$ then $f$ is essentially unary and we are done. Otherwise, observe that since $f$ does
not depend on the $i$-th argument, neither does $g$, and so
there is an $(i-1)$-ary operation $g'$ such that for all $x_1,\dots,x_n \in B$ we have $f(x_1,\dots,x_n)=g(x_1,\dots,x_i)=g'(x_1,\dots,x_{i-1})$, contradicting the choice of $i$.
\end{proof}

Combined with Theorem~\ref{thm:ep}, we obtain for $\omega$-categorical structures a characterization
of the situation where disjunction can be eliminated from existential
positive formulas.

\begin{proposition}\label{prop:unary}
Let $\bB$ be an $\omega$-categorical structure, and let
 $\cC$ be its polymorphism clone. Then the
following are equivalent.
\begin{enumerate}
\item All relations with an existential positive 
definition in $\bB$ also have
a primitive positive definition in $\bB$.
\item The relation $P^B_3$ is contained in $\Inv(\cC)$.
%\item The relation $P^B_4$ is contained in $\Inv(\C)$.
\item All operations in $\cC$ are essentially unary.
%\item The projections and the unary operations in $\cC$ generate $\cC$.
\end{enumerate}
\end{proposition}

\begin{proof}
(1) implies (2). The formula $(x=y) \vee (y=z)$ is existential positive,
and thus has a primitive positive definition in $\bB$; such formulas
are preserved by $\cC$.

(2) implies (3). Follows from Lemma~\ref{lem:unary}.

%(3) implies (4): By definition, every essentially unary operation
%$f\in \C$ is composed out of a projection and a unary operation.

(3) implies (1). Unary operations preserve
all existentially positive formulas. Hence, when $\phi$
is an existential positive formula, then by assumption 
all polymorphisms of
$\bB$ preserve $\phi$, and $\phi$ is equivalent to a
primitive positive formula by Theorem~\ref{thm:ep}.
\end{proof}

If all operations of a clone $\cC$ are essentially unary, 
we say that $\cC$ is \emph{essentially unary}.
% Do we really need this? Don't know, so leave it for now. 

\subsection{Elementary Clones}
If every polymorphism of an $\omega$-categorical structure $\bB$ is 
locally generated by the automorphisms of $\bB$,
then $\bB$ has the remarkable property that every
first-order formula is equivalent to a primitive positive
formula over $\bB$. In this case, the polymorphism clone of $\bB$ is the smallest
element of the lattice of locally closed clones described in Section~\ref{sect:inv-pol}. 
The facts in this section are straightforward combinations
of previous results. We state them for future use. 

%This again demonstrates how purely relational statements
%(item (1)) can be translated into purely operational statement (item (4)), and vice versa.

%A polymorphism $f$ of $\bB$ is said to be \emph{oligopotent}
%if $f(x,\dots,x)$ is locally generated by the automorphisms of $\bB$.

\begin{corollary}\label{cor:elementary}
Let $\bB$ be an $\omega$-categorical structure, and let $\cC$ be
its polymorphism clone. Then the following are equivalent.
\begin{enumerate}
\item Every relation with a first-order definition also has a primitive positive
definition in $\bB$.
\item $\bB$ is a model-complete core, and $P^B_3$ is primitive positive definable in $\bB$.
%\item All operations in $\cC$ are oligopotent and essentially unary.
\item $\cC$ is locally generated by the invertible operations in $\cC$.
\item All operations in $\cC$ are elementary, i.e., preserve all first-order definable relations in $\bB$.
\end{enumerate}
\end{corollary}
\begin{proof}
(1) implies (2).
We assume that every first-order definable relation has a primitive positive definition, and hence is preserved by all polymorphisms
of $\bB$. In particular, the endomorphisms of $\bB$
preserve all first-order definable relations, and hence $\bB$
is a model-complete core.
Moreover,  the relation $P^B_3$ is clearly first-order definable, 
and therefore also primitive positive definable in $\bB$.

%(2) implies (3).
%Straightforward from Proposition~\ref{prop:unary} and Proposition~\ref{prop:mc-core}.

(2) implies (3). Assume (2). Then
Proposition~\ref{prop:unary} implies that all polymorphisms of
$\bB$ are essentially unary. Thus, for every $n$-ary polymorphism $f$ of $\bB$ there is an endomorphism $g$ 
of $\bB$ and an $j \leq n$ such that
$f(x_1,\dots,x_n)=g(x_j)$. 
Since $\bB$ is a model-complete core, and by Theorem~\ref{thm:mc-core}, $g$ is locally generated by the 
automorphisms 
of $\bB$, and in particular by the 
invertible operations in $\cC$. 
Hence, 
$f$ is locally generated by the locally invertible operations in $\cC$, 
which proves (3).

(3) implies (4). Invertible operations of $\cC$ preserve all first-order
definable relations in $\bB$.
Hence, the implication follows from Proposition~\ref{prop:pol-inv}.

(4) implies (1). By Theorem~\ref{thm:inv-pol}.
\end{proof}

We will see in Corollary~\ref{cor:elementary-hard} %in Section~\ref{ssect:pp-interpret} 
that when the equivalent conditions from Corollary~\ref{cor:elementary} apply, 
% Remark: hier brauchen wir unendliche Grundmenge, aber das ist ja implizit
% wenn wir sagen B ist omega-cat.
then $\bB$ has a finite signature reduct $\bB'$ whose CSP is NP-hard.

\subsection{Arity Reduction}
For many combinatorial arguments with oligomorphic clones
it is crucial to have bounds on
the arity of operations with certain properties.
A basic, yet extremely useful observation to obtain such bounds is the following
(which holds for arbitrary structures $\bB$).

\begin{lemma}\label{lem:small-arity}
Let $\bB$ be a relational structure
and let $R$ be a $k$-ary relation 
contained in $m$ orbits of $k$-tuples of
$\aut(\bB)$.  If $\bB$ has a polymorphism $f$ that violates
$R$, then $\bB$ also has an at most $m$-ary polymorphism that violates $R$.
%Then every operation $f$ that violates $R$ 
%generates an at most $m$-ary operation that violates $R$.
\end{lemma}
\begin{proof}
    Let $f'$ be an polymorphism of $\bB$ 
    of smallest arity $l$ that violates $R$.
    Then there are $k$-tuples $t_1,\dots,t_l \in R$
    such that $f'(t_1,\dots,t_l) \notin R$. For $l>m$
    there are two tuples $t_i$ and $t_j$ that
    lie in the same orbit of $k$-tuples,
    and therefore $\bB$ has an automorphism $\alpha$
    such that $\alpha t_j =t_i$. By permuting the arguments of $f'$,
    we can assume that $i=1$ and $j=2$.
    Then the $(l-1)$-ary operation $g$ defined as
    $$g(x_2,\dots,x_l):=f'(\alpha x_2,x_2,\dots,x_l)$$ 
    is also a polymorphism
    of $\bB$, and also violates $R$,
    a contradiction. Hence, $l \leq m$.
\end{proof}

We present applications of Lemma~\ref{lem:small-arity}.
Recall that $r(\cG)$ denotes the \emph{rank} of $\cG$, i.e.,
the number of orbitals of $\cG$ (see Section~\ref{sect:galois}).

\begin{corollary}\label{cor:essential}
Let $\bB$ be a structure with an automorphism group $\cG$.
If $\bB$ has an essential polymorphism, then it must also have
an essential polymorphism of arity at most $2r(\cG)-1$.
\end{corollary}
\begin{proof}
The structure $\bB$ has an essential polymorphism if and only
if it has a polymorphism that violates the relation $P^B_3$, where $B$ is the domain of $\bB$, by Proposition~\ref{prop:unary}. The relation $P^B_3$ consists of at most $2r(\cG)-1$ orbits of triples: there are at most $r(\cG)$ orbits of triples $(t_1,t_2,t_3)$ where $t_1=t_2 \neq t_3$, and at most that many where $t_1 \neq t_2=t_3$. Only the orbit of the tuple where $t_1=t_2=t_3$ is counted twice. The statement follows from Lemma~\ref{lem:small-arity}.
\end{proof}

\begin{corollary}\label{cor:q-endo-viol}
Let $\bB$ be first-order definable in $(\mathbb Q;<)$, and
suppose there is a polymorphism of $\bB$ that violates $<$.
Then there is also an endomorphism of $\bB$ that violates $<$.
\end{corollary}
\begin{proof}
Observe that $<$ consists of a single orbit of pairs in Aut$((\mathbb Q;<))$, and therefore also in $\text{Aut}(\bB)$.
\end{proof}

% Another application: 
%Let $\mathcal S$ be the full symmetric group on $D$, i.e., the
%set of all unary bijective operations in $\O$. 

\begin{corollary}\label{cor:neq-constant}
Let $\cF \subseteq \cO$ 
be a local clone that contains a $2$-transitive permutation group $\cG$.
If there is an $f \in \cF$ that violates $\neq$, then $\cF$
contains a constant operation.
\end{corollary}
\begin{proof}
The relation $\neq$ consists of a single orbit of pairs in $\cG$. 
Hence, there is a unary operation in $\cF$ that violates
$\neq$ by Lemma~\ref{lem:small-arity}. The rest follows by Lemma~\ref{lem:constant}.
\end{proof}

Another application of Lemma~\ref{lem:small-arity} can
be found in Section~\ref{sect:schaefer}, and many applications 
can be found in Chapters~\ref{chap:schaefer} and~\ref{chap:tcsp}.

\subsection{K\'ara's method}
\label{ssect:kara}
We present another method for showing that
an oligomorphic clone with essential operations
must contain a \emph{binary} essential operation.
The method applies in many cases where the arity bounds 
% "bounds": plural ist absicht!
from Corollary~\ref{cor:essential} are too weak.
The idea is taken from~\cite{ecsps}, where it has been stated only 
for structures that are preserved by all permutations of the domain,
and it has been generalized slightly in~\cite{RandomMinOps}.
To state the method in full generality, we introduce the following, 
apparently new, concept.

\begin{definition}\label{def:orbital-ext}
A permutation group $\cG$ on a set $B$ has the
\emph{orbital extension property} 
if there is an orbital $O$ such that 
for all $b_1,b_2 \in B$ there is an element $c \in B$ such that $(c,b_1) \in O$
and $(c, b_2) \in O$.
\end{definition}
Note that every permutation group with the orbital extension property
is transitive, and that every 2-transitive infinite permutation group has
the orbital extension property.
More examples of oligomorphic permutation groups with the orbital extension property are the automorphism group of the Random graph,
$(\mathbb Q;<)$, the universal homogeneous poset, 
%, the dense local order $S(2)$,
% Problem hiermit: dass wir es nicht eingefuehrt haben, weil das CSP nicht unabhaengig
% in der Literatur aufgetaucht ist. Im Buch waere das aber gut, BOOKTD
the universal homogeneous $C$-relation,
and many more. 
An example of a structure without the orbital extension property is $K_{\omega,\omega}$, the complete bipartite graph where both parts are countably infinite. 
An example of an imprimitive oligomorphic permutation 
group with the orbital extension property is the automorphism group
of an equivalence relation on an infinite set with infinitely many infinite classes.

% There is another, similar property:
%A group $\cG$ of permutations of a set $B$ has the
%\emph{orbital extension property} 
%if for all $b_1,b_2 \in B$ there is a $c \in B$ such
%that $(b_1,c)$ and $(b_2,c)$ lie in the same orbital.
%In other words, $\cG$ has the orbital extension property
%if for all $b_1,b_2 \in B$
%there exists a $c \in B$ such that there is an $\alpha \in \cG_c$  with $\alpha(b_1)=b_2$.
% this is incomparable to primitivity: 
 % this is not stronger than primitivitiy: 
% we require a diameter-2 zig-zag condition 
% on orbitals.
% it is weaker, since we only require it for SOME orbital, not for all
% (see omega*K_omega)

% (see the digraph S(3))
% yet another:
% there is an orbital such that for all b_1,b_2 there is a c
% such that (b_1,c) and (b_2,c) are in the same orbit.
%  a stronger requirement. However, for the examples
% the difference doesn't matter. it is still incomparable to primitivity.
% Are the two definitions equivalent?

% true or false: every IMPRIMITIVE transitive 
% permutation group with more
% than two blocks of imprimitivity has the orbital extension property?

\begin{lemma}\label{lem:binary}
Let $\cC$ be a clone that contains a permutation group
$\cG$  with the orbital extension property.
Then, if $\cC$ contains an essential operation, it must also
contain a binary essential operation.
\end{lemma}
\begin{proof}
Let $f$ be an essential operation in $\cC$, and let $k$ be the arity 
of $f$.
Assume without loss of generality that $f$ depends all its arguments
and is at least ternary. In particular, there are $a_1,\dots,a_k$ and $a_1'$ such that
$f(a_1,\dots,a_k) \neq f(a'_1,a_2,\dots,a_k)$.
Let $O$ be the orbital of $\cG$ which exist due to
the orbital extension property of $\cG$. 
We distinguish two cases.

\textbf{Case 1.} There are elements $b_1,\dots,b_k$ such that
$(b_i,a_i) \in O$ for $2 \leq i \leq k$ and $f(b_1,a_2,\dots,a_k) \neq
f(b_1,\dots,b_k)$. Then there are $\alpha_3,\dots,\alpha_k \in \cG$ such that
$\alpha_i(a_2)=a_i$ and $\alpha_i(b_2) =b_i$. We define
$$g(x,y):=
f(x,y,\alpha_3(y),\dots,\alpha_k(y))\; ,$$
which clearly depends on both
arguments.

\textbf{Case 2.} For all $b_1,\dots,b_k$, if $(a_i,b_i) \in O$ for $2
\leq i \leq k$, then $f(b_1,a_2,\dots,a_k)=f(b_1,b_2,\dots,b_k)$.
Since $f$ depends on its second coordinate, there are
$c_1,\dots,c_k$ and $c_2'$ such that
$f(c_1,c_2,c_3,\dots,c_k) \neq f(c_1,c_2',c_3,\dots,c_k)$.
Then $f(c_1,a_2,\dots,a_k)$ can be
equal to either $f(c_1,c_2,c_3,\dots,c_k)$, or to
$f(c_1,c_2',c_3,\dots,c_k)$, but not to both. We assume without loss
of generality that $f(c_1,a_2,\dots,a_k) \neq
f(c_1,c_2,c_3,\dots,c_k)$. 
Since $\cG$ has the orbital extension property, 
 we can choose $d_2,\dots,d_k$ such that for each $2 \leq i \leq k$,
 the pairs $(d_i,a_i)$ all lie in $O$ and the pairs
$(d_i,c_i)$ all lie in $O$. Hence, 
there are $\alpha_3,\dots,\alpha_k \in \cG$ such that
$\alpha_i(c_2)=c_i$ and $\alpha_i(d_2)=d_i$. We claim that the
operation $g$ defined by
$$ g(x,y) := f(x,y,\alpha_3(y),\dots,\alpha_k(y))$$
depends on both arguments.
Indeed, we know
that $g(a_1,d_2)=f(a_1,d_2,\dots,d_k)=f(a_1,\dots,a_k)$, and that
$f(a'_1,d_2)=f(a_1',d_2,\dots,d_k)=f(a'_1,a_2,\dots,a_k)$. By the
choice of the values $a_1,\dots,a_k$ and $a_1'$ these two values
are distinct, and we have that $g$ depends on the first argument.
For the second argument, note that $g(c_1,d_2) = f(c_1,d_2,\dots,d_k) = f(c_1,a_2,\dots,a_k)$ and that $g(c_1,c_2) = f(c_1,c_2,\dots,c_k)$.
Because $f(c_1,a_2,\dots,a_k)$ and $f(c_1,c_2,\dots,c_k)$ are distinct,
the function $g$ also depends on the second argument.
\end{proof}

\begin{corollary}\label{cor:binary}
Let $\bB$ be 2-transitive with an essential polymorphism.
Then $\bB$ also has a binary essential polymorphism.
\end{corollary}

 % ist nur subsection, nicht section.
\subsection{Minimal Operations}
\label{ssect:minimal}
In some classification results, e.g.~in Chapter~\ref{chap:schaefer} and Chapter~\ref{chap:tcsp}
% Unterhalb des Betweenness Klones gibt es nur einen --
% das folgt leider nicht explizit aus der Klassification mit Jan Kara,
it turns out that a \emph{bottom-up} approach works best: for example in Chapter~\ref{chap:schaefer} we first classify all the minimal (with respect to set inclusion) locally closed
clones that strictly contain the automorphism group of the random graph, 
and then the classification argument is organised according to those \emph{minimal} clones.  

\begin{definition}
Let $\cC \subseteq \cO$ be a locally closed clone. We say that 
\begin{itemize} 
\item a local clone $\cD \subseteq \cO$ is \emph{minimal above $\cC$}Ê
if $\cC \subsetneq \cD$, and $\cC \subsetneq \cE \subseteq \cD$ implies $\cE=\cD$ for all locally closed clones $\cE$.
\item a function $f \in \cO$ is \emph{minimal above $\cC$} if $f$ is not from $\cC$, and of minimal arity such that for every $g \notin \cC$ that is locally generated by $\cC \cup \{f\}$ we have that $\cC \cup \{g\}$ locally generates $f$.
\end{itemize}
\end{definition}

The following is obvious from the definitions.

\begin{proposition}\label{prop:minimal-op}
Let $\cC$ be a locally closed clone. Then
every minimal operation above $\cC$ locally generates a
clone that is minimal above $\cC$, and every clone
that is minimal above $\cC$ is locally generated by a minimal
operation above $\cC$.
\end{proposition}

For oligomorphic clones $\cC$, minimality translates into maximality
of $\inv(\cC)$, and we obtain the following.

\begin{proposition}\label{prop:minimal}
Let $\bB$ be an $\omega$-categorical structure,
and let $\bC$ be primitive positive definable in $\bB$.  
Then $\Pol(\bC)$ is minimal
above $\Pol(\bB)$ 
if and only if for every $R \in \langle \bB \rangle_{\text{pp}} \setminus \langle \bC \rangle_{\text{pp}}$ the structure $\bB$ has a primitive positive definition in $(\bC,R)$.
\end{proposition}
\begin{proof}
The equivalence follows from
Proposition~\ref{thm:pp-galois}.
\end{proof}

It is well-known that every clone over a finite domain 
contains a clone $\cD$ which is
minimal above the trivial clone that just contains the projections~\cite{MinClones}.
The same is not true for infinite domains: the clone with
domain $\mathbb N$ 
generated by the operation $x \mapsto x + 1$ does not
contain a clone that is minimal above the set of all projections
over $\mathbb N$.
The situation is again better when $\cC$ is oligomorphic.

\begin{theorem}[from~\cite{OligoClone}]\label{thm:containsminimal}
Let $\bB$ be a finite or $\omega$-categorical structure with a finite
relational signature, and let $\cB$ be its polymorphism clone.
Then any %locally closed: removed in 12/2011, not needed
clone $\cC$ that contains $\cB$ contains a locally closed
clone $\cD$ that is minimal above $\cB$.
\end{theorem}
\begin{proof}
By Proposition~\ref{prop:minimal}, it suffices to show that
there is a structure $\bD$ whose relations are a subset 
of $\inv(\cB)$ such that
for every $R \in \inv(\cB) \setminus \inv(\cC)$
there is a primitive
positive definition of $\bB$ in $(\bD,R)$.

Consider the partially ordered set of all locally closed clones that contain $\cB$ and that are contained in $\cC$, ordered
by inclusion. 
From this poset we remove $\cB$, which is the unique minimal element; the resulting poset will be denoted by $P$.
We claim that in $P$, all descending chains $S_1 \supseteq S_2 \supseteq \cdots$
are bounded, i.e., for all such chains there exists an $S \in P$ such
that $S_i \supseteq S$ for all $i \geq 1$. To see this, observe that $\bigcup_{i \geq 1} \inv(S_i)$ is \emph{closed under primitive positive definability} in the sense that it
can be written as $\langle {\mathfrak S} \rangle_{\text{pp}}$ for some relational
structure ${\mathfrak S}$ (since only a finite number of relations can be mentioned
in a formula, and since each of the $S_i$ is closed under primitive positive definability).

Moreover, there must be a relation $R \in \bB$ that is not contained in
$\bigcup_{i \geq 1} \inv(S_i)$; otherwise, since $\bB$ has
finitely many relations, there is a $j < \omega$ such that $\inv(S_{j})$ contains
all relations from $\bB$, and hence equals $\inv(\cB)$, which is impossible since $\bB$ is not an element of $P$.
Hence, by Theorem~\ref{thm:pp-galois},
the structure ${\mathfrak S}$ has a polymorphism $f$ that is not
from $\cB$. Then $\cB \cup \{f\}$ is contained
in $\bigcap_{i \geq 1} S_i$ and is a lower bound of the descending chain
$(S_i)_{i \geq 0}$. We can thus apply Zorn's lemma and conclude that $P$
contains a minimal element $S$.
Any structure $\bD$ whose relations are exactly the relations from 
$S$ satisfies the initial requirements.
\end{proof}

For essentially unary oligomorpic clones $\cB$, we can bound
the arity of minimal functions above $\cB$; this follows from the following. 

\begin{proposition}[from~\cite{BPT-decidability-of-definability}]\label{prop:finiteMinimalClonesAboveEnd}
    Let $\bB$ be any relational structure with a finite number $p$ of orbitals. Then every minimal clone above $\End(\bB)$ is generated by a function of arity at most $2\cdot p-1$ together with $\End(\bB)$.
\end{proposition}
\begin{proof}
    Let $\cC$ be a minimal clone above $\End(\bB)$. 
    If all the functions in $\cC$ are essentially unary, then $\cC$ is generated by a unary operation together with $\End(\bB)$ and we are done. Otherwise, let $f$ be an essential operation in $\cC$. 
    By Lemma~\ref{lem:unary} the operation $f$ violates $P^B_3$ over the domain $B$ of $\bB$; recall that $P^B_3$ is defined by the formula $(x=y) \vee (y=z)$. 
    The subset of $P^B_3$ that contains all tuples of the form $(a,a,b)$, for $a,b \in B$,
    clearly consists of $p$ orbits in $\bB$. Similarly, the subset of $P^B_3$ 
    that contains all tuples of the form $(a,b,c)$, for $a,b \in B$, consists of the 
    same number of orbits. The intersection of these two relations consists of exactly one orbit (namely, the triples with three equal entries), and therefore $P^3_3$ is the union of
    $2\cdot p-1$ different orbits. The assertion now follows from Lemma~\ref{lem:small-arity}.    
\end{proof}

In Section~\ref{sect:canonization} 
we will see that under further
Ramsey-theoretic assumptions on the structure $\bB$, 
there are only \emph{finitely many} minimal closed clones above the
polymorphism clone of $\bB$. 
 % ist nur subsection, nicht section.
\section{Schaefer's Theorem}
\label{sect:schaefer}
Schaefer's theorem states that every CSP for a 2-element template
is either in P or NP-hard. 
Via the Inv-Pol Galois connection (Section~\ref{sect:inv-pol}), most of the classification arguments in Schaefer's article
follow from earlier work of Post~\cite{Post}, who classified all clones
on a two-element domain. We present a short proof of Schaefer's theorem here, using the results and ideas from Section~\ref{sect:arity}.

The following operations are important in this context. 
A ternary function $f \colon B^3 \rightarrow B$ is called a \emph{minority function} 
if it satisfies $f(x,y,y) = f(y,x,y) = f(y,y,x)=x$ for all $x,y \in B$. 
It is called a \emph{majority function} if $f(x,x,y) = f(x,y,x)=f(y,x,x)=x$ for
all $x,y \in B$. 
Note that on Boolean domains, the given identities determine $f$ uniquely. 
A $k$-ary operation $f$ is called \emph{idempotent} iff $f(x, \dots, x) = x$.
Also recall the definition of $\NAE$ and $\OIT$.
\begin{align*}
\NAE = & \{(0,0,1),(0,1,0),(1,0,0),(1,1,0),(1,0,1),(1,1,0)\}  \\
\OIT = & \{(0,0,1),(0,1,0),(1,0,0)\} 
\end{align*}

%Recall that $\NAE$ denotes the relation $$\{(0,0,1),(0,1,0),(1,0,0),(1,1,0),(1,0,1),(1,1,0)\} \; .$$ 

\begin{lemma}\label{lem:oit-violation}
Let $f$ be an idempotent function on the domain $\{0,1\}$
that violates $\OIT$.
\begin{itemize}
\item If $f$ is binary, then $f$ must be $(x,y) \mapsto \min(x,y)$ or $(x,y) \mapsto \max(x,y)$. 
\item If $f$ is ternary, then $f$ generates $\min$, $\max$, the majority, or the minority operation. 
\end{itemize}
\end{lemma}
\begin{proof}
There are only four binary idempotent operations
on $\{0,1\}$, two of which are projections and therefore preserve $\OIT$. The other two operations are
$\min$ and $\max$. If $f$ is ternary, then the operations defined by $f(x,x,y)$, $f(x,y,x)$, and $f(y,x,x)$ must be projections, or otherwise they generate $\min$ or $\max$ and we are done. So we consider the following eight cases.
\begin{center}
\begin{tabular}{l|llllllll}
Cases & (1) & (2) & (3) & (4) & (5) & (6) & (7) & (8) \\
\hline
$f(x,x,y)$ & $x$ & $x$ & $x$ & $x$ &  $y$ & $y$ & $y$ & $y$ \\
$f(x,y,x)$ & $x$ & $x$ & $y$ & $y$ &  $x$ & $x$ & $y$ & $y$ \\
$f(y,x,x)$ & $x$ & $y$ & $x$ & $y$ &  $x$ & $y$ & $x$ & $y$
\end{tabular}
\end{center}
In case (1) the operation $f$ is a majority, and in case (8) a minority. 
The cases (2), (3), and (5) are impossible since in this case $f$ would preserve $\OIT$.
In the remaining three cases, which are symmetric with respect to permuting arguments
of $f$, the function defined by $f(x,f(x,y,z),z)$ is a majority. 
\end{proof}

The following is well-known; the short proof is taken from~\cite{acm-rendezvous}.

\begin{theorem}\label{thm:minority}
If $R$ is a Boolean relation preserved by the minority operation,
then $R$ has a definition by a conjunction of linear
equalities modulo 2. 
\end{theorem}

\begin{proof}
The proof is by induction on the arity $k$ of $R$.
The statement is clear when $R$ is unary. Otherwise, let $R_0$ be the Boolean relation of arity $k-1$ defined by $R_0(x_2,\dots,x_k) \equiv R(0,x_2,\dots,x_k)$, and let
$R_1 \subseteq \{0,1\}^{k-1}$ be defined by $R_1(x_2,\dots,x_k) \equiv R(1,x_2,\dots,x_k)$. By the inductive assumption, there are conjunctions of linear equalities
$\psi_0$ and $\psi_1$ defining $R_0$ and $R_1$, respectively. If $R_0$ is empty, we may express $R(x_1,\dots,x_k)$ by $x_1=1 \wedge \psi_1$. The case that $R_1$ is empty can be treated analogously. When both $R_0$ and $R_1$ are non-empty,
fix a tuple $(c_2^0,\dots,c_k^0) \in R_0$ and  a tuple $(c_2^1,\dots,c_k^1) \in R_1$. Define $c^0$ to be $(0,c_2^0,\dots,c_k^0)$ and $c^1$ to be $(1,c_2^0,\dots,c_k^0)$. Let $b$ be an arbitrary tuple from $\{0,1\}^{k}$. Observe that if $b \in R$, then $\minority(b,c^0,c^1) \in R$, since $c^0 \in R$ and $c^1 \in R$. Moreover, if
$\minority(b,c^0,c^1) \in R$, then $\minority(\minority(b,c^0,c^1),c^0,c^1) = b \in R$. Thus, $b \in R$ if and only if $\minority(b,c^0,c^1) \in R$. Specializing this to $b_1 = 1$, we obtain 
$$ (b_2,\dots,b_k) \in R_1 \; \Leftrightarrow \; (\minority(b_2,c_2^0,c_2^1), \dots, \minority(b_k,c^0_k,c^1_k)) \in R_0 \; . $$
This implies 
$$ (b_1,\dots,b_k) \in R \; \Leftrightarrow \; (\minority(b_2,c_2^0 b_1,c_2^1 b_1), \dots, \minority(b_k,c_k^0 b_1,c_k^1 b_1)) \in R_0  \; .$$
Thus, $R(x_1,\dots,x_k)$ is defined by the formula 
$$\exists x_2',\dots,x_k'  (\phi_0(x_2',\dots,x_k') \wedge \bigwedge_{i \in \{2,\dots,k\}} (x_i + c_i^0 x_1 + c^1_i x_1 = x_i')) \; .$$
%obtained from $\phi_0(x_2',\dots,x_k')$ by substituting $x_i'$ 
%with $\minority(x_i,c_i^0 x_1,c^1_i x_1)$.
\end{proof}

A binary relation is called \emph{bijunctive} if it can be defined
by a propositional formula that is a conjunction of clauses of size two (aka \emph{2CNF formulas}). 

%When $f: \{0,1\}^k \rightarrow \{0,1\}$ is a function, then the \emph{dual}
%of $f$ is the function $(x_1,\dots,x_k) \mapsto 1-f(1-x_1,\dots,1-x_k)$.

\begin{theorem}[of Post~\cite{Post} and Schaefer~\cite{Schaefer}]
\label{thm:schaefer}
Let $\bB$ be a structure over a two-element universe. 
Then either %$(\{0,1\};\OIT)$ 
$(\{0,1\};\NAE)$ 
has a primitive positive
definition in $\bB$,
and $\Csp(\bB)$ is NP-complete, or 
\begin{enumerate}
\item $\bB$ is preserved by a constant operation.
%, and $\Csp(\bB)$ is (trivially) in P.
\item $\bB$ is preserved by $\min$. In this case, 
every relation of $\bB$
has a definition by a propositional Horn formula.
\item $\bB$ is preserved by $\max$. 
In this case, 
every relation of $\bB$
has a definition by a \emph{dual-Horn} formula, that is,
by a propositional formula in CNF where every clause contains at most one negative literal. 
\item $\bB$ is preserved by the majority operation. In this case, every relation of $\bB$ is bijunctive.
\item $\bB$ is preserved by the minority operation. In this case,
every relation of $\bB$ can be defined by a conjunction of linear equations modulo 2.
\end{enumerate}
In case $(1)$ to case $(5)$, $\Csp(\bB)$ can be solved in polynomial time.
\end{theorem}
\begin{proof}
If the relation $\OIT=\{(0,0,1),(0,1,0),(1,0,0)\}$ has a primitive positive definition in $\bB$, then Lemma~\ref{lem:pp-reduce} shows that the NP-hard problem 
positive 1-in-3-3SAT~\cite{GareyJohnson} (see Example~\ref{expl:sat})
can be reduced to $\Csp(\bB)$. In this case, 
also the relation $\NAE$ 
is primitive positive definable in $\bB$, 
as we have seen in Example~\ref{expl:oit-defines-nai}. 

If $\OIT$ is not primitive positive definable in $\bB$, then by Theorem~\ref{thm:inv-pol} 
there is a polymorphism $f$ of $\bB$ that violates $\OIT$; let $f$ be such an operation 
of minimal arity. Because $\OIT$ consists of three tuples only, Lemma~\ref{lem:small-arity}
asserts that $f$ is at most ternary.

If $f$ is not unary, then $\hat f \colon \{0,1\} \rightarrow \{0,1\}$ defined by 
$x \mapsto f(x,\dots,x)$ must preserve $\OIT$ by the choice of $f$.
Hence, $\hat f$ is the identity and $f$ is idempotent. 
If $f$ is binary, then by Lemma~\ref{lem:oit-violation} it is either $\min$ or $\max$.
In case that $f$ is $\min$, we 
show that all relations in $\bB$ can be defined by propositional Horn formulas. 
It is well-known that positive unit-resolution is a polynomial-time decision
procedure for the satisfiability problem of propositional Horn-clauses~\cite{SchoeningLogic}. The case that $f$ is $\max$ is dual to this case. 

So let $R$ be a Boolean relation preserved by $\min$. Let $\phi$ be a
propositional formula in CNF that defines $\phi$. We can assume without loss of generality that
for all literals in clauses of $\phi$, when we remove this literal from the clause,
the resulting formula is not equivalent to $\phi$. (Otherwise, we keep on removing literals until the formula has the required property.)
Now suppose for contradiction that $\phi$ contains a clause $C$ with 
two positive literals $x$ and $y$. Since $\phi$ is reduced, there is an assignment $s_1$ that satisfies $\phi$ such that $s_1(x)=1$, and such that all other literals of $C$ evaluate to $0$. Similarly, there is a satisfying assignment $s_2$ for $\phi$ such that $s_2(y)=1$ and all other literal s of $C$ evaluate to $0$.
Then $s_0 \colon x \mapsto \min(s_1(x),s_2(y))$ does not satisfy
$C$, and does not satisfy $\phi$, in contradiction to the assumption that $\min$ preserves $R$.

If $f$ is ternary, then $f$ either generates the minority
or the majority operation, by Lemma~\ref{lem:oit-violation} and the choice of $f$.
If $f$ generates the majority operation, we show 
that every relation of $\bB$ is bijunctive. Hence, in this case $\Csp(\bB)$ is
equivalent to the 2SAT problem, and can be solved in linear time~\cite{AspvallPlassTarjan}.
Let again $\phi$ be a reduced definition of a relation from $R$,
and suppose that $\phi$ contains a clause $C$ with three literals $l_1,l_2,l_3$. Since $\phi$ is reduced, there must be satisfying
assignments $s_1,s_2,s_3$ to $\phi$ 
such that under $s_i$ all literals of $C$ evaluate to $0$ except
for $l_i$. Then the mapping $s_0 \colon x \mapsto \majority(s_1(x),s_2(x),s_3(x))$ does not satisfy $C$ and therefore does not satisfy $\phi$, in contradiction to the assumption that 
$\majority$ preserves $R$.

If $f$ generates the minority operation, and $R$ is a relation of $\bB$,
then by Theorem~\ref{thm:minority} the relation 
$R$ has a definition by a conjunction of linear
equalities modulo 2. Then $\Csp(\bB)$ can be solved in polynomial time by Gaussian elimination.

Finally, if $f$ is unary, then $f$ is either constant, and we are
done, or $f$ is the operation $\neg$ defined by $x \mapsto 1-x$. 
If $f$ is $\neg$, then $\NAE$ consists of three orbits of triples.
If $\NAE$ is primitive positive definable in $\Phi$, 
then $\Csp(\bB)$ is NP-hard by reduction from positive not-all-equal-3SAT~\cite{GareyJohnson} (see again Example~\ref{expl:sat}). Otherwise, by Lemma~\ref{lem:small-arity}, there is 
an at most ternary operation $g$ that violates $\NAE$. Since all non-constant unary operations preserve $\NAE$, we can assume
that $g$ is at least binary. 
%Also, 
%$\hat g$ is either the identity or $\neg$.
%In the first case, $g$ is idempotent and violates $\NAE$, and the statement follows from the above proof for $g$ in place of $f$. 
%If $\hat g$ is $\neg$ then $\neg g \in \Pol(\bB)$  violates
%$\NAE$, but is idempotent, and 
If $g$ is binary and violates $\NAE$, then there
are $t_1,t_2 \in \NAE$ such that $t_0 = g(t_1,t_2) \notin \NAE$. 
For $i \in \{1,2\}$, if $t_i \in \OIT$, set $\alpha_i$ to be $\id$, otherwise set $\alpha_i$ to be $\neg$, and
note that $\alpha_i t_i \in \OIT$. Then either $h \colon (x,y) \mapsto g(\alpha_1 x,\alpha_2 y)$
or $h \colon (x,y) \mapsto \neg g(\alpha_1 x,\alpha_2 y)$ is idempotent and violates $\OIT$,
and the statement follows from the above proof when we take $h \in \Pol(\bB)$ in place of $f$.
 The argument for ternary $g$ follows the same lines. 
\end{proof}

Hard Boolean constraint languages can be characterized in many
equivalent ways via Corollary~\ref{cor:elementary}. To see this, we need the 
following proposition.
%Section~\ref{sect:elementary}).

\begin{proposition}\label{prop:hard-boolean-csps}
Let $\bB$ be a structure over a two-element universe. 
Then the following are equivalent.
\begin{enumerate}
\item $(\{0,1\};\NAE)$ has a primitive positive
definition in $\bB$.
\item $\bB$ is neither preserved by $\min$, $\max$, minority, majority, nor the constant operations.
\item The polymorphism clone of $\bB$ either contains only projections,
or is generated by the unary operation $x \mapsto -x$. 
\item In $\bB$ every first-order formula is equivalent to a primitive positive formula.
\end{enumerate}
\end{proposition}
\begin{proof}
The implication from $(1)$ to $(2)$ follows from the fact that 
$\NAE$ is not preserved by $\min$, $\max$, \minority, \majority, and constant operations,
which is straightforward to verify. 
A proof that $(2)$ implies $(3)$ can for instance be found in~\cite{acm-rendezvous}  (Theorem~5.1).
%To prove that $(1)$ implies $(2)$, assume $(1)$, and let $\cC$ be the polymorphism
%clone of $\bB$. 
%We verify $(2)$ using Theorem~\ref{thm:rosenberg} as follows.
%Suppose that $\bC$ contains essential operations.
%The polymorphism clone $\cC := \Pol(\bB)$ contains a minimal clone above 
%the unary clone $\langle\Aut(\bB)\rangle$
%(Theorem~\ref{thm:containsminimal}), which is generated by an operation $g$ that
%is minimal above $\langle\Aut(\bB)\rangle$.
%Since constant polymorphisms violate $\NAE$, the operation $g$ must be essential.
%The domain has just two elements, so every semi-projection of arity three must be a projection. Therefore, by Theorem~\ref{thm:rosenberg}, $g$ must be essentially binary,
%or a ternary Maltsev (which over the Boolean domain is the unique minority operation) 
%or majority operation. 
%It is straightforward to verify that the minority and the majority operation violate $\NAE$. 
%For the binary case, we distinguish whether $\bC$ contains $x \mapsto 1-x$ or not.
%In both cases we see that $g$ is without loss of generality $\min$ or $\max$,
%and both those operations violate $\NAE$.
The implication from $(3)$ to $(4)$ follows from Corollary~\ref{cor:elementary}.
% Maltsev: 001,010,100 mapped to 111
% Majority: 001,010,100 mapped to 000
% Cases: binary: 001,010 maps to 
%observe that all polymorphisms
%of $(\{0,1\};\OIT)$ are projections. This can be verified using Rosenberg's pre-classification of minimal clones (Theorem~\ref{thm:rosenberg}) as follows. Clearly,
%$\OIT$ is not preserved by $x \mapsto 1-x$. Now
%suppose that $\cC := \Pol((\{0,1\};\OIT))$ contains essential operations. Then $\cC$
%contains a minimal clone, generated by a minimal operation $g$. 
%Since the domain
%has just two elements, every semi-projection of arity three must be a projection.
%Therefore, by Theorem~\ref{thm:rosenberg}, $g$ must be binary and idempotent,
%or a ternary Maltsev or majority operation. It is straightforward to verify that all these
%three types of operation do not preserve $\OIT$. 
%
For the implication from $(4)$ to $(1)$, note that $\NAE$ is preserved
by $x \mapsto -x$, and hence preserved by all automorphisms
of $\bB$. 
% BOOKTD: Do we need inv-aut=fo for finite domains?
In particular, $\NAE$ is first-order definable in $\bB$. 
So $(4)$ implies that $\NAE$ also has a primitive positive definition in $\bB$. 
\end{proof}

\section{Pseudo-varieties and Primitive Positive Interpretations}
\label{sect:pseudo-var}
Primitive positive definability is a strong tool to prove that
certain CSPs are hard, but in some cases this tool is not strong enough. In this section we discuss the concept 
of \emph{primitive positive
interpretations}, and the matching universal-algebraic concept, which is the concept of \emph{pseudo-varieties}.

% Achtung, das folgende stimmt nicht: die Atomless
% Boolean algebra mit ihrem binaeren commutativen 
% (modulo automorphismen)
% isomorphismus zu ihrem Quadrat hat ein NP-schweres
% CSP!

%These concepts suffice to perform all hardness proofs
%FOR FINITELY BOUNDED HOMOGENEOUS STRUCTURES in this text purely algebraically. 
%It turns out that
%for the classes of CSPs considered here, hardness
%of $\Csp(\bB)$ can always be shown by giving a primitive positive interpretation
%of $(\{0,1\},\OIT)$ in the model-complete core of $\bB$ (see Example~\ref{expl:sat}), %Section÷\ref{sect:}), 
%or in the expansion of the model-complete core 
%of $\bB$ by finitely many constants.

\subsection{Algebras} 
\label{ssect:algebras}
Algebras have been defined in Chapter~\ref{chap:logic}:
they are simply structures with a purely functional signature.
When $\fA$ is an algebra with signature $\tau$ and domain $A$, we denote by $\Clo(\fA)$
the set of all functions with domain $A$ of the form
$(x_1,\dots,x_n) \mapsto t(x_1,\dots,x_n)$ where $t$ is any
term over the signature $\tau$ whose set of variables is contained in 
$\{x_1,\dots,x_n\}$; clearly, $\Clo(\fA)$ is
closed under compositions, and contains the projections, and therefore forms a clone.
In this section we recall some basic universal-algebraic facts that 
will be used in the following subsections.

When $\cal K$ is a class of algebras of the same signature, then
\begin{itemize}
\item $\PPP({\cal K})$ denotes the class of all products of algebras from $\cal K$.
\item $\PPPfin({\cal K})$ denotes the class of all finite products of algebras from $\cal K$.
\item $\SSS({\cal K})$ denotes the class of all subalgebras of algebras from $\cal K$.
\item $\HHH({\cal K})$ denotes the class of all homomorphic images of algebras from $\cal K$.
\end{itemize}
(Products, subalgebras, and homomorphic images have been defined in Chapter~\ref{chap:logic}.)
Note that closure under homomorphic images implies in particular closure under
isomorphism. For the operators $\PPP$, $\PPPfin$, $\SSS$ and $\HHH$ we often omit the brackets when applying them
to single algebras, i.e., we write $\HHH(\fA)$ instead of $\HHH(\{\bf A\})$.
The elements of $\HS(\bf A)$ are also called the \emph{factors}
of $\bf A$.

A class $\cal V$ of algebras with the same signature $\tau$ is called
a \emph{pseudo-variety} if $\cal V$ contains all homomorphic
images, subalgebras, and direct products
of algebras in $\cal V$, i.e., $\HHH({\cal V})=\SSS({\cal V})=\PPPfin({\cal V})$.
The class $\cal V$ is called a 
\emph{variety} if $\cal V$ also contains all (finite and infinite) 
products of algebras in $\cal V$.
So the only difference between pseudo-varieties
and varieties is that pseudo-varieties need not be closed
under direct products of infinite cardinality.
The smallest pseudo-variety (variety) that contains an algebra $\bf A$
is called the pseudo-variety (variety) \emph{generated} by $\bf A$.

\begin{definition}\label{def:poly-algebra}
Let $\bB$ be a relational structure with domain $B$. An algebra with domain $B$
whose operations are exactly the polymorphisms of $\bB$ is called
a \emph{polymorphism algebra of $\bB$}.
\end{definition}

Note that a relational structure can have many different polymorphism algebras,
since Definition~\ref{def:poly-algebra} does not prescribe how to assign function symbols to the polymorphisms of $\bB$.
In our applications, the precise choice of the signature never plays a role, and therefore we sometimes refer to \emph{the} polymorphism algebra of $\bB$, and denote it by $\Alg(\bB)$. So statements
about \emph{a} polymorphism algebra of $\bB$ (or about \emph{the} polymorphism algebra $\Alg(\bB)$) can typically be translated
to statements that hold for \emph{all} polymorphism algebras of $\bB$ 
(e.g. in~Theorem~\ref{thm:pp-interpret-reduce} below). 

Also note that when $\bB$ is $\omega$-categorical, then the signature of the polymorphism algebra has cardinality $2^\omega$. This follows directly from Theorem~\ref{thm:auto-cardinality} and the remark after Lemma~\ref{lem:interpret}.

% do we really need the following? 
%An algebra is called \emph{weakly oligomorphic} if its operations locally generate an oligomorphic clone. 
%An algebra is called \emph{trivial} (the ones having a constant operation? 1-element algebras.). Don't need this any longer.

\subsection*{Congruences and Quotients} 
\label{ssect:quotients}
When $\mu \colon C \rightarrow D$ is a map, then the \emph{kernel} of
$\mu$ is the equivalence relation $E$ on $C$ where $(c,c') \in E$
if $\mu(c)=\mu(c')$. For $c \in C$, we denote by $c/E$ the equivalence class of $c$ in $E$, and by $C/E$ the set of all equivalence classes of elements of $C$. 
%\begin{definition}
A \emph{congruence} of an algebra $\bf A$ is 
an equivalence relation that is preserved by all operations in $\bf A$.
%\end{definition}
The results in Section~\ref{sect:inv-pol} show that
for $\omega$-categorical structures $\bB$ with polymorphism algebra $\bf B$,
the congruences of $\bf B$ are exactly the primitive positive definable
equivalence relations over $\bB$. 
%Similarly, by the results of Section~\ref{sect:galois}, % from Chapter~\ref{chap:mt},
%the congruences of the automorphism algebra of $\bB$ are exactly the first-order definable equivalence relations over $\bB$ 
%(and this is consistent with the concept of congruences 
%of permutation groups as defined in Section~\ref{sect:galois}).
%from Chapter~\ref{chap:mt}). 

\begin{proposition}[see~\cite{BS}]\label{prop:congruence-homo}
Let $\bf A$ be an algebra. Then $E$ is a congruence of $\bf A$ if and only if
$E$ is the kernel of a homomorphism
from $\bf A$ to some other algebra $\bf B$. 
\end{proposition}

% LATERTD: we could mention that this generalizes simultaneously the concept of 
% a congruence of a permutation group (when we view the permutation group as an algebra of unary functions)
% and kernels of abstract group homomorphisms (when applied to the abstract group)

When $K$ is a congruence of a $\tau$-algebra $\bf A$, then $\fA/K$ denotes $\tau$-algebra with domain $A/K$
where $$ f^{{\bf A}/K}(a_1/K, \dots,a_k/K)=f^{\bf A}(a_1,\dots,a_k)/K$$ where $a_1,\dots,a_k \in A$ and $f \in \tau$
is $k$-ary. This is well-defined since $K$ is preserved by all operations of ${\bf A}$. 
If $K$ is the kernel of $\mu$ then we also write 
$\fA/\mu$ instead of $\fA/K$. 
%When $\bf A$ is a $\tau$-algebra, and $\mu \colon A \rightarrow B$ is a mapping
%such that the kernel $E$ of $\mu$ is a congruence of $\bf A$, we define the \emph{quotient algebra
%${\bf A}/\mu$ of $\bf A$ under $\mu$} to be the algebra with domain $A/E$ where 
%$$ f^{{\bf A}/\mu}(\mu(a_1), \dots,\mu(a_k))=\mu(f^{\bf A}(a_1,\dots,a_k))$$ where $a_1,\dots,a_k \in A$ and $f \in \tau$
%is $k$-ary. This is well-defined since $E$ is preserved by all operations of ${\bf A}$. 
%Note that $\mu$ is a surjective homomorphism from ${\bf A}$ to ${\bf A}/\mu$.
The following is well-known.

\begin{lemma}[The Homomorphism Lemma]\label{lem:hom}
Let $\bf A$ be a $\tau$-algebra, let $K$ be a congruence of ${\bf A}$, 
and let $\mu_1 \colon A \rightarrow B_1$ and $\mu_2 \colon A \rightarrow B_2$ 
be two mappings with kernel $K$. Then ${\bf A}/\mu_1$ is isomorphic to ${\bf A}/\mu_2$.
\end{lemma}

The following is also well known (see e.g. Theorem 6.3 in~\cite{BS}).
%The isomorphism and homomorphism theorems. 
%If $B \subset A$, then $K|_B$ denotes $B^2 \cap K$.
%Let $B^{K}$ be the set $\{ a \in A \; | \; B \cap a/K \neq \emptyset\}$. 
%\begin{theorem}[The third isomorphism theorem; see Theorem 6.18 in~\cite{BS}] If ${\bf B}$ is a subalgebra of ${\bf A}$, and
%$\theta$ a congruence of ${\bf A}$, then ${\bf B}/(\theta |_{\bf B})$ is isomorphic to ${\bf B}^{\theta}/(\theta|_{{\bf B}^\theta}$.
%\end{theorem}
%\begin{proof}
%The mapping that sends $b/(\theta|_B)$ to $b/(\theta|_{B^\theta})$
%is the desired isomorphism.
%\end{proof}

\begin{lemma}%[Theorem 6.3 in~\cite{BS}]
\label{lem:congruences}
Let $\fA$ and $\fB$ be algebras with the same signature, and let $\mu \colon {\bf A} \rightarrow {\bf B}$ be a homomorphism.
Then the image of any subalgebra ${\bf A}'$ of ${\bf A}$ under $\mu$ is a subalgebra of ${\bf B}$, and the preimage of any subalgebra ${\bf B}'$ of ${\bf B}$ under $\mu$ is a subalgebra of ${\bf A}$.
%In both cases, the restriction of $h$ to $A'$ is a homomorphism from ${\bf A}'$ to ${\bf B}'$.
% THIS IS TRIVIAL!
\end{lemma}
\begin{proof}
Let $f \in \tau$ be 
$k$-ary. Then for all $a_1,\dots,a_k \in A'$, 
$$f^{\bf B}(\mu(a_1),\dots,\mu(a_k)) = \mu(f^{\bf A}(a_1,\dots,a_k)) \in h(A') \; ,$$
so $\mu(A')$ is a subalgebra of $\bf C$. Now suppose that $\mu(a_1),\dots,\mu(a_k)$ are in $B'$;
then $f^{\bf B}(\mu(a_1),\dots,\mu(a_k)) \in B'$ and hence $\mu(f^{\bf A}(a_1,\dots,a_k)) \in B'$.
So, $f^{\bf A}(a_1,\dots,a_k)) \in \mu^{-1}(B')$ which shows that $\mu^{-1}(B')$ induces a subalgebra
of $\bf A$.
\end{proof}

% IF WE WOULD MENTION SOMETHING ABOUT 
% VARIETIES AND TERM EQUATIONS, WE COULD 
% DROP A COMMENT THAT the question whether an
% operation e is an automorphism of the underlying
% invariant structure can in fact be answered from the abstract
% clone: there exists an i such that i(e(x))=e(i(x))=x for all x.
% Actually it would be great if this could go into a new section:
% varieties, term equations, abstract clones.
% This section would then more appropriately be called:
% interpretations, pseudo-varieties, and topological clones
% this section could then also contain the fact that whether or not 
% a structure has essentially infinite signature or is finitely homogeneous is a property of the topological clone etc. 

\subsection{Primitive Positive Interpretations}
\label{ssect:pp-interpret}
In Chapter~\ref{chap:mt}, we have seen that
first-order interpretations are a convenient tool
to construct $\omega$-categorical structures from
other $\omega$-categorical structures.
\emph{Primitive positive} interpretations are interpretations $I$
where the domain formula $\delta_I$ and 
all the defining formulas $\phi_I$ are primitive positive.  
As we will see, such interpretations can be used
to study the computational complexity of constraint satisfaction 
problems. 

%The results in this section link 
%the model-theoretic concept of primitive positive interpretations with the universal-algebraic concept of pseudo-varieties.
%To this end, we need the following refined variant
%of interpretations.

\begin{definition}
Let $I$ be an interpretation.
If the domain formula $\delta_I$ 
and the interpreting formulas $\phi_I$ 
are primitive positive,
then we say that $I$ is a \emph{primitive positive interpretation}.
\end{definition}

We first start with a result that is known for finite domain
constraint satisfaction, albeit not using 
the terminology of primitive positive interpretations~\cite{JBK}.
In the present form, it appears first in~\cite{BodirskySurvey}.

\begin{theorem}\label{thm:pp-interpret-reduce}
Let $\bB$ and $\bC$ be structures with finite relational signatures.
If there is a primitive positive interpretation of $\bB$
in $\bC$, then there is a polynomial-time reduction from
$\Csp(\bB)$ to $\Csp(\bC)$. 
\end{theorem}

\begin{proof}
Let $d$ be the dimension of the primitive positive 
interpretation $I$ of the $\tau$-structure $\bB$ in the $\sigma$-structure 
$\bC$, let $\delta_I(x_1,\dots,x_d)$ be the domain formula, 
let $h \colon \delta_I(\bC^d) \rightarrow D(\bB)$ be the coordinate map, and 
let $\phi_I(x_1,\dots,x_{dk})$ be the formula for the $k$-ary relation
$R$ from $\bB$.

Let $\phi$ be an instance of $\Csp(\bB)$ 
with variable set $U = \{x_1,\dots,x_n\}$.
We construct an instance $\psi$ of $\Csp(\bC)$ as follows.
For distinct variables $V := \{y_1^1, \dots, y_n^d\}$, we set 
$\psi_1$ to be the formula
$$ \bigwedge_{1 \leq i \leq n} \delta(y_i^1,\dots,y_i^d) \; .$$
Let $\psi_2$ be the conjunction of
the formulas $\theta_I(y_{i_1}^1,\dots,y_{i_1}^d,\dots,y_{i_k}^1,\dots,y_{i_k}^d)$
over all conjuncts $\theta = R(x_{i_1},\dots,x_{i_k})$ of $\phi$.
By moving existential quantifiers to the front, the sentence 
$$\exists y_1^1, \dots, y_n^d \; (\psi_1 \wedge \psi_2)$$ 
can be re-written to a primitive positive $\sigma$-formula $\psi$, and clearly $\psi$
can be constructed in polynomial time in the size of $\bA$.

We claim that $\phi$ is true in $\bB$
if and only $\psi$ is true in $\bC$. Let $C$ be the domain of $\bC$, $B$ the domain of $\bB$,  and suppose that $f \colon U \rightarrow C$ satisfies all conjuncts of $\psi$ in $\bC$.
Hence, by construction of $\psi$, if $\phi$ has a conjunct 
$\theta = R(x_{i_1},\dots,x_{i_k})$, then 
$$\bC \models \theta_I((f(y_{i_1}^1),\dots,f(y_{i_1}^d)), 
\dots, (f(y_{i_k}^1),\dots,f(y_{i_k}^d))) \; .$$
By the definition of interpretations,
this implies that 
$$\bB \models R(h(f(y_{i_1}^1), \dots, f(y_{i_1}^d)), \dots, h(f(y_{i_k}^1),\dots,f(y_{i_k}^d))) \; .$$
Hence, the mapping $g \colon  U \rightarrow B$ that sends
$x_i$ to $h(f(y_i^1),\dots,f(y_i^d))$ satisfies all conjuncts of $\phi$ in $\bB$.

Now, suppose that $f \colon U \rightarrow B$ 
satisfies all conjuncts of $\phi$ over $\bB$.
Since $h$ is a surjective mapping from $\delta(\bC^d)$
to $B$, there are elements $c_i^1,\dots,c_i^d$ in $\bC$
such that $h(c_i^1,\dots,c_i^d) = f(x_i)$, 
for all $i \in \{1,\dots,n\}$.
We claim that the mapping $g \colon  V \rightarrow C$ that maps
$y_i^j$ to $c_i^j$ is a homomorphism from $\psi$ to $\bC$.
By construction, any constraint in $\psi$ either comes from $\psi_1$ or from $\psi_2$.
If it comes from $\psi_1$ then it must be of the form $\delta_I(y_i^1,\dots,y_i^d)$, and is satisfied since the pre-image of
$h$ is $\delta_I(\bC^d)$.
If the constraint comes from $\psi_2$, then it must be a conjunct of a formula
$\theta_I(y^1_{i_1},\dots,y_{i_1}^d,\dots,y^1_{i_k},\dots,y^d_{i_k})$ that was introduced for a constraint $\theta = R(x_{i_1},\dots,x_{i_k})$ in $\bA$.
It therefore suffices to show that
$$\bC \models \theta_I(g(y^1_{i_1}),\dots,g(y_{i_1}^d),\dots,g(y^1_{i_k}),\dots,g(y^d_{i_k})) \; .$$ 
By assumption, $R(f(x_{i_1}),\dots,f(x_{i_k}))$ holds in $\bB$. 
By the choice of $c^1_1,\dots,c^d_n$, this shows that 
$R(h(c^1_{i_1},\dots,c^d_{i_1}),\dots,h(c^1_{i_k},\dots,c^d_{i_k}))$ holds in $\bC$. By the definition of interpretations,
this is the case if and only if $\theta_I(c^1_{i_1},\dots,c^d_1,$
$\dots,c^1_{i_k}, \dots, c^d_{i_k})$ holds in $\bC$, which is what
we had to show.
\end{proof}

We describe how to compose interpretations, and
observe that compositions of primitive positive interpretations
are again primitive positive.
Note that if $\bC_2$ has a $d_1$-dimensional interpretation $I_1$ in $\bC_1$, and $\bC_3$
has an $d_2$-dimensional interpretation $I_2$ in $\bC_2$,
then $\bC_3$ has a natural $(d_1d_2)$-dimensional interpretation in $\bC_1$, which we denote by $I_2 \circ I_1$.
To formally describe $I_2 \circ I_1$, suppose that the signature of $\bC_i$
is $\tau_i$ for $i = 1,2,3$, and that
$I_1 = (d_1,S_1,h_1)$ and $I_2 = (d_2,S_2,h_2)$.
When $\phi$ is a $\tau_2$-formula, let $\phi_{I_1}$ denote the
$\tau_1$-formula obtained from $\phi$ by replacing each atomic 
$\tau_2$ formula $\psi$ in $\phi$ by the $\tau_1$-formula $\psi_{I_1}$.
Note that when $\phi$ is primitive positive (existential positive), and
the interpreting formulas of $I_1$ are primitive positive (existential positive), then 
$\phi_{I_2}$ is again primitive positive 
(existential positive)\footnote{Note
that this is in general false for \emph{existential} formulas: 
there are existential formulas $\phi$ and existential 
interpretations $I_1$ such that $\phi_{I_1}$ is no longer existential.}.
 
Now the interpretation $I_2 \circ I_1$ is given by $(d_1d_2,S,h)$
where $S := (\delta_{I_2})_{I_1}((\bC_1)^{d_1d_2})$,
and where the coordinate map $h \colon S \rightarrow \bC_3$ is defined by
$$(a^1_1,\dots,a^{d_1}_1,\dots,a^1_{d_2},\dots,a^{d_1}_{d_2}) \; \mapsto \; h_2(h_1(a^1_1,\dots,a^{d_1}_1),\dots,h_1(a^1_{d_2},\dots,a^{d_1}_{d_2})) \; .$$
Observe that when $I_1$ and $I_2$ are primitive positive interpretations, then $I_2 \circ I_1$ is also primitive positive.

%\begin{lemma}\label{lem:pp-interpret-composition}
%When $I_1$ and $I_2$ are primitive positive interpretations, 
%then $I_2 \circ I_1$ is also primitive positive.
%\end{lemma}

In many hardness proofs we use Theorem~\ref{thm:pp-interpret-reduce} in the following way. 
\begin{corollary}\label{cor:pp-interpret-hard}
Let $\bB$ be an $\omega$-categorical relational structure. 
If there is a primitive positive interpretation of
$(\{0,1\}; \OIT)$ or $(\{0,1\}; \NAE)$ in $\bB$, then
$\bB$ has a reduct with finite signature whose $\Csp$ is NP-hard.
\end{corollary}
\begin{proof}
The primitive positive formulas involved in the
primitive positive interpretation
can mention only finitely many relations from $\bB$.
Let $\bB'$ be the reduct of $\bB$ that contains exactly those
relations. Then NP-hardness of $\Csp(\bB')$ follows 
from the NP-hardness of $\Csp((\{0,1\}; \OIT))$ and $\Csp((\{0,1\}; \NAE))$ (see Section~\ref{sect:csp-logical} in Chapter~\ref{chap:intro}, Example~\ref{expl:sat})
via Theorem~\ref{thm:pp-interpret-reduce}.
\end{proof}

We present an application of Theorem~\ref{thm:pp-interpret-reduce} and prove a hardness result that becomes useful at several occasions 
in later sections.

\begin{definition}\label{def:I6}
For any set $B$, we write $I^B_6$ for the relation
 \begin{align*} \big \{ (x_1,x_2,y_1,y_2,z_1,z_2) \in B^6\; | \; &
 (x_1=x_2 \wedge y_1 \neq y_2 \wedge z_1 \neq z_2) \\
   \vee & \; 
 (x_1 \neq x_2 \wedge y_1 = y_2 \wedge z_1 \neq z_2) \\
  \vee & \; (x_1 \neq x_2 \wedge y_1 \neq y_2 \wedge z_1 = z_2) \big \} \; .
 \end{align*}
\end{definition}

\begin{proposition}\label{prop:1in3hard}
For any set $B$ with $|B| \geq 2$, the structure $(\{0,1\}; \OIT)$ 
has a primitive positive interpretation in $(B; I^B_6)$, 
and $\Csp((B; I^B_6))$ is NP-hard.
\end{proposition}
\begin{proof}
Recall that $\OIT$ denotes the boolean relation $\{(1,0,0),(0,1,0),(0,0,1)\}$.
We give a primitive positive interpretation $I$ of the structure
$\bB := (\{0,1\}; \OIT)$ in $(B; I^B_6)$. 
The dimension of $I$ is 2, and the domain formula is $\delta_I := \top$ (for \emph{true}). 
The formula ${\OIT}(x_1,x_2,y_1,y_2,z_1,z_2)_I$ is $I^B_6(x_1,x_2,y_1,y_2,z_1,z_2)$, and the formula $=_I(x_1,x_2,y_1,y_2)$ is 
\begin{align*} \exists a_1,a_2,u_1,u_2,u_3,u_4,z_1,z_2 \; & \big (a_1=a_2 \wedge 
I^B_6(a_1,a_2,u_1,u_2,u_3,u_4) \\
& \wedge I^B_6(u_1,u_2,x_1,x_2,z_1,z_2) 
\wedge I^B_6(u_3,u_4,z_1,z_2,y_1,y_2) \big) \; .
\end{align*}
Note that the primitive positive formula $=_I(x_1,x_2,y_1,y_2)$ is 
equivalent to $x_1=x_2 \Leftrightarrow y_1=y_2$.
The map $h$ maps
$(b_1,b_2) \in B^2$ to $1$ if $b_1=b_2$, and to $0$ otherwise.
NP-hardness of $\Csp((B; I^B_6))$ then
follows from Corollary~\ref{cor:pp-interpret-hard}.
\end{proof}

\begin{corollary}\label{cor:elementary-hard}
Let $\bB$ be an $\omega$-categorical structure where all first-order formulas
are equivalent to primitive positive formulas (or that satisfies some of the other equivalent conditions from Corollary~\ref{cor:elementary}).
Then there is a primitive positive interpretation of $(\{0,1\};\OIT)$ in $\bB$, and
$\bB$ has a finite signature reduct $\bB'$ such that 
$\Csp(\bB')$ is NP-hard.
\end{corollary}
\begin{proof}
Since $I^B_6$ is first-order definable,
it also has a primitive positive definition in $\bB$ by assumption. Proposition~\ref{prop:1in3hard} implies that the structure 
$(\{0,1\};\OIT)$ has a primitive positive interpretation in $\bB$. The last part of the statement follows from Corollary~\ref{cor:pp-interpret-hard}.
\end{proof}

In fact, we could have weakened the assumptions in Corollary~\ref{cor:elementary-hard} 
by only requiring that all polymorphisms of $\bB$ are essentially
unary, and that all endomorphisms of $\bB$ are injective, because it is then easy to see that the relation
$I^B_6$ is preserved by all polymorphisms of $\bB$, and hence primitive positive definable in $\bB$, by Theorem~\ref{thm:inv-pol}. 

There are many situations where Theorem~\ref{thm:pp-interpret-reduce} can be combined with Lemma~\ref{lem:constant-expansion} to prove hardness of CSPs, as described in the following.

\begin{proposition}\label{prop:pp-interpret-with-constants}
Let $\bA$ be a structure with finite relational signature, and
let $\bB$ be a structure with elements $c_1,\dots,c_k$ such that 
\begin{itemize}
\item the orbit of $(c_1,\dots,c_k)$ in $\bB$ is primitive positive definable, and
\item $\bA$ has a primitive positive interpretation in $(\bB,c_1,\dots,c_k)$. 
\end{itemize}
Then there is a finite signature reduct $\bB'$ of $\bB$ and a polynomial-time
reduction from $\Csp(\bA)$ to $\Csp(\bB')$.
\end{proposition}

\begin{proof}
Let $\bC$ denote the expansion of $\bB$ by 
the unary relations $\{c_1\},\dots,\{c_k\}$.
Then the interpretation of $\bA$ in $(\bB,c_1,\dots,c_k)$ shows that there is also a primitive positive interpretation of $\bA$ in $\bC$, and this interpretation mentions only finitely many relations of $\bC$. Let $\bC'$ be the finite signature reduct of $\bC$ that contains exactly those
relations and the relations that are mentioned in the primitive positive definition
of the orbit of $(c_1,\dots,c_k)$. Since $\bC'$ still interprets $\bA$, 
there is a polynomial-time reduction from $\Csp(\bA)$ to $\Csp(\bC')$
by Theorem~\ref{thm:pp-interpret-reduce}. 
Since there is still a primitive positive definition of the orbit of $(c_1,\dots,c_k)$ in $\bC'$,
we can apply Corollary~\ref{cor:mccore-constants} and get a polynomial-time reduction
from $\Csp(\bC')$ to $\Csp(\bB')$, where $\bB'$ is the reduct of $\bB$ that only contains
the relations that are also in $\bC'$; note that $\bB'$ has finite signature. 
Composing reductions, 
we conclude that there is a polynomial-time reduction from $\Csp(\bA)$ to $\Csp(\bB')$.
\end{proof}

Together with Corollary~\ref{cor:mccore-constants} we have the following consequence.

\begin{corollary}\label{cor:pp-interpret-with-constants}
Let $\bB$ be an $\omega$-categorical model-complete core,
and let $\bA$ be a structure with a finite signature and
a hard $\Csp$. 
If $\bA$ has a primitive positive interpretation with parameters 
in $\bB$, then
 $\bB$ has a reduct with finite signature whose $\Csp$ is NP-hard.
\end{corollary}
% TODO: check all applications of this cor

%\begin{proof}
%Let $c_1,\dots,c_k$ be the constants needed to interpret
%$\bA$ in the model-complete core of $\bB$,
%and let $\bC$ denote the expansion of the
%model-complete core of $\bB$ by 
%the relations $\{c_1\},\dots,\{c_k\}$.
%Then there is a primitive positive interpretation of $\bA$ in $\bC$, and this interpretation mentions only finitely many relations of $\bC$. Let $\bC'$ be the finite signature reduct of $\bC$ that contains exactly those
%relations. Since $\bC'$ still interprets $\bA$, 
%$\Csp(\bC')$ is NP-hard by
%Theorem~\ref{thm:pp-interpret-reduce}. 
%We can thus apply Corollary~\ref{cor:mccore-constants} 
%and conclude that there is a finite signature reduct $\bB'$ of $\bB$
%such that $\Csp(\bB')$ is NP-hard.
%\end{proof}

We give an application of this technique in Proposition~\ref{prop:betw-hard} below. Many more applications can be found
in  Section~\ref{sect:endos} and Section~\ref{ssect:tcsp-hard}. We have defined in Example~\ref{expl:betw}
the relation $\Betw$ on ${\mathbb Z}$; 
we use the analogous definition for $\Betw$ over ${\mathbb Q}$, that is, 
$$\Betw := \{(x,y,z) \in {\mathbb Q}^3 \mid x<y<z \vee z<y<x\} \,.$$

\begin{proposition}\label{prop:betw-hard}
The structure $(\{0,1\}; \NAE)$
has a primitive positive interpretation in $({\mathbb Q}; \Betw, 0)$,
and $\Csp(({\mathbb Q}; \Betw))$ is NP-hard.
\end{proposition}
\begin{proof}
Recall that the relation $\NAE$ is $\{0,1\}^3 \setminus \{(0,0,0),(1,1,1)\}$.
The dimension of our interpretation $I$ is one, 
and the domain formula is $\exists z. \, \Betw(x,0,z)$, which is equivalent to $x \neq 0$. The formula $=_I(x_1,y_1)$ is 
$$ \exists z \; \big(\Betw(x_1,0,z) \wedge \Betw(z,0,y_1) \big) \; .$$
Note that $=_I$ 
 is over $({\mathbb Q}; \Betw, 0)$ 
 equivalent to $(x_1>0 \Leftrightarrow y_1 > 0)$.
Finally, the formula ${\NAE}(x_1,y_1,z_1)_I$
is 
$$ \exists u \; \big( \Betw(x_1,u,y_1) \wedge \Betw(u,0,z_1)\big ) \; .$$
The map $h$ maps positive points to $1$, and all other points from ${\mathbb Q}$ to $0$. 

Since the orbit of $0$ is the entire set ${\mathbb Q}$ it is in particular primitive positive
definable, and we can show
NP-hardness of $\Csp\big(({\mathbb Q}; \Betw)\big)$ using Proposition~\ref{prop:pp-interpret-with-constants} and the fact that $\Csp\big((\{0,1\}; \NAE)\big)$ is NP-hard.
\end{proof}

\subsection{Pseudo-varieties}
\label{ssect:pseudo-varieties}
We present the mentioned connection between primitive
positive interpretations and pseudo-varieties.
%, which has
%been published in~\cite{BodirskySurvey}.
%We have not been able to find the following theorem explicitely
%in the literature even in the case of finite algebras,
%and therefore present its proof in full detail.

\begin{theorem}[from~\cite{BodirskySurvey}]\label{thm:pp-interpret}
Let $\bC$ be a finite or $\omega$-categorical relational structure,
and let $\fC$ be a polymorphism algebra of $\bC$. 
Then a structure $\bB$ has a primitive positive interpretation
in $\bC$ if and only if  there is an algebra $\bf B$
in the pseudo-variety generated by ${\bf C}$ 
such that all operations of $\bf B$ are polymorphisms of $\bB$.
\end{theorem}
\begin{proof}
Let $\tau$ be the signature of $\fC$, and let $\cal V$ 
be the pseudo-variety generated by $\fC$. 
Similarly to the famous HSP theorem for varieties (see e.g.~\cite{BS}), 
every algebra in ${\cal V}$ is
the homomorphic image of a subalgebra of a finite direct
product of $\fC$. To see this, we have to verify that 
$\HSPfin(\fC)$ is closed under $\HHH$, $\SSS$, and $\PPPfin$.
It is clear that $\HHH(\HSPfin(\fC))=\HSPfin(\fC)$. 
Lemma~\ref{lem:congruences} implies that $\SSS(\HSPfin(\fC)) \subseteq \HSSP^{\fin}(\fC) = \HSPfin(\fC)$. Finally, 
$$\PPPfin(\HSPfin(\fC)) \subseteq \HHH\PPPfin\SSS\PPPfin(\fC) \subseteq \HSPfin\PPPfin(\fC) = \HSPfin(\fC) \; .$$

First assume that there is an algebra ${\bf B}$
in ${\cal V}$ 
all of whose operations are polymorphisms of $\bB$.
Then there exists a finite number $d \geq 1$,
 a subalgebra $\bf D$ of $\fC^d$,
 and a surjective homomorphism $h$ 
from $\bf D$ to $\bf B$.
We claim that $\bB$ has a first-order interpretation $I$
of dimension $d$ in $\bC$. 
All operations of $\fC$ preserve $D$ (viewed as a 
$d$-ary relation over $\bC$),
since ${\bf D}$ is a subalgebra of $\fC^d$.
By Theorem~\ref{thm:inv-pol}, 
this implies that $D$ has a primitive positive definition 
$\delta(x_1,\dots,x_d)$ in $\bC$, which becomes the 
domain formula $\delta_I$ of $I$.
As coordinate map we choose the mapping $h$.

If $R$ is a $k$-ary relation in $\bB$,
let $R' \subseteq C^{dk}$ be defined by
$$(a^1_1,\dots,a^d_1,\dots, a^1_k,\dots,a^d_k) \in R' \; \Leftrightarrow 
\; (h(a^1_1,\dots,a^d_1),\dots,h(a^1_k,\dots,a^d_k)) \in R \; .$$
%Clearly, $R'$ is preserved by an operation $f$ in ${\bf A}(\bC)$
%if and only if $R$ is preserved by the operation $f$ in ${\bf D}$.
%If $f$ is an operation in ${\bf A}(\bC)$ that violates
%$R'$, then corresponding operation $f$ in ${\bf B}$ violates $R$.
Let $f \in \tau$ be arbitrary. By assumption, $f^\fB$ preserves
$R$. It is easy to verify that then $f^\fC$ 
preserves $R'$. 
Hence, all polymorphisms of $\bC$
preserve $R'$, and because $\bC$ is $\omega$-categorical, the
relation $R'$ has a primitive positive definition  in $\bC$ (Theorem~\ref{thm:inv-pol}), which becomes
the defining formula for $R(x_1,\dots,x_k)$ in $I$.
Finally, since $h$ is an algebra homomorphism, the kernel
$K$ of $h$ is a congruence of ${\bf D}$. 
It follows that $K$, viewed as a $2d$-ary
relation over $C$, is preserved by all operations from $\fC$. 
Theorem~\ref{thm:inv-pol} implies that $K$ has a 
primitive positive definition in $\bC$. This definition becomes 
the formula $=_I$.
It is straightforward to verify that $I$ is a
primitive positive interpretation of $\bB$ in $\bC$.

To prove the opposite direction, 
suppose that $\bB$ has a primitive positive interpretation $I$
in $\bC$.
We have to show that $\cal V$ contains a $\tau$-algebra ${\bf B}$ such
that all operations in ${\bf B}$ are polymorphisms of $\bB$.
Let $d$ be the dimension and $\delta$ be the primitive positive domain formula of $I$. Clearly, the set $\delta(\bC^d)$ is preserved
by all operations in $\fC$, and therefore
induces a subalgebra $\bf D$ of $\fC^d$.

We first show that the kernel $K$ of the coordinate map $h$
of the interpretation is a congruence of ${\bf D}$.
For all
$d$-tuples $\overline a, \overline b \in D$, the $2d$-tuple $(\overline a,\overline b)$
satisfies $=_I$ in $\bC$ if and only if 
$h(\overline a)=h(\overline b)$. Let $S$ be the $2d$-ary relation defined by $=_I$ over
$\bC$. Then $S$ can be viewed as a binary relation over $C^d$, and we have $S \cap D^2 = K$. 
Since $=_I$ is primitive positive
definable in $\bC$, $S$ is preserved
by all polymorphisms of $\bC$. 
To show that $K$ is a congruence of $\bf D$, let $f \in \tau$ be $k$-ary, 
and let $(a^1,b^1), \dots, (a^k,b^k)$ be pairs from $K$.
Let $a = f^{\bf D}(a^1,\dots,a^k)$ and $b = f^{\bf D}(b^1,\dots,b^k)$. We
have to show that $(a,b) \in K$. Since ${\bf D}$ is a subalgebra of $\fC^d$,
$a,b \in D$, and hence it suffices to show that $(a,b) \in S$. 
Recall that $f^{\bf D}$ is defined by applying $f^\fC$ component-wise. 
Since $(a^i,b^i) \in S$ for all $i \leq k$ and $f^\fC$ preserves $S$,
we thus have that $(a,b) \in S$.
Hence, $K$ is a congruence of $\bf D$ and $h$ is a surjective homomorphism from $\bf D$ to ${\bf B} := {\bf D} / h$.

%Let $g$ the natural homomorphism from ${\bf A}(\bC)^d$
%to ${\bf A}(\bC)^d / \theta$. 
%By Lemma~\ref{lem:congruences},
%the image of the subalgebra $\bf D$ of ${\bf A}(\bC)^d$ under $h$ is a subalgebra of 
%${\bf A}(\bC)^d / h$, and will be denoted by $\bf B$ . The restriction of $h$ to $D$ is a homomorphism from $\bf D$
%to $\bf B$, which shows that $K$ is a congruence.

We finally verify that  
every operation in ${\bf B}$ is a polymorphism of $\bB$, i.e.,
for every $f \in \tau$, every relation $R$ of $\bB$ is preserved by $f^{\bf B}$. 
The operation $f^\fB$ preserves $\phi := R(x_1,\dots,x_k)$ 
if and only if 
$f^\fC$ preserves $\phi_I$. Since $f^\fC$
is a polymorphism of $\bC$, and since $\phi_I$
is primitive positive over $\bC$, the operation $f^{\fC}$
indeed preserves $\phi_I$. 
\end{proof}

The proof of Theorem~\ref{thm:pp-interpret} above gives
more information about the link between the algebras
in $\HSPfin(\Alg(\bB))$ and the primitive positive interpretations in $\bB$, and we state it explicitly.  

\begin{theorem}\label{thm:pp-pseudovar-detail}
  Let $\bC$ be a finite or $\omega$-categorical structure, and let $\bB$ be an arbitrary structure. Then the following are equivalent.
  \begin{enumerate}
  \item there is a polymorphism algebra $\fC$ of $\bC$, 
  %(equivalently, for any polymorphism algebra $\fC$ of $\bC$), 
  an algebra $\fS \in \SSS(\fC^d)$ with domain $S$,
  and a surjective homomorphism $h$ from $\fS$ to an algebra $\fB$
  such that $\Clo(\fB) \subseteq \Pol(\bB)$;
 	\item  $\bB$ has the primitive positive interpretation $(d,S,h)$ in $\bC$.
  \end{enumerate}
\end{theorem}

%The following can be shown directly, but it also follows from Theorem~\ref{thm:pp-pseudovar-detail} and the observation that 
%$\fA \in HSP^{d_1}HSP^{d_2}(\fB)$ implies 
%$\fA \in HSP^{d_1+d_2}(\fB)$. 
%\begin{corollary}\label{cor:pp-interpret-composition}
%If $\bC_2$ has a $d_1$-dimensional primitive positive interpretation $I_1$ in $\bC_1$, 
%and $\bC_3$ has an $d_2$-dimensional primitive positive interpretation $I_2$ in $\bC_2$,
%then $\bC_3$ has a $(d_1d_2)$-dimensional primitive positive interpretation in $\bC_1$, which we denote by $I_2 \circ I_1$. 
%\end{corollary}
% This is not unique. But there is a "canonical"
% definition of I_2 \circ I_1. Probably it is better to present 
% the definition of I_2 \circ I_1 from the beginning. 

We return to applications of these concepts to CSPs.
\begin{corollary}\label{cor:pseudovar-hard}
Let $\bB$ be $\omega$-categorical.
If there is an expansion $\bC$ of the model-complete core
of $\bB$ by finitely many constants
such that the pseudo-variety $\V$ generated by $\Alg(\bC)$
contains a 2-element algebra where all operations are projections,
then $\bB$ has a finite signature reduct with an NP-hard $\Csp$.
\end{corollary}
\begin{proof}
Let $\bf D$ be the 2-element algebra in $\V$
where all operations are projections.
All operations of $\bf D$ preserve the relation $\OIT$. 
By Theorem~\ref{thm:pp-interpret},
the structure $(\{0,1\}; \OIT)$ has a primitive positive 
interpretation in $\bC$.
Then Corollary~\ref{cor:pp-interpret-hard} 
shows that $\bB$ has a finite signature reduct $\bB'$
with an NP-hard CSP.
\end{proof}

%All $\omega$-categorical templates known to the author that have an NP-complete $\Csp$ satisfy the condition from Corollary~\ref{cor:pseudovar-hard}. 
% No longer true: there are examples
% of this type that are definable in
% the atomless Boolean algebra. 
% BOOK-TD: present this in detail. 
All templates $\bB$ with a first-order definition in a homogeneous structure with finite relational signature known to the author that have an NP-complete $\Csp$ satisfy the condition from Corollary~\ref{cor:pseudovar-hard}. 
For finite templates $\bB$ there is the 
conjecture (and strong evidence) that $\Csp(\bB)$ is NP-hard if and only if $\bB$ satisfies this condition (see Section~\ref{ssect:tractability}). 

% BOOK-TD: move the first four items to pp interpretations!

\begin{theorem}\label{thm:simulates-tame}
Let $\bB$ be any structure. %model-complete core.
Then the following are equivalent.
\begin{enumerate}
\item \label{enum:oit} there is a primitive positive interpretation of $(\{0,1\};\OIT)$ in $\bB$.
\item \label{enum:nae} 
there is a primitive positive interpretation of $(\{0,1\};\NAE)$ in $\bB$;
% Definitely need in applications the following directions: 
% NAE interpretable -> OIT interpretable
\item \label{enum:fopp} $\bB$ interprets a structure with at least two elements where all first-order formulas are equivalent to primitive positive formulas;
\item \label{enum:all} all finite structures have a primitive positive interpretation in $\bB$.
\setcounter{mycounter}{\value{enumi}}
\end{enumerate}
If $\bB$ is $\omega$-categorical, the following two conditions are equivalent to the conditions above. 
\begin{enumerate}\setcounter{enumi}{\value{mycounter}}
\item \label{enum:allnset} 
the pseudo-variety $\V$ generated by $\Alg(\bB)$
contains for all $n$ an algebra on $n$ elements 
all of whose operations are projections;
\item \label{enum:2set} 
the pseudo-variety $\V$ generated by $\Alg(\bB)$
contains a 2-element algebra all of whose operations are projections.
\end{enumerate}
\end{theorem}
\begin{proof}
The first statement can be shown by proving implications in cyclic order, 
$(1) \Rightarrow (2) \Rightarrow (3) \Rightarrow (4) \Rightarrow (1)$.
Obviously, $(4)$ implies $(1)$. We have given a primitive positive 
definition of $\NAE$ in $(\{0,1\}; \OIT)$ in the proof of Theorem~\ref{thm:schaefer}, 
which implies that $(1)$ implies $(2)$.
The implication from $(2)$ to $(3)$ is by Proposition~\ref{prop:hard-boolean-csps}. 
%$(4)$ obviously implies $(1)$  and $(2)$, so for the first statement it suffices to show that each of $(1)$ and $(2)$ implies $(3)$, and that $(3)$ implies $(4)$. 
%so it remains to show that 
%$(\ref{enum:oit}) \Rightarrow (\ref{enum:nai}) \Rightarrow (\ref{enum:fopp}) \Rightarrow (\ref{enum:all})$. The implication $(\ref{enum:oit}) \Rightarrow (\ref{enum:nai})$ has been shown in the proof of Theorem~\ref{thm:schaefer}. 
%For $(\ref{enum:oit}) \Rightarrow (\ref{enum:fopp})$, observe that all polymorphisms
%of $(\{0,1\};\OIT)$ are projections and thus Theorem~\ref{thm:inv-pol} that all Boolean relations have a primitive positive definition in $(\{0,1\};\OIT)$ (Theorem~\ref{thm:schaefer}). 
%For $(\ref{enum:nae}) \Rightarrow (\ref{enum:fopp})$, note that $(\{0,1\};\NAE)$
%is preserved by $x \mapsto 1-x$, and hence all first-order definable
%relations are preserved by $x \mapsto 1-x$ as well.
%By Corollary~\ref{cor:galois} it therefore suffices
%to show all polymorphisms of $(\{0,1\};\NAE)$ are generated by this operation. 
%This can be done similarly as above, using Therorem~\ref{thm:rosenberg}.

For the implication $(\ref{enum:fopp}) \Rightarrow (\ref{enum:all})$,
let $\bB'$ be the structure that has a primitive positive interpretation in $\bB$, has at least two elements, and where all first-order formulas are equivalent to primitive positive formulas.
Let $\bA$ be a $\tau$-structure with domain $\{1,\dots,n\}$. We prove that $\bA$
has a first-order interpretation in $\bB'$. This yields in fact a primitive positive interpretation since every first-order formula is equivalent to a primitive positive formula in $\bB'$.
The claim then follows by composing the primitive positive interpretation of $\bB'$ in $\bB$ with that of $\bA$ in $\bB'$.
%, via Lemma~\ref{lem:pp-interpret-composition}. 

Our first-order interpretation $I$ of $\bA$ in 
$\bB'$ is $2n$-dimensional. The domain formula 
$\delta_I(x_1,\dots,x_n,x_1',\dots,x_n')$ expresses that for exactly one $i \leq n$ we have $x_i = x_i'$; clearly this is first-order.
Equality is interpreted by the formula 
$$=_I(x_1,\dots,x_n,x_1',\dots,x_n',y_1,\dots,y_{n},y_1',\dots,y_n') :=  \bigwedge_{i=1}^n \big ((x_i=x'_i) \Leftrightarrow (y_i=y_i') \big ) \; .$$ 
Note that the equivalence relation defined by $=_I$ on $\delta((\bB')^{2n})$ 
has exactly $n$ equivalence classes, and the coordinate map sends
$(x_1,\dots,x_n,x_1',\dots,x_n')$ to $i$ if and only if $x_i=x_i'$. 
It is now straightforward to write down first-order formulas $\phi_I$ that interpret atomic $\tau$-formulas $\phi$. When $R \in \tau$ is $k$-ary, then the formula $R(x_1,\dots,x_k)_I$ 
is a disjunction of conjunctions with the $2nk$ variables $x_{1,1},\dots,x_{k,n}$, 
$x'_{1,1},\dots,x'_{k,n}$.
For each tuple $(t_1,\dots,t_k)$ from $R^\bA$ 
the disjunction contains the conjunct $\bigwedge_{i \leq k} x_{i,t_i}=x'_{i,t_i}$. 
%Hence, the coordinate map sends $(x_1,\dots,x_{2n},y_1,\dots,y_{2n}) \in B^4$
%to the same element if
%$\delta(x_1,\dots,x_n,y_1,\dots,y_n$ 
%is $\bigwedge_{i \neq j, 1 \leq i,j \leq n} (x_i \neq x_j \wedge x_i \neq y_$

Now suppose that $\bB$ is $\omega$-categorical. We prove that $(\ref{enum:all})$
implies $(\ref{enum:allnset})$, that $(\ref{enum:allnset})$ implies $(\ref{enum:2set})$,
and that $(\ref{enum:2set})$ implies $(\ref{enum:oit})$. 
For $(\ref{enum:all}) \Rightarrow (\ref{enum:allnset})$, let $\bA$ be the
structure with domain $A = \{1,\dots,n\}$, the relations $P_3^A$, and for
each $i \in \{1,\dots,n\}$ the unary relation $\{i\}$. 
By $(\ref{enum:all})$ there is a primitive positive interpretation of $\bA$
in $\bB$. Hence, Theorem~\ref{thm:pp-interpret} implies that there is an algebra $\fA' \in \V$ such that all operations of $\fA'$ are polymorphisms of $\bA$. But all polymorphisms of $\bA$ are projections (Corollary~\ref{cor:elementary}).
The implication $(\ref{enum:allnset}) \Rightarrow (\ref{enum:2set})$ is trivial.
The implication $(\ref{enum:2set}) \Rightarrow (\ref{enum:oit})$ follows 
from Theorem~\ref{thm:pp-interpret} and the fact that the projections
preserve $\OIT$.
\end{proof}

%\begin{definition}\label{def:simulates}
%Let $\bB$ be an $\omega$-categorical model-complete core, and let $\cC$ be its
%polymorphism clone. 
%If $\bB$ satisfies the equivalent conditions 
%in Theorem~\ref{thm:simulates-tame},
%then we say that $\bB$ \emph{simulates} $(\{0,1\};\OIT)$; otherwise 
%If $\bB$ does not satisfy the equivalent conditions in Theorem~\ref{thm:simulates-tame},
%we say that $\cC$ is \emph{tame}.
%the model-complete core
%of $\bB$ has an expansion $\bC$ by finitely many constants
%such that ${\cal V}(\bf A(\bC))$ contains
%an 2-element algebra where all operations are projections.
%\end{definition}

\subsection{Bi-interpretations and Classification Transfer}
\label{ssect:transfer}
Let $\bC$ be a structure with finite relational signature. By the
\emph{classification project for $\bC$} we mean a complexity
classification for $\Csp(\bB)$ for all first-order 
expansions $\bB$ of $\bC$ that have finite relational signature. For instance, the classification project for the random graph $(\mV;E)$ is treated in Chapter~\ref{chap:schaefer}, and the classification project for
$(\mQ;<)$ is treated in Chapter~\ref{chap:tcsp}.

Sometimes, it is possible to derive the complexity classification
project for $\bC$ from the complexity classification project 
for $\bD$,
for another $\omega$-categorical structure $\bD$. 
For instance, we will show below how to derive the 
classification project for the directed graph 
$$\bC := ({\mathbb N}^2; \{(x,y),(u,v) \; | \; y=u\})$$
from the classification project for
$\bD := ({\mathbb N};=)$ (which will be given in Chapter~\ref{chap:ecsp}); a more advanced application of such a classification transfer can be found in Theorem~\ref{thm:allen-lift} below.

Primitive positive interpretability is a crucial concept for 
the transfer of complexity classifications.
In particular, this section studies \emph{primitive positive bi-interpretations} in this context.
Two interpretations of $\bC$ in $\bD$
with coordinate maps $h_1$ and $h_2$ are called 
\emph{homotopic}\footnote{We are following the terminology from~\cite{AhlbrandtZiegler}.}
%The topological term `homotopic' has been chosen deliberately (as originally inthe term is now standard in model-theory~\cite{Hodges}).}
%and its topological interpretation will be given in Section~\ref{sect:bi-interpret}.
if the relation $\{(\bar x,\bar y) \; | \; h_1(\bar x) = h_2(\bar y) \}$
is first-order definable in $\bD$. If this relation is even primitive positive
definable in $\bD$, we say that the two interpretations are \emph{pp-homotopic}.
The \emph{identity interpretation} of a $\tau$-structure $\bC$
 is the interpretation $I=(1,\text{true},h)$
of $\bC$ in $\bC$ whose coordinate map $h$ is the identity
(note that the identity interpretation is primitive positive). 
Recall that we write
$I_1 \circ I_2$ for the natural composition of two interpretations $I_1$ and $I_2$, defined in Section~\ref{ssect:pp-interpret}.

\begin{definition}
Two structures $\bC$ and $\bD$ with an
interpretation $I$ of $\bC$ in $\bD$ and an interpretation $J$ of $\bC$ in $\bD$ are called \emph{mutually interpretable}. 
If both $I \circ J$ and $J \circ I$ are homotopic
to the identity interpretation (of $\bD$ and of $\bC$, respectively),
then we say that $\bC$ and $\bD$ are \emph{bi-interpretable}. 

When both interpretations $I$ and $J$ are primitive positive, 
then $\bC$ and $\bD$ are called \emph{mutually pp-interpretable}. If moreover $I \circ J$ and $J \circ I$ are pp-homotopic to the identity interpretation,
then $\bC$ and $\bD$ are called \emph{primitive positive bi-interpretable}. 
\end{definition}

\begin{example}\label{expl:lift}
The directed graph $\bC := ({\mathbb N}^2; M)$ where 
$$M := \big \{((u_1,u_2),(v_1,v_2)) \; | \; u_2=v_1 \big \}$$
and the structure $\bD := ({\mathbb N};=)$ 
are primitive positive bi-interpretable. The interpretation $I$ of $\bC$ in $\bD$ is 2-dimensional, the domain formula is true, and the coordinate
map $h$ is the identity. The interpretation $J$ of $\bD$ in $\bC$ is 1-dimensional, the domain formula is true, and the coordinate map $g$
sends $(x,y)$ to $x$. Both interpretations are clearly primitive positive. 

Then $g(h(x,y))=z$ is definable by the formula $x=z$, and hence $I \circ J$ is pp-homotopic to the identity interpretation of $\bD$. 
Moreover, $h(g(u),g(v))=w$ is primitive positive definable
% Formula below corrected 5/3/2013
by $$M(w,v) \wedge \exists p \, (M(p,u) \wedge M(p,w)) \; ,$$ 
so $J \circ I$ is also pp-homotopic to the identity interpretation of $\bC$.
\qed
\end{example}

\begin{example}\label{expl:mutually}
The structures $\bC := \big ({\mathbb N}^2; \{(x,y),(u,v) \; | \; x=u\} \big )$
and $\bD := ({\mathbb N};=)$ are mutually primitive positive interpretable,
but \emph{not} primitive positive bi-interpretable. 
There is a primitive positive interpretation $I_1$
of $\bD$ in $\bC$, and a primitive positive interpretation of 
$\bC$ in $\bD$ such that $I_2 \circ I_1$ is pp-homotopic to the
identity interpretation.
However, 
the two structures are not even \emph{first-order} 
bi-interpretable, as we will see 
in Example~\ref{expl:not-bi-interpret} in Section~\ref{sect:bi-interpret}.
\qed
\end{example}

Here comes the central lemma for complexity classification
transfer.

\begin{lemma}\label{lem:transfer}
Suppose $\bD$ has a primitive positive interpretation $I$ in $\bC$,
and $\bC$ has a primitive positive interpretation $J$ in $\bD$ such that
 $J \circ I$ is pp-homotopic to the identity interpretation of $\bC$.
Then for every first-order expansion $\bC'$ of $\bC$
there is a first-order expansion $\bD'$ of $\bD$
such that $\bC'$ and $\bD'$ are mutually pp-interpretable.
% BOOKTD: aren't they even pp-bi-interpretable?
\end{lemma}
\begin{proof}
Let $I = (c,U,g)$ and $J=(d,V,h)$ be the primitive positive interpretations from the statement,
and let $\bC'$ be 
a first-order expansion of $\bC$.
%and let $\sigma$ be the signature of $\bC$, and $\tau$ be the signature of
Then we set $\bD'$ to be the expansion of $\bD$
that contains for every $k$-ary $R$ in the signature of $\bC'$ 
the $(dk)$-ary relation $S$ defined as follows. 
When $\phi$ is the first-order definition of $R$ in $\bC$, then
$S$ is the relation defined by $\phi_J$ in $\bD$ (see Section~\ref{ssect:oldnew} and Section~\ref{ssect:pp-interpret}).

We claim that $\bC'$ has the primitive positive interpretation 
$(d,V,h)$ in $\bD'$. First note that $V$ is primitive positive definable
in $\bD'$ since $\bD'$ is an expansion of $\bD$.
An atomic formula $\psi$ with free variables $x_1,\dots,x_k$
in the signature of $\bC'$ can be interpreted in $\bD'$ as follows. We replace the relation
symbol in $\psi$ by its definition in $\bC$, and obtain
a formula $\phi$ in the language of $\bC$.
Let $S$ be the symbol in the language of $\bD'$ for the relation
defined by $\phi_{J}(x_1^1,\dots,x_1^d,\dots,x^1_k,\dots,x^d_k)$
 over $\bD'$. Then indeed $S(x_1^1,\dots,x_1^d,\dots,x^1_k,\dots,x^d_k)$ is a defining formula for $\psi$, because 
$$\bC' \models \psi(h(a_1^1,\dots,a_1^d),\dots,h(a^1_k,\dots,a^d_k)) \Leftrightarrow \bD' \models S(a_1^1,\dots,a_1^d,\dots,a^1_k,\dots,a^d_k)$$ for all $a_1,\dots,a_k \in V$.

Conversely, we claim that $\bD'$
has the primitive positive interpretation $(c,U,g)$ in $\bC'$. 
Again, $U$ is primitive positive definable in $\bC'$ since 
$\bC'$ is an expansion of $\bC$. 
Let $\phi$ be an atomic formula in the (relational) signature of $\bD'$.
If the relation symbol in $\phi$ is already in the signature of $\bD$,
then there is a primitive positive interpreting formula in $\bC$ and
therefore also in $\bC'$. Otherwise, by definition of $\bD'$, the relation
symbol in $\phi$ has arity $dk$, and has been introduced for a $k$-ary relation $R$ from $\bC'$. We have to find a defining formula
having $kcd$ variables. 
Let $\theta(x_0,x_{1,1},\dots,x_{1,c},\dots,x_{d,1},\dots,x_{c,d})$ be the primitive positive formula of
arity $cd+1$ that shows 
that $h(g(x_{1,1},\dots,x_{c,1}),\dots,g(x_{1,d},\dots,x_{c,d})) = x_0$ is primitive positive definable in $\bC$. 
Then the defining formula for the atomic formula $\phi(x^1_1,\dots,x^k_d)$ 
%in the language of $\bD'$
has
free variables $x^1_{1,1},\dots,x^k_{c,d}$ %in $\bC'$ 
and equals
$$\exists x^1,\dots,x^k \; 
\big(R(x^1,\dots,x^k) \; \wedge \;  \bigwedge_{i=1}^k \theta(x^i,x^i_{1,1},\dots,x^i_{c,d})\big ) \, .$$
\end{proof}

In particular, when $\bC$, $\bD$, $\bC'$ and $\bD'$ are as in Lemma~\ref{lem:transfer}, 
and $\bC'$ and $\bD'$ have a finite
relational signature, then $\Csp(\bC')$ and $\Csp(\bD')$ have the same computational complexity, by Theorem~\ref{thm:pp-interpret-reduce}. Hence, Lemma~\ref{lem:transfer} shows that the classification project
for $\bC$ can be reduced to the classification project for $\bD$.
%since for every first-order expansion $\bD'$ of $\bD$ there 
%is a first-order expansion of $\bC$ such that $\Csp(\bB)$ and
%$\Csp(\bC
With a slightly stronger assumption we can get the following
consequence. 

\begin{corollary}\label{cor:pp-bi}
Let $\bC$ and $\bD$ be primitive positive bi-interpretable
 $\omega$-categorical structures.
Then every first-order expansion of $\bC$
is primitive positive bi-interpretable with a first-order expansion of $\bD$.
\end{corollary}

More about \emph{first-order} bi-interpretability can be found in Section~\ref{sect:bi-interpret}. 
Let us conclude with a concrete application of Corollary~\ref{cor:pp-bi}.

% THERE IS A GENERAL LEMMA IN THIS CONTEXT: 
% when A has a full interpretation in B, 
% then the orbits of A have a primitive positive
% interpretation in the orbits of B?!

\begin{theorem}
\label{thm:allen-lift}
Let $\bB$ be a 
reduct of Allen's  interval algebra (Example~\ref{example:allen}) 
that contains the relation $m = \big \{((u_1,u_2),(v_1,v_2)) \; | \: u_2 = v_1 \big\}$.
Then $\Csp(\bB)$ is either in P or NP-complete. 
\end{theorem}
\begin{proof}
	We show that the structure $({\mathbb I}; m)$ 
	is primitive positive bi-interpretable with $({\mathbb Q}; <)$. 
	It follows that $\bB$ is primitive positive bi-interpretable with a temporal constraint language $\bB'$, and the result follows by the main result of Chapter~\ref{chap:tcsp} and Corollary~\ref{cor:pp-bi}.
	
	Let $I$ be the $2$-dimensional interpretation of  $({\mathbb I}; m)$ in $({\mathbb Q}; <)$ 
	with domain formula $x<y$, the formula $(y_1=y_2)_I$ is true, 
	and the formula $(m(y_1,y_2))_I$ has variables $x^1_1,x^1_2,x^2_1,x^2_2$ and is
	given by $x^1_2 = x^2_1$. 
	The coordinate map $g$ sends $(x,y) \in {\mathbb Q}^2$ with $x<y$
	to the interval $[x,y] \in {\mathbb I}$.

	Let $J$ be the $1$-dimensional interpretation with domain formula true,
	and where the coordinate map $h$ is $[x,y] \mapsto x$. 
	The formula $(x<y)_I$ is the primitive positive formula $$\exists u,v \, \big(m(u,x_1) \wedge m(u,v) \wedge m(v,x_2)\big) \; .$$
	
	We show that $J \circ I$ and $J \circ I$ are pp-homotopic to the identity interpretation. 
	The relation $\big \{(x_1,x_2,y) \, | \, h(g(x_1,x_2))=y \big \}$ has the primitive positive definition
	$x_1=y$. To see that the relation $R := \big \{(u,v,w) \, | \, g(h(u),h(v))=w \big \}$ has a
	primitive positive definition in $({\mathbb I};m)$, first note that the relation
	$$\big \{(u,v) \; | \; u=[u_1,u_2], v=[v_1,v_2], u_1 = v_1 \big \}$$ has the primitive positive definition
	$\phi_1(u,v) = \exists w \; \big(m(w,u_1) \wedge m(w,u_2)\big)$ in  $({\mathbb I}; m)$. 
	Similarly, $\{(u,v) \; | \; u=[u_1,u_2], v=[v_1,v_2], u_2 = v_2\}$ has a primitive positive definition $\phi_2(u,v)$. Then the formula $\phi_1(u,w) \wedge \phi_2(v,w)$ is equivalent to a primitive
	positive formula over  $({\mathbb I}; m)$, and defines $R$.	
\end{proof}

\section{Varieties}
\label{sect:varieties}
Varieties (which we have introduced briefly in Section~\ref{sect:pseudo-var}) are a fascinatingly powerful concept to study classes of algebras. For a finite structure $\bB$ with finite signature, 
%that contains a relation $\{c\}$ for every element $c$ of its domain, 
the complexity of $\Csp(\bB)$ 
only depends on the variety generated by the polymorphism algebra of $\bB$.
This is in particular related to the fact that a finite algebra is in the variety generated
by a finite algebra $\fB$ if and only if it is in the pseudo-variety generated by $\fB$; % (Theorem~\ref{thm:variety-pseudovariety}); 
the link between the pseudo-variety generated by $\fB$ and the CSP has already been
explained in Section~\ref{sect:pseudo-var}. 

The section has two parts. In Section~\ref{ssect:tractability}
we explain the role of varieties for the study of CSPs with finite templates.
In particular, we present various equivalent forms of the \emph{tractability conjecture}
for finite domain constraint satisfaction.

The second part studies the situation for $\omega$-categorical templates.
It is open whether the complexity of an 
$\omega$-categorical model-complete core only
depends on the variety generated by the polymorphism algebra of $\bB$. 
But it will turn out that the tractability frontier in
the classification results in Chapter~\ref{chap:schaefer} and
Chapter~\ref{chap:tcsp} can be described elegantly using varieties, and
the description is very similar to the tractability conjecture for finite domain constraint satisfaction. In Section~\ref{ssect:canonical} we provide some partial explanation 
for this phenomenon.
%A generalization of the tractability conjecture for
%a large class of $\omega$-categorical structures will be presented 
%in Section~\ref{ssect:inf-tractability}.

\subsection{The Tractability Conjecture}
\label{ssect:tractability}
%As we have mentioned in the introduction, 
%The universal-algebraic approach for CSPs with $\omega$-categorical templates is certainly inspired by the universal-algebraic approach
%for finite domains. 
In this section, we present some classical results
that specifically hold for \emph{finite} algebras and
are relevant to constraint satisfaction. 
We also discuss more recent universal-algebraic results 
about finite algebras.
%, some of which are still
%unmatched for general $\omega$-categorical templates.
%There are many deep results for finite domains,
%some of them answering questions that have been open for more than a decade. 
We cannot cover all recent developments here,
but sketch in this section some of the highlights.
%, in particular those for which we believe that they might provide
%inspiration for pushing further the universal-algebraic approach
%for $\omega$-categorical structures.

We have already mentioned in the introduction the dichotomy conjecture of 
Feder and Vardi~\cite{FederVardi}, which we state here since it is
one of the central stimulating conjectures for finite domain constraint satisfaction.

% BOOKTD: isn't this a bit late? 
\begin{conjecture}[Dichotomy Conjecture~\cite{FederVardi}]\label{conj:dich}
Let $\bB$ be a structure with finite relational signature and finite domain. 
Then $\Csp(\bB)$ is in P or NP-complete.
\end{conjecture}

We will now see a stronger conjecture, due to~\cite{JBK,BulatovJeavons},
 that exactly describes which finite-domain CSPs are NP-hard, and which can be solved in polynomial time.
This conjecture is called the \emph{tractability conjecture},
and it has been confirmed in many important cases, for example for
\begin{itemize}
\item finite structures $\bB$ that contain a unary relation symbol for each subset of the domain of $\bB$, due to~\cite{Conservative} (see also~\cite{Barto-Conservative}),
\item structures over a 3-element domain~\cite{Bulatov}, and
\item digraphs without sources and sinks~\cite{BartoKozikNiven}, and
which includes the case of undirected graphs.
\end{itemize}

The tractability conjecture can be formulated in terms of 
primitive positive interpretability (Section~\ref{sect:pseudo-var}) as follows.

\begin{conjecture}[Tractability Conjecture]\label{conj:tractability}
Let $\bB$ be a finite structure with  finite relational signature, 
and let $\bC$ be 
the core of $\bB$. 
Then $\Csp(\bB)$ is NP-hard if there is a primitive positive interpretation of $(\{0,1\}; \OIT)$ with parameters in $\bC$, and can be solved in polynomial-time otherwise. 
\end{conjecture}

We remark that the first part of this conjecture follows
directly from Corollary~\ref{cor:pp-interpret-with-constants}.
Moreover, we have also seen that
$\Csp(\bB)$ is polynomial-time equivalent to $\Csp(\bC)$, so 
all that remains to be shown is to prove polynomial-time tractability of $\Csp(\bC)$ 
when $\bC$ does not admit a primitive positive 
interpretation of a hard Boolean CSP. 
By the results in Section~\ref{sect:pseudo-var}, this condition 
can be translated into a condition of the pseudo-variety generated
by the polymorphism algebra $\fC$ of $\bC$. 
%The conjecture then says that if the condition given in 
%Corollary~\ref{cor:pseudovar-hard} 
%does not apply, then $\Csp(\bB)$ is in P. 

When $\fB$ is a finite algebra, 
it turns out that 
a finitely generated algebra (and in particular a finite algebra) ${\bf A}$ 
is in the pseudo-variety generated
by ${\bf B}$ if and only if ${\bf A}$ is in the variety generated by ${\bf B}$
(see~\cite{BS}; the claim follows from Exercise 11.5 in combination with the proof of Lemma 11.8 there). Varieties have the advantage that
they can be described by the equations satisfied by its members.

\begin{theorem}[Birkhoff; see e.g.~\cite{HodgesLong} or~\cite{BS}]
\label{thm:birkhoff}
Let $\tau$ be a functional signature, $\bf A$ a $\tau$-algebra, 
and $\C$ be a class of $\tau$-algebras. 
Then the following are equivalent.
\begin{itemize}
\item All universal conjunctive sentences that hold in all members
of $\C$ also
hold in $\bf A$.
\item $\bf A$ is in the variety generated by $\mathcal C$.%${\bf A}$.
\item ${\bf A} \in \HSP(\C)$.
\end{itemize}
%Moreover, when $\bf A$ and $\bf B$ are finite, then we can add the following to the list:
%\begin{itemize}
%\item $\bf B$ is in the pseudo-variety generated by ${\bf A}$. 
%\end{itemize}
\end{theorem}

%Hence, a characterization of interpretability in terms of the variety
%generated by the polymorphism algebra of $\mathfrak A$
%can be used to translate the fact that 
%$\mathfrak A$ does \emph{not}
%interpret a certain structure $\mathfrak B$ 
%into the existence of a polymorphism of $\mathfrak A$
%that satisfies certain universal conjunctive sentences.
Theorem~\ref{thm:birkhoff} is 
important for constraint satisfaction since it can be used to transform the `negative'
statement of not interpreting certain hard boolean CSPs into a `positive'
statement of having polymorphisms satisfying non-trivial identities. 
The following theorem is an application of this philosophy, and
%Theorem~\ref{thm:taylor} 
goes back  to Walter Taylor (Corollary 5.3 in~\cite{Taylor}; 
see also Lemma 9.4 in~\cite{HobbyMcKenzie}).

\begin{theorem}
\label{thm:taylor}
Let $\bB$ be a finite structure,
and suppose that the polymorphism algebra $\bf B$ of $\bB$
is idempotent. Then the following are equivalent.
\begin{itemize}
\item $(\{0,1\}; \OIT)$ does not have a primitive positive
interpretation in $\bB$.
\item every 2-element algebra in the pseudo-variety generated by $\fB$ contains
an essential operation.
\item every 2-element algebra in $\HSP(\bf B)$
contains an essential operation.
%\item every finite algebra in $\HSP(\bf B)$ contains an operation that is not a projection.
%\item every 2-element algebra in $\HS({\bf B})$ (i.e., every factor of $\bf B$) contains an essential operation.
\item $\bf B$ has a \emph{Taylor term}, that is, an $n$-ary operation, for $n \geq 2$, such that 
for every $1 \leq i \leq n$ there are $x_1,\dots,x_n,y_1,\dots,y_n \in \{x,y\}$ such that $(B;f)$ satisfies 
\begin{align*}
\forall x,y. \; & f(x_1,\dots,x_{i-1},x,x_{i+1},\dots,x_n) \\
= & f(y_1,\dots,y_{i-1},y,y_{i+1},\dots,y_n) \; .
\end{align*}
\end{itemize}
\end{theorem}
The equations satisfied by Taylor terms are a special form of a \emph{linear equation}, that is, an equation
of the form 
$$\forall x_1,\dots,x_n. \, f(u_1,\dots,u_n) = f(v_1,\dots,v_n)$$ where
$u_1,\dots,u_n,v_1,\dots,v_n \in \{x_1,\dots,x_n\}$. 

Even though Theorem~\ref{thm:taylor} is of central importance in 
universal algebra, % (see e.g.~\cite{HobbyMcKenzie}),
it was discovered only recently and
under the influence of work in the context of constraint satisfaction
%that this theorem can been improved to the following.
that the existence of Taylor terms is equivalent to the existence of various other terms satisfying stronger conditions,
which are likely to be of greater use in the quest for polynomial-time algorithms for CSPs.
%By inspection of the proof of Theorem~\ref{thm:pp-interpret},
%this is equivalent to saying that 
%when there is a primitive positive interpretation of 
%$(\{0,1\}; \OIT)$ in a finite core structure $\bB$, then there is 
%also a $1$-dimensional interpretation of $(\{0,1\}; \OIT)$ in $\bB$ 

\begin{theorem}[of~\cite{Siggers,Cyclic}]\label{thm:siggers}
Let $\bf B$ be a finite idempotent algebra. 
Then the following are equivalent.
\begin{itemize}
\item $\bf B$ has a Taylor term.
\item $\bf B$ has a \emph{weak near unanimity}, that is, an $n$-ary idempotent operation $f$, for $n \geq 2$, that satisfies 
$$\forall x,y. \; f(x,\dots,x,y)=f(x,\dots,x,y,x)=\dots=f(y,x,\dots,x) \; .$$
\item $\bf B$ has a \emph{Siggers term}\footnote{Originally, Siggers gave equations for a six-ary operation, using the universal-algebraic formulation from~\cite{BulatovHColoring} of the dichotomy theorem for the CSPs for undirected graphs $\bH$ from~\cite{HellNesetril}. This was later improved by an anonymous referee of~\cite{Siggers} to the given identities for a 4-ary operation, using the main result from~\cite{BartoKozikNiven}; see concluding comments in~\cite{Siggers}.}, that is, a four-ary operation $f$ that satisfies 
$$\forall x,y. \; f(y,y,x,x)= f(x,x,x,y) = f(y,x,y,x) \; .$$
\item $\bf B$ has a \emph{cyclic term}, i.e., an $n$-ary operation $f$, for $n \geq 2$, that satisfies $$\forall x_1,\dots,x_n. \; f(x_1,\dots,x_n)=f(x_2,\dots,x_n,x_1) \; .$$
\end{itemize}
\end{theorem}

Note that weak near unanimities, 
Siggers terms and cyclic terms are (special)
Taylor terms, and that cyclic terms are (special) weak near unanimities. 
Also note that the existence of a Siggers term can be decided 
(for an explicitly given finite idempotent algebra $\bf B$),
and hence the condition of the tractability conjecture 
for finite domain constraint satisfaction is decidable.
We would also like to remark that binary \emph{commutative} operations,
that is, operations $f$ satisfying $f(x,y)=f(y,x)$, are Taylor terms, 
and that Conjecture~\ref{conj:tractability} is already open
in this special case. 

%Our optimism about the gain from this insights of 
%Theorem~\ref{thm:siggers}
%with respect to the goal to prove Conjecture~\ref{conj:tractability} is sligthly curbed by the observation that already
%the following special case of Conjecture~\ref{conj:tractability} is open.
%\begin{conjecture}
%Let $\bB$ be a finite core structure with a polymorphism
%that satisfies $f(x,y)=f(y,x)$. Then $\Csp(\bB)$ is in P.
%\end{conjecture} 

Another improvement of Theorem~\ref{thm:taylor} is the 
following result from~\cite{BulatovJeavons} 
(Proposition 4.14). %observation that 
%when the variety generated by a finite idempotent algebra
%$\bB$ contains a 2-element algebra without essential operations,
%then this algebra is already contained in $\HS(\bB)$ (see Proposition 4.14 in~\cite{BulatovJeavons}).

\begin{theorem}[of~\cite{BulatovJeavons}]\label{thm:factors}
Let $\bf B$ be a finite idempotent algebra. Then $\HSP({\bf B})$
contains an algebra without essential operations if and only if
 $\HS({\bf B})$ does.
\end{theorem}

Since all algebras in $\HS({\bf B})$ are smaller than $\bf B$ 
(or isomorphic to $\bf B$), this leads to another algorithm
that decides whether a given structure $\bB$ satisfies
the equivalent conditions in Theorem~\ref{thm:taylor}
(besides the approach via searching for Siggers terms mentioned above).
% mentioned after Theorem~\ref{thm:siggers}).

%This theorem is certainly false for polymorphism algebras $\bf B$
%of $\omega$-categorical structures, even when we assume that
%the structure is a model-complete core. BUT THIS IS NOT FAIR
% since here we make essential use of the idempotency assumption.
% the right generalization would be: 
% 1 \in \HSP(B) iff 1 \in SP(B) for the expansion of B by finitely many
% constants!

\subsection{Canonical Clones}
\label{ssect:canonical}
We cannot offer an full analog of Theorem~\ref{thm:taylor} for $\omega$-categorical
structures $\bB$. This section treats the special case
where the polymorphism algebra of an $\omega$-categorical structure $\bB$ 
resembles a finite algebra in a certain formal sense; in this case, an analog of Theorem~\ref{thm:taylor} can be transferred from the finite.

Let $\bB$ be a structure. Then $f \colon B^k \rightarrow B$
 is called \emph{$m$-canonical} (with respect to $\bB$)
if for all $m$-tuples $t_1,\dots,t_k$, the $m$-type of $f(t_1,\dots,t_k)$
in $\bB$ only depends on the $m$-types of $t_1,\dots,t_k$ in $\bB$. It is called \emph{canonical} if it is $m$-canonical for all finite $m$.
A clone (or an algebra) is called \emph{canonical} if all its operations are canonical (this is still with
respect to some base structure $\bB$). 

\begin{lemma}\label{lem:type-algebra}
Let $\bB$ be a structure with a finite number $q$ of
$m$-types, and let $\fB$ be an algebra with signature $\tau$ such that all
operations of $\fB$ are $m$-canonical with respect to $\bB$. 
Then there exists a $\tau$-algebra $\bf A$ of size $q$ and a surjective
homomorphism $\mu$ from ${\bf B}^m$ to $\bf A$.
\end{lemma}
\begin{proof}
Let $p_1,\dots,p_q$ be the $m$-types of $\bB$.
Define $\mu \colon B^m \rightarrow \{1,\dots,q\}$ by $g(b_1,\dots,b_m)=i$ if 
$(b_1,\dots,b_m)$ has type $p_i$. Since the operations of $\fB$ are $m$-canonical, 
the kernel of $\mu$ is a congruence $K$ of ${\bf B}^m$.
Then ${\bf A} := {\bf B}^m/K$ satisfies the requirements of the statement.
\end{proof}

The algebra $\bf A$ constructed from $\bf B$ in the proof of the previous lemma
will be called the \emph{type algebra} of $\bf B$, denoted by $T_m({\bf B})$.

\begin{lemma}\label{lem:type-idempotent}
Let $\bB$ be an $\omega$-categorical model-complete core, and suppose that all operations of the polymorphism algebra $\bf B$ of $\bB$ are $m$-canonical with respect to $\bB$.  Then $T_m(\bf B)$ is idempotent.
\end{lemma}
\begin{proof}
When $\bB$ is an $\omega$-categorical model-complete core, then
all orbits of $m$-tuples in $\bB$ are preserved by the endomorphisms of $\bB$.
It follows that every operation $f$ of $T_m(\bf B)$
satisfies $\forall x. \, f(x,\dots,x)=x$.
\end{proof}

Note that if $\bB$ is homogeneous in a relational
signature with maximal arity $m$ (or first-order interdefinable
with such a structure), then being $m$-canonical 
implies being $n$-canonical for all $n \geq m$.
In this case, we simply write $T(\bf B)$ instead of $T_m(\bf B)$.

We say that an operation $f \colon B^n \to B$ is \emph{cyclic modulo $e_1,e_2 \colon B \to B$} if  
$$ \forall x_1,\dots,x_n. \; e_1(f(x_1,\dots,x_n)) = e_2(f(x_2,\dots,x_n,x_1)) \; .$$
Similarly, we say that $f \colon B^n \to B$ is
a \emph{weak near unanimity modulo $e_1,\dots,e_n \colon B\to B$}
if the following is satisfied. 
$$ \forall \bar x. \; e_1(f(x,\dots,x,y)) = e_2(f(x,\dots,y,x)) = \cdots = e_n(f(y,x,\dots,x))$$
The same definition can be made for any type of equation, and we therefore also define Taylor operations modulo unary operations and Siggers polymorphisms modulo unary operations analogously.

The idea of the following lemma comes from the proof of Proposition~6.6 in Bodirsky, Pinsker, and Pongracz~\cite{BPP-projective-homomorphisms}, and has been
used in~\cite{Phylo-Complexity}. 
 
\begin{lemma}\label{lem:lift}
Let $\bB$ be $\omega$-categorical, and
$f \in \Pol(\bB)$. 
Suppose that for every finite $A \subset B$ there exists an
$\alpha \in \Aut(\bB)$ such 
that $f(x_1,\dots,x_n) = \alpha f(x_2,\dots,x_n,x_1)$ for all $x_1,\dots,x_n \in A$. 
Then $\bB$ has a cyclic polymorphism modulo 
endomorphisms. 
The analogous statement holds linear equations modulo unary operations in general. 
\end{lemma}
\begin{proof}
We show that there are $e_1,e_2 \in \overline{\Aut(\bB)}$
such that $e_1(f(x_1,\dots,x_n)) = e_2(f(x_2,\dots,x_n,x_1))$ for all $x_1,\dots,x_n$ from the domain $B$ of $\mathfrak B$.  
Construct a rooted tree as follows. 
Each vertex of the tree lies on some level $n \in \mathbb N$.
Let $d_1,d_2,\dots$ be an enumeration
of $B$. Let $F_n$ be the set of partial isomorphisms of
$\bB$ with domain $D_n := \{d_1,\dots,d_n\}$, and define
the equivalence relation $\sim$ on $F_n^2$ as follows:
$(\alpha_1,\alpha_2) \sim (\beta_1,\beta_2)$
if there exists a $\delta \in \Aut(\bB)$
such that $\alpha_i = \delta \circ \beta_i$ for $i \in \{1,2\}$.
Note that for each $n$, the relation $\sim$ has finitely
many equivalence classes on $F_n^2$, by the $\omega$-categoricity of $\bB$ and Theorem~\ref{thm:ryll}. 
The vertices of the tree on level $n$ are precisely 
the equivalence classes $E$ of $\sim$ on $F_n^2$ 
such that for all $(\alpha_1,\alpha_2) \in E$ and $x_1,\dots,x_n \in B$
satisfying $\{f(x_1,\dots,x_n),f(x_2,\dots,x_n,x_1)\} \subseteq D_n := \{d_1,\dots,d_n\}$
we have $\alpha_1(f(x_1,\dots,x_n))=\alpha_2(f(x_2,\dots,x_n,x_1))$.  

The equivalence class of the partial map with the empty domain $D_0$ becomes the root of the tree, on level $n=0$.  
We define adjacency in the tree by restriction as follows: 
when $E$ is a vertex on level $n$, and $E'$ a vertex on level $n+1$,
and $E$ contains $(\alpha_1,\alpha_2)$ and $E'$ contains
$(\alpha'_1,\alpha'_2)$ such that $\alpha_1 = \alpha'_1\upharpoonright_{D_n}$
and $\alpha_2 = \alpha'_2\upharpoonright_{D_n}$, then we make $E$ and $E'$
adjacent in the tree. 
Note that 
the resulting rooted tree is finitely branching. 
By assumption, the tree has vertices on all levels. 
Hence, by K\"onig's tree lemma, there exists an infinite path $E_0,E_1,E_2,\dots$ in the tree, where $E_i$ is from level $i \in \mathbb N$. 

We define $e_1,e_2 \in \overline{\Aut(\bB)}$ as follows. 
Suppose $e_1,e_2$ are already defined on $D_n$ 
such that $\alpha_1 := e_1\upharpoonright_{D_n}$, $\alpha_2 := e_2\upharpoonright_{D_n}$, and 
$(\alpha_1,\alpha_2) \in E_n$. 
We want to define $e_1$ and $e_2$ on $d_{n+1}$, and we will do it in such a way that 
$(e_1\upharpoonright_{D_{n+1}},e_2\upharpoonright_{D_{n+1}}) \in E_{n+1}$. 
Since $E_n$ and $E_{n+1}$ are adjacent, there exist $(\beta_1,\beta_2) \in E_n$ and $(\beta'_1,\beta_2') \in E_{n+1}$ such that $\beta_1 = \beta_1' \upharpoonright_{D_n}$
and $\beta_2 = \beta_2'\upharpoonright_{D_{n}}$. 
By the definition of $\sim$ there
exists a $\delta \in \Aut(\bB)$ such that 
$\alpha_1  =  \delta \circ \beta_1$ and $\alpha_2 =  \delta \circ \beta_2$. 
For $j \in \{1,2\}$, define $\alpha'_j := \delta \circ \beta_j'$ 
so that $(\alpha'_1,\alpha'_2) \in E_{n+1}$
and observe that  
\begin{align*}
\alpha'_j \upharpoonright_{D_n} := & \; (\delta \circ \beta_j') \upharpoonright_{D_n}  =  \delta \circ \beta_j = \alpha_j \; ,
\end{align*}
and hence that $\alpha'_j$ extends $\alpha_j$. 
Define $e_j(d_{n+1}) := \alpha'_j(d_{n+1})$. 
The proof for general linear equations modulo unary operations is analogous. 
\end{proof}

\begin{corollary}\label{cor:type-algebra-lift}
Let $\bC$ be a structure which is homogeneous in a finite relational language, and let $\bB$ be a model-complete core which is first-order interdefinable with $\bC$ such that all the operations of 
the polymorphism algebra $\fB$ of $\bB$ are canonical with respect to $\bB$. Then the following are equivalent:
\begin{enumerate}
\item The type algebra $T(\fB)$ contains a cyclic operation;
\item $\bB$ has a cyclic polymorphism modulo endomorphisms;
\item $\bB$ has a weak near unanimity polymorphism modulo endomorphisms;
\item $\bB$ has a Siggers polymorphism modulo endomorphisms.
\end{enumerate}
\end{corollary}
\begin{proof}
Let $m$ be the maximal arity of $\tau$.
By Lemma~\ref{lem:type-idempotent} the finite algebra $T({\bf B})$ is idempotent. 
Hence, by Theorem~\ref{thm:siggers}
it has a cyclic operation if and only if 
it has a weak near unanimity operation if and only if
it has a Siggers operation. 

Suppose now that 
$\bf B$ has a cyclic polymorphism $g^{\bf B}$ 
modulo endomorphisms $e_1^{\bf B}$ and $e_2^{\bf B}$. Let $\mu$ be the homomorphism from $\bf B$ to $T(\bf B)$. 
, $e_1^{T(\bf B)} = e_2^{T(\bf B)}$ must be the identity, and $g^{T(\bf B)}$ is a cyclic operation. Hence, 2 implies 1. Analogously one can show that 3 implies 1 and that 4 implies 1. 

Conversely, suppose that $T({\bf B})$ has an $n$-ary cyclic operation $f^{T({\bf B})}$. 
We claim that
$f^{\bf B}$ satisfies the assumptions of Lemma~\ref{lem:lift}. 
To show this,
it suffices to prove that for all finite $m$ and all $b^1,\dots,b^m \in B^n$, the $m$-tuples $(f^\fB(b^1),\dots,f^\fB(b^m))$ and 
$(f^\fB(b^1_2,\dots,b^1_{n},b^1_{1})$,
$\dots,f^\fB(b^m_{2},\dots,b^m_{n},b^m_{1}))$ lie in the same orbit of $\bC$.
Indeed, we have
\begin{align*}
\tp^\bC(f^\fB(b^1),\dots,f^\fB(b^m)) \; = & \; f^{T(\fB)}(\tp(b^1_1,\dots,b^m_1),\dots,\tp(b^1_n,\dots,b^m_n)) \\
= & \; f^{T(\fB)}(\tp(b^1_{2},\dots,b^m_{2}),\dots,\tp(b^1_{n},\dots,b^m_{n}),\tp(b^1_{1},\dots,b^m_{1})) \\
= & \; \tp(f^\fB(b^1_{2},\dots,b^1_n,b^1_{1}),\dots,f^\fB(b^m_{2},\dots,b^m_{n},b^m_{1}))
\end{align*}
Lemma~\ref{lem:lift} shows the existence of a
cyclic polymorphism modulo endomorphisms of $\bB$.
Analogously one can show the existence of
weak near unanimity polymorphisms and Siggers polymorphisms modulo endomorphisms. 
\end{proof}

Another important consequence of 
Lemma~\ref{lem:lift} is that the existence
of cyclic polymorphisms modulo
 endomorphisms of an $\omega$-categorical 
 model-complete core
 is inherited by expansions by constants. I am thankful to Trung Van Pham for communicating
 the proof of the following proposition to me
 which strengthens a weaker statement in earlier
 versions of this text. 

\begin{proposition}\label{prop:eq-const-pres}
Let $\bB$ be an $\omega$-categorical model-complete core.
If $\bB$ has an $n$-ary cyclic polymorphism modulo endomorphisms, 
then the expansion of $\bB$ by finitely many constants also has an $n$-ary cyclic polymorphism modulo endomorphisms. Analogous statements hold for weak near unanimity operations modulo unary operations,
and other operations satisfying linear equations modulo unary operations. 
\end{proposition}

\begin{proof}
By the assumption there exist $f\in
\Pol(\bB)$ of arity $k$ and $e_1,e_2\in \End(\bB)$ such that $e_1(f(x_1,x_2,\dots,x_k)) = e_2(f(x_2,x_3,\dots,x_k,x_1))$ 
for all $x_1,\dots,x_n$ from the domain $B$ of $\mathfrak B$. Let
$\hat f \colon B \to B$ be given by 
$\hat f(x):=f(x,x,\dots,x)$ for all $x\in B$.
Clearly, $\hat f$ is an endomorphism of $\bB$.
Let $a =(a_1,\dots,a_n) \in B^n$ be an arbitrary tuple of $n$ constants. Then $a$, $\hat f(a)$, and $e_1(\hat f(a)) = e_2(\hat f(a))$ lie in the same orbit of 
$\Aut(\bB)$
because $\bB$ is a model-complete core. 
Let $\alpha,\beta \in \Aut(\bB)$ be
such that $\alpha e_1(\hat f(a))=a$
and $\beta(\hat f(a)) = a$. 
Let $h_1 := \alpha e_1 \beta^{-1}$ and 
$h_2 := \alpha e_2 \beta^{-1}$,
and $g := \beta f$. 
Clearly,
we have $g(a,\dots,a)=\beta \hat f(a) = a$ by the choice of $\beta$. We will show that
$h_1(a)=a$ and $h_2(a)=a$. We have
$$h_1(a)=h_1(g(a,\dots,a))=\alpha e_1 \beta^{-1}(\beta f(a,\dots,a)) = \alpha e_1(\hat f(a)) = a \, .$$
Similarly one can show that $h_2(a)=a$.
It follows that $h_1,h_2 \in \End(\bB,a_1,a_2,\dots,a_n)$
and that $g \in \Pol(\bB,a_1,a_2,\dots,a_n)$. Moreover, for all
$x_1,x_2,\dots,x_k \in B$ we have that 
\begin{align*}
h_1(g(x_1,x_2,\dots,x_k))= & \; \alpha e_1 \beta^{-1} (\beta f(x_1,x_2,\dots,x_k)) \\
= & \; \alpha e_1(f(x_1,\dots,x_k)) \\
= & \; \alpha e_2(f(x_2,\dots,x_k,x_1)) \\
= & \; \alpha e_2 \beta^{-1}( \beta f(x_2,\dots,x_k,x_1)) \\
= & \; h_2(g(x_2,x_3,\dots,x_k,x_1)) \; .
\end{align*}
This shows that $(\bB,a_1,\dots,a_n)$ has a
cyclic polymorphism modulo endomorphisms. 
The proof for other linear equations modulo endomorphisms is analogous. 
\end{proof}

How strong is the assumption 
that every operation of an oligomorphic clone is canonical
with respect to some homogeneous structure $\mathfrak C$ 
over a finite relational signature? %We do not have a fully satisfying answer to this question.
In Chapter~\ref{chap:ramsey}, we see that under fairly general Ramsey-theoretic assumptions on $\mathfrak C$ we can find canonical operations in a natural way. 
%In the next section we see that there is a more abstract approach to show 
%the existence of operations that satisfy the Taylor term identities modulo the application of an endomorphism. 
Indeed, it turns out that when $\bB$ 
is first-order definable over the random graph $({\mathbb V};E)$
and does not primitively positively interpret $(\{0,1\};\OIT)$,
then $\bB$ is the reduct of a structure where all polymorphisms are canonical, and which still does not primitively positively interpret $(\{0,1\};\OIT)$ (Theorem~\ref{thm:minimalTractableClones}). The same statement
 is not true when we replace the random graph 
 by $({\mathbb Q};<)$. However,
we make the following general conjecture.

\begin{conjecture}\label{conj:ua-dichotomy}
Let $\bB$ be a countable $\omega$-categorical model-complete core. Then either $\bB$ interprets all finite structures primitively positively with parameters, or $\bB$ has a $k$-ary weak near unanimity operation modulo endomorphisms. 
\end{conjecture}

This conjecture has been confirmed for all 
$\bB$ that have a first-order definition over $({\mathbb Q};<)$, and for all $\bB$ that are definable over the random graph. 

Let us remark that in order to `prevent' primitive positive interpretations of all finite structures, it suffices to have a Taylor term $f$ modulo unary operations applied to the arguments of $f$ \emph{and} to the function value of $f$, in the following sense. 

\begin{proposition}\label{prop:deep-taylor-mod-endo-tame}
Let $\bB$ be an $\omega$-categorical structure whose polymorphism algebra $\fB$ contains an $n$-ary $f$ and
unary $a_0,\dots,a_n,b_0,\dots,b_n$
such that for all $i \leq n$
there are $x_1,\dots,x_n$, $y_1,\dots,y_n \in \{x,y\}$ such that $\fB$ satisfies
\begin{align}
\forall \bar x. \, & a_0(f(a_1(x_1),\dots,a_{i-1}(x_{i-1}),a_i(x), a_{i+1}(x_{i+1}),\dots,a_n(x_n)))  \nonumber \\
= \; & b_0(f(b_1(y_1),\dots,b_{i-1}(y_{i-1}),b_i(y),b_{i+1}(y_{i+1}),\dots,b_n(y_n))) \; . \label{eq:taylor-mod-endo}
\end{align}
Then there is no primitive positive interpretation of $(\{0,1\};\OIT)$ in $\bB$. 
\end{proposition}
\begin{proof}
Suppose that $(\{0,1\};\OIT)$ had a primitive positive interpretation
in $\bB$. Then by Theorem~\ref{thm:simulates-tame} there is a 2-element algebra
$\fA$ in the pseudo-variety generated by $\fB$ all of whose operations are projections.
The algebra $\fA$ is in particular in the variety generated by $\fB$, and
satisfies (\ref{eq:taylor-mod-endo}). Since the unary function symbols of $\fB$ must denote the identity in $\fA$, and $\fA$ has more than one element, the $i$-th equation prevents that $f$ is the $i$-th projection. Because we have such
an equation for each argument of $f$, the operation $f$ cannot be a projection.
\end{proof}

In particular, if $\bB$ is a structure with a weak near unanimity polymorphism modulo endomorphisms, then
$(\{0,1\};\OIT)$ is not primitively positively interpretable in $\bB$. Using Proposition~\ref{prop:eq-const-pres},
we even get the following. 

\begin{corollary}\label{cor:ua-dich-disj}
The two cases in Conjecture~\ref{conj:ua-dichotomy}
are indeed disjoint: every $\omega$-categorical
model-complete core with a weak near unanimity 
polymorphism modulo endomorphisms does
not interpret $(\{0,1\};\OIT)$ primitively positively with parameters. 
\end{corollary}

\chapter{Equality Constraint Satisfaction Problems}
\label{chap:ecsp}

%\begin{figure}[h]
\begin{center}
\includegraphics{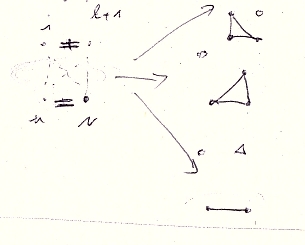} Jan K\'ara, 2005
\end{center}
\vspace{1cm}
%\caption{Drawing by Kan K\'ara, 2005}
%\end{figure}

This section is about structures with a first-order definition in $({\mathbb N}; =)$; such structures will be called \emph{equality constraint languages}. From a model-theoretic perspective, 
equality constraint languages appear to be trivial altogether. However,
the set of all such structures, 
taken up to primitive positive interdefinability and ordered by
inclusion, is a quite complicated object (there are actually $2^\omega$
many equality constraint languages that pairwise do not define each other primitively positively~\cite{BodChenPinsker}). 

By the results of Section~\ref{sect:galois}, % in Chapter~\ref{chap:mt},
a structure $\bB$ is isomorphic to an equality constraint language if and only if
$\bB$ is preserved by all permutations of its domain.
Therefore, this chapter is about locally closed clones that contain
all permutations of the domain. 
On a \emph{finite} domain, such clones have been completely described 
 in \cite{HaddadRosenberg}; it turns out that the number of clones that contain all permutations of a fixed finite domain is finite. 
Clones on infinite sets that contain all permutations 
are of independent interest in universal algebra~\cite{Hei02,Pin05MaxAbovePerm,MP06MinimalAboveS,Pin05NumberOfUnary}.
Local closure is a strong additional assumption, which allows a good understanding of the lattice of all locally closed
clones that contain all permutations~\cite{BodChenPinsker}.

%Thus, in some sense equality constraint languages are those structures that have the largest possible degree of symmetry. 
%When we consider the study of $\omega
%By the theorem of Ryll-Nardzewski (Theorem~\ref{thm:ryll}), $\omega$-categoricity
%itself can be seen as a condition enforcing a high degree of symmetry, and so equality constraint languages appear to be a quite natural class of structures to be looked at. 
%\ignore{Results about primitive positive definability in equality templates can be translated into questions about local generation for clones that
%contain all the permutations of their domain, according to the results in Chapter~\ref{chap:algebra}. It turns out that these questions are
%of a quite fundamental nature for locally closed clones on
%infinite sets; this has been investigated in depth 
%in~\cite{BodChenPinsker}. Since we concentrate on results that are relevant
%for constraint satisfaction, we will here only need a tiny bit of those
%results on locally closed clones containing all the permutations.}

The CSP for a finite equality constraint language is called an \emph{equality constraint satisfaction problem}. Equality CSPs are of fundamental importance in infinite domain constraint satisfaction;
we mention some reasons.
\begin{itemize}
\item NP-hard equality relations are very good candidates for establishing hardness results of other infinite domain CSPs. 
For instance, it follows from the results presented in this section that 
every structure which admits a primitive positive definition of the relation $\{(x,y,z) \; | \; (x=y \neq z) \vee (x \neq y = z)\}$ has an NP-hard CSP.
\item When analyzing an $\omega$-categorical structure $\bB$ with the universal-algebraic approach, 
the question which equality constraint languages can be defined in $\bB$ is of crucial importance, as we will see for instance in
Chapter~\ref{chap:schaefer} and~\ref{chap:tcsp}. 
%For example, if the relation $\{ (x,y,u,v) \; | \; x=y \Rightarrow u=v\}$
For example, if the relation $\{ (x,y,u,v) \; | \; x=y \Leftrightarrow u=v\}$
is primitive positive definable in $\bB$, then every polymorphism of $\bB$ that depends on all its arguments must be injective (Proposition~\ref{prop:all-inj}).
\item Suppose we want to classify the computational complexity of $\Csp(\bB)$ when $\bB$ has a first-order definition in a fixed infinite structure $\bC$; examples of such classifications will be given in Chapter~\ref{chap:schaefer} and~\ref{chap:tcsp}.
%, and more examples are discussed in Chapter~\ref{chap:concl}). 
Then such a classification includes the classification of equality constraint satisfaction problems. 
\end{itemize}

The complexity of equality CSPs has been completely classified~\cite{ecsps}; those problems are in P or NP-hard. 
In this chapter we present a new proof of this result, and show that either the template $\bB$ has a binary injective polymorphism,  
in which case $\bB$ has a quantifier-free Horn definition in $({\mathbb N};=)$, and $\Csp(\bB)$ is in P, 
or $(\{0,1\}; \NAE)$ has a primitive positive interpretation in $\bB$, and $\Csp(\bB)$ is NP-complete.
The new proof has the advantage that it
divides the argument into several steps that each holds
 for a much larger class of structures. Indeed, several results of this chapter turn out to be useful in later
classification arguments, 
in particular in Chapter~\ref{chap:schaefer}.

We would also like to mention that the fact that satisfiability of quantifier-free Horn clauses over $({\mathbb N};=)$ can be decided in polynomial time has already been observed in~\cite{KanellakisKuperRevesz}.
Here, we derive the algorithm from more general principles that will also be important for our algorithmic results in 
Chapter~\ref{chap:schaefer} and Chapter~\ref{chap:tcsp}.

\section{Independence of Disequality}
\label{sect:independence}
The importance of \emph{independence} in constraint satisfaction has been recognized several times;
the first appearance of this concept in the literature
seems to be in~\cite{LassezMcAloon-Indep}, and, 
subsequently, in~\cite{LassezMcAloon-LICS}. 
In this thesis, we focus on \emph{independence of disequality}, which found most applications;
the general definition of independence has been worked out in~\cite{Disj}. Applications of this concept have been studied 
in metric temporal reasoning~\cite{Koubarakis,JonssonBaeckstroem} 
and qualitative reasoning cacluli~\cite{BroxvallJonsson,BroxvallJonssonRenz}; also see~\cite{HornOrFull}.

\begin{definition}[Independence of Disequality]
Let $\bB$ be a structure with relational signature $\tau$. 
Then we say that \emph{$\neq$ is independent from $\bB$} if for all primitive positive $\tau$-formula $\phi$, 
if both $\phi \wedge x \neq y$ and $\phi \wedge u \neq v$
are satisfiable over $\bB$, then $\phi \wedge x \neq y \wedge u \neq v$ is satisfiable over $\bB$ as well.
\end{definition}

In this section we prove that for many $\omega$-categorical structures 
independence of disequality is equivalent to the existence of a binary injective polymorphism.
The following definition comes from~\cite{ecsps}.

\begin{definition}
A relation $R\subseteq B^k$ is called \emph{intersection-closed}
if for all $k$-tuples $(u_1,\ldots,u_k), (v_1,\ldots,v_k) \in R$
there is a tuple $(w_1,\ldots,w_k) \in R$
such that for all $1 \leq i,j \leq k$ we have
$w_i \neq w_j$ whenever $u_i \neq u_j$ or $v_i \neq v_j$.
\end{definition}

\begin{lemma}\label{lem:intersect}
Let $\bB$ be an $\omega$-categorical structure 
where $\neq$ has an 
primitive 
positive definition.
Then the following are equivalent.
\begin{enumerate}
\item Disequality is independent from $\bB$.
\item Every finite induced substructure of $\bB^2$ admits an
injective homomorphism into $\bB$.
\item $\bB$ has a binary injective polymorphism.
\item All primitive positive definable relations in $\bB$ are intersection-closed.
\end{enumerate}
\end{lemma}

\begin{proof}
Throughout the proof,
let  $b_1,b_2, \dots$ be an enumeration
of the domain $B$ of $\bB$.
If $f$ is a binary injective polymorphism of $\bB$,
then clearly every relation in $\bB$ is intersection-closed,
so (3) implies (4).
The implication from (4) to (1) is straightforward as well.

We now show the implication from (1) to (2). Let $\bA$ be a finite induced
substructure of $\bB^2$. Then the domain of $\bA$ is contained in 
$\{b_1,\dots,b_n\}^2$, for sufficiently large $n$. It clearly suffices to show
that the structure induced by $\{b_1,\dots,b_n\}^2$ in $\bB^2$ homomorphically and injectively maps
to $\bB$, so let us assume without loss of generality that the domain of $\bA$ is $\{b_1,\dots,b_n\}^2$.

Consider the formula $\phi$ whose variables $x_1,\dots,x_{n^2}$ are the elements of $\bA$,
$$x_1 := (b_1,b_1),\dots, x_n := (b_1,b_n), \dots, x_{n^2-n+1} := (b_n,b_1) ,\dots, x_{n^2} :=(b_n,b_n) \, ,$$
and which is the conjunction over all
literals $R((b_{i_1},b_{j_1}),\dots,(b_{i_k},b_{j_k}))$ such that
$R(b_{i_1},\dots,b_{i_k})$ and $R(b_{j_1},\dots,b_{j_k})$ hold in $\bB$. So $\phi$ states precisely which relations hold in $\bA$.

Using induction
over the number $m$ of disequalities, we will now show that for any
conjunction $\sigma:=\bigwedge_{1\leq k\leq m} x_{i_k}\neq
x_{j_k}$ with the property that $i_k\neq j_k$ for all $1\leq k\leq
m$, the formula $\phi\wedge \sigma$ is satisfiable over $\bB$. This implies that there exists an $n^2$-tuple $t$ in $\bB$ with pairwise distinct entries
which satisfies $\phi$; the assignment that sends every $x_i$ to $t_i$ is an injective homomorphism from $\bA$ into $\bB$.

For the induction beginning, let $x_i\neq x_j$ be any disequality. Let $r,s$ be the $n^2$-tuples defined as follows.
\begin{align*}
r & := (b_1,\dots,b_1,b_2,\dots,b_2,\dots,b_n,\dots,b_n) \\
s & := (b_1,b_2,\dots,b_n,b_1,b_2,\dots,b_n,\dots,b_1,b_2,\dots,b_n).
\end{align*}
These two tuples satisfy $\phi$, because the projections to
the first and second coordinate, respectively, are homomorphisms
from $\bA$ to $\bB$. Now either $r$ or $s$ satisfies $x_i\neq x_j$, proving that $\phi\wedge x_i\neq x_j$ is satisfiable in $\bB$.

In the induction step, let a conjunction $\sigma:=\bigwedge_{k \in \{1,\dots,m\}} x_{i_k}\neq
x_{j_k}$ be given, where $m\geq 2$.
%and suppose by induction that 
%$\phi \wedge \bigwedge_{k \in \{2,\dots,m\}} x_{i_k}\neq x_{j_k}$ has a solution $s_1 \colon \{x_1,\dots,x_{n^2}\} \to B$ over $\bB$,
%and $\phi \wedge \bigwedge_{k \in \{1,3,\dots,m\}} x_{i_k}\neq x_{j_k}$ has a solution $s_2 \colon \{x_1,\dots,x_{n^2}\} \to B$ over $\bB$. 
 Set $\sigma':=\bigwedge_{3\leq k\leq m} x_{i_k}\neq
x_{j_k}$, and $\phi':=\phi\wedge \sigma'$. Observe that $\phi'$ has a primitive positive definition in $\bB$, as $\phi$ and $\neq$ have such definitions. By induction hypothesis, both $\phi'\wedge x_{i_1}\neq x_{j_1}$ and $\phi'\wedge x_{i_2}\neq x_{j_2}$ are satisfiable in $\bB$. But then $\phi'\wedge x_{i_1}\neq x_{j_1}\wedge x_{i_2}\neq x_{j_2}$,
which is equivalent to $\phi\wedge\sigma$, is satisfiable over $\bB$ as well by (1), concluding the proof.

The implication from (2) to (3) follows from Lemma~\ref{lem:omega-cat-compactness},
because the property that a function is injective can be described by the universal first-order
sentence $\forall \bar x, \bar y \big (\bar x \neq \bar y \Rightarrow f(\bar x) \neq f(\bar y) \big )$.
\end{proof}

Also the situation that the polymorphisms $f$ of an $\omega$-categorical structure are `essentially injective' 
can be characterized using equality relations. 

\begin{proposition}\label{prop:all-inj}
Let $f$ be an operation from $B^k$ to $B$ that depends on all arguments. Then the
following is equivalent.
\begin{enumerate}
\item $f$ is injective.
\item $f$ preserves the relation defined by $x=y \Leftrightarrow u=v$.
\item $f$ preserves the relation defined by $x=y \Rightarrow u=v$.
\end{enumerate}
\end{proposition}
\begin{proof}
For the implication from $(1)$ to $(2)$, suppose that $f$ is injective. 
We check that $f$ preserves  $x=y \Leftrightarrow u=v$.
Let $a,b,c,d$ be elements of $B^k$ such that $a_i = b_i \Leftrightarrow c_i = d_i$ for all $i \leq k$, and let $t$ be the tuple $(f(a),f(b),f(c),f(d))$. 
If $a = b$, we thus have that $c_i=d_i$ for all $i \leq k$, and so $c = d$.
In this case, $t$ satisfies $t_1=t_2$ and $t_3=t_4$, and we are done.
Similarly, if $c = d$ then $a = b$ and we are done.
Otherwise, $a \neq b$ and $c \neq d$, and by injectivity of $f$
we have $t_1 \neq t_2$ and $t_3 \neq t_4$. So we have in all cases that $t_1 = t_2$ 
if and only if $t_3 = t_4$. 

For the implication from (2) to (3), note that $x=y \Rightarrow u=v$ is
equivalent to a primitive positive formula over $(B; R)$ where $R = \{(a,b,c,d) \; | \; a=b \Leftrightarrow c=d \})$. The primitive positive formula is
$$ \exists w \, (R(x,y,u,w) \wedge R(x,y,w,v)) \; .$$

Finally, for the implication from $(3)$ to $(1)$, 
suppose that there are distinct $a,b \in B^k$ such that $f(a)=f(b)$. 
We want to prove that $f$ violates $x=y \Rightarrow u=v$.
Let $I$ be the set of all $i \in \{1,\dots,k\}$ such that $a_i \neq b_i$.
Since $a$ and $b$ are distinct, $I$ is non-empty; let $i \in I$ be arbitrary.
Since $f$ depends on the $i$-th argument, there are $c,d \in D^k$
with $c_j = d_j$ for all $j \neq i$, and $c_i \neq d_i$. 
We claim that $(a, b, c, d)$ shows that $f$ violates $x=y \Rightarrow u=v$. 
First, note that for all $j \in \{1,\dots,k\} \setminus I$, we have that $a_j=b_j$ and $c_j = d_j$. 
Next, note that for all $j \in I$ we have that $a_j \neq b_j$. We conclude that for all $j \in \{1,\dots,k\}$ we have that $a_i = b_i$ implies $c_i=d_i$.
However, $f(a)=f(b)$ and $f(c) \neq f(d)$.
\end{proof}

We close with an application to CSPs. 
When $\bA$ is an instance of $\Csp(\bB)$, then an injective homomorphism from $\bA$ to $\bB$
is also called an \emph{injective solution} for $\bA$. 
%It will in the following be
%more convenient to view instances $\bA$ 
%of $\Csp(\bB)$ as primitive positive sentences $\phi$ (see Section~\ref{sect:csp-logical}); we also use the term \emph{`surjective solution'} for this perspective on the CSP. 

\begin{proposition}\label{prop:injective-sols}
Suppose that $\bB$ has a binary injective polymorphism $h$. 
Then every satisfiable instance $\bA$ of $\Csp(\bB)$ either has an injective solution,
or $\bA$ has two distinct elements $a,a'$ such that $s(a)=s(a')$ in
all solutions $s$ for $\bA$.
\end{proposition}
\begin{proof}
Suppose that $\bA$ has a solution, but no injective solution. 
Let $f$ be a solution such that the cardinality of $f$ is maximal.
Since there is no injective solution, there are two elements $a$, $a'$ of $\bA$
such that $f(a)=f(a')$. We claim that $s(a)=s(a')$ in \emph{all} solutions $s$ of $\bA$. 
Otherwise, if $s(a) \neq s(a')$ for some solution $s$, 
then by the choice of $f$ there must be another
pair $b,b'$ such that $s(b) \neq s(b')$ but $f(b) \neq f(b')$. 
Then the mapping $x \mapsto h(f(x),s(x))$ is also a solution to $\bA$, but
has a strictly larger image than $f$, a contradiction.
\end{proof}

\section{Two-transitive Templates}
We show that two-transitive structures
(in particular, equality constraint languages) with essential but without constant polymorphisms also have
binary injective polymorphisms.
Here we use Corollary~\ref{cor:binary} about existence of binary essential polymorphisms, 
and Lemma~\ref{lem:intersect} about the existence of binary injective polymorphisms.

\begin{theorem}\label{thm:2trans}
Let $\bB$ be a two-transitive structure without a constant polymorphism, but with an essential
polymorphism.
Then $\bB$ has a binary injective polymorphism. 
\end{theorem}
\begin{proof}
Corollary~\ref{cor:binary}
implies that $\bB$ is preserved by a binary essential operation $f$. Since $\bB$ has
no constant polymorphism and is 2-set transitive, Corollary~\ref{cor:neq-constant} implies that all 
polymorphisms of $\bB$ preserve $\neq$, and
hence $\neq$ is primitive positive definable in $\bB$. So we can apply Lemma~\ref{lem:intersect}, and have to show that
 for every primitive positive formula $\phi$ the formula
$\phi \wedge x \neq y \wedge u \neq v$ is satisfiable over $\bB$
whenever 
$\phi \wedge x \neq y$ and $\phi \wedge u \neq v$ are satisfiable over $\bB$.

Let $V$ be the variables of $\phi$, and let $s \colon V \rightarrow B$ 
be a satisfying assignment for $\phi \wedge x \neq y$, 
and $t \colon V \rightarrow B$ 
be a satisfying assignment for $\phi \wedge u \neq v$.
We can assume that $s(u) = s(v)$ and $t(x) = t(y)$,
otherwise we are done. 
Let $k$ be the cardinality of the set 
$\{s(x),s(y),s(u),t(u),t(v),t(x)\}$; note that $4 \leq k \leq 6$.
Suppose that $k=6$, the other cases are simpler.
Since $f$ is essential, it violates the relation $P^B_3$ (Lemma~\ref{lem:unary}). 

Since $f$ preserves $\neq$, we can therefore assume that
there are tuples $(a,a,b)$ for $a \neq b$ and
$(c,d,d)$ for $c \neq d$ such 
that $f(a,c) \neq f(a,d)$ and $f(a,d) \neq f(b,d)$.
By 2-transitivity of $\Aut(\bB)$, there are $\alpha,\beta \in \Aut(\bB)$ such that
$\alpha(s(u),s(x)) = (a,b)$, and $\beta(t(u),t(x))=(c,d)$. Since $s(x) \neq s(y)$ and $t(v) \neq t(u)$,
and $f$ preserves $\neq$, we have $f(\alpha(s(x)),\beta(t(v))) \neq f(\alpha(s(y)),\beta(t(u)))$.
This implies that
$f(\alpha(s(x)),\beta(t(v))) = f(\alpha(s(y)),\beta(t(v)))$ and $f(\alpha(s(y)),\beta(t(v))) = f(\alpha(s(y)),\beta(t(u)))$
cannot both be true. 
By 2-transitivity of $\Aut(\bB)$, there
exist $\alpha',\beta'$ such that $\alpha'(a,\alpha(u)) = (\alpha(u),a)$, and $\beta'(d,\beta(v))= (\beta(v),d)$. 

If $f(\alpha(s(x)),\beta(t(v))) \neq f(\alpha(s(y)),\beta(t(v)))$, then 
$z \mapsto f(\alpha s(z),\beta' t(z))$
is a satisfying assignment for $\phi \wedge x \neq y \wedge u \neq v$.
If $f(\alpha(s(y)),\beta(t(v))) \neq f(\alpha(s(y)),\beta(t(u)))$, then $z \mapsto f(\alpha' s(z),\beta t(z))$
is a satisfying assignment for $\phi \wedge x \neq y \wedge u \neq v$.
\end{proof}

\section{Horn Formulas}\label{sect:horn}
In this section we show that $\bB$ has
certain binary injective polymorphisms if and only if 
all relations in $\bB$ have a quantifier-free Horn definition
over a `base' structure $\bC$; this will often be useful
to design algorithms for $\Csp(\bB)$.

The structure $({\mathbb N};=)$ has quantifier-elimination:
this follows from Lemma~\ref{lem:qe} by the observation that 
every bijection between finite subsets of ${\mathbb N}$ can
be extended to a permutation of ${\mathbb N}$. 
So we could have defined equality constraint languages
as those relational structures where all relations have a \emph{quantifier-free}
definition in $(\mathbb N;=)$. Note that not all equality constraint languages
have quantifier-elimination; however, all equality constraint
languages are model-complete. 

\begin{proposition}
Every equality constraint language $\bB$ is model-complete.
\end{proposition}
\begin{proof}
The permutations of $\mN$ locally generate all injective
self-maps on $\mN$. Hence, the statement 
 follows from Theorem~\ref{thm:mc-omegacat}
by the observation that the embeddings of $\bB$ are locally generated
by automorphisms of $\bB$.
\end{proof}

The following is a simple, but very useful definition to prove
syntactic results.

\begin{definition}[as in~\cite{BodChenPinsker}]\label{def:reduced}
A quantifier-free first-order formula $\phi$ in  
conjunctive normal form is called \emph{reduced} (over a structure $\bB$) if every formula 
obtained from $\phi$ by removing 
a literal or a clause is not equivalent to $\phi$ (over $\bB$).
\end{definition}

%THE FOLLOWING IS NOT TRUE
%a relation R has an existential horn definition in A iff
%R is preserved by all embeddings from $A^2$ into A and from A into A.
% Barny's Counterexample: (Q,\leq,0,1): the relation x=0 or x=1 is
% not Horn, and there is certainly no embedding.

Clearly, every quantifier-free formula is equivalent to 
a reduced formula over $\bB$, 
because we can find one by successively removing literals and clauses from $\phi$.
The following theorem is from~\cite{Maximal} and~\cite{BodHilsMartin-Journal} (stated there for quantifier-free Horn formulas only). 

\begin{theorem}\label{thm:horn}
Let $\bB$ be a structure with an embedding $e$ from $\bB^2$ into $\bB$. Then a relation $R$ with
a quantifier-free definition in $\bB$ has a quantifier-free Horn definition
in $\bB$ if and only if $R$ is preserved by $e$.
\end{theorem}
\begin{proof}
(Backwards.) Let $\delta$ be a quantifier-free 
Horn definition of 
$R$ over $\bB$, written in prenex conjunctive normal form. 
It suffices to demonstrate that $e$ preserves
each clause in $\delta$. Note that
a Horn clause $\psi$ of $\delta$ can always be written in the form $(\phi_1 \wedge \dots \wedge \phi_l) \rightarrow \phi_0$, for atomic $\tau$-formulas $\phi_0,\dots,\phi_l$. Let $V$ be the variables of $\psi$,
and let $s_1,s_2 \colon V \rightarrow \mN$ be two assignments that satisfy the clause.
We claim that $s_3 \colon V \rightarrow \mN$ 
defined by $s_3(x,y) = e(s_1(x),s_2(y))$ satisfies $\psi$. 
There are two cases cases to consider. 
Either there is an $i \leq l$ such that $s_1$ or $s_2$ does not satisfy $\phi_i$.
In this case, since $e$ is an embedding from $\bB^2$ to $\bB$, $s_3$ does not satisfy $\phi_i$, and therefore satisfies $\psi$. Or, if for all $i \leq l$ both $s_1$ and $s_2$ satisfy $\phi_i$, then they also satisfy $\phi_0$. Since $e$ is a polymorphism of $\bB$, it follows that $s_3$ satisfies $\phi_0$, and therefore also $\psi$. 

(Forwards.) Consider a quantifier-free definition $\delta$ of $R$ in $\bB$ such that $\delta$ is in prenex normal form, and that the quantifier-free part $\eta$ of $\delta$ is a 
reduced CNF formula over $\bB$. Assume for contradiction that 
$\delta$ is not Horn, that is, $\eta$ has a clause $\psi = \phi_1 \vee \phi_2 \vee \phi_3 \vee \cdots \vee \phi_l$ where $\phi_1,\phi_2$ are positive literals, and $\phi_3,\dots,\phi_l$ are positive or negative literals. Let $V$ be the variables of $\eta$.
Since $\eta$ is reduced, it has a satisfying assignment
$s_1 \colon V \rightarrow \mN$ such that
$\phi_i$ is false for all $i \leq l$ except for $i=1$;
otherwise, we could remove $\phi_i$ from $\psi$ and
would obtain a formula that is equivalent to $\eta$ over $\bB$,
contradicting the assumption that $\eta$ is reduced.
Similarly, $\eta$ has a satisfying assignment $s_2 \colon V \rightarrow \mN$ such that $\phi_i$ is false for all $i \leq l$ except for $i=2$.
Then $s_3 \colon V \rightarrow \mN$ defined by
$s_3(x,y) = e(s_1(x),s_2(y))$ does not satisfy $\psi$, a contradiction.
\end{proof}

The structure $\bB := ({\mathbb N};=)$ is an obvious example with an embedding from $\bB^2$ into $\bB$. 
When $\bB$ is a relational structure, then $\bB^\neg$ denotes
the expansion of $\bB$ by all relations that are the
complement of a relation from $\bB$. The following is from~\cite{Maximal} (note that we de not assume $\omega$-categoricity of $\bB$ and $\bC$).

\begin{theorem}\label{thm:resolution}
Let $\bC$ be a structure with 
an embedding $e$ from $\bC^2$ into $\bC$.
Let $\bB$ be a relational structure with finite signature $\sigma$
that is preserved by $e$ 
and has a quantifier-free definition in $\bC$. 
Then there is a polynomial-time Turing reduction from
$\Csp(\bB)$ to $\Csp(\bC^\neg)$.
\end{theorem}

\begin{figure*}[h]
\begin{center}
\small
\fbox{
\begin{tabular}{l}
// Input: An instance $\phi$ of $\Csp(\bB)$ \\
// Assumption: $\bB$ has a quantifier-free Horn definition in a $\tau$-structure $\bC$. \\
Replace each constraint $R(x_1,\dots,x_n)$ from $\phi$
by $\delta(x_1,\dots,x_n)$, \\ 
\hspace{.3cm} where $\delta$ is a quantifier-free
Horn definition of $R$ in $\bC$. \\
Let $\psi$ be the resulting $\tau$-sentence, written in prenex conjunctive normal form. \\
Repeat := true \\
While Repeat = true do \\
\hspace{.3cm} Repeat := false \\
\hspace{.3cm} Let $\Psi$ be the set of all singleton clauses in $\psi$. \\
\hspace{.3cm} If $\Psi$ is unsatisfiable over $\bC$ then reject. \\
\hspace{.3cm} For each negative literal $\eta$ of $\psi$ do \\
\hspace{.6cm} If $\Psi \cup \eta$, considered as an instance of $\Csp(\bC^\neg)$, is unsatisfiable \\
\hspace{.9cm} Remove $\eta$ from its clause in $\psi$ \\
\hspace{.9cm} Repeat := true \\
\hspace{.3cm} End for \\
Loop  \\
Accept
\end{tabular}}
\end{center}
\caption{A polynomial-time Turing reduction 
from $\Csp(\bB)$ to $\Csp(\bC^\neg)$ when $\bB$ has a polymorphism that is an embedding of $\bC^2$ into $\bC$.}
\label{fig:alg-Horn}
\end{figure*}

\begin{proof}
We use the algorithm shown in Figure~\ref{fig:alg-Horn}, which is due to~\cite{Disj}.
By Theorem~\ref{thm:horn}, every relation of $\bB$ has
a quantifier-free Horn definition in $\bC$. 
Let $\phi$ be an input instance of $\Csp(\bB)$, and let $\psi$ 
be the sentence in the language of $\bC$ obtained from $\phi$
as described in the algorithm.
Since $\sigma$ is finite and fixed,
and does not depend on the input, 
there is only a linear number of literals that can be deleted from $\psi$ in the course of the algorithm. 
It is thus clear that the algorithm works in polynomial time.

To show that the algorithm is correct, observe that
$\phi$ is false in $\bB$ if and only if $\psi$ is false in $\bC$.
We first show that if the algorithm rejects,
 then $\psi$ is false in $\bB$. 
The reason is that whenever a negative literal $\eta$
is removed from a clause of $\psi$, 
then in fact $\neg \eta$ is implied by the other clauses
in $\psi$, and therefore removing $\eta$ from $\psi$ leads
to an equivalent formula.

Finally, we show that if the algorithm accepts, then $\psi$
is true in $\bC$. Let $B$ be the domain of $\bB$ and $\bC$, and let $V$ be the set of variables of $\psi$. 
%, and $n = |V|$.
Consider the negative 
literals $\eta_1,\dots,\eta_m$ 
that are in clauses of $\psi$ at the final stage of the algorithm.
For all $i \leq m$, let $t_i \colon V \rightarrow B$ be an assignment
that satisfies all clauses of $\psi$ without negative literals, and which also satisfies $\eta_i$. Such an assignment 
must exist, since otherwise $\eta_i$ would have been
false in all solutions, and our algorithm would have
removed $\eta_i$ in the inner loop of the algorithm.
We claim that $s \colon V \rightarrow B$ given by 
$$s(x) = e(t_1(x),e(t_2(x),\dots e(t_{m-1}(x),t_m(x))\dots))$$
 satisfies all clauses of $\psi$.
Negative literals $\eta_k$ are satisfied because
$t_k$ satisfies $\eta_k$, and $e$ is an embedding of $\bC^2$ into $\bC$. 
Positive literals from $\psi$ are satisfied by $s$ because they are satisfied by all the $t_i$,
and since $e$ is a polymorphism of $\bC$. 
\end{proof}

\section{Classification}\label{sect:ecsp-class}
We now finish the complexity classification for $\Csp(\bB)$
where $\bB$ is an equality constraint language,
combining the results from the previous sections of this chapter. 

%Since $\bB$ is two-transitive, 
%the following is a direct consequence 
%of Theorem~\ref{thm:2trans}. 
%\begin{theorem}\label{thm:neq-indep}
%Let $\bB$ be an equality constraint language with an essential
%polymorphism, and suppose that 
%$\neq$ is primitive positive 
%definable in $\bB$.
%Then $\bB$ has a binary injective polymorphism. 
%\end{theorem}

\begin{corollary}\label{cor:ecsp-preclass}
Let $\bB$ be an equality constraint language. Then 
one of the following cases applies.
\begin{enumerate}
\item $\bB$ has a constant polymorphism.
\item $\bB$ has a binary injective polymorphism.
\item In $\bB$ every first-order formula is equivalent to a primitive positive formula.
\end{enumerate}
\end{corollary}
\begin{proof}
Suppose that $\bB$ does not have a constant polymorphism.
Since equality constraint languages have 2-transitive automorphism groups, we can use the contrapositive of Corollary~\ref{cor:neq-constant} to derive that all polymorphisms of $\bB$ must preserve $\neq$. 
The endomorphisms of $\bB$ are therefore injective, and locally generated by the automorphisms of $\bB$. 
If $\bB$ does not have essential polymorphisms, then Corollary~\ref{cor:elementary} shows that all relations that are first-order definable
in $\bB$ are also primitive positive definable in $\bB$, and we are in case (3).
If $\bB$ has an essential polymorphism, then $\bB$ has
a binary injective polymorphism by Theorem~\ref{thm:2trans}.
\end{proof}

We can now give the complexity classification for
equality constraint languages, which confirms Conjecture~\ref{conj:ua-dichotomy} in a special case. 

\begin{theorem}\label{thm:ecsps}
Let $\bB$ be an equality constraint language. Then 
exactly one of the following cases applies.
\begin{itemize}
\item $\bB$ has a polymorphism $f$ and an automorphism $\alpha$ such that 
$$f(x,y)=\alpha f(y,x)$$
for all elements $x$ and $y$ of $\bB$.
In this case, for every
finite reduct $\bB'$ of $\bB$ the problem $\Csp(\bB')$ can be solved in polynomial time.
\item There is a primitive positive interpretation of $(\{0,1\};\OIT)$ in $\bB$. In this case, there is a finite reduct $\bB'$ of $\bB$ such that $\Csp(\bB')$ is NP-complete.
\end{itemize}
\end{theorem}
\begin{proof}
By Proposition~\ref{prop:deep-taylor-mod-endo-tame}, the two cases are disjoint. 
%It is also clear that the two cases are disjoint, since
%if $(\{0,1\};\OIT)$ has a primitive positive interpretation in 
% not have a primitive positive
%interpretation in $\bB$ since $(\{0,1\};\OIT)$ doesn't have a constant polymorphism.
If $\bB$ has a constant polymorphism, 
then clearly there are $f$ and $\alpha$ such that $f(x,y)=\alpha f(y,x)$ for all $x,y \in B$. 
The claim for finite reducts of $\bB$ follows from Proposition~\ref{prop:const-core}.

Now suppose that $\bB$ has a binary injective polymorphism $f$. 
Such an operation
is an isomorphism between $({\mN};=)^2$ and $({\mN};=)$,
and we can find a permutation $\alpha$ of $B$ such that $f(x,y)=\alpha f(y,x)$ 
for all $x,y \in \mN$. 
Since every relation of $\bB$ has a quantifier-free definition
over $({\mathbb N};=)$,
Theorem~\ref{thm:horn}  
shows that every relation of $\bB$ even has a quantifier-free Horn
definition over $\bB$.  
By Theorem~\ref{thm:resolution}, the CSP for every finite signature reduct of $\bB$
can be reduced to $\Csp((\mN; =,\neq))$ in polynomial time. 
Tractability of $\Csp(({\mN}; =,\neq))$
has been shown in Section~\ref{sect:homo}.
By Corollary~\ref{cor:ecsp-preclass}, the only remaining case is that
over $\bB$ all first-order formulas are equivalent to primitive positive formulas.
In this case the claim follows from Corollary~\ref{cor:elementary-hard}.
\end{proof}

\chapter{Topology}
\label{chap:topology}
\vspace{-.5cm}
\begin{center}
\includegraphics{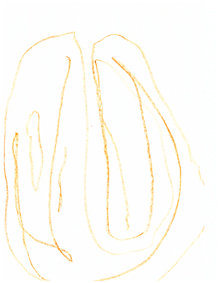}
\end{center}
\vspace{1cm}

Several important properties of $\omega$-categorical structures only depend
on their automorphism group considered as a \emph{topological group},
that is, on their automorphism group viewed as an abstract group, with the topology
of pointwise convergence on the group elements. This is in particular the 
case for certain Ramsey properties that become important in the next chapter.
We therefore give a self-contained introduction to basic topological background,
with a focus on the topics that become relevant for our applications to automorphism groups of $\omega$-categorical structures. 

\section{Topological Spaces}
\label{sect:topological-spaces}
% define neighbourhood? 
A \emph{topological space} is a set $S$ together with a collection 
of subsets of $S$, called the \emph{open} sets of $S$, such that
\begin{enumerate}
\item the empty set and $S$ are open;
\item arbitrary unions of open sets are open;
\item the intersection of two open sets is open.
\end{enumerate}
Complements of open sets are called \emph{closed}.
%The smallest closed set that contains a set $E$ is called the \emph{closure} of $E$ (in $S$), and denoted by $\bar E$.  
%Would have to formally do this with infinite intersection, may be more standard to do it as follows:
For $E \subseteq S$, the \emph{closure} of $E$ is the set of all points $x$ such that every open set in $S$ that contains $x$ also contains a point from $E$. 
Clearly, the closure of $E$ is a closed set. 
%Proof: For every element not in E there is an open set around it that doesnt interset E, so we can take the union of all those
%opens, and define the complement of the closure of E in this way.
A subset $E$ of $S$ is called \emph{dense (in $S$)} if its closure is the full space $S$.
%if it has non-empty intersection with every nonempty open set of $S$.
The \emph{subspace of $S$ induced on $E$} is the topological space on $E$ 
where the open sets are exactly the intersections
of $E$ with the open sets of $S$.

\begin{definition}
A mapping between two topological spaces is called \emph{continuous} if the pre-images of open sets are open, and \emph{open} if images of open sets are open.
A bijective open and continuous map is called a \emph{homeomorphism}.
\end{definition}

A \emph{basis} of $S$ is a collection of open subsets of $S$ such that every open set in $S$ is the union of sets from the collection.
%contains an open set from this collection. 
For $s \in S$, a collection of open subsets of $S$ is called a \emph{basis at $s$} if each set from the collection contains $s$,
and every open set containing $s$ also contains an open set from the collection. 
For a sequence $(s_n)_{n \geq 1}$ of elements of $S$, we write $\lim_{n \to \infty} s_n = s$, and say that $s_n$ \emph{converges} (against the \emph{limit} $s$) if for every open set $U$ of $S$ that contains $s$ there exists an $n_0$ such that $s_n \in U$ for all $n \geq n_0$. A topological space $S$ is called 
\begin{itemize} 
\item \emph{discrete} if every subset of $S$ is open (and hence also closed);
\item \emph{compact} if for an arbitrary
collection $\{U_i\}_{i \in A}$ of open subsets of $S$
with $S = \bigcup_{i \in A} U_i$ there is a finite
subset $B$ of $A$ such that $S = \bigcup_{i \in B} U_i$;
\item \emph{Hausdorff} if for any two distinct 
points $u,v$ of $S$ there are disjoint open sets $U$ and $V$ that
contain $u$ and $v$, respectively;
\item \emph{first-countable} if for all $s \in S$ there exists a countable basis at $s$.
%collection $U_n$ of open sets containing $x$ so
%that any open $V$ that contains $x$ also contains some $U_n$. 
\item \emph{separable}  if there is a countable dense set;
\end{itemize} 

The following equivalent characterization of continuity of maps from a first-countable space $S$ to a topological space $T$ is often easier to work with.
For $x \in S$, we say that $f$ is \emph{continuous at $x$} if for every open $V \subseteq T$ containing $f(x)$ there is an open $U \subseteq S$ containing $x$ whose image $f(U)$ is contained in $V$.

\begin{proposition}\label{prop:continuity}
Let $S$ be a first-countable and $T$ an arbitrary topological space. Then for every $f \colon S \rightarrow T$ the following are equivalent.
\begin{enumerate}
\item $f$ is continuous.
\item For all $s_n$, if $\lim_{n \to \infty} s_n = s$ then $\lim_{n \to \infty} f(s_n) = f(s)$. 
\item $f$ is continuous at every $x \in S$. 
\end{enumerate}
\end{proposition}
\begin{proof}
The implication from (1) to (2) is true even without the assumption that $S$ is first-countable. Let $(s_n)_{n \geq 1}$ be such that $\lim_{n \to \infty} s_n = s$, and let $V$ be open so that $f(s) \in V$. Then $U := f^{-1}(V)$ is open, and $s \in U$. 
So there exists an $n$ with $s_n \in U$. For this $n$, $f(s_n) \in V$. So $\lim_{n \to \infty} f(s_n) = f(s)$.

For the implication from (2) to (3), we show the contraposition. Suppose that $f$ is not continuous at some $s \in S$. 
That is, there exists an open set $V$ containing $f(s)$ such that 
all open sets $U$ that contain $x$ %are not contained in $f^{-1}(V)$. 
have an image that is not contained in $V$. 
Since $S$ is first-countable, there exists a countable collection $U_n$ of open sets containing $x$ so
that any open $V$ that contains $x$ also contains some $U_n$. Replacing $U_n$ by $\cap_{k=1}^n U_k$ where necessary, we
may assume that $U_1 \supset U_2 \supset \cdots$. If $U_n \subseteq f^{-1}(V)$, then $f(U_n) \subseteq V$, in contradiction
to our assumption; so we can pick an $x_n \in U_n \setminus f^{-1}(V)$ for all $n$, and obtain a sequence that converges to $x$. But $s_n \notin f^{-1}(V)$ for all $n$, and so $f(s_n)$ does not converge to $f(s) \in V$. 

Finally, the implication from $(3)$ to $(1)$ again holds in arbitrary topological spaces. Let $V \subseteq T$ be open.
We want to show that $U := f^{-1}(V)$ is empty. 
%If $f^{-1}(V)$ is empty there is nothing to show. 
%Otherwise, let $s \in f^{-1}(V)$ be arbitrary. 
When $s$ is a point from $U$, then because $f$ is continuous at $s$, and $V$ contains $f(s)$ and is open,
there is an open set $U_s \subseteq S$ containing $s$ whose image $f(U_s)$ is contained in $V$. 
Then $\bigcup_{s \in U} U_s = U$ is open as a union of open sets. 
\end{proof}

Important examples of topologies come from metric spaces. 
A sequence $(s_n)_{n \in \mathbb N}$ 
of elements of a metric space $(S;d)$ is called a \emph{Cauchy sequence} if for every $\epsilon>0$ 
there is an $n_0 \in \mathbb N$ such that for all $n,m > n_0$ we have that $d(s_n,s_m) < \epsilon$. A topological space $S$ is called 
\begin{itemize}
\item \emph{metrizable} if there exists a metric $d$ on $S$
which is \emph{compatible}, i.e., the open sets are unions of sets of the form $\{y \in S \; | \; d(c,y) \leq r\}$, for $x \in S$, $0 \leq r \in \mathbb R$;
\item \emph{completely metrizable} if it has a compatible \emph{complete} metric $d$, i.e., 
a metric $d$ on $S$ where every Cauchy sequence
converges against
an element of $S$;
\item \emph{Polish} if $S$ is separable and completely metrizable.
\end{itemize}

% Conversely, suppose that $f$ is not continuous. Then there exists a point x and an open set V containing f(x) such that
% f^-1(V) does not contain any open set containing x. Since $S$ is first-countable, there is a
% sequence s_n with s_n -> x. 

The \emph{product} $\prod_{i \in I} S_i$ of a family of topological
spaces $(S_i)_{i \in I}$ is the topological space on
the cartesian product $\times_{i \in I} S_i$ where the open
sets are unions of sets of the form $\times_{i \in I} U_i$ where $U_i$
is open in $S_i$ for all $i \in I$, and $U_i = S_i$ for all but finitely many $i \in I$. 
When $I$ has just two elements, say $1$ and $2$, we also write $S_1 \times S_2$ for the product 
(this operation is clearly associative and commutative). 
We denote by $S^k$ for
the $k$-th power $S \times \dots \times S$ of $S$, equipped
with the product topology as described above.

We also write $S^I$ to a $|I|$-th power of $S$, where the factors are 
indexed by the elements of $I$. In this case, 
we can view each element of $T := S^I$ as a function from $I$ to $S$ in the obvious way.
The product topology on $T$ is also called the \emph{topology of pointwise convergence}, due to the following.

\begin{proposition}\label{prop:pointwise-convergence}
Let $S$ be a topological space, and $I$ be a set. 
Let $(f_n)_{n \in \mathbb N}$ be a sequence of elements
of the product space $T := S^I$.
Then $\lim_{n \to \infty} f_n = f$ if and only if $\lim_{n \to \infty} f_n(j) = f(j)$ in $S$ for all $j \in I$. 
\end{proposition}
\begin{proof}
Suppose first that $\lim_{n \to \infty} f_n = f$ in $T$. Let $j \in I$ be arbitrary and let $V$ be an open set that contains
$f(j)$. 
%$U := \prod_{1 \leq i < j} S \times V \times \prod_{j < i \leq \kappa} S$ 
Then the set $U:=\prod_{i \in I} T_i$ where
$T_i = V$ if $i = j$, and $T_i = S$ otherwise,  
is open in $T$ and contains
$f$. So there is an $n_0$ such that $f_n \in U$ for all $n \geq n_0$. But then $f_n(j) \in V$ for all $n \geq n_0$,
and so $\lim_{n \to \infty} f_n(j) = f(j)$. 

Now suppose that $\lim_{n \to \infty} f_n(j) = f(j)$ in $S$ for all $j \in I$, and let $V$ be an open set of $T$ that contains $f$. 
Then there exists a finite $J \subseteq I$
and open subsets $(V_j)_{j \in J}$ of $S$ 
such that $f \in \prod_{i \in I} T_i$ where
$T_i = V_i$ if $i \in J$ and $T_i = S$ otherwise. For each $j \leq J$ there exists an $n_j$ so that $f_n(j) \in V_j$ for all $n \geq n_j$. 
Then $f_n \in V$ for all $n \geq \max_{j \in J} n_j$, and hence $\lim_{n \to \infty} f_n = f$. 
\end{proof}

\begin{example}\label{expl:baire-space}
When we equip the natural numbers $\mathbb N$ with the discrete topology, then ${\mathbb N}^{\mathbb N}$ with the topology of pointwise convergence is called the \emph{Baire space}. The open sets are exactly 
the unions of sets of the form $\{ g \in {\mathbb N} \rightarrow {\mathbb N} \; | \; g(\bar a) = \bar b\}$ for some $\bar a, \bar b \in \mN^k$, $k \in \mN$. \qed
\end{example}

%Since a topological space is Hausdorff if and only if limits of converging sequences are unique, 
%t follows from Proposition~\ref{prop:pointwise-convergence} that products of Hausdorff spaces are Hausdorff. 
%The following is more substantial.

% NOT TRUE:
%Let U be open in R iff R\U is countable or U is \emptyset. If U,V are
%open then U\cap V is not empty so this space is not Hausdorff, and in
%this space no sequence converges.

\begin{theorem}[Tychonoff; see e.g.~\cite{Jaenich}]\label{thm:tychonoff}
Products of compact spaces are compact. 
\end{theorem}

\section{Topological Groups}
\label{sect:topo-groups}
%In Section~\ref{sect:galois}, we considered permutation groups as sets of permutations of some domain $D$.
%We will see in the following that many fundamental properties of an $\omega$-categorical structure $\bB$ only depend
%on the automorphism group of $\bB$ considered as a \emph{topological group}, which is a concept that we define now. 

A \emph{topological group} is an (abstract) group $\bf G$ together with a topology on the elements $G$ of $\bf G$ such that 
$(x,y) \mapsto xy^{-1}$ is continuous from $G^2$ to $G$. In other words, 
we require that the binary group operation and the inverse function are continuous.  
Two topological groups are said to be \emph{isomorphic}
if the groups
are isomorphic, and the isomorphism is a homeomorphism between the respective topologies. 

\begin{example}\label{expl:sym-topology}
The elements of the group $\Sym({\mathbb N})$ form a  (non-closed) subset of the Baire space ${\mathbb N}^{\mathbb N}$ (Example~\ref{expl:baire-space}),
and the topology induced by the Baire space on $\Sym({\mathbb N})$ is also called the \emph{topology of pointwise convergence}. 
The open sets are 
the unions of sets of the form $\{ g \in \Sym({\mathbb N}) \; | \; g(\bar a) = \bar b\}$ for some finite tuples $\bar a,\bar b$ over ${\mathbb N}$. \qed
\end{example}

An action of a topological group $\bf G$ on a topological space
 $S$ is \emph{continuous} if it is continuous as a function from $G \times S$ into $S$. 
 A continuous action of $\bf G$ on $S$ gives rise 
 to a homomorphism
 from $\bf G$ into the group of all homeomorphisms of $S$.
An action is faithful if this homomorphism
is injective. 
If $\bf G$ is a subgroup of $\Sym({\mathbb N})$ and the space $S$ is countable and equipped with the discrete topology,
it makes sense to call a such a homomorphism \emph{topologically faithful} if additionally the 
homomorphism is a homeomorphism whose
image is closed in $\Sym(S)$ (equipped with the product topology). 
 
An important example of a continuous action of topological groups is the following. 
A \emph{left coset} of a subgroup $\bf V$ of $\bf G$ is a set of the form
$\{h g \; | \; g \in V\}$ for $h \in G$, also written $hV$. Clearly, the set of all left cosets of $\bf G$ partitions $G$, and is denoted
by ${\bf G}/{\bf V}$. The cardinality of ${\bf G}/{\bf V}$ is the \emph{index} of $\bf V$ in $\bf G$. 
The set ${\bf G}/{\bf V}$ can be viewed a topological space where a set of left-cosets is open if their union is open in 
$\bf G$. We can define a continuous action of $\bf G$ on ${\bf G}/{\bf V}$ by setting $g \cdot hV = g h V$.  This action is also called the \emph{action of $\bf G$ on ${\bf G}/{\bf V}$ by left translation}. 
Analogously we define the space ${\bf G} \backslash {\bf V}$ of all \emph{right-cosets} $Vh$, and the action of ${\bf G}$ on ${\bf G} \backslash {\bf V}$ by \emph{right translation}. 

Every open subgroup $\bf H$ of $\bf G$ is closed,
 since the complement of $H$ in $G$ is the open set given by the union of open sets $gH$ for $g \in G \setminus H$.
% Cauchy sequences in topological groups:
%A sequence  in a topological group  is a Cauchy sequence if for every open neighbourhood  of the identity in  there exists some number  such that whenever  it follows that . As above, it is sufficient to check this for the neighbourhoods in any local base of the identity in .
% BOOK-TD: 
% THIS WOULD BE A GREAT PLACE FOR THE LOGIC ACTION!
A topological group $\bf G$ is
\begin{itemize} 
\item \emph{Hausdorff (metrizable, Polish)} if the topology of $\bf G$ is Hausdorff (metrizable, Polish, respectively);
\item \emph{first-countable} if it has a countable basis at the identity.
%\item \emph{Polish} if the topology of $\bf G$ is Polish.
\item \emph{non-archimedian} if it has a basis at the identity consisting of open subgroups. %~\cite{Tsankov}.
\end{itemize}
%It is easy to see that every metrizable topological group is Hausdorff and first-countable. Quatsch!
We also recall the following. 

\begin{proposition}[Proposition 13 and Proposition 14 in~\cite{Bourbaki}]
Let ${\bf G}$ be a topological group, and let ${\bf H}$ be a
subgroup of ${\bf G}$. Then 
\begin{itemize}
\item ${\bf G}/{\bf H}$ is discrete if and only if $H$ is open in $G$;
\item ${\bf G}/{\bf H}$ is Hausdorff if and only if $H$ is closed in $G$.
\end{itemize}
\end{proposition}

The following is sometimes useful to verify that an action
is topologically faithful. 

\begin{proposition}[Proposition 2.2.1 in~\cite{Gao}]
\label{prop:Polish}
Let ${\bf G}$ be a Polish group and ${\bf H}$ a subgroup
of ${\bf G}$ with the subspace topology. Then $H$
is Polish if and only if $H$ is closed in $G$.
\end{proposition}

The \emph{topological automorphism group} of a structure $\bB$ with domain $B$ is a topological group obtained from 
the abstract automorphism group ${\bf G}$ of $\bB$ 
(see Section~\ref{ssect:products}) % of Chapter~\ref{chap:mt}) 
by equiping the elements $G$ of ${\bf G}$ 
with the topology of pointwise convergence, that is, 
the topology induced on $G$ by the one 
on $\Sym(B)$ as given in Example~\ref{expl:sym-topology}.

%\begin{definition}
%Let $G$ be a set of permutations of a set $D$.
%The \emph{topology of pointwise convergence} on $G$ is 
%the topology where the open sets are precisely the unions of left cosets of pointwise stabilizers $G_{\bar a}$, for finite tuples $\bar a$ from $D$.
%\end{definition}
% Hence, the topology of pointwise convergence on a set of permutations $G$ of a set $D$
%is exactly the subspace of $\Sym(D)$ induced on $G$. 

\begin{proposition} 
%Let $G$ be a set of permutations of a set $D$. Then the following are equivalent.
%\begin{itemize}
%\item $G$ is closed in the topology of pointwise convergence;
%\item $G$ is a closed subspace of $\Sym(D)$;
%\item $G$ is locally closed as in Definition~\ref{def:lc};
%\item $G$ is the automorphism group of a relational structure. 
%\end{itemize}
%%%Note that the closure of a subset of $\cG$ in this topology 
%%%is the same as local closure for permutation groups, as defined in Section~\ref{sect:galois}. 
A set $\cB$ of permutations of a set $B$ is a closed subset of $\Sym(B)$ if and only if it is
locally closed as defined in Definition~\ref{def:lc}.
\end{proposition}
\begin{proof}
The set of operations $\cB$ is \emph{not} closed in the topology of pointwise convergence if and only if there exists a permutation $g \in \Sym(B) \setminus \cB$ such that every open set containing $g$ also contains an element of $\cB$. This is the case if and only that for every tuple $\bar a$, 
$\cB$ contains an operation $h$ such that the restriction of $g$ to the elements of $\bar a$ equals the restriction of $h$ to those elements.
According to Definition~\ref{def:lc}, this is exactly the case when $g$ is in the local closure by $\cB$. 
\end{proof}

% Brauchen wir doch nicht:
%Lascar showed the following.
%\begin{theorem}[Corollary 2.8 in~\cite{Lascar}]\label{thm:homeo-from-cont}
%Every continuous isomorphism between closed
%subgroups of $\Sym({\mathbb N})$ is a homeomorphism.
%\end{theorem}

In the following, let $\bf G$ be a topological group that is the automorphism group of a relational
structure $\bB$, and let $G$ be its domain (equipped with the topology of pointwise convergence). 
Note that if $G$ is compact then all orbits of $G$ must be finite. 
% if there is an infinite orbit, we can choose as cover of S just
% stabilizers for all tuples in this orbit. Then no finite subset will
% do it. The converse should also hold, but is not needed here.
Hence, when $\bB$ is $\omega$-categorical,
$G$ cannot be compact.
%It can be checked easily that the topology of $\bf G$ is Hausdorff. % follows from metrizability
It is clear that $\bf G$ is non-archimedian. 
The topology on $G$ has the following compatible metric $d$.
When $b_1,b_2,\dots$ is an enumeration of the domain $B$ of $\bB$,
then for elements $f,g \in G$ we define $d(f,g) = 1/2^{n+1}$ where
$n$ is the least natural number such that $f(b_n) \neq g(b_n)$.
In fact, $d$ is an \emph{ultrametric}, that is, it satisfies $d(x,z) \leq max(d(x,y),d(y,z))$ for all $x,y,z$. 
Moreover, $d$ is \emph{left-invariant}, i.e., $d(gx,gy) = d(x,y)$ for all $g,x,y \in G$. 
% TODO: verify this (shows en passant that choice of enumeration is not important).
This metric is not complete: to see this, let $f$ be an arbitrary injective
non-surjective mapping from $B \rightarrow B$. For each $n$, there exists
a permutation $h_n$ of $B$ such that $h_n(b_i)=f(b_i)$ for all $i \leq n$.
Hence, the sequence $(h_n)_{n \geq 1}$ is Cauchy, but it does not converge to a permutation.

The topology on $G$ is also completely metrizable.
%also has the following compatible \emph{complete} metric $d$.
To see this, we define  
a compatible complete 
metric $d'$ by setting $d'(f,g) = 1/2^{n+1}$ for elements $f,g \in G$ where
$n$ is the least natural number such that $f(b_n) \neq g(b_n)$ or $f^{-1}(b_n) \neq g^{-1}(b_n)$. 
Alternatively, and more generally, when
$d$ is a compatible left-invariant metric, then 
$d(x,y) + d(x^{-1},y^{-1})$ defines a compatible complete metric (see~\cite{BeckerKechris}).

Finally, $\bf G$ is separable: for all finite tuples $\bar a$, $\bar b$ that lie in the same orbit
we fix an element of $G$ that maps $\bar a$ to $\bar b$; the (countable) set of all the selected
elements of $G$ is clearly dense in $G$. 

In this thesis, we will be exclusively interested in topological groups that arise as automorphism groups of countable structures. Those groups can be characterised in topological terms, as demonstrated in Proposition~\ref{prop:topo-auto-groups} below. 

\begin{proposition}[Section 1.5 in~\cite{BeckerKechris}; also see Theorem~2.4.1 and Theorem 2.4.4 in~\cite{Gao}]
\label{prop:topo-auto-groups}
Let $\bf G$ be a topological group. Then the following are equivalent.
\begin{enumerate}
\item $\bf G$ is isomorphic to 
the topological automorphism group of a countable relational structure.  
%THIS is the actual source of interest
\item $\bf G$ is isomorphic to a closed subgroup of $\Sym(\mathbb N)$. 
%Have already seen this equivalence
\item $\bf G$ is Polish and admits a compatible left-invariant ultrametric.
%Find this most intuitive of all the characterizations
\item $\bf G$ is Polish and non-archimedian. 
%This is the formulation that Todor and Solecki seem to prefer
%\item $\bf G$ is Polish and has a countable neighbourhood basis 
%of $e^{\bf G}$ consisting of open subgroups. 
% This is what the proof gives us, and what we find in Gao; omit it
\item ${\bf G}$ is Polish and has an at most countable basis \emph{closed under left multiplication}, that is, an at most countable basis $\mathcal B$ of $\bf G$ so that for any $U \in \mathcal B$ and $g \in G$ we have $gU \in \mathcal B$. 
% this helps to give intuition how to ``go back''.
\end{enumerate}
\end{proposition}
\begin{proof}
The equivalence of (1) and (2) has been shown in Proposition~\ref{prop:loc-clos-group}. 
The implication from (1) to (3) has been explained in the paragraphs preceding the statement of the proposition.
So it suffices to show (3) $\Rightarrow$ (4) $\Rightarrow$ (5) $\Rightarrow$ (2).

For the implication from (3) to (4), let $d$ be a left-invariant ultrametric on $G$. 
Let $U_n = \{x \in G \; | \; d(x,1)<2^{-n} \}$, for $n \in \mN$. We claim that the set of all those $U_n$ forms a basis at the identity consisting of open subgroups. Since $d$ is a left-invariant ultra-metric, for $x,y \in G$ we have 
$$d(x^{-1}y,1) = d(y,x) \leq \max(d(y,1),d(1,x))$$
and thus $U_n$ is a indeed a subgroup. 

For the implication from (4) to (5), assume (4). 
Let $\{U_1,U_2,\dots\}$ be an at most countable basis at the identity (which exists since $G$ is metrizable). 
Each $U_i$ has an open subset $V_i$ which is a subgroup, 
since $\bf G$ has a basis at the identity consisting of open subgroups. Then $\{V_1,V_2,\dots\}$ is a countable 
basis of the identity consisting of open subgroups. 
Each $V_i$ has at most countably many cosets since $G$ is separable. 
So the set of all cosets of those groups gives an at most countable basis that is closed under left multiplication. 

Finally, we show that (5) implies (2). 
Let $\mathcal B = \{U_1,U_2,\dots\}$ be an at most countable basis closed under left multiplication. 
If $\mathcal B$ is infinite,
we define the map $\xi \colon G \rightarrow \Sym({\mathbb N})$ by setting
$$\xi(g)(n)=m \; \Leftrightarrow \; g U_n = U_m \; .$$
If $|\mathcal B|=n_0$ is finite, we define 
the map $\xi \colon G \rightarrow \Sym({\mathbb N})$ 
similarly, but set $\xi(g)(n)=n$ for all $n > n_0$.
It is straightforward to verify that $\xi(f g) = \xi(f) \xi(g)$.
% phi(f)\phi(g)(n) = m iff f(g U_n) = U_m iff fg U_n = U_m iff \phi(gf)(n)=m
The mapping $\xi$ is injective: 
when $f$ and $g$ are distinct, then there are disjoint open subsets $U$ and $V$ with
$f \in U$ and $g \in V$, because the topology is Hausdorff; 
since $\mathcal B$ is a basis, we can assume that $U=U_{n_1}$ and $V=U_{n_2}$, for some $n_1,n_2 \geq 1$. If $f U_{n_1} = g U_{n_1}$, then $g \in U_{n_1}=U$ since $f \in U_{n_1}$, contradicting the assumption that $U$ and $V$ are disjoint. Hence,
 $\xi(f)(n_1) \neq \xi(g)(n_1)$, and so $\xi(f) \neq \xi(g)$. 
Since bijective 
algebra homomorphisms are isomorphisms, $\xi$ is an isomorphism between $\bf G$ and a subgroup of $\Sym({\mN})$. To verify that $\xi$ is continuous,
let $g \in G$ be arbitrary, and let $V \subseteq \Sym({\mathbb N})$ be an open set containing $\xi(g)$. 
Then $V$ is a union of basic open sets of the form
$V_{\bar a,\bar b} := \{f \in \Sym({\mathbb N}) \; | \; f(\bar a) = \bar b\}$ for some $\bar a,\bar b \in {\mathbb N}^n$. The preimage of $V_{\bar a,\bar b}$ under $\xi$ is $\{g \in G \; | \; g U_{a_1} = U_{b_1} \wedge \dots \wedge g U_{a_n} = U_{b_n} \}$. 
Since multiplication in $G$ is continuous, this set is open. Hence the preimage of $V$ is a union of open sets and therefore open as well, which concludes the proof that
$\xi$ is continuous. 

It can also be verified that $\xi$ is open; for the details of this last step, we refer to~\cite{Gao} (Theorem 2.4.4). 
Therefore, $\xi$ is a homeomorphism between $\bf G$ 
and its image $\xi(G)$, which is therefore also Polish, and a subgroup of the Polish group $\Sym({\mathbb N})$.
By Proposition~\ref{prop:Polish}, $\xi(G)$ is a \emph{closed} subgroup of $\Sym({\mathbb N})$.
\end{proof}

%When we later refer to topological groups that satisfy the equivalent conditions in 
% the previous proposition, we will refer to those groups as "closed subgroups of S_omega"
% since it is simultaneously short and informative.

%A basis at the identity is a collection of open sets containing 1 (the identity of the group) such that every open set containing 1 contains also an element of the basis.
%Suppose now that G has a basis of open subgroups. As G is Polish, you can assume this basis is countable. Now G acts on the countable set of all left cosets of all subgroups in the basis and it is easy to check that this action gives a topological embedding of G in the group of all permutations of this set. As G is Polish, the image of the embedding has to be closed.

A subgroup ${\bf N}$ of ${\bf G}$ with domain $N$ is called \emph{normal} 
if $g N = N g$ for all elements $g$ of $\bf G$. 
%The subgroup of ${\bf G}$ consisting of the identity element only is called
%\emph{trivial}, and subgroups of ${\bf G}$ that are distinct from ${\bf G}$
%are called \emph{proper}.
Recall the following equivalent characterizations of 
normality of subgroups, which can be seen as a refinement of Proposition~\ref{prop:congruence-homo} for the
case of groups.

% The third items appears in the preservation theorem for extreme amenability the second is conceptually important. So lets keep it. 
\begin{proposition}\label{prop:normal-subgroups}
Let ${\bf G}$ be a group, and ${\bf N}$ be a subgroup of ${\bf G}$.
Then the following are equivalent.
\begin{enumerate}
\item ${\bf N}$ is normal.
\item $\bf G$ has the congruence $E = \{(a,b) \; | \; a b^{-1} \in N\}$. %$\{(a,b) \; | \; \exists v \in N. \; ab=va\}$. 
\item There is a homomorphism $h$ from ${\bf G}$ to some group such that $N = h^{-1}(0)$.
\item For every $g \in G$ and every $v \in N$ we have $g v g^{-1} \in N$.
\end{enumerate}
\end{proposition}
\begin{proof}
(1) $\Rightarrow$ (2): to verify that $E$ is a congruence, we have to show that for all $(a_1,b_1), (a_2,b_2) \in E$,
$(a_1a_2,b_1b_2) \in E$. Indeed, $(a_1a_2)(b_1b_2)^{-1} = a_1 (a_2 b_2^{-1}) b_1^{-1} \in a_1 N b_1^{-1} = N a_1 b_1^{-1} \subseteq NN = N$. 

(2) $\Rightarrow$ (3):  follows from Proposition~\ref{prop:congruence-homo}:  
$g \mapsto gN$ is a group homomorphism from ${\bf G}$ to ${\bf G}/{\bf N}$. 

(3) $\Rightarrow$ (4): For $g \in G$ and $v \in h^{-1}(0)$, we must show that $gvg^{-1} \in h^{-1}(0)$. Indeed, 
$h(gvg^{-1}) = h(g) h(v) h(g)^{-1} = h(g) 0 h(g)^{-1} = 0$. 

(4) $\Rightarrow$ (1): assume that $gNg^{-1} \subseteq N$ for all $g \in G$. Let $a \in G$ be arbitrary. 
Applying the assumption for $g=a$ we find that $aN \subseteq Na$. 
Applying the assumption for $g=a^{-1}$ we find that $a^{-1}N(a^{-1})^{-1} = a^{-1}Na \subseteq N$, and hence $Na \subseteq aN$. 
We conclude that $aN=Na$. 
\end{proof}

When ${\bf G}$ is an automorphism group, %closed subgroup of $\Sym({\mathbb N})$,
then \emph{closed normal subgroups} of ${\bf G}$ typically
arise as the subgroups consisting of those elements of ${\bf G}$
that fix the equivalence classes of a congruence relation on the elements of ${\bf G}$.
This can be made precise as follows\footnote{I thank Todor Tsankov for pointing this out to me.}.

\begin{proposition}\label{prop:closed-normal-subgroups}
Let $\bf G$ be the automorphism group of a relational structure $\bB$
with domain $B$. If $E$ is a $\bf G$-invariant equivalence relation on $B^n$, for some $n$, then the subgroup of $\bf G$ that preserves each equivalence class of $E$ is closed and normal. Conversely, every closed normal subgroup of $\bf G$ is 
the intersection of closed normal subgroups that arise in this way.
%, that is, 
%that are the automorphism group of a structure that contains as relations the equivalence classes of a $\bf G$-invariant equivalence relation on $B^n$, for some $n$.
%Conversely, 
%when $\bf G$ has a 
%non-trivial proper 
%closed normal subgroup ${\bf N}$, then there is a non-trivial
%proper congruence of the component-wise action 
%of $\bf G$ on $B^n$ for some $n$.
% What was the point of Dugald?
\end{proposition}

\begin{proof}
Let $\bC$ be the expansion of $\bB$ by a unary relation
for each equivalence class of $E$. Then $\Aut(\bC)$ is closed by Proposition~\ref{prop:loc-clos-group}, 
and it is a normal subgroup of $\Aut(\bB)$:
when $g \in \Aut(\bB)$ and $h \in \Aut(\bC)$, then $g \circ h \circ g^{-1}$ preserves each equivalence class of $E$, 
and thus is an automorphism of $\bC$. Normality of $\Aut(\bC)$ follows from Proposition~\ref{prop:normal-subgroups}. 

For the second part, suppose that ${\bf G}$ has a closed normal subgroup
${\bf N}$. Consider the relation
$$R_n := \{(x,y) \; | \; x,y \in B^{n} \text{ and there is } h \in N \text{ such that }
h(x)=y \} \; .$$
This relation is obviously an equivalence relation, and it is preserved by 
all the automorphisms of $\bB$. For this, we have
to show that for all $g \in G$ and all $(x,y) \in R_n$
we have that $(g(x),g(y)) \in R_n$. So suppose that $x,y \in B^n$ such 
that $h(x)=y$ for some $h \in N$. Then $g(y)=g(h(x)) \in (gN)(x) = (Ng)(x) = N(g(x))$
by normality of ${\bf N}$. Hence there exists an $h' \in N$
such that $h'(g(x)) = g(y)$, which shows that $(g(x),g(y)) \in R_n$.

Let $\bC$ be the structure that contains for all $n$ the $n$-ary relations given by the 
equivalence classes of the relations $R_n$ for all $n \geq 0$.
We claim that ${\bf N}$ is precisely the automorphism group of $\bC$.
As in the first part we can verify that every $h \in N$ is an automorphism of $\bC$. The converse follows by local closure as follows.
Let $g$ be an automorphism of $\bC$, and let $x,y$ be from $B^n$ so that $g(x) = y$. 
Since $g$ preserves the equivalence classes of
$R_n$, there exists an $h \in N$ such that $h(x) = y$. 
Hence, $g$ lies in the closure of ${\bf N}$,
which implies that $g$ is from ${\bf N}$ since ${\bf N}$ is closed.
%When for all $n$, the blocks of $R_n$ are first-order definable in $\bB$, it follows that ${\bf N} = {\bf G}$, in contradiction
%to the assumption that ${\bf N}$ is a proper normal subgroup. 
%So choose $n$ such that the blocks of $R_n$ are not first-order definable,
%and note that in particular $R_n$ 
%is not the full equivalence relation on $B^n$, and therefore a \emph{proper}
%congruence. 
%Since ${\bf N}$ is non-trivial, there are distinct 
%$a,b \in B$ such that $h(a)=b$ for some $h \in N$,
%and therefore also $\bar a, \bar b \in B^n$ such that $h(\bar a)=\bar b$.
%This shows that $R_n$ is non-trivial. 
\end{proof}

%"Your second example is a special case of the first. In general, if H is a normal closed subgroup of the oligomorphic G = Aut M, you can define for every n an equivalence relation E_n on M^n by
%a E_n b  <->  \exists h \in H  h.a = b
%(i.e. just the H-orbits) As H is normal, this equivalence relation is G-invariant and thus definable. Then H is exactly the stabilizer of all classes of all the E_n's. In your second example, it is enough to consider just E_2 to recover H."

%There are a few more things you can say, for example, H acts oligomorphically on M iff G/H is compact.
% Proof? Do we need normality here?
% Suppose H acts oligomorphically on M. Then we can
% simply factor the larger group and get compactness 
% -- here we don't need normality. 
% Conversely, if G/H is compact, then we learn 
% from compactness in how many ways an orbit
% of $G$ splits into orbits of $H$. 
% also here I guess that we don't need normality. 

\begin{example}
The automorphism group ${\bf G}$ of the structure
$\bB = ({\mathbb Q}; \Betw)$, 
where $\Betw = \big \{(x,y,z) \; | \; (x<y<z) \vee (z<y<x) \big \}$, is 2-transitive
and therefore primitive.  
However, the relation $\big \{((x_1,x_2),(y_1,y_2)) \; | \: (x_1 < x_2 \wedge y_1 < y_2) \vee (x_1 > x_2 \wedge y_1 > y_2) \vee (x_1=x_2 \wedge y_1 = y_2) \big \}$ is a $\bf G$-invariant equivalence relation on ${\mathbb Q}^2$. And indeed, 
${\bf G}$ has a closed normal subgroup ${\bf N}$ that is isomorphic
to the automorphism group of $({\mathbb Q}; <)$, and ${\bf G}/{\bf N}$ has two elements, corresponding
to the automorphisms that reverse the order $<$, and the automorphisms that preserve the order. \qed
\end{example}

\section{Oligomorphic Groups}
\label{sect:roelcke}
In the last section we have seen conditions that describe when a topological group is the automorphism
group of a countable structure. 
In this section, we see conditions that describe when a topological group is the automorphism group of a countable \emph{$\omega$-categorical} structure.
As a permutation group, we have seen that
these groups are precisely 
the closed \emph{oligomorphic} permutation groups (as we have seen in Section~\ref{sect:galois}); we therefore
call a topological group \emph{oligomorphic} if it
is isomorphic to an oligomorphic subgroup of $\Sym({\mathbb N})$. In fact, Theorem~\ref{thm:tsankov} 
below shows that 
the information whether a topological group ${\bf G}$ is oligomorphic can be expressed quite naturally in terms
of the open subgroups of ${\bf G}$ 
without referring to any particular oligomorphic action
of ${\bf G}$. 

A topological group $\bf G$ is called \emph{Roelcke precompact} if for every  open set $U \subseteq G$ that contains the identity there exists a finite set $F \subseteq G$ such that $G=UFU$. 
The following theorem is essentially from Tsankov~\cite{Tsankov};
there, the focus has been a characterization of 
Roelcke precompact groups in terms of oligomorphic groups. Here, on the other hand, the focus will be the
characterization of oligomorphic groups in terms of Roelcke precompact ones, and this motivates the following formulation of Tsankov's theorem\footnote{I am grateful to Todor Tsankov for his help with the presented reformulation of his result.}. 

% Corrected version; 
% The problem was that there are Roelcke precompact
% groups that do no appear as automorphism groups
% of omega-categorical structures. 

\begin{theorem}[of Tsankov~\cite{Tsankov}]\label{thm:tsankov}
Let ${\bf G}$ be isomorphic to a closed subgroup of $\Sym({\mathbb N})$. Then the following are equivalent.
\begin{enumerate}
\item ${\bf G}$ is the automorphism group of a countably infinite $\omega$-categorical structure.
\item ${\bf G}$ is Roelcke precompact,
and ${\bf G}$ has  an open subgroup $V$ of countably infinite index such that
for all open subgroups $U$ of  ${\bf G}$ there are $g_1,\dots,g_n \in G$
such that $\bigcap_{i \leq n} g_i V g_i^{-1} \subseteq U$. 
\item For every open subgroup $U$ of ${\bf G}$ 
the set $\{UfU \;| \; f \in G\}$ is finite, and ${\bf G}$ has an open subgroup $V$ of countably infinite index such that
for all open subgroups $U$ of  ${\bf G}$ there are $g_1,\dots,g_n \in G$
such that $\bigcap_{i \leq n} g_i V g_i^{-1} \subseteq U$. 
\item ${\bf G}$ has a topologically 
faithful transitive action on a countably infinite set with the discrete topology, and every such action of $\bf G$ is oligomorphic. 
\item ${\bf G}$ is the automorphism group of a countably infinite $\omega$-categorical structure with only one orbit.
\end{enumerate}
\end{theorem}
\begin{proof}
The implication from $(5)$ to $(1)$ is trivial, and we prove
$(1) \Rightarrow (2) \Rightarrow (3) \Rightarrow (4) \Rightarrow (5)$. 
For the implication from $(1)$ to $(2)$, suppose 
that $\bf G$ is the automorphism group
of an $\omega$-categorical structure $\bB$, and let $G$ be the domain of $\bf G$, which is a set of permutations
of the domain $B$ of $\bB$. 
Since $\bB$ is $\omega$-categorical, it has a finite
number $k$ 
of orbits by Theorem~\ref{thm:ryll}; choose orbit representatives $b_1,\dots,b_k \in B$, and write
$\bar b$ for $(b_1,\dots,b_k)$. Then the stabilizer 
$V := G_{\bar b}$ is an open subgroup of ${\bf G}$ of countably infinite index. 
Let $U$ be an arbitrary open subgroup of ${\bf G}$. 
Then $U$ contains $G_{\bar a}$ for some $\bar a \in B^n$.
For $j \leq n$, let $g_j \in G$ be such that $g_j(a_j) = b$ where $b \in \{b_1,\dots,b_k\}$ is from the same orbit as $a_j$. We claim that $K := \bigcap_{j \leq n} g_j^{-1} V g_j \subseteq U$. To see this, let $h \in K$ be arbitrary. 
Since $h \in g_j^{-1} V g_j$ we find that $h(a_j) = a_j$. 
Hence, $h \in G_{\bar a} \subseteq U$.

To show that $\bf G$ is 
Roelcke precompact, let $U \subseteq G$ be open with $1 \in U$. 
Then there exists an $n$ such that $U$ contains the stabilizer $G_{\bar a}$ for an $n$-tuple $\bar a$ of elements of $B$. % Is clear; but still would be better
% to have a lemma for this
It suffices to show the existence
of a finite number of elements $g_1,\dots,g_k$ of $G$
such that $G = \bigcup_{i \leq k} G_{\bar a} g_i G_{\bar a}$.
By Theorem~\ref{thm:ryll}, $G$ has finitely many 
orbits of $2n$-tuples; so let $(\bar a,g_1 \cdot \bar a),\dots,(\bar a,g_k \cdot \bar a)$
be a complete list of representatives for those orbits
of $2n$-tuples that are contained in $G \cdot \bar a \times G \cdot \bar a$. 
We claim that 
$G_{\bar a} g_1 G_{\bar a} \cup \cdots \cup  
G_{\bar a} g_k G_{\bar a} = G$. 
Let $f \in G$ be arbitrary. Let $i \leq k$ be such that
$(\bar a,f \cdot \bar a)$ and $(\bar a,g_i \cdot \bar a)$ lie in the same orbit
of $n$-tuples in $G_{\bar a}$. 
So there exists
an $h \in G_{\bar a}$ such that $f \cdot \bar a = h g_i \cdot \bar a$. 
Then $f^{-1} \circ h \circ g_i$ lies in $G_{\bar a}$,
so $f \in G_{\bar a} g_i G_{\bar a}$ as required.  

For the implication (2) implies (3), 
let $U$ be an open subgroup of $\bf G$.
Since $\bf G$ is Roelcke precompact there exists
a finite set $F \subseteq G$ such that $G = UFU$. 
Then $|F|$ bounds 
the sets of the form $\{UfU \; | \; f \in G\}$ 
because those sets partition $G$.

(3) implies (4). Since $V$ is open, 
${\bf G}/{\bf V}$ has the discrete topology, 
and the action of $\bf G$ on the countably 
infinite set ${\bf G}/{\bf V}$ 
by left translation is continuous and transitive.
We show that this action, as a map $\xi$ 
from ${\bf G}$ to $\Sym({\bf G}/{\bf V})$, is open. 
Let $U \subset G$ be open. 
By assumption, there are $g_1,\dots,g_n \in G$
such that $K := \bigcap_{i \leq n} g_i V g_i^{-1} \subseteq U$. Note that every $h \in K$ fixes $g_i V$ for all $i \leq n$.
Hence, $\xi(U)$ contains the stabilizer 
of finitely many elements, and hence is open. 
It also follows that the action is faithful:
to see this, let $f,g \in G$ be distinct.
We have to show that $\phi(f g^{-1}) \neq 1$. 
Since $f g^{-1} \neq 1$ there is an open subgroup $U$ that contains $1$ but not $fg^{-1}$. 
Since $\xi(U)$ is open, there are finitely
many $h_1,\dots,h_m$ such that 
$$U' := \{ h \in U \; | \; h h_1 V = h_1 V \wedge \cdots \wedge h h_m V = h_m V\} \subseteq U \; .$$
Then $U'$ still contains $1$ and not $fg^{-1}$, so assume
in the following that $U'=U$. 
Since the sets of the form $g V g^{-1}$ are precisely
the point stabilizers of ${\bf G}/{\bf V}$,
we have that the kernel $\phi^{-1}(1)$ of $\xi$
can be expressed as $\phi^{-1} = \bigcap_{g \in G} gVg^{-1}$. Since 
$fg^{-1} \notin U 
= \bigcap_{g \in \{h_1,\dots,h_m\}} h V h^{-1}$, it follows in particular
that $fg^{-1} \notin \phi^{-1}(1)$, which is what we wanted to show. 

Now $\xi$ is a homeomorphism between the Polish group $\bf G$ and its image ${\bf H} := \xi(U)$
and hence ${\bf H}$ is Polish as well. We can 
apply Proposition~\ref{prop:Polish} to the subgroup ${\bf H}$
of the Polish group $\Sym({\bf G}/{\bf V})$,
and conclude that $H$ is closed in
$\Sym({\bf G}/{\bf V})$. Hence, the action is topologically faithful. 

We show by induction on $n$ that this action has only finitely many orbits of $n$-tuples,
for all $n$. Since the action is transitive, this is true 
for $n=1$. For the induction step, fix $\bar a = (a_1,\dots,a_n) \in D^n$, 
and let $c$ be an arbitrary element
from $D \setminus \{a_1,\dots,a_n\}$. 
By Roelcke precompactness of $\bf G$, 
there exists a finite set
$\{f_1,\dots,f_k\} \subseteq G$ such that $G = G_{\bar ac} f_1 G_{\bar ac} \cup \dots \cup G_{\bar ac} f_k G_{\bar ac}$.
Let $B(\bar a)$ be $\{f_1 \cdot c,\dots,f_k \cdot c\}$.

{\bf Claim 1.} For every $d \in D \setminus \{a_1,\dots,a_n\}$ there is an $h \in G_{\bar a}$ and $b \in B(\bar a)$ such that $d=h \cdot b$. By transitivity of $G$, there is a $g \in G$ so that $d=g \cdot c$, 
for arbitrary $d \in D \setminus \{a_1,\dots,a_n\}$.
Let $i,h_1,h_2$ be such that $h_1,h_2 \in G_{\bar ac}$ and $g=h_1 f_i h_2$. Then $d=gc=h_1 f_i h_2 \cdot c = h_1 f_i \cdot c$, proving Claim 1. 

{\bf Claim 2.} When $\{\bar a_1,\dots,\bar a_s\}$ is a complete set of representatives for the orbits of $n$-tuples of the permutation group $G$, then 
$$\{(\bar a_i,b) \; | \; i \leq s, b \in B(\bar a_i)\}$$
is a complete set of representatives for the orbits of $(n+1)$-tuples.
Let $(\bar c,d) \in D^{n+1}$. By assumption there exists $g \in G$ such that $g \cdot \bar a_i = \bar c$. Find $h \in G_{\bar a_i}$ and $b \in B(\bar a_i)$ such that $g^{-1} \cdot d=h \cdot b$. Then one has $$gh \cdot (\bar a_i,b) = g \cdot (\bar a_i,h \cdot b) = (\bar c,d) \; .$$
This shows that $G$ has finitely many orbits of $(n+1)$-tuples, and concludes the induction step.

The implication from (4) to (5) follows from
Corollary~\ref{cor:galois}: let 
$D$ be the countably infinite set on which ${\bf G}$
acts continuously, transitively, and topologically faithfully. Then
the set of permutations of $D$ induced by this action is a closed oligomorphic permutation group, and hence 
the automorphism group of an
$\omega$-categorical relational structure with domain $D$. Since the action is transitive, the structure has only one orbit.
\end{proof}

Note that the groups from Theorem~\ref{thm:tsankov}
must always have continuum cardinality; this follows from the
following and the remarks after Lemma~\ref{lem:interpret}. 

\begin{theorem}[Corollary 4.1.5 in~\cite{Hodges}]
\label{thm:auto-cardinality}
Let $\bB$ be a countable structure. Then the following are equivalent.
\begin{enumerate}
\item $|\Aut(\bB)| \leq \omega$
\item $|\Aut(\bB)| < 2^\omega$
\item There is a finite tuple $\bar a$ in $\bB$ such that $|\Aut((\bB,\bar a))|=1$ 
\end{enumerate}
\end{theorem}

% FROM Kaye, Macpherson in Auto book
% If G and H are Polish groups, $G$ has the small index
% property, and phi: G \rightarrow H is a homomorphism of 
% abstract groups, then phi is continuous
% Proof: If K < H is open, it has index at most alpha_0 in H,
% so its inverse image phi^{-1} H has index at most alpha_0 
% in G and hence is open. 

% Does something like this work for clones? What might be the concept of "index" there? Something that
% * transfers under homomorphism
% * implies openness of pre-images of homomorphisms

\section{Bi-interpretations}
\label{sect:bi-interpret}
When two $\omega$-categorical structures share the same topological automorphism group, then
the relationship between the two structures can be described model-theoretically. 

\begin{theorem}[Ahlbrandt and Ziegler~\cite{AhlbrandtZiegler}]\label{thm:bi-interpret}
Two $\omega$-categorical structures $\bB$ and $\bC$ are bi-interpretable if and only if $\Aut(\bB)$ and $\Aut(\bC)$ are isomorphic as topological groups.
\end{theorem}

The subgroup of ${\bf G}$ consisting of the identity element only is called
\emph{trivial}, and subgroups of ${\bf G}$ that are distinct from ${\bf G}$ are called \emph{proper}.

\begin{example}\label{expl:not-bi-interpret}
The structures $\bC := ({\mathbb N}^2; \{(x,y),(u,v) \; | \; x=u\})$
and $\bD := ({\mathbb N};=)$ are mutually primitive positive interpretable,
but \emph{not} bi-interpretable. To see this, observe that $\Aut(\bC)$ has
a proper non-trivial closed normal subgroup $\fN$
such that $\Aut(\bC)/\fN$ is isomorphic to $\Aut(\bD)$ 
(see Proposition~\ref{prop:closed-normal-subgroups}), 
whereas $\Aut(\bD)$, 
the symmetric permutation group of a countably infinite set, has 
no proper non-trivial closed normal
subgroups (it has exactly three proper non-trivial normal subgroups~\cite{SchreierUlam}, none of which is closed). \qed
\end{example}

Theorem~\ref{thm:bi-interpret} has many 
consequences. For instance, it shows
in combination with Theorem~\ref{thm:tsankov}
that every
$\omega$-categorical structure is bi-interpretable
with an $\omega$-categorical structure that has
only one orbit. 
Ahlbrandt and Ziegler also showed the following. 

\begin{theorem}[From~\cite{AhlbrandtZiegler}; also see Theorem 5.3.5 and 7.3.7 in~\cite{HodgesLong}]
\label{thm:az}
  Let $\bC$ be an $\omega$-categorical structure with at least two
  elements. Then a structure
  $\bB$ has a first-order interpretation in $\bC$
  if and only if $\bB$ is the reduct of a structure $\bB'$ 
  such that there is a surjective continuous
  group homomorphism $f \colon \Aut(\bC) \to \Aut(\bB')$. 
  %such that
  %the image of $f$ has finitely many orbits in its action on $\bB$.
\end{theorem}

% Book-TD. THIS WOULD BE A GOOD PLACE FOR AUTOMATIC CONTINUITY?
% 1 -- small index prop, examples
% 2 -- reconstruction up to bi-interpret from small index prop
% 3 -- phenomenon of automatic continuity
% 4 -- set theoretic assumptions to get automatic continuity

Several fundamental properties of $\omega$-categorical structures $\bB$ are preserved by bi-interpretability, and therefore, by Theorem~\ref{thm:bi-interpret}, only depend on the topological automorphism group of $\bB$. 
As we will see in Chapter~\ref{chap:ramsey}, this is for instance the
case for the property whether an ordered homogeneous structure
has the Ramsey property. We give another property of this type.

%BOOKTD: the following facts have primitive positive analogs!

\begin{definition}
Let $\bB$ be an $\omega$-categorical structure.
Then $\bB$ has \emph{essentially infinite signature}\ if every relational structure $\bC$ that is interdefinable with $\bB$ (equivalently, has the same set of automorphisms as $\bB$, by Proposition~\ref{prop:inv-aut-omega-cat}) has an infinite signature. 
%\item $\bB$ is \emph{homogeneous in a finite signature}
%if $\bB$ is interdefinable with a homogeneous
%structure $\bB$ of finite (but not necessarily relational) signature.
%\end{itemize}
\end{definition}

%The proofs of the following statements are easy and similar to the proof of Lemma~\ref{lem:transfer}, and left to the reader. 

We show that the property to have 
essentially infinite language is preserved by
bi-interpretability. 

\begin{proposition}\label{prop:finite-languages}
Let $\bB$ and $\bC$ be $\omega$-categorical structures that are first-order bi-interpretable. 
%\begin{itemize}
%\item 
Then $\bB$ has essentially infinite signature if and only if $\bC$ has.
%\item If $\bB$ is homogeneous in a finite signature, then so is $\bC$.
% PROOF WAS FALSE, don't know whether
% this is true.
%\item If the number of orbits of $n$-subsets in $\bB$ is bounded by $O(2^{P(n)})$ for a polynomial $P$, then so is the number of orbits of $n$-subsets in $\bC$. 
% This already follows from mutual interpretability, is less interesting!
%\end{itemize}
\end{proposition}

\begin{proof}
Let $\tau$ be the signature of $\bB$.
We have
to show that if $\bC$ has finite signature,
then $\bB$ is interdefinable with a structure $\bB'$ with a finite signature.
Let $\sigma \subseteq \tau$ be the set of all relation symbols that
appear in all the formulas of the interpretation of $\bC$ in
$\bB$. Since the signature of $\bC$ is finite, the
cardinality of $\sigma$ is finite as well.

We will show that there is a first-order definition of $\bB$
in the $\sigma$-reduct $\bB'$ of $\bB$. Suppose that the interpretation
$I_1$ of $\bC$ in $\bB$ is $d_1$-dimensional, and that
the interpretation
$I_2$ of $\bB$ in $\bC$ is $d_2$-dimensional. 
Let $\theta(x,y_{1,1},\dots,y_{d_1,d_2})$ be the formula 
that shows that $I_2 \circ I_1$ is homotopic to the identity interpretation.
That is, $\theta$ defines in $\bB$ the $(d_1d_2+1)$-ary relation 
that contains a tuple $(a,b_{1,1},\dots,b_{d_1,d_2})$ iff 
$$a=h_2(h_1(b_{1,1},\dots,b_{1,d_2}),\dots,
h_1(b_{d_1,1},\dots,b_{d_1,d_2})) \; .$$

Let $\phi$ be an atomic $\tau$-formula with $k$ free variables $x_1,\dots,x_k$. We will specify a 
%primitive positive 
$\sigma$-formula that
is equivalent to $\phi$ over $\bB'$. 
%Let $\psi$ be the 
%is interpreting $R$ over $\bC$ ($\phi_R$ has arity $kd_2$). Then the relation $(\phi_R)^{I_1}$ has arity $kd_1d_2$. We claim that the formula 
\begin{align*}
\exists y_{1,1}^1,\dots,y_{d_1,d_2}^k \; \big( & \bigwedge_{i \leq k} \theta(x_i,y^i_{1,1},\dots,y^i_{d_1,d_2}) \\
& \wedge \; \phi_{I_1 I_2} (y_{1,1}^1,\dots,y_{1,d_2}^k,y_{2,d_2}^1,\dots,y_{2,d_2}^k,\dots,y_{d_1,d_2}^k) \big )
\label{eq:bi-interpret}
\end{align*}
is equivalent to $\phi(x_1,\dots,x_k)$ over $\bB$. 
Indeed, by surjectivity of $h_2$, for every
element $a_i$ of $\bB$ there
are elements $c^i_{1},\dots,c^i_{d_2}$
of $\bC$ such that $h_2(c^i_{1},\dots,c^i_{d_2})=a_i$, and by surjectivity of $h_1$, for every
element $c^i_j$ of $\bC$ there are elements
$b^i_{1,j},\dots,b^i_{d_1,j}$ of $\bB$ such that
$h_1(b^i_{1,j},\dots,b^i_{d_1,j})=c^i_j$.
Then 
\begin{align*}
\bB \models R(a_1,\dots,a_k) \; \Leftrightarrow \; & 
\bC \models \phi_{I_2}(c^1_1,\dots,c_{d_2}^1,\dots,c_1^{k},\dots,c_{d_2}^k) \\
\Leftrightarrow \; & \bB' \models \phi_{I_1I_2}(b_{1,1}^1,\dots,b_{1,d_2}^k,b_{2,d_2}^1,\dots,b_{2,d_2}^k,\dots,b_{d_1,d_2}^k) 
\end{align*}
\end{proof}

Note that if $\bB$ and $\bC$ are
 $\omega$-categorical structures that are even primitive positive bi-interpretable, then 
 the above proof even shows that 
 $\bB$ is primitively positive interdefinable with
 a structure with a finite domain if and only
 if $\bC$ is.

Proposition~\ref{prop:finite-languages} shows via Theorem~\ref{thm:bi-interpret} that 
having essentially infinite signature only depends on the topological automorphism group of $\bB$. But unlike the property of $\omega$-categoricity and Theorem~\ref{thm:tsankov}, we are not aware
of any elegant characterization of those properties that is directly stated in terms of the topological group. 

%We remark that it is not true that being homogeneous in a finite \emph{relational}Êsignature is a property 
%that does \emph{not}Êonly
%depend on the topological automorphism group. (...)

%\section{Topological Clones}

\chapter{Ramsey Theory}
\label{chap:ramsey}
\begin{center}
\includegraphics[scale=0.4]{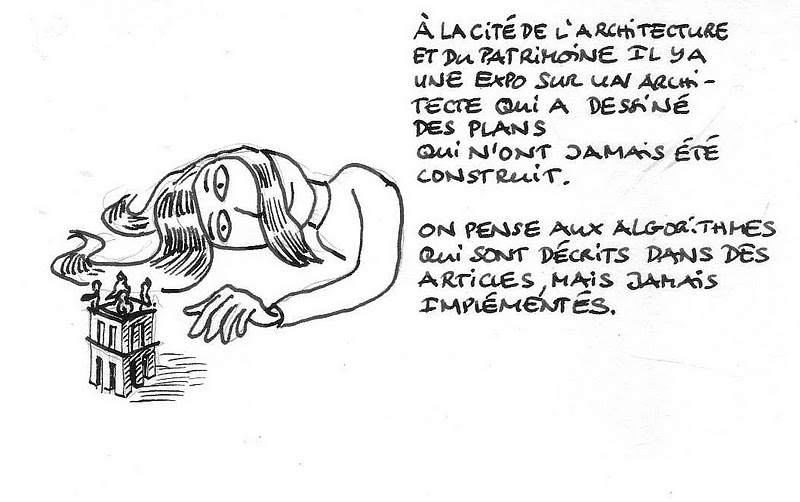}
\end{center}

\vspace{-.5cm}
The application of Ramsey theory to study the expressive power
of constraint languages via polymorphisms is one of the central contributions of this thesis. 
The idea is that polymorphisms
must behave in a regular way on large parts of their domain. 
This also leads us to decidability results for several
meta-questions about the expressive power of constraint
languages. The same idea can be used to show statements
of the type `every polymorphism that violates $R$ must locally generate $g$', for certain fixed operations $g$ with good properties.
Such statements will be crucial in the classification projects in Chapters~\ref{chap:schaefer} and~\ref{chap:tcsp}.

In this chapter we first revisit classical concepts and results from structural Ramsey theory in Section~\ref{sect:ramsey-classes}. 
In order to apply Ramsey theory to analyze polymorphism clones, we need the product Ramsey theorem, but also
other fundamental facts from Ramsey theory, some of which appear to be new (such as Corollary~\ref{cor:Ramsey-constants}). 
Those facts will be derived from a recently discovered fundamental connection between Ramsey theory and topological dynamics
due to Kechris, Pestov, and Todorcevic~\cite{Topo-Dynamics}. 
This connection allows a more systematic understanding of Ramsey-theoretic principles, and we present it in Section~\ref{sect:extr-amen}.  The way in which we apply Ramsey theory to polymorphisms is described in Section~\ref{sect:canonization}. 
We close with an application of our technique in Section~\ref{sect:decidability}, and prove 
the decidability of various \emph{meta-problems} concerning constraint satisfaction problems,
that is, problems where the input is a description of a template $\bB$, and the question is whether
the corresponding CSP has certain properties (for instance, whether certain relations are primitive positive definable).

Some of the results presented here have been published
in~\cite{BPT-decidability-of-definability}; there is also
a survey article~\cite{BP-reductsRamsey} that additionally covers the applications 
of our technique for the classification of `the first-order reducts' of a given homogeneous structure $\bC$, that is, the structures that are first-order definable in $\bC$.

\section{Ramsey Classes}
\label{sect:ramsey-classes}
This section is about classes $\C$ of finite structures that 
satisfy the following \emph{Ramsey-type} property:
for all $\bA,\bB \in \C$ there exists a $\bC \in \C$ such that $\bB$ embeds into $\bC$, and
when we assign finitely many `colors' to the substructures of $\bC$ that are isomorphic to $\bA$, then we can find a `monochromatic' copy of $\bB$ in $\bC$, 
i.e., an induced substructure of $\bC$ that is isomorphic to $\bB$ and in which all copies of $\bA$ in this substructure have the same color.
Before we formalize this in detail, we give the classical result of Ramsey, which provides a prototype of a class with the Ramsey property.

From now on, we denote the set $\{1,\dots,n\}$ also by $[n]$.
Subsets of a set of cardinality $m$ will be called $m$-subsets in the following. 
Let ${S \choose m}$ denote the set of all $m$-subsets of $S$. We also refer to mappings $f \colon {S \choose m} \rightarrow [r]$ as a \emph{coloring}
of $S$ (with the \emph{colors} $[r]$).

\begin{theorem}[Ramsey's theorem]\label{thm:inf-ramsey}
Let $B$ be a countably infinite set, and let $m,r$ be finite integers. 
%When $\chi$ is a mapping from 
%the $m$-element subsets of $B$ into $[r]$, then
For every $\chi \colon {B \choose m} \rightarrow [r]$
there exists an infinite $P \subseteq B$ such that $\chi$ is constant
on all $m$-element subsets of $P$.
\end{theorem}

A proof of Theorem~\ref{thm:inf-ramsey} can be found in~\cite{Hodges} (Theorem 5.6.1); for a broader introduction to Ramsey theory see~\cite{GrahamRothschildSpencer}.
It is easy to derive the following finite version of Ramsey's theorem
from Theorem~\ref{thm:inf-ramsey} via a compactness argument.

\begin{theorem}[Finite version of Ramsey's theorem]\label{thm:fin-ramsey}
For all positive integers $r,m,k$ there is a positive integer $l$ such that for every $\chi \colon {[l] \choose m} \rightarrow [r]$ 
there exists a $k$-subset $S$ of $[l]$ such that $\chi$ is constant
on ${S \choose m}$.
\end{theorem}
\begin{proof}
A proof by contradiction: suppose that there are positive integers $r,m,k$ 
such that for all positive integers $l$ there is a $\chi \colon {[l] \choose m} \rightarrow [r]$
such that for all $k$-subsets $S$ of $[l]$ the mapping $\chi$ is not constant
on ${S \choose m}$. Since the property that for all $k$-subsets $S$ of $[l]$ the mapping $\chi$ is not constant
on ${S \choose m}$ is universal first-order, and since the image of $\chi$ is finite,
we can apply Lemma~\ref{lem:omega-cat-compactness} and get the existence
of a mapping $\chi$ with the same property but defined on all integers. 
This contradicts Theorem~\ref{thm:inf-ramsey}.
\end{proof}

We write ${\bf R}(r,m,k)$ for the \emph{smallest} number $l$
whose existence is asserted by Theorem~\ref{thm:fin-ramsey}.

More generally, when $\bA$ and $\bB$ are $\tau$-structures, we write
${\bB \choose \bA}$ for the set of all induced substructures of $\bB$
that are isomorphic to $\bA$. 
When $\bA,\bB,\bC$ are $\tau$-structures, and
$r \geq 1$ is finite, 
we write $$\bC \rightarrow (\bB)^{\bA}_r$$ if for all 
$\chi \colon {\bC \choose \bA} \rightarrow [r]$ 
there exists $\bB' \in {\bC \choose \bB}$ such that $\chi$ is constant
on ${\bB' \choose \bA}$.

\begin{definition}
A class of relational structures that is closed
under isomorphisms and induced substructures
is called \emph{Ramsey} if for every $\bA,\bB \in \cal C$ and for every finite $k \geq 1$ there exists a $\bC \in \cal C$ such that $\bB$ embeds into $\bC$, and $\bC \rightarrow (\bB)^{\bA}_k$. 
\end{definition}

Our first example of a Ramsey class is the class of all finite
structures over the empty signature; this is an immediate consequence of Theorem~\ref{thm:fin-ramsey}. We also observe the following. Recall that the \emph{age}
of a $\tau$-structure $\bB$ is the class of all finite $\tau$-structures that embed into $\bB$. 
%Theorem~\ref{thm:fin-ramsey} also implies the following.

\begin{corollary}
The age of $({\mathbb Q};<)$ is Ramsey.
\end{corollary}
\begin{proof}
This is again a direct consequence of Theorem~\ref{thm:fin-ramsey},
since whether or not an $m$-element substructure is isomorphic
to an $n$-element substructure of $({\mathbb Q};<)$ only depends
on $n$ and $m$. 
\end{proof}

We will now present further examples of Ramsey classes;
the proofs are non-trivial and fall out of the scope 
of this thesis, but we provide references.

\begin{example}\label{expl:atomless}
The class of all finite Boolean algebras 
$\bB = (B;\sqcup,\sqcap,c,{\bf 0},{\bf 1})$ has the Ramsey property. 
This is explicitly stated in~\cite{Topo-Dynamics}, page 147, line 3ff (see also page 112, line 9ff), where it is observed that this follows from a result by Graham and Rothschild~\cite{GrahamRothschild}. \qed
\end{example}

It might be instructive to also see an example of a class of structures that
is \emph{not} Ramsey. Typical examples come from 
classes 
that contain structures with non-trivial automorphism groups, 
as in the following.

\begin{example}\label{expl:graphs}
The class of all finite graphs is not a Ramsey class.
To see this, let $\bA$ be the (undirected) graph $\big (\{0,1,2\}; \{\{0,1\},\{1,2\}\} \big)$
(since the edge relation is symmetric we write edges as 2-element subsets of the vertices)
with three vertices and two edges, and let $\bB$ be $C_4$,
that is, the undirected four-cycle 
$$\big (\{0,1,2,3\}; \{\{0,1\},\{1,2\},\{2,3\},\{3,0\}\} \big) \; .$$ 
Let $\bC$ be an arbitrary graph. 
We want to show that there is a way to color the copies of $\bA$
in $\bC$ without producing a monochromatic copy of $\bB$.
For that, fix an arbitrary linear 
order $<$ on the vertices of $\bC$. 
We define a coloring $\chi \colon {\bC \choose \bA} \rightarrow \{0,1\}$ 
as follows. If there is an embedding $h$ of $\bA$ into $\bC$
such that $h(0)<h(1)<h(2)$, 
then we color the corresponding copy of $\bA$ in $\bC$ with $0$;
all other copies of $\bA$ in $\bC$ are colored by $1$.
We claim that any copy of $\bB$ in $\bC$ contains a copy
of $\bA$ that is colored by $1$, and one that is colored by $0$.
The reason is that for any ordering of the vertices of $\bB$
there is an embedding $h'$
of $\bA$ into $\bB$ such that $h'(0)<h'(1)<h'(2)$, and an embedding
$h''$ of $\bA$ into $\bB$ such that not $h''(0)<h''(1)<h''(2)$. 
Hence, $\bC \not\rightarrow (\bB)^\bA_2$. \qed
\end{example}

Frequently, a class without the Ramsey property
can be made Ramsey by expanding its members appropriately 
with a linear ordering. %(Section~\ref{ssect:kpt} will explain this phenomenon). 
We will see several examples.

\begin{example}\label{expl:all-structs-Ramsey}
\Nesetril\ and R\"odl~\cite{NesetrilRoedlOrderedStructures} and independently Abramson and Harrington~\cite{AbramsonHarrington} showed that for any relational signature $\tau$, the class $\cal C$ of all finite \emph{ordered} $\tau$-structures is 
a Ramsey class. That is, the members of $\cal C$ are finite structures 
$\bA = (A; <, R_1,R_2,\dots)$ for some fixed signature $\tau = \{<,R_1,R_2,\dots\}$, where $<$ is a total linear order of $A$.

A shorter and simpler proof of this substantial result can 
be found in~\cite{NesetrilRoedlPartite} and \cite{NesetrilSurvey};
the proof there uses the \emph{partite method},
which uses amalgamation to reduce the statement to proving the so-called \emph{partite lemma}; the proof of the partite lemma relies on the Hales-Jewett theorem from Ramsey theory (see~\cite{GrahamRothschildSpencer}).\qed
\end{example} 

\begin{example}\label{expl:ordered-c-relation}
Recall the homogeneous structure $\bB = (B; |)$ carrying a $C$-relation, 
introduced in Section~\ref{ssect:c-relation}. %of Chapter~\ref{chap:mt}, 
%In the following, we use $uv|w$ as a shortcut for $w|uv$.
We consider the expansion of $\bB$ by a linear
order, defined as follows. 
It is easy to see that for every finite tree $\bT$ there is an ordering
$<$ on the leaves $L$ of $\bT$ such that for all $u,v,w \in L$ with
$u<v<w$ we have either $u|vw$ or $uv|w$ (recall Definition~\ref{def:bar}).
This can be seen from the obvious existence of an
embedding of $\bT$ on the plane so that all leaves lie on a line and
none of the edges cross, and take the linear order induced by the line. 
We call such an ordering of $L$ \emph{compatible} with $\bT$.
% BOOKTD: wondering whether it makes sense to rather call it "convex"
The class of all finite substructures $\bC$ of $\bB$ expanded by 
a compatible ordering of the underlying tree of $\bC$
is a Ramsey class; this follows from more general results by Milliken
(Theorem 4.3 in~\cite{Mil79}, building
on work in~\cite{DeuberTreeRamsey}), and has been shown explicitly in~\cite{BodirskyPiguet}. \qed
%The class of all finite substructures $\bC$ of $\bB$, expanded by 
%a compatible ordering of the underlying tree of $\bC$,
%is again an amalgamation class, and the \Fresse-limit of this class
%is an expansion of $\bB$. 
% LATERTD: the second claim should follow from
% a general "amalgamation-expansion lemma"? 
% (an inverse of a lemma on reasonable order classes 
% from KPT)
\end{example}

\begin{example}\label{expl:irreducible-forbidden-Ramsey}
The Ramsey classes from Example~\ref{expl:all-structs-Ramsey} have been further generalized by \Nesetril\ and R\"odl as follows~\cite{NesetrilRoedlOrderedStructures}.
Suppose that $\mathcal N$ is a (not necessarily finite) class of %relational 
structures 
with finite 
% LATER-TD Check: discuss infinite signatures here? 
relational signature $\tau$ 
whose Gaifman graph (Definition~\ref{def:gaifman}) is a clique -- such structures
have been called \emph{irreducible} in the Ramsey theory literature.
It can be readily verified that $\C := \Forb(\mathcal N)$ is an amalgamation class. Then the class of all
expansions of the structures in $\C$ by a linear order is a Ramsey class;
again, this can been shown by the partite method~\cite{NesetrilRoedlPartite}.
This example is indeed a generalization of Example~\ref{expl:all-structs-Ramsey} since we obtain the previous result by taking ${\mathcal N} = \emptyset$. \qed
\end{example}

%\begin{example} %hmh, to be worked out
%Finite semi-linear orders with a linear extension and a 
%compatible $C$-relation.
%\end{example}

%\begin{example}
%Finite posets with a linear extension.
%\end{example}
% Don't have a reference at hand;
% didn't check the proof;
% don't need it right now since the corresponding CSPs have
% not been mentioned.

%\begin{definition}
%Let $\mathcal C$ be a class of $\tau$-structures, for a finite
%relational signature $\tau$. Then $\mathcal C$ is called
%a \emph{truncated Ramsey class} if there exists a relational
%signature $\sigma$ that contains $\tau$, and a class $\mathcal D$

The fact that all the above Ramsey classes could be described
as the age of a homogeneous structure is not a coincidence.
We have the following (the proof is from~\cite{RamseyClasses}, 
and presented here for the convenience of the reader).

\begin{theorem}[of~\cite{RamseyClasses}]\label{thm:ramsey-amalgamation}
Let $\tau$ be a relational signature, and let $\mathcal C$ be a class of ordered
finite $\tau$-structures that is closed under induced substructures, isomorphism, 
and has the joint embedding property (see Section~\ref{sect:fraisse}). 
If $\mathcal C$ is Ramsey,
then it has the amalgamation property.
\end{theorem}
\begin{proof}
Let $\bA,\bB_1,\bB_2$ be members of $\mathcal C$
such that $\bA$ is an induced substructure of both $\bB_1$ and
$\bB_2$. Since $\cal C$ has the joint embedding property, 
there exists a structure $\bC \in \cal C$
with embeddings $e_1,e_2$ of $\bB_1$ and $\bB_2$ into $\bC$.
If $e_1,e_2$ have the same restriction to $\bA$, then we are done, so
assume otherwise. 

Let $\bD \in \mathcal C$ be such that $\bD \rightarrow (\bC)^{\bA}_2$. 
Define a coloring $\chi \colon {\bD \choose \bA} \rightarrow \{1,2\}$ as follows. When $\bA' \in {\bD \choose \bA}$, and $f \colon \bA \rightarrow
\bA'$ is an isomorphism, then $\chi(\bA')=1$ if and only if there
 is an embedding $h$ of $\bC$ into $\bD$
such that $f = h \circ e_1$. 

Since $\bD \rightarrow (\bC)^{\bA}_2$, there exists $\bC' \in {\bD \choose \bC}$, witnessed by an embedding $h'$ of $\bC$ into $\bD$
such that $\chi$ is constant on ${\bC' \choose \bA}$. Now any copy of $\bC$ in $\bD$ contains a copy $\bA'$ of $\bA$ with $\chi(\bA')=1$.
Thus $\chi$ is constant $1$ on ${\bC' \choose \bA}$. 
%Let $h'$ be the isomorphism between $\bC$ and $\bC'$. 

Consider the embedding $h' \circ e_2$ of $\bA$ into
$\bD$; as we have seen above, the corresponding copy of $\bA$
in $\bD$ is colored $1$. Thus there exists an embedding $h''$ of
$\bC$ into $\bD$ such that $f=h'' \circ e_1 = h' \circ e_2$ (here we use the assumption
that the structure $\bA$ is ordered). This shows that $\bD$ together with the embeddings 
$h'' \circ e_1 \colon B_1 \rightarrow D$ and $h' \circ e_2 \colon B_2 \rightarrow D$ is the amalgam
of $\bB_1$ and $\bB_2$ over $\bA$.
\end{proof}

It is often convenient to work with the \Fresse-limit
of a Ramsey class rather than the class itself. 
Indeed, we have the following.
% BOOKTD: reduce to the 2-color case!

\begin{proposition}
Let $\mathcal C$ be an amalgamation class, and let
$\bC$ be the \Fresse-limit of $\mathcal C$. Then $\mathcal C$ has the Ramsey property if and only if $\bC \rightarrow (\bB)^\bA_k$
for all $\bA,\bB \in \mathcal C$, and $k \geq 2$.
\end{proposition}
\begin{proof}
Let $\bA,\bB \in \mathcal C$, and $k \geq 2$ an integer. When $k$ is the cardinality of ${\bB \choose \bA}$, then for any structure $\bD$ 
the fact that $\bD \rightarrow (\bB)^\bA_r$ can equivalently be expressed in terms of $r$-colorability of a certain $k$-uniform hypergraph, defined as follows. Let $\bG=(V;E)$ be the structure whose vertex set $V$ is ${\bD \choose \bA}$,
and where $(\bA_1,\dots,\bA_k) \in E$ if there exists a
$\bB' \in {\bD \choose \bB}$ such that ${{\bB' \choose \bA}} = \{\bA_1,\dots,\bA_k\}$.
Let $\bH = ([r];E)$ be the structure where $E$ contains all tuples except the tuples $(1,\dots,1),\dots,(r,\dots,r)$. 
Then $\bD \rightarrow (\bB)^\bA_r$ if and only if $\bG$ homomorphically maps to $\bH$.
By Lemma~\ref{lem:infinst}, this is the case if and only if all finite substructures of $\bG$ homomorphically map to $\bH$.
Thus, $\bC \rightarrow (\bB)^\bA_r$ if and only if 
$\bC' \rightarrow (\bB)^\bA_r$ for all finite substructure $\bC'$ of $\bC$.
\end{proof}

%We say that a homogeneous structure is \emph{Ramsey} if it is the \Fresse-limit of a Ramsey class.
When $\bB$ is a homogeneous structure with a finite relational signature whose age is a Ramsey class, 
then this fact is useful for studying which relations are primitive positive definable in $\bB$, 
as we will see for instance in Section~\ref{sect:decidability}.
In fact, for those applications of Ramsey theory 
 it suffices that $\bB$ can be expanded
to a structure $\bC$ that is homogeneous, Ramsey, and has a finite relational signature. The following question has been asked.
%We make the following conjecture. 

\begin{question}\label{quest:expand}
Let $\bB$ be a homogeneous structure with a finite relational
signature. Then there is a homogeneous 
expansion $\bC$ of $\bB$ with finite relational signature
whose age has the Ramsey property.
\end{question}

% several variations possible, but this one seems to make
% most sense. One variation would be to define "truncated Ramsey
% classes" (the AGE can be expanded by finitely many relations
% so that we have an Ramsey class after that (each structure might be expanded in several ways))
% another variant is to allow infinite signature: but then we should
% require omega-categoricity, it is no longer implied.
% Without omega-categoricity it is trivially true: make the 
% automorphism group trivial, and we have extreme amenability for 
% trivial reasons. 
% The good thing about this formulation:
% it puts rather strong assumptions (homogeneous in finite
% language) that is still rich seen from a CSP complexity perspective,
% at the same time we require something that will definitely be useful
% for the universal algebraic approach.)

\section{Extremely Amenable Groups}
\label{sect:extr-amen}
This section presents a link between Ramsey classes and Polish groups that are \emph{extremely amenable}. The link rests on the theorem of Kechris, Pestov, and Todorcevic 
that characterizes those ordered homogeneous structures that are Ramsey
in terms of their topological automorphism group, and will be presented
in Section~\ref{ssect:kpt}.
In Section~\ref{ssect:cont-hom}, \ref{ssect:products}, and~\ref{ssect:open}, we use this result to obtain a more systematic understanding of Ramsey classes.

\subsection{Extreme Amenability}
\label{ssect:kpt}
The Ramsey property for ordered homogeneous
structures $\bB$ has an elegant characterization in terms of the topological automorphism group of $\bB$: %, due to Kechris, Pestov, and Todorcevic:
the age of $\bB$ is Ramsey if and only if the topological automorphism group of $\bB$ is extremely amenable. Extreme amenability is a concept from group theory studied since the 60s~\cite{Granirer}.

\begin{definition}
    A topological group is \emph{extremely amenable} iff every continuous action of the group on a compact Hausdorff space has a fixed point.
\end{definition}

%Let $\bB$ be a homogeneous relational structure. 
%When $O_1$ and $O_2$ are orbits of finite subsets of $B$,
%we write $O_1 \preceq O_2$ if there are $F_1 \in O_1$ and 
%$F_2 \in O_2$ such that $F_1 \subseteq F_2$.

The following is the combination of Proposition~4.2, Proposition~4.3, Theorem~4.5,
and Theorem~4.7 from~\cite{Topo-Dynamics}.

\begin{theorem}[Kechris, Pestov, Todorcevic~\cite{Topo-Dynamics}]\label{thm:kpt}
    Let $\bB$ be a homogeneous relational structure, 
    and let $\bf G$ be the topological automorphism group of $\bB$. 
    Then the following are equivalent. 
    \begin{enumerate}
    \item The age of $\bB$ has the Ramsey property, and 
    $G$ preserves a linear order $<$ on $B$. 
    \item The age of $\bB$ only contains rigid structures, and has the Ramsey property.
    \item \emph{(a)} For any finite subset $F$ of $B$, the substructure induced 
    by $F$ in $\bB$ is rigid, and
    \emph{(b)} for all orbits $O_1,O_2$ of finite subsets in $\bB$, and 
    for every $\chi \colon O_1 \rightarrow [r]$ there is an $i \leq r$
    and an $F \in O_2$ such that $\chi(F')=i$ for all $F' \subseteq F$ where $F' \in O_1$. 
    \item For any open subgroup $\bf V$ of $\bf G$, every $\chi \colon {\bf G}/{\bf V} \rightarrow [r]$, and every finite $A \subseteq {\bf G}/{\bf V}$ there is $g \in G$ and $1 \leq i \leq r$ such that $\chi(ga)=i$ for all $a \in A$.  
    \item $\bf G$ 
    is extremely amenable.
\end{enumerate}
\end{theorem}

For structures $\bB$ that are not homogeneous
but $\omega$-categorical, the equivalence between (3), (4), and (5) remains valid, since every
$\omega$-categorical structure has a homogeneous expansion by first-order definable relations -- and such an expansion has the same automorphism group as $\bB$; we can then apply Theorem~\ref{thm:kpt} to the expansion. 
It therefore makes sense to call an $\omega$-categorical structure $\bB$ \emph{Ramsey}
if the age of the expansion of $\bB$ by all first-order definable relations
is Ramsey. 
There are also interesting applications
of Theorem~\ref{thm:kpt} when $\bB$ is \emph{not} $\omega$-categorical, see~\cite{Topo-Dynamics}.

%The assumption that 
%$\bB$ has a relation that denotes the total order is
%relatively natural. In fact, we have have the following. 
%\begin{proposition}
%Let $\mathcal C$ be a Ramsey class of rigid structures.
%Then the \Fresse-limit of $\mathcal C$ has a first-order definable
%linear order.
%\end{proposition}
%\begin{proof}
%By Theorem~\ref{thm:kpt}, the automorphism group $\cC$
%of the \Fresse-limit of $\mathcal C$ is extremely amenable. 
%Let $\bB$ be the \Fresse-limit of $\mathcal C$. 
%The statement now 
%follows from Proposition 4.3. in~\cite{Topo-Dynamics}, which says that for any closed permutation group $\cG$ the following are equivalent:
%\begin{itemize}
%\item $G$ is extremely amenable;
%\item For an finite non-empty subset $F$ of the domain of $\cG$, 
%we have $\cG_{(F)} = \cG_F$, and for any two orbits
%\end{itemize}
%\end{proof}

We point out a remarkable 
consequence of Theorem~\ref{thm:kpt} in combination with Proposition~\ref{prop:inv-aut-omega-cat}.

% Relevant example in this context: the binary branching C-relation freely jointed with the homogeneous universal
% tournament. The finite substructures here are rigid, but I see no way to define an order. 
% This is not a problem for the statement, since this structure apparently does not have the Ramsey property. 

\begin{corollary}\label{cor:rigid-order}
Suppose that $\bB$ is an $\omega$-categorical Ramsey structure where all finite induced substructures are rigid. Then a linear order is first-order definable in $\bB$.
\end{corollary}

\ignore{
\begin{proof}
We claim that $\bB$ satisfies in particular property (3) (b) in Theorem~\ref{thm:kpt}.
The statement then follows from the implication (2) $\Rightarrow$ (1) in Theorem~\ref{thm:kpt}.
So suppose that $O_1$ and $O_2$ are orbits of finite subsets in $\bB$, and $\chi \colon O_1 \rightarrow [r]$ is 
any mapping. 
Let $F_2$ be from $O_2$; order the elements of $F_2$ in an arbitrary way, 
say $e_1 < \dots < e_n$, $F_2 = \{e_1,\dots,e_n\}$. 
If there is no $F_1 \subseteq F_2$ with $F_1 \in O_1$, there is nothing to show. 
Otherwise, consider the structure induced by $F_1$ in $\bB$.
This structure is ordered by restricting the order of $F_2$ to $F_1$. Moreover, it is by assumption rigid,
and hence we obtain the same ordering $f_1 < \dots < f_m$ for all subsets $F_1 = \{f_1,\dots,f_m\}$ of $F_2$ that are in $O_1$. 
Let $P_1$ be the orbit of $(f_1,\dots,f_m)$ in $\bB$. 
Define $\chi' \colon P_1 \rightarrow [r]$ by $\chi'((f_1',\dots,f_m'))=\chi(\{f_1',\dots,f_m'\})$. 
Since $\bB$ is Ramsey, we find an $n$-tuple $(e_1',\dots,e_n')$ in the same orbit as $(e_1,\dots,e_n)$ so that 
$\chi'$ is constant on
all the $m$-tuples from $P_1$ with entries from $(e'_1,\dots,e_n')$.
This shows that $\chi$ is constant for all $F_1 \subseteq F_2$ where $F_1 \in O_1$. 
\end{proof}
}

When a Ramsey structure $\bB$ has substructures with non-trivial automorphisms, Theorem~\ref{thm:kpt} is therefore
not directly applicable. For the applications of Ramsey theory we have in mind, 
though, it is sufficient
to know that $\bB$ has a first-order definition in an ordered Ramsey structure that is homogeneous in a finite relational signature (see Section~\ref{sect:canonization} and~\ref{sect:decidability}).

The choice of the order is not arbitrary, but plays an important role when
we want to preserve the Ramsey property.
To give another example, consider again 
the countable atomless Boolean algebra, which is an example of an $\omega$-categorical structure that is 
\emph{not} homogeneous in a finite relational signature (Corollary~\ref{cor:not-finitely-homogeneous}). 
In this case an order expansion with an extremely amenable automorphism group has
been specified in~\cite{Topo-Dynamics}, and can be found below. 
%(the class of all arbitrarily ordered finite Boolean algebras is \emph{not} a Ramsey class).  % REFERENCE? PROOF?

\begin{example}
Let $\bB = (B;\sqcup,\sqcap,c,{\bf 0},{\bf 1})$ be a finite Boolean algebra and $A$ its set of atoms (see Example~\ref{expl:atomless} in Section~\ref{ssect:mt-set-constraints}).  Then every ordering $a_1 < \dots < a_n$ of $A$ gives an ordering of $B$ as follows (we follow~\cite{Topo-Dynamics}).
For $x,y \in B$, we set $x < y$ if there exists an $i_0 \in \{1,\dots,n\}$ such that
\begin{itemize}
\item for all $i \in \{1,\dots, i_0-1\}$ we have that $a_i \sqcap x = a_i \sqcap y$, and
\item $x \sqcap a_{i_0} = {\bf 0}$ and $y \cap a_{i_0} \neq {\bf 0}$.
\end{itemize}
Such an ordering of the elements of $\bB$ is called a \emph{natural ordering}.
It can be shown that the class $\mathcal C$ of all naturally ordered finite atomless Boolean algebras has the Ramsey property 
(see the comments preceding Theorem 6.14 in~\cite{Topo-Dynamics}, and Proposition 5.6 in~\cite{Topo-Dynamics}).
By Theorem~\ref{thm:ramsey-amalgamation}, $\mathcal C$ is an amalgamation class.
The reduct of the \Fresse-limit of $\mathcal C$ with signature $\{\sqcup,\sqcap,c,{\bf 0},{\bf 1}\}$ is the atomless Boolean algebra (Propositions 5.2 and 6.13 in~\cite{Topo-Dynamics}),
so we indeed found an extremely amenable order expansion of the atomless Boolean algebra. \qed
\end{example}

The main focus of the article by Kechris, Pestov, and Todorcevic~\cite{Topo-Dynamics}
is the application of Theorem~\ref{thm:kpt}
to prove that certain groups are extremely amenable,
using known and deep Ramsey results. 
Here we are rather interested in the opposite direction: we are applying Theorem~\ref{thm:kpt} in the following sections
to obtain a more systematic understanding of which classes 
of structures have the Ramsey property. 

%\subsection{Surjective continuous homomorphisms}
%\label{ssect:surj-hom}
\subsection{Continuous Homomorphisms}
\label{ssect:cont-hom}
Interestingly, whether an $\omega$-categorical 
structure $\bB$ is Ramsey only depends on the automorphism
group of $\bB$ viewed as a \emph{topological} group. For this observation we do not need the full power 
of Theorem~\ref{thm:kpt}: the equivalence of (1) and (3) suffices.
More generally, we have the following.

\begin{proposition}
\label{prop:cont-hom}
Let ${\bf G}$ be an extremely amenable group, and let
${\bf H}$ be a Polish group. If there is a continuous
homomorphism $\xi \colon {\bf G} \rightarrow {\bf H}$ such that $\xi(G)$ is dense in $H$, 
then ${\bf H}$ is also
extremely amenable.
\end{proposition}
\begin{proof}
Let $a \colon H \times S \rightarrow S$ be a continuous action of
${\bf H}$ on a compact Hausdorff space $S$. 
Then $b \colon G \times S \rightarrow S$ given by $(g,s) \mapsto a(\xi(g),s)$
is a continuous action of ${\bf G}$ on $S$. Since ${\bf G}$ is extremely amenable,
$b$ has a fixed point $s_0$. Now let $h \in H$ be arbitrary. Since $\xi(G)$ is dense in $H$,
there exists a sequence $(g_i)_{i \geq 1}$ of elements of $G$ such that $\lim_i \xi(g_i)$
converges against $h$ in $H$. 
Therefore there exists an $i_0$ such that
for all $j \geq i_0$ we have that 
$$a(h,s_0) = a(\xi(g_j),s_0) = b(g_j,s_0) = s_0$$
and hence $s_0$ is also a fixed point under $a$.
\end{proof}

\begin{example}
Recall Example~\ref{example:allen}, where we introduced an $\omega$-categorical structure $\bA$ with binary relations
and a two-dimensional first-order interpretation over $({\mathbb Q}; <)$ -- the corresponding relation algebra is also called 
\emph{Allen's Interval Algebra} (see Section~\ref{ssect:allen}). 
% of Chapter~\ref{chap:intro}). %, and this is why we also refer to
%$\bA$ by this term. 

It follows from the proof of Theorem~\ref{thm:allen-lift} that $\bA$ is first-order bi-interpretable
with $({\mathbb Q};<)$. Since the automorphism group of $({\mathbb Q};<)$ is extremely amenable, 
Theorem~\ref{thm:kpt} therefore shows that the automorphism group of $\bA$ is extremely amenable,
and that $\bA$ is Ramsey. Since $\bA$ is homogeneous, 
we also have the Ramsey property for the age of $\bA$. \qed
\end{example}

Unfortunately, Theorem~\ref{thm:kpt} leaves Question~\ref{quest:expand}
%Section~\ref{sect:ramsey-classes} 
unresolved.
An important variant of Question~\ref{quest:expand}
with a reformulation in
terms of topological groups turned out to be false. 

\begin{theorem}[Evans~\cite{Evans-extremely}]
There exists a closed oligomorphic permutation group 
without a closed oligomorphic extremely amenable
subgroup. 
Equivalently, there exists an $\omega$-categorical structure without an ordered 
$\omega$-categorical Ramsey expansion. 
\end{theorem}

% BOOK-TD: add something about wreath products here
\subsection{Products}
\label{ssect:prt}
% BOOK-TD version: do it also with direct limits and infinite products. 
In this section we present an important tool to build new extremely amenable groups, Theorem~\ref{thm:new-extr-amen} below.
Recall the definition of direct products of two groups from Section~\ref{ssect:products}. 
The direct product of two \emph{topological
groups} $\bf G_1$ and $\bf G_2$ is the direct product of the respective abstract groups,
together with the product topology on the group elements.

\begin{theorem}[Proposition 6.7 in~\cite{Topo-Dynamics}]\label{thm:new-extr-amen} 
Let ${\bf G}$ be a topological group. 
\begin{enumerate}
\item Let ${\bf N}$ a %closed: don't need!!
normal subgroup
of ${\bf G}$. If both ${\bf N}$ and ${\bf G}/{\bf N}$ are extremely amenable, then so is ${\bf G}$. 
\item  
When ${\bf G}$ is a finite direct product of extremely amenable groups, then ${\bf G}$ is extremely amenable. 
\end{enumerate}
\end{theorem}
Item (1) has been stated in~\cite{Topo-Dynamics} under the additional
assumption that ${\bf N}$ is \emph{closed}; however, as we see in the proof, this assumption is not necessary.
Item (2) can be generalized to infinite products of groups, but we do not need this here, and refer to~\cite{Topo-Dynamics} instead.

\begin{proof}
We first show (1).
Suppose ${\bf G}$ acts continuously on a compact Hausdorff space $S$. 
Let $S_{\bf N}$ be the subspace of $S$ induced on $\{ x \in S \; | \: h \cdot x = x \text{ for all } h \in N \}$, which is clearly also Hausdorff.
Moreover, it is closed. To see this, let $x \in S$ and $f \in N$ be such that $f \cdot x \neq x$. 
Then there exists an open set $V(x,f) \subseteq S$ that contains $x$ such that $f \cdot y \neq y$ for all $y \in V(x,f)$. 
Otherwise, there exists a sequence $(x_n)_{n \geq 1}$ with $x_n \rightarrow x$ such that $f(x_n)=x_n$ for all $n \geq 1$. 
But $f \cdot x = f \cdot \lim x_n = \lim f \cdot x_n = \lim x_n = x$, a contradiction.
Then $\bigcup_{f \in N, x \in S} V(f,x)$ defines the complement of $S_{\bf N}$ and is open.
So $S_{\bf N}$ is closed, and also compact since $S$ is compact and closed subsets of compact spaces are compact. 

As $\bf N$ is extremely amenable, $S_{\bf N}$ is non-empty. 
The set $S_{\bf N}$ is preserved by the action of $\bf G$ on $S$: when $x \in S_{\bf N}$ and $g \in G$, then for any $h \in N$
$$ h \cdot (g \cdot x) = hg \cdot x = g \cdot (g^{-1}hg) \cdot x = g \cdot x$$
(where the last equality is by normality of $N$ and Proposition~\ref{prop:normal-subgroups}),
and so $g\cdot x \in S_{\bf N}$. 

Now, consider the action of ${\bf G}/{\bf N}$ on $S_{\bf N}$ defined by $(gN) \cdot x = g \cdot x$, which is clearly well-defined.
To verify continuity of this action with Proposition~\ref{prop:continuity},
%we have to verify that the set $\{(g,x) \; | \; gx \in U\}$ is open for any open subset $U$ of $S_{\bf N}$. 
let $(g_nN,s_n) \rightarrow (gN,s)$. 
Then $(g_n N) \cdot s_n = g_n \cdot s_n \rightarrow g \cdot s = (gN) s$ 
since the action of ${\bf G}$ on $S$ is continuous.

By extreme amenability of ${\bf G}/{\bf N}$ there is a point $p \in S_{\bf N}$ such that $f \cdot p = p$ for all $f \in {\bf G}/{\bf N}$.
But then, $p$ is also a fixed point for the action of $\bf G$ on $S$, since
$g \cdot p = (gN) \cdot p = p$ for any $g \in G$.
 %$p$ and $g \cdot p$ are in the same 
 %left coset of $\bf N$ in $\bf G$, and so there is an $h \in {\bf N}$ such that 
 %$h \cdot 
 % $p$ = \dots = p$.
 
(2) follows from (1): suppose that ${\bf G} = {\bf H}_1 \times {\bf H}_2$, and that ${\bf H}_1$ and ${\bf H}_2$
are extremely amenable. Then ${\bf H}_1$ is a normal subgroup of $\bf G$, and ${\bf G}/{\bf H_1}$ is isomorphic to ${\bf H_2}$,
so ${\bf G}$ is extremely amenable by (1).  
The statement for ${\bf H}_1 \times \dots \times {\bf H}_n$ follows by induction on $n$.
\end{proof}

When we later apply Ramsey theory to analyze polymorphism clones in Section~\ref{sect:canonization} of this chapter,
and when the polymorphisms are higher-ary, it will be crucial to apply 
the so-called \emph{product Ramsey theorem}.
For every ordered Ramsey class there is a corresponding product Ramsey theorem (which usually has various slightly different formulations). This can be shown either directly, or by applying the general results from topological dynamics.

For illustration, we give a direct proof for the class of all finite linear orders (Theorem~\ref{thm:product-ramsey}, which will be used extensively for $d=m=2$ in Chapter~\ref{chap:tcsp}). 
For concreteness, we give specialized terminology in this case. 
If $S_1, \dots, S_d$ are sets, we call a set of the
form $S_1 \times \dots \times S_d$ a \emph{grid}, and also write $S^d$ for a
product of the form $S \times \dots \times S$ with $d$ factors.  A
\emph{$[k]^d$-subgrid} of a grid $S_1 \times \dots \times S_d$ is a subset of
$S_1 \times \dots \times S_d$ of the form $S_1' \times \dots \times S_d'$, where
$S_i'$ is a $k$-element subset of $S_i$. 

\begin{theorem}[Product Ramsey Theorem]\label{thm:product-ramsey}
For all positive integers $d$, $r$, $m$, and $k \geq m$, there is a positive integer
$L = {\bf R}(d,r,m,k)$ such that for every coloring of the
$[m]^d$ subgrids of $[L]^d$ with $r$ colors
there exists a \emph{monochromatic}
$[k]^d$ subgrid $G$ of $[L]^d$, i.e., $G$ is such that all its $[m]^d$ subgrids have the same color.
\end{theorem}
\begin{proof}
Let $d$, $r$, $m$, and $k \geq m$ be positive integers.
We claim that we can choose
$L = {\bf R}(d,r,m,k)$ to be ${\bf R}(r,dm,dk)$.
To verify this, let $\chi$ be a coloring of the $[m]^d$ subgrids
of $[L]^d$ with $r$ colors. We have to find a monochromatic
subgrid of $[L]^d$.

We use $\chi$ to define an $r$-coloring $\xi$ of the $dm$-subsets of 
$[L]$ as follows. Let $S = \{s_1,s_2,\dots,s_{dm}\}$
be a $dm$-subset of $[L]$, with $s_1<Ês_2 < \dots < s_{dm}$.
Then define $$\xi(S) = \chi(\{s_1,\dots,s_m\} \times \dots \times \{s_{m(d-1)+1},\dots,s_{dm}\}) \; .$$
By Theorem~\ref{thm:fin-ramsey},
there is a $dk$-subset $\{t_1, t_2, \dots, t_{dk} \}$ of $[L]$ such that $\xi$ is constant
on the $dm$-element subsets of $\{t_1,\dots,t_{dk}\}$. 
Suppose that $t_1 < t_2 < \dots < t_{dk}$. 
Then $G=\{t_1,\dots,t_k\} \times \dots \times \{t_{k(d-1)+1},\dots,t_{dk}\}$ is a subgrid of $[L]^d$ that is monochromatic with respect to $\chi$.
\end{proof}

We next present a formulation of the product Ramsey theorem for arbitrary ordered Ramsey structures.
The proof uses Theorem~\ref{thm:kpt} and Theorem~\ref{thm:new-extr-amen}.

\begin{theorem}\label{thm:prt}
Let $\bB_1,\dots,\bB_d$ be $\omega$-categorical ordered Ramsey structures.
Then $\bP := \bB_1 \boxtimes \cdots \boxtimes \bB_d$ is Ramsey.
\end{theorem}
\begin{proof}
By Theorem~\ref{thm:kpt}, the automorphism groups ${\bf G}_1,\dots,{\bf G}_d$ of $\bB_1,\dots,\bB_d$ are extremely amenable,
and it suffices to show that the automorphism group $\bf G$ of $\bP$ is extremely amenable. 
The group $\bf G$ is given by the product action of ${\bf G}_1 \times \dots \times {\bf G}_d$ on $B_1,\dots,B_d$ (see Section~\ref{sssect:product-action}). % of Chapter~\ref{chap:mt}).
Hence, extreme amenability of $\bf G$ follows from Theorem~\ref{thm:new-extr-amen}.
\end{proof}

Theorem~\ref{thm:prt} indeed generalizes Theorem~\ref{thm:product-ramsey}, which can be seen as follows. 
Let $r,d,m,k$ be positive integers. 
We consider the $\omega$-categorical ordered Ramsey structure $({\mathbb Q};<)$, 
and apply Theorem~\ref{thm:prt} where $d$ in Theorem~\ref{thm:prt} equals the $d$ given above. 
Let $\bA$ be the structure induced in $\bP := ({\mathbb Q};<)^{[d]}$ by some (equivalently, every) $[m]^d$ subgrid of ${\mathbb Q}^d$,
and let $\bB$ be the structure induced in $\bP$ by some (equivalently, every) $[k]^d$ subgrid of ${\mathbb Q}^d$.
Since $\bC$ is Ramsey, there exists an induced substructure $\bC$ of $\bP$ such that $\bC \rightarrow (\bB)^\bA_r$. 
If $\bC$ is not induced by an $[L]^d$ subgrid of ${\mathbb Q}^d$, for some large enough $L$, we can
clearly choose a larger substructure $\bC$ with this property, such that still $\bC \rightarrow (\bB)^\bA_r$.
%By the relations $=_1,\dots,=_d$ of $\bP$, GEHT AUCH OHNE WEIL ES EINE ORDNUNG GIBT
The occurrences of $\bA$ in $\bC$ correspond precisely to the
$[m]^d$-subgrids of $[L]^d$, which proves the claim.

\subsection{Open Subgroups}
\label{ssect:open}
In this section we show that open subgoups of extremely amenable groups are again extremely amenable. 
This fact will be important in Section~\ref{ssect:canonical-violation} and~\ref{sect:decidability} when it comes to the applications for analyzing polymorphism clones.
We first show the following basic fact. 

\begin{proposition}\label{prop:actions-on-product-space}
Let $X$ be a topological space, and $Y$ be any set.
Let ${\bf G}$ be a topological group
that acts on $X^Y$. Then the action is continuous if %and only if 
for every $y \in Y$, 
the map $f_y: {\bf G} \times X^Y \rightarrow X$ given by $f_y(g,\xi) := (g \cdot \xi)(y)$ is continuous.
\end{proposition}
\begin{proof}
Suppose that $\lim_{n \to \infty} (g_n,\xi_n) = (g,\xi)$. Then by Proposition~\ref{prop:pointwise-convergence} we have $\lim_{n \to \infty} g_n = g$ and $\lim_{n \to \infty} \xi_n(y) = \xi(y)$ for all $y \in Y$.
Since $f_y$ is continuous and by Proposition~\ref{prop:continuity} 
$$\lim_{n \to \infty} (g_n \cdot \xi_n)(y) = \lim_{n \to \infty} f_y(g_n,\xi_n) = f_y(g,\xi) = (g \cdot \xi)(y)$$ 
for all $y \in Y$. We again apply
Proposition~\ref{prop:pointwise-convergence} and obtain that $\lim_{n \to \infty} (g_n \cdot x_n) = g \cdot \xi$,
which implies continuity of the action of ${\bf G}$, again using Proposition~\ref{prop:continuity}.
% Opposite direction is missing.
\end{proof}

\begin{proposition}[from~\cite{BPT-decidability-of-definability}]\label{prop:open-subgroups}
Let ${\bf G}$ be an extremely amenable group, and let 
${\bf H}$ be an open subgroup of ${\bf G}$. Then ${\bf H}$
is also extremely amenable. 
\end{proposition}
In the proof, it is not essential but technically more convenient to use right cosets instead of left cosets. 
\begin{proof}
Let ${\bf H}$ act continuously on a compact space $X$; we will show that
this action has a fixed point. 
%Denote by $H \backslash G$ the set of
%right cosets of $H$ in $G$ (i.e. $H \backslash G = \{Hg : g \in G\}$). 
Denote by
$\pi \colon {\bf G} \to {\bf H} \backslash {\bf G}$ the quotient map and let $s \colon
{\bf H} \backslash {\bf G} \to {\bf G}$ be a section for $\pi$ (i.e., a mapping
satisfying $\pi \circ s = \mathrm{id}$) such that $s(H) = 1$.
Let $\alpha$ be
the map from ${\bf H} \backslash {\bf G} \times {\bf G} \to {\bf H}$ defined by
\[ \alpha(w, g) = s(w) g s(wg)^{-1}  \; .\]
For $w \in {\bf H} \backslash {\bf G}$ and $g \in {\bf G}$,
note that $s(w)g$ and $s(wg)$ lie
in the same right coset of ${\bf H}$, namely $wg$, and hence
the image of $\alpha$ is $H$.
The map $\alpha$ satisfies
\begin{align*}
\alpha(w,g_1g_2) & = s(w)g_1 g_2 (s(w g_1 g_2))^{-1} \\
& = s(w) g_1 s(w g_1) s(w g_1)^{-1} g_2 (s(w g_1 g_2))^{-1} \\
& = \alpha(w,g_1) \alpha(wg_1,g_2) \; .
\end{align*}

As $H$ is open, ${\bf H} \backslash {\bf G}$ is discrete. Hence, $s$ is
continuous, and therefore $\alpha$ is continuous as a composition of
continuous maps. 

Now consider the product space $X^{{\bf H} \backslash {\bf G}}$ 
which is Hausdorff and compact by Theorem~\ref{thm:tychonoff}.
The \emph{co-induced action} of ${\bf G}$ on $X^{{\bf H} \backslash {\bf G}}$ is defined by
\[ (g \cdot \xi)(w) = \alpha(w, g) \cdot \xi(wg). \]
We claim that this action is continuous. 
By Proposition~\ref{prop:actions-on-product-space}, it suffices to verify that the
map $(g, \xi) \mapsto (g \cdot \xi)(w)$ is a continuous map from $G \times X^{H
\backslash G} \to X$ for every fixed $w \in H \backslash G$.
We already know that $\alpha$ is continuous and that the action of ${\bf H}$ on $X$
 is continuous. To see that $(g, \xi) \mapsto \xi(wg)$ is continuous, suppose that $\lim_{n \to \infty} (g_n, \xi_n) = (g, \xi)$. Let $w=Hk$. As
$\lim_{n \to \infty} g_n = g$ and $k^{-1}Hk$ is open, we will have that eventually $g_ng^{-1}
\in k^{-1}Hk$, giving that $kg_n (kg)^{-1} \in H$, or, which is the same, $Hkg_n = Hkg$.
We obtain that for sufficiently large $n$, $wg_n = wg$. Therefore $\lim_{n \to \infty} \xi_n(w g_n) = \xi(wg)$.

By the extreme amenability of ${\bf G}$, the co-induced action has a fixed point $
\xi_0$. Now we check that $\xi_0(H) \in X$ is a fixed point of the
action $H \curvearrowright X$. Indeed, for any $h \in H$, $h \cdot
\xi_0 = \xi_0$ and we have
\begin{align*}
\xi_0(H) = & (h \cdot \xi_0)(H) \\
= & \alpha(H, h) \cdot \xi_0(Hh) \\
= & s(H) h s(Hh)^{-1} \xi_0(H) \\
= & h \cdot \xi_0(H),
\end{align*}
finishing the proof.
\end{proof}

Proposition~\ref{prop:open-subgroups} can be applied to provide a short and elegant proof of the following. 

\begin{corollary}[from~\cite{BPT-decidability-of-definability}]\label{cor:Ramsey-constants}
    Let $\bB$ be ordered homogeneous Ramsey, and let $c_1$, \ldots, $c_n$ be elements of $\bB$. Then $(\bB,c_1,\ldots,c_n)$ is ordered homogeneous Ramsey as well.
\end{corollary}
\begin{proof}
It is easy to see that the expansion of any homogeneous structure $\bB$ by finitely many constants is again homogeneous. 
When $\bB$ is additionally ordered Ramsey, then $\Aut(\bB)$ is extremely amenable. The automorphism group of $(\bB,c_1,\ldots,c_n)$ is an open subgroup of $\Aut(\bB)$. The statement thus follows directly from Proposition~\ref{prop:open-subgroups} and Theorem~\ref{thm:kpt}. 
\end{proof}

Note that in order to preserve homogeneity,
we have to add constants $c$ as unary 
function symbols, and not
as unary singleton relations $\{c\}$. 
Consider for example the homogeneous structure $(\mV;E)$, and let $u \in \mV$ (see Example~\ref{expl:random}). Then 
$(\mV;E,\{u\})$ is \emph{not} homogeneous, 
since there are no automorphisms that map neighbours of $u$ to non-neighbours.

\section{Canonization}
\label{sect:canonization}
% Central differences to the texts with Michael: 
% * dont use the ugly terminology with "arbitrary large"
% * there is no point in treating the unary case differently
In this section we apply Ramsey theory to analyse 
endomorphism monoids and polymorphism clones of Ramsey structures $\bB$.
The central idea is that for arbitrary finite substructures $\bC$ of 
$\bB$, any mapping from $B^k \rightarrow B$ 
must `behave canonically' on a copy of $\bC$ in $\bB$. 
We first consider the more general case of functions between two possibly distinct structures, 
and introduce a refinement of the notion of canonicity from Section~\ref{ssect:canonical}.
% in Chapter~\ref{chap:algebra}.

\begin{definition}\label{def:canonical}
Let $\bC$ be a structure with domain $C$, and $S$ a subset of $C$. 
Let $\bB$ be a structure with domain $B$, and let $f \colon C \rightarrow B$ be
a function. We say that $f$ is \emph{canonical on $S$ as a map from $\bC$ to $\bB$} if 
for all $n$ and every $n$-tuple $t$ over $S$ 
the $n$-type of $f(t)$ in $\bB$ only depends on the $n$-type of $t$ in $\bC$. 
%We say that $f$ is \emph{canonical on $S$}, for a subset $S$ of $B$, if the restriction of $f$ to $S$ is canonical.
\end{definition}

The basic lemma to apply Ramsey theory in the analysis of functions is the following. 

% BOOKTD: provide second proof by topological dynamics, using that Functions/Aut(C)
% is compact. (Should then be able to get rid of the assumption that B is omega-cat.)

\begin{lemma}\label{lem:canonize}
Let $\bC$ be an $\omega$-categorical 
ordered Ramsey structure with domain $C$ and finite relational signature $\tau$, 
let $\bB$ be an $\omega$-categorical structure with domain $B$, and let 
$f \colon C \rightarrow B$ be an operation. Then for all finite subsets $S$ of $C$ there
is an automorphism $\alpha$ of $\bC$ so that the operation 
$x \mapsto f(\alpha x)$ is canonical on $S$ as a map from $\bC$ to $\bB$.
\end{lemma}
\begin{proof}
Let $m$ be the maximal arity of the relations in $\tau$,
and $<$ be the linear order in the signature of $\bC$.
Let $\bC'$ be the homogeneous $\omega$-categorical expansion of $\bC$ by all relations that are
first-order definable in $\bC$. Then the age of $\bC'$
is a Ramsey class.  
Let $\bS$ be the substructure induced by $S$ in $\bC'$, and $n := |S|$.

When $\bA$ is a substructure of $\bC'$ of size $m$, then $a^\bA$ is the tuple $(a_1,\dots,a_m)$ such that
$\{a_1,\dots,a_m\}$ are the elements of $\bA$ and $t_1<\dots<t_m$. 

Let $\bA_1,\dots,\bA_k$ be a 
list all non-isomorphic substructures of $\bS$ of cardinality $m$.
Since $\bC'$ is Ramsey, there is a substructure $\bC_1$ of $\bC'$ such that 
$\bC_1 \rightarrow (\bS)^{\bA_1}_r$. 
%whenever all tuples of type $t_1$ are colored with 
%$l$ colors, then there exists a substructure $\bH_1$ of $\bC_1$ isomorphic to $\bS$ on which the coloring is constant. 
Further, there is a substructure $\bC_2$ of $\bC'$ such that 
$\bC_2 \rightarrow (\bC_1)^{\bA_2}_r$.
%whenever all tuples of type $t_2$ in $\bC_2$ are colored with $l$
%colors, then there exists a substructure $\bH_2$ of $\bC_2$ isomorphic to $\bC_1$ on which the coloring is constant. 
We iterate this $k$ times, arriving at a structure $\bC_k$. 
For each $i \leq k$,
the operation $f$ defines 
a coloring $\chi_i$ of ${\bC_k \choose \bA_i}$ 
with finitely many colors as follows: 
the color of a copy $\bA$ of $\bA_i$ 
is just the type of $f(a^\bA)$ in $\bB$; 
since $\bB$ is $\omega$-categorical, 
the number $r$ of $m$-types in $\bB$ is finite.

Now going back the argument, we find that $\bC_k$ contains a copy of $\bS$ on which all colorings $\chi_1,\dots,\chi_k$ are constant. Since $\bC'$ is homogeneous, 
there exists an automorphism $\alpha$ of $\bC'$ that sends $S$ to this copy. 
Then $x \mapsto f(\alpha x)$ is canonical on $S$ as a map 
from $\bC$ to $\bB$.
\end{proof}

Note that the assumption that $\bB$ is ordered is necessary in Lemma~\ref{lem:canonize}:
for instance, if $f$ is an injective function from $X \rightarrow (\mQ;<)$, then
$f$ is not canonical as a map from $(X;=)$ to $({\mathbb Q};<)$ on any two-element subset of $X$. 

\subsection{Multivariate functions}
%The multivariate case is analogous; its proof relies on the product Ramsey theorem, Theorem~\ref{thm:prt}. 
%We now show how to apply Lemma~\ref{lem:canonize} to prove the basic theorem to analyze polymorphisms 
%of $\omega$-categorical Ramsey structures. 
The appropriate generalisation of canonicity for multivariate 
functions is the following.

\begin{definition}\label{def:multivariate-canonical}
Let $\bB$ be a structure with domain $B$, and $\bC$ a structure with domain $C$.
When $f \colon C^d \rightarrow B$ is a function, 
and $S$ is a subset of $C^d$
we say that $f$ is \emph{canonical on $S$}
if for all $n$ and all $n$-tuples $t^1,\dots,t^d$ where $(t_i^1,\dots,t_i^d) \in S$ for all $i \leq n$ 
the $n$-type of $f(t^1,\dots,t^d)$ in $\bB$ only depends on the $n$-types of $t^1,\dots,t^d$ in $\bC$.
We say that $f$ is \emph{canonical (as a $d$-ary map from $\bC$ to $\bB$)} if $f$ is canonical on $B^d$.
\end{definition}

When $\bB=\bC$ in the definition above, 
we say that \emph{$f$ is 
canonical on $\bB$} 
when it is canonical as a $d$-ary map from
$\bB$ to $\bB$.

\begin{example} 
Let \lex\ be a binary operation on $\mathbb{Q}$ such that
$\lex(a,b) < \lex(a',b')$ if either $a < a'$, or $a=a'$ and $b<b'$.
Clearly, such an operation exists. 
Note that $\lex$ is injective, that it preserves $<$, and that it is canonical as a binary polymorphism of $({\mathbb Q};<)$.
%(Definition~\ref{def:multivariate-canonical}). 
\end{example}

In the proof of the following we use the product Ramsey theorem, Theorem~\ref{thm:prt}.

\begin{theorem}\label{thm:poly-canonize}
Let $\bB$ be an $\omega$-categorical ordered Ramsey structure with finite relational signature and domain $B$, and let 
$f \colon B^d \rightarrow B$ be any operation. Then for all finite subsets $S_1,\dots,S_d$ of $B$ there
are automorphisms $\alpha_1,\dots,\alpha_d$ of $\bB$ so that the operation 
$(x_1,\dots,x_d) \mapsto f(\alpha_1 x_1,\dots,\alpha_d x_d)$ is canonical on $S_1 \times \dots \times S_d$.
\end{theorem}
\begin{proof}
By Theorem~\ref{thm:prt}, the structure $\bB^{[d]}$ is Ramsey. Hence, Lemma~\ref{lem:canonize}
shows the existence of an automorphism $\alpha$ of $\bB^{[d]}$ such that $x \mapsto f(\alpha x)$ is canonical on 
$S_1 \times \cdots \times S_d$ as a map
from $\bB^{[d]}$ to $\bB$. 

Let $\bf G$ be the topological automorphism group of $\bB$.
Since the automorphism group of $\bB^{[d]}$ is induced by the product action
of ${\bf G}^k$ 
on $B^k$, there are group elements $a_1,\dots,a_d$ of $\bG$ so that $\alpha(x_1,\dots,x_d)) = (\alpha_1 x_1,\dots,\alpha_d x_d)$.
Now clearly the function $(x_1,\dots,x_d) \mapsto f(\alpha_1 x_1,\dots,\alpha_d x_d)$ is canonical on $S_1 \times \dots \times S_d$
as a multivariate function on $\bB$. 
\end{proof}

% BOOK-TD: add counterexample for when $\bB$ is not ordered?

\subsection{Interpolation modulo automorphisms}
\label{sect:interpolation-modulo-autos}
One of the central questions when analysing a polymorphism of a structure $\bB$ 
is to find out what functions it generates (since those functions will also be polymorphisms of $\bB$, 
see Section~\ref{sect:inv-pol}). % of Chapter~\ref{chap:algebra}).
Theorem~\ref{thm:poly-canonize} can be used for this purpose; to illustrate this,
we present the following. 

%We present an application of Theorem~\ref{thm:poly-canonize}. 

\begin{corollary}\label{cor:inj-can}
Let $\bB$ be an $\omega$-categorical ordered Ramsey structure with finite relational signature. Then every injective operation 
$f \colon B^k \rightarrow B$ together with the automorphisms of $\bB$ locally generates a canonical injective operation $g$.
\end{corollary}

This corollary follows in a straightforward way from Theorem~\ref{thm:poly-canonize} and a compactness argument,
which we do not present here since we will present a proper generalisation of it in full detail, Theorem~\ref{thm:interpolate-modulo-autos}.

Note that when we drop the injectivity assumption for $f$ in the statement of Corollary~\ref{cor:inj-can} then the statement of the corollary becomes trivially true, since every operation locally generates
the projections, which are canonical on the entire domain. We therefore need a concept that is weaker than
local closure, but stronger than interpolation, to turn the idea of Corollary~\ref{cor:inj-can} into a meaningful statement for all
functions from $B^k$ to $B$.

\begin{definition}
Let $\bB$ be an $\omega$-categorical structure with domain $B$, and $f,g \colon B^d \rightarrow B$ be functions.
Then $f$ \emph{interpolates $g$ modulo automorphisms of $\bB$} if for every finite $S \subseteq B$ there are automorphisms
$\alpha_0,\alpha_1,\dots,\alpha_d$ of $\bB$ such that $g(x_1,\dots,x_d)=\alpha_0(f(\alpha_1 x_1,\dots,\alpha_d x_d))$. 
\end{definition}

\begin{theorem}\label{thm:interpolate-modulo-autos}
Let $\bB$ be an ordered $\omega$-categorical Ramsey structure with finite relational signature and domain $B$, and $f \colon B^d \rightarrow B$ any operation.
Then there is a canonical operation $g \colon B^d \rightarrow B$ that is interpolated by $f$ modulo automorphisms.
\end{theorem}

%To derive Theorem~\ref{thm:interpolate-modulo-autos} 
%from Lemma~\ref{lem:infinst}, and 
Theorem~\ref{thm:interpolate-modulo-autos} is still not in its most general and most useful form. 
%To state a useful generalization of Theorem~\ref{thm:interpolate-modulo-autos},
For this, we need a further generalisation of the notion of interpolation 
modulo automorphisms to the situation where $f$ is a function from a structure $\bC$ to a different structure $\bB$. 

\begin{definition}
Let $\bB$, $\bC$ be structures with domains $B$ and $C$, and let $f,g \colon C \rightarrow B$ be functions.
Then $f$
\emph{interpolates $g$ modulo automorphisms of $\bC$} if for every finite $S \subseteq C$ there is an $\alpha \in \Aut(\bC)$
and a $\beta \in \Aut(\bB)$ such that $g(x)=\beta(f(\alpha x))$. 
\end{definition}

The following is the central statement about Ramsey structures and interpolation modulo automorphisms. 
% Absolut unwichtiger Kommentar: 
%we do not have to require that $\bC$ is $\omega$-categorical, and we do not have to require
%that $\bB$ has the Ramsey property.

\begin{theorem}\label{thm:general-interpolate-modulo-autos}
Let $\bC$ be $\omega$-categorical ordered Ramsey with finite relational signature and domain $C$, 
and let $\bB$ be $\omega$-categorical with domain $B$.  
Then every $f \colon C \rightarrow B$ interpolates a canonical operation $g \colon C \rightarrow B$ modulo automorphisms 
of $\bC$.
\end{theorem}
\begin{proof} 
By Lemma~\ref{lem:omega-cat-compactness}, it suffices to show that
for every finite subset $C'$ of $C$ there is a function from $C \rightarrow B$
that is canonical on $C'$ and interpolated by $f$ modulo automorphisms of $\bC$, since the property to be canonical on $C'$ is a universal first-order statement
about $f$. This follows from Lemma~\ref{lem:canonize}.
\end{proof}

\begin{proof}[Proof of Theorem~\ref{thm:interpolate-modulo-autos}]
We apply Theorem~\ref{thm:general-interpolate-modulo-autos} to the structure $\bC := \bB^{[d]}$, expanded 
by the first-order definable lexicogrpahic ordering on $\bB^{[d]}$, 
which is 
Ramsey when $\bB$ is Ramsey, by Theorem~\ref{thm:prt}.
As in the proof of Theorem~\ref{thm:poly-canonize}, canonicity of a function $g$ from $\bB^{[d]}$ to $\bB$ translate
into canonicity of $g$ as a $d$-ary function on $\bB$, and interpolation operations modulo automorphisms of
$\bB^{[d]}$ and $\bB$ corresponds to interpolation of $d$-ary functions modulo automorphisms of $\bB$.
\end{proof}

Note that Theorem~\ref{thm:interpolate-modulo-autos} is indeed a generalization of Corollary~\ref{cor:inj-can}, since
clearly operations that are interpolated by injective operations modulo automorphisms are again injective.
%Theorem~\ref{thm:interpolate-modulo-autos} 

`Canonization' of operations as exhibited in Theorem~\ref{thm:general-interpolate-modulo-autos}  becomes particularly powerful when we combine it with expansions by constants. 
The following theorem has numerous applications.

\begin{theorem}[fom~\cite{BPT-decidability-of-definability}]\label{thm:orderedCanonical}
% BOOK-TD: check whether we need all those assumptions, add topo dynamics proof
Let $\bB$ be an $\omega$-categorical ordered Ramsey structure with domain $B$ and finite relational signature. 
Let $c_1,\ldots,c_m \in B$, and let $f \colon B^k\To B$ be any function. Then $\{f\} \cup \Aut(\bB)$ locally generates a function which is canonical as a function from $(\bB,c_1, \ldots, c_m)^k$ to $\bB$, and which equals $f$ on all tuples containing only values $c_i$.
\end{theorem}
\begin{proof}
By Corollary~\ref{cor:Ramsey-constants}, also the structure 
$(\bB,c_1,\dots,c_m)$ is ordered Ramsey.
By Theorem~\ref{thm:prt}, the structure
$\bC := (\bB,c_1,\dots,c_m)^{[k]}$ is ordered Ramsey (and still $\omega$-categorical). 
Let $d_1,\dots,d_n$ be an enumeration
of the image of the restriction of $f$ to $\{c_1,\dots,c_m\}$. 
The structure $(\bB,d_1,\dots,d_n)$ is still
$\omega$-categorical. 
Then Theorem~\ref{thm:general-interpolate-modulo-autos} shows 
that $f$ interpolates modulo automorphisms of $\bC$ and $(\bB,d_1,\dots,d_n)$ an operation which is canonical as a function to $(\bB,d_1,\dots,d_n)$ and therefore also as a function to $\bB$. 
In particular, $g$ is locally generated by 
$\{f\} \cup \Aut(\bB)$, and the restrictions of
$f$ and $g$ to $\{c_1,\dots,c_m\}$ are equal. 
\end{proof}

\subsection{Behavior of Operations}
\label{ssect:canonical-behavior}
It is sometimes important to work with operations that exhibit a `behavior' that is only partially canonical. 
The following definition from~\cite{BP-reductsRamsey} gives us some flexibility in specifying such functions.

\begin{definition}
    Let $\bC$ and $\bB$ be structures with domains $C$ and $B$, and let $k\geq 1$. An \emph{($n$-)type condition} between $\bC$ and $\bB$ is a $k+1$-tuple $(t^1,\ldots,t^d,s)$, where each $t_i$ is an $n$-type in $\bC$, and $s$ is an $n$-type in $\bB$. A $d$-ary function $f \colon C^d \To B$ \emph{satisfies an $n$-type condition $(t^1,\ldots,t^d,s)$ on $S \subseteq C^d$} if for all $n$-tuples $a^i$ of type $t^i$  in $\bC$ with $(a^i_1,\dots,a^i_d) \in S$ for all $i \leq d$,
    the $n$-tuple $(f(a_1^1,\ldots,a_1^d),\ldots,f(a_n^1,\ldots,a_n^d))$ is of type $s$ in $\bB$. 
    
A \emph{behavior} between two structures $\bC$ and $\bB$ is a set $\Lambda$ of type conditions. 
A function $f \colon C^d \To B$ \emph{has behavior $\Lambda$
on $S \subseteq C^d$} 
if it satisfies all the type conditions of $\Lambda$ on $S$.
We say that $f$ \emph{has behavior $\Lambda$} 
if it has behavior $\Lambda$ 
on all of $C^d$.
\end{definition}

Note that a $d$-ary operation $f \colon \bC^d \To \bB$ is canonical if for all $n \geq 1$ and all $d$-tuples $(t^1,\ldots,t^d)$ of types of $n$-tuples in $\bC$ there exists a type $s$ of an $n$-tuple in $\bB$ such that $f$ satisfies the type condition $(t^1,\ldots,t^d,s)$. 
When $\bB$ is homogeneous in a relational signature with maximal arity $n$, then
already the $n$-type conditions determine the behavior of functions over $\bB$.

When $\bB$ is $\omega$-categorical then
the clone generated by $\Aut(\bB)$ and a canonical $d$-ary function $f$ over $\bB$ 
is completely described by the behavior of $f$. 
In fact, when 
$f,g \colon B^d \rightarrow B$ are functions with the same behavior,
then $\{f\} \cup \Aut(\bB)$ generates $g$,
and $\{g\} \cup \Aut(\bB)$ generates $f$, by local closure.

\begin{lemma}\label{lem:behavior-generates}
Let $\bB$ be ordered $\omega$-categorical with domain $B$. 
Let $\Lambda$ 
be a behaviour for functions from $\bB^k$ to $\bB$,
and let $g \colon B^k \rightarrow B$ be arbitrary. If for every finite substructure $\bA$ of $\bB$ there are copies $\bA_1,\dots,\bA_k$ of $\bA$ in $\bB$ such that $g$ has behavior 
$\Lambda$ on $\bA_1 \times \cdots \times \bA_k$
then $\{g\} \cup \Aut(\bB)$ locally generates a function
$f$ of behaviour $\Lambda$. 
\end{lemma}
\begin{proof}
A direct consequence of Lemma~\ref{lem:omega-cat-compactness}. 
\end{proof}

An orbit of $(\bB,c_1,\dots,c_m)$ is called \emph{full} if it contains copies of all finite substructures of $\bB$. 
The following follows from Lemma~\ref{lem:behavior-generates}. As we will see in Chapter~\ref{chap:schaefer}
and Chapter~\ref{chap:tcsp}, it becomes important in the context of canonization after expansions by constants (Theorem~\ref{thm:orderedCanonical}).

\begin{lemma}\label{lem:full-orbits-generation}
Let $\bB$ be $\omega$-categorical, and let $c_1,\dots,c_m$ be elements from $\bB$. When $\bA$ is the substructure induced in $\bB$ 
by a full orbit $O$ of $(\bB,c_1,\dots,c_m)$, and $f$ is a function from
$(\bB,c_1, \ldots, c_m)^k$ to $\bB$ with behaviour $\Lambda$ on $O$,
%Then $f$ interpolates modulo automorphisms of 
%$(\bB,c_1, \ldots, c_m)^k$ and $\bB$ a function with
%behavior $B$ from $\bB^k$ to $\bB$. 
then $\{f\} \cup \Aut(\bB)$ locally generates
a function from $\bB^k$ to $\bB$ with behaviour $\Lambda$.
\end{lemma}

\begin{figure}
\begin{center}
\includegraphics[scale=.5]{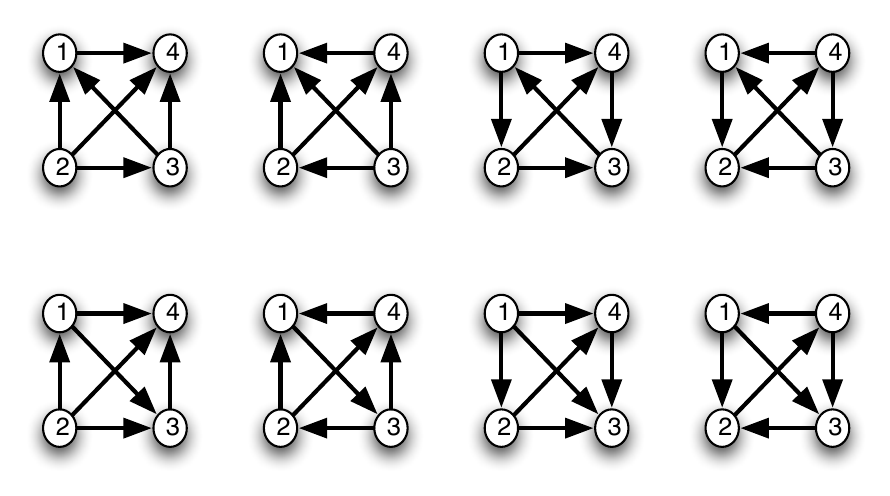}
\caption{Canonical behavior on $[2]^2$ grids.}
\label{fig:order-types}
\end{center}
\end{figure}

\begin{figure}
\begin{center}
\includegraphics[scale=.5]{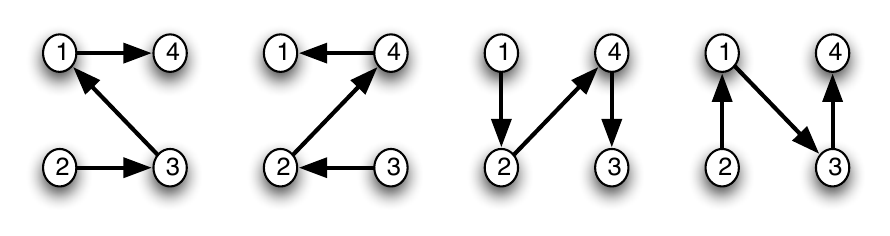}
\caption{When $f$ is a canonical binary injective polymorphism of $({\mathbb Q}; <)$, then there is
one linear order of a $[2]^2$ grid as depicted here such that all $[2]^2$ subgrids of ${\mathbb Q}^2$ are linearly ordered in this way.}
\label{fig:lex-types}
\end{center}
\end{figure}

Not every behavior $\Lambda$ between $\bC$ and $\bB$ is \emph{realized} by a function in the sense that there exists a function from $C \rightarrow B$ that has behavior $\Lambda$. 
We give an example.
There are eight distinct candidates for canonical behavior of injective
maps from $(\mQ;<)^2$ to $(\mQ;<)$;
they are illustrated in Figure~\ref{fig:order-types}. However, only four of those canonical behaviors are realized by
binary injective polymorphisms of $({\mathbb Q};<)$; those are illustrated in Figure~\ref{fig:lex-types}. The others would imply
the existence of three points $x,y,z$ in the image such that $x<y$ and $y<z$ and $z<x$, which is impossible 
over $({\mathbb Q};<)$. The not necessarily injective case can be analyzed similarly, 
and we get the following.

\begin{lemma}\label{lem:lex}
Let $f$ be a canonical binary polymorphism of $({\mathbb Q}; <)$.
Then $f$ has the same behavior as one out of the following seven operations.
\begin{itemize}
\item $\lex(x,y)$ or $\lex(y,x)$;
\item $\lex(x,-y)$ or $\lex(y,-x)$.
\item $(x,y) \mapsto x$ or $(x,y) \mapsto y$. 
\item a constant operation
\end{itemize}
\end{lemma}

Together with Corollary~\ref{cor:inj-can}, we find that every binary injective polymorphism of $({\mathbb Q};<)$ locally
generates $\lex(x,y)$, $\lex(y,x)$, $\lex(x,-y)$, or $\lex(y,-x)$.

\subsection{Canonical Violation}
\label{ssect:canonical-violation}
% important design question: what is easier: deriving finitely many minimal clones above finite signature structures
% or the violation lemma, or are there distinct things anyway?
Let $\bC$ be an $\omega$-categorical ordered Ramsey structure, 
and let $\bB$ be a structure with a first-order definition in $\bC$. 
Suppose that a relation $R$ does not have a primitive positive definition
in $\bB$. We wish to show that then there exists a polymorphism of $\bB$ that violates $R$ and is canonical 
as a function over $\bC$. Boldly stated like this, this cannot hold true. 
However, the results from Section~\ref{ssect:open} show us how to fix the statement.
To illustrate the basic idea, we first discuss the unary case, 
with existential positive definability instead of primitive positive definability.

\begin{theorem}\label{thm:unary-canonical-violation}
Let $\bC$ be an $\omega$-categorical ordered 
Ramsey structure, 
% Don't we need homogeneous in a finite language? No.
% We can even get rid of omega-cat, but in earlier proofs we had it so we keep it also here.
and let $\bB$ be a structure with a first-order definition in $\bC$.
Suppose that the $k$-ary relation $R$ does not have an existential positive definition in $\bB$. Then there exists
an endomorphism $e$ of $\bB$ and a $k$-tuple $t = (t_1,\dots,t_k) \in R$ such that 
\begin{itemize}
\item $e(t) \notin R$
\item $e$ is canonical as a map from $(\bC,t_1,\dots,t_k)$ to $\bC$.
\end{itemize}
\end{theorem}
\begin{proof} 
The structure $\bB$ is $\omega$-categorical.
If $R$ does not have an existential definition, 
then by Theorem~\ref{thm:ep} there is an endomorphism $e'$ of $\bB$ which violates $R$,
that is,  there is a $k$-tuple $t = (t_1,\dots,t_k) \in R$ such that $e'(t) \nin R$. 
By Theorem~\ref{thm:orderedCanonical}, $\{e'\} \cup \Aut(\bB)$ locally generates an operation $e$ that is canonical as a function
from $(\bB,t_1,\dots,t_k)$ to $\bB$ that has the same 
restriction to $\{t_1,\dots,t_k\}$ as $e'$, and $e$ has the required properties from the statement of the theorem.
\end{proof}

Here comes the multivariate analog of Theorem~\ref{thm:unary-canonical-violation}, whose proof
is analogous to the proof of the previous theorem.

\begin{theorem}\label{thm:multivariate-canonical-violation}
Let $\bC$ be an $\omega$-categorical ordered Ramsey structure, let $\bB$ be a structure with a first-order definition in $\bC$, and suppose that
the $k$-ary relation $R$ does not have a primitive positive definition in $\bB$. Then there exists a finite $d$, a $d$-ary polymorphism $f$ of $\bB$, and $k$-tuples $t^1,\dots,t^d \in R$ such that 
\begin{itemize}
\item $f(t^1,\dots,t^d) \notin R$
\item $f$ is canonical as a map from 
%$(\bC^{[d]},(t^1_1,\dots,t^d_1),\dots,(t_k^1,\dots,t_k^d))$ to $\bC$.
$(\bC,t^1) \boxtimes \cdots \boxtimes (\bC,t^d)$ to $\bC$. 
\end{itemize}
\end{theorem}
\begin{proof}
If $R$ does not have a primitive positive definition in $\bB$, then since $\bB$ is $\omega$-categorical, by Theorem~\ref{thm:inv-pol} there is a polymorphism $f'$ of $\bB$ which violates $R$. 
By Lemma~\ref{lem:small-arity}, we can assume that the arity of $f'$ equals the number $d$ of orbits of $k$-tuples contained in $R$, which is bounded by $o^\bB(k)$.
So there are $k$-tuples $t^1,\ldots,t^d \in R$ such that $f'(t^1,\ldots,t^m)\nin R$. 
By Corollary~\ref{cor:Ramsey-constants}, for all $i \leq d$ the structure $(\bC,t^i)$ is Ramsey,
%$(\bC^{[d]},(t^1_1,\dots,t^d_1),\dots,(t_k^1,\dots,t_k^d))$ 
and by Theorem~\ref{thm:prt} the structure
$(\bC,t^1) \boxtimes \cdots \boxtimes (\bC,t^d)$ is Ramsey.
Then Theorem~\ref{thm:general-interpolate-modulo-autos} shows that $f'$ interpolates modulo automorphisms of 
%$(\bC^{[d]},(t^1_1,\dots,t^d_1),\dots,(t_k^1,\dots,t_k^d))$ 
$(\bC,t^1) \boxtimes \cdots \boxtimes (\bC,t^d)$
a canonical operation $f$, and $f(t^1,\dots,t^d) = f'(t^1,\dots,t^d) \notin R$. 
Since $f$ is in particular locally generated by polymorphisms of $\bB$, it is itself an polymorphism of $\bB$.
\end{proof}

We are now in the situation to prove the following, which has been announced already in Section~\ref{ssect:minimal}. 

\begin{theorem}[from~\cite{BPT-decidability-of-definability}]\label{thm:climbing-up}
Let $\bB$ be a structure with finite relational signature, and with a first-order definition in an ordered  homogeneous Ramsey structure $\bC$ with a relational signature of maximal arity $k$.  Then there are finitely many minimal closed clones above $\Pol(\bB)$.
\end{theorem}
\begin{proof}
Every minimal closed clone above $\Pol(\bB)$ is locally generated by a minimal operation $f$ (Proposition~\ref{prop:minimal}),
and by Theorem~\ref{thm:pp-galois} there must be a relation $R$ in $\bB$ that is violated by $f$, that is, 
there are $t^1,\dots,t^d \in R$ such that $f(t^1,\dots,t^d) \notin R$. 
Since $f$ is a minimal operation, Theorem~\ref{thm:multivariate-canonical-violation} implies that $f$ must be canonical as a map from $(\bC,t^1) \boxtimes \cdots \boxtimes (\bC, t^d) \rightarrow \bC$. But since $\bC$ is homogeneous is a finite relational signature, and $\bB$ has finite relational signature,
there are only finitely many canonical behaviors of such operations; since two minimal operations with the same
behavior locally generate the same closed clone above $\Pol(\bB)$, we are done.
\end{proof}

\section{Decidability Results for Meta-Problems}
\label{sect:decidability}
We turn to another application of the ideas of the previous sections. For a fixed structure $\bC$
with a finite relational signature $\tau$ and domain $C$, 
consider the following computational problem. 

\cproblem{$\Exprfo(\bC)$}
{Quantifier-free first-order $\tau$-formulas $\phi_0,\ldots,\phi_n$ defining the relations $R_0,\ldots,R_n$ over $\bC$.}
{Is there a first-order definition of $R_0$ in $(C; R_1,\ldots,R_n)$?}

We are also interested in the variants of this problem where we replace first-order definability by other syntactically restricted
versions of definability, in particular by primitive positive definability. The corresponding computational problem for primitive positive definability is denoted by $\Exprpp(\bC)$ (and the problem for existential and existential positive definability by 
$\Exprex(\bC)$ and $\Exprep(\bC)$, respectively). 

For \emph{finite} structures $\bC$ the problem $\Exprpp(\bC)$ is in co-NEXPTIME (and in particular decidable).
For the variant where the finite structure
$\Gamma$ is part of the input, the problem has recently shown to be also co-NEXPTIME-hard~\cite{Willard-cp10}.
An algorithm for  $\Exprpp(\bC)$ has theoretical and practical consequences in the study of the computational complexity of CSPs for structures that are first-order definable in $\bC$,
as illustrated in the following examples.

\begin{example}
We can use an algorithm for $\Exprpp(\bC)$ 
to decide whether all polymorphisms of a structure $(C;R_1,\dots,R_n)$, given by $\tau$-formulas $\phi_1,\dots,\phi_n$ that define $R_1,\dots,R_n$ over $\bC$, are essentially unary. For that, we simply apply the algorithm to $x \neq y \vee y \neq z, \phi_1,\dots,\phi_n$, for each $i$; here we use Proposition~\ref{prop:unary}.
\end{example}

\begin{example}
To decide whether a structure $(C;R_1,\dots,R_n)$, again given by $\tau$-formulas $\phi_1,\dots,\phi_n$ that define $R_1,\dots,R_n$ over $\bC$, is a core, we apply the algorithm for $\Exprpp(\bC)$ to $\neg \phi_i, \phi_1,\dots,\phi_n$, 
for each $i$. Additionally, we apply the algorithm to $x \neq y, \phi_1,\dots,\phi_n$. 
The structure  $(C;R_1,\dots,R_n)$ is a core if and only if
none of those calls reports
false, that is, all the relations defined by $\neg \phi_i$ or by $x \neq y$ are primitive positive definable in $(C;R_1,\dots,R_n)$.
\end{example}

\begin{example}
We can use an algorithm for $\Exprpp(\bC)$ to effectively test the hardness condition for $\Csp(\bB)$ given in Proposition~\ref{prop:1in3hard} for
structures $\bB$ with a first-order definition in $\bC$ and a finite relational signature.
\end{example} 

The main result of this section is the decidability of $\Exprpp(\bC)$ for a certain class of
structures $\bC$. Even for the simplest of countable structures, namely the structure $(X;=)$ having no relations but equality, the decidability of $\Exprpp(\Gamma)$ is not obvious (see~\cite{BodChenPinsker}). Recall the concept of \emph{finitely bounded} structures $\bC$ (Definition~\ref{def:finitely-bounded}):
we require that the age of $\bC$ is given by a finite set of finite forbidden induced substructures.

\begin{theorem}[from~\cite{BPT-decidability-of-definability}]\label{thm:pp-decidability}
    Let $\bC$ be of finite relational signature, and first-order definable over a structure $\bD$ which is homogeneous, ordered, Ramsey, finitely bounded, and with finite relational signature. Then $\Exprpp(\bC)$ is decidable.
\end{theorem}

% Design decision in the following proof: choose CSP perspective since
% 1 pushes forward to CSP application perspective
% 2 clearly indicates the implementation on a computer +  can use existing solvers
% 3 the implementation will in this way be more rapid using CSP solver technology
% 4 provides also formally a greater degree of detail

% A possible simplification would be to set n to be even larger than the size of the largest obstruction, in which case
% we would only have to check for compatibility, but not for realizability! Realizability constraints would be implicit from the 
% n-type selection. But from practical perspectives, for consistency with past versions, and also for conceptual reasons
% I prefer to keep the realizability item. The same applies to the arity of $R_0$, and also the arity of $R_1,\dots,R_n$:
% we would simply make n larger than all of those; it would be simpler, but uglier.

\begin{proof}%[Proof of Theorem~\ref{thm:pp-decidability}]
Let $D$ be the domain of $\bD$ and $\bC$.
The input consists of formulas $\phi_0,\phi_1,\ldots,\phi_k$ in the signature of $\bC$. 
Those formulas define the relations $R_0,R_1,\ldots,R_k$ over $\bC$.
Set $\bB$ to be the structure $(D;R_1,\ldots,R_k)$.
We will decide whether there is a primitive positive definition of $R_0$ in $\bB$. We can without loss of generality assume that in each formula $\phi_0,\dots,\phi_k$, the variables
are called $x_1,\dots,x_p$, for some $p$. 

By Theorem~\ref{thm:multivariate-canonical-violation}, if $R_0$ is $m$-ary and does not have a primitive positive definition in $\bB$, then there exists a finite $d$, a $d$-ary polymorphism $f$ of $\bB$,
and $m$-tuples $t^1,\dots,t^d \in R_0$ such that $f(t^1,\dots,t^d) \notin R_0$, and 
$f$ is canonical as a map from $(\bD,t^1) \boxtimes \cdots \boxtimes (\bD,t^d)$ to $\bD$. % Really want to take $\bD$ here on both sides since we want to work with HOMOGENEOUS base structure
Such a polymorphism of $\bB$ will be called a \emph{witness} at $t^1,\dots,t^d$ (for the fact that $R_0$ is not primitive positive definable in $\bB$).
The question whether such a witness exists for a specific choice of tuples $t^1,\dots,t^d$ does of course only depend
on the orbits of $t^1,\dots,t^d$ in $\bD$, and by $\omega$-categoricity of $\bD$ there are only finitely many such orbits. 
Moreover, by homogeneity of $\bD$, the orbits of $n$-tuples are in one-to-one correspondence to the $n$-element induced substructures of $\bD$, which can be effectively stored and enumerated on a computer.
So it suffices in the following to consider the case where $t^1,\dots,t^d$ are fixed, and to show
how to decide whether a witness exists at this choice of $t^1,\dots,t^d$.

Since expansions of homogeneous structures by finitely many constants are homogeneous, the one-to-one correspondence between orbits of $l$-tuples, maximal $l$-types, and induced $l$-element substructures of $\bD$ extends to the structures $(\bD,t^i)$.
%Also note that since the signature of $\bC$ is finite,
%$l$-element substructures of $\bC$ (and of the structures $(\bC,t^i)$)
%are a convenient way to represent orbits of $l$-tuples and $l$-types of $\bC$ on a computer.
In the following, let $n$ be $\max(3,n'+1)$ where $n'$ is the maximal arity of the relations in $\bD$.
Then by homogeneity of $\bD$ 
the behavior of $f$ is determined by the $n$-type conditions that are satisfied by $f$ (for this property we only need that $n \geq n'$;  the requirement 
 $n \geq 3$ is motivated by the way how we treat equality in our approach, as we will see later). When $f$ is canonical, then the set $\Lambda$ of  
$n$-type conditions can be viewed as a function from $S^{(\bD,t^1)}_n \times \cdots \times S^{(\bD,t^d)}_n$ to $S^\bD_n$.
By $\omega$-categoricity of $\bD$ and of $(\bD,t^i)$ there are only finitely
many such functions $\Lambda$. 

We decide the existence of a witness by reduction to a finite-domain 
constraint satisfaction problem. The domain of the CSP 
is the set of all $n$-types of $\bD$. The instance of the CSP has a variable for every
$d$-tuple $(S_1,\dots,S_d)$ where $S_i$ is an $n$-type of $(\bD,t^i)$; in fact, we identify the variables of the instance and those $d$-tuples of $n$-types.
The constraints are described below. The idea is that the solutions to this CSP are exactly the
functions $\Lambda$ for witnesses as described above. 

To implement this in detail, it will be convenient to make the assumption that $\mathcal N$ is \emph{minimal} in the sense
that it does not contain
structures $\bA_1, \bA_2$ such that $\bA_1$ is an induced substructure of $\bA_2$; this assumption is without loss of
generality since otherwise we remove $\bA_2$ from $\mathcal N$, and find that
the resulting set of structures still bounds $\bD$.

\begin{itemize}
\item (Compatibility.) 
       %If $\Lambda$ is a behavior of a witness, then for all $1 \leq l \leq n$ 
        %it must also be extendible to a function from $S^{(\bD,t^1)}_l\mult\cdots\mult S^{(\bD,t^d)}_l$ to $S^\bD_l$. 
        Note that every behavior of a witness must have an extension to 
        a function from $S^{(\bD,t^1)}_l\mult\cdots\mult S^{(\bD,t^d)}_l$ to $S^\bD_l$, for all $1 \leq l \leq n$.        
	Hence, when $(S_1,\dots,S_d),(T_1,\dots,T_d) \in S^{(\bD,t^1)}_n\mult\cdots\mult S^{(\bD,t^d)}_n$,
	and $I \subset [n]$, and if for all $i \leq d$ the subtype of $S_i$ induced by $I$ and the subtype
	of $T_i$ induced by $I$ coincide, then we impose the binary constraint that $I$ induces the same subtype in 
	$\Lambda(S_1,\dots,S_d) \in S^\bD_n$ and in $\Lambda(T_1,\dots,T_d) \in S^\bD_n$. 
\item (Realizability.) 
	We also want to make sure that the behavior $\Lambda$ can be realized by an operation 
	(recall the example given in Section~\ref{ssect:canonical-behavior}).
	The idea is that when $\bD$ is finitely bounded, then $\Lambda$ 
	should not force the existence of one of the forbidden substructures in the image, since in this case
	it would be impossible to find an operation with image in $\bD$ whose behavior is $\Lambda$.
	As we will see, it suffices here to consider structures $\bA \in \mathcal N$ whose number of elements $s$ exceeds $n$. 
	
	For each structure $\bA \in \mathcal N$ with $s > n$ elements and each sequence $S_1,\dots,S_d$ with
	$S_i \in S_s^{(\bD;t^i)}$ for all $i \leq d$ we have a constraint of arity $r := {s \choose n}$. 
	Let $a_1,\dots,a_s$ be the elements of $\bA$. 
	Observe that for every subset $I \subseteq [s]$ with $| I | = n$ the structure induced by $\{ a_i \; | \: i \in I\}$
	in $\bA$ is an induced substructure of $\bD$, by the minimality assumption on $\mathcal N$. 
	Let $\phi^\bA[I]$ be the formula with variables $x_1,\dots,x_n$ 
	that contains for $i_1,\dots,i_m \in I$ the conjunct $R(x_{i_1},\dots,x_{i_m})$ if and only if $(a_{i_1},\dots,a_{i_m}) \in R^\bA$.
	By the observation we just made, $\phi^\bA[I]$ 
	is contained in a unique $n$-type of $\bD$. 
	
	The constraint of arity $r$ requires 
	that for some $I \subseteq [s]$ with $| I | = n$ the subtype of $\Lambda(S_1,\dots,S_d)$ induced by $I$
	does not contain $\phi^\bA[I]$. 
\item (Violation.) We want that $\Lambda$ is the behavior of an
	operation that violates the $m$-ary relation $R_0$. 
	For simplicity of presentation, we assume that $m \geq n$;
	this is without loss of generality, 
	since we can otherwise add dummy variables to $\phi_0$.
	
	For $t = (t_1,\dots,t_m)$ and $i_1,\dots,i_n \in [m]$ with $i_1 < \dots < i_n$, 
	denote by $t[\{i_1,\dots,i_n\}]$ the tuple $(t_{i_1},\dots,t_{i_n})$. 
	When $I \subseteq [m]$ we denote by $\phi_0[I]$ the subtype of $\{\phi_0\}$ induced by $I$; here, $\{\phi_0\}$ 
	is viewed as a type over the empty theory. 
	We add the ${m}\choose{n}$-ary constraint that 
	for some $I \subset [m]$ of cardinality $n$ the type
	$\Lambda(\tp^{(\bD,t^1)}(t^i[I]),\dots,\tp^{(\bD,t^d)}(t^i[I]))$ does not contain $\phi_0[I]$. 
\item (Preservation.) We also want that $\Lambda$ is the behavior of an
	operation that preserves $\bB$. 
	Let $j \leq k$, and suppose that the relation $R_j$ of $\bB$ defined by $\phi_j$ is $p$-ary. 
	For simplicity of presentation, 
	we assume that $p \geq n$, otherwise we add dummy arguments to $R_j$.
	 For every list $S_1,\dots,S_d$ such that $S_i \in S^{(\bD,t^i)}_p$ contains $\phi_j$ for all $i \leq d$, 
	 we impose the following constraint of arity $q =  {p \choose n}$.
	For all $I \subseteq [p]$ with $|I| = n$, let $S^I_i$ be the subtype of $S_i$ induced by $I$,
	and let $S^I_0$ be the subtype of $\{\phi_j\}$ (of the empty theory) induced by $I$.
	We add the constraint that $\Lambda(S^I_1,\dots,S^I_d)$ contains $S^I_0$ for all $I \subseteq [p]$.
\end{itemize}

We now prove that there is a witness $f$ at $t^1,\dots,t^d$ for the fact that $R_0$ is not primitive positive definable in $\bB$ 
if and only
if the described CSP instance has a satisfying assignment, which concludes the proof. 
For the easy direction, suppose that there exists such a witness, and let $\Lambda$ be its behavior.
Then $\Lambda$ clearly satisfies compatibility, realizability, violation, and preservation constraints. 

For the opposite direction, suppose that $\alpha$ is a solution
to the described CSP, i.e., a mapping from $S^{(\bD,t^1)}_n \times \cdots \times S^{(\bD,t^d)}_n$ to $S^\bD_n$ that satisfies
compatibility, realizability, violation and preservation constraints. 
We show the existence of a witness $f$ at $t^1,\dots,t^d$ in three steps.

We first construct an infinite structure $\bE$ with domain $D^d$ of the same signature $\tau$ as $\bD$
as follows. When $a^1,\dots,a^d \in D^n$ are such that $a^i \in S^{(\bD,t^i)}_n$, 
then for $R \in \sigma$ the relation %$R(a^1,\dots,a^d)$ holds in the substructure of $\bE$ induced by the $n$-tuple
$R((a^1_1,\dots,a^d_1),\dots,(a^1_n,\dots,a^d_n))$ holds in $\bE$ if and only if
$R(x_1,\dots,x_n)$ is contained in $\Lambda \big(\tp^{(\bD,t^1)}(a^1) \times \dots \times \tp^{(\bD,t^d)}(a^d)\big)$.
This is well-defined by the compatibility constraints.

Next, we consider the relation $\sim$ on the domain of $\bE$ such that $a_1 \sim a_2$ for $a_1,a_2 \in D^d$ if and only if
there exist $a_3,\dots,a_n \in D^d$ such that 
the subtype of 
$\Lambda(\big(\tp^{(\bD,t^1)}((a^1_1,\dots,a^1_n)) \times \dots \times \tp^{(\bD,t^d)}(a^d_1,\dots,a^d_n)\big)$ 
induced by $\{1,2\}$ contains $x_1=x_2$. 
Note that since $n \geq 3$, the relation $\sim$ must be an equivalence relation (since the properties of an equivalence relation can be formulated with a universal formula with three variables).  
Then the quotient structure $\bE/_{\sim}$ is defined to be the $\tau$-structure 
whose elements are the equivalence classes $E/_{\sim}$ of $\sim$, and
where $R(E_1,\dots,E_p)$ holds for a $p$-ary $R \in \tau$ and $E_1,\dots,E_p \in E/_{\sim}$ if and only if there are $b_1 \in E_1,\dots,b_p \in E_p$ such that
$R(b_1,\dots,b_p)$ holds in $\bE$. 

The final step is to show that there exists an embedding $f$ of $\bE/_{\sim}$ into $\bD$.
By $\omega$-categoricity of $\bD$ and Lemma~\ref{lem:infinst},
it suffices to show every finite substructure $\bA$ of $\bE/_{\sim}$ embeds into $\bD$. 
Since $\bD$ is finitely bounded by $\mathcal N$, we thus have to show
that no structure in $\mathcal N$ embeds into $\bA$. 
Suppose to the contrary that there is an embedding $e$ of $\bF \in \mathcal N$ into $\bA$. Let $u_1,\dots,u_s$ be the elements of $\bF$. 
Pick any elements $v_1,\dots,v_s$ from the equivalence classes 
$e(u_1),\dots,e(u_s)$, respectively. 
The rest of the paragraph is devoted to the argument that
 the mapping $e'$ that maps $u_i$ to $v_i$ is an embedding of
$\bF$ into $\bE$, which contradicts the realizability constraints. 
% LATERTD: do we have to expand this part?
It is obvious that $e'$ is injective. To see that it is a strong
homomorphism, let $R$ be an $(n-1)$-ary symbol from $\tau$;
the case that $R$ has a smaller arity can be dealt with by adding dummy variables.
Note that in the following we use the assumption that $n$ is strictly larger
than the maximal arity of $\bD$; intuitively, we implement Leibniz' law for equality.  
We have $R(e(u_{i_1}),\dots,e(u_{i_p}))$ if and only if
there are $v'_{i_1} \in e(u_{i_1}),\dots,v'_{i_p} \in e(u_{i_p})$ 
so that $R(v'_{i_1},\dots,v'_{i_p})$ holds in $\bE$.  
Since the formula $R(x_1,\dots,x_p) \wedge x_p = x_{p+1}$ is contained in 
$$\Lambda \big(\tp^{(\bD,t^1)}(v'_{i_1}[1],\dots,v'_{i_p}[1],v_{n}[1]),\dots,\tp^{(\bD,t^d)}(v'_{i_1}[d]),\dots,v'_{i_p}[d],v_{i_n}[d]) \big) \; ,$$
this type of $\bD$ must also contain $R(x_1,\dots,x_{p-1},x_{p+1})$. In this way we argue successively for all arguments of $R$, 
and finally obtain that $R(v_{i_1},\dots,v_{i_p})$ holds in $\bE$. The argument can be reverted, 
and we have that $e'$ is a strong homomorphism. 
We conclude that $e'$ is an embedding. 

Observe that the mapping $g$ from $D^d$ to $D$ that maps $\bar u$ to $f(\bar u/_{\sim})$ is a polymorphism of $\bB$, 
by the preservation constraints, and it is canonical by construction.
By the violation constraints, $g$ violates $R_0$, and hence is the desired witness.
\end{proof}

Analogously to the proof of this theorem, one can show the following.

\begin{theorem}[from~\cite{BPT-decidability-of-definability}]\label{thm:ep-decidability}
    Let $\bC$ be with finite relational signature, and first-order definable over a structure $\bD$ which is ordered, homogeneous, Ramsey, finitely bounded, and with finite relational signature. 
    Then $\Exprex(\bC)$ and $\Exprep(\bC)$ are decidable.
\end{theorem}

An important open problem is whether the method can be extended to show
decidability of $\Exprfo(\bC)$, under the same assumptions on $\bC$ as in Theorems~\ref{thm:pp-decidability} and~\ref{thm:ep-decidability}.
By the theorem of Ryll-Nardzewski, first-order definability
is characterized by preservation under automorphisms, i.e., surjective self-embeddings.
But the requirement of surjectivity is difficult to deal with in our approach.

\begin{question}\label{quest:fo-decidability}
Let $\bB$ be with finite relational signature, and definable in a structure $\bC$ which is ordered, homogeneous, Ramsey, finitely bounded, and with finite relational signature.  
Is $\Exprfo(\bB)$ decidable?
\end{question}

While the conditions of Theorem~\ref{thm:pp-decidability} 
might appear rather restrictive at first sight, they actually are quite general: we want
to point out that we only require that $\bC$ is first-order definable over an ordered, homogeneous, Ramsey, and finitely bounded structure,
rather than requiring that $\bC$ itself to have these properties. We do not know 
of a single homogeneous structure $\bC$ with finite relational signature which does not satisfy the conditions of Theorems~\ref{thm:pp-decidability} and~\ref{thm:ep-decidability}. Examples of structures $\bC$ that do satisfy the assumptions, and the corresponding references, see Section~\ref{sect:ramsey-classes}.

%Examples of structures $\bD$ that satisfy the assumptions
%of Theorem~\ref{thm:pp-decidability} are $({\mathbb Q};<)$,
%the Fra\"{i}ss\'{e} limit of ordered finite graphs (or tournaments~\cite{RamseyClasses}), the Fra\"{i}ss\'{e} limit of
%finite partial orders with a linear extension~\cite{RamseyClasses}, and the homogeneous universal `naturally ordered' $C$-relations (for definition and basic properties of $C$-relations, see~\cite{AdelekeNeumann}, in particular Theorem~14.7. The fact that the homogeneous universal naturally ordered $C$-relations have the Ramsey property follows from Theorem 4.3 in~\cite{Mil79}; an
% explicit and elementary verification of the Ramsey property for the binary
% branching case can be found in~\cite{BodirskyPiguet}). 
% Note that our result shows that it is decidable whether a given relation 
% from Allen's Interval Algebra~\cite{Allen,RandomReducts} is primitive positive definable 
% in a given fragment of Allen's Interval Algebra.

We finally show that the assumption in Theorem~\ref{thm:pp-decidability} of $\bC$ being finitely bounded is necessary.

\begin{proposition}
There exists a homogeneous ordered Ramsey structure $\bC$ with finite relational signature such that $\Exprfo(\bC)$, $\Exprpp(\bC)$, $\Exprep(\bC)$, and $\Exprex(\bC)$ is undecidable.
\end{proposition}
\begin{proof}
Recall the definition of \emph{Henson digraphs} from Example~\ref{expl:Henson}.
When $\C'$ is the age of a Henson digraph $\bC'$, then the class $\C$ consisting of all structures obtained from
the digraphs in $\C'$ by adding an arbitrary linear order on the vertices, is again an amalgamation class.
In fact, it is even a Ramsey class by the results described in Example~\ref{expl:irreducible-forbidden-Ramsey}.
Let  $\bC$ denote its \Fresse-limit. 

We first show that non-isomorphic Henson digraphs $\bC_1$ and $\bC_2$ have distinct $\Exprpp$ problems.
In fact, we show the existence of a first-order formula $\phi_1$ over digraphs such that the input with
$\phi_0 := E(x,y)$ and $\phi_1$ is a yes-instance of $\Exprpp(\bC_1)$ and a no-instance of $\Exprpp(\bC_2)$, or vice versa. Since there are uncountably many Henson digraphs, but only countably many
algorithms, this clearly shows the existence of  Henson digraphs $\bC'$ such that
$\Exprpp(\bC')$ is undecidable. It follows that for the ordered Ramsey structure
$\bC$ described above the problem $\Exprpp(\bC)$ is undecidable as well. 

Since $\bC_1$ and $\bC_2$ are non-isomorphic, there must be a structure $\bA$ that
embeds to $\bC_1$ but not to $\bC_2$, or that embeds to $\bC_2$ but not to $\bC_1$.
Assume the former is the case; in the latter, simply exchange $\bC_1$ and $\bC_2$.
Let $s$ be the number of elements of $\bA$, and denote the elements by $a_1,\dots,a_s$. 
Let $\psi$ be the formula with variables $x_1,\dots,x_s$ that has for distinct $i,j \leq s$
a conjunct $E(x_i,x_j)$ if $E(a_i,a_j)$ holds in $\bA$,
and a conjunct $\neg E(x_i,x_j) \wedge x_i \neq x_j$ otherwise. 
Let $\phi$ be the formula $\psi \Rightarrow E(x_{s+1},x_{s+2})$. 

Let $C_1$ be the domain of $\bC_1$, and
consider the relation $R_1 \subseteq (C_1)^{s+2}$ defined by $\phi$ in $\bC_1$. 
Let $R$ be a relation symbol of arity $s+2$,
 and $\bB$ be the structures with signature $\{R\}$, domain $C_1$,
 and where $R$ denotes the relation $R_1$. 
It is clear that 
$\exists x_1,\dots,x_s. \, R(x_1,\dots,x_s,x,y)$ is a primitive positive definition of $E(x,y)$ in $\bB$. 

Now consider the relation $R_2$ defined by $\phi$ in $\bC_2$ over the domain $C_2$.
Since $\bA$ does not embed into $\bC_2$, the precondition of $\phi$ is never satisfied, and
the relation $R_2$ is empty. 
Hence, the structure $(C_2;R_2)$ is preserved by all permutations.
But the relation $E(x,y)$ is certainly not first-order definable over a structure
that is preserved by all permutations. 
\end{proof}

\chapter{Schaefer's Theorem for Graphs}
\label{chap:schaefer}

%\begin{figure}[h]
\begin{center}
\includegraphics{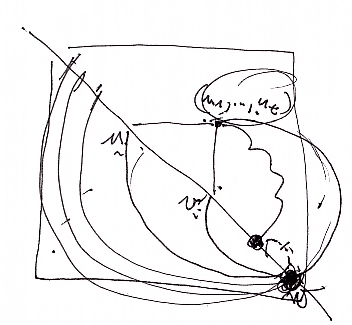} Jaroslav Ne\v{s}et\v{r}il, 2004
\end{center}
\vspace{1cm}
%\caption{Drawing by Jaroslav Ne\v{s}et\v{r}il, 2004}
%\end{figure}

This chapter is based 
on results from~\cite{RandomMinOps,
BodPin-Schaefer-Both,RandomReducts}.

\section{Motivation and the Result}
In an influential paper in 1978, Schaefer~\cite{Schaefer} classified
the computational complexity of $\Csp(\bB)$ for all structures
$\bB$ over a two-element universe; the result and a simple proof have been presented in Section~\ref{sect:schaefer}. % of Chapter~\ref{chap:mt}.
Schaefer's theorem can be viewed as a complexity classification 
for systematic syntactic restrictions of the Boolean
satisfiability problem, as reflected in the following formulation of the result.

Let $\Psi = \{\psi_1,\dots,\psi_n\}$ be
a finite set of propositional (Boolean) formulas.

\cproblem{Boolean-SAT$(\Psi)$}
{Given a finite set of variables $W$ and a propositional formula of the form
$\Phi = \phi_1 \wedge \dots \wedge \phi_l$ where each $\phi_i$ for $1 \leq i \leq l$ is obtained from one of the formulas $\psi$ in $\Psi$ by substituting the variables of $\psi$ by variables from $W$.}
{Is there a satisfying Boolean assignment to the variables of $W$ (equivalently, those of $\Phi$)?}

Schaefer's theorem (Theorem~\ref{thm:schaefer}) states that Boolean-SAT$(\Psi)$ can be solved in polynomial time
if and only if $\Psi$ is a subset of one of six sets of Boolean formulas (called \emph{0-valid, 1-valid, Horn, dual-Horn, affine}, and \emph{bijunctive}),
and is NP-complete otherwise.

We prove a similar classification result, but for the first-order logic
of graphs instead for propositional logic. More precisely, let $E$ be a relation symbol which denotes an antireflexive and symmetric binary relation and hence stands for the edge relation of a (simple, undirected) graph. We consider formulas that are constructed from atomic formulas of the form $E(x,y)$ and $x=y$ by the usual boolean connectives (negation, conjunction, disjunction), and call formulas of this form \emph{graph formulas}. A graph formula $\Phi(x_1,\ldots,x_m)$ is \emph{satisfiable} if there exists a graph $H$ and an $m$-tuple $a$ of elements in $H$ such that $\Phi(a)$ holds in $H$.

The problem to decide whether a given graph formula is satisfiable can be very difficult.
For example the question whether
the Ramsey number $R(5,5)$ is larger than $43$ (which is an open question, see e.g.~\cite{Exoo}) can be easily formulated 
in terms of satisfiability of a single graph formula.
Recall that $R(5,5)$ is the least number $k$ such
that every graph with at least $k$ vertices either contains
a clique of size $5$ or an independent set of size $5$. 
So the question whether $R(5,5)$ is greater than $43$
can be formulated 
using $43$ variables $x_1,\dots,x_{43}$ by imposing 
the constraints that all variables denote a different vertex in the graph,
and by imposing for every subset of the variables of cardinality five a constraint that
forbids that the corresponding five variables form a clique or an independent set.
If there is a solution to this instance, then this implies that $R(5,5) > 43$,
and otherwise $R(5,5) \leq 43$.

Similarly as in Schaefer's theorem, 
we systematically investigate restrictions of the satisfiability problem for graph formulas
that can be solved in polynomial time.
Let $\Psi = \{\psi_1,\dots,\psi_n\}$ be a finite set of graph formulas.
Then $\Psi$ gives
rise to the following computational problem.

\cproblem{Graph-SAT$(\Psi)$}
{Given a set of variables $W$ and a graph formula of the form
$\Phi = \phi_1 \wedge \dots \wedge \phi_l$ where each $\phi_i$ for $1 \leq i \leq l$ is obtained from one of the formulas $\psi$ in $\Psi$ by substituting the variables from $\psi$ by variables from $W$.}
{Is $\Phi$ satisfiable?}

\begin{example}
Let $\Psi$ be the set that just contains the formula
\begin{align}
 & (E(x,y) \wedge \neg E(y,z) \wedge \neg E(x,z))  \nonumber \\
\vee\; &
 (\neg E(x,y) \wedge E(y,z) \wedge \neg E(x,z)) \nonumber \\ %\label{eq:rel-hard} \\
 \vee \;  & (\neg E(x,y) \wedge \neg E(y,z) \wedge E(x,z)) \; . \nonumber
\end{align}
Then Graph-SAT$(\Psi)$ is the problem of deciding whether there
exists a graph such that certain prescribed subsets of its vertex set of cardinality at most three induce subgraphs with exactly one edge. The problem Graph-SAT$(\Psi)$ is NP-complete. 
\end{example}

\begin{example}
There are also many interesting tractable Graph-SAT problems,
for instance when $\Psi$ consists of the formula 
\begin{align}
& (E(x,y) \wedge \neg E(y,z) \wedge \neg E(x,z)) \nonumber \\
 \vee \;  & (\neg E(x,y) \wedge E(y,z) \wedge \neg E(x,z)) \nonumber \\
 %label{eq:rel-tractable} \\
 \vee \; & (\neg E(x,y) \wedge \neg E(y,z) \wedge E(x,z)) \nonumber \\
 \vee \; & (E(x,y) \wedge E(y,z) \wedge E(x,z)) \; . \nonumber
\end{align}
\end{example}

It is obvious that the problem Graph-SAT$(\Psi)$ is
for all $\Psi$ contained in NP. 
The goal of this chapter is to prove the following dichotomy result.

\begin{theorem}\label{thm:gsat-dichotomy}
For all $\Psi$, the problem $\Gsat(\Psi)$ is NP-complete or in P. Moreover, the problem
to decide for given $\Psi$ whether $\Gsat(\Psi)$ is NP-complete or in P is decidable.  
\end{theorem}

%In Theorem~\ref{thm:minimalTractableClones} 
%and Theorem~\ref{thm:sigger-terms} we will see exactly which sets $\Psi$ correspond to tractable problems and which to hard ones.
We establish our result by translating Graph-SAT problems
into CSPs. %In the formulation as a CSP, we can make 
%use of the techniques from Chapter~\ref{chap:algebra}
%and Chapter~\ref{chap:ramsey}.
More specifically, for every set of formulas $\Psi$ we present
a relational structure $\bB_\Psi$ such that
Graph-SAT$(\Psi)$ is equivalent to $\Csp(\bB_\Psi)$. 
The relational structure $\bB_\Psi$
has a first-order definition in the random graph $(\mV;E)$ introduced in Chapter~\ref{chap:mt}.
This perspective allows us to use polymorphisms
to classify the computational complexity of Graph-SAT problems as outlined in Chapter~\ref{chap:algebra}. 
Our proof also relies crucially on
strong results from structural Ramsey theory. Following the technique from
Chapter~\ref{chap:ramsey},
we use such results to find regular patterns in the behavior of polymorphisms of structures on $(\mV;E)$, which in turn allows us to find analogies with polymorphisms of structures on Boolean domains. Our dichotomy result
can be stated as follows.

\begin{theorem}\label{thm:gcsp-tractability}
Let $\bB$ be a relational structure with a first-order definition in $(\mV;E)$. Then exactly
one of the following two statements applies. 
\begin{enumerate}
\item there is a primitive positive interpretation of 
all finite structures in the model-complete core of $\bB$. In this case, $\bB$ has a finite-signature reduct with an NP-hard CSP, by Corollary~\ref{cor:pp-interpret-hard}.
\item $\bB$ has a cyclic polymorphism $f$ modulo an endomorphism, i.e., there are $f \in \Pol(\bB)$ and 
$e \in \End(\bB)$ satisfying 
$$ f(x_1,\dots,x_n) = e(f(x_2,\dots,x_n,x_1))$$
for all $x,y \in \mV$. In this case, every finite-signature reduct of $\bB$ has a polynomial-time tractable CSP. 
\end{enumerate}
\end{theorem}

%It turns out that there is one class of reducts $\bB$ for which $\Csp(\bB)$ is in P for trivial reasons; further, there are
%16 classes of structures $\bB$ for which $\Csp(\bB)$
%(and the corresponding Graph-SAT problems)
%can be solved by
%non-trivial algorithms in polynomial time.

%It turns out that every structure $\bB$ with a first-order
%definition in $(\mV;E)$ either has a 4-ary polymorphism $f$ which is canonical as a function over $(\mV;E)$ and satisfies 
%$$ \forall x,y \, f(y,y,x,x) = a_1 ( f(x,x,x,y)) = a_2 ( f(y,x,y,x)) \, , $$
%or else there is a primitive positive interpretation of the structure $(\{0,1\};\NAE)$ in $\bB$. In the latter case, CSP$(\Gamma)$
%is NP-hard by Theorem~\ref{thm:pp-interpret-reduce}.
%In the first case, CSP$(\bB)$ 

The proof of this theorem can be found at the end of Section~\ref{sect:g-classification}.
The algorithmic part of Theorem~\ref{thm:gcsp-tractability} 
is obtained by combinations of ideas from 
Section~\ref{sect:horn} 
and reductions to the tractable cases of Schaefer's theorem (Theorem~\ref{thm:schaefer}). 

In the remainder of this chapter, $\bB$ denotes a 
relational structure with a first-order definition in the random graph 
$(\mV;E)$. 
Since all the polymorphism clones of this chapter contain the automorphism group $\Aut(\mV;E)$ of the random graph, we also make the following convention, which exclusively holds for this chapter. For a set of functions $F$ and a function $g$ on the domain $\mV$, we say that $F$ \emph{generates} $g$
when $F \cup \Aut(\mV;E)$ locally generates
$g$; also, we say that a function $f$ \emph{generates} $g$ if $\{f\}$ generates $g$. That is, in this paper we consider the automorphisms of $(\mV;E)$ be present in all sets of functions when speaking about the local generating process.

\section{Endomorphisms}\label{sect:endos}
The goal of this section is the proof of Proposition~\ref{prop:endos}, which allows us to reduce the classification task for $\bB$
to the classification of those structures $\bB$ where the 
$\Aut(\bB)$ is dense in $\End(\bB)$; moreover, 
we give a description of the five possible automorphism
groups that can appear, due to Thomas~\cite{RandomReducts}. 

We write $N$ for the relation $\{(x,y) \in \mV^2 \mid x \neq y \wedge \neg E(x,y)\}$. 
Note that $(\mV;N)$ is the complement of the graph $(\mV;E)$, and that $(\mV;N)$ is  
isomorphic to $(\mV;E)$ (it is straightforward to verify the extension
property). Let $-$ be such an isomorphism. 
For any finite subset $S$ of $\mV$, if we flip edges and non-edges
between $S$ and $\mV \setminus S$ in $(\mV;E)$, then the resulting graph is
isomorphic to $(\mV;E)$ (it is straightforward to verify the extension
property). For any non-empty set $S$, we write $i_S$ for such an isomorphism.
Note that when $S$ and $T$ are two finite non-empty subsets of $\mV$, then
$i_S$ and $i_T$ generate one another.
We also write \sw\ for $i_{\{0\}}$, where $0 \in \mV$ is any
fixed element of $\mV$.

\begin{definition}[$R^{(k)}$ and $S^{(k)}$]
Let $R^{(k)}$ ($S^{(k)}$) be the $k$-ary relation that holds on $x_1,\dots,x_k \in \mV$ if $x_1,\dots,x_k$ are pairwise distinct, and the number of edges
between these $k$ vertices is odd (even).
\end{definition}

Observe that 
$R^{(3)}$ and $R^{(4)}$ are preserved by $\sw$, 
that $R^{(4)}$ and $R^{(4)}$ are preserved by $-$,
and that $R^{(5)}$ and $S^{(5)}$ are preserved by $-$ and by
$\sw$, but not by all permutations of $\mV$.

\begin{proposition}\label{prop:endos}
Let $\bB$ be first-order definable in $(\mV;E)$. Then at least one of the following holds.
\begin{enumerate}
\item[(a)] The endomorphisms of $\bB$ are generated by $\Aut(\mV;E)$. 
\item[(b)] $\bB$ has a constant endomorphism. In this case $\Csp(\bB)$ is trivial.
\item[(c)] $\bB$ is homomorphically equivalent to a countably infinite structure that is preserved by all permutations of its domain. Such structures have been classified in Chapter~\ref{chap:ecsp}.
\item[(d)] The endomorphisms of $\bB$ are precisely the functions generated by $\sw$; equivalently, $\End(\bB) = \End(\mV;R^{(3)},S^{(3)})$. 
\item[(e)] The endomorphisms of $\bB$ are precisely the functions generated by $-$; equivalently, $\End(\bB) = \End(\mV;R^{(4)},S^{(4)})$. 
\item[(f)] The endomorphisms of $\bB$ are precisely the functions generated by $\{-,\sw\}$; equivalently, $\End(\bB) = \End(\mV;R^{(5)},S^{(5)})$. 
\end{enumerate}
\end{proposition}

To prove the proposition, we first cite a result about
structures definable over the random graph due to Thomas~\cite{Thomas96};
the formulation of the result presented here first appeared in~\cite{RandomMinOps}. The graph $(\mV;E)$ contains all countable graphs as
induced subgraphs. In particular, it contains an infinite complete
subgraph, denoted by $K_\omega$.
It is clear that
any two injective operations from $\mV \rightarrow \mV$
whose images induce
$K_\omega$ in $(\mV;E)$  generate one another.
Let $e_E$ be one such operation.
Similarly,
$(\mV;E)$ contains an infinite independent set, denoted by $I_\omega$.
Let $e_N$ be an injective operation from $\mV \rightarrow \mV$ whose image induces $I_\omega$ in $(\mV;E)$.

\begin{theorem}[of~\cite{Thomas96,RandomMinOps}]\label{thm:g-endos}
Let $\bB$ be first-order definable in $(\mV;E)$. Then 
one of the following cases applies.
\begin{enumerate}
\item $\bB$ has a constant endomorphism.
\item $\bB$ has the endomorphism $e_E$.
\item $\bB$ has the endomorphism $e_N$.
\item The endomorphisms of $\bB$ are locally generated by the automorphisms of $\bB$.
\end{enumerate}
\end{theorem}

\begin{corollary}[from~\cite{RandomMinOps}]\label{cor:g-mc}
    All relational structures $\bB$ with a first-order definition in $(\mV;E)$ are model-complete.
\end{corollary}
\begin{proof}
    By Theorem~\ref{thm:mc-omegacat}, an $\omega$-categorical structure $\bB$ 
    is model-complete if and only if $\Aut(\bB)$ is dense in the monoid $\cM$ 
    of self-embeddings of $\bB$. 
    We apply Theorem~\ref{thm:g-endos} to $\cM$, which, 
    as a closed monoid containing $\Aut((\mV;E))$, 
    is also an endomorphism monoid of a structure
    $\bB'$ with a first-order definition in $(\mV;E)$. 
    Clearly, $\bB'$ and $\bB$ have the same automorphisms, 
    namely those permutations in $\cM$ whose inverse is also in $\cM$. 
    Therefore we are done if the last case of 
    Theorem~\ref{thm:g-endos} holds. 
    Note that $\cM$ cannot contain a constant operation as all its operations are injective. 
    So suppose
    that $\cM$ contains $e_N$ -- the argument for $e_E$ is analogous. Let $R$ be any relation of $\bB$, and $\phi_R$ be its defining quantifier-free formula; $\phi_R$ exists since $(\mV;E)$ has quantifier-elimination. Let $\psi_R$ be the formula obtained by replacing all occurrences of $E$ by \emph{false}; so $\psi_R$ is a formula over the empty language. Then a tuple $a$ satisfies $\phi_R$ in $(\mV;E)$ iff $e_N(a)$ satisfies $\phi_R$ in $(\mV;E)$ (because $e_N$ is an embedding) if and only if $e_N(a)$ satisfies $\psi_R$ in $(\mV;E)$ (as there are no edges on $e_N(a)$) if and only if $e_N(a)$ satisfies $\psi_R$ in the substructure induced by $e_N[\mV]$ (since $\psi_R$ does not contain any quantifiers). Thus, $\bB$ is isomorphic to the structure on $e_N[\mV]$ which has the relations defined by the formulas $\psi_R$. Therefore, $\bB$ is isomorphic to a structure with a first-order definition over the empty signature. This structure has, of course, all injections as self-embeddings, and all permutations as automorphisms, and hence is model-complete; 
    the same is true for $\bB$.
\end{proof}

The last case in Theorem~\ref{thm:g-endos} 
splits into five sub-cases, corresponding to the five
locally closed permutation groups that contain $\Aut((\mV;E))$ exhibited by Thomas~\cite{RandomReducts}.

%We can now describe the structures with a first-order definition in the random graph up to first-order interdefinability. 

\begin{definition}
    For all $k\geq 3$, let $P^{(k)}$ denote the $k$-ary relation that holds on $x_1,\dots,x_k \in \mV$
    if $x_1,\dots,x_k$ are pairwise distinct, and the graph induced by $\{x_1,\ldots,x_k\}$ in $(\mV;E)$ is neither an independent set nor a clique.
\end{definition}

%  (see Section~\ref{ssect:inv-aut}). %in Chapter~\ref{chap:mt}). 

\begin{theorem}[of~\cite{RandomReducts}]\label{thm:reducts}
Let $\bB$ be a relational structure with a first-order definition in $(\mV;E)$. Then 
exactly one of the following is true.
\begin{enumerate}
\item $\bB$ is first-order interdefinable with $(\mV;E)$; equivalently, $\Aut(\bB)=\Aut((\mV;E))$.
\item $\bB$ is first-order interdefinable with $(\mV;R^{(4)})$; equivalently, $\bB$ is preserved by $-$, but not by $\sw$.
\item $\bB$ is first-order interdefinable with $(\mV;R^{(3)})$; equivalently, $\bB$ is preserved by $\sw$, but not by $-$.
\item $\bB$ is first-order interdefinable with $(\mV;R^{(5)})$; equivalently, $\bB$ is preserved by $-$ and by $\sw$, but not by all permutations of $\mV$.
\item $\bB$ is first-order interdefinable with $(\mV;=)$; equivalently, $\bB$ is preserved by all permutations of $\mV$.
\end{enumerate}
\end{theorem}

We are now ready to prove Proposition~\ref{prop:endos}.

\begin{proof}[Proof of Proposition~\ref{prop:endos}]
%Suppose $E$ does not have a primitive positive definition in $\bB$. We have that $E$ consists of just one orbit of pairs in $(\mV;E)$, and thus, since $\Aut(\bB)\supseteq\Aut((\mV;E))$, also in $\bB$. Hence,
%Lemma~\ref{lem:small-arity} shows the existence of an endomorphism $e$ of $\bB$ that violates $E$.
Theorem~\ref{thm:g-endos} states that 
$\bB$ has a constant
endomorphism, or the endomorphism $e_E$, or the endomorphism $e_N$, or all endomorphisms of $\bB$ are generated by its automorphisms. If $\bB$ has a constant endomorphism we are in Case~(b) and done. If $\bB$ has the endomorphisms $e_E$ or $e_N$, then we
are in Case~(c) since $e_E[\mV]$ and $e_N[\mV]$ induce structures in $(\mV;E)$ which are invariant under all permutations of their domain.
So assume in the following that $\bB$ has neither
$e_E$, nor $e_N$, nor a constant as an endomorphism, and that all endomorphisms of $\bB$ are generated by $\Aut(\bB)$. 
The statement now follow from Theorem~\ref{thm:reducts}. 
\end{proof}

Proposition~\ref{prop:endos} allows us to focus
in the following on the situation where $\Aut(\bB)$ is dense in $\End(\bB)$. Moreover, there are only five
possibilities for 
$\Aut(\bB)$, and the case $\Aut(\bB) = \Aut(\mV;=)$ has already been solved.

%\section{Higher Arity Polymorphisms}\label{sect:higher-arity}
% The four subsections of this
%section are dedicated to the remaining four cases,
%which are first-order expansion of $(\mV;R^{(3)})$, 
%$(\mV;R^{(4)})$, or $(\mV;R^{(5)})$, respectively. 

\section{First-order Expansions of $(\mV;E,N)$}
\label{sect:higherArity}

%contains the relations $E$, $N := \{(u,v) \in \mV^2 \; | \; \neg E(u,v) \wedge u \neq v\}$,  and $\neq$. While in the last section, we only dealt with endomorphisms and automorphisms of structures, 
%the remaining cases require the study of higher arity polymorphisms of $\bB$. 

We remark that since $(\mV;E)$ has only binary relations, a function $f \colon \mV^k\To \mV$ is canonical if and only if
it is $2$-canonical (Section~\ref{ssect:canonical}). The polymorphisms that imply tractability of $\Csp(\bB)$ will be canonical with respect to $(\mV;E)$. 
We now define different behaviors that some of these canonical functions might have. For $Q_1,\ldots,Q_k\in\{E,N,=,\neq\}$, we will in the following write $Q_1\cdots Q_k$ for the binary relation on $\mV^k$ that holds between two $k$-tuples $x,y\in \mV^k$ iff $Q_i(x_i,y_i)$ holds for all $1\leq i\leq k$.

\begin{definition}
    We say that a binary injective operation $f \colon \mV^2\To \mV$ is
    \begin{itemize}
        \item \emph{balanced in the first argument} if for all $u,v\in \mV^2$ we have that $\EEQ(u,v)$ implies $E(f(u),f(v))$ and $\NEQ(u,v)$ implies $N(f(u),f(v))$;
        \item \emph{balanced in the second argument} if
        $(x,y) \mapsto f(y,x)$ is balanced in the first argument;
        \item \emph{balanced} if $f$ is balanced in both arguments, and
        \emph{unbalanced} otherwise;
        \item \emph{$E$-dominated ($N$-dominated) in the first argument} if for all $u,v \in \mV^2$ with $\NEQEQ(u,v)$
        we have that $E(f(u),f(v))$ ($N(f(u),f(v))$);
        \item \emph{$E$-dominated ($N$-dominated) in the second argument} if
        $(x,y) \mapsto f(y,x)$ is $E$-dominated ($N$-dominated) in the first argument;
        \item \emph{$E$-dominated ($N$-dominated)} if it is $E$-dominated ($N$-dominated) in both arguments;
        \item \emph{of type $\mini$} if for all $u,v\in \mV^2$ with $\NEQNEQ(u,v)$ we have
        $E(f(u),f(v))$ if and only if $\EE(u,v)$;
        \item \emph{of type $\maxi$} if for all $u,v\in \mV^2$ with $\NEQNEQ(u,v)$ we have
        $N(f(u),f(v))$ if and only if $\NN(u,v)$;
        \item \emph{of type $p_1$} if for all $u,v \in \mV^2$ with $\NEQNEQ(u,v)$ we have
        $E(f(u),f(v))$ if and only if $E(u_1,v_1)$;
        \item \emph{of type $p_2$} if $(x,y) \mapsto f(y,x)$ is of type $p_1$;
        \item \emph{of type projection} if it is of type $p_1$ or $p_2$.
    \end{itemize}
\end{definition}

Note that, for example, being of type $\maxi$ is a behavior of binary functions that does not force a function to be canonical, since the condition only talks about certain types of pairs in $\mV^2$, but not all such types; however, being of type $\maxi$ and balanced does mean that a function is canonical. The next definition contains some important behaviors of ternary functions.

\begin{definition}
    An injective ternary function $f \colon \mV^3\To \mV$ is of type
     \begin{itemize}
        \item \emph{majority} if for all $u,v\in \mV^3$ we have that $E(f(u),f(v))$ if and only if $\EEE(u,v)$, $\EEN(u,v)$, $\ENE(u,v)$, or $\NEE(u,v)$;

        \item \emph{minority} if for all $x,y \in \mV^3$ we have $E(f(x),f(y))$ if and only if $\EEE(u,v)$, $\NNE(u,v)$, $\NEN(u,v)$, or $\ENN(u,v)$.
     \end{itemize}
\end{definition}

While the polynomial-time tractability results of this section will be shown by means of a number of different canonical functions, all hardness cases will be established by the following single relation.

\begin{definition}
We define the 6-ary relation $H_1(x_1,y_1,x_2,y_2,x_3,y_3)$ on $\mV$ by
\begin{align}
& \bigwedge_{i,j \in \{1,2,3\}, i \neq j, u \in \{x_i,y_i\},v \in \{x_j,y_j\}} N(u,v) \nonumber \\
\wedge & \; \big((E(x_1,y_1) \wedge N(x_2,y_2) \wedge N(x_3,y_3))
\nonumber \\ 
& \vee \; (N(x_1,y_1) \wedge E(x_2,y_2) \wedge N(x_3,y_3)) \label{eq:rel} \\
& \vee \; (N(x_1,y_1) \wedge N(x_2,y_2) \wedge E(x_3,y_3)) \big)\; . \nonumber
\end{align}
\end{definition}

The goal of this section is to prove the following proposition.

\begin{proposition}\label{prop:higherArity}
    Let $\bB = (\mV; E,N,\neq,\dots)$ be first-order definable in $(\mV;E)$. Then at least one of the following holds:
    \begin{enumerate}
\item[(a)] There is a primitive positive definition of $H_1$ in $\bB$. In this case, $\Csp(\bB)$ is NP-complete.
\item[(b)] $\bB$ has a canonical polymorphism of type minority, as well as a canonical binary injection which of type $p_1$ and $E$-dominated or $N$-dominated in the second argument. In this case, $\Csp(\bB)$ is polynomial-time tractable.
\item[(c)] $\bB$ has a canonical polymorphism of type majority, as well as a canonical binary injection which of type $p_1$ and $E$-dominated or $N$-dominated in the second argument. In this case, $\Csp(\bB)$ is polynomial-time tractable.
\item[(d)] $\bB$ has a canonical polymorphism of type minority, as well as a canonical binary injection which is balanced and of type projection. In this case, $\Csp(\bB)$ is polynomial-time tractable.
\item[(e)] $\bB$ has a canonical polymorphism of type majority, as well as a canonical binary injection which is balanced and of type projection. In this case, $\Csp(\bB)$ is polynomial-time tractable.
\item[(f)] $\bB$ has a canonical polymorphism of type $\maxi$ or $\mini$. In this case, $\Csp(\bB)$ is polynomial-time tractable.
\end{enumerate}
\end{proposition}

The remainder of this section contains the proof of Proposition~\ref{prop:higherArity}, except for the polynomial-time tractability 
proofs, which will be given in Section~\ref{sect:g-algs}.
For the other statements of Proposition~\ref{prop:higherArity}, we proceed as follows. In Section~\ref{ssect:hardnessOfH}, we show that the relation $H_1$ is hard. In particular,  a structure $\bB=(\mV;E,N,\neq,\ldots)$ with a first-order definition in $(\mV;E)$ must have an essential polymorphism, or has a finite reduct with an NP-hard CSP. In Section~\ref{ssect:g-bin-inj} we show
when $\bB$ has an essential polymorphism, it must 
also have a binary injective polymorphism. 
We finally prove in Section~\ref{ssect:producingCanonical} 
%that if $H$ does not have a primitive positive definition in a reduct $\bB$ as in Proposition~\ref{prop:higherArity}, then 
that $\bB$ has one of the polymorphisms listed in cases~(b) to~(f) of the proposition. 

%Tractability of cases~(b) and (c) is shown in Section~\ref{ssect:edgeMinMajUnbalanced}, tractability of case~(d) in Section~\ref{ssect:edgeMinorityBalanced}, of case~(e) in Section~\ref{ssect:edgeMajorityBalanced}, and finally tractability of case~(f) in Section~\ref{ssect:maxMin}.

\subsection{Hardness of $H_1$}\label{ssect:hardnessOfH}
This section is devoted to case~(a) of Proposition~\ref{prop:higherArity}.

\begin{proposition}\label{prop:h-is-hard}
    There is a primitive positive interpretation of
    $(\{0,1\}; \OIT)$ in $(\mV;H_1)$, and $\Csp(\mV; H_1)$ is NP-hard.
\end{proposition}
\begin{proof}
We give a 2-dimensional interpretation $I$ of $(\{0,1\}; \OIT)$
in $\bB$. The domain formula is \emph{true}. The formula
$=_I(x_1,x_2,y_1,y_2)$ is
\begin{align*}
\exists z_1,z_2,u_1,u_2,v_1,v_2 \, & \big (H_1(x_1,x_2,u_1,u_2,z_1,z_2) \wedge N(u_1,u_2) \\
\wedge \; & H_1(z_1,z_2,v_1,v_2,y_1,y_2) \wedge N(v_1,v_2) \big )
\end{align*}
This formula is equivalent to a primitive positive formula
over $\bB$ since $N(x,y)$ is primitive positive definable
by $H_1$. The formula $\OIT_I(x_1,x_2,y_1,y_2,z_1,z_2)$
is
\begin{align*}
\exists x_1',x_2',y_1',y_2',z_1',z_2' \, &
\big (H_1(x_1',x_2',y_1',y_2',z_1',z_2') \\
\wedge \; & =_I(x_1,x_2,x_1',x_2')
\wedge =_I(y_1,y_2,y_1',y_2')
\wedge =_I(z_1,z_2,z_1',z_2') \big )
\end{align*}
The coordinate map sends a tuple $(x_1,x_2)$ to $1$ if $E(x_1,x_2)$
and to $0$ otherwise. The second part of the statement follows
from Corollary~\ref{cor:pp-interpret-hard}.
\end{proof}

\subsection{Producing binary injections}
\label{ssect:g-bin-inj}
We now show that if a structure $\bB=(\mV;E,N,\neq,\ldots)$ with a first-order definition in $(\mV;E)$
has an essential polymorphism, then $\bB$ must also have a binary injective polymorphism. This is in particular the case when
the relation $H_1$ from the previous section is not primitive positive
definable in $\bB$: since $E$ and $N$ are among the relations of $\bB$, and since any essentially unary polymorphism preserving both $E$ and $N$ preserves all relations with a first-order definition in $(\mV;E)$, we have that the polymorphism violating $H_1$ must be essential. 

\begin{theorem}[from~\cite{RandomMinOps}]\label{thm:g-bin-inj}
Let $\bB = (\mV; E,N,\neq,\dots)$ be first-order definable in $(\mV;E)$,
and suppose that $\bB$ has an essential polymorphism.
Then $\bB$ also has a binary injective polymorphism.
\end{theorem}
\begin{proof}%[Proof of Theorem~\ref{thm:g-bin-inj}]
Let $f \colon \mV^k\To \mV$ be an essential polymorphism 
of $\bB$ of minimal arity. By Lemma~\ref{lem:binary}, $f$ 
must be binary. Hence, we may apply
Lemma~\ref{lem:intersect} to $\bB$, and in order
to show that $\bB$ is preserved by a binary injection, it
suffices to show that if $\phi$ is a primitive positive formula over
$\bB$ such that both $\phi \wedge x \neq y$ and $\phi \wedge s
\neq t$ are satisfiable over $\bB$, then $\phi \wedge x \neq y
\wedge s \neq t$ is satisfiable over $\bB$ as well. The proof follows the idea of the proof of Theorem~\ref{thm:2trans}.

Let $\phi$ be a primitive positive formula
over the signature of $\bB$ such that
\begin{itemize}
\item there is a tuple $t_1$ that satisfies
$\phi \wedge x \neq y$
\item there is a tuple $t_2$ that satisfies
$\phi \wedge s \neq t$.
\end{itemize}
Let $a_1,a_2,a_3,a_4$ and $b_1,b_2,b_3,b_4$ be the values for
$x,y,s,t$ in $t_1$ and $t_2$, respectively. We have $a_1\neq a_2$
and $b_3\neq b_4$. We want to show that $\phi \wedge x \neq y \wedge
s \neq t$ is satisfiable over $\bB$. Thus, if $a_3\neq a_4$ or
$b_1\neq b_2$, there is nothing to show, and so we assume that
$a_3=a_4$ and $b_1= b_2$.

We claim that there are automorphisms $\alpha,\beta$ of $(\mV;E)$ such
that in the tuple $t_3 := f(\alpha(t_1),\beta(t_2))$ the value of
$x$ is different from the value of $y$, and the value of $s$ is
different from the value of $t$. Then, since $f$ preserves $\bB$,
the tuple $t_3$ shows that $\phi \wedge x \neq y \wedge s \neq t$ is
satisfiable over $\bB$, and concludes the proof.

To prove the claim, we will find tuples $c:=(c_1,c_2,c_3,c_4)$ and
$d:=(d_1,d_2,d_3,d_4)$ of the same type as $(a_1,a_2,a_3,a_4)$ and
$(b_1,b_2,b_3,b_4)$, respectively, such that the tuple $e:=f(c,d)$
satisfies $e_1\neq e_2$ and $e_3\neq e_4$. Then, by the homogeneity
of $(\mV;E)$, we can find automorphisms $\alpha$ and $\beta$ of $(\mV;E)$ sending $a$ to $c$ and $b$ to $d$, which suffices for the proof of
our claim.

In the sequel, we will assume that $X(a_1,a_2)$ and $Y(b_3,b_4)$,
where $X,Y\in\{E,N\}$.\\

\textbf{Case 1.} Suppose first that $a_3=a_4\in\{a_1,a_2\}$ and
$b_1=b_2\in\{b_3,b_4\}$; without loss of generality, $a_3=a_2$ and $b_1=b_3$.

\textbf{Case 1.1} There exists $u\in \mV$ such that for all $p,v\in \mV$
with $(u,v) \in Y$ we have $f(p,u)=f(p,v)$. Then, because $f$ preserves
$\neq$, we have $f(p,u)\neq f(q,u)$ for all $p\neq q$. 
If for all $p,v \in \mV$ we have that $f(p,u) = f(p,v)$, then this implies 
that for all $p,v,v' \in \mV$ we have that $f(p,v') = f(p,v)$, contradicting our assumption
that $f$ is essential. So there are $p,v\in \mV$ such that $f(p,u)\neq f(p,v)$. Pick
$w\in \mV$ such that $(w,u), (w,v)\in Y$. Pick moreover $q\in \mV$ such that
$(p,q)\in X$. We have $f(p,v)\neq f(p,u)=f(p,w)$. Moreover,
$f(p,w)=f(p,u)\neq f(q,u)=f(q,w)$. Hence, the tuples $c:=(q,p,p,p)$
and $d:=(w,w,w,v)$ prove the claim.

\textbf{Case 1.2} For all $u\in \mV$  there exist $p,v\in \mV$ with
$(u,v) \in Y$ such that $f(p,u)\neq f(p,v)$. Pick $m,n, u\in \mV$ with
$(m,n)\in X$ and $f(m,u)\neq f(n,u)$. Pick $p,v\in \mV$ such that $(u,v) \in
Y$ and $f(p,u)\neq f(p,v)$. If we can pick $p$ in such a way that
$(p,m), (p,n) \in X$, then since either $f(m,u)\neq f(p,u)$ or $f(n,u)\neq
f(p,u)$ we have that either $(m,p,p,p)$ or $(n,p,p,p)$ proves the
claim together with the tuple $(u,u,u,v)$. So suppose that this is
impossible. Then for any $q\in \mV$ with $(q,m), (q,n)\in X$ we have
$f(q,u)=f(q,v)\neq f(p,u)$, so we have that $(q,p,p,p)$ and
$(u,u,u,v)$ satisfy the claim.\\

\textbf{Case 2.} Now suppose that $a_3=a_4\in\{a_1,a_2\}$ and
$b_1=b_2\notin\{b_3,b_4\}$; wlog $a_3=a_2$. Write $(b_1,b_3)\in Q_3$
and $(b_1,b_4)\in Q_4$, where $Q_3,Q_4\in\{E,N\}$.

\textbf{Case 2.1} There exists $u\in \mV$ such that for all $p,v,r$
with $(v,r)\in Y$, $(u,v) \in Q_3$ and $(u,r)\in Q_4$ we have $f(p,v)=f(p,r)$.
Then one easily concludes that for all $p\in \mV$ and all $v,v'\in \mV$
with $v,v'\neq u$ we have $f(p,v)=f(p,v')$. This implies that
$f(p,v)\neq f(q,v)$ whenever $p\neq q$ and $v\neq u$. Since $f$ is
essential, there exist $p,v\in \mV$ with $(u,v)\in Y$ such that
$f(p,u)\neq f(p,v)$. Now pick $w,q\in \mV$ such that $(w,u)\in Q_3$,
$(w,v)\in Q_4$, and $(q,p)\in X$. Then $f(p,w)\neq f(q,w)$, and so the
tuples $(q,p,p,p)$ and $(w,w,u,v)$ prove the claim.

\textbf{Case 2.2} For all $u$ there exist $p,v,r$ with $(v,r)\in Y$,
$(u,v)\in Q_3$, $(u,r)\in Q_4$ and $f(p,v)\neq f(p,r)$. Pick $m,n, u$ with
$(m,n)\in X$ and $f(m,u)\neq f(n,u)$. Pick $p,v,r\in \mV$ such that
$(v,r)\in Y$, $(u,v)\in Q_3$, $(u,r)\in Q_4$ and $f(p,v)\neq f(p,r)$. If we
can pick $p$ in such a way that $(p,m), (p,n) \in X$, then either
$(m,p,p,p)$ and $(u,u,v,r)$ or $(n,p,p,p)$ and $(u,u,v,r)$ prove the
claim. So suppose that this is impossible. Then for any $q$ with
$(q,m), (q,n)\in X$ and all $v,r\in \mV$ with  $(v,r)\in Y$, $(u,v)\in Q_3$,
$(u,r)\in Q_4$ we have $f(q,v)= f(q,r)$. This implies that for all such
$q$ and all $v,v'\neq u$ we have $f(q,v)=f(q,v')$. Pick $w$ such
that $(w,v)\in Q_3$, $(w,r)\in Q_4$. Pick $q$ such that $(q,p)\in X$. We have
$f(q,w)\neq f(p,w)$, and so $(q,p,p,p)$ and $(w,w,v,r)$ prove the
claim. \\

\textbf{Case 3.} To finish the proof, suppose that
$a_3=a_4\notin\{a_1,a_2\}$ and $b_1=b_2\notin\{b_3,b_4\}$. Write
$(a_3,a_1)\in P_1$, $(a_3,a_2)\in P_2$, $(b_1,b_3)\in Q_3$ and $(b_1,b_4)\in
Q_4$, where $P_i, Q_i\in\{E,N\}$.

\textbf{Case 3.1} There exists $u$ such that for all $p,v,r$ with
$(v,r)\in Y$, $(u,v)\in Q_3$ and $(u,r)\in Q_4$ we have $f(p,v)=f(p,r)$. Then
one easily concludes that for all $p\in \mV$ and all $v,v'\in \mV$ with
$v,v'\neq u$ we have $f(p,v)=f(p,v')$. This implies that $f(p,v)\neq
f(q,v)$ whenever $p\neq q$ and $v\neq u$. 
We claim that
there exist $p,v$ with $(u,v)\in Y$ such that $f(p,u)\neq f(p,v)$.
Otherwise, if $f(p,u)=f(p,v)$ for all $p$, $\mV$, then $f(p,v)=f(p,v')$ for all $p,v,v'$,
and $f$ depends only on its first variable, contradicting the assumption that $f$ is essential.
Now
pick $w,m,n$ such that $(w,u)\in Q_3$, $(w,v)\in Q_4$, $(m,n)\in X$, $(m,p)\in
P_1$, and $(n,p)\in P_2$. Then the tuples $(m,n,p,p)$ and $(w,w,u,v)$
prove the claim.

\textbf{Case 3.2} For all $u$ there exist $p,v,r$ with $(v,r)\in Y$,
$(u,v)\in Q_3$, $(u,r)\in Q_4$ and $f(p,v)\neq f(p,r)$. Pick $m,n, u$ with
$(m,n)\in X$ and $f(m,u)\neq f(n,u)$. Pick $p,v,r$ such that $(v,r)\in Y$,
$(u,v)\in Q_3$, $(u,r)\in Q_4$ and $f(p,v)\neq f(p,r)$. If we can pick $p$
in such a way that $(p,m)\in P_1$ and $(p,n) \in P_2$, then $(m,n,p,p)$
and $(u,u,v,r)$ prove the claim, so suppose that this is impossible.
Then for any $q$ with $(q,m)\in P_1$ and $(q,n)\in P_2$ and all $v,r$ with
$(v,r)\in Y$, $(u,v)\in Q_3$, $(u,r)\in Q_4$ we have $f(q,v)= f(q,r)$. This
is easily seen to imply that for all such $q$ and all $v,v'\neq u$
we have $f(q,v)=f(q,v')$. Pick $w$ such that $(w,v)\in Q_3$, $(w,r)\in
Q_4$, and $w\neq u$. Pick $q,q'$ such that $(q,q')\in X$, $(q,p)\in P_1$ and
$(q',p)\in P_2$. We have $f(q,w)\neq f(q',w)$, and thus
$(q,q',p,p)$
and $(w,w,v,r)$ prove the claim.
\end{proof}

\subsection{Minimal Binary Functions}
\label{ssect:min-bin-inj}
Let $\cC$ be the clone generated by $\Aut(\mV;E)$. 
We know from Theorem~\ref{thm:g-endos} and Theorem~\ref{thm:g-bin-inj} that all essential functions that are minimal above $\cC$
are binary, injective, and preserve both $E$ and $N$. It is the goal of this section to determine these binary minimal functions.
To state the main result, we define the \emph{dual} of an operation $f$ on $(\mV;E)$, which can be imagined as the function obtained from $f$ by exchanging the roles of $E$ and $N$.

\begin{definition}\label{defn:dual}
    The \emph{dual} of a function $f(x_1,\ldots,x_n)$ on $(\mV;E)$ is the function $-f(-x_1,\ldots,-x_n)$.
\end{definition}

\begin{theorem}[from~\cite{RandomMinOps}]\label{thm:minimal-ops}
    If $\bB=(\mV;E,N,\neq,\ldots)$ is first-order definable in $(\mV;E)$ and has an essential polymorphism, it must also
    have at least one of the following binary injective canonical polymorphisms.
    \begin{itemize}
    \item a balanced operation of type $p_1$;
    \item a balanced operation of type $\maxi$;
    \item an $E$-dominated operation of type $\maxi$;
    \item an $E$-dominated operation of type $p_1$;
    \item a binary operation of type $p_1$ that is balanced in the first and
    $E$-dominated in the second argument;
    \end{itemize}
    or one of the duals of the last four operations (the first operation is self-dual).
\end{theorem}

Our proof of Theorem~\ref{thm:minimal-ops} makes essential use of the Ramsey techniques from Chapter~\ref{chap:ramsey}.
As we have seen in Example~\ref{expl:graphs}, the class of all finite graphs is \emph{not} a Ramsey class.
However, the class of all finite \emph{ordered} graphs \emph{is} a Ramsey class (see Example~\ref{expl:all-structs-Ramsey} for a more general result).
This class is clearly an amalgamation class, and we denote its \Fresse-limit by $(\mV;E,<)$. 
Note that the reduct of this structure without the order has the extension property, and hence is isomorphic to the random graph.
It therefore makes sense to use the same symbol $\mV$ for the elements of $(\mV;E,<)$ and the elements of the random graph. Also note that the reduct $(\mV;<)$ is isomorphic to $({\mathbb Q};<)$. 
By the argument above, the following is a direct consequence of Corollary~\ref{cor:inj-can}.

\begin{corollary}
Every essential function that is minimal above the clone generated by $\Aut(\mV;E)$ is a binary injection that is
canonical as a function from $(\mV;E,<)^2$ to $(\mV;E,<)$. 
\end{corollary}

In the rest of this section, \emph{canonical} means canonical as a function from $(\mV;E,<)^2$ to $(\mV;E,<)$,
and \emph{minimal} means minimal as an operation above $\cC$. The following behavior of functions from $(\mV;E,<)^2 \To (\mV;E,<)$ is useful to
describe canonical functions. 

\begin{definition}
    Let $f \colon \mV^2 \rightarrow \mV$, and let $\{R_1,R_2\} = \{E,N\}$. 
    If for all $(x_1,x_2)$, $(y_1,y_2)\in \mV^2$ with $x_1< y_1$, $x_2< y_2$, $R_1(x_1,y_1)$, and $R_2(x_2,y_2)$ we have
    \begin{itemize}
        \item $N(f(x_1,x_2),f(y_1,y_2))$, then we say that \emph{$f$ behaves like $\min$ on input $(<,<)$}.
        \item $E(f(x_1,x_2),f(y_1,y_2))$, then we say that \emph{$f$ behaves like $\max$ on input $(<,<)$}.
        \item $R_1(f(x_1,x_2),f(y_1,y_2))$, then we say that \emph{$f$ behaves like $p_1$ on input $(<,<)$}.
        \item $R_2(f(x_1,x_2),f(y_1,y_2))$, then we say that \emph{$f$ behaves like $p_2$ on input $(<,<)$}.
    \end{itemize}
    Analogously, we define behavior on input $(<,>)$ using pairs $(x_1,x_2), (y_1,y_2)\in \mV^2$ with $x_1 < y_1$ and $x_2> y_2$.
\end{definition}

Of course, we could also have defined ``behavior on input $(> ,> )$'' and ``behavior on input $(> ,< )$''; however, behavior on input $(> ,> )$ equals behavior on input $(< ,< )$, and behavior on input $(> ,< )$ equals behavior on input $(<,>)$ since graphs are symmetric. 
Thus, there are only two kinds of inputs to be considered, namely ``straight input" $(<,<)$ and ``twisted input"  
$(<,>)$.

\begin{proposition}\label{prop:binaryBehaviour}
Let $f \colon \mV^2 \rightarrow \mV$ be injective and canonical, and suppose it preserves $E$ and $N$. Then it behaves like $\min$, $\max$, $p_1$ or $p_2$ on input $(<,<)$. Moreover, it behaves like on $\min$, $\max$, $p_1$ or $p_2$ on input $(<,>)$.
\end{proposition}
\begin{proof}
    By canonicity it suffices to check the statement for all possible types of pairs $x,y\in \mV^2$.
\end{proof}

We remark that the four possibilities correspond to the four binary operations $g$ on the two-element domain $\{E,N\}$ that are \emph{idempotent}, i.e., that satisfy that $g(E,E)=E$ and $g(N,N)=N$.

\begin{definition}
    If $f \colon \mV^2 \rightarrow \mV$ behaves like $X$ on input $(<,<)$ and like $Y$ on input $(<,>)$, where $X,Y\in\{\max,\min, p_1,p_2\}$, then we say that $f$ is of \emph{type $X / Y$}.
\end{definition}

Fix an automorphism $\Neg$ of the graph $(\mV;E)$ that reverses the order on $\mV$; such an automorphism clearly exists
since $(\mV;E,<)$ and $(\mV;E,>)$ are isomorphic. 
Lemma~\ref{lem:lex} shows that any canonical binary injective polymorphism of $(\mV;<)$ has the same behavior as $\lex(x,y)$, $\lex(x,\Neg y)$, $\lex(y,x)$, or $\lex(y,\Neg x)$. 
If $f$ is any function that is canonical as a map from $(\mV;<;E)^2$ to $(\mV;<)$, and does not preserve $<$, then
$(x,y) \mapsto \Neg f(x,y)$ preserves $<$. Moreover, by passing from $f$ to $(x,y) \mapsto f(\Neg x,y)$ or $(x,y) \mapsto f(\Neg x,y)$
we can assume that $f$ behaves either like $\lex(x,y)$ or $\lex(y,x)$. 

We will now prove that minimal binary canonical injections are never of mixed type, i.e., they have to behave the same way on straight and twisted inputs.

\begin{lemma}
    Suppose that $f \colon \mV^2 \rightarrow \mV$ is injective and canonical, and suppose that it is of type $\max / p_i$ or of type $p_i / \max$, where $i\in\{1,2\}$. Then $f$ is not minimal.
\end{lemma}
\begin{proof}
     We prove that $f$ generates a binary injective canonical function $g$ which is of type $\max / \max $. Clearly, all binary injective canonical functions generated by $g$ then are also of type $\max / \max $, so $g$ cannot generate $f$, proving the lemma.

    Assume without loss of generality that $f$ is of type $\max / p_i$, and note that we assume that $f$ behaves like $\lex(x,y)$. Set $h(u,v):=f(u,\Neg v)$. Then $h$ behaves like $p_i$ on input
    $(< ,< )$ and like $\max$ on input $(< ,> )$; moreover, $f(x_1,x_2)< f(y_1,y_2)$ iff $h(x_1,x_2)< h(y_1,y_2)$, for all
     $x_1\neq y_1$ and $x_2\neq y_2$. We then have that $g(u,v):=f(f(u,v),h(u,v))$ is of type $\max / \max$, finishing the proof.
\end{proof}

\begin{lemma}
    Suppose that $f \colon \mV^2 \rightarrow \mV$ is injective and canonical, and suppose that it is of type $\min / p_i$ or of type $p_i / \min$, where $i\in\{1,2\}$. Then $f$ is not minimal.
\end{lemma}
\begin{proof}
    The dual proof works.
\end{proof}

\begin{lemma}
    Suppose that $f \colon \mV^2 \rightarrow \mV $ is injective and canonical, and suppose that it is of type $\max / \min$ or of type $\min / \max$. Then $f$ is not minimal.
\end{lemma}
\begin{proof}
    Assume without loss of generality that $f$ is of type $\max / \min$, and recall that we assume that $f$ behaves like $\lex(x,y)$. Consider $h(u,v):=f(f(u,v),\Neg v)$. Then $h$ is of type $p_2 / p_2$, so it cannot reproduce $f$.
\end{proof}

\begin{lemma}
    Suppose that $f \colon \mV^2 \To \mV $ is injective and canonical, and suppose that it is of type $p_1 / p_2$ or of type $p_2 / p_1$. Then $f$ is not minimal.
\end{lemma}
\begin{proof}
    If $f$ is of type $p_1 / p_2$, then $h(u,v):=f(f(u,v),\Neg v)$ is of type $p_2 / p_2$ and cannot reproduce $f$. If $f$ is of type $p_2 / p_1$, then $g(u,v):=f(u,\Neg v)$ is of type $p_1 / p_2$ and still behaves like $\lex(x,y)$; hence, we are back in the first case.
\end{proof}

We have seen that actually no ``mixed'' types appear for minimal functions. 
In other words, minimal functions that are canonical as functions from $(\mV;E,<)^2 \To (\mV;E,<)$
are also canonical as functions from $(\mV;E)^2 \To (\mV;E)$. This motivates the following definition.

\begin{definition}
    Let $f \colon \mV^2 \To \mV$. We say that $f$ \emph{behaves like $\min$ ($\max$, $p_1$, $p_2$)
    on input $(\neq,\neq)$} iff it behaves like $\min$ ($\max$, $p_1$, $p_2$) both on input $(< ,< )$ and on input $(<,> )$. We also say that $f$ is \emph{of type $\min$ ($\max$, $p_1$, $p_2$)}. If $f$ is of type $p_1$ or $p_2$ then we also say that $f$ is \emph{of type projection}.
\end{definition}

Our observations so far can be summarized as follows.

\begin{proposition}\label{prop:essentialMinimalIndependentOfOrder}
Let $f \colon \mV^2 \To \mV $ be essential and minimal. Then it is injective, canonical as a function from $(\mV;E)^2 \To (\mV;E)$, and behaves like $\min$, $\max$, $p_1$ or $p_2$ on input $(\neq,\neq)$.
\end{proposition}

In the following, we consider further types of tuples $x,y\in \mV^2$. So far, we did not consider the case where $x_1=y_1$ or $x_2=y_2$.

\begin{definition}
    Let $f \colon \mV^2 \To \mV $. We say that $f$ \emph{behaves like $e_E$ ($e_N$, $\id$, $-$) on input $(\neq,=)$}
    iff for every fixed $c\in \mV$, the function $g(x):=f(x,c)$ behaves like $e_E$ ($e_N$, $\id$, $-$). Similarly we define behavior on input $(=,\neq)$.
\end{definition}

If $f$ is canonical and injective, then it behaves like one of the mentioned functions on input $(\neq,=)$ and $(=,\neq)$, respectively. This motivates the following. 

\begin{definition}
We say that $f \colon \mV^2\To \mV$ is \emph{of type  $E/N$} iff $f$ behaves like $e_E$ on input $(\neq,=)$ and like $e_N$ on input $(=,\neq)$. Similarly we define the types $E/E$, $N/E$, $E/\id$, $E/-$, etc. Moreover, we say that $f$ is \emph{balanced} iff it is of type $\id/\id$, we say it is \emph{$E$-dominated} iff it is of type $E/E$, and we say it is \emph{$N$-dominated} iff it is of type $N/N$.
\end{definition}

In the following theorem, we finally characterize those canonical behaviors that yield minimal functions.

\begin{theorem}\label{thm:minimalFunctions}
    The minimal polymorphisms of $(\mV;E,N)$ are precisely the binary
    injective canonical operations of the following types:
       \begin{enumerate}
        \item Projection and balanced.
        \item $\max$ and balanced.
        \item $\min$ and balanced.
        \item $\max$ and $E$-dominated.
        \item $\min$ and $N$-dominated.
        \item Projection and $E$-dominated.
        \item Projection and $N$-dominated.
        \item $p_2$ and $E/\id$, or $p_1$ and $\id/E$.
        \item $p_2$ and $N/\id$, or $p_1$ and $\id/N$.
   \end{enumerate}
   Moreover, these 9 different kinds of minimal functions do not generate one another, and any two functions in the same group do generate one another.
\end{theorem}

If $e$ is essential, then it must be binary and injective by Theorem~\ref{thm:g-bin-inj}.
The rest of the theorem follows from Proposition~\ref{prop:essentialMinimalIndependentOfOrder} and the following lemmas.
By the homogeneity of $(\mV;E)$ and local closure,
it is easy to see that a binary canonical injection in one of the classes of Theorem~\ref{thm:minimalFunctions}  generates all other functions in
    the same class.
The verification of Lemmas~\ref{lem:minimalAnfang} to ~\ref{lem:minimalEnde}  is left to the reader; the proof always uses induction over terms.

\begin{lemma}\label{lem:minimalAnfang}
    Any binary essential function generated by a binary canonical injection of type $\min$, $\max$, or projection, respectively, is of the same type.
\end{lemma}

\begin{lemma}\label{lem:balancedStaysBalanced}
    Any binary essential function generated by a binary canonical injection that is balanced and preserves $E$ and $N$ is balanced.
\end{lemma}

We thus have that the first three classes of functions of Proposition~\ref{thm:minimalFunctions} are indeed minimal. The following lemma proves minimality for items (4) and (5).

\begin{lemma}\label{lem:EdomAndMaxIsEdom}
    Any binary essential function generated by an $E$-dominated binary canonical injection of type $\max$ is $E$-dominated. Dually, any binary essential function generated by an $N$-dominated binary canonical injection of type $\min$ is $N$-dominated.
\end{lemma}

The following lemma proves minimality for items (6) and (7).

\begin{lemma}\label{lem:minimalEnde}
    Any binary essential function generated by an $E$-dominated binary canonical injection of type projection is $E$-dominated. Dually, any binary essential function generated by an $N$-dominated binary canonical injection of type projection is $N$-dominated.
\end{lemma}

It remains to prove minimality for items (8) and (9), which is achieved in the following lemma.
\begin{lemma}\label{lem:skewMinimal}
    Any binary essential function generated by a binary canonical injection of type $E / \id$ and $p_2$ is either of the same type or of type $\id / E$ and $p_1$. Dually, any binary essential function generated by a binary canonical injection of type $N / \id$ and $p_2$ generates is either of the same type or of type $\id / N$ and $p_1$.
\end{lemma}
\begin{proof}
    Let $f(u,v)$ be of type  $E/\id$ and $p_2$. $f(v,u)$ is of type $\id/E$ and $p_1$. Both $f(u,f(u,v))$ and $f(v,f(u,v))$ are of type $E/\id$ and $p_2$. So is $f(f(u,v),v)$. The function $f(f(u,v),u)$ is of type $\id/E$ and $p_1$. Finally, $f(f(u,v),f(v,u))$ also is of type $\id/E$ and $p_1$, so $f$ cannot generate any new behaviors.
\end{proof}

Next we claim that no other functions except for those listed in Theorem~\ref{thm:minimalFunctions} are minimal. This will be achieved in the following lemmas.

\begin{lemma}
    Let $f$ be a binary canonical injection of type $\max$. If $f$ is not balanced or $E$-dominated, then $f$ is not minimal.
\end{lemma}
\begin{proof}
    If $f$ is of type $E/\id$, then $g(x,y):=f(f(x,y),x)$ is $E$-dominated. By Lemma~\ref{lem:EdomAndMaxIsEdom}, $g$ cannot reproduce $f$. If $f$ is of type $E/N$, then $g$ is $E$-dominated as well. So it is if $f$ is of type $E/-$.

    If $f$ is of type $N/\id$, then $g(x,y):=f(x,f(x,y))$ is balanced, so $f$ is not minimal by Lemma~\ref{lem:balancedStaysBalanced}. If $f$ is of type $N/-$, then $g$ is balanced as well.

    If $f$ is of type $\id/-$ or of type $-/-$, then $g(x,y):=f(x,f(x,y))$ is of type $E/\id$, which we have already shown not to be minimal.

    By symmetry, if we switch the arguments in a type of $f$, e.g., if $f$ is of type $\id/E$, then $f$ is not minimal either. We have thus covered all possible types.
\end{proof}

    Analogously, we find that every minimal binary injection of type $\min$ is balanced or $N$-dominated.
    
\begin{lemma}
    Let $f$ be a binary canonical injection of type $p_1$. If $f$ is not balanced, $E$-dominated, $N$-dominated, of type $\id/E$, or of type $\id/N$, then $f$ is not minimal.
\end{lemma}
\begin{proof}
    If $f$ is of type $E/\id$, $E/-$, $-/\id$, or $-/-$, then $g(x,y):=f(x,f(x,y))$ is balanced and cannot reproduce $f$. If it is of type $E/N$ or $\id/-$, then $g$ is of type $E/\id$, and we are back in the preceding case. Dually, if $f$ is of type $N/\id$ or $N/-$, then $g$ is balanced. If it is of type $N/E$, then $g$ is of type $N/\id$, bringing us back to the preceding case. If it is of type $-/E$, then $g$ is of type $\id/E$ and $p_1$, and hence cannot reproduce $f$ by Lemma~\ref{lem:skewMinimal}. The dual argument works if $f$ is of type $-/N$.
\end{proof}

Analogously, we find that every minimal binary injection of type $p_2$ is balanced, $E$-dominated, $N$-dominated, or
of type $E/\id$. 

\subsection{Producing functions that are not of type projection}
\label{ssect:producingCanonical}
Theorem~\ref{thm:minimal-ops} and the following proposition together imply that indeed, if case~(a) of Proposition~\ref{prop:higherArity} does not apply, then one of the other cases does.

\begin{proposition}\label{prop:maxMinMajorityMonority}
    Suppose that $f$ is an operation on $\mV$ that preserves the relations $E$ and $N$ and violates the relation $H_1$.
    Then $f$ generates a binary injective canonical operation of type $\min$ or $\max$, or a ternary injective canonical operation of type minority or majority.
\end{proposition}

We first prove the following.

\begin{lemma}\label{lem:atMostTernary}
 Let $f$ be an operation on $(\mV;E)$ which preserves $E$ and $N$ and violates $H_1$. Then $f$ generates a binary or ternary injection which shares the same properties.
\end{lemma}
\begin{proof}
Since the relation $H_1$ consists of three orbits of 6-tuples, 
by Lemma~\ref{lem:small-arity} $f$ generates an at most ternary function that violates $H_1$,
    and hence we can assume without loss of generality that $f$ itself is at most ternary. The operation
    $f$ must certainly be essential, since essentially unary operations that
    preserve $E$ and $N$ also preserve $H_1$. Applying Theorem~\ref{thm:minimal-ops}, we get that $f$ generates a binary injective canonical function of type $\min$, $\max$, or $p_1$. In the first two cases we are done, since binary injections of type $\min$ and $\max$ violate $H_1$. So 
    consider the last case and denote the function of type $p_1$ by $g$.
    By adding a dummy variable, we may assume that $f$ is ternary. Now consider $$h(x,y,z):= g(g(g(f(x,y,z),x),y),z)\; .$$ 
Then $h$ is clearly injective, and still violates $H_1$ -- the latter can easily be verified combining the facts that $f$ violates $H_1$, $g$ is of type $p_1$, and all tuples in $H_1$ have pairwise distinct entries.
\end{proof}

It will turn out that just as in the proof of Lemma~\ref{lem:atMostTernary}, there are two cases for $f$ in the proof of Proposition~\ref{prop:maxMinMajorityMonority}: either all binary canonical injections generated by $f$ are of type projection, and $f$ generates an edge majority or an edge minority, or $f$ generates a binary canonical injection of type $\min$ or $\max$. We start by considering the first case, which is combinatorially less involved.

\begin{proposition}\label{prop:generatesMajority}
	Let $f$ be an operation on $(\mV;E)$ which preserves $E$ and $N$ and violates $H$. Suppose moreover that all binary injections  generated by $f$ are of type $p_1$ or $p_2$. Then $f$ generates a canonical ternary injection of type majority or minority.
\end{proposition}
\begin{proof}
By Lemma~\ref{lem:atMostTernary}, we can assume that $f$ is a binary or ternary injection; 
if it was binary, it would be of type projection and thus preserve $H_1$, so it must be ternary. 
Because $f$ violates $H_1$, there are $x^1,x^2,x^3 \in H$ such that
$(f(x_1),\dots,f(x_6)) \notin H$, where $x_i := (x^1_i,x^2_i,x^3_i)$ for $1 \leq i \leq 6$.

If there was an automorphism $\alpha$ such that $\alpha x^i = x^j$ for $i \neq j \leq 3$,
then $f$ generates a binary injection that still violates $H_1$, which contradicts the 
assumption that all binary injections generated by $f$ are of type projection. 
By permuting arguments of $f$ if necessary, we can therefore 
assume without loss of generality that 
\begin{align*}
& E(x^1_1,x^1_2), N(x^1_3,x^1_4), N(x^1_5,x^1_6), \\
& N(x^2_1,x^2_2), E(x^1_3,x^1_4), N(x^1_5,x^1_6), \\ 
& N(x^3_1,x^3_2), N(x^1_3,x^1_4), E(x^1_5,x^1_6). 
\end{align*}
We set $$ S := \{ y \in \mV^3 \; | \; \NNN(x_i,y) \text{ for all } i \leq 6 \} \; .$$
Consider the binary relations $Q_1Q_2Q_3$ on $\mV^3$, 
where $Q_i \in \{E,N\}$ for $1 \leq i \leq 3$; each of these
relations defines a 2-type in $(\mV;E)^{[3]}$. 
We claim that for every $2$-type $s$ defined by one of those relations
there is a 2-type $s'$ of $(\mV;E)$ such that $f$ satisfies the type condition 
$(s,s')$ on $S$.
To prove the claim, fix a relation $Q_1Q_2Q_3$ and let $u,v \in S$
be such that $Q_1Q_2Q_3(u,v)$ holds; we must show that 
whether $E(f(u),f(v))$ or $N(f(u),f(v))$ depends only on 
$Q_1Q_2Q_3$ (and not on $u,v$). We go through all possibilities of
$Q_1Q_2Q_3$. 
\begin{enumerate}
\item $Q_1Q_2Q_3 = \ENN$. Let $\alpha \in \Aut(\mV;E)$ 
be such that $\alpha(x^2_1,x^2_2,u_2,v_2)=(x^3_1,x^3_2,u_3,v_3)$; such an automorphism exists since $\NNN(x_1,u)$, $\NNN(x_1,u)$,
$\NNN(x_2,u)$, $\NNN(x_2,v)$
and since $(x^2_1,x^2_2)$ has the same type as
$(x^3_1,x^3_2)$, and $(u_2,v_2)$ has the same type as $(u_3,v_3)$.
By assumption, the operation $g$ defined by $g(x,y) := f(x,y,\alpha y)$ must
be of type projection. Hence, $E(g(u_1,u_2),g(v_1,v_2))$ iff
$E(g(x_1^1,x_1^2),g(x_2^1,x^2_2))$. Combining this with the
equations $(f(u),f(v)) = (g(u_1,u_2),g(v_1,v_2))$ and 
$(g(x^1_1,x_1^2),g(x_2^1,x_2^2)) = (f(x_1),f(x_2))$, we 
get that $E(f(u),f(v))$ iff $E(f(x_1),f(x_2))$, and so we are done.
\item $Q_1Q_2Q_3=\NEN$ or $Q_1Q_2Q_3=\NNE$. These cases are analogous to the previous case.
\item $Q_1Q_2Q_3=\NEE$. Let $\alpha$ be defined as in the first case. 
By assumption, the operation defined by $f(x,y,\alpha y)$ must
be of type projection. Reasoning as above, one gets that $E(f(u),f(v))$ iff $N(f(x_1),f(x_2))$.
\item $Q_1Q_2Q_3=\ENE$ or $Q_1Q_2Q_3=\EEN$. These cases are analogous to the previous case.
\item $Q_1Q_2Q_3= \EEE$ or $Q_1Q_2Q_3=\NNN$. These cases are trivial since $f$ preserves $E$ and $N$.
\end{enumerate}

To show that $f$ generates an operation of type majority or minority, 
it suffices to prove that $f$ generates a function of type majority or minority on $S$ (that is, has on $S$ the same behavior as a function of type majority), since $S$ contains copies of arbitrary finite products of substructures of $(\mV;E)$, and by Lemma~\ref{lem:behavior-generates}. We prove this by another case distinction, based on the fact that $(f(x_1),\dots,f(x_6)) \notin H$. 
\begin{enumerate}
\item Suppose that $E(f(x_1),f(x_2)), E(f(x_3),f(x_4)),E(f(x_5),f(x_6))$. 
Then $f$ itself is of type minority on $S$.
\item Suppose that $N(f(x_1),f(x_2)),N(f(x_3),f(x_4)),N(f(x_5),f(x_6))$. 
Then $f$ itself is of type majority on $S$.
\item Suppose that $E(f(x_1),f(x_2)),E(f(x_3),f(x_4)),N(f(x_5),f(x_6))$. 
Let $e$ be a self-embedding of $(\mV;E)$ such that for all $w \in V$ and all $i \leq 6$ 
we have that $N(x_i,e(w))$. 
% TODO:
Then $(u_1,u_2,e(f(u_1,u_2,u_3))) \in S$ for all 
$(u_1,u_2,u_3) \in S$. 
%  END TODO
Hence, by the above, the ternary operation defined by $f(x,y,e(f(x,y,z)))$ is of type majority on $S$.
\item Suppose that $E(f(x_1),f(x_2))$, $N(f(x_3),f(x_4))$, and $E(f(x_5),f(x_6))$, or that
$N(f(x_1),f(x_2))$, $E(f(x_3),f(x_4))$, and $E(f(x_5),f(x_6))$. 
These cases are analogous to the previous case.
\end{enumerate}
Let $h(x,y,z)$ be a ternary injection of type majority or minority generated by $f$;  it remains to make $h$ canonical. By Theorem~\ref{thm:minimal-ops}, $f$ generates a binary canonical injection $g(x,y)$, which is of type projection by our assumption on $f$. Assume without loss of generality that it is of type $p_1$ and set $t(x,y,z):=g(x,g(y,z))$. Then the function $h(t(x,y,z),t(y,z,x),t(z,x,y))$ is still of type majority or minority and canonical; we leave the straightforward verification to the reader.
\end{proof}

In order to obtain a full proof 
of Proposition~\ref{prop:maxMinMajorityMonority}, 
it remains to show the following proposition.

\begin{proposition}\label{prop:nonProjGeneratesMin}
    Let $f \colon \mV^2 \rightarrow \mV$ be a binary injection preserving $E$ and $N$ that is neither of type $p_1$ nor of type $p_2$. Then $f$ generates a binary injection of type $\min$ or of type $\max$.
\end{proposition}

In the remainder of this section we will show this by a 
Ramsey theoretic analysis of $f$.
The global strategy behind what follows now is to take a binary injection $f$ and fix finitely many constants $\bar c$ from $\mV^2$ on which it can be seen that $f$ is not of type projection. 
Then, using Theorem~\ref{thm:orderedCanonical}, we generate a binary canonical function which is identical with $f$ on all tuples with elements from $\bar c$; this canonical function then still is not of type projection, and can be handled more easily because it is canonical. To reduce the number of cases that we have to consider, we rule out some behaviors of canonical functions already before introducing the constants. 

\begin{lemma}\label{lem:mixtyp:maxp}
    Suppose that $f \colon \mV^2 \rightarrow \mV$ is injective and canonical, and suppose that it is of type $\max / g$ or of type $g / \max$, where $g \in \{\min,p_1,p_2\}$. Then $f$ generates a binary injection of type $\max$.
\end{lemma}
\begin{proof}
    Assume without loss of generality that $f$ is of type $\max / g$ (when $f$ is of type $g / \max$, replace $f$ by $f(x,\Neg y)$, which is of type $g / \max$).
   We also assume that $f$ obeys $p_1$ for the order (otherwise, continue with
   $f(y,x)$ instead of $f(y,x)$). 
   
   Set $h(u,v):=f(u,\Neg v)$. Then $h$ behaves like $g$ on input
    $(<,<)$ and like $\max$ on input $(<,>)$; moreover, $f(x_1,x_2)<f(y_1,y_2)$ iff $h(x_1,x_2)<h(y_1,y_2)$, for all
     $x_1\neq y_1$ and $x_2\neq y_2$. We then have that $f(f(u,v),h(u,v))$ is of type $\max / \max$, which means that it is of type $\max$.
\end{proof}

\begin{lemma}\label{lem:mixtyp:minp}
    Suppose that $f \colon \mV^2 \rightarrow \mV$ is injective and canonical, and suppose that it is of type $\min / p_i$ or of type $p_i / \min$, where $i\in\{1,2\}$. Then $f$ generates a binary injection of type $\min$.
\end{lemma}
\begin{proof}
    The dual proof works.
\end{proof}

\begin{lemma}\label{lem:mixtyp:maxmin}
    Suppose that $f \colon V^2 \rightarrow V $ is injective and canonical as a function from $(V;E,\prec)^2$ to $(V;E,\prec)$, and suppose that it is of type $\maxi / \mini$ or of type $\mini / \maxi$. Then $f$ generates a binary injection of type $\maxi$ (and by duality, a binary injection of type $\mini$).
\end{lemma}
\begin{proof}
    Assume without loss of generality that $f$ is of type $\maxi / \mini$, and remember that we may assume that $f$ obeys $p_1$ for the order. Then $g(x,y):=f(x,f(x,y))$ is of type $\maxi / p_1$ and generates a binary injection of type $\maxi$ by Lemma~\ref{lem:mixtyp:maxp}.
\end{proof}

We next consider the last remaining mixed behavior, $p_1/p_2$, by combining operational with relational arguments.

\begin{lemma}\label{lem:min-relationally}
Let $\bB$ be a structure that is first-order definable in $(\mV;E)$, contains the relations $E$, $N$, $\neq$, and is preserved by a binary injection of type $p_1$. 
Then the following are equivalent.
\begin{enumerate}
\item $\bB$ has a binary injective polymorphism of behavior $\min$.
\item For every primitive positive formula $\phi$ over $\bB$, if 
$\phi \wedge N(x_1,x_2) \wedge \bigwedge_{1\leq i<j\leq 4} x_i \neq x_j$ and $\phi \wedge N(x_3,x_4) \wedge \bigwedge_{1\leq i<j\leq 4} x_i \neq x_j$ 
are satisfiable over $\bB$, then $\phi \wedge N(x_1,x_2) \wedge N(x_3,x_4)$ is satisfiable over $\bB$ as well. 
\item For every finite $F \subseteq \mV^2$ there exists a binary injective polymorphism of $\bB$ which behaves like $\min$ on $F$.
\end{enumerate}
\end{lemma}

\begin{proof}
The implication from (1) to (2) follows directly by applying a binary injective polymorphism of behavior $\min$ to tuples $r, s$ satisfying $\phi \wedge N(x_1,x_2) \wedge \bigwedge_{1\leq i<j\leq 4} x_i \neq x_j$ and $\phi \wedge N(x_3,x_4) \wedge \bigwedge_{1\leq i<j\leq 4} x_i \neq x_j$, respectively.

To prove that (2) implies (3), assume (2) and let $F \subset \mV^2$
be finite. Without loss of generality we can
assume that $F$ is of the form
$\{e_1,\dots,e_n\}^2$, for sufficiently large $n$. Let $\bA$ be the structure induced by
$F$ in $\bB^2$. We construct an injective homomorphism $h$ from 
$\Delta$ to $\bB$; every homomorphism can clearly be extended to a binary polymorphism of $\bB$, for example inductively by using universality of $(\mV;E)$. We construct $h$ in such a way
that the extension behaves as $\min$ on $F$.

To construct $h$, 
consider the formula $\phi_0$ 
with variables $x_{i,j}$ for $1\leq i,j\leq n$ which is the conjunction over all
literals $R(x_{i_1,j_1},\dots,x_{i_k,j_k})$ such that $R$ is a relation in $\bB$ and 
$R(e_{i_1},\dots,e_{i_k})$ and $R(e_{j_1},\dots,e_{j_k})$ hold in $\bB$. So $\phi_0$ states precisely which relations hold in $\bB^2$ on elements from $F$.
Since $\bB$ is preserved by a binary injection, we have that
$\phi_1 := \phi_0 \wedge \bigwedge_{1\leq i,j,k,l \leq n, (i,j) \neq (k,l)} x_{i,j} \neq x_{k,l}$ 
is satisfiable.

Let $P$ be the set of pairs of the form $((i_1,i_2),(j_1,j_2))$
with $i_1,i_2,j_1,j_2 \in \{1,\dots, n\}$, $i_1 \neq j_1$, $i_2 \neq j_2$,
and where $N(e_{i_1},e_{j_1})$ or $N(e_{i_2},e_{j_2})$.
We show by induction on the size of $I \subseteq P$
that the formula $\phi_1 \wedge \bigwedge_{((i_1,i_2),(j_1,j_2)) \in I} N(x_{i_1,i_2},x_{j_1,j_2})$ is satisfiable over $\bB$. Note that this statement
applied to the set $I = P$ gives us the a
homomorphism $h$ from $\bA$ to $\bB$ such that 
for all $a,b \in F$ we have $N(h(a),h(b))$ whenever
$\EN(a,b)$ or $\NE(a,b)$ by setting $h(e_i,e_j):= s(x_{i,j})$, where $s$ is the satisfying assignment for $\phi_1 \wedge \bigwedge_{((i_1,i_2),(j_1,j_2)) \in P} N(x_{i_1,i_2},x_{j_1,j_2})$.

For the induction beginning, let $p = ((i_1,i_2),(j_1,j_2))$ be any
element of $P$. 
 Let $r,s$ be the $n^2$-tuples defined as follows.
\begin{align*}
r & := (e_1,\dots,e_1,e_2,\dots,e_2,\dots,e_n,\dots,e_n) \\
s & := (e_1,e_2,\dots,e_n,e_1,e_2,\dots,e_n,\dots,e_1,e_2,\dots,e_n)
\end{align*}
In the following we use double indices for the entries of $n^2$-tuples;
for example, $r=(r_{1,1},\dots,r_{1,n},r_{2,1},\dots,r_{n,n})$. 
The two tuples $r$ and $s$ satisfy $\phi_0$. To see this, observe that by definition of $\phi_0$ the tuple $$((e_1,e_1),\dots,(e_1,e_n),(e_2,e_1),\ldots,(e_n,e_n))$$ satisfies $\phi_0$ in $\bB^2$; since $r$ and $s$ are projections of that tuple onto the first and second coordinate, respectively, and projections are homomorphisms, $r$ and $s$ satisfy  $\phi_0$ as well. Let $g$ be a binary injective polymorphism of $\bB$ which is of type $p_1$, 
and set $r' := g(r,s)$ and $s' := g(s,r)$. 
Then $r'$ and $s'$ satisfy $\phi_1$ since $g$ is injective. 
Since $p \in P$, 
we have that $N(e_{i_1},e_{j_1})$ or $N(e_{i_2},e_{j_2})$. 
Assume that $N(e_{i_1},e_{j_1})$;
the other case is analogous. Since $r_{i_1,i_2} = e_{i_1}$, 
$r_{j_1,j_2}=e_{j_1}$,
$r' := g(r,s)$, and $g$ is of type $p_1$,
we have that $N(r'_{i_1,i_2},r'_{j_1,j_2})$,
proving that $\phi_1 \wedge N(x_{i_1,i_2},x_{j_1,j_2})$ is satisfiable in $\bB$.

In the induction step, 
let $I \subseteq P$ be a set of cardinality $n \geq 2$, and assume that the
statement has been shown for subsets of $P$ of cardinality $n-1$. Pick any distinct $q_1,q_2\in I$.
We set $$\psi := \phi_1 \wedge \bigwedge_{((i_1,i_2),(j_1,j_2)) \in I \setminus \{q_1,q_2\}} N(x_{i_1,i_2},x_{j_1,j_2})$$
and observe that $\psi$ is a primitive positive formula over $\bB$
(here we use the assumption that $\bB$ contains the relations $N$ and $\neq$).
Write $q_1 = ((u_1,u_2),(v_1,v_2))$ and 
$q_2 = ((u'_1,u'_2),(v'_1,v'_2))$. 
Then the inductive assumption shows that each of
$\psi \wedge N(x_{u_1,u_2},x_{v_1,v_2})$
and
$\psi \wedge N(x_{u'_1,u'_2},x_{v'_1,v'_2})$
is satisfiable in $\bB$. Note that
$\psi$ contains in particular conjuncts that state that
the four variables $x_{u_1,u_2}$,
$x_{v_1,v_2}$,
$x_{u'_1,u'_2}$,
$x_{v'_1,v'_2}$ denote distinct elements. Hence, by (2), the formula $\psi \wedge N(x_{u_1,u_2},x_{v_1,v_2}) \wedge N(x_{u'_1,u'_2},x_{v'_1,v'_2})$ is satisfiable over $\bB$ as well, which is what we had to show.

The implication from (3) to (1) follows from Lemma~\ref{lem:omega-cat-compactness}.
\end{proof}

\begin{lemma}\label{lem:generatingMin:1}
Let $f \colon \mV^2 \rightarrow \mV$ be a binary injection of behavior $p_1/p_2$ which preserves $E$ and $N$. 
Then $f$ generates a binary injection of type $\min$ and a binary injection of type $\max$.
\end{lemma}
\begin{proof}
By Theorem~\ref{thm:minimal-ops}, $f$ generates a binary injection of type $\max$, $\min$, or $p_1$.
Suppose first that it does not generate a binary injection of type $\max$ or $\min$; we will lead this to a contradiction. 
Let $\bB$ be the structure with domain $\mV$ that contains all relations that are first-order definable in $(\mV;E)$ and that are preserved by $f$. 
Since $f$ generates a binary injection of type $p_1$, 
we may apply implication (2) $\rightarrow$ (1) from
Lemma~\ref{lem:min-relationally}. Let $\phi$ be a primitive positive formula with variable set $S$, $\{x_1,\dots,x_4\} \subseteq S$,
such that the formulas 
$\phi \wedge N(x_1,x_2) \wedge \bigwedge_{i<j\leq 4} x_i \neq x_j$ 
and $\phi \wedge N(x_3,x_4) \wedge \bigwedge_{i<j\leq 4} x_i \neq x_j$ 
have in $\bB$ the satisfying assignments $r$ and $s$ from $S \rightarrow \mV$, respectively.

We can assume without loss of generality that $r(x_1) < r(x_2)$ and $r(x_3) < r(x_4)$;
otherwise, since $r(x_1),\dots,r(x_4)$ must be pairwise distinct, 
we can apply an automorphism of $(\mV;E)$ to $r$ such that the resulting map has the required property.
Similarly, by applying an automorphism of $(\mV;E)$ to $s$,
we can assume without loss of generality that 
$s(x_1) < s(x_2)$ and $s(x_3) > s(x_4)$.
Then the mapping $t \colon S \rightarrow \mV$ defined by $t(x) = f(r(x),s(x))$ 
shows that $\phi \wedge N(x_1,x_2) \wedge N(x_3,x_4)$ 
is satisfiable in $\bB$:
\begin{itemize}
\item The assignment $t$ satisfies $\phi$ since $f$ is a polymorphism of $\bB$.
\item We have that $N(t(x_1),t(x_2))$ since $r(x_1) < r(x_2)$, $s(x_1) < s(x_2)$,
 $f$ is of type $p_1$ on input $(<,<)$, and $N(r(x_1),r(x_2))$. 
\item We have that $N(t(x_3),t(x_4))$ since $r(x_3)< r(x_4)$, $s(x_3)> s(x_4)$,
$f$ is of type $p_2$ on input $(<,>)$, and $N(s(x_3),s(x_4))$. 
\end{itemize}
By Lemma~\ref{lem:min-relationally}, we conclude that $\bB$ is preserved by a binary injection of type $\min$, and consequently $f$ generates a binary injection of type $\min$ -- a contradiction.

Therefore, $f$ generates a binary injection of type $\max$ or $\min$. Since the assumptions of the lemma are symmetric in $E$ and $N$, we infer \emph{a posteriori} that $f$ generates both a binary injection of type $\max$ and a binary injection of type $\min$.
\end{proof}

Having ruled out some behaviors without constants, we now examine behaviors when we add constants to the language. 
In the sequel, we will also say that a function $f \colon \mV^2\To \mV$ has behavior $B$ \emph{between two points $x,y\in \mV^2$} if it has behavior $B$ on $\{x,y\}$.

\begin{lemma}\label{lem:generatingMin:2}
   Let $u \in \mV^2$, and set $U:= (\mV\setminus\{u_1\})\times (\mV\setminus\{u_2\})$. Let $f \colon \mV^2 \To \mV$ be a binary injection which preserves $E$ and $N$, 
behaves like $p_1$ between all points $v, w\in U$, and which behaves like
$p_2$ between $u$ and all points in $U$. Then $f$ generates a binary injection of type $\min$ as well as a binary injection of type $\max$.
\end{lemma}

\begin{proof}
Let $\bB$ be the structure with domain $\mV$ that contains all relations
that are first-order definable in $(\mV; E)$ and that are preserved by $f$. Since $U$ contains copies of products of arbitrary finite graphs, $f$ behaves like $p_1$ on arbitrarily large finite substructures of $(\mV; E)^2$, and hence generates a binary injection of type $p_1$ by Lemma~\ref{lem:behavior-generates}. 
Hence $\bB$ is also preserved by such a function, and we may apply the implication from (2) to (1) in Lemma~\ref{lem:min-relationally} to $\bB$. 

Let $\phi$ be a primitive positive formula with variable
set $S$, $\{x_1,\dots,x_4\} \subseteq S$, such that
$\phi \wedge N(x_1,x_2) \wedge \bigwedge_{1\leq i<j\leq 4} x_i \neq x_j$ and $\phi \wedge N(x_3,x_4) \wedge \bigwedge_{1\leq i<j\leq 4} x_i \neq x_j$ 
are satisfiable over $\bB$, witnessed by satisfying assignments
$r,s \colon S \rightarrow \mV$, respectively.

Let $\alpha$ be an automorphism of $(\mV; E)$ that maps $r(x_3)$ to $u_1$,
and let $\beta$ be an automorphism of $(\mV; E)$ that maps $s(x_3)$ to $u_2$. Then $(\alpha r(x_3),\beta s(x_3))=u$, 
and $v:=(\alpha r(x_4),\beta s(x_4)) \in U$ since $\alpha r(x_4) \neq \alpha r(x_3) = u_1$ and $\beta s(x_4) \neq \beta s(x_3) = u_2$. Thus, $f$ behaves
like $p_2$ between $u$ and $v$, and since $s$ satisfies $N(x_3,x_4)$,
we have that $t \colon S \rightarrow \mV$ defined by 
$$t(x) = f(\alpha x,\beta x)$$
satisfies $N(x_3,x_4)$, too. Since $\alpha,\beta,f$ are polymorphisms of 
$\bB$, the assignment $t$ also satisfies $\phi$.
To see that $N(t(x_1),t(x_2))$, observe that
$\alpha r(x_1) \neq \alpha r(x_3)$ and $\beta s(x_1) \neq \beta s(x_3)$, and hence $p:=(\alpha r(x_1),\beta s(x_1)) \notin U$. 
Similarly, $q:=(\alpha r(x_2),\beta s(x_2)) \notin U$. 
Hence, $f$ behaves as $p_1$ between $p$ and $q$, and
since $N(r(x_1),r(x_2))$, so does $t$. 

By Lemma~\ref{lem:min-relationally} we conclude that $\bB$ is
preserved by a binary injection of type $\min$, and
consequently $f$ generates a binary injection of type $\min$.

Since our assumptions on $f$ were symmetric in $E$ and $N$, it follows that $f$ also generates a binary injection of type $\max$.
\end{proof}

\begin{lemma}\label{lem:generatingMin:3}
   Let $u \in \mV^2$, and let $f \colon \mV^2 \To \mV$ be a binary injection that
behaves like $p_1$ between all points $v, w\in U :=
(\mV\setminus\{u_1\})\times (\mV\setminus\{u_2\})$, and which behaves like
$\min$ between $u$ and all points in $U$. Then $f$ generates a binary injection of type $\min$.
\end{lemma}
\begin{proof}
	The proof is identical with the proof in the preceding lemma; note that our assumptions on $f$ here imply more deletions of edges as the assumptions in that lemma, so it can only be easier to generate a binary injection of type $\min$.
\end{proof}

\begin{lemma}\label{lem:generatingMin:4}
   Let $u, v\in \mV^2$ such that ${\neq}{\neq}(u,v)$ and set $W:=(\mV\setminus\{u_1,v_1\})\times (\mV\setminus\{u_2, v_2\})$. Let $f \colon \mV^2\To \mV$ be a binary injection that 
   \begin{itemize}
   \item preserves $E$ and $N$ 
   \item behaves like $p_1$ between all points $w,r\in W$
   \item behaves like $p_1$ between $u$ and all points $w\in W$
   \item behaves like $p_1$ between $v$ and all points $w\in W$
   \item  does not behave like $p_1$ between $u$ and $v$.
   \end{itemize}
    Then $f$ generates a binary
injection of type $\min$ as wella s a binary injection of type $\max$.
\end{lemma}
\begin{proof}
We have to consider the case that $\EN(u,v)$ and $N(f(u),f(v))$,
and the case that $\NE(u,v)$ and $E(f(u),f(v))$. In the first case we prove that
$f$ generates a binary injection of type $\min$; it then follows by duality that in the second case, $f$ generates a binary injection of type $\max$.

As in Lemma~\ref{lem:generatingMin:2}, we apply the implication (2) $\rightarrow (1)$ from Lemma~\ref{lem:min-relationally}. Let $\bB$, $\phi$, $x_1,\dots,x_4$, and $S$ be as in the proof of Lemma~\ref{lem:generatingMin:2}; by the same argument as before, $\bB$ is preserved by a binary injection of type $p_1$. If $N(r(x_3),r(x_4))$, then the assignment $r$ shows that  $\phi \wedge N(x_1,x_2) \wedge N(x_3,x_4)$ is satisfiable and we are done. Otherwise, since $r(x_3) \neq r(x_4)$, we have 
$E(r(x_3),r(x_4))$. Therefore, there is an $\alpha \in \Aut(\mV;E)$
such that $(\alpha r(x_3),\alpha r(x_4))=(u_1,v_1)$. Similarly, 
since $N(s(x_3),s(x_4))$ and $N(u_2,v_2)$,
there is a $\beta \in \Aut(\mV;E)$ such that $(\beta s(x_3),\beta s(x_4))=(u_2,v_2)$. 
We claim that the map $t \colon S \rightarrow \mV$ defined by 
$$t(x) = f(\alpha x,\beta x)$$
is a satisfying assignment for $\phi \wedge N(x_1,x_2) \wedge N(x_3,x_4)$.
The assignment $t$ satisfies $\phi$ since $\alpha,\beta$ and $f$ are polymorphisms of $\bB$.
Then $N(t(x_3),t(x_4))$ holds because $(\alpha r(x_3),\beta s(x_3))=u$
and $(\alpha r(x_4),\beta s(x_4))=v$, and $N(f(u),f(v))$. 
To prove that $N(t(x_1),t(x_2))$ holds, observe that $r(x_1) \neq r(x_3)$
and $r(x_1) \neq r(x_4)$, and hence
$$\alpha r(x_1) \notin \{\alpha r(x_3),\alpha r(x_4)\} = \{u_1,v_1\} \; .$$ 
Similarly, $\beta s(x_1) \notin \{\beta s(x_3),\beta s(x_4)\} = \{u_2,v_2\}$. 
Hence, $(\alpha r(x_1),\beta s(x_1) \in W$. A similar argument for
$x_2$ in place of $x_1$ shows that $(\alpha r(x_2),\beta s(x_2) \in W$.
Since $f$ behaves like $p_1$ between all points of $W$,
and since $r$ satisfies $N(x_1,x_2)$, we have proved the claim. 
This shows that $\bB$ is preserved by a binary injection of type $\min$, and hence $f$ generates such a function.

By symmetry of our assumptions on $f$ in $E$ and $N$, it follows that $f$ generates a binary injection of type $\min$ if and only if it generates a binary injection of type $\max$.
\end{proof}

We are now set up to prove Proposition~\ref{prop:nonProjGeneratesMin}, and hence complete the proof of Proposition~\ref{prop:maxMinMajorityMonority}.
\begin{proof}[Proof of Proposition~\ref{prop:nonProjGeneratesMin}]
    Let $f$ be given. By Theorem~\ref{thm:minimal-ops}, $f$ generates a binary canonical injection $g$ of type projection, $\min$, or $\max$. In the last two cases we are done, so consider the first case. We claim that $f$ also generates a (not necessarily canonical) binary injection $h$ of type $\min$ or $\max$. Then $h(g(x,y),g(y,x))$ is still of type $\min$ or $\max$ and in addition canonical, and the proposition follows.
    
    To prove our claim, fix a finite set $C:=\{c_1,\ldots, c_m\}\subseteq \mV$ such that the fact that $f$ does not behave like a projection is witnessed on $C$. Invoking Theorem~\ref{thm:orderedCanonical}, we may henceforth assume that $f$ is canonical as a function from $(\mV;E,<, c_1,\ldots,c_m)^2$ to $(\mV;E,<)$ (and hence also to $(\mV;E)$ since tuples of equal type in $(\mV;E,<)$ have equal type in $(\mV;E)$). It is clear that this new $f$ must be injective.
    
In the following we consider orbits of elements in the structure $(\mV;E,<, c_1,\ldots,c_m)$. The infinite orbits are precisely the sets of the form $$\{v\in \mV \; | \; Q_i(v,c_i) \text{ and } R_i(v,c_i) \text{ for all } 1\leq i\leq m\},$$ for $Q_1,\ldots,Q_m\in\{E,N\}$, and $R_1,\ldots,R_m\in\{<,>\}$. The finite orbits are of the form $\{c_i\}$ for some $1\leq i\leq m$. Each infinite orbit of $(\mV;E,<, c_1,\ldots,c_m)$ contains copies of arbitrary linearly ordered finite graphs, and in particular, forgetting about the order, of all finite graphs. 
    
Therefore, if $f$ behaves like $\min$ or $\max$ on an infinite orbit of $(\mV;E,<, c_1,\ldots,c_m)$, then by Lemma~\ref{lem:behavior-generates} it generates a function which behaves like $\min$ or $\max$ everywhere, and we are done. Moreover, if $f$ is of mixed type on an infinite orbit, then, again by Lemma~\ref{lem:behavior-generates}, $f$ generates a canonical function which has the same mixed behavior everywhere. But then we are done by Lemmas~\ref{lem:mixtyp:maxp}, \ref{lem:mixtyp:minp}, 
%\ref{lem:mixtyp:maxmin}, 
and \ref{lem:generatingMin:1}. Hence, we may henceforth assume that $f$ behaves like a projection on every infinite orbit. Fix in the following an infinite orbit $O$ and assume without loss of generality that $f$ behaves like $p_1$ on $O$.
 
Now suppose that there exists an infinite orbit $W$ such that $f$ behaves like $p_2$ between all points $u\in O^2$ and $v\in W^2$ for which $u_1< v_1$ and $u_2< v_2$. Then fix any $v\in W^2$, and set $O_1:=\{o\in O \; | \;  o< v_1 \}$ and $O_2:=\{o\in O \; | \; o< v_2 \}$. Set $O_1':=O_1\cup\{v_1\}$ and $O_2':=O_2\cup\{v_2\}$. We then have that $f$ behaves like $p_2$ between $v$ and any point $u$ of $(O_1'\setminus\{v_1\})\times (O_2'\setminus\{v_2\})$, and like $p_1$ between any two points of $(O_1'\setminus\{v_1\})\times (O_2'\setminus\{v_2\})$. Since $(O_i';E,v_i)$ contains copies of all finite substructures of $(\mV;E,v_i)$, for $i\in\{1,2\}$, by Lemma~\ref{lem:behavior-generates} we get that $f$ generates a function which behaves like $p_2$ between $v$ and any point $u$ of $(\mV\setminus\{v_1\})\times (\mV\setminus\{v_2\})$, and which behaves like $p_1$ between any two points of $(\mV\setminus\{v_1\})\times (\mV\setminus\{v_2\})$. Then Lemma~\ref{lem:generatingMin:2} implies that $f$ generates a binary injection of type $\min$ and we are done.

This argument is easily adapted to any situation where there exists an infinite orbit $W$ such that $f$ behaves like $p_2$ between all points $u\in O^2$ and $v\in W^2$ with $R_1(u_1, v_1)$ and $R_2(u_2, v_2)$, for $R_1, R_2\in\{<,>\}$.

When there exists an infinite orbit $W$ such that $f$ behaves like $\min$ between all points $u\in O^2$ and $v\in W^2$ with $R_1(u_1, v_1)$ and $R_2(u_2, v_2)$, then we can argue similarly, invoking Lemma~\ref{lem:generatingMin:3} at the end. Replacing $\min$ by $\max$ we can use the dual argument, with the notable difference that $f$ generates a binary injection of type $\max$ rather than $\min$.

Since $f$ is canonical, one of the situations described so far must occur. Putting this together, we conclude that for every infinite orbit $W$ and all points $u\in O^2$ and $v\in W^2$, $f$ behaves like $p_1$ between $u$ and $v$. Having that, suppose that for an infinite orbit $W$, $f$ behaves like $p_2$ on $W$. Then exchanging the roles of $O$ and $W$ and of $p_1$ and $p_2$ above, we can again conclude that $f$ generates a binary injection of type $\min$. We may thus henceforth assume that $f$ behaves like $p_1$ between all points $u,v\in (\mV \setminus C)^2$.

Pick any $u\in C^2$. Suppose that there exists $v\in (\mV \setminus C)^2$ such that $f$ does not behave like $p_1$ between $u$ and $v$; say without loss of generality that $\EN(u,v)$ and $N(f(u),f(v))$. Let $O_i$ be the (infinite) orbit of $v_i$, for $i\in\{1,2\}$. Then for all $v\in O_1\times O_2$ we have $\EN(u,v)$ and $N(f(u),f(v))$ since $f$ is canonical. Now let $w\in O_2\times O_1$. We distinguish the two cases $E(f(u),f(w))$ and $N(f(u),f(w))$. In the first case, $f$ behaves like $p_2$ between $u$ and all $v\in (O_1\cup O_2)^2$. We can then argue as above and are done. In the second case, $f$ behaves like $\min$ between $u$ and all $v\in (O_1\cup O_2)^2$, and we are again done by the corresponding argument above. We conclude that we may assume that for all $u\in C^2$ and all $v\in (\mV \setminus C)^2$, $f$ behaves like $p_1$ between $u$ and $v$ as well.

Now pick $u,v\in C^2$ such that $f$ does not behave like $p_1$ between $u$ and $v$, say without loss of generality $\EN(u,v)$ and $N(f(u),f(v))$; this is possible since the fact that $f$ does not behave like $p_1$ everywhere is witnessed on $C$. Pick any 16 infinite orbits $O_1,\ldots,O_{16}$ such that for all $Q_1,Q_2,R_1,R_2\in\{E,N\}$ there exists $w\in (O_1\cup\cdots\cup O_{16})^2$ with $Q_1Q_2(u,w)$ and $R_1R_2(v,w)$. Set $S_1:=\{u_1,v_1\}\cup O_1\cup\cdots\cup O_{16}$ and $S_2:=\{u_2,v_2\}\cup O_1\cup\cdots\cup O_{16}$. Then $S_i$ contains copies of all finite substructures of $(\mV;E,u_i,v_i)$, for $i\in\{1,2\}$, and hence applying Lemma~\ref{lem:behavior-generates} to functions from $(\mV;E,u_1,v_1)\times (\mV;E,u_2,v_2)$ to $(\mV;E)$ we see that 
 $f$ generates a function which behaves like $p_2$ between $u$ and $v$, like $p_1$ between $u$ and all points $w\in (\mV\setminus\{u_1,v_1\})\times (\mV\setminus\{u_2,v_2\})$, like $p_1$ between $v$ and all points $w\in (\mV\setminus\{u_1,v_1\})\times (\mV\setminus\{u_2,v_2\})$, and like $p_1$ between any two points $w, r\in (\mV\setminus\{u_1,v_1\})\times (\mV\setminus\{u_2,v_2\})$. But then we are done by Lemma~\ref{lem:generatingMin:4}.
\end{proof}

\section{First-order Expansions of $(\mV;R^{(3)},S^{(3)})$}
\label{sect:r3}
The structure of this section will be similar
to the one of Section~\ref{sect:higherArity}, but $R^{(3)}$ will take the role of $E$, and $S^{(3)}$ will take the role of $N$. The relation $H_1$ will be replaced by the following relation.

\begin{definition}\label{defn:H2}
Let $H_2$ be the smallest $9$-ary relation that
is preserved by $\{\sw\}$ and contains all tuples
$(x_1,y_1,z_1,x_2,y_2,z_2,x_3,y_3,z_3) \in \mV^9$ such that
\begin{align}
& \bigwedge_{i,j \in \{1,2,3\}, i \neq j, u \in \{x_i,y_i,z_i\}, v \in \{x_j,y_j,z_j\}} N(u,v) \nonumber \\
\wedge & \; \big((R^{(3)}(x_1,y_1,z_1) \wedge S^{(3)}(x_2,y_2,z_2) \wedge S^{(3)}(x_3,y_3,z_3))
\nonumber \\ 
& \vee \; (S^{(3)}(x_1,y_1,z_1) \wedge R^{(3)}(x_2,y_2,z_2) \wedge S^{(3)}(x_3,y_3,z_3)) \nonumber\\  
& \vee \; (S^{(3)}(x_1,y_1,z_1) \wedge S^{(3)}(x_2,y_2,z_2) \wedge R^{(3)}(x_3,y_3,z_3)) \big)\; . \nonumber
\end{align}
\end{definition}

\begin{proposition}\label{prop:sw}
Let $\bB$ be a reduct of $(\mV;E)$ whose endomorphisms are precisely the unary functions generated by $\{\sw\}$. 
Then either $H_2$ is primitive positive definable
in $\bB$, or $\bB$ satisfies item (b) or (d)
of Proposition~\ref{prop:higherArity}. 
%Then at least one of the following holds:
%\begin{itemize}
%\item[(a'')] $H_2$ is primitive positive definable
%in $\bB$.
%, and $\Csp(\bB)$ is NP-complete.
%\item[(b'')] $\Pol(\bB)$ contains
%a canonical ternary injection of type minority, as well as a canonical binary injection which is balanced and of type projection. 
%\item[(d'')] $\Pol(\bB)$ contains
%a canonical ternary injection of type minority, as well as a canonical binary injection which is of type $p_1$ and either $E$-dominated or $N$-dominated in the second argument. 
% TODO: check ob diese unterscheidung ueberhaupt noch Sinn macht. 
%\end{itemize}
\end{proposition}

\begin{proposition}\label{prop:h2}
There is a primitive positive interpretation of
    $(\{0,1\}; \OIT)$ in $(\mV;H_2)$, and $\Csp(\mV; H_2)$ is NP-hard.
\end{proposition}
\begin{proof}
This can be shown analogously to Proposition~\ref{prop:h-is-hard}, 
but this time we represent 
1 by triples from $R^{(3)}$ instead of pairs that satisfy $E$, and 0 by triples from $S^{(3)}$, and then use
$H_2$ analogously as we have used $H_1$ in the proof 
of Proposition~\ref{prop:h-is-hard}. 
\end{proof} 

\subsection{Producing canonical functions of type projection} 
As in Section~\ref{ssect:g-bin-inj}, we show that
if $\bB$ has an essential polymorphism $f$, then it must also contain a binary injective polymorphism. 
Every binary injective function generates
a binary injective canonical function, and
those can be classified similarly as in Section~\ref{ssect:min-bin-inj}. Luckily, even though
we do not work in this section under the assumption that $E$ and $N$ are preserved by $f$, we are able to
reduce to this case in our argument.

\begin{proposition}\label{prop:binary-without-EN}
Suppose that $\bB$ has an essential polymorphism. Then $\bB$ is preserved by a constant function, 
$e_E$, $e_N$, or by a canonical binary injection of type $\mini$, 
$\maxi$, or $p_1$. 
\end{proposition}
\begin{proof}
If there is a primitive positive definition of $E$ and $N$, then the statement follows from Theorem~\ref{thm:minimal-ops}. 
So suppose that this is not that case; also suppose that
$\bB$ is not preserved by $e_E$, $e_N$, or a constant function. Then $\Aut(\bB)$ is dense in $\End(\bB)$ by Proposition~\ref{prop:endos}, and so they must violate $E$ and $N$ as otherwise these relations would have a primitive positive definition. By Theorem~\ref{thm:reducts}, we then see that $\Aut(\bB)$ is 2-transitive. 
By Theorem~\ref{thm:2trans}, 
    $\bB$ has a binary injective polymorphism $g$. 
    Since $(\mV;E,\prec)$ is Ramsey (Example~\ref{expl:all-structs-Ramsey}), we can apply 
    Corollary~\ref{cor:inj-can} and obtain that 
      $g$ generates a binary injective function $h$ which is canonical as a function from $(\mV;E,\prec)^2$ to $(\mV;E,\prec)$. The function $x\mapsto h(x,x)$ either preserves $E$ and $N$, or behaves like $-$, $e_E$ or $e_N$. We can assume that it does not behave like $e_E$ or $e_N$, and if it behaves like $-$, we can replace $h$ by $-h$ and assume that 
      $x \mapsto h(x,x)$ preserves $E$ and $N$. Now consider the function  $x\mapsto h(x,\alpha(x))$, where $\alpha\in\Aut(\mV;E)$ reverses $\prec$. Again, we may exclude the possibility that it behaves like $e_E$ or $e_N$. But then the function $(x,y)\mapsto h(h(x,y),h(y,x))$ preserves $E$ and $N$ and we can apply Theorem~\ref{thm:minimal-ops} to conclude that it generates a binary  injection which is canonical as a function from $(\mV;E)^2$ to $(\mV;E)$ and of type $\mini$, $\maxi$, or $p_1$. 
\end{proof}

%We write $S^{(3)}$ for the relation that contains
%all triples of pairwise distinct elements of $\mV$ that
%are not in $R^{(3)}$. 

\begin{corollary}\label{cor:r3-p1}
Let $\bB=(\mV;R^{(3)},S^{(3)},\ldots)$ be first-order definable over $(\mV;E)$ with an essential polymorphism. 
Then $\bB$ is preserved by a binary canonical injection of type $p_1$.
\end{corollary}
\begin{proof}
Since $e_N$ and functions of type $\mini$ 
do not preserve $R^{(3)}$
and $e_E$ and functions of type $\maxi$ do not preserve $S^{(3)}$, 
Proposition~\ref{prop:binary-without-EN}
implies that $\bB$ is preserved by a binary canonical injection of type $p_1$.
\end{proof}

% wo verwendet?
%\begin{proposition}\label{prop:r3-binaryBehaviour}
%Let $f \colon \mV^2 \rightarrow \mV$ be injective and canonical as a function from $(\mV;E,\prec)^2$ to $(\mV;E,\prec)$, and suppose it preserves $R^{(3)}$ and $S^{(3)}$. Then it behaves like $p_1$ or $p_2$ on input $(\prec ,\prec )$ (and similarly on input $(\prec ,\succ)$).
%\end{proposition}
%\begin{proof}
%By definition of the term canonical; one only needs to enumerate all possible types of pairs $(u,v)\in \mV^2$,
%and noting that no function of type $\mini$, $\maxi$, 
%and no function that generates $-$, $e_E$, and $e_N$ is a polymorphism of $(\mV;R^{(3)},S^{(3)})$. 
%\end{proof}

\subsection{Eliminating mixed behavior} 
\begin{lemma}\label{lem:r3-nomixed}
Let $f \colon \mV^2 \to \mV$ be a binary injection that preserves $R^{(3)}$ and $S^{(3)}$. Then $f$ is not of type 
$p_1/p_2$. 
\end{lemma}
\begin{proof}
Suppose for contradiction that $f$ does have the behavior $p_1/p_2$. 
Let $u_1,u_2,u_3 \in \mV$ with $u_1 \prec u_2 \prec u_3$,
$E(u_1,u_2)$, $N(u_2,u_3)$, and $N(u_1,u_3)$.
Let $v_1,v_2,v_3 \in \mV$ with $v_1 \prec v_2 \prec v_3$
and $N(v_1,v_2),E(v_2,v_3),N(v_1,v_3)$. 
Then $E(f(u_1,v_1),f(u_2,v_3))$ and $N(f(u_1,v_1),f(u_3,v_2))$ since $f$ behaves
like $p_1$ on input $(\prec,\prec)$. Moreover, we have
$E(f(u_2,v_3),f(u_3,v_2))$ since $f$ behaves 
like $p_2$ on input $(\prec,\succ)$. 
Then $(u_1,u_2,u_3) \in R^{(3)}$ 
and $(u_1,u_2,u_3)\in R^{(3)}$,
but $(f(u_1,v_2),f(u_2,v_3),f(u_3,v_2)) \notin R^{(3)}$, in contradiction to our assumptions. 
\end{proof}

\subsection{Behaviors relative to vertices}

\begin{lemma}\label{lem:r3-constants}
   Let $u \in \mV^2$, and set 
   $U := (\mV \setminus\{u_1\})\times (\mV \setminus\{u_2\})$. 
   Let $f \colon \mV^2 \to \mV$ be a binary injection which 
behaves like $p_1$ on $U$,
%between all points $v, w\in U$,
 and which behaves like
$p_2$ or $\maxi$ between $u$ and all points in $U$. Then $f$ does not preserve $R^{(3)}$. 
\end{lemma}
\begin{proof}
Let $v,w \in U$ be such that $\NE(u,v)$, $\EN(v,w)$, and $\NN(u,w)$. Then we have $E(f(u),f(v))$, 
$E(f(v),f(w))$, and $N(f(u),f(w))$. Hence,
$R^{(3)}(u_i,v_i,w)$ for $i\in\{1,2\}$, but $S^{(3)}(f(u),f(v),f(w))$. 
\end{proof}

\begin{definition}
We say that a binary injective function $f \colon \mV^2 \to \mV$ is
\begin{itemize}
\item of type $R^{(3)}$-$p_i$, for $i \in \{1,2\}$, 
iff for all $u,v,w \in \mV^2$ with $\NEQNEQ(u,v)$,
$\NEQNEQ(v,w)$, and $\NEQNEQ(u,w)$ we have $R^{(3)}(f(u),f(v),f(w))$ if and only if $R^{(3)}(u_i,v_i,w_i)$.
\item of type $R^{(3)}$-projection iff it is
of type $R^{(3)}$-$p_1$ or of type $R^{(3)}$-$p_2$. 
\end{itemize}
\end{definition}

%Clearly, when $f$ is of type $\xor$, then $(x,y,z) \mapsto f(f(x,y),z)$ is of type minority. 

\begin{proposition}\label{prop:r3-projection}
    Suppose that $f \colon \mV^2 \rightarrow \mV$ preserves $R^{(3)}$ and $S^{(3)})$.
    Then $f$ is of type $R^{(3)}$-projection.
\end{proposition}
\begin{proof}
The proof is similar to the proof of Proposition~\ref{prop:nonProjGeneratesMin}. 
 Fix a finite set $C:=\{c_1,\ldots, c_m\}\subseteq \mV$ such that the fact that $f$ is not of type $R^{(3)}$-projection is witnessed on $C$. Invoking Theorem~\ref{thm:orderedCanonical}, we may henceforth assume that $f$ is canonical as a function from $(\mV;E,\prec, c_1,\ldots,c_m)^2$ to $(\mV;E,\prec)$.
    
In the following we consider orbits in the structure ${(\mV;E,\prec, c_1,\ldots,c_m)}$. The infinite orbits are precisely the sets of the form 
$$\{v\in \mV\; |\; Q_i(v,c_i) \text{ and } R_i(v,c_i) \text{ for all } 1\leq i\leq m\},$$ 
for $Q_1,\ldots,Q_m\in\{E,N\}$, and $R_1,\ldots,R_m\in\{\prec,\succ\}$. The finite orbits are of the form $\{c_i\}$ for some $1\leq i\leq m$. 
Each infinite orbit of $(\mV;E,\prec, c_1,\ldots,c_m)$ 
induces in $(\mV;E,\prec)$ a structure isomorphic to $(\mV;E,\prec)$. Lemma~\ref{lem:full-orbits-generation} implies that if $f$ has a certain behaviour on such an infinite orbit, then it generates a canonical function which has the same behaviour everywhere. 
Therefore we have for all infinite orbits $O$ that $f$
\begin{itemize}
\item cannot be of type $\mini$ or $\maxi$ on $O$ since it preserves $R^{(3)}$ and $S^{(3)}$;
\item cannot have behaviour $\maxi/p_i$ or $p_i/\maxi$ for $i \in \{1,2\}$ on $O$, by Lemma~\ref{lem:mixtyp:maxp};
\item cannot have behaviour $\mini/p_i$ or $p_i/\mini$ for $i \in \{1,2\}$ on $O$, by \ref{lem:mixtyp:minp};
\item it cannot have behaviour $\maxi/\mini$ or $\mini/\maxi$ on $O$, by Lemma~\ref{lem:mixtyp:maxmin};
\item it cannot have behavior $p_1/p_2$ or $p_2/p_1$ on $O$, by Lemma~\ref{lem:r3-nomixed}. 
\end{itemize}
Hence, we may assume that $f$ behaves like a projection on every infinite orbit. Fix in the following an infinite orbit $O$ and assume without loss of generality that $f$ behaves like $p_1$ on $O$.

Let $W$ be any infinite orbit. 
Then since $f$ is canonical, 
it behaves like $p_1$, $p_2$, $\mini$, or $\maxi$
between all $u,v$ with $u \in O^2$, $v  \in W^2$
and $u_1 \prec v_1$ and $u_2 \prec v_2$. 
Consider the case where there exists an infinite orbit $W$ such that $f$ behaves like $p_2$ or $\maxi$ between all points $u\in O^2$ and $v\in W^2$ for which $u_1\prec v_1$ and $u_2\prec v_2$. Then fix any $v\in W^2$, and set $O_1:=\{o\in O\; |\;  o\prec v_1 \}$ and $O_2:=\{o\in O\; |\; o\prec v_2 \}$. Set $O_1':=O_1\cup\{v_1\}$ and $O_2':=O_2\cup\{v_2\}$. We then have that $f$ behaves like $p_2$ or $\maxi$ between $v$ and any point $u$ of $(O_1'\setminus\{v_1\})\times (O_2'\setminus\{v_2\})$, and like $p_1$ between any two points of $(O_1'\setminus\{v_1\})\times (O_2'\setminus\{v_2\})$. Since $(O_i';E,v_i)$ is isomorphic to $(\mV;E,v_i)$, for $i\in\{1,2\}$, by Lemma~\ref{lem:full-orbits-generation} we get that $f$ generates a function which behaves like $p_2$ or $\maxi$ between $v$ and any point $u$ of $(\mV \setminus\{v_1\})\times (\mV \setminus\{v_2\})$, and which behaves like $p_1$ between any two points of $(\mV \setminus\{v_1\})\times (\mV \setminus\{v_2\})$. This is impossible by Lemma~\ref{lem:r3-constants}.
This argument is easily adapted to any situation where there exists an infinite orbit $W$ such that $f$ behaves like $p_2$ between all points $u\in O^2$ and $v\in W^2$ with $R_1(u_1, v_1)$ and $R_2(u_2, v_2)$, for $R_1, R_2\in\{\prec,\succ\}$.
When there exists an infinite orbit $W$ such that $f$ behaves like $\mini$ between all points $u\in O^2$ and $v\in W^2$ with $R_1(u_1, v_1)$ and $R_2(u_2, v_2)$, then we can argue similarly.

Since $f$ is canonical, one of the situations described so far must occur. Putting this together, we conclude that for every infinite orbit $W$ and all points $u\in O^2$ and $v\in W^2$, $f$ behaves like $p_1$ between $u$ and $v$. Having that, suppose that for an infinite orbit $W$, $f$ behaves like $p_2$ on $W$. Then exchanging the roles of $O$ and $W$ and of $p_1$ and $p_2$ above, we again arrive at a contradiction. We may thus henceforth assume that $f$ behaves like $p_1$ on $(\mV \setminus C)^2$. 
%between all points $u,v\in (V\setminus C)^2$. 

Pick any $u\in C^2$. Suppose that there exists $v\in (\mV \setminus C)^2$ such that $f$ does not behave like $p_1$ between $u$ and $v$. Assume first that $\EN(u,v)$ and $N(f(u),f(v))$. Let $O_i$ be the (infinite) orbit of $v_i$, for $i\in\{1,2\}$. Then for all $v\in O_1\times O_2$ we have $\EN(u,v)$ and $N(f(u),f(v))$ since $f$ is canonical. Now let $w\in O_2\times O_1$. We distinguish the two cases $E(f(u),f(w))$ and $N(f(u),f(w))$. In the first case, $f$ behaves like $p_2$ between $u$ and all $v\in (O_1\cup O_2)^2$. We can then argue as above and are done. In the second case, $f$ behaves like $\mini$ between $u$ and all $v\in (O_1\cup O_2)^2$, and we are again done by the corresponding argument above. The dual argument works when $\NE(u,v)$ and $E(f(u),f(v))$. Now assume that $\EE(u,v)$ and $N(f(u),f(v))$. We claim that $\EE(u,v')$ implies $N(f(u),f(v'))$ and $\NN(u,v')$ implies $E(f(u),f(v'))$ for all $v'\in (\mV \setminus C)^2$. Suppose that $v'\in (\mV \setminus C)^2$ is a counterexample. We can find $v''\in (\mV \setminus C)^2$ such that $v_1',v_1''$ and $v_2',v_2''$ belong to the same orbit  and such that $R^{(3)}(u_i,v_i,v_i'')$ for $i\in\{1,2\}$. But then $S^{(3)}(f(u),f(v),f(v''))$, a contradiction. By applying a version of $\sw$ which switches edges and non-edges with respect to $f[C^2]$ to $f$ from the left, we may assume that $f$ behaves like $p_1$ between all $u\in C^2$ and all $v\in (\mV \setminus C)^2$

Since $f$ does not behave like $R^{(3)}$-$p_1$ on $C^2$, in particular it does not behave like $p_1$ on $C^2$. Pick $u,v\in C^2$ witnessing this. Then $f$ behaves like $p_1$ between any point in $\{u,v\}$ and any point in $(\mV \setminus C)^2$. Since $(\mV \setminus C)\cup\{u_i,v_i\}$ induces an isomorphic copy of the random graph for $i\in\{1,2\}$, we can refer to Lemma~\ref{lem:generatingMin:4} to arrive at a contradiction: $f$ generates $e_E$, $e_N$, or a binary injection of type $\mini$ or $\maxi$, all of which violate either $R^{(3)}$ or $S^{(3)}$.
\end{proof}

\begin{definition}
We say that a ternary injective function $f \colon \mV^3 \to \mV$ is
\begin{itemize}
\item \emph{of type $R^{(3)}$-majority}
iff for all $u,v,w \in \mV^3$ with $\NEQNEQNEQ(u,v)$, $\NEQNEQNEQ(u,w)$, $\NEQNEQNEQ(v,w)$ we have $R^{(3)}(f(u),f(v),f(w))$ if and only if $R^{(3)}R^{(3)}R^{(3)}(u,v,w)$, $R^{(3)}R^{(3)}S^{(3)}(u,v,w)$, $R^{(3)}S^{(3)}R^{(3)}(u,v,w)$, or $S^{(3)}R^{(3)}R^{(3)}(u,v,w)$. 
\item \emph{of type $R^{(3)}$-minority} iff 
for all $u,v,w \in \mV^3$ with $\NEQNEQNEQ(u,v)$, $\NEQNEQNEQ(u,w)$, $\NEQNEQNEQ(v,w)$ we have $R^{(3)}(f(u),f(v),f(w))$ if and only if $R^{(3)}R^{(3)}R^{(3)}(u,v,w)$, $R^{(3)}S^{(3)}S^{(3)}(u,v,w)$, $S^{(3)}R^{(3)}S^{(3)}(u,v,w)$, or $S^{(3)}S^{(3)}R^{(3)}(u,v,w)$. 
\end{itemize}
\end{definition}

\begin{lemma}\label{lem:r3-majority}
Functions $f \colon \mV^3 \to \mV$ of type $R^{(3)}$-majority do not preserve $R^{(3)}$.
\end{lemma}
\begin{proof}
Let $u^1,u^2,u^3 \in \mV^4$ be such that 
\begin{itemize}
\item $E(u^1_1,u^1_2)$ and $N(u^1_i,u^1_j)$ for all pairs
$(i,j)$ of distinct elements from $\{1,\dots,4\}$ that
are distinct from $(1,2)$. 
\item $E(u^2_2,u^2_3)$ and $N(u^1_i,u^1_j)$ for all pairs $(i,j)$ of distinct elements from $\{1,\dots,4\}$ that
are distinct from $(2,3)$.  
\item $E(u^3_1,u^3_3)$ and $N(u^3_i,u^3_j)$ for all pairs $(i,j)$ of distinct elements from $\{1,\dots,4\}$ that are distinct from $(1,3)$.  
\end{itemize}
Since $f$ is of type $R^{(3)}$-majority $S^{(3)}(f(u_1),f(u_2),f(u_4))$,
$S^{(3)}(f(u_1),f(u_3),f(u_4))$, and $S^{(3)}(f(u_2),f(u_3),f(u_4))$. Since for all four-element subsets 
of $\mV$ there must always be an
even number of three-element subsets in 
$R^{(3)}$, we have $S^{(3)}(f(x_1),f(x_2),f(x_3))$,
and hence $f$ does not preserve $R^{(3)}$. 
\end{proof}

\begin{lemma}\label{lem:r3-minority}
Let $f \colon \mV^3 \to \mV$ be of type $R^{(3)}$-minority. Then $\{f,\sw\}$ generates a function of
type minority.
\end{lemma}
\begin{proof}
Let $g$ be any ternary injection of type minority, and let $u,v,w \in \mV^3$ with $\NEQNEQNEQ(u,v),\NEQNEQNEQ(u,w),\NEQNEQNEQ(v,w)$ be given. We will show that $R^{(3)}(g(u),g(v),g(w))$
if and only if $R^{(3)}(f(u),f(v),f(w))$. 
Recall that $R^{(3)}(f(u),f(v),f(w))$ 
if and only if
\begin{align*}
& R^{(3)}S^{(3)}S^{(3)}(u,v,w), \\
& S^{(3)}R^{(3)}S^{(3)}(u,v,w), \\
& S^{(3)}S^{(3)}R^{(3)}(u,v,w), \\
\text{or } & R^{(3)}R^{(3)}R^{(3)}(u,v,w) \, .
\end{align*}
This is in turn the case if and only if the cardinality of the set $$E \cap \bigcup_{i \in \{1,2,3\}}\{(u_i,v_i),(u_i,w_i),(v_i,w_i)\}$$ is odd, which 
is the case if and only if $E \cap \{(g(u),g(v)),(g(u),g(w)),(g(v),g(w))\}$ is odd, which is the case if and only if $R^{(3)}(g(u),g(v),g(w))$ holds. 

By Corollary~\ref{cor:r3-p1}, $f$ generates a binary canonical injection $s(x,y)$ of type $p_1$. Set $t(x,y,z):=s(x,s(y,z))$. As in the proof of 
 Proposition~\ref{prop:generatesMajority}
 the function $p(x,y,z):=f(t(x,y,z),t(y,z,x),t(z,x,y))$ is still of type $R^{(3)}$-minority, and the function $q(x,y,z):=g(t(x,y,z),t(y,z,x),t(z,x,y))$ is still of type minority. Moreover, by the above we have $R^{(3)}(p(u),p(v),p(w))$ if and only if $R^{(3)}(q(u),q(v),q(w))$ for all $u,v,w \in \mV^3$, since $t$ is injective. Therefore, the homogeneity of $(\mV;R^{(3)})$ implies that for all finite $S\subseteq \mV^3$ there exists a unary operation $a$
generated by $\{\sw\}$ such that the ternary function $a(p(x,y,z))$ agrees with $q(x,y,z)$ on $S$. By local closure, $q$ is thus generated by $\{f,\sw\}$.
\end{proof}

\begin{lemma}\label{lem:Ternary2}
Let $\bB=(\mV;R^{(3)},S^{(3)},\ldots)$ be a reduct of $(\mV;E)$ 
such that $H_2$ is not primitive positive definable.
Then $\bB$ has a ternary injective polymorphism which violates $H_2$. 
\end{lemma}
\begin{proof} 
Since the relation $H_2$ consists of three orbits of 9-tuples in $\Aut(\mV;R^{(3)})$, 
Lemma~\ref{lem:small-arity} implies that $f$ generates an at most ternary function that violates $H_2$, and hence we can assume that $f$ itself is at most ternary; by adding a dummy variable if necessary, we may assume that $f$ is actually ternary. Moreover, 
    $f$ must certainly be essential, since essentially unary operations that
    preserve $R^{(3)}$ and $S^{(3)}$ are generated by $\{\sw\}$ and hence
    also preserve $H_2$.  
    Corollary~\ref{cor:r3-p1} implies that 
       $\bB$ is preserved by a binary canonical injection $g$ of type $p_1$. Consider $$h(x,y,z):= g(g(g(f(x,y,z),x),y),z)\; .$$ 
Then $h$ is clearly injective, and still violates $H_2$ -- the latter can easily be verified combining the facts that $f$ violates $H_2$, $g$ is of type $p_1$, and all tuples in $H_2$ have pairwise distinct entries.
\end{proof}

\begin{proposition}\label{prop:violatesh2}
Let $f$ be an operation on $(\mV;E)$ that preserves
$R^{(3)}$ and $S^{(3)}$ and violates $H_2$.
%Suppose moreover that all binary injections
%generated by $\{f,\sw\}$ are of type $R^{(3)}$-projection. 
Then $\{f,\sw\}$ generates a ternary canonical injection of type minority.
\end{proposition}
\begin{proof}
The proof is similar to the proof of Proposition~\ref{prop:generatesMajority}.
By Lemma~\ref{lem:Ternary2}, we can 
assume that $f$ is a ternary injection. 
Because $f$ violates $H_2$, there are $x^1,x^2,x^3 \in H_2$ such that $f(x^1,x^2,x^3) \notin H_2$. In the following, we will write 
$x_i := (x_i^1,x_i^2,x_i^3)$ for $1 \leq i \leq 9$. 
So $(f(x_1),\dots,f(x_9)) \notin H_2$. If there were a
map $a$ generated by $\sw$ such that 
 $a(x^i) = x^j$ for $1\leq i \neq j \leq 3$,
then $\{f,\sw\}$ would generate a binary injection that still violates $H_2$. 
Proposition~\ref{prop:r3-projection} asserts that all binary injections generated 
by $\{f,\sw\}$ are of type $R^{(3)}$-projection, so we have reached a contradiction since 
operations of type $R^{(3)}$-projection preserve $H_2$. By permuting arguments of $f$ if necessary, 
we can therefore 
assume without loss of generality that 

\begin{align*}
	R^{(3)}S^{(3)}S^{(3)}(x_1,x_2,x_3),\, S^{(3)}R^{(3)}S^{(3)}(x_4,x_5,x_6),\,\text{and } S^{(3)}S^{(3)}R^{(3)}(x_7,x_8,x_9).
\end{align*}

We set $$S := \{ y \in \mV^3 \; | \; \NNN(x_i,y) \text{ for all } 1\leq i \leq 9 \} \; .$$
Consider the ternary relations $Q_1Q_2Q_3$ on $\mV^3$, where $Q_i \in \{R^{(3)},S^{(3)}\}$ for $1\leq i\leq 3$; each of these relations defines a 3-type in $(\mV;R^{(3)})$. 
We claim that for fixed  $Q_1Q_2Q_3$, whether or not $R^{(3)}(f(u),f(v),f(w))$ holds for $u,v,w\in S$ with $Q_1Q_2Q_3(u,v,w)$ does not depend on $u,v,w$. We go through all possibilities of $Q_1Q_2Q_3$.
\begin{enumerate}
\item $Q_1Q_2Q_3=R^{(3)}S^{(3)}S^{(3)}$. Let $\alpha \in \Aut(\mV;R^{(3)})$ be such that the tuple $(x^2_1,x^2_2,x^2_3,u_2,v_2,w_2)$ is mapped
to $(x^3_1,x^3_2,x^3_3,u_3,v_3,w_3)$; such an automorphism exists since $\NNN(x_1, u)$, 
$\NNN(x_1, v)$, $\NNN(x_1,w)$,
$\NNN(x_2, u)$, $\NNN(x_2, v)$, $\NNN(x_2, w)$ and since the tuple $(x^2_1,x^2_2,x^2_3)$ has the same type as
$(x^3_1,x^3_2,x^3_3)$, and $(u_2,v_2,w_2)$ has the same type as $(u_3,v_3,w_3)$ in $(\mV;R^{(3)})$.
By Proposition~\ref{prop:r3-projection}, the operation $g$ defined by $g(x,y):=f(x,y,\alpha(y))$ must
be of type $R^{(3)}$-projection. Hence, $R^{(3)}(g(u_1,u_2),g(v_1,v_2),g(w_1,w_2))$ iff $R^{(3)}(g(x_1^1,x_1^2),g(x_2^1,x_2^2),g(x_3^1,x_3^2))$. Combining this with the equations 
\begin{align*}
(f(u),f(v),f(w))=& \; (g(u_1,u_2),g(v_1,v_2),g(w_1,w_2)) \text{ and } \\(g(x_1^1,x_1^2),g(x_2^1,x_2^2),g(x_3^1,x_3^2))= &  \;(f(x_1),f(x_2),f(x_3))
\end{align*}
we get that $R^{(3)}(f(u),f(v),f(w))$ iff $R^{(3)}(f(x_1),f(x_2),f(x_3))$, and so we are done.
\item $Q_1Q_2Q_3=S^{(3)}R^{(3)}S^{(3)}$ or $Q_1Q_2Q_3=S^{(3)}S^{(3)}R^{(3)}$. These cases are analogous to the previous case.
\item $Q_1Q_2Q_3=S^{(3)}R^{(3)}R^{(3)}$. Let $\alpha$ be defined as in the first case. 
By Proposition~\ref{prop:r3-projection}, the operation defined by $f(x,y,\alpha(y))$ must
be of type projection. Reasoning as above, one obtains that $R^{(3)}(f(u),f(v),f(w))$ iff $S^{(3)}(f(x_1),f(x_2),f(x_3))$.
\item $Q_1Q_2Q_3=R^{(3)}S^{(3)}R^{(3)}$ or $Q_1Q_2Q_3=R^{(3)}R^{(3)}S^{(3)}$. These cases are analogous to the previous case.
\item $Q_1Q_2Q_3= R^{(3)}R^{(3)}R^{(3)}$ or $Q_1Q_2Q_3=S^{(3)}S^{(3)}S^{(3)}$. These cases are trivial since $f$ preserves $R^{(3)}$ 
and $S^{(3)}$.
\end{enumerate}

To show that $f$ generates an operation of type minority, by Lemma~\ref{lem:full-orbits-generation} 
it suffices to prove that $f$ generates a function 
of type minority on  
$S$, since $S$ is the product of isomorphic copies of $(\mV;E)$. We show this by another case distinction, based on the fact that 
$(f(x_1),\dots,f(x_9)) \notin H_2$. 
\begin{enumerate}
\item Suppose that $R^{(3)}(f(x_1),f(x_2),f(x_3))$, 
$R^{(3)}(f(x_4),f(x_5),f(x_6))$
and that $R^{(3)}(f(x_7),f(x_8),f(x_9))$. 
By the above, note that $R^{(3)}(f(u),f(v),f(w))$ for $u,v,w \in S$
if and only if
$R^{(3)}S^{(3)}S^{(3)}(u,v,w)$,  
$S^{(3)}R^{(3)}S^{(3)}(u,v,w)$,
$S^{(3)}S^{(3)}R^{(3)}(u,v,w)$, or
$R^{(3)}R^{(3)}R^{(3)}(u,v,w)$.
Hence, $f$ behaves like an $R^{(3)}$-minority on $S$,
and we are done by Lemma~\ref{lem:r3-minority}. 
\item Suppose that 
$S^{(3)}(f(x_1),f(x_2),f(x_3))$, 
$S^{(3)}(f(x_4),f(x_5),f(x_6))$,
and that $S^{(3)}(f(x_7),f(x_8),f(x_9))$.
Then $f$ behaves like an $R^{(3)}$-majority on $S$,
which is impossible by Lemma~\ref{lem:r3-majority}. 
\item Suppose that 
$R^{(3)}(f(x_1),f(x_2),f(x_3))$,
$R^{(3)}(f(x_4),f(x_5),f(x_6))$,
and that $S^{(3)}(f(x_7),f(x_8),f(x_9))$. 
Let $e$ be a self-embedding of $(\mV;E)$ such that for all $w \in \mV$, all $1\leq j\leq 3$, and all $1\leq i \leq 9$ 
we have that $N(x_i^j,e(w))$. 
Then $(u_1,u_2,e(f(u_1,u_2,u_3))) \in S$ for all 
$(u_1,u_2,u_3) \in S$. 
Hence, by the above, 
the ternary operation defined by $f(x,y,e(f(x,y,z)))$ is of type $R^{(3)}$-majority on $S$; but this is impossible by Lemma~\ref{lem:r3-majority}.
\item Suppose that 
$R^{(3)}(f(x_1),f(x_2),f(x_3))$, 
$S^{(3)}(f(x_4),f(x_5),f(x_6))$, and that $R^{(3)}(f(x_7),f(x_8),f(x_9))$. Analogous to the previous case.
\item Suppose that $S^{(3)}(f(x_1),f(x_2),f(x_3))$, $R^{(3)}(f(x_4),f(x_5),f(x_6))$, and that $R^{(3)}(f(x_7),f(x_8),f(x_9))$. 
Analogous to the previous case.
\end{enumerate}
Let $h(x,y,z)$ be a ternary injection of type minority generated by $f$; 
 it remains to make $h$ canonical. 
 By Corollary~\ref{cor:r3-p1}, $f$ generates a binary canonical injection $g(x,y)$ of type $p_1$. Set $t(x,y,z):=g(x,g(y,z))$. As in the proof of 
 Proposition~\ref{prop:generatesMajority}
 the function $h(t(x,y,z),t(y,z,x),t(z,x,y))$ is still of type minority and canonical.
\end{proof}

\begin{proof}[Proof of Proposition~\ref{prop:sw}]
Assume that $H_2$ is not primitive positive definable; by Theorem~\ref{thm:inv-pol} there exists a polymorphism $f$ of $\bB$ that violates $H_2$. 
Since $\Aut(\bB)$ contains $\sw$, the relations $R^{(3)}$ and $S^{(3)}$ consist of only one orbit of triples in $\bB$. Therefore, since they are preserved by all endomorphisms of $\bB$, it follows by Theorem~\ref{thm:inv-pol} and Lemma~\ref{lem:small-arity} that
these relations are primitive positive definable in $\bB$. 

We can now apply Proposition~\ref{prop:violatesh2}
and obtain that $\{f,\sw\}$ generates a ternary injection of type minority which is canonical as a function from $(\mV;E)$ to $(\mV;E)$. 
Corollary~\ref{cor:r3-p1} implies that $\bB$
is preserved by a binary injection of type $p_1$ which is canonical as a function from $(\mV;E)$ to $(\mV;E)$, and the statement follows from Theorem~\ref{thm:minimal-ops}.
\end{proof}

\section{First-order Expansions of $(\mV;R^{(4)},S^{(4)})$}
\label{sect:r4}
We assume that the endomorphisms of $\bB$ are exactly the functions generated by $\{-\}$. In particular, $\Aut(\bB)$ contains $-$ but not $\sw$, and
the automorphisms of $\bB$ generate
its endomorphisms.

\begin{definition}\label{def:H1p}
Let $H_1'$ be the smallest $6$-ary relation that
is preserved by $\{-\}$ and contains $H_1$.
\end{definition}

\begin{proposition}\label{prop:h1prime}
   There is a primitive positive interpretation of
    $(\{0,1\}; \NAE)$ in $(\mV;H'_1)$, and $\Csp(\mV; H'_1)$ is NP-hard.
\end{proposition}
\begin{proof}
Similar to the proof of Proposition~\ref{prop:h-is-hard}.
\end{proof}

The following is an analog of Proposition~\ref{prop:higherArity} for the situation of this section.
 
\begin{proposition}\label{prop:minus}
Let $\bB$ be a reduct of $(\mV;E)$ whose endomorphisms are precisely the unary functions generated by $\{-\}$. Then either $H_1'$ is primitive positive definable in $\bB$, or one of the cases (b)-(e)
of Proposition~\ref{prop:higherArity} applies.
%Then at least one of the following holds:
%\begin{itemize}
%\item[(a')] $H_1'$ is primitive positive definable
%in $\bB$.
%\item[(b')] $\Pol(\bB)$ contains a canonical ternary injection of type minority, as well as a canonical binary injection which is of type $p_1$ and either $E$-dominated or $N$-dominated in the second argument.
%\item[(c')] $\Pol(\bB)$ contains
%a canonical ternary injection of type majority, as well as a canonical binary injection which is balanced and of type projection. 
%\item[(d')] $\Pol(\bB)$ contains a canonical ternary injection of type minority, as well as a canonical binary injection which is balanced and of type projection.
%\item[(e')] $\Pol(\bB)$ contains
%a canonical ternary injection of type majority, as well as a canonical binary injection which is of type $p_1$ and either $E$-dominated or $N$-dominated in the second argument. 
%\end{itemize}
\end{proposition}
% TODO: check whether all four cases
% can really appear!
\begin{proof}
Note that $H_1'$ consists of three orbits
of 6-tuples in $\Aut(\bB)$,
and hence, if $H_1'$ is not primitive positive
definable in $\bB$, then
there exists by
Theorem~\ref{thm:inv-pol}
and Lemma~\ref{lem:small-arity} a ternary polymorphism $f$ of
$\bB$ that violates $H_1'$.
That is, there are $t^1,t^2,t^3 \in H_1'$ such
that $f(t^1,t^2,t^3) \notin H_1'$. Note that
for each $t^j$, either $t^j$ or $-t^j \in H_1$.
In the first case we set $g_j$ to be the identity function on $\mV$, in the second
case we let $g_j$ be the operation $-$.
Now consider the function $f'$ defined by
$f'(x_1,x_2,x_3) := f(g_1(x_1),g_2(x_2),g_3(x_3))$.
We have that $s^{j} := g_j^{-1}(t^j) \in H_1$, but $f'(s^1,s^2,s^3) =
f(t^1,t^2,t^3)$ is not in $H_1'$. Consider the function $h(x):=
f'(x,x,x)$; since the endomorphisms of $\bB$ are generated by
$\{-\}$, $h$ either preserves $E$ and $N$, or it flips them. By
replacing $f'$ by $-(f')$ in the latter case we may assume that $h$ preserves
$E$ and $N$. Note that we still have that $f'(s^1,s^2,s^3)$ is not in
$H_1'$, and therefore not in $H_1$ either.  Hence,
$f'$ violates $H_1$.

Now suppose that $f'$ violates $E$ or $N$; we will derive a
contradiction. Say without loss of generality that there are $u,v\in
\mV^3$ with $\EEE(u,v)$ such that $E(f'(u),f'(v))$ does not hold. Pick distinct 
$a,b,c,d\in \mV$ such that $\{a,b,c,d\}$ induces a clique in $(\mV;E)$, and such that each element is connected to all entries on $u,v$ by an edge. Pick then
$\alpha_1,\alpha_2,\alpha_3\in\Aut(\mV;E)$ such that $\alpha_i(a)=u_i$ and
$\alpha_i(b)=v_i$ for all $i\in\{1,2,3\}$, and such that
$\alpha_1({c})=\alpha_2({c})=\alpha_3({c})=c$ and
$\alpha_1({d})=\alpha_2({d})=\alpha_3({d})=d$. We then have that the
function $x \mapsto f'(\alpha_1(x),\alpha_2(x),\alpha_3(x))$ maps
$(c,d)$ to an edge since $h(x)$ preserves $E$, but it does not map
$(a,b)$ to an edge, by our assumption on $u$ and $v$. This is,
however, impossible, since  the function must be generated by $\{-\}$.

Therefore, $f'$ preserves $E$ and $N$. Then
Proposition~\ref{prop:higherArity} implies that $f'$
generates functions with the desired properties, or a binary canonical injection 
of type $\maxi$ or $\mini$.
A binary canonical injection of type $\maxi$
together with $\{-\}$ generates a binary canonical injection of type $\mini$, and vice
versa. Then $$\maxi(\mini(x,y),\mini(y,z),\mini(x,z))$$ 
is a ternary canonical injection of type majority with the desired properties, and we are also
done in this case, since identifying two of its variables yields a binary canonical injection of type projection.
\end{proof}

\section{First-order Expansions of $(\mV;R^{(5)},S^{(5)})$}
\label{sect:r5}
We assume that the endomorphisms of $\bB$ are precisely the unary functions generated by $\{-,\sw\}$. In particular, $\Aut(\bB)$ contains $-,\sw$, and
the automorphisms of $\bB$ generate
its endomorphisms. 
%The proof for this case is similar to that for Case~({c}) of Proposition~\ref{prop:endos}, presented in Section~\ref{sect:R4}. 

\begin{definition}\label{def:H2p}
Let $H_2'$ be the smallest $9$-ary relation that
is preserved by $-$ and contains $H_2$.
\end{definition}

\begin{proposition}\label{prop:h2prime}
   There is a primitive positive interpretation of
    $(\{0,1\}; \NAE)$ in $(\mV;H_2')$, and $\Csp(\mV; H_2')$ is NP-hard.
\end{proposition}
\begin{proof}
Similar to Proposition~\ref{prop:h2}. 
%If $H_2'$ is primitive positive definable in 
%$\bB$, then 
%one can show similarly 
%as in the proof of Proposition~\ref{prop:h2-is-hard} 
%that $\Csp(\bB)$ is NP-hard, 
%by reduction from positive Not-all-three-equal-3SAT instead of positive 1-in-3-3SAT, 
%and by simulating $1$ with $R^{(3)}$ instead of $E$,
%and $0$ with $S^{(3)}$ instead of $N$.
\end{proof}

The following is an analog of Proposition~\ref{prop:higherArity} for the situation of this section.

\begin{proposition}\label{prop:minus-sw}
Let $\bB$ be a reduct of $(\mV;E)$ whose endomorphisms are precisely the unary functions generated by $\{-,\sw\}$. 
Then $H_2'$ is primitive positive definable
in $\bB$, or (b) or (d) from Proposition~\ref{prop:higherArity} applies. 
% TODO: check whether all four cases
% can really appear!
%, and $\Csp(\bB)$ is NP-complete.
%\item[(b'')] $\Pol(\bB)$ contains a canonical ternary injection of type minority, as well as a canonical binary injection which is of type $p_1$ and either $E$-dominated or $N$-dominated in the second argument.
%\item[(c''')] $\Pol(\bB)$ contains
%a canonical ternary injection of type majority, as well as a canonical binary injection which is balanced and of type projection. 
%\item[(d''')] $\Pol(\bB)$ contains a canonical ternary injection of type minority, as well as a canonical binary injection which is balanced and of type projection.
%\item[(e''')] $\Pol(\bB)$ contains
%a canonical ternary injection of type majority, as well as a canonical binary injection which is of type $p_1$ and either $E$-dominated or $N$-dominated in the second argument. 
%\end{itemize}
\end{proposition}
\begin{proof}
Note that $H_2'$ consists of three orbits
of 9-tuples in $\Aut(\bB)$, 
and hence, if $H_2'$ is not primitive positive
definable in $\bB$, then
there exists by 
Theorem~\ref{thm:inv-pol} 
and Lemma~\ref{lem:small-arity} a ternary polymorphism $f$ of
$\bB$ that violates $H_2'$.  
That is, there are $t^1,t^2,t^3 \in H_2'$ such
that $f(t^1,t^2,t^3) \notin H_2'$. Note that 
for each $t^j$, either $t^j$ or $-t^j \in H_2$. 
In the first case we set $g_j$ to be the identity function on $\mV$, in the second
case we let $g_j$ be the operation $-$. 
Now consider the function $f'$ defined by
$f'(x_1,x_2,x_3) := f(g_1(x_1),g_2(x_2),g_3(x_3))$.
We have that $s^{j} := g_j^{-1}(t^j) \in H_2$, but $f'(s^1,s^2,s^3) = f(t^1,t^2,t^3)$ is not in $H_2'$, and therefore not in $H_2$ either.  Hence,
$f'$ violates $H_2$. 
The function 
$h(x):=f'(x,x,x)$ is generated by $\{-,\sw\}$, and hence $h$ either preserves $R^{(3)}$ and $S^{(3)}$,
or it flips them. Since $f'(s^1,s^2,s^3)$ is not in $H_2'$, neither is $-f'(s^1,s^2,s^3)$, and in particular not in $H_2$, so also $-f'$ violates $H_2$.  
Hence, by replacing $f'$ with $-f'$ if necessary, 
we may assume 
that $h$ preserves $R^{(3)}$ and $S^{(3)}$. 

We claim that $f'$ preserves $R^{(3)}$ and $S^{(3)}$. 
Suppose for contradiction that there are $u,v,w \in \mV^3$ with $R^{(3)}(u_i,v_i,w_i)$ for all $i \in \{1,2,3\}$ such that
$R^{(3)}(f'(u),f'(v),f'(w))$ does not hold;
the case where $f'$ violates $S^{(3)}$ can be treated similarly. 
If $(u_1,v_1,w_1)$, $(u_2,v_2,w_2)$, and $(u_3,v_3,w_3)$ all lie in the same orbit of triples in $(\mV;E)$, then we choose 
$a,b,c \in \mV$ with $R^{(3)}(a,b,c)$
such that $N(x,y)$ for $x \in \{a,b,c\}$ 
and $y \in \{u_1,v_1,w_1,u_2,v_2,w_2,u_3,v_3,w_3\}$. Then by the homogeneity of $(\mV;E)$ there is for each $i \in \{2,3\}$ a unary operation 
$\alpha_i \in \Aut(\mV;E)$ such that
%$\alpha_i(u_1) = u_i$, $\alpha_i(v_1) = v_i$, $\alpha_i(w_1) = w_i$, and $\alpha_i(a) = 
$\alpha_i(u_1,v_1,w_1,a,b,c) = (u_i,v_i,w_i,a,b,c)$. 
We then have that the unary function $g(x):= f'(x,\alpha_2(x),\alpha_3(x))$ maps $(u_1,v_1,w_1) \in R^{(3)}$ to $(f'(u),f'(v),f'(w)) \notin R^{(3)}$. But $g$ and the function $h$ above agree on $\{a,b,c\}$, and hence $g$ preserves $R^{(3)}$ on $\{a,b,c\}$, but violates it on $\{u_1,v_1,w_1\}$. 
This contradicts the assumption that $g$ is generated by $\{-,\sw\}$. 

So suppose in the following that $R^{(3)}(f'(u),f'(v),f'(w))$ for all $u,v,w \in \mV^3$ with $R^{(3)}(u_i,v_i,w_i)$ for all $i \in \{1,2,3\}$ such that $u,v,w$ belong to the same orbit of triples in $(\mV;E)$. We now show that $R^{(3)}(f'(u),f'(v),f'(w))$ for all $u,v,w \in \mV^3$ with $R^{(3)}(u_i,v_i,w_i)$ for all $i \in \{1,2,3\}$. 
To this end, note that for each $i \in \{2,3\}$ there is a subset $S_i$ of $\{u_i,v_i,w_i\}$
such that $(\sw_{S_i}(u_i),\sw_{S_i}(v_i),\sw_{S_i}(w_i))$ and $(u_1,v_1,w_1)$ belong to the same orbit in $(\mV;E)$.  
%$E(\sw_{S_i}(a),\sw_{S_i}(b))$, 
%$E(\sw_{S_i}(b),\sw_{S_i}(c))$, and
%$E(\sw_{S_i}(a),\sw_{S_i}(c))$. 
Hence, there is $\beta_i \in \Aut(\mV;E)$ such that
$\beta_i(\sw_{S_i}(u_1)) = u_i$,
$\beta_i(\sw_{S_i}(v_1)) = v_i$, and
$\beta_i(\sw_{S_i}(w_1)) = w_i$. 
Pick $a,b,c \in V \setminus \bigcup_{i \in \{1,2,3\}} 
\{u_i,v_i,w_i\}$. % with $E(a,b)$, $E(b,c)$, $E(a,c)$. 
Note that for both $i \in \{2,3\}$ we have that the triples 
$(a,b,c)$ and $(\sw_{S_i}(a),\sw_{S_i}(b),\sw_{S_i}(c))$ lie in the same orbit. 
We then have that the function $x \mapsto f'(x,\beta_2(\sw_{S_2}(x)),\beta_3(\sw_{S_3}(x)))$ maps $(u_1,v_1,w_1) \in R^{(3)}$ to $(f'(u),f'(v),f'(w)) \notin R^{(3)}$. But the same unary function also maps $(a,b,c) \in R^{(3)}$ to a tuple in $R^{(3)}$ 
since $f'$ by assumption preserves $R^{(3)}$ on tuples 
$R^{(3)}$ that lie in the same orbit, and indeed we have
that for $i \in \{2,3\}$ the triples
$(a,b,c)$ and $(\beta_i(\sw_{S_i}(a)),\beta_i(\sw_{S_i}(b)),\beta_i(\sw_{S_i}(c)))$ lie in the same orbit.  
This again contradicts the assumption that the unary function is generated by
$\{-,\sw\}$. 

We therefore have that $f'$ preserves $R^{(3)}$ and $S^{(3)}$. Since it violates $H_2$, 
Proposition~\ref{prop:h2} implies that $\{f',\sw\}$ generates a ternary canonical injection of type minority,
and we are done. 
\end{proof}

\section{Algorithms for Graph-SAT problems}
\label{sect:g-algs}
Throughout this section we assume that $\bB$ 
is a structure with finite relational signature $\tau$ and a first-order definition in $(\mV;E)$.

\subsection{The unbalanced case}\label{ssect:unbalanced}
We now prove tractability of the CSP for templates $\bB$ as in cases~(b) and~(c) of Proposition~\ref{prop:higherArity}, that is, for structures with a first-order definition in $(\mV;E)$
that have 
\begin{itemize}
\item a ternary polymorphism of type majority or minority, and 
\item a binary polymorphism of type $p_1$ which is either $E$-dominated or $N$-dominated in the second argument. 
\end{itemize}
By duality, we may assume that the polymorphism of type $p_1$ is $E$-dominated in the second argument.

It turns out that for such templates $\bB$ we can reduce $\Csp(\bB)$ to the CSP of a structure that we call the \emph{injectivization} of $\bB$. This implies in turn that the CSP can be reduced to a CSP over a Boolean domain.

\begin{definition}\label{def:injective}
	A tuple is called \emph{injective} if all its entries have pairwise distinct entries.
    A relation is called \emph{injective} if all its tuples are injective. 
    A structure is called \emph{injective} if all its relations are injective. 
\end{definition}

\begin{definition}\label{def:inj}
    We define \emph{injectivizations} for relations, atomic formulas, and structures.
    \begin{itemize}
        \item Let $R$ be any relation. Then the \emph{injectivization of $R$}, denoted by $\inj(R)$, is the (injective) relation consisting of all injective tuples of $R$.
        \item Let $\phi(x_1,\ldots,x_n)$ be an atomic formula in the language of $\bB$, where $x_1,\ldots,x_n$ is a list of the variables that appear in $\phi$. Then
        the \emph{injectivization of $\phi(x_1,\dots,x_n)$} is the formula $R^{\inj}_\phi(x_1,\ldots,x_n)$, where $R^{\inj}_\phi$ is a relation symbol which stands for the injectivization of the relation defined by $\phi$.
        \item The \emph{injectivization} of a relational structure $\bB$, denoted by $\inj(\bB)$, is the relational structure $\bC$ with the same domain as $\bB$ whose relations are the injectivizations of the atomic formulas over $\bB$, i.e., the relations $R^{\inj}_\phi$.
    \end{itemize}
\end{definition}

Note that $\inj(\bB)$ also contains the injectivizations of relations that are defined by atomic formulas in which one variable might appear several times. In particular, the injectivization of an atomic formula $\phi$ might have smaller arity than the relation symbol that appears in $\phi$.

To state the reduction to the CSP of an injectivization, we also need the following operations on instances of $\Csp(\bB)$.
Here, it will be convenient to view instances of $\Csp(\bB)$ as primitive positive $\tau$-sentences (see Section~\ref{sect:csp-logical}).

\begin{definition}
    Let $\phi$ be an instance of $\Csp(\bB)$. Then
    the \emph{injectivization of $\phi$}, denoted by $\inj(\phi)$, is the instance $\psi$
    of $\Csp(\inj(\bB))$ obtained from $\phi$ by replacing each conjunct
    $\phi(x_1,\dots,x_n)$ of $\phi$ 
    by $R^{\inj}_\phi(x_1,\ldots,x_n)$.
\end{definition}

We say that a constraint in an instance of $\Csp(\bB)$ \emph{is false} if it defines an empty relation in $\bB$.
Note that a constraint
 $R(x_1,\dots,x_k)$ might be false
even if the relation
 $R$ is non-empty (simply because some of the variables from $x_1,\dots,x_k$
 might be equal).

\begin{figure*}[t]
\begin{center}
\small
\fbox{
\begin{tabular}{l}
{\rm // Input: An instance $\phi$ of CSP$(\bB)$ with variables $W$}
 \\
While $\phi$ contains a constraint that implies $x=y$ for $x,y \in W$ do \\
\hspace{.5cm} Replace each occurrence of $x$ by $y$ in $\phi$.  \\
\hspace{.5cm} If $\phi$ contains a false constraint then reject \\
Loop \\
Accept if and only if $\inj(\phi)$ is satisfiable in $\inj(\bB)$.
\end{tabular}}
\end{center}
\caption{Algorithm for $\Csp(\bB)$ when $\bB$ is preserved by an unbalanced binary injection, using an algorithm for $\inj(\bB)$.}
\label{fig:alg-unbalanced}
\end{figure*}

\begin{proposition}\label{prop:inj-reduction}
    Let $\bB$ be preserved by a
    binary injection $f$ of type $p_1$ that is $E$-dominated in the second argument.
    Then the algorithm shown in Figure~\ref{fig:alg-unbalanced}
    is a polynomial-time reduction of $\Csp(\bB)$ to $\Csp(\inj(\bB))$.
\end{proposition}
\begin{proof}
    In the main loop, when the algorithm detects a constraint that is false and therefore rejects, then $\phi$ cannot hold in $\bB$, because
     the algorithm only contracts variables $x$ and $y$
    when $x=y$ in all solutions to $\phi$  -- and contractions are the
    only modifications performed on the input formula $\phi$.
    So suppose that the algorithm does not reject, and let $\psi$ be
    the instance of $\Csp(\bB)$ computed by the
    algorithm when it reaches the final line of the algorithm.

    By the observation we just made it suffices to show that
    $\psi$ holds in $\bB$
    if and only if $\inj(\psi)$ holds in $\inj(\bB)$.
    It is clear that when $\inj(\psi)$ holds
    in $\inj(\bB)$ then $\psi$ holds in $\bB$ (since the constraints in $\inj(\psi)$ have been made stronger).
    We now prove that if $\psi$ has a solution $s$ in $\bB$,
    then there is also a solution for $\inj(\psi)$ in $\inj(\bB)$.

    Let $s'$ be any mapping from the variable set $W$ of $\psi$ 
    to $\mV$ such that for all distinct $x,y \in W$ we have that
    \begin{itemize}
    \item if $E(s(x),s(y))$ then $E(s'(x),s'(y))$;
    \item if $N(s(x),s(y))$ then $N(s'(x),s'(y))$;
    \item if $s(x)=s(y)$ then $E(s'(x),s'(y))$.
    \end{itemize}
    By universality of $(\mV;E)$, such a mapping exists. We claim that $s'$ is a solution to $\psi$
    in $\bB$. Since $s'$ is injective, it is then clearly
    also a solution to $\inj(\psi)$.
     To prove the claim, let $\gamma$ be a constraint of $\psi$ on the variables
    $x_1,\dots,x_k \in W$. Since we are at the final stage of the algorithm, we can conclude that
    $\gamma(x_1,\dots,x_k)$ does not imply equality of any of the variables $x_1,\dots,x_k$,
    and so there is for all $1 \leq i < j \leq k$ a tuple $t^{(i,j)}$ such that $R(t^{(i,j)})$ and
    $t_i \neq t_j$ hold. Since $\gamma(x_1,\ldots,x_k)$ is preserved by a binary injection, it is also preserved by injections of arbitrary arity (it is straightforward to build such terms from a binary injection). Application of an injection of arity $\binom{k}{2}$ to the tuples $t^{(i,j)}$ shows that $\gamma(x_1,\ldots,x_k)$ is satisfied by an injective tuple $(t_1,\dots,t_k)$.

    Consider the mapping $r \colon \{x_1,\dots,x_k\} \rightarrow \mV$ 
    given by $r(x_l) := f(s(x_l),t_l)$.
    This assignment has the property that  for all $i,j \in S$
    if $E(s(x_i),s(x_j))$, then $E(r(x),r(y))$,
    and if $N(s(x_i),s(x_j))$ then $N(r(x_i),r(x_j))$, because $f$ is of type $p_1$.
    Moreover, if $s(x_i)=s(x_j)$ then $E(r(x_i),r(x_j))$ because $f$ is $E$-dominated in the second argument.
    Therefore, $(s'(x_1),\dots,s'(x_n))$ and $(r(x_1),\dots,r(x_n))$ have the same type in $(\mV;E)$.
    Since $f$ is a polymorphism of $\bB$, we have that $(r(x_1),\dots,r(x_n))$ satisfies the constraint $\gamma(x_1,\ldots,x_n)$. Hence, $s'$ satisfies
    $\gamma(x_1,\ldots,x_n)$ as well.
    We conclude that $s'$ satisfies all the constraints of $\psi$, proving our claim.
\end{proof}

To reduce the CSP for injective structures to Boolean CSPs,
we make the following definition.

\begin{definition}
    Let $t$ be a $k$-tuple of distinct vertices of $(\mV;E)$, and let $q$ be ${k}\choose{2}$.
     Then $\Bool(t)$ is the $q$-tuple $(a_{1,2},a_{1,3},\dots,a_{1,k}$,
    $a_{2,3},\dots,a_{k-1,k}) \in \{0,1\}^q$
    such that $a_{i,j}=0$ if $N(t_i,t_j)$
     and $a_{i,j} = 1$ if $E(t_i,t_j)$.
    If $R$ is a $k$-ary injective relation, then $\Bool(R)$ is the $q$-ary Boolean relation $\{ \Bool(t) \; | \;  t \in R \}$.
    If $\phi$ is a formula that defines an injective relation
    $R$ over $(\mV;E)$, then we also write $\Bool(\phi)$ instead of $\Bool(\inj(R))$. Finally, for injective $\bB$, we write $\Bool(\bB)$ for the structure over a Boolean domain with the relation $\Bool(R)$ for each relation $R$ of $\bB$.
\end{definition}

\begin{proposition}\label{prop:injectiveToBooleanReduction}
    Let $\bB$ be injective. Then there is a polynomial-time reduction from $\Csp(\bB)$ to $\Csp(\Bool(\bB))$.
\end{proposition}
\begin{proof}
    Let $\phi$ be an instance of $\Csp(\bB)$ with variable set $W$.
    We create an instance $\psi$ of $\Csp(\Bool(\bB))$ as follows.
    The variable set of $\psi$ is the set of unordered pairs of variables from $\phi$.
    When $\gamma = R(x_1,\dots,x_k)$ is a constraint in $\phi$, then $\psi$ contains the constraint $\Bool(R)(x_{1,2},x_{1,3},\dots,x_{1,k},x_{2,3},\dots,x_{k-1,k})$.
    It is straightforward to verify that $\psi$ can be computed from $\phi$
    in polynomial time, and that $\phi$ is a satisfiable instance of $\Csp(\bB)$ if and only if $\psi$ is a satisfiable instance of $\Csp(\Bool(\bB))$.
\end{proof}

% \emph{Boolean majority operation} is the unique ternary function $f$ on a Boolean domain satisfying $f(x,x,y)=f(x,y,x)=f(y,x,x)=x$. The \emph{Boolean minority operation} is the unique ternary function $f$ on a Boolean domain satisfying $f(x,x,y)=f(x,y,x)=f(y,x,x)=y$.

The following proposition, together with Propositions~\ref{prop:inj-reduction} and~\ref{prop:injectiveToBooleanReduction}, solves the case where $\Pol(\bB)$ contains a ternary injection of type minority or majority as well as one of the functions of Theorem~\ref{thm:minimal-ops} which are unbalanced and of type projection. It thus shows tractability of cases~(b) and~(c) of Proposition~\ref{prop:higherArity} given that none of the other cases applies.

\begin{proposition}\label{prop:edgeMajorityImpliesBooleanMajority}
    Let $\bB$ be injective, and suppose it has an polymorphism of type minority (majority). Then $\Bool(\bB)$ has a minority (majority) polymorphism, and $\Csp(\Bool(\bB))$ can be solved in polynomial time.
\end{proposition}
\begin{proof}
    It is straightforward to show that $\Bool(\bB)$ has a minority (majority) polymorphism. We have seen in Theorem~\ref{thm:schaefer}
    that $\Csp(\Bool(\bB))$ can then be solved in polynomial time.
\end{proof}

\subsection{Tractability for type minority}
\label{ssect:edgeMinorityBalanced}
We show tractability of $\Csp(\bB)$ when $\bB$ has a polymorphism of type minority as well as a binary canonical injection of type $p_1$ which is balanced. We start by proving that in this case the relations of $\bB$ can be defined in $(\mV;E)$ by first-order formulas of a certain restricted syntactic form; this normal form will later be essential for our algorithm.

Recall that a Boolean relation $R$ is affine
if it can be defined by a conjunction of linear equations modulo 2,
which is the case if and only if $R$ is preserved
by the Boolean minority operation (see Theorem~\ref{thm:schaefer}).
In the following, we denote the Boolean exclusive-or connective (xor) by $\oplus$.

\begin{definition}\label{defn:edgeAffine}
    A graph formula is called \emph{edge affine} if it is
    a conjunction of formulas of the form
    \begin{align*}
    x_1 \neq y_1 \; \vee \; & \dots \; \vee \; x_k \neq y_k \\
    \vee \; & \big (u_1 \neq v_1 \wedge \dots \wedge u_l \neq v_l \\
    & \quad \wedge E(u_1,v_1) \oplus \dots \oplus E(u_l,v_l) = p \big) \\
    \vee \; & (u_1=v_1 \wedge \dots \wedge u_l=v_l) \; ,
    \end{align*}
    where $p \in \{0,1\}$, variables need not be distinct, and each of $k$ and $l$ can be $0$.
\end{definition}

\begin{definition}\label{defn:balanced}
    A ternary operation $f \colon \mV^3 \rightarrow \mV$
    is called \emph{balanced} if for every $c \in \mV$,
    the binary operations $(x,y) \mapsto f(x,y,c)$,
    $(x,z) \mapsto f(x,c,z)$, and $(y,z) \mapsto f(c,y,z)$ are balanced
    injections of type $p_1$.
\end{definition}

Observe that the existence of balanced operations and even balanced minority operations follows from the fact that $(\mV;E)$ contains all countable graphs as induced subgraphs.

\begin{proposition}\label{prop:syntax-graph-affine}
    Let $R$ be a relation with a first-order definition in $(\mV;E)$.
    Then the following are equivalent:
    \begin{enumerate}
        \item $R$ can be defined by an edge affine formula;
        \item $R$ is preserved by every injection of type minority which is balanced;
        \item $R$ is preserved by an injection of type minority, and a balanced binary injection of type $p_1$.
    \end{enumerate}
\end{proposition}

\begin{proof}
    We first show the implication from 1 to 2, that every $n$-ary relation $R$ defined by an edge affine formula $\psi(x_1,\dots,x_n)$ is preserved by balanced functions $f$ of type minority.  We verify that each clause $\phi$ from $\psi$
    is preserved by $f$. 
       By injectivity of $f$, it is easy to see that we only have to show this for the case that $\phi$ does not contain disequality disjuncts
    (i.e., for the case $k=0$). In this case $\phi$ is of the following form, for $p \in \{0,1\}$ and $u_1,\dots,u_l,v_1,\dots,v_l \in \{x_1,\dots,x_n\}$.
   
    \begin{align*}
    \phi =  \; & \big (u_1 \neq v_1 \wedge \dots \wedge u_l \neq v_l \\
    & \quad \wedge \; (E(u_1,v_1) \oplus \dots \oplus E(u_l,v_l) = p) \big ) \\
    & \vee \; (u_1=v_1 \wedge \dots \wedge u_l=v_l) 
    \end{align*}
	In the following, it will sometimes be notationally convenient to consider
    tuples in $(\mV;E)$ satisfying a formula as mappings from the variable set of the formula to $\mV$.
    Let $t_1,t_2,t_3 \colon \{x_1,\dots,x_n\} \rightarrow \mV$ be three mappings that satisfy $\phi$. We have to show that
     the mapping $t_0 \colon \{x_1,\dots,x_n\} \rightarrow \mV$ defined by  $t_0(x) = f(t_1(x),t_2(x),t_3(x))$ satisfies
    $\phi$.

    Suppose first that each of $t_1,t_2,t_3$ satisfies $u_1 \neq v_1 \wedge \dots \wedge u_l \neq v_l$. In this case,
    $t_0(u_1) \neq t_0(v_1) \wedge \dots \wedge t_0(u_l) \neq t_0(v_l)$, since $f$ preserves $\neq$.
     Note that $E(t_0(u_i),t_0(v_i))$, for $1 \leq i \leq l$,
     if and only if $E(t_1(u_i),t_1(v_i)) \oplus E(t_2(u_i),t_2(v_i)) \oplus E(t_3(u_i),t_3(v_i)) = 1$.
     Therefore, since each $t_1,t_2,t_3$ satisfies
     $E(u_1,v_1) \oplus \dots \oplus E(u_l,v_l) = p$,
     we find that $t_0$ also satisfies $E(u_1,v_1) \oplus \dots \oplus E(u_l,v_l) = p \oplus p \oplus p = p$.

    Next, suppose that one of $t_1,t_2,t_3$ satisfies $u_i = v_i$
    for some (and therefore for all) $1\leq i \leq l$.
    By permuting arguments of $f$, we can assume
    that $t_1(u_i)=t_1(v_i)$ for all $i \in \{1, \dots, l\}$.
    Since the function $f$ is balanced, the operation $g \colon (y,z) \mapsto f(t_1(u_i),y,z)$ is a balanced injection of type $p_1$.
    Suppose that $t_2(u_i)=t_2(v_i)$.
    Then
    $E(t_0(u_i),t_0(v_i))$ if and only if $E(t_3(u_i),t_3(v_i))$, since
    $g$ is balanced.
    Hence, $t_0$ satisfies $\phi$.
    Now suppose that $t_2(u_i) \neq t_2(v_i)$. Then $E(t_0(u_i),t_0(v_i))$
    if and only if $E(t_2(u_i),t_2(v_i))$, since $g$ is of type $p_1$.
    Again, $t_0$ satisfies $\phi$.
    This shows that $f$ preserves $\phi$.

    The implication from 2 to 3 is trivial, since every balanced function of type minority generates
    a balanced binary injection of type $p_1$ by identification of two of its variables. It is also here that we have to check the existence of balanced  injections of type minority; as mentioned above, this follows easily from the universality of $(\mV;E)$.

    We show the implication from 3 to 1 by induction on the arity
    $n$ of the relation $R$.
    Let $g$ be the balanced binary injection of type $p_1$,
    and let $h$ be the operation of type minority.
    For $n=2$ the statement of the theorem holds, because all binary
    relations with a first-order definition in $(\mV;E)$ can be defined over $(\mV;E)$
    by expressions as in Definition~\ref{defn:edgeAffine}:
    \begin{itemize}
        \item For $x \neq y$ we set $k=1$ and $l=0$.
        \item For $\neg E(x,y)$ we can set $k=0, l=1,p=0$.
        \item For $\neg N(x,y)$ we can set $k=0, l=1,p=1$.
        \item Then, $E(x,y)$ can be expressed as $(x \neq y) \wedge \neg N(x,y)$.
        \item $N(x,y)$ can be expressed as $(x \neq y) \wedge \neg E(x,y)$.
        \item $x=y$ can be expressed as $\neg E(x,y) \wedge \neg N(x,y)$.
        \item The empty relation can be expressed as $E(x,y) \wedge N(x,y)$.
        \item Finally, $\mV^2$ can be defined by the empty conjunction.
    \end{itemize}

    For $n>2$, we construct a formula $\phi$ that defines
    the relation $R(x_1,\dots,x_n)$ as follows.
    If there are distinct $i,j \in \{1,\dots,n\}$ such that for all  tuples
    $t$ in $R$ we have $t_i=t_j$, consider the relation
    defined by $\exists x_i. R(x_1,\dots,x_n)$. This relation is also
    preserved by $g$ and $h$, and by
    inductive assumption has a definition $\psi$ as required.
    Then the formula $\phi := (x_i = x_j) \wedge \psi$ 
    proves the claim.
    So let us assume that for all distinct $i,j$ 
    there is a tuple $t \in R$
    where $t_i \neq t_j$. Note that since $R$ 
    is preserved by the binary injective operation $g$, 
    this implies that $R$ also contains an injective tuple.

    Since $R$ is preserved by an operation of type minority,
    the relation $\Bool(\inj(R))$ 
    is preserved by the Boolean minority
    operation, and hence has a definition by a conjunction of
    linear equations modulo 2 (Theorem~\ref{thm:schaefer}).
    From this definition it is straightforward
    to obtain a definition $\psi(x_1,\dots,x_n)$ of $\inj(R)$ which is the conjunction
    of $\bigwedge_{i < j \leq n} x_i \neq x_j$ and of
    formulas of the form
    $$E(u_1,v_1) \oplus \dots \oplus E(u_l,v_l) = p\; ,$$
    for $u_1,\dots,u_l,v_1,\dots,v_l \in \{x_1,\dots,x_n\}$.
    It is clear that we can assume that none of the formulas of the form $E(u_1,v_1) \oplus \dots \oplus E(u_l,v_l) = p$ in $\psi$
    can be equivalently replaced by a conjunction
    of shorter formulas of this form.

    For all $i,j \in \{1,\dots,n\}$ with $i<j$,
    let $R_{i,j}$ be the relation that holds for the tuple $(x_1,\ldots,x_{i-1},x_{i+1},\ldots,x_n)$ iff $R(x_1,\dots,x_{i-1},x_j,x_{i+1},\dots,x_n)$ holds. Because $R_{i,j}$ is preserved by $g$ and $h$, but has arity $n-1$, it has a definition
    $\psi_{i,j}$ as in the statement by inductive assumption.
    We call the conjuncts of $\psi_{i,j}$ also the \emph{clauses}
    of $\psi_{i,j}$.
    %We add to each clause of $\psi_{i,j}$ a disjunct $x_i \neq x_j$.

    Let $\phi$ be the conjunction composed of conjuncts from the following two groups:
    \begin{enumerate}
        \item $\gamma \vee (x_i \neq x_j)$ for all $i < j \leq n$ and each clause $\gamma$ of $\psi_{i,j}$;
        \item when $\eta = (E(u_1,v_1) \oplus \dots \oplus E(u_l,v_l) = p)$ is a conjunct
        of $\psi$, then
        $\phi$ contains the formula
        \begin{align*} & (u_1 \neq v_1 \wedge \dots \wedge u_l \neq v_l \;\wedge \; \eta) \\
        \vee \; & (u_1=v_1 \wedge \dots \wedge u_l=v_l) \; .
        \end{align*}
    \end{enumerate}
    Obviously, $\phi$ is a formula of the required form.
    We have to verify that $\phi$ defines $R$.

    Let $t$ be an $n$-tuple such that $t \notin R$.
    If $t$ is injective, then
    $t$ violates a formula of the form
    $$E(u_1,v_1) \oplus \dots \oplus E(u_l,v_l) = p$$
    from the formula $\psi$ defining $\inj(R)$, and hence it violates a conjunct of $\phi$ of the second group.
    If there are $i,j$ such that $t_i=t_j$
    then the tuple $t^i:=(t_1,\dots,t_{i-1},t_{i+1},\dots,t_n) \notin R_{i,j}$.
    % REASON:
    % R_{i,j} precisely contains
    % the tuples such that there 
    % EXISTS an extension to a 
    % tuple in R, an this extension
    % actually is unique (we work
    % with the equality relation)
    Therefore some conjunct $\gamma$ of
    $\psi_{i,j}$ is not satisfied by $t^i$, and $\gamma \vee (x_i \neq x_j)$
    is not satisfied by $t$. Thus, in this case $t$ does not satisfy $\phi$ either.

    It remains to verify that all $t \in R$ satisfy $\phi$.
    Let $\gamma \vee (x_i \neq x_j) $ be a conjunct of $\phi$ created from some clause in $\psi_{i,j}$.
    If $t_i \neq t_j$, then $t$ satisfies $x_i \neq x_j$. 
    If $t_i = t_j$, 
    then $(t_1,\dots,t_{i-1},t_{i+1},\dots,t_n) \in R_{i,j}$ 
    and thus this tuple satisfies $\psi_{i,j}$. 
    This also implies that $t$ satisfies $\phi$.
    Now, let $\eta$ be a conjunct of $\phi$ from the second group.
    We distinguish three cases.
    \begin{enumerate}
    \item For all $1 \leq i \leq l$ we have 
    that $t$ satisfies $u_i = v_i$. In this case we are clearly
    done since $t$ satisfies the second disjunct of $\eta$.
    \item For all $1 \leq i \leq l$ we have that $t$ satisfies $u_i \neq v_i$.
    Suppose for contradiction that $t$ does not satisfy $E(u_1,v_1) \oplus \dots \oplus E(u_l,v_l)=p$. Let $r\in R$ be injective, and consider the tuple $s:=g(t,r)$. Then $s\in R$, and $s$ is injective since the tuple $r$ and the function $g$ are injective. However, since $g$ is of type $p_1$, we have $E(s(u_i),s(v_i))$ if and only if $E(t(u_i),t(v_i))$, for all $1\leq i\leq l$. Hence, $s$ violates the conjunct $E(u_1,v_1) \oplus \dots \oplus E(u_l,v_l)=p$ from $\psi$, a contradiction since $s \in \inj(R)$.
    \item The remaining case is that there is a proper non-empty subset $S$ of $\{1,\dots,l\}$ such that $t$ satisfies $u_i=v_i$ for all $i \in S$ and $t$ satisfies $u_i \neq v_i$ for all
    $i \in \{1,\dots,n\} \setminus S$. 
    We claim that this case cannot occur.
    Suppose that all tuples $t'$ from $\inj(R)$ satisfy that
    $\bigoplus_{i \in S} E(u_i,v_i) = 1$. 
    In this case we could have
    replaced $E(u_1,v_1) \oplus \dots \oplus E(u_l,v_l)=p$ 
    by the two shorter formulas
    $\bigoplus_{i \in S} E(u_i,v_i) = 1$ and 
 $\bigoplus_{i \in [n] \setminus S} E(u_i,v_i) = p \oplus 1$, in contradiction to our assumption on $\psi$.
    Hence, there is a tuple $s \in \inj(R)$ where $\bigoplus_{i \in S} E(u_i,v_i) = 1$.
    Now, for the tuple $g(t,s)$ we have
    \begin{align*}
    \bigoplus_{i \in [n]} E(u_i,v_i) = & \; \bigoplus_{i \in S} E(u_i,v_i) \oplus \bigoplus_{i \in [n] \setminus S} E(u_i,v_i) \\
    = & \; 1 \oplus p \\
    \neq & \; p
    \end{align*}
    which is a contradiction since $g(t,s) \in \inj(R)$.
    \end{enumerate}
    Hence, all $t \in R$ satisfy all conjuncts of $\phi$. We conclude
    that $\phi$ defines $R$.
\end{proof}

We now present a polynomial-time algorithm for $\Csp(\bB)$
for the case that $\bB$ has finitely many relations that are all edge affine.

\begin{definition}
    Suppose all relations of $\bB$ are edge affine, 
    and let $\phi$ be an instance of $\Csp(\bB)$.
    Then the \emph{graph of $\phi$} is the (undirected) graph
    whose vertices are unordered pairs 
    of distinct variables of $\phi$, and which
    has an edge between distinct sets $\{a,b\}$ and $\{c,d\}$ 
    if $\phi$ contains a
    constraint whose definition 
    as in Definition~\ref{defn:edgeAffine}
    has a conjunct of the form
    \begin{align*}
    & \big (u_1 \neq v_1 \wedge \dots \wedge u_l \neq v_l
    \wedge (E(u_1,v_1) \oplus \dots \oplus E(u_l,v_l) = p) \big) \\
    \vee \; & (u_1=v_1 \wedge \dots \wedge u_l=v_l)
    \end{align*}
    such that $\{a,b\}=\{u_i,v_i\}$ and $\{c,d\} = \{u_j,v_j\}$ for
    some $i,j \in \{1,\dots,l\}$.
\end{definition}

It is clear that for $\bB$ with finite signature,
the graph of an instance $\phi$ of $\Csp(\bB)$ can be computed
in linear time from $\phi$.

\begin{definition}
    Let $\bB$ only have edge affine relations, 
    and let $\phi$ be an instance of $\Csp(\bB)$.
    For a set $C$ of 2-element subsets of variables of $\phi$, we
    define $\inj(\Phi,C)$ to be the following affine Boolean
    formula.
    The set of variables of $\inj(\phi,C)$ is
    $C$.
    The constraints of $\inj(\phi,C)$ are obtained from
    the constraints $\gamma$ of $\phi$ as follows.
    If $\gamma$ has a definition as in Definition~\ref{defn:edgeAffine} with a clause of the form
    \begin{align*}
    & \big (u_1 \neq v_1 \wedge \dots \wedge u_l \neq v_l
    \wedge (E(u_1,v_1) \oplus \dots \oplus E(u_l,v_l) = p) \big) \\
    \vee \; & (u_1=v_1 \wedge \dots \wedge u_l=v_l)
    \end{align*}
    where all pairs $\{u_i,v_i\}$ are in $C$,
    then $\inj(\phi,C)$ contains the conjunct
    $\{u_1,v_1\} \oplus \dots \oplus \{u_l,v_l\} = p$.
\end{definition}

\begin{figure*}[t]
\begin{center}
\small
\fbox{
\begin{tabular}{l}
    {\rm // Input: An instance $\phi$ of $\Csp(\bB)$} \\
    Repeat \\
    \hspace{.3cm} For each connected component $C$ of the graph of $\phi$ do \\
    \hspace{.6cm} Let $\psi$ be the affine Boolean formula $\inj(\phi,C)$. \\
    \hspace{.6cm} If $\psi$ is unsatisfiable then \\
    \hspace{.9cm} For each $\{x,y\} \in C$ do \\
    \hspace{1.2cm} Replace each occurrence of $x$ by $y$ 
    in $\phi$. \\
    \hspace{.9cm} If $\phi$ contains a false constraint then reject \\
    \hspace{.3cm} Loop \\
    Until $\inj(\phi,C)$ is satisfiable for all components $C$ \\
    Accept
\end{tabular}}
\end{center}
\caption{A polynomial-time algorithm
for $\Csp(\bB)$ when $\bB$ is preserved by a balanced operation of type minority.}
\label{fig:balanced-minority}
\end{figure*}

Tractability of case~(d) of Proposition~\ref{prop:higherArity} now follows from the following proposition and Proposition~\ref{prop:syntax-graph-affine}.

\begin{proposition}\label{prop:alg-minority}
Let $\bB$ be a structure with a first-order definition in $(\mV;E)$ and a finite signature, and suppose that $\bB$ is preserved by a balanced injection 
    of type minority. Then the algorithm shown 
    in Figure~\ref{fig:balanced-minority} 
    solves $\Csp(\bB)$ in polynomial time.
\end{proposition}

\begin{proof}
    We first show that when the algorithm detects a constraint that
    is false and therefore rejects in the innermost loop, 
    then $\phi$ must be unsatisfiable. 
    Since variable contractions are the
    only modifications performed on the input formula $\phi$,
    it suffices to show that the algorithm only
     equates variables $x$ and $y$
    when $x=y$ in all solutions to $\phi$.
    To see that this is true, assume that $\psi := \inj(\phi,C)$
    is an unsatisfiable Boolean formula 
    for some connected component $C$. 
    Hence, in any solution $s$ to $\phi$ there must be a pair
    $\{x,y\}$ in
    $C$ such that $s(x)=s(y)$. 
    It follows immediately from the definition
    of the graph of $\phi$ that then $s(u)=s(v)$ for all
    $\{u,v\}$ adjacent to $\{x,y\}$ in the graph of $\phi$.
    By connectivity of $C$, we have that $s(u)=s(v)$ 
    for all $\{u,v\} \in C$.
    Since this holds for any solution to $\phi$, 
    the contractions in the
    innermost loop of the algorithm preserve satisfiability.

    So we only have to show that when the algorithm accepts,
    there is indeed a solution to $\phi$. 
    When the algorithm accepts,
    we must have that inj$(\phi,C)$ 
    has a solution $s_C$ for all components
    $C$ of the graph of $\phi$. 
    Let $s$ be a mapping from the variables of $\phi$ 
    to the $\mV$ such that $E(x_i,x_j)$ if $\{x_i,x_j\}$ 
    is in component $C$ of the graph of $\phi$ 
    and $s_C(\{x_i,x_j\})=1$, and $N(x_i,x_j)$ otherwise.
    It is straightforward to verify that this assignment satisfies all of the constraints.
\end{proof}

\subsection{Tractability for type majority}\label{ssect:edgeMajorityBalanced}
We turn to case~(e) of Proposition~\ref{prop:higherArity}, i.e., the case where $\bB$ has ternary injection of type majority and a binary canonical injection of type $p_1$ which is balanced.

 Recall that a Boolean relation is called \emph{bijunctive}
if it can be defined by a conjunction of clauses of size at most two.
It is well-known that a Boolean relation is bijunctive if and only if
it is preserved by the Boolean majority operation (see Section~\ref{sect:schaefer}).

\begin{definition}
A formula is called \emph{graph bijunctive}
iff it is a conjunction of \emph{graph bijunctive clauses}, i.e., formulas of the form
\begin{align*}
x_1 \neq y_1 & \vee \cdots \vee x_k \neq y_k \vee \phi
\end{align*}
where $\phi$ is of one of the following forms
\begin{enumerate}
\item[(i)] $u_1 = v_1$;
\item[(ii)] $L_1(u_1,v_1)$;
\item[(iii)] $L_1(u_1,v_1) \vee L_2(u_2,v_2)$; 
\item[(iv)] $L_1(u_1,v_1) \vee u_1 = v_1$;
\item[(v)] $(L_1(u_1,v_1) \vee u_1=v_1 \vee L_2(u_2,v_2)) \wedge (u_1 \neq v_1 \vee L_2(u_2,v_2) \vee u_2 = v_2)$;
\end{enumerate}
for $L_1,L_2 \in \{E,N\}$, and $k \geq 0$. 
\end{definition}

Note that when $M_1,M_2$ are such that $\{L_i,M_i\} = \{E,N\}$ for $i \in \{1,2\}$, then the graph bijunctive clause in Item (v) can be equivalently written in the form 
$$(M_1(u_1,v_1) \Rightarrow L_2(u_2,v_2)) \wedge (u_1=v_1 \Rightarrow \neg M_2(u_2,v_2)) \; .$$ 

\ignore{
\noindent {\bf Example.} Note that the formula $x_1 \neq y_1 \vee \cdots \vee x_k \neq y_k \vee x_0 = y_0$ is equivalent to the graph bijunctive formula
\begin{align*}
\big (x_1 \neq y_1 \vee \cdots \vee x_k \neq y_k \vee \neg E(x_0,y_0))
\wedge (x_1 \neq y_1 \vee \cdots \vee x_k \neq y_k \vee \neg N(x_0,y_0)) \; .
\end{align*}
}

\ignore{%not needed anymore
\noindent {\bf Example.}
For $L \in \{E,N\}$, the formula
\begin{align*}
\big (u_1 \neq v_1 \Rightarrow L(u_2,v_2) \big) & \wedge \big (u_1 = v_1 \Rightarrow (L(u_2,v_2) \vee u_2 = v_2)\big) 
\end{align*}
is equivalent to the following graph bijunctive formula. 
\begin{align*}
& \big (E(u_1,v_1) \vee u_1 = v_1 \vee L(u_2,v_2) \big ) \wedge (u_1 \neq v_1 \vee L(u_2,v_2) \vee u_2 = v_2)  \\
\wedge \quad & \big (N(u_1,v_1) \vee u_1 = v_1 \vee L(u_2,v_2)\big) \wedge (u_1 \neq v_1 \vee L(u_2,v_2) \vee u_2 = v_2)  
\end{align*}
}

\begin{theorem}\label{thm:syntax}
Let $R$ be a relation with a first-order definition over $(\mV;E)$. Then the following are equivalent.
\begin{enumerate}
\item $R$ can be defined by a graph bijunctive formula;
\item $R$ is preserved by every ternary injection which is of type majority and balanced; 
\item $R$ is preserved by some 
ternary injection of type majority and some binary balanced injection of type $p_1$.
\end{enumerate}
\end{theorem}
\begin{proof}
We show the equivalence of (1) and (2);
the equivalence between (2) and (3) is easy and is to be added later.

For the implication $(1) \Rightarrow (2)$, 
let $\psi$ be a graph bijunctive clause.
It suffices to show that $\psi$ is preserved by every balanced injection $f$ of type majority. 
Let $t_1,t_2,t_3$ be three tuples that satisfy $\psi$.
If $\psi$ contains 
an inequality disjunct $x_i \neq y_i$,
and one of $t_1,t_2,t_3$ satisfies $x_i \neq y_i$,
then by injectivity of $f$ we have that $t_0 = f(t_1,t_2,t_3)$ satisfies $x_i \neq y_i$ and therefore also $\psi$. So we can focus on the case $k = 0$, i.e.,  
$\psi$ does not contain any inequality disjunct. 
If $\psi$ is of the form $u_1=v_1$, $\psi$ is clearly preserved. If $\psi$ is of the form $L_1(u_1,v_1)$ or 
of the form $\neg L_1(u_1,v_1)$, then $f$ preserves $\psi$ since it is of type majority and balanced. Suppose now that $\psi$ is of the form $L_1(u_1,v_1) \vee L_2(u_2,v_2)$ for $L_1,L_2 \in \{E,N\}$. 
Then at least two of $t_1,t_2,t_3$ satisfy $L_1(u_1,v_1)$, or at least two of $t_1,t_2,t_3$ satisfy $L_2(u_2,v_2)$. In the former case, $t_0$ satisfies $L_1(u_1,v_1)$, in the latter case $t_0$ satisfies $L_2(u_2,v_2)$, since $f$ is of type majority and balanced. 

Finally, suppose that $\psi$ is as in item (v) of the definition of graph bijunctive formulas.
If $t_0 = f(t_1,t_2,t_3)$ satisfies $\neg M_1(u_1,v_1) \wedge u_1 \neq v_1$, then $t_0$ satisfies
both conjuncts of $\psi$ and we are done. We thus may assume that $t_0$ satisfies either $u_1 = v_1$ or $M_1(u_1,v_1)$.
If $t_0$ satisfies $u_1 = v_1$,
then $t_0$ satisfies the first conjunct of $\psi$.
By injectivity of $f$ we must have that all of 
$t_1,t_2,t_3$ satisfy $u_1=v_1$, and therefore
all three tuples satisfy $L_2(u_2,v_2) \vee u_2=v_2$. Since $f$ is of type majority and balanced, also $t_0$ satisfies $L_2(u_2,v_2) \vee u_2=v_2$, which is the second conjunct of $\psi$, and we are done also in this case.

Suppose now that $t_0$ satisfies $M_1(u_1,v_1)$.
Since $f$ is of type majority and balanced, either 
\begin{enumerate}
\item[(a)] at least two out of $t_1,t_2,t_3$ satisfy $M_1(u_1,v_1)$, or 
\item[(b)] $t_1$ satisfies $M_1(u_1,v_1)$ and exactly one out
of $t_2,t_3$ satisfy $u_1=v_1$, or
\item[(c)] $t_1$ satisfies $u_1=v_1$ and $t_2$ satisfies $M_1(u_1,v_1)$.
\end{enumerate}
If at least two tuples out of $t_1,t_2,t_3$ satisfy
$M_1(u_1,v_1)$, then they also satisfy
$L_2(u_2,v_2)$, and
so does $t_0$ since $f$ is of type majority and balanced. We conclude that $t_0$ satisfies $\psi$. 
Now assume (b). Then $t_1$ satisfies $M_1(u_1,v_1)$, and therefore also satisfies $L_2(u_2,v_2)$. 
Moreover, one of $t_2,t_3$ satisfies $u_1 = v_1$, and therefore also
$L_2(u_2,v_2) \vee u_2 = v_2$. Since $f$ is balanced and of type majority we have that 
$t_0$ satisfies $L_2(u_2,v_2)$, and therefore also $\psi$.
Suppose finally that (c) holds, i.e., $t_1$ satisfies $u_1=v_1$ and $t_2$ satisfies $M_1(u_1,v_1)$.
In this case $t_1$ satisfies $L_2(u_2,v_2) \vee u_2 = v_2$ and $t_2$ satisfies
$L_2(u_2,v_2)$. Again, since $f$ is balanced and of type majority, we have that 
$t_0$ satisfies $L_2(u_2,v_2)$, and therefore also $\psi$.\\

We next show the implication $(2) \Rightarrow (1)$. 
Let $R$ be a relation preserved by a ternary injection $f$ which is of type majority and balanced.
Let $\Phi$ be a formula in CNF 
that defines $R$ over $(\mV;E,N)$ such that 
all literals of $\Phi$ are of the form $E(x,y)$, $N(x,y)$,
$x \neq y$, or $x=y$. This can be achieved by replacing literals of the form $\neg L(x,y)$ by $M(x,y) \vee x=y$, for $M$ such that $\{L,M\} = \{E,N\}$. 
Also suppose that 
$\Phi$ is \emph{minimal} in the sense that no clause $\phi$ of $\Phi$ can be replaced by a set of clauses such that
\begin{enumerate}
\item each replacing clause has fewer literals of the form $L(x,y)$ for $L \in \{E,N\}$ than $\phi$, or
\item each replacing clause has the same number of literals of the form $L(x,y)$, but fewer
literals of the form $x=y$ than $\phi$, or
\item each replacing clause has the same number of literals of the form $L(x,y)$ and of the form $x=y$, but fewer literals of the form $x \neq y$ than $\phi$. 
\end{enumerate}

Let $\Psi$ be the set of all graph bijunctive clauses that are implied by $\Phi$. To prove  {(2) $\Rightarrow$ (1)}, it suffices to show that  $\Psi$ implies
all clauses $\phi$ of $\Phi$. Let $\phi$ such a clause. 
In the entire proof we make the convention
that $L_1,\dots,L_n$ denote elements of 
$\{E,N\}$, and
$M_1,\dots,M_n$ are such that 
$\{L_i,M_i\} = \{E,N\}$, for all $i \leq n$.\\

{\it Observation 1: } The clause $\phi$ cannot contain two different literals of the form $x_1=y_1$ 
and $x_2=y_2$. 
Otherwise,
since $\Phi$ is minimal, the formula obtained by removing $x_1=y_1$ from $\phi$ is inequivalent to $\Phi$, and hence there exists a tuple $t_1$ that 
satisfies $\Phi$, and none of the literals in $\phi$ except for $x_1=y_1$. Similarly, there exists
a tuple $t_2$ that satisfies $\Phi$, and none of the literals in $\phi$ except for $x_2 = y_2$. 
By the injectivity of $f$, the tuple 
$t_0 = f(t_1,t_2,t_2)$ satisfies $x_1 \neq y_1$
and $x_2 \neq y_2$. 
Moreover, $t_0$ does not satisfy any other literal of $\phi$
because the fact that it is of type majority and balanced implies that $f$ preserves the negations of all literals of the form $x = y$, 
$E(x,y)$, $N(x,y)$, and $x \neq y$.
Therefore, $t_0$ satisfies
none of the literals in $\phi$, 
contradicting
the assumption that $f$ preserves $\Phi$. \\

{\it Observation 2: } The clause $\phi$ contains at most two literals of the form $L(x,y)$, where $L\in\{E,N\}$. Suppose to the contrary that $\phi$ contains three different literals of the form
$L_1(x_1,y_1)$, $L_2(x_2,y_2)$, 
and $L_3(x_3,y_3)$. Let $\theta$
be the clause obtained from $\phi$ by
removing those three literals from $\phi$.
Note that it is impossible that $\Phi$ has
satisfying assignments $t_1,t_2,t_3$ with
\begin{align*}
t_1 \models & M_2(x_2,y_2) \wedge M_3(x_3,y_3) \wedge \neg \theta \\
t_2 \models & M_1(x_1,y_1) \wedge M_3(x_3,y_3) \wedge \neg \theta \\
t_3 \models & M_1(x_1,y_1) \wedge M_2(x_2,y_2) \wedge \neg \theta \; .
\end{align*}
%such that $t_1,t_2,t_3$ satisfy none of the other literals of $\phi$:
Otherwise, $t_0 = f(t_1,t_2,t_3)$ satisfies 
$M_1(x_1,y_1) \wedge M_2(x_2,y_2) \wedge M_3(x_3,y_3)$ since $f$ is of type majority and balanced. Moreover, $t_0$ satisfies $\neg \theta$, since $f$ preserves the negations of
literals of the form $x = y$, 
$E(x,y)$, $N(x,y)$, and $x \neq y$. 
Therefore, $t_0$ does not satisfy 
$\phi$, in contradiction to the assumption that $f$ preserves $\Phi$.

Suppose without loss of generality that there 
is no satisfying assignment $t_1$ as above. 
In other words,
$\Phi$ implies the clause 
\begin{align} \theta \vee L_2(x_2,y_2) \vee (x_2 = y_2) \vee L_3(x_3,y_3) \vee (x_3 = y_3) \; . \label{eq:no-good}
\end{align}
Note that $\Phi$ also implies the clauses
\begin{align} \theta \vee L_1(x_1,y_1) \vee L_2(x_2,y_2) \vee (x_3 \neq y_3)
\label{eq:eq1}
\end{align} 
\begin{align} \theta \vee L_1(x_1,y_1) \vee (x_2 \neq y_2) \vee L_3(x_3,y_3)
\label{eq:eq2}
\end{align} 
since they are obvious weakenings of $\phi$.
We claim that the clauses in (\ref{eq:no-good}), (\ref{eq:eq1}), and  (\ref{eq:eq2}) together imply $\phi$.
To see this, suppose they hold for a tuple $t$ which does not satisfy $\phi$. Then $t$ satisfies neither $\theta$ nor any of the $L_i$, and hence it satisfies both $(x_2 \neq y_2)$ and $(x_3 \neq y_3)$, by (\ref{eq:eq1}) and  (\ref{eq:eq2}). On the other hand, in this situation (\ref{eq:no-good}) implies $x_2=y_2\vee x_3=y_3$, a contradiction. Hence $\phi$ is equivalent to the conjunction of these three clauses. Now replacing $\phi$ by this conjunction in $\Phi$, we arrive at a contradiction to the minimality of $\Phi$.\\

Taking the two observations together, we conclude that $\phi$ contains at most one literal of the form $x=y$, and at most two literals of the 
form $L(x,y)$. If it has no literal of the form $x=y$ or no literal of the form $L(x,y)$ then it is itself graph bijunctive and hence an element of $\Psi$, and we are done. So assume henceforth that  
$\phi$ contains a literal $x_1=y_1$
and a literal of the form $L_2(x_2,y_2)$. It may or may not contain at most one more literal $L_3(x_3,y_3)$; all other literals of $\phi$ are of the form $x \neq y$.

Let us first consider the case where $\phi$ does not
contain the literal $L_3(x_3,y_3)$. 
Let $\theta$ be the clause obtained from $\phi$ by removing $x_1=y_1$ and $L_2(x_2,y_2)$; all
literals in $\theta$ are of the form $x \neq y$. 
We claim that $\Phi$ implies the following formula.
\begin{align}
\theta \vee (x_1 \neq y_1) \vee L_2(x_2,y_2) \vee (x_2=y_2)
\label{eq:important}
\end{align}
To show the claim, suppose for contradiction
that there is a tuple $t_1$ that
satisfies $\Phi \wedge \neg \theta \wedge (x_1 = y_1) \wedge M_2(x_2,y_2)$. By minimality of $\Phi$, there is also a
tuple $t_2$ that satisfies $\Phi \wedge \neg \theta \wedge (x_1 \neq y_1) \wedge L_2(x_2,y_2)$. Then 
$f(t_1,t_1,t_2)$ satisfies $\Phi \wedge \neg \theta \wedge x_1 \neq y_1 \wedge M_2(x_2,y_2)$ since $f$ is of type majority and balanced; but this is a contradiction
since such a tuple does not satisfy $\phi$. 
We next show that $\Phi$ implies the graph bijunctive formulas
\begin{align}
& \theta \vee (E(x_1,y_1) \vee x_1 = y_1 \vee L_2(x_2,y_2)) \wedge (x_1 \neq y_1 \vee L_2(x_2,y_2) \vee x_2 = y_2) 
\label{eq:case1-1} \\
& \theta \vee (N(x_1,y_1) \vee x_1 = y_1 \vee L_2(x_2,y_2)) \wedge (x_1 \neq y_1 \vee L_2(x_2,y_2) \vee x_2 = y_2) \; .
\label{eq:case1-2}
\end{align}
Since $\Phi$ implies~(\ref{eq:important}),
it suffices to show that $\Phi$ implies $\theta \vee E(x_1,y_1) \vee (x_1 = y_1) \vee L_2(x_2,y_2)$
and $\theta \vee N(x_1,y_1) \vee (x_1 = y_1) \vee L_2(x_2,y_2)$. But this is clear since those formulas are weakenings of $\phi$.
%Formula~(\ref{eq:case1-1}) and~(\ref{eq:case1-2})
%are both in the form of item (5) 
%in the definition of graph bijunctive formulas, 
%and therefore in $\Psi$. 
Hence, the formulas~(\ref{eq:case1-1}) and~(\ref{eq:case1-2}) are in $\Psi$.
As
$E(x_1,y_1) \vee (x_1 = y_1) \vee L_2(x_2,y_2)$
and 
$N(x_1,y_1) \vee (x_1 = y_1) \vee L_2(x_2,y_2)$
implies $(x_1 = y_1) \vee L_2(x_2,y_2)$,
the formulas~(\ref{eq:case1-1}) 
and (\ref{eq:case1-2}) 
imply $\phi$, and therefore $\Psi$ implies $\phi$. 

Finally, we consider the case where $\phi$ also contains a literal $L_3(x_3,y_3)$. 
Let $\theta$ be the clause obtained from $\phi$ by removing $x_1=y_1$, $L_2(x_2,y_2)$, and $L_3(x_3,y_3)$; all literals of $\theta$ are of the form $x \neq y$.  If $\Phi$ implies $\theta \vee \neg M_2(x_2,y_2)$, then we could have replaced
$\phi$ by the two clauses $\theta \vee L_2(x_2,y_2) \vee (x_2 = y_2)$ and 
$\theta \vee (x_1 = y_1) \vee (x_2 \neq y_2) \vee L_3(x_3,y_3)$ 
which together imply $\phi$, in contradiction to the minimality of $\Phi$. The same argument shows that $\Phi$ does not imply $\theta \vee \neg M_3(x_3,y_3)$.
Now observe that $\Phi$ implies the following.

\begin{align}
& \theta \vee x_1 = y_1 \vee x_2 \neq y_2 \vee x_3 \neq y_3 \label{eq:easy1} \\
& \theta \vee \neg M_2(x_2,y_2) \vee \neg M_3(x_3,y_3) \label{eq:easy2} \\
& \theta \vee x_2 \neq y_2 \vee L_3(x_3,y_3) \vee x_3 = y_3 \label{eq:guard1} \\
& \theta \vee x_3 \neq y_3 \vee L_2(x_2,y_2) \vee x_2 = y_2\; .
\label{eq:guard2}
\end{align}
This is obvious for~(\ref{eq:easy1}). For~(\ref{eq:easy2}), assume otherwise that there is an assignment $t$ satisfying $\Phi \wedge \neg \theta \wedge M_2(x_2,y_2) \wedge M_3(x_3,y_3)$. By minimality of
$\Phi$ there is also an assignment $t'$ satisfying 
$\Phi \wedge \neg \theta \wedge (x_1 \neq x_2)$. 
Then $f(t,t,t')$ satisfies none of the literals of $\phi$,
a contradiction. We now show that~(\ref{eq:guard1})
is implied; the proof for~(\ref{eq:guard2}) is symmetric.
Assume otherwise that $t$ satisfies 
$\Phi \wedge \neg \theta \wedge (x_2 = y_2) \wedge M_3(x_3,y_3)$.
There also exists a tuple $t'$
that satisfies $\Phi \wedge \neg \theta \wedge M_2(x_2,y_2)$ 
since $\Phi$ does not imply $\theta \vee \neg M_2(x_2,y_2)$ as we have observed above. Then $f(t,t,t')$ satisfies
$\neg \theta \wedge M_2(x_2,y_2) \wedge M_3(x_3,y_3)$,
which contradicts~(\ref{eq:easy2}).

We now claim that $\Phi$ also implies at least one of the following two formulas. 
\begin{align}
& \theta \vee L_2(x_2,y_2) \vee x_2=y_2 \vee L_3(x_3,y_3)
%\wedge (x_2 \neq y_2 
%\vee L_3(x_3,y_3) \vee x_3 = y_3) \big)
\label{eq:first} \\
& \theta \vee L_3(x_3,y_3) \vee x_3=y_3 \vee L_2(x_2,y_2)\; .
%\wedge (x_3 \neq y_3 
%\vee L_2(x_3,y_3) \vee x_2 = y_2) \big)
 \label{eq:second}
\end{align}
Otherwise, there would be a tuple $t$ satisfying
$\Phi \wedge \neg \theta \wedge M_2(x_2,y_2) \wedge \neg L_3(x_3,y_3)$ and a tuple $t'$ satisfying
$\Phi \wedge \neg \theta \wedge \neg L_2(x_2,y_2) \wedge M_3(x_3,y_3)$. Then $f(t,t',t')$ would satisfy
$\neg \theta \wedge M_2(x_2,y_2) \wedge M_3(x_3,y_3)$, which is impossible by~(\ref{eq:easy2}). Suppose without loss of generality that
$\Phi$ implies $\theta \vee L_2(x_2,y_2) \vee (x_2=y_2) \vee L_3(x_3,y_3)$. Since $\Phi$
also implies~(\ref{eq:guard1}),
%$\theta \vee (x_2 \neq y_2) \vee L_3(x_3,y_3) \vee (x_3 = y_3)$, 
we have that 
$\Psi$ contains the graph bijunctive formula
\begin{align}
\theta \vee \big ((L_2(x_2,y_2) \vee x_2=y_2 \vee L_3(x_3,y_3)) \wedge (x_2 \neq y_2 \vee L_3(x_3,y_3) \vee x_3 = y_3) \big) \; . \label{eq:last}
\end{align}
We finally show that $\Psi$ implies $\phi$.
Let $t$ be a tuple that satisfies $\Psi$.
If $t$ satisfies $\theta \vee (x_1 = y_1)$ there is nothing to show, so suppose otherwise. 
Then~(\ref{eq:easy1}), which is graph bijunctive and thereofore in $\Psi$, implies that either $x_2 \neq y_2$ or $x_3 \neq y_3$. If $x_2 \neq y_2$, then
by the first conjunct in~(\ref{eq:last}) we have
that $L_2(x_2,y_2)$ or $L_3(x_3,y_3)$, in which case $t$
satisfies $\phi$ and we are done. Otherwise, suppose
that $x_2 = y_2$. Then $x_3 \neq y_3$ as we have seen above. 
But then the second conjunct in~(\ref{eq:last}) implies 
that $L_3(x_3,y_3)$, and we are again done. 
\end{proof}

\begin{proposition}
Let $\bB$ be a reduct of $(\mV;E)$ with finite
relational signature, and suppose that $\bB$
has a balanced ternary polymorphism of type majority. 
Then $\Csp(\bB)$ can be solved in polynomial time.
\end{proposition}
\begin{proof}
Let $\Phi$ be an instance of $\Csp(\bB)$ with variables $S$, and let
$\Psi$ be the set of clauses obtained from $\Phi$ by replacing each constraint by its graph bijunctive definition over $(\mV;E,N)$ which exists by Theorem~\ref{thm:syntax}. 
Clearly, $\Phi$ is satisfiable in $\bB$ if and only if $\Psi$ is satisfiable in $(\mV;E,N)$. 

We associate to $\Psi$ a 2SAT instance $\psi = \psi(\Psi)$ as follows. For each unordered pair $\{u,v\}$ 
of distinct variables $u,v$ of $\Psi$ we have a
variable $x_{\{u,v\}}$ in $\psi(\Psi)$. Then
\begin{itemize}
\item if $\Psi$ contains the clause $E(u,v)$ or the clause $E(u,v) \vee u=v$ then $\psi(\Psi)$ contains the clause $\{x_{\{u,v\}}\}$;
\item if $\Psi$ contains the clause $N(u,v)$ or the clause $N(u,v) \vee u=v$ then $\psi(\Psi)$ contains the clause $\{\neg x_{\{u,v\}}\}$;
\item if $\Psi$ contains the clause $N(a,b) \vee E(c,d)$ then $\psi(\Psi)$ contains the clause $\{\neg x_{\{a,b\}},x_{\{c,d\}}\}$. Clauses of the form $L_1(a,b) \vee L_2(c,d)$ are translated correspondingly for all $L_1,L_2 \in \{E,N\}$;
\item if $\Psi$ contains the clause 
$(N(a,b) \vee a=b \vee E(c,d)) \wedge (a \neq b \vee E(c,d) \vee c = d)$
 then $\psi(\Psi)$ contains the clause $\{\neg x_{\{a,b\}}, x_{\{c,d\}}\}$. Clauses of the form $(L_1(u_1,v_1) \vee u_1=v_1 \vee L_2(u_2,v_2)) \wedge (u_1 \neq v_1 \vee L_2(u_2,v_2) \vee u_2 = v_2)$ are translated correspondingly for all $L_1,L_2 \in \{E,N\}$.
\end{itemize}
All other clauses of $\Psi$ are ignored for the definition of $\psi(\Psi)$. 
%We call $\psi$ the \emph{2SAT instance associated} to $\Psi$. 

We recall an important and well-known 
concept to decide 
satisfiability of 2SAT instances $\psi$.
If $\psi$ contains clauses of size one,
we can reduce to the case where all clauses have size two by replacing the clause $\{x\}$ by $\{x,x\}$. 
 The
\emph{implication graph} $G_\psi$ of a conjunction $\psi$ of propositional clauses of size two is the directed
graph whose vertices $T$ are the variables $x,y,z,\dots$ of $\psi$, and the negations 
$\neg x, \neg y, \neg z$ of the variables. 
The edge set of $G_\psi$ contains $(x,x') \in V^2$ 
if $\psi$ contains the clause $\{\neg x, x'\}$ 
(here we identify $\neg ( \neg x)$ with $x$).
It is well-known that $\psi$ is unsatisfiable if and only if there exists $x \in T$ such that
$x$ and $\neg x$ belong to the same strongly connected component (SCC) of $G_\psi$. 

\begin{figure*}[h]
\begin{center}
\small
\fbox{
\begin{tabular}{l}
{\rm // Input: A set of graph bijunctive clauses $\Psi$} 
\\
Do \\
\hspace{0.5cm} While $\Psi$ contains a clause of the form $u=v$ do  \\
\hspace{1cm} Replace each occurrence of $v$ by $u$ in $\Psi$.  \\
\hspace{1cm} Remove literals of the form $E(u,u)$, $N(u,u)$, and $u \neq u$ from $\Psi$. \\
\hspace{1cm} If $\Psi$ contains an empty clause then reject. \\
\hspace{.5cm} Loop. \\
%\hspace{0.5cm} Let $\Theta$ be the set of unit clauses in $\Psi$ (none of them of the form $u = v$). \\
\hspace{.5cm} Compute the 2SAT instance $\psi = \psi(\Psi)$, and the graph $G_\psi$. \\
%\hspace{.5cm} Compute $G_\phi$. \\
\hspace{.5cm} If $G_\psi$ contains $x_{\{u,v\}}$ such that $x_{\{u,v\}}$ and $\neg x_{\{u,v\}}$ are in the same SCC then \\
%\hspace{2cm} are in the same strong component, then \\
\hspace{1cm} Replace each occurrence of $v$ by $u$ in $\Psi$.  \\
\hspace{1cm} Remove literals of the form $E(u,u)$, $N(u,u)$, and $u \neq u$ from $\Psi$. \\
\hspace{1cm} If $\Psi$ contains an empty clause then reject. \\
Loop until $\Psi$ does not change any more. \\
Accept.
\end{tabular}}
\end{center}
\caption{Polynomial-time algorithm to test satisfiability of a given set of graph bijunctive clauses.}
\label{fig:alg-bijunctive}
\end{figure*}

Now consider the algorithm displayed in Figure~\ref{fig:alg-bijunctive}. We make the following claims. 
\begin{enumerate}
\item Whenever the algorithm replaces all occurrences of a variable $v$ in $\Psi$ by a variable $u$, then $u$ and $v$ must have the same value in all solutions of $\Psi$.
\item When the algorithm rejects an instance, then $\Psi$ is unsatisfiable. 
\item When the algorithm accepts, then
the input formula indeed is indeed satisfiable. 
\end{enumerate}
The first claim can be shown inductively over
the execution of the algorithm as follows.
When the algorithm replaces all occurrences of $v$ by $u$ in line 4 of the algorithm, the first claim is trivially true. The only other variable contraction can
be found in line 10 of the algorithm. 

So let $\Psi$ be the set of graph bijunctive clauses when we reach line 10, and 
suppose that $x_{\{u,v\}}$ and $\neg x_{\{u,v\}}$
lie in the same SCC
of $G_\psi$.
Since $x_{\{u,v\}}$ and $\neg x_{\{u,v\}}$ belong to the same SCC, there is a path $x_{\{u,v\}}=x_0,x_1,\dots,x_n=\neg x_{\{u,v\}}$ from $x_{\{u,v\}}$ 
to $\neg x_{\{u,v\}}$, and a path $\neg x_{\{u,v\}}=y_0,y_1,\dots,y_m=x$ from $\neg x_{\{u,v\}}$ to $x_{\{u,v\}}$.

Suppose that $\Psi$ has a solution $s \colon S \rightarrow V$. We have to show that $s(u)=s(v)$. 
Suppose otherwise that $s(u) \neq s(v)$;
without loss of generality, $E(s(u),s(v))$ holds. 
Let $\{u_i,v_i\}$ be the pair of variables of $\Phi$ that corresponds to $x_i$. We show by induction on $i$ that if $x_i$ is positive, then $E(s(u_i),s(v_i))$, and if $x_i$ is negative
then $N(s(u_i),s(v_i))$. 
Suppose without loss of generality that $x_i$
is positive, and suppose inductively that $E(s(u_i),s(v_i))$. 
There is a clause in
$\Psi$ that contributed the edge 
$(x_i,x_{i+1})$ to $G_\psi$. If $x_{i+1}$ is a positive literal, then this clause is
 either of the form $N(u_i,v_i) \vee E(u_{i+1},v_{i+1})$,
 or of the form
$$(N(u_i,v_i) \vee u_i=v_i \vee E(u_{i+1},v_{i+1})) \wedge (u_i \neq v_i \vee E(u_{i+1},v_{i+1}) \vee u_{i+1} = v_{i+1}) \; .$$
 In both cases, the clause together with
 $E(s(u_i),s(v_i))$ implies that $E(s(u_{i+1}),s(v_{i+1}))$. The argument in the case
 that $x_{i+1}$ is a negative literals is similar. 
 For $i+1=n$ we obtain that
 $N(s(u),s(v))$, in contradiction to our assumption.
 Therefore, 
we conclude that $s(u)=s(v)$, which concludes the proof of the first claim. 

Since the only modifications to $\Psi$ are variable contractions, the first claim implies that when at some stage during the execution of the algorithm the formula $\Psi$ contains an empty clause, then there is indeed no solution to the original input formula; this proves the second claim. 

To prove the third claim, suppose that 
the algorithm accepts. Let $\psi = \psi(\Psi)$ be
the 2SAT instance in the final round of the main loop of the algorithm, and let $T$ be the set of variables of $\psi$. 
The 2SAT formula $\psi$ must have a solution, since
otherwise the algorithm would have changed $\Phi$,
in contradiction to our assumptions.
From a solution 
$t \colon T \to \{0,1\}$ for $\psi$ we obtain a solution
$s \colon S \to V$ for the clause set $\Psi$ at the end of the execution of the algorithm by assigning distinct vertices of $V$ to every variable of $\Psi$ such that
$(s(u),s(v)) \in E$ 
if and only if $s(x_{\{u,v\}})=0$. We also get a solution to the originally given set of clauses (before contractions of variables) by setting contracted variables to the same value. 

The three claims show the correctness of the algorithm. It is easy to see that the algorithm can be
implemented in polynomial (in fact, in quadratic) time in the input size. 
\end{proof}

\subsection{Tractability of types max and min}\label{ssect:maxMin}

We are left with the case where $\bB$ has a canonical binary injective polymorphism of type $\maxi$ or $\mini$, which corresponds to case~(f) of Proposition~\ref{prop:higherArity}.

We claim that we can assume that this polymorphism is either balanced, or of type $\maxi$ and $E$-dominated, or of type $\mini$ and $N$-dominated.

\begin{proposition}\label{prop:maxMinStandardizing}
    If $\bB=(\mV;E,N,\neq,\ldots)$ is first-order definable in $(\mV;E)$ and has a canonical binary injective polymorphism of type $\maxi$ ($\mini$), then it also has a canonical binary injective polymorphism of type $\maxi$ which is balanced or $E$-dominated ($N$-dominated).
\end{proposition}
\begin{proof}
    We prove the statement for type $\maxi$ (the situation for $\mini$ is dual). Let $p$ be the polymorphism of type $\maxi$.
    Then $h(x,y):=p(x,p(x,y))$ is not $N$-dominated in the first argument; this is easy to see. But then $p(h(x,y),h(y,x))$ is either balanced or $E$-dominated, and still of type $\maxi$.
\end{proof}

We apply Theorem~\ref{thm:resolution} to our setting as follows.

\begin{proposition}\label{prop:horn}
    Let $\bB$ be a structure with a first-order definition in $(\mV;E)$ and a finite relational signature, and 
    suppose $\mB$ is preserved by a binary canonical injection which is of type $\maxi$ and balanced or $E$-dominated, or of type $\mini$ and balanced or $N$-dominated. 
    Then $\Csp(\bB)$ can be solved in polynomial time.
\end{proposition}

\begin{proof}
    First note that
    $$\Csp(\mV;E,\neg E,N,\neg N,=,\neq)$$ can be solved in polynomial time. 
    One way to see this is to verify that all relations are preserved by a balanced polymorphism of type majority, and to use the algorithm presented in Section~\ref{ssect:edgeMajorityBalanced}. We observe the following. 
    \begin{itemize}
        \item A canonical binary injection which is of type $\mini$ 
        and $N$-dominated is an embedding 
        of $(\mV;E,=)^2$ into $(\mV;E,=)$.
        \item A canonical binary injection which is of type $\maxi$ 
        and $E$-dominated is a an embedding of $(\mV;N,=)^2$ 
        into $(\mV;N,=)$.
        \item A canonical binary injection which is of type $\maxi$ 
        and balanced is an embedding of $(\mV;\neg E,=)^2$
        into $(\mV;\neg E,=)$.
        \item A canonical binary injection which is of type $\mini$ 
        and balanced is an embedding of $(\mV;\neg N,=)^2$
        into $(\mV;\neg N,=)$.
    \end{itemize}
    In each case, polynomial-time tractability of $\Csp(\bB)$ follows from Theorem~\ref{thm:resolution}.
\end{proof}

This completes the proof of Proposition~\ref{prop:higherArity}.

\section{Classification}
\label{sect:g-classification}
%Theorem~\ref{thm:gsat-dichotomy} shows that the CSP for every template with a first-order definition in $(\mV;E)$ 
%is NP-hard, or polynomial-time tractable. 
In this section we present a refined description of
the polymorphisms of structures with a first-order definition in $(\mV;E)$ that imply tractability. This leads to a  dichotomy result 
%for structures $\bB$ with a first-order definition in $(\mV;E)$ 
that has already been stated in Theorem~\ref{thm:gcsp-tractability}, and which holds without any complexity-theoretic assumptions: either 
\begin{enumerate}
\item there is a primitive positive interpretation of 
%$(\{0,1\};\OIT)$ 
all finite structures in the model-complete core of $\bB$,
or 
\item $\bB$ has a cyclic polymorphism modulo an endomorphism. 
\end{enumerate}
It follows from Proposition~\ref{prop:deep-taylor-mod-endo-tame}
that $(1)$ and $(2)$ are indeed disjoint cases.
In order to prove that every $\bB$ satisfies $(1)$ or $(2)$ above,
we first determine a list of 17 
operations with the following properties: 
\begin{itemize}
\item[(a)] 
every structure
$\bB$ with a first-order definition in $(\mV;E)$ either 
interprets $(\{0,1\};\OIT)$ or $(\{0,1\};\NAE)$
or is preserved by one of those 17 operations; and
\item[(b)] 
the list is minimal, that is, if any operation is removed
from the list, then the list looses property~(a).
\end{itemize}

Our next step (Proposition~\ref{prop:g-cyclic}) will be the verification that each of the 17 operations generates an operation that is cyclic modulo an endomorphism of $(\mV;E)$. 
It will also turn out that $\Csp(\bB)$ can be solved in polynomial time if $\bB$ has
one of those operations as a polymorphism (Proposition~\ref{prop:g-tract}).

\begin{definition}
Let $B$ be a behaviour for binary functions on $(\mV;E)$. 
A ternary injection $f\colon \mV^3\To \mV$ is \emph{hyperplanely of type $B$} 
if the binary functions $(x,y) \mapsto f(x,y,c)$, $(x,z) \mapsto f(x,c,z)$, 
and $(y,z) \mapsto f(c,y,z)$ have behaviour $B$ for all $c\in \mV$.
\end{definition}

We have already met a special case of this concept in Definition~\ref{defn:balanced} of Section~\ref{ssect:edgeMinorityBalanced}: a ternary function is balanced if and only if it is hyperplanely balanced of type $p_1$.
The following behaviors of binary functions 
appear hyperplanely in ternary functions of our classification
result.

\begin{definition}
    A binary injection $f\colon \mV^2\To \mV$ is of type
    \begin{itemize}
        \item \emph{$E$-constant} if the image of $f$ is a clique;
        \item \emph{$N$-constant} if the image of $f$ is an independent set;
        \item \emph{$\xnor$} if for all $u,v\in \mV^2$ with $\NEQNEQ(u,v)$ the relation $E(f(u),f(v))$ holds if and only if $\EE(u,v)$ or $\NN(u,v)$ holds;
        \item \emph{$\xor$} if for all $u,v\in \mV^2$ with $\NEQNEQ(u,v)$ the relation $E(f(u),f(v))$ holds if and only if neither $\EE(u,v)$ nor $\NN(u,v)$ hold.
    \end{itemize}
\end{definition}

Recall from Definition~\ref{def:I6}
that $I_6 := I^{\mV}_6$ denotes the 6-ary relation defined by
\begin{align*} \{(x_1,x_2,y_1,y_2,z_1,z_2) \in \mV^6\; | \; &
 (x_1=x_2 \wedge y_1 \neq y_2 \wedge z_1 \neq z_2) \\
  & \vee  \;
 (x_1 \neq x_2 \wedge y_1 = y_2 \wedge z_1 \neq z_2) \\
 & \vee  \; (x_1 \neq x_2 \wedge y_1 \neq y_2 \wedge z_1 = z_2) \} \; .
 \end{align*}
%It is easy to see that the polymorphisms of $I_6$ are precisely the essentially unary operations which after deletion of all dummy variables are injective.
Similarly, we define relations $E_6$ and $N_6$ by altering the above definition and replacing all occurrences of $\neq$ by $E$ and $N$, respectively.

\begin{theorem}\label{thm:minimalTractableClones}
Let $\bB$ be a structure with a first-order definition in $(\mV;E)$.
Then either one of the following relations
is primitive positive definable in $\bB$: 
 $I_6$, $E_6$, $N_6$, $H_1$, $H_1'$,  $H_2$, $H_2'$, 
and $(\{0,1\};\OIT)$ or $(\{0,1\};\NAE)$ are primitive positive interpretable in $\bB$, or $\bB$ has a polymorphism of one of the following types. 
    \begin{enumerate}
        \item \label{mt:const} A constant operation,
        \item \label{mt:maxBalanced} a balanced binary injection of type $\maxi$,
        \item \label{mt:maxEdominated} an $E$-dominated binary injection of type $\maxi$,
        \item \label{mt:majHpProjBalanced} a function of type majority which is hyperplanely of type projection and balanced,
         \item \label{mt:majHpEfanatic} a function of type majority which is hyperplanely $E$-constant,
\item \label{mt:majHpMaxEdominated}
a function of type majority which is hyperplanely of type $\max$ and $E$-dominated. 
       \item \label{mt:minHpProjBalanced} a function of type minority which is hyperplanely of type projection and balanced,
         \item \label{mt:minHpProjEdominated} a function of type minority which is hyperplanely of type projection and $E$-dominated,
        \item \label{mt:minHpXorBalanced} a function of type minority which is hyperplanely of type $\xnor$ and balanced,
         \item \label{mt:Efanatic} a binary injection which is $E$-constant,
     \end{enumerate}
    or the dual of one of the last seven operations.
\end{theorem}
\begin{proof}
If $\bB$ has a constant endomorphism, 
then we are in case~$(\ref{mt:const})$, so we may assume that this is not the case. 

First consider the case where all polymorphisms
of $\bB$ are essentially unary. Then either 
$I_6$, $N_6$, or $E_6$ is preserved by 
all polymorphisms of $\bB$, and hence primitive positive definable in $\bB$, and we are done.
So we assume that $\bB$ has an essential operation. 
By Lemma~\ref{lem:binary}, we even
have a binary essential polymorphism $f$. 

Consider now the case that $e_E \in \Pol(\bB)$.  
Then consider the structure $\bD$ induced in $\bB$ on the image $D := e_E[\mV]$. 
This structure $\bD$ is preserved by all permutations of its domain, and hence is first-order definable in $(D;=)$. 
It follows from Corollary~\ref{cor:ecsp-preclass} that $\bD$ either has a 
constant polymorphism, or a binary injection, or all polymorphisms of $\bD$ are essentially unary. 
The structure $\bD$ cannot have a constant endomorphism as otherwise also $\bB$ has a constant polymorphism by composing the constant with $e_E$. 
Suppose that $f(a,a)=f(a,b)$ for all $a,b \in \mV$ with $E(a,b)$. 
We claim that $f(u,u)=f(u,v)$ for \emph{every} $u,v \in \mV$. To see this,
let $w \in \mV$ be such that $E(u,w)$ and $E(v,w)$. 
Then $f(u,u)=f(u,w)=f(u,v)$, as required. 
It follows that $f$ does not depend
on its first variable, a contradiction. Hence, 
there exist $a,b \in \mV$ such that $E(a,b)$ and $f(a,a) \neq f(a,b)$. 
Similarly, there exist $c,d \in \mV$ such that $E(c,d)$ and $f(c,c) \neq f(d,c)$. 
Let $T$ be an infinite clique adjacent to $a,b,c,d$. 
Then $f$ is either essential on $T \cup \{a,b\}$ or on $T \cup \{c,d\}$, both cliques. Suppose without loss of generality
that $f$ is essential on $C = T \cup \{a,b\}$. Since
all operations with the same behaviour as $e_E$ generate each other,
we can also assume that the image of $e_E$ is $C$.
Then the 
restriction $f'$ of $(x_1,x_2) \mapsto e_E(f(x_1,x_2))$ to $e_E[\mV]$
is an essential polymorphism of $\bD$. 
Hence, Corollary~\ref{cor:ecsp-preclass} implies that
$\bD$ has a binary injective polymorphism $h'$. 
Then $h(x,y):=h'(e_E(x),e_E(y))$ is a polymorphism of $\bB$. 
But $h$ is a binary canonical injection which is $E$-constant, 
and so $\bB$ has a polymorphism from Item~$(\ref{mt:Efanatic})$ of our list. 
When $\bB$ is preserved by $e_N$ the dual argument works.

Hence, by Theorem~\ref{thm:g-endos} it remains to consider the case where the 
endomorphisms of $\bB$ are generated by the automorphisms of $\bB$,
that is, $\bB$ is a model-complete core (Theorem~\ref{thm:mc-core}). 
By Theorem~\ref{thm:reducts} there are five possibilities for $\End(\bB)$. Suppose first that $\bB$ is preserved by all permutations on $\mV$. Then by Corollary~\ref{cor:ecsp-preclass}, 
 $\bB$ is preserved by all binary injections, 
 and in particular $\bB$ is preserved by, say, a balanced binary canonical injection operation of type $\maxi$. 

We can therefore assume that $\End(\bB)$ does not contain all permutations. 
We consider the case where the
automorphism of $(\mV;E)$ are dense in $\End(\bB)$. 
If $H_1$ is primitive positive definable in 
$\bB$, then we have found a primitive positive interpretation
of $(\{0,1\};\OIT)$ in $\bB$ by Proposition~\ref{prop:h2}.  
Otherwise, 
Proposition~\ref{prop:higherArity} applies, 
and $\Pol(\bB)$ contains 
a binary canonical injection of type $\maxi$ or $\mini$, or a function of type minority or majority. 
If it contains a canonical injection of type $\maxi$ or $\mini$, then $\bB$ is preserved by some of the operations from case~$(\ref{mt:maxBalanced})$, 
$(\ref{mt:maxEdominated})$, or their duals, by Proposition~\ref{prop:maxMinStandardizing}. Otherwise, $\Pol(\bB)$ contains a
ternary injection $t$ of type minority or majority, and one of the binary canonical injections of type projection listed in Theorem~\ref{thm:minimal-ops}, which we denote by $p$. 
Set $s(x,y,z):=t(p(x,y),p(y,z),p(z,x))$ and $w(x,y,z):=s(p(x,y),p(y,z),p(z,x))$. Then $w$ has the same behavior as one of the operations from case~$(\ref{mt:majHpProjBalanced})$ to case~$(\ref{mt:minHpXorBalanced})$ --- see Figure~\ref{fig:classification-correspondence}; we leave the verification to the reader.

Now suppose that $\End(\bB)$ is the monoid generated by $\sw$.
If $H_2$ is primitive positive definable in $\bB$,
then we have a primitive positive interpretation
of $(\{0,1\};\OIT)$ in $\bB$ by Proposition~\ref{prop:h2}.  
Otherwise we are done by Proposition~\ref{prop:sw},
since the functions that appear in  
Proposition~\ref{prop:sw} are a subset 
of the function that appear in Proposition~\ref{prop:higherArity} and that we have treated above. 

Next consider the case where $\End(\bB)$ is the monoid generated by $-$. 
If the relation $H_1'$ is primitively positively definable in $\bB$,
then $(\{0,1\};\NAE)$ has a primitive positive interpretation 
in $\bB$ by Proposition~\ref{prop:h1prime}. Otherwise we are 
done by Proposition~\ref{prop:minus}. 

If $\End(\bB)$ is the monoid generated by $\{\sw,-\}$
and $H_2'$ is primitive positive definable in $\bB$,
then we have a primitive positive interpretation
of $(\{0,1\};\NAE)$ in $\bB$ by Proposition~\ref{prop:h2}.  
Otherwise we are done by Proposition~\ref{prop:minus-sw}. 
\end{proof}

\begin{proposition}\label{prop:g-cyclic}
    The 17 operations listed in Theorem~\ref{thm:minimalTractableClones} 
    generate 17 distinct clones, 
    each containing an operation which is
    cyclic modulo a unary operation. 
\end{proposition}
\begin{proof}
% Todo: incomparability
Observe that all of the 17 operations listed in Theorem~\ref{thm:minimalTractableClones} are canonical as functions over $(\mV;E)$.
Let $g$ be one of them. By Lemma~\ref{lem:type-algebra} there is a homomorphism $\mu$ from
$(\mV;g)^2$ to an algebra ${\bf A} = (\{=,E,N\},f)$ (where $=$, $E$, and $N$ are the image of $\mu$ for the pairs $(x,y) \in \mV^2$ such that $x=y$, $E(x,y)$, or $N(x,y)$, respectively).

In 
 case~(\ref{mt:const}), case~(\ref{mt:Efanatic}), and its dual, 
the algebra ${\bf A}$ is not 
idempotent. In case~(\ref{mt:const}), when $g$ is constant, then $g$ is in particular a cyclic polymorphism. 
In case~$(\ref{mt:Efanatic})$, 
the image of $g$ induces an infinite clique in $(\mV;E)$. 
As in the proof of Corollary~\ref{thm:mc} we see that 
$\bB$ is preserved by all permutations of its domain. 
Also note that $\bf A$ has a congruence with the congruence classes $\{=\}$
and $\{E,N\}$ (see Proposition~\ref{prop:all-inj}),
and in the corresponding quotient algebra $g$ denotes $\max$ with respect to 
the order $\{=\} < \{E,N\}$. It is then easy to see that $f\colon (x,y,u,v) \mapsto g(g(g(x,y),u),v)$ is a cyclic operation modulo an endomorphism of $(\mV;E)$.
The dual proof works for the dual of case~$(\ref{mt:Efanatic})$.

By Theorem~\ref{thm:factors} in combination with Theorem~\ref{thm:taylor} and Theorem~\ref{thm:siggers}, 
every finite idempotent algebra either 
\begin{itemize}
\item has a 2-element factor all of whose
operations are projections, or 
\item has a cyclic term. 
\end{itemize}
If the second case applies, then Corollary~\ref{cor:type-algebra-lift} shows that $\bB$ has a cyclic polymorphism 
modulo an endomorphism. Therefore, 
it suffices to show in the remaining 14 cases that 
all factors of $\fA$ contain operations that are not projections. 
Since in those 14 cases both $E$ and $N$ are preserved by $g$, the
relation $\neq$ is also preserved, and $\{E,N\}$ induce a subalgebra of $\fA$.
In this subalgebra and if the operation $g$ is ternary, it either acts as a majority (that is, $g(x,x,y)=g(x,y,x)=g(y,x,x)=x$), or as a minority (that is, $g(x,x,y)=g(x,y,x)=g(y,x,x)=y$).
If $f$ is binary, it satisfies $g(x,y)=g(y,x)$ in this subalgebra. In all cases,
$f$ does not act as a projection.
Four out of the 14 remaining operations are balanced, which is equivalent to saying that
both $\{E,=\}$ and $\{N,=\}$ induce a subalgebra ${\bf B}$ in ${\bf A}$. In this case it is easy to check from the description of the balanced
operations in Theorem~\ref{thm:minimalTractableClones} that
\begin{align}
g(x,y) \text{ satisfies } g(x,y)=g(y,x) & \text{ if $f$ is binary, and} \label{eq:commutative}
\\ 
h(x,y) := g(x,x,y) \text{ satisfies } h(x,y)=h(y,x) 
& \text{ if $g$ is ternary.} \label{eq:almost-commutative}
\end{align}
So $g$ is not a projection in those factors as well. 
For five of the remaining non-balanced operations we have that
$\{E,=\}$ induces a subalgebra of ${\bf A}$. Again, $g$ satisfies
the condition in (\ref{eq:commutative}).
For the other five remaining operations, the set $\{N,=\}$ induces a subalgebra, and the argument that the operation $f$ is not a projection in those algebras is analogous.

Finally, we have to argue that the operation $g$ is in none of the 2-element homomorphic 
images of ${\bf A}$ a projection.
Since all of the 14 remaining operations are injective, they 
have a congruence with the classes $\{E,N\}$ and $\{=\}$ (Proposition~\ref{prop:all-inj}). 
Then the operation
$g$ satisfies (\ref{eq:commutative}) in the corresponding factor.
It can be verified that from all 14 operations, only
\begin{itemize}
\item the balanced operation of type max,
\item the N-dominated operation of type min, 
\item and the edge majority that is hyperplanely of type
min and N-dominated 
\end{itemize}
preserve the relation $E(x,y) \Leftrightarrow E(u,v)$.
In those cases,  the algebra ${\bf A}$ has a congruence with the 
classes $\{E\}$ and $\{N,=\}$. 
For the balanced operation of type max, and the N-dominated operation of type min,
in the corresponding quotient the operation $g$ satisfies
the condition in (\ref{eq:commutative}).
For the edge majority that is hyperplanely of type min and N-dominated, 
the condition in (\ref{eq:almost-commutative}) applies. 
Congruences of ${\bf A}$ with the classes $\{N\}$ and $\{E,=\}$ can be checked analogously.
%We conclude that all operations in the list generate tame clones.
% Proof: 
% balanced of type max
% (N,=),(E,E) -> (E,E);      E <-> E wird vermutlich erhalten
% (E,=),(N,N) -> (E,N);     zerstoert N <-> N
% E-dominated of type max
% (N,=),(N,N) -> (N,E);     zerstoert E <-> E
% (E,=),(E,E) -> (E,E) ;     N <-> N wird vermutlich erhalten.
%
% edge majority hp balanced of tp p_1
% (N,=),(E,E),(=,N) -> (N,E) zerstoert E<->E 
% (E,=),(N,N),(=,E) -> (E,N) zerstoert N<->N 
% edge majority hp E-fanatic
% (N,=),(E,E),(N,N) -> (N,E) zerstoert E<->E
% (E,=),(N,N),(N,N) -> (N,E) zerstoert N<->N
% Edge majority hp tp max E-dom
% (N,=),(N,=),(E,N) -> (N,E) zerstoert E<->E
% (E,=),(E,E),(N,N) -> (E,E) koennte N<->N erhalten!!!
% (E,=),(=,E),(N,N) -> (E,E) koennte N<->N erhalten!!!
% (E,=),(E,=),(N,N) -> (E,E) koennte N<->N erhalten!!!
% edge minority hp balanced of tp p_1
% (N,=),(N,N),(E,E) -> (E,N) zerstoert E<->E
% (E,=),(E,E),(N,N) -> (N,E) zerstoert N<->N
% edge minority hp p1 and E-dominated
% analogous to the previous case
% edge minority hp match-rewarding E-dominated:
% (N,=),(N,N),(N,N) -> (N,E)   zerstoert E <-> E
% (N,=),(N,=),(N,N) -> (N,E)   zerstoert N <-> N 
\end{proof}

\begin{figure*}
\begin{center}
{\small
 \begin{tabular}{|l|l|l|}
    \hline
    Binary injection type $p_1$ & Type majority & Type minority\\
    \hline
    Balanced & Hp.\  balanced, type $p_1$ & Hp.\  balanced, type $p_1$\\
    $E$-dominated & Hp.\  $E$-constant & Hp.\  type $p_1$, $E$-dominated\\
    $N$-dominated & Hp.\  $N$-constant & Hp.\  type $p_1$, $N$-dominated\\
    Balanced in 1st, $E$-dom. in 2nd arg.& Hp.\  type $\maxi$, $E$-dom. & Hp.\  type $\xnor$, balanced. \\
    Balanced in 1st, $N$-dom. in 2nd arg.& Hp.\  type $\mini$, $N$-dom. & Hp.\  type $\xor$, balanced. \\
    \hline
 \end{tabular}
 }
\end{center}
\caption{Minimal tractable canonical functions of type majority and minority, and their corresponding canonical binary injections of type projection.}\label{fig:classification-correspondence}
\end{figure*}

\begin{lemma}\label{prop:g-tract}
Suppose that $\Pol(\bB)$ contains one of the 17 operations from Theorem~\ref{thm:minimalTractableClones}. Then 
every finite signature reduct $\bB'$ of
$\bB$ has a polynomial-time tractable CSP. 
\end{lemma}
\begin{proof} 
For the constant operation 
this is Proposition~\ref{prop:const-core}.
Cases~(\ref{mt:maxBalanced}) and (\ref{mt:maxEdominated}) are tractable by case~(f) of Proposition~\ref{prop:higherArity}. In all cases, the duals can be solved analogously.
The functions of type majority or minority are tractable by cases~(b) 
to (e) of Proposition~\ref{prop:higherArity}: in those cases, certain binary canonical injections of type projection are required -- these are obtained by identifying any two variables of the function of type majority / minority; 
Figure~\ref{fig:classification-correspondence} shows which function of type majority / minority yields which type of binary injection.
 We leave the verification to the reader. Finally, let $f(x,y)$ be an $E$-constant binary injection (case (10)), and denote the reduct corresponding to this clone by $\bB$. Then $g(x):=f(x,x)$ is a homomorphism from $\bB$ to the structure $\bC$ induced by the image $g[\mV]$ in $\bB$. This structure $\bC$ is invariant under all permutations of its domain, and hence is definable in $(g[\mV];=)$; such structures have been treated in Chapter~\ref{chap:ecsp}. 
 The structure $\bC$ has a binary injection among 
 its polymorphisms (namely, the restriction of $f$ to $\bC$). It then follows from Theorem~\ref{thm:ecsps} that $\Csp(\bC)$ is tractable. Hence, $\Csp(\bB)$ is tractable as well, since $\bB$ and $\bC$ are homomorphically equivalent. 
 \end{proof}

\begin{figure*}
\begin{center}
\includegraphics[scale=0.5]{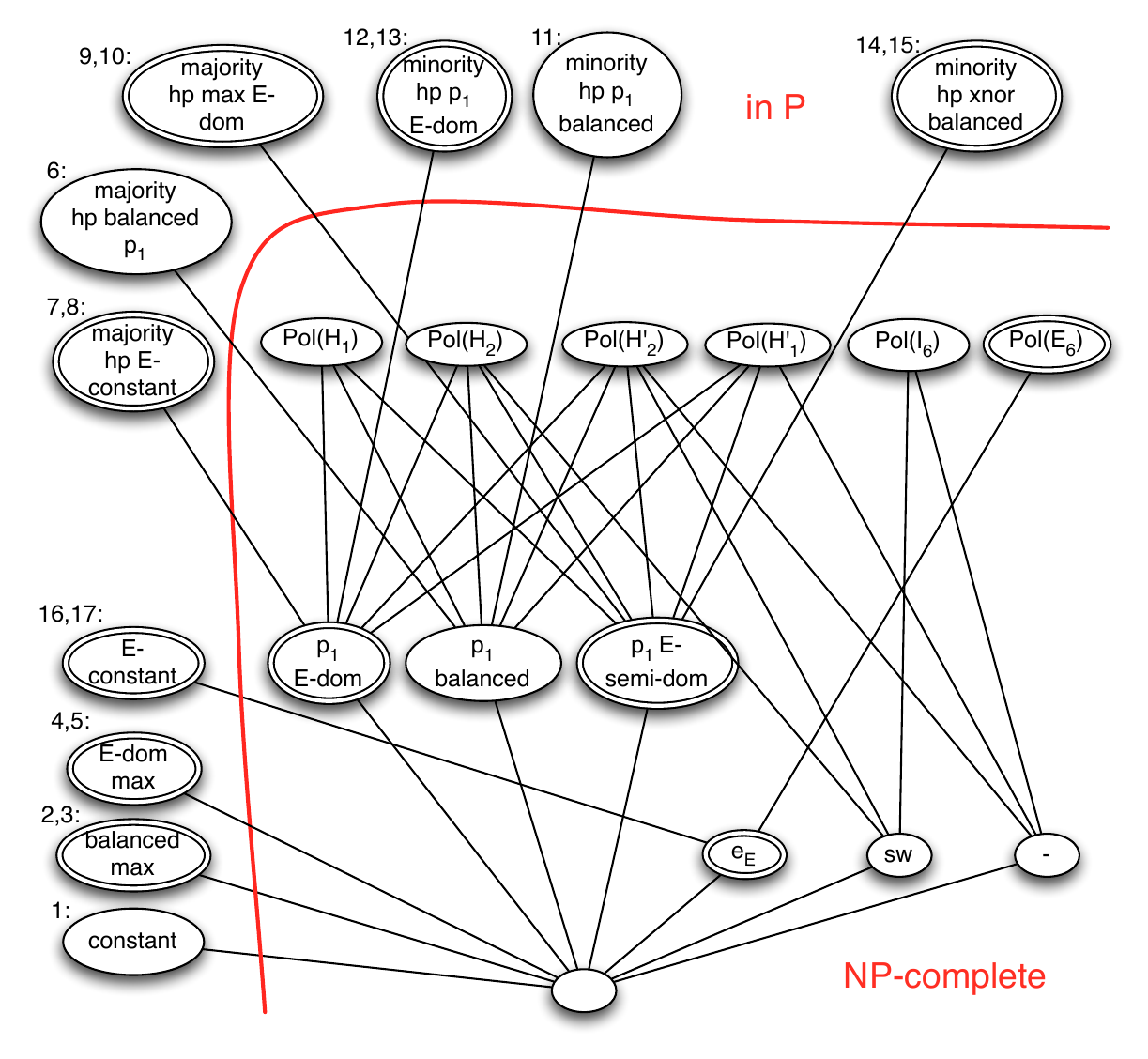}
\end{center}
\caption{The border: Minimal tractable and maximal hard clones containing $\Aut((\mV;E))$.}\label{fig:gcsp-main}
\end{figure*}

Figure~\ref{fig:gcsp-main} shows the border between the hard and the tractable clones. The picture contains all minimal tractable clones as well as all maximal hard clones, plus some other clones that are of interest in this context. 
Lines represent containment of clones, but edges that are implied by transitivity of containment are not drawn. Note that lines do not mean to imply that there are no other clones between them which are not shown in the picture. Clones are symbolized with a double border when they have a dual clone (generated by the dual function in the sense of Definition~\ref{defn:dual}, whose behavior is obtained by exchanging $E$ with $N$, $\maxi$ with $\mini$, and $\xnor$ with $\xor$). Of two dual clones, only one representative (the one which has $E$ and $\maxi$ in its definition) is included in the picture. The numbers of the minimal tractable clones refer to the numbers in Theorem~\ref{thm:minimalTractableClones}. ``$E$-semidom'' refers to ``balanced in the first and $E$-dominated in the second argument''. 

\begin{proof}[Proof of Theorem~\ref{thm:gcsp-tractability}]
By Theorem~\ref{thm:minimalTractableClones},
either there is a primitive positive interpretation of $(\{0,1\};\OIT)$ in $\bB$,
and the statement follows from Corollary~\ref{cor:pp-interpret-hard}, 
or $\bB$ is preserved by one out of the 17 canonical  operations
listed in Theorem~\ref{thm:minimalTractableClones}.
By Proposition~\ref{prop:g-cyclic}, these
operations are cyclic modulo an endomorphism of $\bB$, and polynomial-time tractability for the CSP of
finite-signature reducts of $\bB$ follows from Proposition~\ref{prop:g-tract}. 
\end{proof}

%By inspection of the primitive positive interpretations used in the classification proof, we also obtain the following 
%(the relation $E^{\mV}_6$ has been introduced in Definition~\ref{def:e6}).
%\begin{corollary}\label{cor:relational-main}
%$\bB$ primitively positively interprets $\OIT$ if and only if
%one $E^{\mV}_6$, $T$, $H$, or $P^{(3)}$ has a 
%primitive positive definition in $\bB$.
%\end{corollary}

\chapter{Temporal Constraint Satisfaction Problems}
\label{chap:tcsp}

%\begin{figure}[h]
\begin{center}
\includegraphics{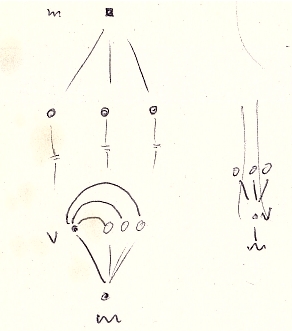} Martin Kutz, 2001
\end{center}
\vspace{1cm}
%\caption{Drawing by Martin Kutz, 2001}
%\end{figure}

This chapter contains results from~\cite{tcsps-journal,ll,BP-reductsRamsey}.

\section{Introduction}
\label{sect:tcsp-intro}
A \emph{temporal relation} is a relation $R \subseteq {\mathbb Q}^k$,
for some finite $k$, with a first-order definition in $(\mathbb Q;<)$, the ordered rational numbers (which can be thought of \emph{time points}). 
A \emph{temporal constraint language} is a set of temporal relations,
and will be treated here as a relational structure with a
first-order definition in $(\mathbb Q;<)$.
Constraint satisfaction problems for
temporal constraint languages will be called \emph{temporal CSPs}
in the following.
We have already discussed some temporal constraint languages in 
Section~\ref{sect:csp-examples}: %of Chapter~\ref{chap:intro}: 
for instance and/or precedence constraints
from scheduling, and Ord-Horn constraints on time points.

There are also several famous NP-complete temporal CSPs. 
For example the Betweenness Problem~\cite{GareyJohnson},
which has been introduced in Example~\ref{expl:betw} as a CSP with domain $\mathbb Z$, can also be formulated as
$\Csp(({\mathbb Q}; \btw))$ where $\btw$ is the ternary relation 
$\btw = \{ (x,y,z) \in {\mathbb Q}^3 \;  | \; (x < y < z) \vee (z < y < x) \}$. We have seen in Proposition~\ref{prop:betw-hard}
that this CSP is NP-hard. 
Similarly, the Cyclic Ordering Problem~\cite{GareyJohnson} can be formulated as the CSP for $({\mathbb Q}; \{(x,y,z) \; | \; (x<y<z) \vee (y<z<x) \vee (z<x<y)\})$, and is also NP-complete~\cite{GalilMegiddo} (a hardness proof using primitive positive interpretations can be found in Section~\ref{ssect:tcsp-hard}).

A subclass of temporal
CSPs called \emph{ordering CSPs} has been introduced in~\cite{GM2}. An ordering CSP is a temporal CSP where the constraint language only contains relations where the arguments are pairwise distinct (thus, 
CSP$(({\mathbb Q}; \leq,\neq))$ is \emph{not} an ordering CSP). Satisfiability thresholds  for \emph{random instances} of ordering CSPs have been studied in~\cite{Goerdt}. 
\emph{Approximability} of ordering CSPs has been studied in~\cite{ApproxOrderingCSP}.

The class of temporal constraint languages is
of fundamental importance for infinite domain constraint satisfaction,
since CSPs for such languages appear as important special cases in several other classes of
CSPs that have been studied, e.g.,
constraint languages about branching time, partially ordered time, spatial reasoning, and set constraints~\cite{DrakengrenJonssonSets,RCC5JD,BroxvallJonsson}.
Moreover, several polynomial-time solvable classes of constraint
languages on \emph{time intervals}~\cite{Nebel,KrokhinAllen,DrakengrenJonssonMetric} can be solved
by translation into polynomial-time solvable temporal constraint languages.

In this chapter we prove a complete classification of the
computational complexity of $\Csp(\bB)$ when $\bB$ is a temporal constraint language.

\begin{theorem}\label{thm:tcsp-main}
Let $\bB$ be a structure with a first-order definition in $({\mathbb Q};<)$, and let
$\bC$ be the model-complete core of $\bB$. 
Then exactly one of the following two cases is true.
\begin{itemize}
\item $\bC$ has an (at most ternary) weak near unanimity
polymorphisms modulo endomorphisms.
In this case, $\Csp(\bB')$ is in P for every finite reduct $\bB'$ of $\bB$. 
\item All finite structures have a primitive positive interpretation with parameters in $\bC$.
In this case, 
$\Csp(\bB)$ is NP-hard by Corollary~\ref{cor:pp-interpret-hard}. 
\end{itemize}
\end{theorem}

Our classification proof  is based on the universal-algebraic
approach and Ramsey theory as described in Chapter~\ref{chap:algebra}
and Chapter~\ref{chap:ramsey}.

\section{Preliminaries}
\subsection{Cameron's theorem}
\label{ssect:tcsp-aut}
In this subsection we recall the classical result of Cameron~\cite{Cameron5} that describes temporal constraint languages
up to first-order interdefinability. 
For $x_1,\ldots,x_n\in \mathbb{Q}$ write $\overrightarrow{x_1\cdots x_n}$ when $x_1<\cdots<x_n$.

\begin{theorem}[Relational version of Cameron's theorem; see e.g.~\cite{JunkerZiegler}]\label{thm:order-reducts}
Let $\bB$ be a temporal constraint language.
Then $\bB$ is first-order interdefinable with exactly 
one out of the following five homogeneous structures.
\begin{itemize}
\item The dense linear order $(\mathbb{Q};<)$ itself,
\item The structure $(\mathbb{Q}; \btw)$, where $\btw$ is the ternary relation 
$$\big \{(x,y,z) \in {\mathbb Q}^3 \; | \; \overrightarrow{xyz} \, \vee \,
\overrightarrow{zyx} \big \} \, ,$$
\item The structure $(\mathbb{Q}; \Cycl)$, where $\Cycl$ is the ternary
  relation 
   $$\big \{(x,y,z) \; | \; \overrightarrow{xyz} \vee \overrightarrow{yzx} \vee \overrightarrow{zxy} \big \} \, ,$$
\item The structure $(\mathbb{Q}; \Sep)$, where $\Sep$ is the 4-ary relation 
\begin{align*}
 \big \{(x_1,y_1,x_2,y_2) \;  | \; &
 \overrightarrow{x_1x_2y_1y_2} \vee \overrightarrow{x_1y_2y_1x_2} \vee
  \overrightarrow{y_1x_2x_1y_2} \vee \overrightarrow{y_1y_2x_1x_2} \\
  \vee \; & \overrightarrow{x_2x_1y_2y_1}  \vee
  \overrightarrow{x_2y_1y_2x_1} \vee \overrightarrow{y_2x_1x_2y_1} \vee \overrightarrow{y_2y_1x_2x_1} \big \} \, ,
  \end{align*}
\item The structure $(\mathbb{Q};=)$.
\end{itemize}
\end{theorem}
The relation $\Sep$ is the so-called \emph{separation relation};
note that $\Sep(x_1,y_1,x_2,y_2)$ holds for elements $x_1,y_1,x_2,y_2 \in \mathbb Q$ iff all four points $x_1,y_1,x_2,y_2$ are distinct
and the smallest interval over $\mathbb Q$ 
containing $x_1,y_1$ properly overlaps with the smallest interval containing $x_2,y_2$
(where properly overlaps means that the two intervals have a non-empty intersection, but none of the intervals contains the other).

The next theorem is also due to Cameron~\cite{Cameron5}, and was his original motivation for the investigation of structures with a first-order definition in $(\mathbb Q; <)$. 
It is not used for our results; however, we would like to state
it here because it provides a fundamentally different characterization
of the class of temporal constraint languages.

\begin{theorem}\label{thm:set-trans}
A relational structure $\bB$ is highly set-transitive if and only if it is a temporal constraint language.
\end{theorem}

\subsection{Polymorphisms of Temporal Constraint Languages}
\label{sect:tpols}
For this chapter only, we make the following convention.
We say that a set of operations $\F$ {\em generates} an operation $g$ if $\F$ together
with all automorphisms of $(\mathbb{Q};<)$ locally generates $g$.
In case that $\F$ contains just one operation $f$, we also say that
$f$ generates $g$. 

A $k$-ary operation $f$ on $\mathbb{Q}$ defines a weak linear
order $\preceq$ on $\mathbb{Q}^k$, as follows: for 
$x,y \in {\mathbb Q}^k$, let $x \preceq y$ iff $f(x) \leq f(y)$.
The following observation follows straightforwardly from Proposition~\ref{prop:pol-inv}. 

\begin{observation}\label{obs:f-gen-g-order}
Let $f$ and $g$ be two $k$-ary operations that define the same weak linear order on $\mathbb{Q}^k$. Then $f$ generates $g$ and $g$ generates $f$.
\end{observation}

We now define fundamental operations on $\mathbb Q$.
The unary operation $\Neg$ is defined as ${\Neg}(x):=-x$ 
in the usual sense.
Let $c$ be any irrational number, 
and let $e$ be any order-preserving bijection between $(-\infty,c)$
and $(c,\infty)$. Then the operation $\Cyc$ is defined by $e(x)$ for
$x<c$ and by $e^{-1}(x)$ for $x>c$.
With these operations and the notion of generation, Cameron's theorem can be rephrased as follows.

\begin{theorem}[Operational version of Camerons theorem; see e.g.~\cite{JunkerZiegler}]\label{thm:cameron}
Let $\bB$ be a temporal constraint language.
Then exactly one of the following holds.
\begin{itemize}
\item $\Aut(\bB)$ equals $\Aut((\mathbb{Q};<))$;
\item The automorphisms of $\bB$ are the permutations generated by $\Neg$;
\item The automorphisms of $\bB$ are the permutations generated by $\Cyc$;
\item The automorphisms of $\bB$ are the permutations generated by $\Neg$ and $\Cyc$;
\item $\Aut(\bB)$ equals $\Sym({\mathbb Q})$.
\end{itemize}
\end{theorem}

If $f$ is a $k$-ary operation on $\mathbb Q$, then the operation $\Neg f(\Neg x_1,\dots,\Neg x_k)$ 
is called the \emph{dual} of $f$. Note that if $f$ preserves an $m$-ary relation $R$, then the dual of $f$ preserves the relation $\Neg R$, which
is defined to be the relation $\{ (\Neg a_1,\dots, \Neg a_m) \; | \; (a_1,\dots,a_m) \in R \}$. 
Clearly, 
CSP$(({\mathbb Q}; R_1,\dots,R_k))$ and 
CSP$(({\mathbb Q}; \Neg R_1,\dots, \Neg R_k))$ are exactly the same computational problem.

\subsection{Hard temporal CSPs}
\label{ssect:tcsp-hard}
In this subsection we discuss various important NP-complete 
temporal constraint satisfaction problems. 
We have already
mentioned in the introduction that the Betweenness and
the Cyclic Ordering Problem in~\cite{GareyJohnson} can
be formulated as temporal CSPs, and that these problems are NP-complete. 
The corresponding relations
$\Betw$ and $\Cycl$ re-appeared in Cameron's theorem (Theorem~\ref{thm:order-reducts}).
Another important relation for our classification is the relation $T_3$,
defined as follows.

\begin{definition}\label{def:lcsp-smin-smax}
Let $T_3$ be 
the ternary relation 
$$ \{ (x,y,z) \in \mathbb Q^3\; | \; (x=y<z) \vee (x=z<y) \}$$ 
\end{definition}

\begin{proposition}\label{prop:Shard}
The structure $(\{0,1\}; \OIT)$ has a primitive positive interpretation in $({\mathbb Q}; T_3,0)$.
The problem $\Csp(({\mathbb Q}; T_3))$ is NP-hard.
\end{proposition}
\begin{proof}
We give a 2-dimensional primitive positive interpretation $I$ of the structure 
$(\{0,1\}; \OIT)$ in $({\mathbb Q}; T_3,0)$.
The domain formula 
$\delta_I(x_1,x_2)$ is $T_3(0,x_1,x_2)$;
the formula $\OIT_I(x_1,x_2,y_1,y_2,z_1,z_2)$ is
% equivalent to (x_1=0 \wedge y_2=0 \wedge z_2 = 0) \vee ...
$$ \exists u \, (T_3(u,x_1,y_1) \wedge T_3(0,u,z_1));$$
%
%\item For each clause $Z$ of $F$ with variables $x_i$, $x_j$, $x_k$, we add a new
%variable $v_C$ and introduce the constraints
%$S(v_C,v_i^1,v_j^1)$ and $S(a,v_C,v_k^1)$.
the formula $=_I(x_1,x_2,y_1,y_2)$ is $T_3(0,x_1,y_2)$.
% equivalent to $(x_1=y_1=0) \vee (x_2=y_2=0)$;
The coordinate map $h\colon \delta_I(\bB^2) \rightarrow \{0,1\}$ 
is defined as follows. 
Let $(b_1,b_2)$ be a pair of elements
of $\bB$ that satisfies $\delta_I$. 
Then exactly one of $b_1, b_2$ must have value $0$, 
and the other element is strictly greater than $0$. 
We define $h(b_1,b_2)$ to be $1$ if $b_1=0$,
and to be $0$ otherwise.

To see that this is the intended interpretation, let
$(x_1,x_2),(y_1,y_2),(z_1,z_2) \in \delta_I(\bB^2)$, and 
suppose that 
$t:=(h(x_1,x_2),h(y_1,y_2),h(z_1,z_2))=(1,0,0) \in \OIT$.
We have to verify that $(x_1,x_2,y_1,y_2,z_1,z_2)$ satisfies $\OIT_I$ in $\bB$.
Since $h(x_1,x_2)=1$, we have $x_1 = 0$, and similarly we get
that $y_1,z_1 > 0$. We can then set
$u$ to $0$ and have $T_3(u,x_1,y_1)$ since $0=u=x_1<y_1$,
and we also have $T_3(0,u,z_1)$ since $0=u<z_1$. 
The case that $t=(0,1,0)$ is analogous. Suppose now that 
$t=(0,0,1) \in \OIT$.
Then $x_1,y_1 > 0$, and $z_1=0$. We can then set $u$ to
$\min(x_1,y_1)$, and therefore have $T_3(u,x_1,y_1)$,
and $T_3(0,u,z_1)$ since $0=z_1<u$.
Conversely, suppose that $(x_1,x_2,y_1,y_2,z_1,z_2)$ satisfies $\OIT_I$ in $\bB$. Since $T_3(0,u,z_1)$, 
exactly one out of 
$u,z_1$ equals $0$. When $u=0$, then because of
$T_3(u,x_1,y_1)$ exactly one out of $x_1,y_1$ equals $0$,
and we get that $(h(x_1,x_2),h(y_1,y_2),h(z_1,z_2)) \in \{(0,1,0),(1,0,0)\} \subseteq \OIT$. When $u>0$, then $x_1>0$ and $y_1>0$,
and so $(h(x_1,x_2),h(y_1,y_2),h(z_1,z_2)) = (0,0,1) \in \OIT$.

Since the orbit of $0$ is primitive positive definable, 
NP-hardness of $\Csp(({\mathbb Q};T_3))$ follows from the NP-hardness of
$\Csp((\{0,1\};\OIT))$ via Proposition~\ref{prop:pp-interpret-with-constants}.
\end{proof}

We will see in Theorem~\ref{thm:pre-main} 
that if no relation among $\Betw$, $\Cycl$, $\Sep$, $T_3$, $\Neg T_3$,
or $E_6$ is primitive positive definable in a temporal
constraint language $\bB$, then $\Csp(\bB)$ is tractable.
In fact, when $\bB$ is $({\mathbb Q}; R)$ for one of the relations $R$ above, 
then we give primitive positive interpretations of
$(\{0,1\}; \OIT)$ 
with finitely many constants in $\bB$. 
Thus, hardness of temporal CSPs can always be shown with 
Proposition~\ref{prop:pp-interpret-with-constants}.
We have already seen this for $\Betw$, $T_3$ (and thus $\Neg T_3$), and $E_6$,
and close by showing it for $\Cycl$ and $\Sep$. 
We thank Trung Van Pham for pointing
out a simpler proof for $\Cycl$ than our original
proof which was inspired by the 
NP-hardness proof of~\cite{GalilMegiddo} 
for the `Cyclic ordering problem' (see~\cite{GareyJohnson}).

\begin{theorem}\label{thm:cycl-hard}
The structure $(\{0,1\}; <,\Betw)$ has a primitive positive 
interpretation in $({\mathbb Q}; \Cycl,0,1)$. 
The structure $(\{0,1\}; \OIT)$ has a
primitive positive interpretation with parameters
in $(\mQ; \Cycl)$, and $\Csp((\mQ; \Cycl))$ is NP-hard.
\end{theorem}
\begin{proof}
Our interpretation of $(\mQ;<,\Betw)$
in $(\mQ; \Cycl,0,1)$ is $1$-dimensional. 
The domain formula $\delta(x)$ is 
$\Cycl(0,x,1)$, and defines the open interval $(-1,1) \subseteq \mQ$. 
The coordinate map $c$ is any isomorphism
between $(\mQ;<)$ and the substructure induced by these numbers.
The interpreting relation for $x<y$ is
$\Cycl(0,x,y)$. 
It is easy to verify that
the relation $\Cycl$ is not preserved by any of the
relations listed in Lemma~\ref{lem:lcsp-lex-or-pp}. 
Hence, $\Betw$ has a primitive positive definition in $({\mathbb Q};\Cycl,<)$, which is the interpreting formula for $\Betw$ in the interpretation. 

Since $(\mQ;\Betw)$ can in turn interpret primitively positively $(\{0,1\};\OIT)$ with parameters by Proposition~\ref{prop:betw-hard}, the desired interpretation can be obtained 
by composing interpretations (see Section~\ref{ssect:transfer}). 
We can then apply Proposition~\ref{prop:pp-interpret-with-constants}, 
and the NP-hardness of $\Csp((\mQ;\Cycl))$ 
follows from the NP-hardness of $(\{0,1\};\OIT)$. 
\end{proof}

Another relation that appeared
in Theorem~\ref{thm:order-reducts} is the separation relation $\Sep$. The corresponding
CSP is again NP-complete.

\begin{proposition}\label{prop:lcsp-sep-hard}
There is a primitive positive interpretation 
of $(\mQ; \Betw)$ in 
$(\mQ;\Sep,-1,1,2)$, 
and therefore also a primitive positive
interpretation of $(\{0,1\};\OIT)$ in $(\mQ; \Sep)$
with parameters.   
The problem $\Csp((\mQ; \Sep))$ is NP-hard.
\end{proposition}
\begin{proof}
Our interpretation of $(\mQ; \Betw)$
in $(\mQ; \Sep,-1,1,2)$ is $1$-dimensional. 
The domain formula $\delta(x)$ is 
$\Sep(-1,1,x,2)$, and defines the open interval $(-1,1) \subseteq \mQ$.
The coordinate map $c$ is any isomorphism
between $(\mQ;<)$ and the substructure induced by these numbers. 
Then the formula $\Sep(x,z,y,1)$ interprets $\Betw$: $x$ and $y$ must satisfy $\delta$,
and so $-1<x,y<1$. Therefore, 
\begin{align*}
\Betw(c(x),c(y),c(z)) & \Leftrightarrow -1<x<y<z<1 \text{ or } -1<z<y<x<1 \\
& \Leftrightarrow  \Sep(x,z,y,1) 
\end{align*}

A primitive positive interpretation of $(\{0,1\};\OIT)$ can be obtained as follows. 
The argument above shows that the structure $(\mQ; \Sep)$ can interpret primitively positively $(\mathbb Q; \Betw,0)$ with parameters, 
which in turn can interpret primitively positively $(\{0,1\};\OIT)$ by Proposition~\ref{prop:betw-hard}. 
Then the desired interpretation can be obtained 
by composing interpretations (see Section~\ref{ssect:transfer}). 
We can then apply Proposition~\ref{prop:pp-interpret-with-constants}, 
and the NP-hardness of $\Csp((\mQ;\Sep))$ 
follows from the NP-hardness of $(\{0,1\};\OIT)$. 
\end{proof}

\section{Endomorphisms}
\label{sect:tcsp-endos}
In this section we study the endomorphisms of temporal constraint languages. As an application, we obtain a reduction of the complexity classification 
for temporal constraint satisfaction problems to the classification
for those languages that admit a primitive positive definition of the binary relation $<$.

\begin{theorem}\label{thm:t-endo}
Let $\bB$ be a temporal constraint language. 
Then exactly one of the following cases applies.
\begin{enumerate}
\item $\bB$ has a constant endomorphism;
\item All endomorphisms of $\bB$ preserve $<$;
\item $\End(\bB)$ equals the set of unary operations generated by $\Neg$;
\item $\End(\bB)$ equals the set of unary operations generated by $\Cyc$;
\item $\End(\bB)$ equals the set of unary operations 
generated by $\Neg$ and $\Cyc$;
\item $\End(\bB)$ equals the set of all injective unary operations.
\end{enumerate}
\end{theorem}
\begin{proof}
First note that all the cases are indeed disjoint:
a constant endomorphism violates $<$, and cannot be generated
by a set of injective unary operations; this shows that the first case is distinct from all others. %The second case is distinct from all others because $\Neg$ and $\Cyc$ do not preserve $<$. 
Disjointness of the remaining cases follows from Theorem~\ref{thm:cameron}.
%The third, fourth, and fifth case are distinct because neither $\Neg$ generates $\Cyc$ nor $\Cyc$ generates $\Neg$. Finally, the last case is distinct because
%$\Neg$ and $\Cyc$ together preserve the relation \Sep and hence

If $\bB$ has a non-injective endomorphism, then
Corollary~\ref{cor:q-endo-viol} shows that there is also a constant
endomorphism.
Otherwise all endomorphisms of $\bB$ are injective.
We show that then all endomorphisms $e$ of $\bB$ are 
locally invertible: for 
any $a_1,\dots,a_l \in \mathbb Q$ there exists a self-embedding 
$f$ of $\bB$ into $\bB$ such that $f(e(a_i)) = a_i$ for all $i \in \{1,\dots,l\}$.
Because $e$ is injective,
there is an $\alpha \in \AQ$ 
such that $\alpha e(\{a_1,\dots,a_l\}) = \{a_1,\dots,a_l\}$. 
Then $(\alpha e)^{l!}$, i.e., the composition of $(\alpha e) \dots (\alpha e) $ with $l$-factorial many terms of the form $(\alpha e)$,
maps $a_i$ to itself for all $1 \leq i \leq l$.
Then $(\alpha e)^{l!-1} \alpha$ is also an endomorphism of $\bB$,
and we have $\big ((\alpha e)^{l!-1} \alpha\big ) (e(a_1),\dots,e(a_l)) = (\alpha e)^{l!}(a_1,\dots,a_l) = (a_1,\dots,a_l)$.
This proves that $e$ is locally invertible.

Theorem~\ref{thm:mc-omegacat}
shows that the endomorphisms of $\bB$ are generated by the automorphisms of $\bB$. The claim of the statement follows directly from Theorem~\ref{thm:cameron}.
\end{proof}

The following theorem shows that we can focus on constraint languages
where $<$ is primitive positive definable.

\begin{theorem}\label{thm:cores}
Let $\bB$ be a  temporal constraint language. Then it satisfies at least one of the following:
\begin{enumerate}
\item[(a)] There is a primitive positive definition of $\Cycl$, $\Betw$, or $\Sep$ in $\bB$.
\item[(b)] $\pol(\bB)$ contains a constant operation.
\item[(c)] ${\it Aut}(\bB)$ contains all permutations of $\mathbb{Q}$.
\item[(d)] There is a primitive positive definition of $<$ in $\bB$.
\end{enumerate}
\end{theorem}
\begin{proof}
If there is a pp definition of $\Betw$ in $\bB$ we are in case (a). Otherwise, since $\Betw$ consists of two orbits of triples of the automorphism group of $({\mathbb Q}; <)$, Lemma~\ref{lem:small-arity} shows that $\bB$ has a binary polymorphism that violates $\Betw$. 
If there is a pp definition of $<$ in $\bB$, we are in case (d). Otherwise, again by Lemma~\ref{lem:small-arity}, there is a unary polymorphism of $\bB$ that violates $<$. Proposition~\ref{thm:t-endo} shows
that $\bB$ is preserved by a constant, $-$, or $\cyc$.
For each of these three operations we show 
the claim of the statement separately in the following three paragraphs. 

If $\bB$ is preserved by a constant we are in case (b), 
so we assume in the following 
that $\bB$ is not preserved by a constant.

If $\bB$ is preserved by $-$, the relation $\Betw$ consists of only one orbit of triples, and Lemma~\ref{lem:small-arity}
shows that there is an endomorphism
that violates $\Betw$. Proposition~\ref{thm:t-endo} then implies
that $\bB$ is also preserved by $\cyc$.
Thus, the relation $\Sep$ consists of only one orbit of $4$-tuples.
Again, either $\Sep$ has a pp definition, and we are in case (a), or there is an endomorphism that violates
$\Sep$. Proposition~\ref{thm:t-endo} now shows that 
$\bB$ is preserved by all injective unary operations and we are in case (c).

If $\bB$ is preserved by $\cyc$, then 
the relation $\Cycl$ consists of only one orbit of triples.
If $\Cycl$ has a pp definition in $\bB$, we are in case (a). Otherwise, 
Lemma~\ref{lem:small-arity}
shows that there is an endomorphism
that violates $\Cycl$. Proposition~\ref{thm:t-endo} then shows
that $\bB$ is also preserved by $-$. But the statement of the lemma has already been shown in the case that
$\bB$ is preserved by both $-$ and $\cyc$ in the previous paragraph, so we are done.
\end{proof}

In case (a), there is a finite signature reduct $\bB'$ of $\bB$ such that $\Csp(\bB')$ is NP-hard, as we have seen in Section~\ref{ssect:tcsp-hard}. 
In case (b), for all finite signature reducts $\bB'$ of $\bB$ the problem $\Csp(\bB)$
is trivially in P (see Proposition~\ref{prop:const-core}). In case (c) the complexity of $\Csp(\bB)$ has been classified in Chapter~\ref{chap:ecsp}.
In the following, we therefore study only those temporal constraint languages where $<$
is pp definable.

\section{Lex-closed Constraints}
\label{sec:lcsp-classif}
\begin{figure}
\begin{center}
\includegraphics[scale=0.5]{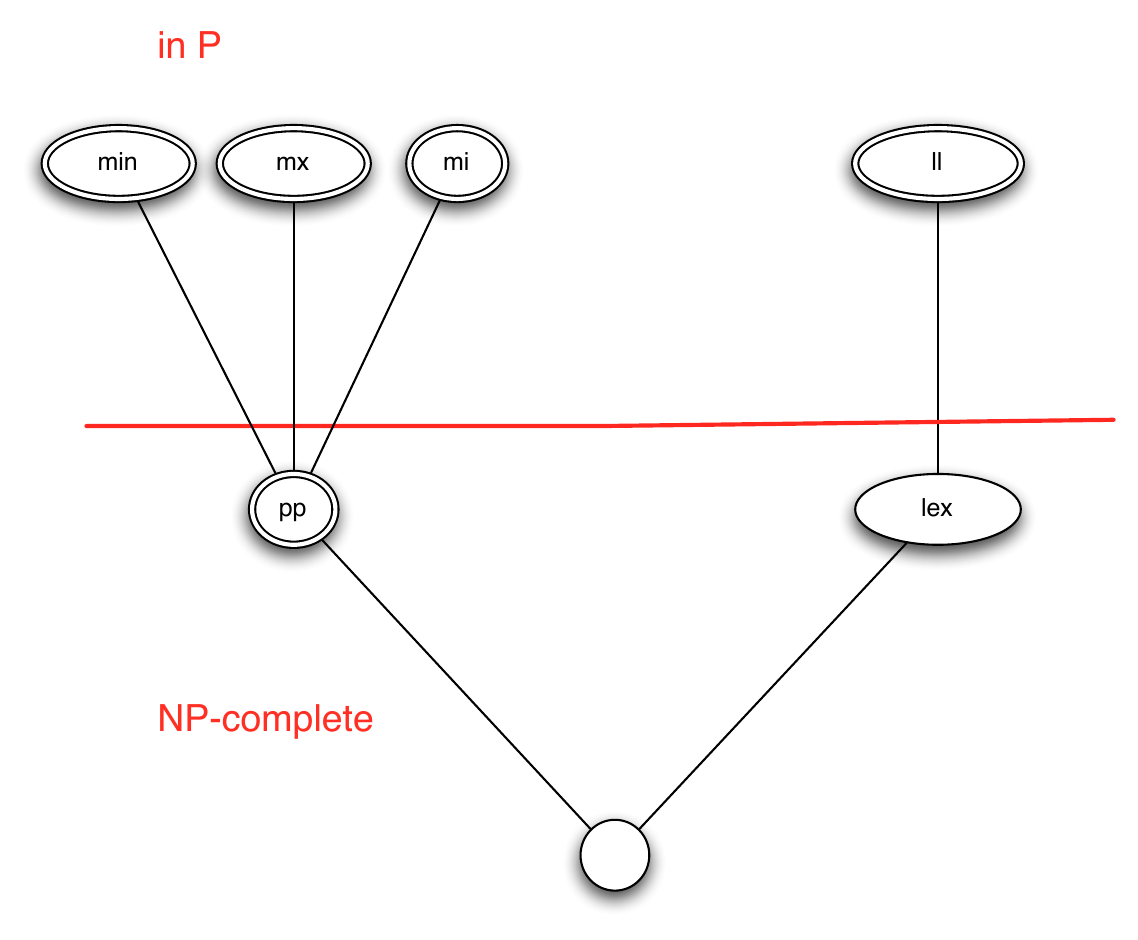}
\caption{An illustration of the classification result for temporal constraint languages that contain $<$. 
Double-circles mean that the corresponding operation has a dual generating a distinct clone which is not drawn in the figure.}
\label{fig:simple-main}
\end{center}
\end{figure}

First-order expansions of $({\mathbb Q};<)$ can be divided into four (non-disjoint) groups: those where the betweenness relation is primitive positive definable, 
those that are preserved by an operation called \pp, an operation called \dpp, or by the 
binary injective operation called \lex\ that we have already encountered in Section~\ref{ssect:canonical-behavior}.
None of the three polymorphisms \pp, \dpp, and \lex\ alone guarantees tractability of the CSP. An illustration of the complexity classification result for first-order expansions of $({\mathbb Q};<)$ can be found in Figure~\ref{fig:simple-main}.

\subsection{The operations \lex\ and \lele}
\label{ssect:ll}
An important class of  temporal constraint languages
are the languages preserved by the operation $\lex$, 
introduced in Section~\ref{ssect:canonical-behavior}.
Recall that \lex\ is a binary injective operation on $\mathbb{Q}$ such that
$\lex(a,b) < \lex(a',b')$ if either $a < a'$, or $a=a'$ and $b<b'$.
By Observation~\ref{obs:f-gen-g-order}, all such operations generate the same clone. 
We also write 
\begin{itemize}
\item $\lex_{y,x}$ for the operation $(x,y) \mapsto \lex(y,x)$, 
\item $\lex_{y,-x}$ for the operation $(x,y) \mapsto \lex(y,-x)$, 
\item $\lex_{x,-y}$ for the operation $(x,y) \mapsto \lex(x,-y)$,   
\item $\lex_{x,y}$ for the operation $(x,y) \mapsto \lex(x,y)$,
\item $p_x$ for the operation $(x,y) \mapsto x$, and
\item $p_y$ for the operation $(x,y) \mapsto y$.
\end{itemize}

\begin{figure}[h]
\begin{center}
\includegraphics[scale=.5]{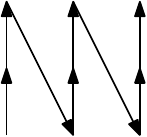} \quad \quad
\includegraphics[scale=.5]{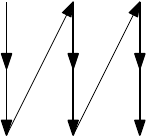} \quad \quad
\includegraphics[scale=.5]{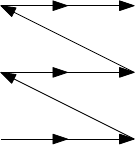}\quad \quad
\includegraphics[scale=.5]{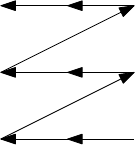}\quad \quad
\includegraphics[scale=.5]{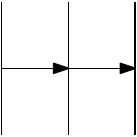}\quad \quad
\includegraphics[scale=.5]{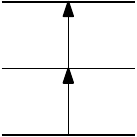}
\caption{Illustrations of the six basic operations $\lex_{x,y}$, 
$\lex_{x,-y}$, $\lex_{y,x}$, $\lex_{y,-x}$,  $p_x$, $p_y$.}
\label{fig:six-types}
\end{center}
\end{figure}

A $k$-ary operation $f \colon \mathbb Q^k \rightarrow \mathbb Q$
is \emph{dominated} by the $i$-th argument when for all $\bar a, \bar b \in \mQ^k$
it holds that
$f(a_1,\dots,a_k) \leq f(b_1,\dots,b_k)$ if and only if $a_i \leq b_i$.
Examples of operations dominated by the first argument
are $p_x$, $\lex_{x,y}$, and $\lex_{x,-y}$, and examples of operations dominated by the second argument are $p_y$, $\lex_{y,x}$, $\lex_{y,-x}$.

It is easy to see that the relation $\btw$ is preserved by $\lex$, and more generally by all operations that are dominated by one argument.
Therefore, we are interested in further restrictions
of languages preserved by $\lex$ that imply tractability of the 
corresponding CSP. 

A large tractable temporal constraint language has been introduced 
in~\cite{ll}. The language is defined in terms of a binary polymorphism, denoted by $\lele$,
and it has a dual version, which is tractable as well. 
We will see in Proposition~\ref{prop:contains} that this language contains the class of Ord-Horn constraints (Section~\ref{ssect:ord-horn}).

\begin{definition}
Let $\lele \colon \mQ^2 \to \mQ$ be
such that $\lele(a,b) < \lele(a',b')$ if 
\begin{itemize}
\item $a \leq 0$ and $a<a'$, or 
\item $a \leq 0$ and $a=a'$ and $b<b'$, or
\item $a,a' > 0$ and $b<b'$, or 
\item $a > 0$ and $b=b'$ and $a<a'$.
\end{itemize} 
\end{definition}

\begin{figure}[h]
\begin{center}
\includegraphics[scale=0.8]{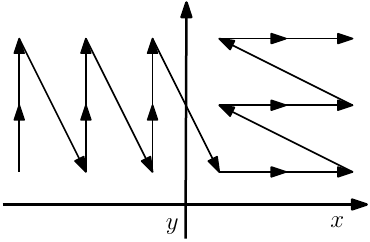}\hskip1cm\includegraphics[scale=0.8]{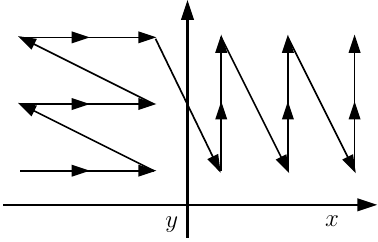}
\caption{A visualization of \lele\ (left) and \dll\ (right).}
\label{fig:ll_dualll}
\end{center}
\end{figure}

All operations satisfying these conditions are by
definition injective, and they all generate the same clone. 
For an illustration of \lele\ and its dual, see Figure~\ref{fig:ll_dualll}.
It is easy to see that
\lele\ generates \lex. 

All Ord-Horn
relations (Section~\ref{ssect:ord-horn}) are preserved by ll. 

\begin{proposition}\label{prop:contains}
All relations in Ord-Horn are preserved by ll and dual ll.
\end{proposition}
\begin{proof}
We give the argument for ll only; the argument for
dual ll is analogous.
It suffices to show that every relation 
that can be defined by a formula $\phi$ 
of the form $(x_1=y_1 \wedge \dots \wedge x_{k-1}=y_{k-1}) \rightarrow x_k \; O \; y_k$ is preserved by ll, where $O \in \{=,<,\leq,\neq\}$.
Let $t_1$ and $t_2$ be two $2k$-tuples that satisfy
$\phi$. Consider a $2k$-tuple $t_3$ obtained by applying
ll componentwise to $t_1$ and $t_2$. Suppose first that 
there is an $i \leq k-1$ such that one of the tuples 
does not satisfy $x_i = y_i$. Then $x_i = y_i$ is not satisfied
in $t_3$ as well, by injectivity of ll, 
and therefore the tuple $t_3$ satisfies $\phi$.
Now consider the case that 
$x_i=y_i$ holds for all $i\leq k-1$ in both tuples $t_1$ and $t_2$. 
Since $t_1$ and $t_2$ satisfy $\phi$, the literal $x_k O y_k$ holds 
in both $t_1$ and $t_2$. Because ll preserves all relations in $\{=,<,\leq,\neq\}$, 
the literal $x_k O y_k$ holds in $t_3$, and
therefore $t_3$ satisfies $\phi$ as well.
\end{proof}

Since the relation $R^{min}$ defined by $(x > y) \vee (x > z)$ (see Section~\ref{ssect:and-or}) is preserved by ll but not by dual ll, the class of ll-closed constraints is \emph{strictly} larger than Ord-Horn. 

\subsection{Operations generating \lele, \dll, or \lex}
\label{sec:lcsp-lex-classif}
In this section we present operations that generate \lele, \dll, or \lex.
We again use the concept of a \emph{behavior} of operations over a relational structure;
note that a $k$-ary operation
$f$ behaves like a $k$-ary operation $g$ on $S = S_1 \times \cdots \times S_k$
if for all $t,t' \in S$ we have $f(t) \leq f(t')$ iff $g(t) \leq g(t')$. That is, the weak linear order induced by $f$ on the tuples from $G$ (in the sense of Observation~\ref{obs:f-gen-g-order})
is the same as the weak linear order induced on these tuples by $g$.
Let $\mQ^+$ denote the set of all positive rational numbers, and let $\mQ^-_0$ denote $\mQ \setminus \mQ^+$.

\begin{definition}\label{def:merge}
Let $f,g$ be from $\mQ^2 \rightarrow \mathbb Q$. 
Then $[f|g]$ denotes an arbitrary operation  from $\mathbb Q^2 \rightarrow \mathbb Q$ with the following
properties. For all $x,x',y,y' \in \mathbb Q$,
\begin{itemize} 
\item if $x \leq 0$ and $x' > 0$ then $[f|g](x,y) < [f|g](x',y')$;
\item %if $x,x' \leq 0$ then $[f|g](x,y) \leq [f|g](x',y')$ if and only if $f(x,y)<f(x',y')$;
$[f|g]$ behaves like $f$ on $\mathbb Q^-_0 \times \mathbb Q$;
\item %if $x,x' > 0$ then $[f|g](x,y) \leq [f|g](x',y')$ if and only if
%$g(x,y) \leq f(x',y')$.
$[f|g]$ behaves like $g$ on $\mathbb Q^+ \times \mathbb Q$;
\end{itemize}
\end{definition}

For example, if $f=lex_{x,y}$ and $g=\lex_{y,x}$, then 
$[f|g]$ behaves like $ll$.

\begin{lemma}\label{lem:genl-l}
Let $f,g \in \{\lex_{x,y}, \lex_{x,-y}, \lex_{y,x},\lex_{y,-x},p_x,p_y\}$, 
and let $f'$ ($g'$) be $\lex_{x,y}$ if $f$ ($g$) is dominated by the first argument, and $\lex_{y,x}$ otherwise.
Then $\{\lex,[f|g]\}$ generates $[f'|g'](x,y)$. 
\end{lemma}

\begin{proof}
By Proposition~\ref{prop:pol-inv} it suffices to show that every relation $R$ preserved by $\lex$ and $[f|g]$
is preserved by $[f'|g']$. So let
$R$ be an arbitrary relation preserved by $\lex$ and $[f|g]$, let $k$ denote its arity, and let $t_1,t_2$ be $k$-tuples from $R$. 
We have to show that $t_3:=[f'|g'](t_1,t_2)$ is in $R$.

Let $\alpha \in \AQ$ be such that  for each entry $x$ of $t_1$ and for each entry $y$ of $t_2$, the value of $\alpha \lex(x,y)$ is negative 
when $x \leq 0$, and positive otherwise. We will show that there is
an automorphism of $(\mathbb Q;<)$ that maps the tuple
$$s := [f|g](\alpha \lex(t_1,t_2),\lex(t_2,t_1))$$ to $t_3$, 
which proves that $t_3$ is in $R$.
It suffices to show for $j_1,j_2 \in [k]$ that
\begin{align}
s[j_1] \leq s[j_2] \; \text{ if and only ifÊ} \; t_3[j_1] \leq t_3[j_2] \; . \label{eq:flip}
\end{align}

We can assume that
$t_1[j_1] \leq t_1[j_2]$ by exchanging the name of $j_1$ and $j_2$ if
necessary, and distinguish three cases:
\begin{itemize}
\item $t_1[j_1] \leq 0$, $t_1[j_2] > 0$. Then $t_3[j_1] < t_3[j_2]$ by definition of $[f'|g']$. 
Since for $j \in [k]$, the value of $\alpha \lex(t_1[j],t_2[j])$
is positive if and only if the value of $t_1[j]$ is positive, 
we have $s[j_1] < s[j_2]$ by definition of $[f|g]$. Thus we have verified (\ref{eq:flip}) in this case.
\item $t_1[j_2] \leq 0$.  %Suppose that $f$ is dominated by the first argument; the case that $f$ is dominated by the second argument
%is analogous. 
%Note that then $f(\lex(x,y),y)$ behaves like $\lex(x,y)$.
Note that $f(\lex(x,y),\lex(y,x))$ behaves like $f'(x,y)$.
%$\lex(x,y)$ when $f$
%is dominated by the first argument, and like $\lex(y,x)$, when 
Thus, writing $a[j]$ for $\lex(t_1[j],t_2[j])$ and $b[j]$ for $\lex(t_2[j],t_1[j])$, we have the following equivalences.
\begin{align*}
t_3[j_1] \leq t_3[j_2] \quad \text{iff} & \quad f'(t_1[j_1],t_2[j_1]) \leq f'(t_1[j_2],t_2[j_2]) \\
%\; \text{iff} & \quad \lex(t_1[j_1],t_2[j_1]) \leq  \lex(t_1[j_2],t_2[j_2])\\
%\; \text{iff} & \quad \lex(a[j_1],b[j_1]) \leq  \lex(a[j_2], b[j_2])\\
\; \text{iff} & \quad f(a[j_1],b[j_1]) \leq  f(a[j_2], b[j_2])\\
\; \text{iff} & \quad f(\alpha a[j_1],b[j_1]) \leq  f(\alpha a[j_2], b[j_2])\\
\; \text{iff} & \quad s[j_1] \leq s[j_2]
\end{align*}
\item $t_1[j_1] > 0$. This case is analogous to the previous one and left to the reader.
\end{itemize}
\end{proof}

\begin{lemma}\label{lem:gen}
For $f,g \in \{p_y,\lex_{y,x}\}$ the operation $[f|g]$ generates
$[\lex_{x,y}|g]$.
\end{lemma}

\begin{figure}
\begin{center}
\includegraphics[scale=.9]{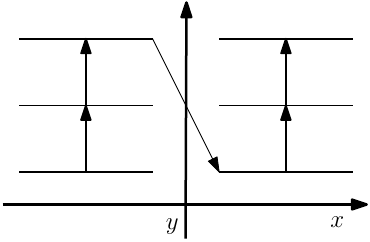}
\includegraphics[scale=.9]{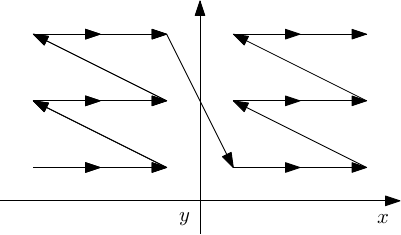}
\caption{An illustration of the operation $[p_y|p_y]$ (on the left) and the operation $[\lex_{y,x}|\lex_{y,x}]$ (on the right). }
\label{fig:wl-wl}
\end{center}
\end{figure}

In particular, for $f=g=lex_{y,x}$ the lemma shows
that $[f|g]$ generates ll. For
$f=g=p_y$, the lemma shows that $[f|g]$ generates $[\lex_{x,y}|p_y]$ and in particular $\lex_{x,y}$. See Figure~\ref{fig:wl-wl} for illustrations of those two cases.

%\begin{lemma}\label{lem:gen-lex}
%The operation $[p_y|p_y]$ generates $\lex$.
%\end{lemma}

\begin{proof}[Proof of Lemma~\ref{lem:gen}]
We show that every relation $R$ preserved by $[f|g]$ 
is preserved by $[\lex_{x,y}|g]$, and conclude 
by Proposition~\ref{prop:pol-inv} that $[f|g]$ generates $[\lex_{x,y}|g]$. So let
$R$ be an arbitrary relation preserved by $[f|g]$, let $k$ denote its arity, and let $t_1,t_2$ be $k$-tuples from $R$. 
We have to show that $t_3:=[\lex_{x,y}|g](t_1,t_2)$ is in $R$.

Let $l$ denote the number of non-positive values in $t_1$.
We take $\alpha_1,\dots,\alpha_l$ from $\AQ$ such that $\alpha_i$ maps all but the $i$ smallest values in $t_1$ to positive values. 
We define a sequence of tuples $s_1, \dots, s_l$ as follows: $s_1=t_2$, and
for $i \geq 2$
\begin{align*}
s_i & := [f|g](\alpha_i t_1, s_{i-1})  \; .
\end{align*}

Clearly, for all $i\in[l]$ the tuple $s_i$ is in $R$. 
We will show that there is an automorphism of $(\mathbb Q;<)$ that maps $s_l$ to $t_3$,
which proves that $t_3$ is also in $R$.
By symmetry it is enough to
show for $j_1,j_2\in[k]$ with $t_1[j_1]\leq t_1[j_2]$ that
\begin{align}
s_l[j_1]\leq s_l[j_2] \text{ if and only if } t_3[j_1]\leq t_3[j_2]\; .  
\label{eq:grid-calc}
\end{align}
We distinguish three cases:
\begin{itemize}
%\item $t_1[j_1] = t_1[j_2]$ - this is the same as the case 2 of Lemma 44.
\item $t_1[j_1]=t_1[j_2] \leq 0$. Since $\alpha_i t_1[j_1]=\alpha_i t_1[j_2]$ for all $i \in [l]$, we have $s_l[j_1] \leq s_l[j_2]$ if and only if $s_1[j_1]\leq s_1[j_2]$. Since $s_1=t_2$ and $t_1[j_1] \leq 0$, and because $f$ is dominated by the second argument, 
%$s_1[j_1]\leq s_1[j_2]$ if and only if 
$s_1[j_1] \leq s_1[j_2]$ if and only if $t_3[j_1]\leq t_3[j_2]$, which proves (\ref{eq:grid-calc}).
%\item $t_1[j_1]\leq 0$ - this is the same as the case 1 of Lemma 44.
\item $t_1[j_1] < t_1[j_2]$, $t_1[j_1] \leq 0$.
Let $i\in[l]$ be such that
$\alpha_i t_1[j_1] \leq 0$ and $\alpha_i t_1[j_2] > 0$.
By definition of $[f|g]$ we see that $s_i[j_1]<s_i[j_2]$. Because 
$\alpha_i t_1[j_1]<\alpha_i t_1[j_2]$ for all $i \in [l]$, and because
$[f|g]$ preserves $<$, by induction on $i' \geq i$ we have 
that $s_{i'}[j_1] < s_{i'}[j_2]$. In particular, $s_l[j_1] < s_l[j_2]$.
On the other hand, $t_3[j_1]<t_3[j_2]$ 
by definition of $\lex_{x,y}$ and $[\lex_{x,y}|g]$, 
and so (\ref{eq:grid-calc}) also holds in this case.
\item $t_1[j_1]>0$. Observe that by the choice of $l$ we have $\alpha_i t_1[j_1] > 0$ for all $i \in [l]$. Thus (\ref{eq:grid-calc}) holds, because
both $[f|g]$ and $[\lex_{x,y}|g]$ behave like $g$ on $\mathbb Q^+\times \mathbb Q$.
\end{itemize}
\end{proof}

\subsection{A syntactic description of ll-closed constraints}
\label{ssect:ll-syntax}
In this section we present a syntactic characterisation of ll-closed
relations. As a consequence, we also obtain a
better understanding of the clone generated
by $\lele$. 

\begin{definition}\label{def:ll-horn}
A formula is called \emph{ll-Horn} if it is a conjunction of formulas
of the following form %(slightly abusing terminology, we call 
%these formulas the \emph{clauses} of the ll-Horn formula)
\begin{align*}
(x_1 = y_1 \wedge \dots \wedge x_k = y_k) & \Rightarrow (z_1 < z_0\vee \dots \vee z_l < z_0)
\; \hbox{, or}\\
(x_1 = y_1 \wedge \dots \wedge x_k = y_k) & \Rightarrow (z_1 < z_0 \vee \dots \vee z_l < z_0 \vee (z_0 = z_1 = \dots = z_l)) 
\end{align*}
where $0 \leq k,l$.
\end{definition}

Note that $k$ or $l$ might be $0$: if $k=0$, we obtain
a formula of the form $z_1 < z_0 \vee \dots \vee z_l < z_0$ or $(z_1 < z_0 \vee \dots \vee z_l < z_0 \vee (z_0 = z_1 = \dots = z_l))$,
and if $l=0$ we obtain a disjunction of disequalities.
Also note
that the variables $x_1,\dots,x_k,y_1,\dots,y_k,z_0,$
$\dots,z_l$ need not be pairwise distinct. 
On the other hand, the clause $z_1 < z_2 \vee z_3 < z_4$ is an example of a formula that is \emph{not} ll-Horn.

The following result is from~\cite{ll}, but Antoine
Mottet found a mistake in the proof presented there; 
the new proof
presented below is also due to him. 

\begin{proposition}\label{prop:ll-horn}
A temporal relation is preserved by $\lele$ if and only if it can be defined by an ll-Horn formula.
\end{proposition}
\begin{proof}
The proof that every relation defined by an ll-Horn formula is ll-closed
is similar to the proof of Proposition~\ref{prop:contains}.
We just need to additionally check that the relation 
defined by 
$z_1 < z_0 \vee \dots \vee z_l < z_0$
and the relation defined by $z_1 < z_0 \vee \dots \vee z_l < z_0 \vee 
(z_0 = \dots = z_l)$ are preserved by ll.
So let $s$ and $t$ be two assignments
that satisfy $\phi := z_1 < z_0 \vee \dots \vee z_l < z_0$, and let $r := \lele(s,t)$. 
Let $i \in \{1,\dots,l\}$ be such that
$s(z_i) = \min(s(z_1),\dots,s(z_l))$.
Note that $s(z_i)<s(z_0)$. 
Let $j \in \{1,\dots,l\}$ be such that
$t(z_j) < t(z_0)$. 
\begin{itemize}
\item If $s(z_i) \leq 0$ then $\lele(s(z_i),t(z_i)) < \lele(s(z_0),s(z_0))$ since $s(z_i) < s(z_0)$,
and hence $r$ satisfies
$\phi$. 
\item If $s(z_i) > 0$, then $s(z_0) > s(z_i) > 0$
and $s(z_j) > s(z_i) > 0$, 
and hence $\lele(s(z_j),t(z_j)) < \lele(s(z_0),s(z_0))$ since $t(z_j) < t(z_0)$,
and hence $r$ satisfies $\phi$.
\end{itemize}
When $t_1$ and $t_2$ are satisfying
assignments of $z_1<z_0 \vee \dots \vee z_l <z_0 \vee (z_0 = \dots = z_l)$ where one of
the assignments satisfies the last clause,
then the statement follows 
from the fact that $\lele$ is injective and preserves $\leq$.

Let $R$ be a temporal relation, and
let $\phi$ be a quantifier-free formula in CNF that
defines $R$ over $({\mathbb Q};<)$. 
In this formula, we replace literals of the form
$\neg (y < x)$ by $x < y \vee x = y$, and we use
$x \leq y$ as shortcut for those two literals. For reasons that will
become clear later, we additionally allow that clauses contain `clustered equations' which are
expressions of the form $x_1 = x_2 = \cdots = x_n$
and which stand for $x_1 = x_2 \wedge \cdots \wedge x_1 = x_n$; such an expression will be treated as one literal. 

We describe four  rewriting rules that yield a formula $\psi$ that also defines $R$ over $({\mathbb Q};<)$ 
such that $R$ is preserved by $\lele$ if and only if $\psi$ is ll-Horn. 
\begin{enumerate}
% do we really need this one?
\item \label{rule:x} If $\phi$ implies $x=y$ for distinct variables $x,y$ of $\phi$, replace all occurrences of $u$ in $\phi$
by $x$ and add the clause $x=y$. 
\item \label{rule:a} Suppose that $\phi$ 
contains a clause $\theta$ of the form 
$$ x < y \vee u < v \vee \theta' \, ,$$ let $\phi'$ be the other clauses of $\phi$, and suppose that
\begin{align*}
(\phi' \wedge \neg \theta' \wedge x < y) & \text{ implies } 
(u \leq v \vee x \leq v) \\
 \text{ and } (\phi' \wedge \neg \theta' \wedge u < v) & 
 \text{ implies }
(x \leq y \vee u \leq y) \; .
\end{align*}
Then replace $\theta$ by 
\begin{align*}
 & (u \leq v \vee x \leq v \vee \theta') \wedge (u \neq v \vee x < y \vee \theta')  \\
\wedge \; & (x \leq y \vee u \leq y \vee \theta') \wedge (x \neq y \vee u < v \vee \theta') \, .
\end{align*}
\item \label{rule:b} Suppose that $\phi$ contains a clause $\theta$
of the form $$x < y \vee u < v \vee \theta' \, .$$ 
Let $\phi'$ be the other clauses of $\phi$, and suppose that 
$$(\phi' \wedge \neg \theta' \wedge x < y) \text{ implies \ } u \leq v \, .$$ 
Then replace $\theta$ by 
\begin{align*}
 (u \leq v \vee \theta') 
\wedge (x < y \vee u \neq v \vee  \theta').
\end{align*}
\item \label{rule:c} Suppose that $\theta$ is a clause of $\phi$ of the form $$x_1 \neq y_1 \vee \cdots \vee x_k \neq y_k \vee z_1 < z_0 \vee \cdots \vee z_l < z_0 \vee u = v \, ,$$ 
let $\phi'$ be the other clauses of $\phi$, 
%and that $\phi \wedge u=v$ implies that $z_0 = z_1 = \cdots = z_l$. 
and that 
\begin{align*}
& \phi' \wedge x_1 = y_1 \wedge \cdots \wedge x_k = y_k \wedge z_0 \leq z_1 \wedge \cdots \wedge z_0 \leq z_l \wedge u = v 
\end{align*}
implies that $z_0 = z_1 = \cdots = z_l$.
Then replace $\theta$ by 
\begin{align*}
& (x_1 \neq y_1 \vee \cdots \vee x_k \neq y_k \vee z_0 \neq z_1 \vee \cdots \vee z_0 \neq z_l \vee u = v) \\
\wedge \; & (x_1 \neq y_1 \vee \cdots \vee x_k \neq y_k \vee z_1 < z_0 \vee \cdots \vee z_l < z_0 \vee z_0 = z_1 = \cdots = z_l) \; . 
\end{align*}
%\item If $\phi$ contains a clause $\theta$
%and the remaining clauses imply $\theta$, remove $\theta$
\item \label{rule:d}
If $\phi$ contains a literal such that removing this
literal from $\phi$ results in an equivalent formula, then
remove the literal. 
\end{enumerate}
We claim that for each of the four rewriting rules, the resulting formula
$\psi$ is equivalent to $\phi$. This is obvious for rules~(\ref{rule:x}) and (\ref{rule:d}). To see that $\phi$ implies 
the new clauses in rule (\ref{rule:a}), let $s$ be a satisfying
assignment to $\phi$. If $s$ satisfies
$\theta'$, then $s$ also satisfies the new clauses, 
so let us assume that $\theta'$ is false. Then
$s$ satisfies $x<y$ or $u<v$. The two cases are
symmetric, so we only treat the case that $s$ satisfies
$x<y$ in the following. By assumption,
$s$ must then satisfy $u \leq v \vee y \leq v$,
and hence the first new clause is satisfied by $s$.
Since $x<y$, the other new clauses are satisfied, too. 

Now suppose conversely that $s$ is a solution
to $\phi'$ and the four new clauses, and suppose
for contradiction
that $\theta$ does not hold. Because of the second and fourth new clause, we then must have $u \neq v$ and $x \neq y$. Then the first new clause implies that $x \leq v$
and the third new clause implies that $u \leq y$. 
But then $x \leq v \leq u \leq y \leq x$, a contradiction
to $x \neq y$. 

For rule~(\ref{rule:b}), let $s$ be a solution to $\phi$.
Then $s$ obviously satisfies the first
new clause if $u < v$ or $\theta'$ holds; otherwise, 
$s$ must satisfy $x < y$ because of $\theta$. But then  $u \geq v$ by assumption and hence the first new clause also holds in this case. The second new clause
is weaker then $\theta$, so it is also satisfied by $s$.
Now suppose conversely that $s$ satisfies $\phi'$ and the two new clauses, and suppose for contradiction that $\theta$ does not hold. Then in particular $v \leq u$ holds and the first new clause implies that $u = v$, and hence $x < y$ because of the second new clause, contradiction to the assumption that $\theta$ does not hold. 

Finally, for the fourth rule, the first new clause
is a weakening of $\theta$, and the second new clause is a consequence of $\phi$ by assumption. 
Conversely, suppose that $s$ satisfies all clauses
of $\phi$ except for $\theta$ which is not satisfied. 
Then the first new clause implies that $z_1 < z_1 \vee \cdots \vee z_l < z_0$, and thus the second new clause implies that $u=v$, and hence $\theta$ holds, contradiction. Hence, $\psi$ is indeed equivalent to $\phi$. 

%A \emph{bad pair} is a pair of literals of the same clause of one of the following two forms:
%\begin{itemize}
%\item the first literal is of the form 
%$x>y$ and the second of the form $u>v$ where $x$ and $u$ are distinct variables;
%\item the first literal is of the form $x>y$ and the second of the form $u=v$. 
%\end{itemize}

Note that rules~(\ref{rule:a}) and~(\ref{rule:b}) strictly reduce the number of pairs of literals $x<y$ and $u<v$ in the same clause where $y$ and $v$ are distinct variables. 
Rule~(\ref{rule:c}) leaves this number invariant, 
but strictly reduces the number of literals of the form $u = v$
or of the form $u < v$ in the clause (here, we do not count complex equations). Rules~(\ref{rule:x})
and~(\ref{rule:d}) do not increase these numbers,
and strictly reduce the number of variables that occur more than once, or strictly reduce the total number
of literals. Hence, when we repeatedly 
apply these rules, the procedure will eventually terminate. 

{\bf Claim 1.} The formula 
$\psi$ cannot contain a clause $\theta$ of the form 
$x < y \vee u < v \vee \theta'$ where $x$ and $u$
are distinct variables. Since rule~(\ref{rule:a})  
is not applicable, there must 
exist a solution $s$ to
$\phi' \wedge \neg \theta' \wedge x < y \wedge v < u \wedge v < x$
or to
$\phi' \wedge \neg \theta' \wedge u < v \wedge y < x \wedge y < u$. 
Suppose the former is the case, since
the latter case can be treated similarly. 
Since rule~(\ref{rule:b}) is not applicable, there exists a solution $t$ 
to $\phi' \wedge \neg \theta' \wedge u<v \wedge y < x$. Let $\alpha \in \Aut(\mQ;<)$ be such that $\alpha s(v) = 0$. We claim that
$r = \lele(\alpha s,t)$ does not satisfy 
$\theta$: 
\begin{itemize}
\item we have $r(y) < r(x)$ since $0<s(x),s(y)$ and $t(y) < t(x)$; 
\item we have $r(v) < r(u)$ since $s(v)=0$
and $s(u) > 0$; 
\item finally, $r$ does not satisfy $\theta'$
since neither $s$ nor $t$ satisfy $\theta'$. 
\end{itemize}
Hence, $r$ does not satisfy $\psi$,
in contradiction to the assumption that 
$\lele$ preserves $R$. 

{\bf Claim 2.} The formula $\psi$ cannot contain
a clause with two distinct literals $x = y$ and
$u = v$. This is because
rule~(\ref{rule:d}) and since $\phi$ is preserved
by the injective function $\lele$. 

{\bf Claim 3.} If $\psi$ contains a clause
with a literal 
$z_1 < z_0$ and a literal $u = v$, 
then $\{u,v\} = \{x,y\}$. This is because of
Claim 1 and Claim 2, any such clause must be of the form $x_1 \neq y_1 \vee \cdots \vee x_k \neq y_k \vee z_1 < z_0 \vee \cdots \vee z_l < z_0 \vee u = v$. Since rule~(\ref{rule:c}) does
not apply, there exists a solution $s$ to
\begin{align*}
& \phi' \wedge x_1 = y_1 \wedge \cdots \wedge x_k = y_k \wedge z_0 \leq z_1 \wedge \cdots \wedge z_0 \leq z_l \wedge u = v \\
\wedge \; &  z_0 \neq z_1 \vee \cdots \vee z_0 \neq z_l \, .
\end{align*}
Hence, there exists an $i \in \{1,\dots,l\}$
such that $s(z_0) \neq s(z_i)$. 
Because the literal $z_i < z_0$ cannot
be removed from $\psi$ with rule~(\ref{rule:d}),
there exists a solution $t$ to $\phi$
such that $z_i < z_0$ is the only
literal in $\theta$ satisfied by $t$. 
Let $\alpha \in \Aut({\mathbb Q};<)$
be such that $\alpha t(z_i) = 0$.
Then $r := \lele(\alpha t,s)$ does not satisfy 
$\theta$: 
\begin{itemize}
\item $r$ satisfies $\theta' \wedge x_1 = y_1 \wedge \cdots \wedge x_k = y_k$ since
both $t$ and $s$ satisfy this formula. 
\item $r(z_i) < r(z_0)$ since $0 = \alpha t(z_i) < \alpha t(z_0)$. 
\item $r(z_j) \leq r(z_0)$ for all $i \in \{1,\dots,k\} \setminus \{i\}$ since 
$t(z_0) \leq t(z_j)$ and $s(z_0) \leq s(z_j)$.  
\item $r(u) \neq r(v)$ since $t(u) \neq t(v)$
and $\lele$ is injective. 
\end{itemize}
The three claims imply that each of the  clauses 
of $\psi$ must be logically equivalent 
to an implication as in Definition~\ref{def:ll-horn}, and this concludes the proof. 
\end{proof}
 
% commented out: these equations are not stable under
% expansion with constants

\subsection{Weak Commutativity}
%For a uniform presentation of the tractability results in Section~\ref{sect:tcsp-classification}, 
In this section we present a different description of ll-closed temporal constraint languages. Let $c_1,\dots,c_n \in \mQ$ be arbitrary. 
Partition the rationals $\mQ = Q_1 \uplus Q_1$ such that each of $Q_1$ and $Q_2$ is dense in ${\mathbb Q}$, and let $f$ be any injective binary operation that preserves $<$ and $\leq$ such that
\begin{itemize}
\item $f(c_i,c_i) = c_i$ for all $i \leq n$;
\item for $x \in Q_1$, we have that $f(x,y_1) > f(y_2,x)$ for all $y_1,y_2 > x$;
\item for $x \in Q_2$, we have that $f(x,y_1) < f(y_2,x)$ for all $y_1,y_2 > x$. 
\end{itemize}
We even assume that $f$ is bijective:
since the image of any such function will induce a dense
linear order without endpoints in $(\mQ;<)$,
the existence of such a function follows from
$\omega$-categoricity of $(\mQ;<)$.

\begin{figure}
\begin{center}
\includegraphics[scale=.6]{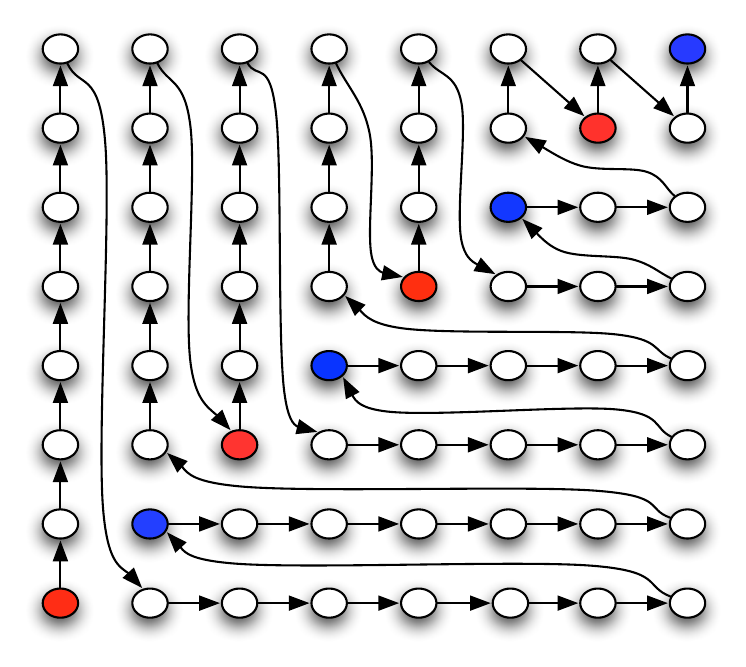}
\caption{Illustration of an operation $f$ generated by $\lele$ that
satisfies $f(x,y)=\alpha f(\beta(x),\beta(y))$.}
\label{fig:good-ll}
\end{center}
\end{figure}

For an illustration of $f$, see Figure~\ref{fig:good-ll}.
The red vertices are the elements of $\{(x,x) \; | \; x \in Q_1\}$, 
and the blue blue vertices the elements of $\{(x,x) \; | \; x\in Q_2\}$.
Observe that $f(x,y)<f(z,z)$ for all $x,y,z$
such that $x<z$ or $y<z$. 

\begin{proposition}\label{prop:good-ll}
There are automorphisms $\alpha,\beta$ of $(\mQ;<,c_1,\dots,c_n)$ such that $f(x,y) = \alpha f( \beta y, \beta x)$ holds 
for all $x,y \in \mQ$.
\end{proposition}
\begin{proof}
Let $\beta$ be an automorphism of $(\mQ;<,c_1,\dots,c_n)$ 
that maps $Q_1 \setminus \{c_1,\dots,c_n\}$ to $Q_2  \setminus \{c_1,\dots,c_n\}$ and $Q_2 \setminus \{c_1,\dots,c_n\}$ to $Q_1  \setminus \{c_1,\dots,c_n\}$; such an automorphism can easily be constructed by going back-and-forth.  
To define $\alpha \in \Aut((\mQ;<,c_1,\dots,c_n))$, 
let $d \in \mQ$
be arbitrary. Since $f$ and $\beta$ are bijective,
$f' \colon (x,y) \mapsto f(\beta y, \beta x)$ is bijective as well,
so there exists
a unique pair $(a,b) \in \mQ^2$ such that
$f'(a,b)=d$. Define $\alpha(d) = f(a,b)$. 
Then by definition $\alpha f( \beta y, \beta x) = f(x,y)$ holds for all $x,y \in \mQ$, and it is straightforward to verify that
$\alpha \in \Aut((\mQ;<,c_1,\dots,c_n))$. 
\ignore{ (we omit this horrible calculation, it will 
be of no help to the reader):
To see this, let $p_1,q_1,p_2,q_2 \in \mQ$ be arbitrary. 
If $p_1 \leq p_2$ and $q_1 \leq q_2$, or if
$p_1 \geq p_2$ and $q_1 \geq q_2$, 
then $f(p_1,p_2)<f(q_1,q_2)$ and $f'(p_1,p_2)<f'(q_1,q_2)$. 
since
both $f$ and $f'$ are injective and preserve $\leq$,
so there is nothing to show.
Otherwise, either $p_1 < p_2$ and $q_1 > q_2$,
or $p_1 > p_2$ and $q_1 < q_2$. 
Again, the situation
is symmetric and we assume in the following that 
$p_1 < q_1$ and $p_2 < q_2$.
If $p_1 < q_1$ and $p_2 < q_2$ then
again $f(p_1,p_2)<f(q_1,q_2)$ and $f'(p_1,p_2)<f'(q_1,q_2)$.
Similarly, we are done if $p_1 > q_1$ and $p_2 > q_2$. 
So either $p_1 < q_1$ and $p_2 \geq q_2$,
or $p_1 \geq q_1$ and $p_2 < q_2$. 
Assume that $p_1 < q_1$ and $p_2 \geq q_2$. 
If $p_1 < q_2$ then $f(p_1,q_1) < f(q_2,q_2) 
\leq f(p_2,q_2)$, and similarly $f'(p_1,q_1) < f'(p_2,q_2)$. 
If $p_1 > q_2$ then $f(p_2,q_2) < f(p_1,q_1)$ and
$f'(p_2,q_2) < f'(p_1,q_1)$.
The remaining case is that $p_1 = q_2 =: x$. 
Since $x < q_1,p_2$, if $x \in Q_1$ then $f(x,q_1) > f(p_2,x)$.
So we have $f(p_1,q_1) > f(p_2,q_2)$,
and have to show that $f'(p_1,q_1) > f'(p_2,q_2)$.
Since $b(x) \in Q_2$, and $b(x) < b(q_1),b(p_2)$
we have
$f(b(x),b(p_2)) > f(b(q_1),b(x))$,
and therefore also $a(f(b(q_2),b(p_2)) > a(f(b(q_1),b(p_{i+q})))$, 
which is what we had to show. 
The situation that $x \in Q_2$ is analogous. 
}
\end{proof}

\begin{proposition}
The operation $f$ defined above is generated by $\lele$, and generates $\lele$.
\end{proposition}
\begin{proof}
It is easy to see that $f$ interpolates $\lele$.
For the converse, 
it suffices to verify that $f$ preserves all ll-Horn formulas, by Proposition~\ref{prop:ll-horn}. 
Since $f$ is injective, it suffices
to show that $f$ preserves formulas of the form
$$(z_0 > z_1) \vee \cdots \vee (z_0 > z_l)$$
and formulas of the form
$$(z_0 > z_1) \vee \cdots \vee (z_0 > z_l) \vee (z_0 = z_1 = \cdots = z_l) \; .$$
Preservation of formulas of the latter type reduces to the
former type, since $f$ preserves $\leq$ and is injective binary.
Now suppose that $\bar a = (a_0,a_1,\dots,a_l)$ and $\bar b = (b_0,b_1,\dots,b_l)$ 
are two tuples that satisfy $z_1 < z_0 \vee \cdots \vee z_l < z_0$.
Assume that $a_0 \leq b_0$.
Let $i$ be such that $a_i = \min(a_1,\dots,a_l)$. Then $a_i < a_0$, and we get that $f(a_0,b_0) \geq f(a_0,a_0) >  f(a_i,b_i)$ by the properties of $f$. Therefore,
$\min(f(a_1,b_1),\dots,f(a_l,b_l)) < f(a_0,b_0)$.
We can argue analogously in the case that $a_0 \geq b_0$. 
\end{proof}

\subsection{Weak near-unanimity modulo endomorphisms}
For a uniform presentation of the classification result in Section~\ref{sect:tcsp-classification}, we need yet another description of 
the clone generated by $\lele$.

%\begin{proposition}\label{prop:ll-wnu-def}
%There exists an injective function $f \colon {\mQ}^3 \to \mQ$ such that 
%$f(x,y,z) \leq f(x',y',z')$ if and only if
%$(\min(x,y,z),\max(\min(x,y),\min(x,z),\min(y,z)),x,y,z)$ is lexicographically smaller
%than 
%$(\min(x',y',z'),\max(\min(x',y'),\min(x',z'),\min(y',z')),x',y',z')$.
%\end{itemize}
%\end{proposition}
%\begin{proof}
%The given conditions on $f$
%clearly define a linear order on ${\mathbb Q}^3$, and we know that all countable linear
%orders embed into $({\mathbb Q};<)$. 
%\end{proof}

We write $\lex(x_1,\dots,x_n)$ as a shortcut
for $\lex(x_1,\lex(x_2,\dots \lex(x_{n-1},x_n) \dots ))$. 

\begin{proposition}\label{prop:ll-wnu}
%Let $f \colon {\mathbb Q}^3 \to {\mathbb Q}$
%be any function with the properties from
%Proposition~\ref{prop:ll-wnu-def}. 
There are $a,b,c \in \End({\mathbb Q};<)$
such that the ternary function $f \colon {\mathbb Q}^3 \to {\mathbb Q}$ defined by 
$$ f(x,y,z) = \lex(\min(x,y,z),\max(\min(x,y),\min(x,z),\min(y,z)),x,y,z)$$
satisfies for all $x,y \in \mQ$ 
\begin{align*}
a(f(x,x,y)) = b(f(x,y,x)) = c(f(y,x,x)) \, .
\end{align*}
That is, $f$ is a weak near unanimity 
modulo endomorphisms of $({\mathbb Q};<)$. 
\end{proposition}
\begin{proof}
By Lemma~\ref{lem:lift}, it suffices to show
that for every finite $S \subset \mQ$
there are $\alpha, \beta \in \Aut(\mQ;<)$ such that for all $x,y \in S$
\begin{align*}
f(y,x,x) = \alpha_1 f(x,y,x) = \alpha_2 f(x,x,y)  \, .
\end{align*}
By the properties of $f$ we have that $f(y,x,x) \leq f(y',x',x')$
if and only if one of the following holds:
\begin{itemize}
\item $\min(x,y)<\min(x',y')$;
\item $\min(x,y) = \min(x',y')$ and $x<x'$;
\item $\min(x,y) = \min(x',y')$, $x=x'$, and $y<y'$;
\item $x=x'$ and $y=y'$. 
\end{itemize}
Note that this is the case if and only if
$f(x,y,x) < f(x',y',x')$, and if and only if
$f(x,x,y) < f(x',x',y')$. Hence, the
existence of $\alpha_1$ and $\alpha_2$ follows from the homogeneity of $({\mathbb Q};<)$. 
\end{proof}

Note that the function $f$ defined in Proposition~\ref{prop:ll-wnu} is injective and 
preserves $\leq$. 

\begin{theorem}\label{thm:ll-wnu}
Let $R \subseteq {\mathbb Q}^n$ be first-order definable over $({\mathbb Q};<)$. 
Then the following are equivalent.
\begin{enumerate}
\item $R$ is preserved by the operation $f$ as defined in Proposition~\ref{prop:ll-wnu} (a weak near unanimity modulo endomorphisms). 
\item $R$ is preserved by $\lele$. 
\item $R$ has an ll-Horn definition. 
\end{enumerate}
\end{theorem}
\begin{proof}
The implication from (1) to (2) follows
from the observation that
$(x,y) \mapsto f(x,x,y)$ interpolates $\lele$.
The implication from (2) to (3) is Lemma~\ref{prop:ll-horn}. 
For the implication from (3) to (1), 
it suffices to verify that $f$ preserves all ll-Horn formulas. Since $f$ is injective, it suffices
to show that $f$ preserves formulas $\phi$ 
of the form
$$(z_1 < z_0) \vee \cdots \vee (z_l < z_0)$$
and of the form
\begin{align*}
(z_1 < z_0) \vee \cdots \vee (z_l < z_0) \vee (z_0 = z_1 = \cdots = z_l) \; .
%\label{eq:second-ll}
\end{align*}
Suppose that $s_1,s_2,s_3$ 
are assignments that 
 satisfy $\phi$; we have to show that the assignment $s$ defined by $s(x) := f(s_1(x),s_2(x),s_3(x))$ satisfies $\phi$. 
Let $j \in \{1,2,3\}$ be such that 
$s_j(z_0) = \min(s_1(z_0),s_2(z_0),s_3(z_0))$. 

 Suppose first that $s_j$ satisfies 
 $(z_1 < z_0 \vee \cdots \vee z_l < z_0)$. 
 %Suppose first that 
%$\phi = 
%all of $s_1,s_2,s_3$ satisfy
%$(z_1 < z_0 \vee \cdots \vee z_l < z_0)$.
Let $i$ be such that $s_j(z_i) = \min(s_j(z_1),\dots,s_j(z_l))$. 
Then $s_j(z_i) < s_j(z_0)$ by assumption, 
and hence 
$$\min(s_1(z_i),s_2(z_i),s_3(z_i)) <
\min(s_1(z_0),s_2(z_0),s_3(z_0)) \, .$$
Therefore, 
$f(s_1(z_i),s_2(z_i),s_3(z_i)) <
f(s_1(z_0),s_2(z_0),s_3(z_0))$ by the properties of $f$, and $s$ satisfies 
$(z_1 < z_0 \vee \cdots \vee z_l < z_0)$.

Otherwise, 
%$\phi$ is of the form
%$(\ref{eq:second-ll})$ and that 
%there is exactly one $j \in \{1,2,3\}$ 
%such that 
$s_j$ must satisfy 
$z_0=z_1=\cdots = z_l$.
%and the other solutions from $\{s_1,s_2,s_3\}$ 
%satisfy $(z_1 < z_0) \vee \cdots \vee (z_l < z_0)$.
Let $a,b$ be such that $a < b$ and 
$\{a,b\} = \{1,2,3\} \setminus \{j\}$. 
%\begin{itemize}
%\item Suppose 
We next consider the case 
that there exists 
$c \in \{a,b\}$ and $p \in \{1,\dots,l\}$ such that 
$s_j(z_0) > s_{c}(z_p)$. 
Let $d \in \{a,b\} \setminus \{c\}$. 
Note that 
\begin{align*}
\min(s_{c}(z_p), s_{d}(z_p)) & \leq
 s_{c}(z_p) < 
s_j(z_p)  = s_j(z_0) 
= \min(s_{c}(z_0), s_{d}(z_0)) \\
\min(s_{j}(z_p),s_{c}(z_p)) & = s_{c}(z_p) < 
s_j(z_p) = s_j(z_0) = \min(s_{c}(z_0), s_{j}(z_0)) \\
\min(s_{j}(z_p), s_{c}(z_p)) & = s_{c}(z_p) < 
s_j(z_p)  = s_j(z_0) = \min(s_{c}(z_0), s_{j}(z_0))
\end{align*}
and hence 
\begin{align*}
& \max(\min(s_{c}(z_p),s_{d}(z_p)),\min(s_{j}(z_p),s_{c}(z_p)),\min(s_{j}(z_p),s_{d}(z_i))) \\
< \; & \max(\min(s_{c}(z_0),s_{d}(z_0)),\min(s_{j}(z_0),s_{c}(z_0)),\min(s_{j}(z_0),s_{d}(z_i))) \, .
\end{align*}
Thus, by the definition of $f$, we have 
$s(z_p) < s(z_0)$ and $s$ satisfies $\phi$. 

Otherwise, 
$s_j(z_0) \leq \min(s_{a}(z_1),\dots,s_{a}(z_l))$ and 
$s_j(z_0) \leq \min(s_{b}(z_1),\dots,s_{b}(z_l))$. For all $i \in \{0,1,\dots,l\}$ we have $s_j(z_i) = s_j(z_0)$ and hence 
\begin{align*}
& \max(\min(s_{a}(z_i),s_{b}(z_i)),\min(s_{j}(z_i),s_{a}(z_i)),\min(s_{j}(z_i),s_{b}(z_i))) \\
= \; & \min(s_{a}(z_i),s_{b}(z_i)) \geq s_{j}(z_i) \, .
\end{align*}
The definition of $f$ then implies that 
$f(s_1(z_i),s_2(z_i)$, $s_3(z_i)) <
f(s_1(z_0),s_2(z_0),s_3(z_0))$ 
if and only if $s_a(z_i) < s_a(z_0)$. 
If there exists an $i \in \{1,\dots,l\}$
such that $s_k(z_i) < s_k(z_0)$, we therefore
have $s(z_i) < s(z_0)$ and 
$s$ satisfies $(z_1 < z_0) \vee \cdots \vee (z_l < z_0)$. 
Otherwise, we must have that
$$s_a(z_0) = s_a(z_1) = \cdots = s_a(z_l)$$
If also $s_b(z_0) = s_b(z_1) = \cdots = s_b(z_l)$
then $s$ satisfies $z_0=z_1=\cdots = z_l$, too. So suppose that there exists a $p \in \{1,\dots,l\}$ such that $s_b(z_p) < s_b(z_0)$. 
Since $s_j(z_p) = s_j(z_0)$ and 
$s_a(z_p) = s_a(z_0)$ we then have $s(z_p) < s(z_0)$ since $s$ is injective
and preserves $\leq$. 
Hence, $s$ satisfies $\phi$ also in this case. 
\end{proof}

\subsection{An Algorithm for ll-closed Constraints}
\label{sect:alg}
\newcommand{\KK}{\cal{K}}
\newcommand{\CC}{\cal{C}}
% Alternative: use the syntax result to set up a retraction algorithm
% in the style of the gcsp paper.
% in any case, rewrite the algorithm box in the following style:
In this section we present an algorithm for ll-closed constraints.  
One of the underlying ideas of the algorithm is to use a subroutine
that tries to find a solution where every variable
has a different value. If this is impossible, the subroutine must
return a set of at least two variables that denote the same
value in all solutions -- since the constraints are preserved by a binary
injective operation, such a set must exist (Proposition~\ref{prop:injective-sols}). 

The $i$-th entry in a $k$-tuple $t$ is called {\em minimal} 
if $t[i]\leq t[j]$ for every $j\in[k]$. 
It is called {\em strictly minimal} if $t[i]<t[j]$ for every $j\in[k]
\setminus\{i\}$.

\begin{definition}\label{def:min-set}
Let $R$ be a $k$-ary temporal relation. A set $S \subseteq [k]$ is
called a \emph{min-set for the $i$-th entry in $R$} if there exists a tuple
$t\in R$ such that the $i$-th entry is minimal in $t$, 
and for all $j \in [k]$ it holds that $j \in S$ if and only if $t[i]=t[j]$. 
We say that $t$ is a \emph{witness} for this min-set.
\end{definition}

Let $R$ be a $k$-ary relation 
that is preserved by \lex\ 
(recall that ll-closed constraints are preserved by \lex\ as well),
and suppose that the $i$-th entry has 
the min-sets $S_1,\ldots,S_l$, for $l \geq 1$, 
with the corresponding witnesses $t_1,\dots,t_l$. Consider the tuple
$t:=\lex(t_1,\lex(t_2,\ldots$ $\lex(t_{l-1},t_l)))$. Since the entry $i$ 
is minimal in every tuple $t_1,\dots,t_l$, and since \lex\ preserves 
both $<$ and $\leq$, it is also minimal in $t$. 
Because \lex\ is injective, we have that $t[i]=t[j]$ 
if and only if these two entries are equal in each tuple $t_1,\dots,t_l$. 
Hence, the min-set for the $i$-th entry in $R$ witnessed by
the tuple $t$ is a subset of every other min-set $S_1,\ldots,S_l$.
We then call this set the {\em minimal min-set} for the $i$-th entry in $R$.

\begin{lemma}
\label{lem:lex-free-set-equal}
Let $R$ be a $k$-ary relation preserved by \lex, 
and let $S$ be the minimal min-set for the $i$-th entry in $R$.
If $t \in R$ is such that $t[j] \geq t[i]$ for every $j \in S$,
then $t[i]=t[j]$ for every $j\in S$.
\end{lemma}
\begin{proof}
Let $t' \in R$ be the tuple that witnesses the minimal min-set $S$. 
Let $t \in R$ be such that not all entries in $S$ are equal
(in particular, $|S| > 1$). 
Consider the tuple $s:=lex(t',t)$. By the properties of \lex\ 
it holds that $s[i] < s[j]$ 
for every $j\in[k]\setminus S$. Furthermore, 
$s[i]\leq s[j]$
for $j\in S$ if and only if $t[i]\leq t[j]$. Thus, unless $s$ witnesses
a smaller min-set for $i$ in $R$ (which would be a contradiction), we have 
that $s[i]>s[j]$ for some $j\in S$.
\end{proof}

To develop our algorithm, we use a specific notion of \emph{constraint graph} 
of a temporal CSP instance, defined as follows.

\begin{definition}
The \emph{constraint graph} $G_\phi$ of a temporal CSP instance $\phi$ is a
directed graph $(V;E)$ defined on the variables $V$ of $\phi$.
For each constraint of the form $R(x_1,\dots,x_k)$ from $\phi$ we add
a directed edge $(x_i,x_j)$ to $E$ if in every tuple from $R$ where the $i$-th entry is minimal, the $j$-th entry is minimal as well.
\end{definition}

%\begin{lemma}
%\label{lem:lex-free-set-equal}
%Let $R$ be a $k$-ary relation preserved by \lex,
%and let $S_i$ be the set of indices such that 
%$G_{\{R(x_1,\dots,x_k)\}}$ has an edge from $x_i$ to $x_j$.
%Then in any tuple $t$ from $R$, either $t[i]=t[j]$ for every $j\in S$,
%or there is a $j\in S$ such that $t[j]<t[i]$.
%\end{lemma}

\begin{definition}
If an instance of a temporal CSP 
contains a constraint $\phi$ imposed on $y$
such that $\phi$ does not admit a solution where $y$ denotes the minimal value,
the we say that $y$ is \emph{blocked (by $\phi$)}.
\end{definition}

We can easily determine for each constraint which variables are blocked by this
constraint: For a constraint represented by weak linear orders we just check all
weak linear orders and build a set of variables that are not minimal in any of them.
%For a constraint represented by an ll-Horn formula, 
%a variable $x_i$ is blocked if and only if the formula is of the form 
%$x_i > z_1 \vee \dots \vee x_i > z_l$.
Thus, by
inspecting all the constraints it is possible to compute the blocked variables
in linear time in the input size. We want to use the constraint graph to
identify variables that have to denote the same value in all solutions, and
therefore introduce the following concepts.

\begin{definition} 
A strongly connected component $K$ of the constraint graph $G_\phi$ for a
temporal CSP instance $\phi$ is called a \emph{sink component} if no edge in
$G_\phi$ leaves $K$, and no variable in $K$ is blocked.  
A vertex of $G$ that
belongs to a sink component of size one is called a \emph{sink}.
\end{definition}

%The following lemma shows an important consequence of \lex-closure of 
%constraints.
\begin{lemma}
\label{lem:component-contract}
Let $\bB$ be a \lex-closed temporal constraint language. 
Let $\phi$ be an instance of $\Csp(\bB)$ with variables $V$, 
and let $K \subseteq V$ be a sink component of the graph $G_\phi$. Then in every solution of $\phi$ all variables from $K$ must have equal values.
\end{lemma}

\begin{proof}
We assume that $\phi$ has a solution $s\colon V \rightarrow \mQ$, and that $K$ has at least two vertices;
otherwise the statement is trivial.
Define $M := \{ x \in K \; | \; s(x) \leq s(y) \text{ for all } y \in K\}$. 
We want to show that $M=K$. 
Otherwise, because $K$ is a strongly connected component, 
there is an edge in $G_\phi$ from
some vertex $u \in M$ to some vertex $v \in K \setminus M$. 
By the definition of $G_\phi$, 
there is a constraint $\psi$ in $\phi$ such that 
whenever $u$ denotes the minimal value 
of a solution of $\psi$, then $v$ has to denote the minimal value as well.
By permuting arguments, we can assume without loss of generality
that $\psi$ is of the form $R(w_1,\dots,w_k)$ where $w_1=u$. % and $w_2=v$.
Because $K$ is a sink component, the variable $u$ cannot be blocked,
and hence there is a minimal min-set $S$ for the first entry in $R$.
%Therefore, $2 \in S$, because $v$ is the second argument of $\phi$.

Note that $G_\phi$ contains an edge from $u$ to $w_i$ for all
$i \in S$. Since $K$ is a strongly connected component, 
all these variables $w_i$ are in $K$. Because
$s(u) \leq s(y)$ for all $y \in K$,
there is no variable $w_i$, $i \in S$, such that $s(w_i) < s(u)$. 
This contradicts Lemma~\ref{lem:lex-free-set-equal},
because $s(u) \neq s(v)$.
\end{proof}

Lemma~\ref{lem:component-contract} immediately implies that we can add
constraints of the type $x=y$ for all variables $x,y$ from the same sink 
component $K$. Equivalently, we can consider the CSP instance where
all the variables in $K$ are \emph{contracted}, i.e., where
all variables from $K$ are replaced by the same variable.
When $\phi = \exists x_1,\dots,x_n \, (\phi_1 \wedge \dots \wedge \phi_m)$ 
is an instance of a $\Csp(\bB)$, and $x_i \in V := \{x_1,\dots,x_n\}$,
then we write $\phi[V \setminus \{x_i\}]$ for the formula $\exists x_1,\dots,x_{i-1},x_{i+1},\dots,x_n \big(  (\exists x_i. \phi_1) \wedge \cdots \wedge (\exists x_i. \phi_m) \big )$.
Note that if $\bB$ contains all primitive positive definable relations
whose arity is bounded by the maximal arity of the relations in $\bB$,
then $\phi[V \setminus \{x_i\}]$ can be viewed as an instance of $\Csp(\bB)$.

\begin{lemma}
\label{lem:injective-extension}
Let $\bB$ be an ll-closed temporal constraint language. 
Let $\phi$ be an instance of $\Csp(\bB)$ with variables $V$, and 
let $x$ be a sink in $G_\phi$. If $\phi[V \setminus \{x\}]$ 
has an injective solution, then $\phi$ has an injective solution as well.
\end{lemma}
\begin{proof}
Let $s\colon V \rightarrow \mQ$ be an injective solution to $\phi[V \setminus \{x\}]$. 
We claim that any extension $r$ of $s$ to $x$
such that $r(x) < s(y)$ for all $y \in V \setminus \{x\}$ is injective and satisfies $\phi$.
If $x$ appears in no constraint in $\phi$, the statement is trivial. 
Consider a constraint $\psi=R(x_1,\dots,x_k)$ from $\phi$ 
that is imposed on $x$, and let $S \subseteq [k]$ be such that $i \in S$ if and only if
$x=x_i$. By the definition of $\phi[V \setminus \{x\}]$, the mapping $s$ has an extension
$s'$ that is also defined on $x$ such that $(s'(x_1),\dots,s'(x_k)) \in R$. 
Because $x$ is a sink, there is tuple
$t \in R$ such that $S$ is the minimal min-set for the $i$-th entry
of $R$ for each $i \in S$. 
Let $t'$ be the tuple $(s'(x_1),\dots,s'(x_k))$,  and 
let $\alpha \in \AQ$ be such that $\alpha s'(x)=0$.
Then $r:=ll(\alpha t', t) \in R$. 
Note that for $i,j \in [k] \setminus S$, we have that $r[i] \leq r[j]$ if and only if $r(x_i) \leq r(x_j)$. Hence, $r$ satisfies all constraints from $\phi$, which is what we had to show. 
\end{proof}

Our algorithm for ll-closed constraints can be found in Figure~\ref{alg:main};
we are now ready to prove its correctness. 

\begin{figure*}[t]
\begin{center}
\small
\fbox{
\begin{tabular}{l}
Spec$(\phi)$ \\
// Input: An instance $\phi$ of $\Csp(\bB)$ with variables $V$. \\
// Output: If algorithm returns \false\ then $\phi$ has no solution. \\
// If $\phi$ has an injective solution, then algorithm returns \true. \\
// Otherwise return $S \subseteq V$, $|S| \geq 2$, such that \\
//  for all $x,y \in S$ we have $x=y$ in all solutions to $\phi$. \\
%Let $G$ be the graph of $\Phi$ \\
Set $X := \emptyset$ \\ % , $\phi' := \phi$ \\ %$G' = G$ \\
While $G_{\phi}$ contains a sink $s$ \\
\hspace{.3cm} $X := X \cup \{s\}$ \\
%\text{projection of } \Phi' \text{ to } V \setminus X$. \\
%\hspace{.3cm} $G' := \text{ReconstructGraph}(\Phi[X \setminus Y])$ \\
\hspace{.3cm} If $X=V$ then return \true \\
\hspace{.3cm} else $\phi := \phi[V \setminus X]$ \\
If $G_{\phi}$ has sink component $S$ return $S$ \\
\hspace{.3cm} else return \false \\
end if
\end{tabular}}
\end{center}
\caption{A polynomial-time algorithm for $\Csp(\bB)$ when $\bB$ is ll-closed: the sub-procedure Spec.}
\end{figure*}

\begin{figure*}[t]
\begin{center}
\small
\fbox{
\begin{tabular}{l}
Solve$(\Phi)$ \\
// Input: An instance $\phi$. \\
// Output: accept if $\phi$ is true, reject otherwise. \\
$S := \text{Spec}(\phi)$ \\
If $S=\false$ then reject \\
else if $S = \true$ then accept \\
else \\
\hspace{.3cm} Let $\phi'$ be contraction of $S$ in $\phi$. \\
\hspace{.3cm} Return Solve$(\phi')$. \\
end if
\end{tabular}}
\end{center}
\caption{A polynomial-time algorithm for $\Csp(\bB)$ when $\bB$ is ll-closed: the main procedure.}
\label{alg:main}
\end{figure*}

\begin{theorem}\label{thm:ll-alg}
The procedure Solve$(\phi)$ in Algorithm~\ref{alg:main} 
decides whether a given set of ll-closed constraints $\phi$ has a solution. There is an implementation of the algorithm that runs 
in time $O(nm)$, where $n$ is the number of variables
of $\phi$ and $m$ is the size of the input.
\end{theorem}

\begin{proof}
The correctness of the procedure Spec immediately implies the correctness 
of the procedure Solve.
In the procedure Spec, after iterated deletion of sinks in $G_\phi$, 
we have to distinguish three cases. 

First, consider the case $V=X$. In this case it follows by a straightforward induction from
Lemma~\ref{lem:injective-extension} that $\phi$ has an injective solution.
Otherwise, consider the case that $G_\phi$ contains a sink component $S$ with $|S| \geq 2$. 
We claim that for all
variables $x,y \in S$ we have $x=y$ in all solutions to $\phi$.
Lemma~\ref{lem:component-contract}
applied to $\phi[V \setminus X]$ 
implies that all variables in the same sink component must have the same value in every
solution, and hence the output is correct in this case as well.

In the third case we have
$X \neq V$ but $G_\phi$ does not contain a sink component.
Note that in every solution to $\phi$ some variable must take
the minimal value. However, since each strongly connected component
without outgoing edges contains a blocked vertex, there is no variable
that can denote the minimal element, and hence $\phi$ has no solution.
Because $\phi$ is at all times of the execution of the algorithm implied
by the original input constraints, the algorithm correctly rejects.

Since in each recursive call of Solve the instance
in the argument has at least one variable less, Solve is executed at most $n$
times. It is not difficult to implement the algorithm such that
the total running time is cubic in the input size.
However, it is possible to implicitly represent the constraint 
graph and to implement all sub-procedures
such that the total running time is in $O(nm)$; for the details, we refer to~\cite{ll}.
\end{proof}
\section{Shuffle-closed Constraints}
\label{sect:shuffle}
An important subclass of temporal constraint languages
are \emph{shuffle-closed} constraint languages. 
As we will see, there are NP-complete shuffle-closed constraint languages.
However, in this section we present three additional restrictions
for shuffle-closed constraint languages that imply that the corresponding CSPs can be
solved in polynomial time. 

\subsection{Shuffle closure}
We define shuffle closure, and show how shuffle closure 
can also be described by 
a certain binary operation on $\mathbb Q$.

\begin{definition}
A $k$-ary relation $R$ is called 
 \emph{shuffle-closed} 
iff for all tuples $t_1,t_2 \in R$ and all indices $l \in [k]$
there is a tuple $t_3 \in R$ such that for all $i,j \in [k]$ we have
$t_3[i] \leq t_3[j]$ iff 
\begin{itemize}
\item $t_1[i] \leq t_1[l]$ and $t_1[i] \leq t_1[j]$, or
\item $t_1[l] < t_1[i] $, $t_1[l] < t_1[j]$, and $t_2[i] \leq t_2[j]$.
\end{itemize}
\end{definition}

\begin{figure}
\begin{center}
\includegraphics[scale=.8]{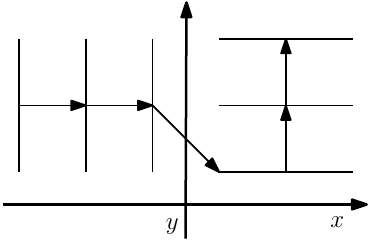}\hskip1cm\includegraphics[scale=.8]{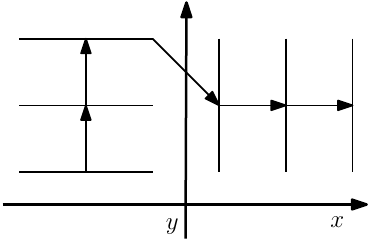}
\caption{A visualization of \pp\ (left) and \dpp\ (right).}
\label{fig:pp_dualpp}
\end{center}
\end{figure}

Let \pp\ be an arbitrary binary operation on $\mathbb{Q}$ such that $\pp(a,b)\leq
\pp(a',b')$ iff one of the following cases applies:
\begin{itemize}
\item $a\leq 0$ and $a \leq a'$
\item $0 < a$, $0 < a'$, and $b\leq b'$.
\end{itemize}
Clearly, such an operation exists. 
For an illustration, see the left diagram in Figure~\ref{fig:pp_dualpp}.
In diagrams for binary operations $f$ as in Figure~\ref{fig:pp_dualpp},
we draw a directed edge from $(a,b)$ to $(a',b')$ if $f(a,b) < f(a',b')$. 
Unoriented lines in rows and columns 
of the picture for an operation $f$ relate pairs of values that get the same value under $f$.
The right diagram of Figure~\ref{fig:pp_dualpp} is  an illustration of the \dpp\ operation. The name of the operation $\pp$ is derived from the word \emph{`projection-projection'},
since the operation behaves as a projection to the first argument
for negative first argument, and a projection to the second argument for positive first argument.

%The following is straightforward.
\begin{proposition}\label{prop:shuffle}
A temporal relation is shuffle-closed if and only if it is preserved by \pp.
\end{proposition}
\begin{proof} 
Let $R$ be a shuffle-closed relation, and let $t_1$ and $t_2$ be
tuples from $R$. We want to show that $t_3 = \pp(t_1,t_2) \in R$.
If $t_1$ only contains positive values, then there clearly
exists an $\alpha \in \AQ$ such that $t_3 = \alpha t_2$,
and since $R$ is preserved
by the automorphisms of $(\mathbb Q;<)$, we are done.
Otherwise, let $l \in[k]$ be an index such that $t_1[l]$ is the largest 
entry in $t_1$ that is not positive.
Because $R$ is shuffle-closed, we know that
there exists a tuple $t_3' \in R$ such that $t_3'[i] \leq t_3'[j]$
iff ($t_1[i] \leq t_1[l]$ and $t_1[i] \leq t_1[j]$) 
or ($t_1[l] < t_1[i]$, $t_1[l] < t_1[j]$, and $t_2[i] \leq t_2[j]$)
for all $i,j \in [k]$.
By the definition of \pp, and the choice of $l$,
the tuple $t_3$ satisfies the same property, and therefore
there exists $\beta \in \AQ$
such that $t_3=\beta t_3'$, and hence $t_3 \in R$.

For the opposite direction, we assume that $R$ is preserved
by $\pp$, and have to show shuffle closure of $R$.
Let $t_1,t_2$ be tuples in $R$, and let $l \in [k]$.
Choose $\gamma \in \AQ$ such that $\gamma$
maps $t_1[l]$ to $0$. Then $t_3=\pp(\gamma t_1,t_2)$
is a tuple that satisfies the conditions specified in the definition 
of shuffle-closure.
\end{proof}

Due to Proposition~\ref{prop:shuffle}, we 
use 
the phrase `$\bB$ is shuffle-closed' interchangeably with 
`$\bB$ is preserved by \pp' .
The following lemma states an important property of shuffle-closed
languages that will be used several times in the next subsections.
%sometimes
%say that $\Gamma$ is shuffle-closed, and sometimes

\begin{lemma}
\label{lem:pp-combination}
Let $t_1,\dots,t_l$ be tuples from a $k$-ary shuffle-closed relation $R$, and let $M_1,\ldots,M_l\subset[k]$ be disjoint sets of indices such that
$\bigcup_{i=1}^l M_i = [k]$
and such that 
for all $i,j \in [l]$ with $i<j$ and for all $i' \in M_i$, $j' \in M_j$ it
holds that $t_i[i']<t_i[j']$. Then there is a tuple $t\in R$ such
that 
\begin{itemize}
\item $t[i']<t[j']$ for all $i,j\in[l]$ with $i<j$ and for all $i'\in M_i, j'\in M_j$;
\item $t[i']\leq t[i'']$ iff $t_i[i']\leq t_i[i'']$ for all $i \in [l]$ and all $i',i''\in M_i$.
\end{itemize}
\end{lemma}
\begin{proof}
Let $\beta_1,\dots,\beta_{l-1} \in \AQ$ be such that
$\beta_i$ maps $max\{t_i[i'] | i'\in M_i\}$ to $0$. We set
$$t:=\pp(\beta_1 t_1,\pp(\beta_2 t_2,\dots,\pp(\beta_{l-1} t_{l-1},t_l)\dots)) \; .$$
The tuple $t$ clearly belongs to $R$.

We prove by induction on $l$ that $t$ satisfies the other conditions of the
lemma.  Observe that $\beta_1$ maps all the entries of $t_1$ at $M_1$ to
non-positive values. Thus for $l=2$, it is easy to check from the properties of
\pp\ that for each $i\in M_1$ and $i'\in M_2$ we have $t[i]<t[i']$ as required by the
statement of the lemma. Also the second condition is immediate. For
$l>2$ let $t'$ be defined by
$$t':=\pp(\beta_2 t_2,\pp(\beta_3 t_3,\dots,\pp(\beta_{l-1} t_{l-1},t_l)\dots)) \; .$$
Then we have $t=\pp(\beta_1 t_1,t')$. Now we apply the same argument as for $l=2$.
Because the order on $[k]\setminus M_1$ is preserved by the application
of \pp, we know that the conditions are satisfied for the sets $M_2,\dots,M_l$.
The argument also shows that  the entries at $M_1$ are smaller than the entries
at $[k]\setminus M_1$ and that their order is the same as in $t_1$.
\end{proof}

The following lemma is a simple criterion for showing that certain operations generate $\pp$.

\begin{lemma}\label{lem:gen-pp}
Let $f$ be a binary operation preserving $<$ such that for some $\alpha,\beta \in \Aut((\mathbb Q; <))$ we have
$f(x,y)=\alpha x$ for all $x\leq -1$, $0<y<1$, and $f(x,y)=\beta y$ for all $x>1$,
$0<y<1$. Then $f$ generates $\pp$.
\end{lemma}
\begin{proof}
It suffices to show that every relation preserved by $f$
is also preserved by $\pp$. Let $R$ be preserved by $f$, and let $t_1,t_2$ be two tuples from $R$.
%Let $m$ be the maximal value from all components of $t_2$,
Let $\gamma_1 \in \AQ$ be such that $\gamma x=x+1$ for all positive entries $x$ of $t_2$
and $\gamma_1 x=x-1$ for all other entries $x$ of $t_2$. Let $\gamma_2 \in \AQ$ be such 
that all entries of $\gamma_2 t_2$ are larger than $0$ and smaller than $1$. Then $f(\gamma_1 t_1,\gamma_2 t_2)$ is in the same orbit as $\pp(t_1,t_2)$, which is what we wanted to show.
\end{proof}

It is easy to verify that the relation $T_3$, defined in Section~\ref{ssect:tcsp-hard},
%of this chapter,
is shuffle-closed. Proposition~\ref{prop:Shard} shows that CSP$(({\mathbb Q}; S))$ 
is NP-complete, and thus the property of shuffle-closure
is not strong enough to guarantee tractability.

\subsection{Min-union closure}
\label{ssect:min}
\begin{figure}
\begin{center}
\includegraphics{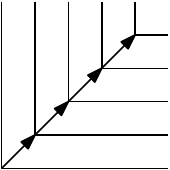}
\caption{Illustration of the operation \min.}
\label{fig:min}
\end{center}
\end{figure}

This section introduces and studies a stronger property than shuffle-closure, namely
preservation under the binary operation $\min$ 
that maps two values $x$ and $y$ to the smaller of the two values; see Figure~\ref{fig:min} for an illustration
of the operation \min.
We also present a sufficient condition that implies that a temporal
constraint language is preserved by $\min$.

For constraint languages over a finite domain, \min-
and \max-closed relations were studied in~\cite{Ordered}.
An equivalent clausal description
of such constraints is known; however, the equivalence only holds
for \emph{finite} domains. The tractability
of the CSP where the constraint language has such a clausal description has also been shown for infinite domains~\cite{CJJK}. But the algorithm presented in~\cite{CJJK} cannot be applied to all $\min$-closed constraint languages over 
an infinite domain; it is already not clear how to adapt this approach
to deal with the relation $\{(x,y,z) \; | \; x>y \vee x >z \}$, 
which is preserved by $\min$.
In Section~\ref{ssect:alg-min} we describe an algorithm 
that efficiently 
solves the CSP for temporal constraint languages that are preserved by $\min$. 

% BOOK-TD: give pointer to paper with Johan

\begin{definition}
Let $t$ be from $\mathbb Q^k$. 
The set of indices $$\{i \in [k] \; | \; t[i]\leq t[j] \text{ for all } j\in[k]\}$$
is called the \emph{min-set of $t$}, and denoted by $M(t)$.
\end{definition}

\begin{definition}\label{def:min-union}
A relation is called \emph{min-union closed}
if for all tuples $t_1$, $t_2$ in $R$
there exists a tuple $t_3$ in $R$ such that $M(t_3) = M(t_1) \cup M(t_2)$.
\end{definition}

We now want to link min-union closure of the relations
in the constraint language to the existence of certain polymorphisms.
%Min-union closure of all relations in a temporal constraint language is 
%only a necessary condition for preservation under $\min$.

\begin{definition}\label{def:prov-min-union}
Let $f$ be a binary operation preserving $<$. We say that $f$ {\em provides min-union closure} if $f(0,0)=f(0,x)=f(x,0)$ for all integers $x>0$. 
\end{definition}

The operation \min\ is an example of an operation providing min-union closure. The following lemma connects Definition~\ref{def:min-union} 
and Definition~\ref{def:prov-min-union}.

\begin{lemma}\label{lem:mu-mu}
Let $R$ be a temporal relation preserved by 
an operation $f$ providing min-union closure. 
Then $R$ is min-union closed.
\end{lemma}

\begin{proof}
Let $t_1$ and $t_2$ be tuples in $R$,
and let $a_1$ and $a_2$ be the minimal values
among the entries of $t_1$ and $t_2$, respectively.
Then there are $\alpha_1, \alpha_2 \in \AQ$ such that
$\alpha_1 a_1=\alpha_2 a_2=0$, and such that $\alpha_1$ and $\alpha_2$ map all other entries of $t_1$ and $t_2$ to integers.
Observe that all entries at $M(t_1) \cup M(t_2)$ in the tuple
 $t_3 = f(\alpha_1 t_1,\alpha_2 t_2)$ have the same value. 
Because $f$ preserves $<$,
  this value is strictly smaller than the values at all other entries in $t_3$.
Hence, $M(t_3) = M(t_1) \cup M(t_2)$.\end{proof}

The following proposition implies that $\{f,\pp\}$
generates $\min$ for every
operation $f$ that provides min-union closure.

\begin{proposition}\label{prop:min}
A temporal relation $R$ is preserved by \pp\ and an operation
providing min-union closure if and only if $R$ is preserved by
\min.
\end{proposition}
\begin{proof}
Clearly, $\min$ provides min-union closure. Also observe that $\min$ satisfies the conditions of Lemma~\ref{lem:gen-pp}, and hence generates $\pp$.

For the opposite direction, suppose that $R$ is $k$-ary
and preserved by \pp\ and an operation $f$ providing min-union
closure. We show that for any two tuples $t_1,t_2\in R$ the tuple
$t_3=\min(t_1,t_2)$ is in $R$ as well.  Let $l$ be the number of distinct
values in $t_3$ and $v_1<v_2<\dots<v_l$ be these values.  We define $M_i$,
$i \in [l]$, to be the set of indices of $t_3$ with the $i$-th lowest value,
i.e., $M_i=\{j \in [k] \; | \; t_3[j] = v_i\}$.

Now let $\alpha_1,\dots,\alpha_l \in \AQ$ be such that
$\alpha_i v_i=0$ and such that the entries of $\alpha_i t_1$ and $\alpha_i t_2$ are integers. Using these automorphisms we define the tuples
$s_1,\dots,s_l$ by $s_i=f(\alpha_i t_1,\alpha_i t_2)$. Clearly, these tuples
belong to $R$. It also holds that $s_i$ is constant at $M_i$ because for each
$j\in M_i$ at least one of the entries $t_1[j], t_2[j]$ is equal to $v_i$
(the other one can be only greater) which
is subsequently mapped to 0 by $\alpha_i$ and $f$ maps all such pairs to the
same value.  Furthermore, for each $j'\in M_{i'}$ for $i<i'\leq l$
we have that $s_i[j']$ is greater than the value of $s_i$ at $M_i$, because
$\min(t_1[j'],t_2[j'])=v_{i'}$ is greater than $v_i$ and $f$ preserves $<$.

Now we can apply Lemma~\ref{lem:pp-combination} to the obtained tuples
$s_1,\dots,s_l$ and the corresponding sets $M_1,\ldots,M_l$. The lemma gives us
some tuple $t'_3$ from $R$ which is constant at each set $M_i$, 
$i\leq [l]$, and
such that for each $i < j \leq l$ the value of $t'_3$ at $M_i$ is lower than
the value of $t'_3$ at $M_j$. Thus $t'_3$ has the same order of entries
as $t_3$ which shows that $t_3$ is in $R$ as well.
\end{proof}

\subsection{Min-intersection closure}\label{ssect:mi}
In this section, we study a different restriction of shuffle-closed constraint languages.

\begin{definition}\label{def:min-intersection}
A relation $R$ is called min-intersection closed
if for all tuples $t_1$, $t_2$ in $R$, 
if $M(t_1) \cap M(t_2) \neq \emptyset$,
then there exists a tuple $t_3$ in $R$ 
such that $M(t_3) = M(t_1) \cap M(t_2)$.
\end{definition}

\begin{definition} 
Let $f$ be a binary operation preserving $<$. We say that $f$ {\em provides min-intersection closure} if $f(0,0)<f(0,x)$ and $f(0,0)<f(x,0)$ for all integers $x>0$. 
\end{definition}

\begin{lemma}\label{lem:mi-mi}
Let $R$ be a temporal relation that is preserved by an operation $f$ that provides min-intersection closure. Then $R$ is min-intersection closed.
\end{lemma}
\begin{proof}
Let $t_1$ and $t_2$ be two tuples in $R$ such that
$M(t_1) \cap M(t_2)$ is non-empty, that is, it 
contains an index $i$.
Choose $\alpha_1, \alpha_2 \in \AQ$ such that 
$\alpha_1 t_1[i] = \alpha_2 t_2[i] = 0$, and such that $\alpha_1$ and $\alpha_2$ map all other entries of $t_1$ and $t_2$ to integers.
Consider the tuple $t_3 = f(\alpha_1 t_1,\alpha_2 t_2)$.
Because at the entries from $M(t_1)$ (from $M(t_2)$) the tuple $\alpha_1 t_1$ ($\alpha_2 t_2$) equals 0, and because 
$f(0,0) < f(0,x)$ and $f(0,0) < f(x,0)$ for all positive integers
$x$, it follows that in $t_3$ all entries at $M(t_1) \cap M(t_2)$
have a strictly smaller value than all values at the symmetric difference
$M(t_1) \bigtriangleup M(t_2)$. Because $f$ preserves $<$,
it also follows that all entries at $M(t_1) \cap M(t_2)$ 
have a strictly smaller value than the entries not at $M(t_1) \cup M(t_2)$.
We conclude that $M(t_3)=M(t_1) \cap M(t_2)$.
\end{proof}

\begin{figure}[h]
\begin{center}
\includegraphics{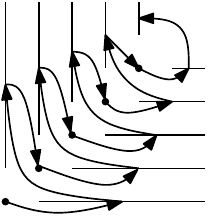}
\caption{Illustration of the operation \mi.}
\label{fig:lcsp-min-int-closure}
\end{center}
\end{figure}

An example of an operation that
provides min-intersection closure is the operation $\mi$, defined by
$$
\mi(x,y) := \left\{\begin{array}{l l l}
a(x) & \text{if } x < y \\
b(x) & \text{if } x=y \\
c(y) & \text{if } x > y 
\end{array} \right.
$$
where $a,b,c$ are unary operations that preserve $<$ such that
$$b(x) < c(x) < a(x) < b(x+\varepsilon)$$
 for all $x\in\mathbb{Q}$ and all
$0<\varepsilon\in\mathbb{Q}$ 
(see Figure~\ref{fig:lcsp-min-int-closure}). Operations $a,b,c$ with these properties can be constructed as follows.
Let $q_1,q_2,\dots$ be an enumeration of $\mathbb Q$. 
%Set $\beta(a_1)$, $\gamma(a_1)$, $\alpha(a_1)$ to arbitrary values from $\mathbb Q$ such that
%$\beta(a_1) < \beta(a_1) < \alpha(a_1)$. 
Inductively assume that 
we have already defined $a,b,c$ 
on $\{q_1,\dots, q_n\}$ such that 
%\begin{align}Ê\label{eq:midef}
$b(q_i) < c(q_i) < a(q_i) < b(q_j)$
%\end{align} 
whenever $q_i < q_j$, for $i,j \in [n]$.
Clearly, this is possible for $n=1$. 
If $q_{n+1} > q_i$ for all $i \in [n]$, let $q_j$ be the maximum
of $\{q_1,\dots,q_n\}$,  and define $a(q_j) < b(q_{n+1}) < c(q_{n+1}) < a(q_{n+1})$.
In the case that $q_{n+1} < q_i$ for all $i \in [n]$ we proceed analogously. Otherwise, let $i,j \in [n]$ such that
$q_i$ is the largest possible and $q_j$ is smallest possible 
such that $q_i<q_{n+1}<q_j$. In this case, define $a(q_i) < b(q_{n+1}) < c(q_{n+1}) < a(q_{n+1}) < b(q_j)$.
In this way we define unary operations
$a,b,c$ on all of $\mQ$ with the desired properties.

In fact, the operation \mi\ will be of special importance, because
the following proposition shows that \pp\ together with any operation providing min-intersection closure
generates the operation \mi.

\begin{proposition}\label{prop:mi}
A temporal relation $R$ is preserved by \pp\ and an operation $f$
providing min-intersection closure if and only if $R$ is
preserved by \mi.
\end{proposition}

% proof for mi

\begin{proof}
It is clear that $\mi$ provides min-intersection closure, and
Lemma~\ref{lem:gen-pp} shows that $\mi$ generates $\pp$.

For the opposite direction, suppose $R$ is $k$-ary and preserved by \pp\ and an operation $f$ providing
min-intersection closure. We show that for any two tuples $t_1,t_2\in R$ the
tuple $t_3=\mi(t_1,t_2)$ is in $R$ as well. Let 
$a,b,c$ be the mappings from
the definition of the operation $\mi$. Let $v_1<\dots<v_l$ be the minimal-length sequence of rational numbers such that for each $i'\in[k]$ it holds that 
$t_3[i']\in \bigcup_{j \in [l]}
\{a(v_j),
b(v_j),c(v_j) \}$. Let $M_i$ be 
$$ \big \{ i' \in [k] \; | \; t_3[i'] \in \{a(v_i),b(v_i),c(v_i)\} \big \} \; . $$
Observe that for each $i'\in M_i$ at least one of $t_1[i']$ and $t_2[i']$ is equal
to $v_i$ and the other value is greater or equal to $v_i$. 
Let $M_i^a$
be the set of those $i'\in M_i$ where $v_i=t_1[i']<t_2[i']$, 
$M_i^b$
the set of those $i'\in M_i$ where $v_i=t_1[i']=t_2[i']$,
and $M_i^c$
the set of those $i'\in M_i$ where $v_i=t_2[i']<t_1[i']$.

Let $\alpha_1,\dots,\alpha_l \in \AQ$ be such that $\alpha_i$
maps $v_i$ to $0$ and such that the entries of $\alpha_i t_1$
and $\alpha_i t_2$ are integers.
Let $\beta \in \AQ$ be such that $\beta f(0,0)=0$.
For each $ i \in [l]$ we define

\begin{align} 
s_i:=\pp(\beta f(\alpha_i t_1,\alpha_i t_2)),\pp(\alpha_i t_2,t_1) \; .
\label{eq:mi}
\end{align}

We verify that for all $i\in[l]$ the tuple $s_i$ is constant on each of the sets $M_i^a, M_i^b, M_i^c$, the value at $M_i^b$ is lower than the value at
$M_i^c$ which is lower than the value at $M_i^a$. Furthermore, for
each $j\in[l], j>i$, and each $i'\in M_i,j'\in M_j$, it holds that
$s_i[i']<s_i[j']$.  Having this, we can apply Lemma~\ref{lem:pp-combination}
and obtain a tuple from $R$ with the same ordering of entries as in $t_3$, which proves the lemma.

Because $\alpha_i$ maps $v_i$ to 0, the properties of \pp\ 
imply that the tuple $t'_i=pp(\alpha_i t_2,t_1)$ is constant at
$M_i^b\cup M_i^c$ and at $M_i^a$, 
and the value at the first set
is smaller than the value at the second set. 
Because the values of $t_2$ at
$M_i^a \cup \bigcup_{j=i+1}^l M_j$ 
are greater than $v_i$ and the values
of $t_1$ at $\bigcup_{j=i+1}^l M_j$ 
are also greater than $v_i$ 
(recall that for each $j\in [l]$, $j' \in M_j$ 
it holds that $\min(t_1[j'],t_2[j'])=v_j$) we conclude that
the values of $t'_i$ at $\bigcup_{j=i+1}^l M_j$ 
are greater than those at $M_i$. 

The application of $f$ in (\ref{eq:mi}) yields a tuple which is constant on $M_i^b$
and its value there (which is consequently mapped to 0 by $\beta$) is smaller
than the values at $M_i^a \cup M_i^c \cup \bigcup_{j=i+1}^l M_j$.
Thus it is easy to verify from the properties of \pp\ that the outer
application of \pp\ in (\ref{eq:mi}) yields a tuple with the desired properties.
\end{proof}

%We say that $f$ {\em provides min-union closure} if $f(0,0)=f(0,x)=f(x,0)$ for all integers $x>0$. We say that $f$ {\em provides min-xor closure} 
%if $f(0,0)>f(0,x)=f(x,0)$ for all integers $x>0$.

%\paragraph{Example.}
\begin{example}
An interesting example of a relation that is preserved by \mi\ but not by \min\ is the 
$4$-ary relation $I$ defined 
as follows.
\begin{align*}
	I(a,b,c,d) \; \equiv \; & (a=b \wedge b<c \wedge c=d) \\
	\vee \; & (a=b \wedge b>c \wedge c=d) \\
	\vee \; & (a=b \wedge b<c \wedge c<d) \\
	\vee \; & (a>b \wedge b>c \wedge c=d)
\end{align*}

To see that $I$ is preserved by $\mi$, let $t_1$ and $t_2$ be
two tuples from $I$. We have to show that $t_3 := \mi(t_1,t_2) \in I$.
First note that $I(a,b,c,d)$ is equivalent to $$(a \geq b) \wedge (b \neq c) \wedge (c \leq d) \wedge (a=b \vee b > c) \wedge (b < c \vee c=d) \; ,$$
and that $\mi$ preserves $\leq$ and $\neq$.

We distinguish the following cases.
\begin{enumerate}
\item $t_1[2] < t_1[3]$ and $t_2[2]<t_2[3]$.
Then $t_1[1] = t_1[2]$ and $t_2[1] = t_2[2]$, and hence $t_3[1]=t_3[2]$.
Since $\mi$ preserves $<$, we have $t_3[2]<t_3[3]$.
Since $\mi$ preserves $\leq$, we have that $t_3[3] \leq t_3[4]$,
and hence $t_3[1]=t_3[2]<t_3[3]<t_3[4]$ or $t_3[1]=t_3[2]<t_3[3]=t_3[4]$, which proves the claim.
\item $t_1[2] < t_1[3]$ and $t_2[2] > t_2[3]$. Then 
$t_1[1]=t_1[2]$ and $t_2[3]=t_2[4]$. 
We verify that $t_3$ satisfies the equivalent characterization of $I$
given above; since $\mi$ preserves $\leq$ and $\neq$, 
this amounts to proving that $t_3$ satisfies the two clauses
$(a=b \vee b > c) \wedge (b < c \wedge c=d)$.

The first sub-case we consider is $t_3[2]<t_3[3]$. Then by the assumptions on $t_1$ and $t_2$ and
by definition of \mi\ we have that $t_1[2]<t_2[2]$.
Therefore, $t_1[1]=t_1[2]<t_2[2] \leq t_2[1]$
and thus $t_3[1]=t_3[2]$ again by the properties of $\mi$; we see that both clauses are satisfied.
The second sub-case is that $t_3[2]>t_3[3]$. Then by the assumptions on $t_1$ and $t_2$ and by definition
of \mi\ we have that $t_1[4]\geq t_1[3]>t_2[3]=t_2[4]$. Thus
$t_3[3]=t_3[4]$ and again both clauses are satisfied.

% ZU SCHWACH:
%Suppose that $t_3[1] \neq t_3[2]$. By definition
%of $\mi$, we must have that $t_2[2] \leq t_1[2]$. Hence, 
%$$t_2[4] = t_2[3] < t_2[2] \leq t_1[2] < t_1[3] \leq t_1[4]$$
%and in particular $t_2[3]<t_1[3]$ and $t_2[4]<t_1[4]$.
%Thus $t_3[3]=t_3[4]$ by definition of $\mi$. Therefore, $t_3[1] = t_3[2]$ or $t_3[3] = t_3[4]$.
\item $t_1[2] > t_1[3]$ and $t_2[2] > t_2[3]$. This is analogous to the first case.
\item $t_1[2] > t_1[3]$ and $t_2[2] < t_2[3]$. This is analogous to the second case.
\end{enumerate}

The relation $I$ is not preserved by $\min$ since $(0,0,1,2) \in I$ and
$(2,1,0,0) \in I$ but $\min((0,0,1,2),$
$(2,1,0,0))=(0,0,0,0) \notin I$.
\qed \end{example}

%\vspace{.4cm}
%\paragraph{Example.}
\begin{example}
The following ternary temporal relation $U$ is preserved by \min\ (we omit the easy proof),
but not preserved by $\mi$. 

\begin{align*}
	U(x,y,z) \; \equiv \; & (x=y \wedge y<z) \\
	\vee \; & (x=z \wedge z<y) \\
	\vee \;  & (x=y \wedge y=z)
\end{align*}

To see that $U$ is not preserved by $\mi$,  note that the tuple $\mi((0,0,1),(0,1,0))$ has
three distinct values and hence is not in $U$, but $(0,0,1),(0,1,0) \in U$. An algorithm that solves temporal constraint languages preserved by 
\mi\  can be found in Section~\ref{ssect:alg-mi}.
\qed \end{example}

\subsection{Weak near-unanimity modulo endomorphisms}
For a uniform presentation of the classification
result in Section~\ref{sect:tcsp-classification}, 
we need the following alternative description of the clone generated by $\mi$.
When $A,B$ are two subsets of ${\mathbb Q}^3$, 
and $f \colon {\mathbb Q}^3 \to {\mathbb Q}$, 
we write $A <_f B$ if for all $(x,y,z) \in A$
and $(x',y',z') \in B$ we have $f(x,y,z) < f(x',y',z')$. 

\begin{figure}[h]
\begin{center}
\includegraphics[scale=.4]{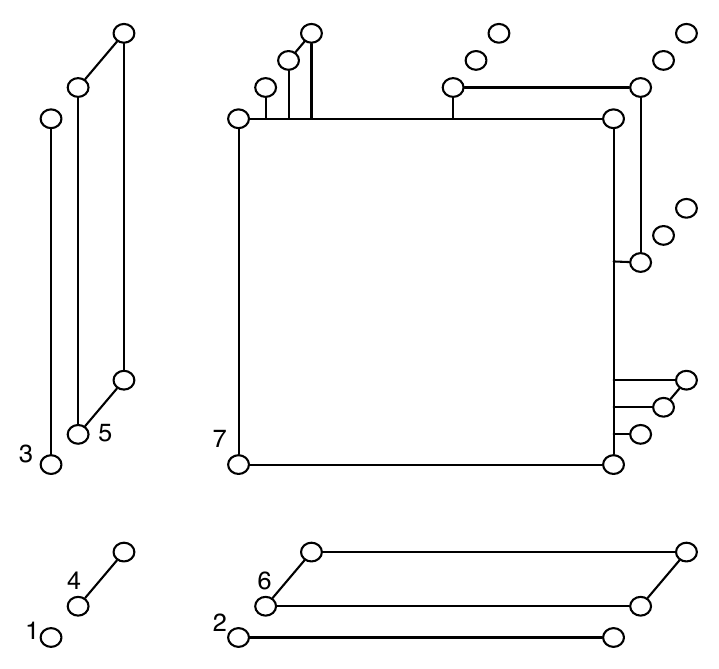}
\caption{Illustration of the function $f$
 from Proposition~\ref{prop:wnu-mi}.}
\label{fig:wnu-mi}
\end{center}
\end{figure}

\begin{proposition}\label{prop:better-mi}
There exists a function $f \colon {\mathbb Q}^3 \to {\mathbb Q}$ 
whose kernel has the following classes: for each $u \in {\mathbb Q}$ %, the class of all $x,y,z \in {\mathbb Q}^3$ such that 
\begin{enumerate}
\item $x(u) := \{(a,b,c) \; | \; u = b = c, a>c\}$;
% x-Balken
\item $y(u) := \{(a,b,c) \; | \; u = a = c, b>a\}$;
% y-Balken
\item $z(u) := \{(a,b,c) \; | \; u = a = b, c>a\}$; 
% z-Balken
\item $X(u) := \{(a,b,c) \; | \; u = a, b > a, c > a\}$; 
% x-Platt
\item $Y(u) := \{(a,b,c) \; | \; u = b, a > b, c > b\}$;
% y-Platte
\item $Z(u) := \{(a,b,c) \; | \; u = c, a > c, b > c\}$; 
% z-Platte
\item $D(u) := \{(u,u,u)\}$. 
% Diagonale
\end{enumerate}
Moreover, for $u<v$, we have 
$$ D(u) <_f x(u) <_f y(u) <_f z(u) <_f Z(u) <_f  Y(u) <_f X(u) <_f D(v) \; .$$
\end{proposition}
\begin{proof}
The specified countable family of 
subsets of ${\mathbb Q}^3$ indeed forms 
a partition of ${\mathbb Q}^3$. 
To see this, note that we distinguish 
which entries of the tuple are
equal to the minimum $u$ of the entries of the tuple. 
This splits ${\mathbb Q}^3$ into seven different classes
for a given $u$, all of them pairwise disjoint. 
See Figure~\ref{fig:wnu-mi}. 
Note that $<_f$ defines a linear order on 
this countable family, and since $({\mathbb Q};<)$
embeds all countable linear orders, the existence
of such a function $f$ follows. 
\end{proof}

\begin{proposition}\label{prop:wnu-mi}
Let $f \colon {\mathbb Q}^3 \to {\mathbb Q}$ 
be any
function with the properties in Proposition~\ref{prop:better-mi}. 
Then there are $a,b,c \in \End({\mathbb Q};<)$
such that for all $x,y \in {\mathbb Q}$
$$a(f(y,x,x)) = b(f(x,y,x)) = c(f(x,x,y)) \, .$$
That is, $f$ is a weak near unanimity 
modulo endomorphisms of $({\mathbb Q};<)$. 
\end{proposition}
\begin{proof}
By Lemma~\ref{lem:lift}, it suffices
to show that for all finite $S \subset \mQ$
there are $\alpha,\beta \in \AQ$ such that
for all $x,y \in S$
\begin{align}
f(y,x,x) & \; = \alpha(\mi(y,x)) \label{eq:yxx} \\
f(x,y,x) & \; = \beta(\mi(y,x)) \label{eq:xyx} \\
f(x,x,y) & \; = \gamma(\mi(y,x)) \label{eq:xxy}
\end{align}
Observe that for all $u,v,u',v' \in \mQ$ we have
$f(v,u,u) \leq f(v',u',u')$ iff one of the following
cases applies:
\begin{itemize}
% 1
\item $(v,u,u) \in D(u)$, $(v',u',u') \in D(u') \cup x(u') \cup X(u')$, and $u \leq u'$;
% 2
\item $(v,u,u) \in x(u)$, $(v',u',u') \in x(u') \cup X(u')$, and $u \leq u'$;
% 3
\item $(v,u,u) \in x(u)$, $(v',u',u') \in D(u')$, and $u < u'$;
% 4
\item $(v,u,u) \in X(u)$, $(v',u',u') \in X(u')$, and $u \leq u'$;
% 5
\item $(v,u,u) \in X(u)$, $(v',u',u') \in D(u') \cup x(u')$, and $u < u'$.
\end{itemize}
Note that this is the case if and only if
\begin{itemize}
\item $u=v \leq u' = v'$; % 1
\item $u<v$, $u' \neq v'$, $u \leq u'$, % 2
\item $u<v$, $u' = v'$, $u < u'$, % 3
\item $v<u$, $v' < u'$, $u \leq u'$, % 4
\item $v<u$, $u' = v'$, $u < u'$. % 5
\end{itemize}
This in turn is the case if and only if $\mi(v,u) \leq \mi(v',u')$. Then the statement for (\ref{eq:yxx}) 
 follows from homogeneity of $(\mQ;<)$. 
\ignore{
For $f(x,y,x)$, the proof is similar:
again, let $u,v,u',v' \in \mQ$ be arbitrary,
and observe that 
$f(u,v,u) \leq f(u',v',u')$ iff one of the following
cases applies: % WORKING HERE
\begin{itemize}
% 1
\item $(u,v,u) \in D(u)$, $(u',v',u') \in D(u') \cup x(u') \cup X(u')$, and $u \leq u'$;
% 2
\item $(u,v,u) \in y(u)$, $(u',v',u') \in y(u') \cup Y(u')$, and $u \leq u'$;
% 3
\item $(u,v,u) \in y(u)$, $(u',v',u') \in D(u')$, and $u < u'$;
% 4
\item $(u,v,u) \in Y(u)$, $(u',v',u') \in Y(u')$, and $u \leq u'$;
% 5
\item $(u,v,u) \in Y(u)$, $(u',v',u') \in D(u') \cup x(u')$, and $u < u'$. 
\end{itemize}
Note that this is the case if and only if
\begin{itemize}
\item $u=v \leq u' = v'$; % 1
\item $u<v$, $u' \neq v'$, $u \leq u'$, % 2
\item $u<v$, $u' = v'$, $u < u'$, % 3
\item $v<u$, $v' < u'$, $u \leq u'$, % 4
\item $v<u$, $u' = v'$, $u < u'$. % 5
\end{itemize}
This in turn is the case if and only if $\mi(v,u) \leq \mi(v',u')$. This concludes the
proof of (\ref{eq:xyx}). }
The proof for (\ref{eq:xyx}) and for (\ref{eq:xxy}) is analogous. 
\end{proof}

The addition of the 
third item in the following proposition is a result of Micha\l\ Wrona~\cite{WronaMi}.

\begin{theorem}\label{thm:wnu-mi}
Let $R \subseteq {\mathbb Q}^n$ be first-order definable over $({\mathbb Q};<)$. Then the following are equivalent.
\begin{enumerate}
\item $R$ is preserved by the operation $f$ as defined in Proposition~\ref{prop:better-mi} 
(a weak near unanimity modulo endomorphisms of $({\mathbb Q};<)$ by Proposition~\ref{prop:wnu-mi}).  
\item $R$ is preserved by $\mi$.
\item $R$ can be defined by a conjunction of formulas of the form
\begin{align}
& (x_1 \neq z) \vee \cdots \vee (x_k \neq z) \nonumber \\ 
\vee \; &  (y_1<z) \vee \cdots \vee (y_{l}<z) \label{eq:mi-syntax} \\ 
\vee \; &  (y_0 \leq z) \; . \nonumber
\end{align}
\end{enumerate}
\end{theorem}
\begin{proof}
The implication from (1) to (2)
follows from the observation that $(x,y) \mapsto f(x,x,y)$ induces the same weak
linear order on $\mQ^2$ as $\mi$ and hence generates $\mi$. 
The implication from (2) to (3) is an unpublished result from~\cite{WronaMi}. 
For the implication from (3) to (1) we verify that $f$ preserves
formulas $\phi$ of the form as in (\ref{eq:mi-syntax}). 
%Let $t_1 := (c,a_1,\dots,a_k,b_0,b_1,\dots,b_l)$
%and $t_2 := (c',a'_1,\dots,a'_k,b'_0,b_1',\dots,b'_l)$ be two tuples that satisfy $\phi(z,x_1,\dots,x_k,y_0,y_1,\dots,y_l)$. 
Let $t_1,t_2,t_3$ be assignments that
satisfy $\phi$. 
Suppose 
for contradiction that $t_0 := f(t_1,t_2,t_3)$ 
does not satisfy $\phi$.
In particular, $t_0(z)=t_0(x_1)=\dots=t_0(x_k)$.  
%$t_0(z) \leq \min(t_0(y_1),\dots,t_0(y_l))$,
%and $t_0(z) < t_0(y_0)$.
By the definition of $f$, there exists a $u \in \mQ$ such that $$D := \big \{(t_1(z),t_2(z),t_3(z)),(t_1(x_1),t_2(x_2),t_3(x_3)),\dots,(t_1(x_k),t_2(x_k),t_3(x_k)) \big \}$$ 
is contained
in one of $x(u)$, $y(u)$, $z(u)$, $X(u)$, 
$Y(u)$, $Z(u)$, or $D(u)$. 
It follows that there exists an $i \in \{1,2,3\}$
such that $t_i(z)=t_i(x_1) = \cdots = t_i(x_k)=u$. Suppose without loss
of generality that $i=1$. 
Since $t_1$ satisfies $\phi$, there must 
be a $j \in \{0,1,\dots,l\}$ such 
that $t_1(y_j) < t_1(z)=u$ 
or $t_1(y_0) = t_1(z)=u$.
\begin{itemize}
\item If $t_1(y_0) = u$ and 
$t_2(y_0) < t_1(y_0) = u$ 
or $t_3(y_0) < t_1(y_0) = u$ then 
$t_0(y_0) < t_0(z)$.
\item If $t_2(y_0) > t_1(y_0) = u$ and 
$t_3(y_0) > t_1(y_0) = u$ then 
$t_0(y_0) = t_0(z)$.
\item If $t_1(y_j) < t_1(z)=u$, 
then $t_0(y_j) < t_0(z)$. 
\end{itemize}
In each of the three cases, we have reached
a contradiction to the assumption that
$t_0$ does not satisfy $\phi$. 
\end{proof}

\subsection{Min-xor closure}
\label{ssect:mx}
We now introduce the last of the three mentioned closure conditions.

\begin{definition}\label{def:min-xor}
A relation is called min-xor closed
if for all tuples $t_1$, $t_2$ in $R$ where the symmetric difference
$M(t_1) \bigtriangleup M(t_2)$ is nonempty
there exists a tuple $t_3$ in $R$ 
such that $M(t_3) = M(t_1) \bigtriangleup M(t_2)$.
\end{definition}

\begin{definition} 
Let $f$ be a binary operation preserving $<$. 
We say that $f$ {\em provides min-xor closure} 
if $f(0,0)>f(0,x)=f(y,0)$ for all integers $x,y>0$.
\end{definition}

\begin{figure}
\begin{center}
\includegraphics{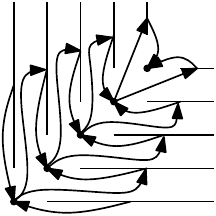}
\caption{Illustration of the operation \mx.}
\label{fig:mx}
\end{center}
\end{figure}

For an example of a 
binary operation that provides min-xor closure,
consider the following binary operation, which we denote by $\mx$.

$$
\mx(x,y) := 
\left\{\begin{array}{l l l}
a(min(x,y)) & \text{if } x \neq y \\
b(x) & \text{if } x=y
\end{array} \right.
$$
where $a$ and $b$ are unary operations that preserve $<$ such that
$a(x) < b(x) < a(x+\varepsilon)$ for all $x\in\mathbb{Q}$ and all $0<\varepsilon\in\mathbb{Q}$ 
(see Figure~\ref{fig:mx}).  Similarly as in the definition of $\mi$,
such operations $a,b$ can be easily constructed. It is easy to see that the operation $\mx$ neither preserves the relation $I$ nor the relation $U$ introduced in Section~\ref{ssect:mi}.

\begin{lemma}\label{lem:mx-mx}
Let $R$ be a temporal relation that is preserved by
an operation $f$ providing min-xor closure. Then $R$
is min-xor closed.
\end{lemma}
\begin{proof}
Let $t_1$ and $t_2$ be tuples in $R$, and
suppose that the symmetric difference $M(t_1) \bigtriangleup M(t_2)$
of $M(t_1)$ and $M(t_2)$ is non-empty.
Let $v_1$ and $v_2$ be the minimal values 
of the entries of $t_1$ and of $t_2$, respectively.
%contains an index $i$.
Then there are $\alpha_1,\alpha_2 \in \AQ$ such that
$\alpha_1 v_1=0$ and $\alpha_2 v_2=0$ and such that $\alpha_1$ and $\alpha_2$ map all other entries of $t_1$ and $t_2$ to integers.
Consider the tuple
$t_3 = f(\alpha_1 t_1,\alpha_2 t_2)$. 
Because $\alpha_1 t_1$ is $0$ for all entries at $M(t_1)$,
$\alpha_2 t_2$ is $0$ for all entries at $M(t_2)$, 
and $f(0,0)>f(0,x)=f(y,0)$ for
all $x,y > 0$, it follows that in $t_3$ all entries at $M(t_1) \cap M(t_2)$ have a strictly larger value than all entries at $M(t_1)\bigtriangleup M(t_2)$, which all have the same
value.  Because $f$ preserves $<$, all entries of $t_3$ at
$M(t_1) \cap M(t_2)$ have a smaller value
than all entries not at $M(t_1) \cup M(t_2)$. 
We conclude that the tuple $t_3 \in R$ satisfies
$M(t_3) = M(t_1) \bigtriangleup M(t_2)$.
\end{proof}

The following lemma implies that $\{f,\pp\}$ generates
$\mx$ for any operation $f$ that provides min-xor closure.

\begin{proposition}\label{prop:mx}
A temporal relation $R$ is preserved by $\pp$ and
an operation $f$ providing min-xor closure if and only if $R$
is preserved by $\mx$.
\end{proposition}
% proof for mx
\begin{proof}
Clearly, $\mx$ provides min-xor closure.
Lemma~\ref{lem:gen-pp} shows that $\mx$ generates $\pp$.

For the opposite direction, suppose that
$R$ is $k$-ary and preserved by \pp\ and an operation $f$ providing
min-xor closure. We show that for any two tuples $t_1,t_2\in R$ the
tuple $t_3=\mx(t_1,t_2)$ is in $R$ as well. Let $a,b$ be the mappings as in the definition of the operation $\mx$.
Let $v_1<\dots<v_l$ be minimal
set of rational numbers such that $t_3[i]\in \bigcup_{j\in[l]} \{ a(v_j),
b(v_j)\}$ for all $i\in[k]$, and let 
 $M_i$ be the set of indices 
 $\{i' \in[k] | t_3[i'] \in\{a(v_i),b(v_i)\}\}$.
Observe that for each $i'\in M_i$ at least one of $t_1[i']$ and $t_2[i']$ is equal
to $v_i$ and the other value is greater or equal to $v_i$. Let $M_i^a$
be the set of those $i'\in M_i$ where $t_1[i']\neq t_2[i']$ and $M_i^b$
the set of those $i'\in M_i$ where $v_i=t_1[i']=t_2[i']$.

Let $\alpha_1,\dots,\alpha_l \in \AQ$ be
such that $\alpha_i$
maps $v_i$ to 0 and such that the entries of $\alpha_i t_1$ and $\alpha_i t_2$ are integers. For each $i\in[l]$ we define
$s_i:=f(\alpha_i t_1,\alpha_i t_2)$.
It is easy to see from the choice of $\alpha_i$ and properties of $f$
that for each $i\in[l]$ the tuple $s_i$ is constant at $M_i^a,
M_i^b$, and that the value at $M_i^a$ is lower than the value at $M_i^b$.
Furthermore, because $f$ preserves $<$, because the values of $t_1$ at
$\bigcup_{j=i+1}^l M_j$ are greater than $v_i$, and because the values of
$t_2$ at $\bigcup_{j=i+1}^l M_j$ are greater than $v_i$, we see that
for each $j\in[l], j>i$ and each $i'\in M_i,j'\in M_j$, it holds
that $s_i[i']<s_i[j']$.  Having this, we can apply
Lemma~\ref{lem:pp-combination} and obtain a tuple from $R$ with the same
ordering of entries as in $t_3$, which proves the lemma.
\end{proof}

\begin{example}
An interesting example of a temporal relation that is preserved
by \mx\  is the ternary relation $X$ defined as follows.

\begin{align*}
	X(x,y,z) \; \equiv \; & (x=y \wedge y<z) \\
	\vee \; & (x=z \wedge z<y) \\
	\vee \; & (y=z \wedge y<x)
\end{align*}

The relation is not preserved by \min\ and by \mi:
the tuples $t_1 = (0,0,1)$, $t_2 = (0,1,0)$ are in $X$, but
$\min(t_1,t_2)=(0,0,0) \notin R$, and $\mi(t_1,t_2)$ has three distinct entries
and hence is not in $X$ as well.
\qed \end{example}

An algorithm that solves constraint languages preserved by \mx\ can
 be found in Section~\ref{ssect:alg-mx}.

\subsection{Operations generating \min, \mi, \mx}
\label{ssec:lcsp-classif-pp}
As we have seen in Proposition~\ref{prop:Shard}, 
if the relation $T_3$ has a primitive positive
definition in $\bB$, then CSP$(\bB)$ is NP-hard.
We show that if a temporal constraint language is shuffle-closed 
and does not admit a primitive positive definition of $T_3$,
then it is preserved by \min, \mi, or \mx.

%\begin{theorem}\label{thm:lcsp-pp-hard}
%Let $\bB$ be a temporal constraint language that is able to %express the
%relation $S(x,y,z)$ from Definition~\ref{def:lcsp-smin-smax}. Then
%\CSP$(\bB)$ is NP-complete.
%\end{theorem}
%\begin{proof}
%\end{proof}

If the relation $T_3$ does not have a primitive
positive definition in $\bB$, then Theorem~\ref{thm:inv-pol} 
implies that $\bB$ has a polymorphism that does not preserve $T_3$. By Theorem~\ref{thm:cores}, it suffices to consider operations that preserve $<$. 
%\subsection{Languages expressing inequality}
%In the first and most important case, we consider languages $\bB$
%such that $<$ has a primitive positive definition in $\bB$.
We start with a sequence of auxiliary lemmas.

\begin{lemma}
\label{lem:lcsp-pp-neg-inc}
Let $f$ be a binary operation preserving $<$,
and suppose that there is an infinite sequence $x_1<x_2<\ldots$ of elements of $\mathbb Q$ 
and $y_1 \in \mathbb Q$ such that $f(x_1,y_1)\geq f(x_2,y_1)<f(x_i,y_1)$ for
all $i>2$. Then $f$ generates an operation providing
min-intersection closure.
\end{lemma}
\begin{proof}
Because $f$ preserves $<$, we have that for any infinite sequence
$y_1<y_2<\ldots$ it holds that $f(x_2,y_i)>f(x_1,y_1)$. Hence,
the binary operation defined by $f(\alpha(x),\beta(y))$ provides
min-intersection closure, where $\alpha \in \AQ$ maps
$0,1,\ldots$ to $x_2,x_3,\ldots$ and $\beta \in \AQ$ maps 
$0,1,2,\ldots$ to $y,y_1,y_2,\ldots$
\end{proof}

\begin{lemma}\label{lem:mi-koenig}
Suppose $f$ preserves $<$ and generates a sequence of operations $f_1,f_2,\ldots$ 
such that for each $f_k$ it holds that $f_k(0,0)<f_k(x,0)$ and
$f_k(0,0)<f_k(0,x)$ for all integers $x \in [k]$. 
Then $f$ generates an operation $g$ providing min-intersection closure.
\end{lemma}
\begin{proof}
A direct consequence of Lemma~\ref{lem:behavior-generates}.
\end{proof}

\begin{lemma}
\label{lem:lcsp-pp-neg-dec}
Let $f$ be a binary operation preserving $<$ such that there is an infinite
sequence $x_1<x_2<\ldots$ and $y_1\in\mathbb{Q}$ satisfying
$f(x_i,y_1)>f(x_j,y_1)$ for all $1\leq i<j$. 
Then $\{f,\pp\}$ generates an operation providing min-intersection closure.
\end{lemma}
\begin{proof}
By Lemma~\ref{lem:mi-koenig}, it suffices to show that there is a sequence of operations $f_1,f_2,\dots,$ 
generated by $\{f,\pp\}$ such that $f_k(0,0)<f_k(x,0)$ and $f_k(0,0)<f_k(0,x)$
for all $k \geq 1$ and all $x \in [k]$.
%Then we use this sequence to
%generate an operation $g$ providing min-intersection closure.

So let $k \geq 0$ be a fixed integer, and $y_1<y_2<\dots$ be an arbitrary infinite sequence.
Let $\alpha_k$ be from \AQ such that
%$\{f(x_1,y_i), f(x_i,y_1) \,|\, $
%$1\leq i\leq k\}$  
 $$\alpha \big \{f(x_1,y_i) \, \big | \, 1 \leq i \leq k\} \cup \{f(x_i,y_1) \; | \; 1 \leq i \leq k \big \} \subseteq \{x_2,\dots,x_{2k}\}$$
and $\beta_1,\beta_2\in\AQ$ such that $\beta_1$ maps $0,1,2,\dots$ to
$x_1,x_2,x_3,\dots$ and $\beta_2$ maps $0,1,2,\dots$ to $y_1,y_2,y_3,\dots$
We define 
$$f_k(x,y) := f(\alpha_k f(\beta_1 x,\beta_2 y),\beta_2 y) \; ,$$
and show that $f_k$ has the required properties. It follows from the assumptions on $f$ 
that for all positive integers $x$ we have $f(\beta_1 0,\beta_20) = f(x_1,y)>f(\beta_1 x,y_1) = f(\beta_1 x,\beta_2 0)$, and due to the properties of $\alpha_k$ it holds that $f_k(0,0)<f_k(x,0)$ for all
integers $x \in [k]$. 

We also have for every $x \in [k]$ 
that $\beta_2 x>y_1$
and $\alpha_k f(\beta_1 0,\beta_2 x) > x_1$. Because $f$ preserves
$<$, this shows that $f_k(0,x)=f(\alpha_k f(\beta_1 0,\beta_2 x)),\beta_2 x)>f(x_1,y_1)$. Moreover,
$f_k(0,0)=f(\alpha_k f(x_1,y_1),y_1) < f(x_1,y_1)$ by the assumptions
on $f$. Hence, $f_k(0,x) > f(x_1,y_1) > f_k(0,0)$ for all $x \in [k]$.
\end{proof}

The following lemma applies (a special case of) Ramsey's theorem; more substantial applications of Ramsey theory can be found
in Section~\ref{sect:tcsp-classification}. 

\begin{figure}
\begin{center}
\includegraphics[scale=.8]{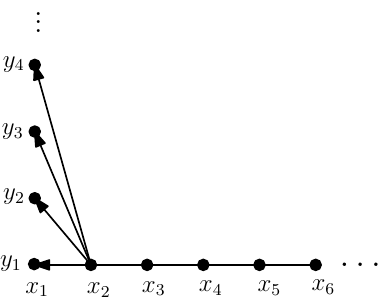}
\caption{Illustration for Case 3 of Lemma~\ref{lem:lcsp-pp-neg-equal}.}
\label{fig:lcsp-pp-intersection}
\end{center}
\end{figure}

\begin{lemma}
\label{lem:lcsp-pp-neg-equal}
Let $f$ be a binary operation preserving $<$ such that there is an infinite
sequence $x_1<x_2<\ldots$ and $y_1\in\mathbb{Q}$ satisfying
$f(x_1,y_1)>f(x_i,y_1)=f(x_j,y_1)$ for all $1<i<j$. 
Then $\{f,\pp\}$ generates an operation providing min-intersection or
min-xor closure.
\end{lemma}
\begin{proof}
By the infinite pigeon-hole principle 
there must be an infinite sequence $y_2 < y_3 < \dots$ of elements of $\mathbb Q$ larger than $y_1$ such that
\begin{enumerate}
\item $f(x_2,y_1) = f(x_1,y_i)$  for all $i \geq 2$, or
\item $f(x_2,y_1) > f(x_1,y_i)$ for all $i \geq 2$, or
\item $f(x_2,y_1) < f(x_1,y_i)$ for all $i \geq 2$.
\end{enumerate}
In case 1, $f$ generates an operation providing min-xor closure and we are done. In case 2, we apply Ramseys theorem (Theorem~\ref{thm:inf-ramsey})
in the special case of $m=2$, $r=3$ as follows. Let $D$ be $\{y_1,y_2,\dots\}$.
For $i<j$, define $\chi(\{y_i,y_j\}) = 1$ if $f(x_1,y_i) = f(x_1,y_j)$,
$\chi(\{y_i,y_j\}) = 2$ if $f(x_1,y_i) > f(x_1,y_j)$, and
$\chi(\{y_i,y_j\}) = 3$ if $f(x_1,y_i) < f(x_1,y_j)$.
Then Theorem~\ref{thm:inf-ramsey} applied to $\chi$ shows that 
there exists an infinite subsequence $z_1 < z_2 < \dots$
of $y_1 < y_2 < \dots$ such that 
\begin{enumerate}
\item[2a.] $f(x_1,z_i) = f(x_1,z_j)$ for all $1 \leq i<j$, or
\item[2b.] $f(x_1,z_i) > f(x_1,z_j)$ for all $1 \leq i<j$, or
\item[2c.] $f(x_1,z_i) < f(x_1,z_j)$ for all $1 \leq i<j$.
\end{enumerate}
In case 2a, we swap arguments of $f$ and
proceed as in case 3. In case 2b, we swap arguments of $f$, apply
Lemma~\ref{lem:lcsp-pp-neg-dec}, and conclude that $f$ generates an operation providing min-intersection closure.
In case 2c, note that $f(x_1,y_1) > f(x_2,y_1) > f(x_1,y_i)$ for all $i \geq 2$,
and thus we can apply Lemma~\ref{lem:lcsp-pp-neg-inc}
to conclude that $f$ generates an operation providing min-intersection closure.

In case 3, we show that similarly as in Lemma~\ref{lem:lcsp-pp-neg-dec} there
is a sequence of operations $f_1,f_2,\ldots$ generated by $\{f,\pp\}$ such that for each $f_k$ it holds that $f_k(0,0)<f_k(x,0)$ and $f_k(0,0)<f_k(0,x)$ for
all integers $x \in [k]$, and conclude by application
of Lemma~\ref{lem:mi-koenig}. See 
Figure~\ref{fig:lcsp-pp-intersection} for an illustration.

Let $\alpha_k$ be from \AQ such that $\alpha_k f(x_2,y_1) = x_1$ and
$\alpha \{f(x_1,y_i)\, | \, 1\leq i\leq k\} \subseteq \{x_2,\dots,x_{k+1}\}$. Furthermore let
$\beta_1,\beta_2\in\AQ$ be such that $\beta_1$ maps $0,1,2,\dots$ to
$x_1,x_2,x_3,\dots$ and $\beta_2$ maps $0,1,2,\dots$ to $y_1,y_2,y_3,\dots$
We define 
$$f_k(x,y) := f(\alpha_k f(\beta_1 x,\beta_2 y),\beta_2 y) \; .$$  
Then for all integers $x>0$
\begin{align*}
f_k(0,0) & = f(\alpha_k f(x_1,y_1),y_1)=f(x_2,y_1) \;, \quad \text{and} \\
f_k(x,0) & = f(\alpha_k f(\beta_1 x,y_1),y_1)=f(x_1,y_1) \; .
\end{align*}
Hence
$f_k(0,0)<f_k(x,0)$ for all integers $x>0$. Finally, since $\beta_2 x>y_1$ and
$\alpha_k f(x_1,\beta_2 x)>x_1$ for all integers $x>0$, we have that
$f_k(0,x)>f(x_1,y_1)>f(x_2,y_1)=f_k(0,0)$.
\end{proof}

The previous two lemmas are combined in the following result.

\begin{lemma}\label{lem:lcsp-pp-pres-leq}
Let $f$ be a binary operation that preserves $<$ and violates the relation
$\leq$. Then $\{f,\pp\}$ generates an operation providing min-intersection or min-xor closure.
\end{lemma}
\begin{proof}
As $f$ violates $\leq$, we can without loss of generality assume that there is $y\in\mathbb{Q}$ and
$x_1,x_2\in\mathbb{Q}$, $x_1 < x_2$, such that $f(x_1,y)>f(x_2,y)$. 

We claim that there are only three possibilities:
\begin{enumerate}
\item[a)] There is an infinite sequence $x_3<x_4<\dots$ such that $x_2<x_3$ and
$f(x_i,y)>f(x_2,y)$  for all $i>2$. 
\item[b)] There is an infinite sequence $x_3<x_4<\dots$ such that $x_2<x_3$ and
$f(x_i,y) > f(x_j,y)$ for all $2\leq i<j$. 
\item[c)] There is an infinite sequence $x_3<x_4<\dots$ such that $x_2<x_3$ and
$f(x_i,y) = f(x_2,y)$ for all $i>2$. 
\end{enumerate}
To show this claim, observe that by the infinite pigeon-hole principle
there is an infinite sequence $x_3<x_4<\dots$ with $x_2<x_3$ such that 
$f(x_i,y)>f(x_2,y)$ for all $i > 2$, $f(x_i',y) = f(x_2,y)$ for all $i > 2$, 
or $f(x_i,y) < f(x_2,y)$ for all $i > 2$.
In the first and the second case the claim holds. In the third case, 
we repeat the argument with $x_2<x_3$ instead of $x_1 < x_2$. Again,
we distinguish three cases, and as before in two of them we are immediately done. In the third case, we repeat again. 
If we repeat this for infinitely many times we obtain a sequence
$x_3=x_3' < x'_4 < \dots$ such that $x_2<x_3$ and
$f(x_i',y) > f(x_j',y)$ for all $2\leq i<j$.

In a) the conditions of Lemma~\ref{lem:lcsp-pp-neg-inc}
are satisfied and we conclude that $\{f,\pp\}$ generates an operation providing min-intersection closure. 
In b)
Lemma~\ref{lem:lcsp-pp-neg-dec} shows that $\{f,\pp\}$ generates an operation providing min-intersection closure.
In c) we apply Lemma~\ref{lem:lcsp-pp-neg-equal}
and conclude that $\{f,\pp\}$ generates an operation providing min-intersection 
or min-xor closure.
\end{proof}

The following is the main result of this subsection.
Recall that the relation $T_3$ was defined in Definition~\ref{def:lcsp-smin-smax} to be
$$\{(x,y,z) \in {\mathbb Q}^3\; | \; (x = y < z) \; \vee (x=z < y) \; \} \; .$$

\begin{lemma}\label{lem:lcsp-pp-one-closure}
Let $f$ be a binary operation that preserves $<$ and violates the relation $T_3$.
Then $\{f,\pp\}$ generates \min, \mi, or \mx.
\end{lemma}
\begin{proof}
By Proposition~\ref{prop:min}, \ref{prop:mi}, and \ref{prop:mx},
it suffices to show that $\{f,\pp\}$ generates
an operation providing min-intersection, min-union,
or min-xor closure.
If $f$ violates $\leq$, then we are immediately done by
Lemma~\ref{lem:lcsp-pp-pres-leq}. So we further assume that $f$ preserves
$\leq$.

Because $f$ preserves $<$ and violates $T_3$, we can assume without loss
of generality (possibly after swapping arguments) that there are
$x_1,x_2,y_1,y_2\in\mathbb{Q}$ such that $x_1<x_2$, $y_1<y_2$ and
$t:=(f(x_1,y_1),f(x_2,y_1),f(x_1,y_2))\not\in T_3$. Because $f$ preserves $\leq$
we have that $f(x_1,y_1)\leq f(x_2,y_1)$ and $f(x_1,y_1)\leq f(x_1,y_2)$.
Since $t\not\in T_3$, there are only two possibilities:
\begin{enumerate}
\item $t[1]<t[2]$ and $t[1]<t[3]$. In this case, choose infinite sequences $x_3<x_4<\dots$ and $y_3<y_4<\dots$ such that $x_2<x_3$,
$y_2<y_3$. Because $f$ preserves $\leq$, we have for all $i>1$
that $f(x_2,y_1) \leq f(x_i,y_1)$ and $f(x_1,y_2) \leq f(x_1,y_i)$.
Since $t[1]=f(x_1,y_1)<t[2]=f(x_2,y_1)$ we have 
that $f(x_1,y_1)<f(x_i,y_1)$ for all $i>1$,
and since $t[1]=f(x_1,y_1)<t[3]=f(x_1,y_2)$
we have that $f(x_1,y_1)<f(x_1,y_i)$ for all $i>1$.
Hence, $f$ provides min-intersection closure.
\item $t[1]=t[2]=t[3]$. In this case we can choose infinite sequences
$x'_2<x'_3<\dots$ and $y'_2<y'_3<\dots$ such that $x_1<x'_2$, $y_1<y'_2$,
and for all $i>1$, $x'_i<x_2$ and $y'_i<y_2$. As $f$ preserves $\leq$, we
see that $f(x'_i,y_1)=f(x_1,y_1)=f(x_1,y'_i)$ for all $i>1$ and thus $f$
provides min-union closure.
\end{enumerate}
\end{proof}

\subsection{Algorithms for shuffle-closed languages}
\label{sect:shuffle-algs}
In this section we present three algorithms, for the languages
preserved by $\mi$, by $\min$, and by $\mx$, respectively.
All three algorithms follow a common strategy.
They are searching 
for a subset of the variables that can have the minimal value in a solution. 
If they have found such a a subset, $S$, the algorithms add equalities and inequalities that are implied
by all constraints under the assumption that the variables in $S$ denote the minimal
value in \emph{all} solutions. 
Next, the algorithms recursively solve the instance consisting of the 
projections of all constraints
to the variables that do not denote the minimal value in all solutions.
We later show that for languages preserved by $\pp$ it is true that 
if the instance has a solution, it also has a
solution that satisfies all the additional constraints.

Throughout this section we assume that $\bB$ is a structure with
 a first-order definition in $(\mQ;<)$ and a finite relational signature.
For the formulation of the algorithms and their correctness proofs
it will be convenient to work with an expanded constraint
language, that contains the binary relation $=$ for the equality relation.
We also add to the temporal constraint language $\bB$ 
several other temporal relations that are primitive positive definable in $\bB$.

\begin{definition}
Let $R$ be an $n$-ary temporal relation and $L = \{p_1,\dots,p_k\} \subseteq [n]$ where $p_1 < \dots < p_k$. Let $\{q_1,\dots,q_l\}$ be $[n] \setminus L$.
Then the {\em ordered projection of $R$ to $L$} is the $k$-ary
relation $R'$ with the primitive positive definition 
$$R'(x_{p_1},\dots, x_{p_k}) \; \equiv \; \exists x_{q_1},\dots,x_{q_l}. R(x_1,\dots,x_n) \wedge \bigwedge_{i \in [n] \setminus L, \; j \in L} x_i < x_j  \, .$$
\end{definition}

Note that if $\bB$ is a finite temporal constraint language,
then there are only finitely many projections and ordered projections of
relations in $\bB$. 
In case that there is a primitive positive definition of $<$ in $\bB$,
ordered projections are primitive positive definable.
By Lemma~\ref{lem:pp-reduce}, we can assume in this case that
$\bB$ contains all relations that can be defined by ordered projections from relations in $\bB$.

To formally introduce our algorithms, we also need the concept of an
ordered projection of \emph{instances} of the CSP. 

\begin{definition}
Let $\bB$ be a temporal constraint language that contains all ordered projections of relations
from $\bB$.  Let $\Phi$
be an instance of $\Csp(\bB)$ and $X \subseteq V(\Phi)$. 
Then the {\em ordered projection of $\Phi$ to $X$} is the 
instance of $\Csp(\bB)$ that contains
for each constraint $R(x_1,\dots,x_n)$ in $\Phi$, with not necessarily distinct variables $x_1,\dots,x_n$, 
the constraint
$R'(x_{k_1},\dots,x_{k_l})$ where
$k_1 < \dots < k_l$
are such that 
$\{k_1,...,k_l\} = \{k \in [n] \; | \; x_k\in X\}$,
and $R'$ is the ordered projection of $R$ to $\{k_1,\dots,k_l\}$.
\end{definition}

Let $\Phi$ be an instance of a temporal CSP.

\begin{definition}
If $\psi=R(x_1,\dots,x_k)$ is a constraint from $\Phi$,
then a subset $X$ of the variables
of $\psi$ is called a min-set (of $\psi$) if there exists
a $k$-tuple $t$ satisfying $\psi$ such that $x \in X$ iff the value
for $x$ in $t$ is the minimum of all entries of $t$.
A set of variables $S \subset V(\Phi)$ is called {\em free} 
iff it is non-empty and for all constraints $R(x_1,\dots,x_k)$ in $\Phi$
the set $S \cap \{x_1,\dots,x_k\}$ is either empty or
a min-set of $R$.
\end{definition}

We will show how to use the concept of freeness
to solve instances of $\Csp(\bB)$ for 
shuffle closed temporal constraint languages.

\begin{lemma}\label{lem:lcsp-pp-compose}
Let $\Phi$ be an instance of $\Csp(\bB)$
for some shuffle closed $\bB$, and let $S$ be a free set of variables of $\Phi$. Then $\Phi$ 
has a solution if and only if the ordered projection $\Phi'$ of
$\Phi$ to $V(\Phi) \setminus S$ has a solution.
\end{lemma}

\begin{proof}
First suppose $\Phi'$ has a solution $s'$.
Let $\psi = R(x_1,\dots,x_m)$ be a constraint of $\Phi$ such that
$V(\psi) \cap S = \{x_{p_1},\dots,x_{p_k} \} \neq \emptyset$.
Let $\{x_{q_1},\dots,x_{q_l} \} = V(\psi) \setminus S$ for
$q_1 < \dots < q_l$. 
By the definition of an ordered projection, 
there is a tuple $t_1 \in R$ such that $s'(x_i)=t_1[i]$
for all $i \in \{q_1,\dots,q_l\}$.
Since $V(\psi) \cap S$ is a min-set of $R$, 
there is a tuple $t_2 \in R$ such that $M(t_2) = \{p_1,\dots,p_k\}$. 
Let $\alpha \in \AQ$ be such that $\alpha$ maps the minimal value 
of $t_2$ to $0$. 
Because $R$ is preserved by \pp, 
the tuple $t_3:=\pp(\alpha(t_2),t_1)$ 
is in $R$.
It is easy to verify that $M(t_3)=\{p_1,\dots,p_k\}$ and that
there is $\beta \in \AQ$ such that $\beta t_3[i]=s'(x_i)$
for $i \in \{q_1, \dots, q_l \}$. Because 
we can find such a tuple for all the constraints $\psi$ in
$\Phi$ where $V(\psi) \cap S \neq \emptyset$, 
we conclude that a solution $s'$
of $\Phi'$ can be extended to a solution $s$ of $\Phi$ by setting all the
variables in $S$ to some value that is smaller than 
the smallest value in $\{ s'(x)  \; | \; x \in V(\Phi') \}$. 
Clearly,
all the constraints $\psi$ in $\Phi$ with $V(\psi) \cap S = \emptyset$ or $V(\psi) \subset S$ are satisfied by $s$ as well.

Now suppose that $\Phi$ has a solution $s$. 
Let $x_1,\dots,x_n$ be the variables of $\Phi$,
and let $\{x_{r_1},\dots,x_{r_{|S|}}\}$ be $S$. 
Let $s'$ be a mapping from $V(\Phi)$ to $\mathbb Q$ such that 
$M((s'(x_1),\dots,s'(x_n)))=\{r_1,\dots,r_{|S|}\}$, and $s'(x)=s(x)$
for $x \in V(\Phi) \setminus S$. 
We claim that $s'$ is a solution for $\Phi'$.
Let $\psi = R(y_1,\dots,y_m)$ be a constraint of $\Phi$ such that 
$V(\psi) \cap S \neq \emptyset$.
Clearly, $t_1 := (s(y_1),\dots,s(y_m))$ is in $R$ 
since $s$ is a solution of $\Phi$. 
Let $\{y_{p_1},\dots,y_{p_l}\}$ be $S \cap \{y_1,\dots,y_m\}$.  
Since $\{y_{p_1},\dots,y_{p_l}\}$ is a min-set of $R$, 
there is a tuple $t_2 \in R$ such that $M(t_2) = \{p_1,\dots,p_l\}$.
Let $\alpha \in \AQ$ be such that
$\alpha$ maps the minimal value of $t_2$ to $0$. 
Because $R$ is preserved by \pp, 
the tuple $t_3:=\pp(\alpha t_2, t_1)$ is in $R$.
It is easy to verify that $M(t_3) = \{p_1,\dots,p_l\}$, and
that there is an automorphism $\beta$ such that $\beta t_3[i] = s(y_i)$
for $i \in [m] \setminus \{p_1,\dots,p_l\}$.
Clearly, the restriction of $s'$ to $V(\Phi) \setminus S$ 
is a solution to the ordered projection $\Phi'$ of $\Phi$ to
$V(\Phi) \setminus S$ since $s'$ also satisfies all the inequalities imposed by the ordered projection. 
Therefore $\Phi'$ is satisfied by $s'$.
\end{proof}

\begin{figure}
\begin{center}
\small
\fbox{
\begin{tabular}{l}
Solve($\Phi$) \\
{\rm // Input: An instance $\Phi$ of $\Csp(\bB)$} \\
{\rm // for a shuffle closed temporal language $\bB$} \\
{\rm // Output: A solution $s$ to $\Phi$, or reject if there is no solution.} \\
$i$ := $0$ \\
while $V(\Phi) \neq \emptyset$ do begin \\
\hspace{.5cm}     $S$ := FindFreeSet($\Phi$) \\
\hspace{.5cm}    if $S=\false$ then reject \\
\hspace{.5cm}    for each $x \in S$ do $s(x)$ := $i$ \\
\hspace{.5cm}    $i$ := $i+1$ \\
\hspace{.5cm}    $\Phi$ :=\textrm{ ordered projection of $\Phi$ to $V (\Phi) \setminus S$} \\
end \\
return $s$ 
\end{tabular}}
\end{center}
\caption{An algorithm that efficiently solves instances of a shuffle closed constraint language if free sets can be computed efficiently.}
\label{fig:alg-pp}
\end{figure}

The above lemma asserts that if we are able to identify a free set for instances of $\Csp(\bB)$ for a shuffle-closed 
temporal language $\bB$ in polynomial time, then
we also have a
polynomial time algorithm that solves $\Csp(\bB)$. 
%More precisely, we have the following.
%\begin{lemma}\label{lem:pp-alg}
%Suppose there is an algorithm that computes for a given
%instance $\Phi$ a free set, or returns $\false$, if no such 
The running time of the algorithm is $O(n \cdot (m+t(n,m)))$,
where $n = |V|$, $m$ is the number of constraints in $\Phi$,
and $t(n,m)$ is the running time of the procedure that
computes the free set of an instance with $n$ variables and $m$
constraints.

\subsection*{An algorithm for languages preserved by \min}
\label{ssect:alg-min}
Now, we concentrate on the problem to find a free set of $\Phi$
if $\bB$ is preserved by the operation $\min$. 

Let $\psi=R(x_1,\dots,x_k)$ be a constraint where $R$ 
is from $\bB$ 
and let $L$ be a subset of $\{x_1,\dots,x_k\}$. Let $A_1,\ldots,A_l$ be all min-sets of $\psi$  that are contained in $L$.
When $l \geq 1$, i.e., when such min-sets exist, there is a unique set $A_j$, $j \in [l]$, with the property that $A_i\subseteq A_j$ for all $i \in [l]$, because $R$ is preserved by $\min$, and thus min-union
  closed by Lemma~\ref{lem:mu-mu}.
We call this min-set the {\em maximal min-set of $\psi$ contained in $L$}. Note that for some $L$ it could be that $l=0$, i.e.,
$L$ does not contain min-sets of $R$.

Figure~\ref{fig:alg-min} shows our procedure for finding a free set
for a min-union closed constraint language.
It is straightforward to check that the procedure {\tt FindFreeSetUC} has a running time $O(nm)$, where $n$ is
the number of variables and $m$ is the number of constraints of $\Phi$. 

\begin{figure}
\begin{center}
\small
\fbox{
\begin{tabular}{l}
FindFreeSetUC($\Phi$) \\
{\rm // Input: An instance $\Phi$ of $\Csp(\bB)$ with variables $V$} \\
{\rm //  for a temporal constraint language $\bB$ preserved by \min.} \\
{\rm // Output: A free set $S \subseteq V$ of $\Phi$, or reject.} \\
{\rm // If the algorithm rejects, $\Phi$ is unsatisfiable} \\
%{\rm //   If there is no such set $S$, return \false.} \\
$S$ := $V$ \\
$recheck$ := \true \\
while $recheck$ do begin \\
\hspace{.5cm}    $recheck$ := \false \\
\hspace{.5cm}    for all $\psi \in \Phi$ do begin \\
\hspace{1cm}      if $S\cap V(\psi)\neq\emptyset$ then begin \\
\hspace{1.5cm}        $S$ := $(S \setminus V(\psi))\;\cup$
 the maximal min-set of $\psi$ contained in $S\cap V(\psi)$ \\
\hspace{1.5cm}        if $S$ \textrm{changed} then $recheck$ := \true \\
\hspace{1cm}      end \\
\hspace{.5cm}    end \\
end \\
if $S\neq\emptyset$ then return $S$ \\
else reject \\ 
end
\end{tabular}}
\end{center}
\caption{A polynomial time algorithm that computes free sets for constraint languages preserved by \min.}
\label{fig:alg-min}
\end{figure}

\begin{lemma}\label{lem:lcsp-pp-uc-sub-correct}
The procedure {\tt FindFreeSetUC} in
Figure~\ref{fig:alg-min} returns a free set of $\Phi$, or rejects.
If it rejects, $\Phi$ is unsatisfiable.
\end{lemma}
\begin{proof}
Suppose that the algorithm returns a (non-empty) set $S$. 
Then $recheck$ must be
 set to \false. Therefore, for all constraints $R(x_1,...,x_k)$ of $\Phi$ such that $S\cap \{x_1,\dots,x_k\} \neq\emptyset$
the maximal min-set of $\psi$ contained in $S$ equals $S\cap \{x_1,\dots,x_k\}$.
We conclude that $S$ is a
free set of $\Phi$.

We now have to argue that in case that $\Phi$ is satisfiable,
the algorithm does not reject (i.e., it finds a free set).
If $\Phi$ has a solution, there is some set $S'$ of variables 
that have the minimal value in this solution. At the beginning of the procedure, $S$ is set to $V$ and
therefore $S' \subseteq S$. We show that $S' \subseteq S$ during the entire execution of the procedure. 
Let $\psi = R(x_1,\dots,x_k)$ be a constraint from $\Phi$. Because
$S' \cap \{x_1,\dots,x_k\}$ 
is a min-set of $\psi$ that is contained in $S$, 
the maximal min-set of $\psi$
added to $S \setminus \{x_1,\dots,x_k\}$ 
certainly contains $S' \cap \{x_1,\dots,x_k\}$. 
Therefore, after
the modification to $S$ it still holds that $S\supseteq S'$. 
When the procedure terminates, it returns the set $S$,
because $\emptyset \neq S'\subseteq S$.
\end{proof}

\begin{theorem}\label{thm:min-alg}
If $\bB$ is preserved by $\min$ there is an algorithm solving 
$\Csp(\bB)$ in time $O(n^2m)$.
\end{theorem}
\begin{proof}
We use the procedure
{\tt FindFreeSetUC} in Figure~\ref{fig:alg-min}
for the subroutine {\tt FindFreeSet} 
in Figure~\ref{fig:alg-pp}.
Then Lemma~\ref{lem:lcsp-pp-compose} 
and Lemma~\ref{lem:lcsp-pp-uc-sub-correct} 
imply the correctness of the resulting algorithm. 
\end{proof}

\subsection*{An algorithm for languages preserved by \mi}
\label{ssect:alg-mi}
In this section we describe how to find free sets in instances
of $\Csp(\bB)$ for 
languages $\bB$ that are preserved by $\mi$.
We define the notion of a minimal
min-set: Let $\psi=R(x_1,\dots,x_k)$ be a constraint 
from an instance $\Phi$ of $\Csp(\bB)$,
and let $L \subseteq \{x_1,\dots,x_k\}$.
Let $A_1,\dots,A_l$ be all min-sets of $\psi$ that contain $L$. Because
$R$ is preserved by $\mi$, and thus is min-intersection
  closed by Lemma~\ref{lem:mi-mi},
there is a min-set $A_j$ of $\psi$ that is a subset of every 
min-set containing $L$. We call $A_j$ 
the {\em minimal min-set of $R$ containing $L$}.

The procedure for finding a free set
for min-intersection closed constraint languages
is given in Figure~\ref{fig:alg-mi}.
It is straightforward to verify that the 
above algorithm runs in time $O(n^2m)$
where $n$ is the number of variables and $m$ is the number
of constraints in $\Phi$.

\begin{figure}
\begin{center}
\small
\fbox{
\begin{tabular}{l}
FindFreeSetIC($\Phi$) \\
{\rm // Input: An instance $\Phi$ of $\Csp(\bB)$ where $\bB$ is preserved by $\mi$}  \\
{\rm // Output: A free set $S \subseteq V(\Phi)$ of $\Phi$, or reject} \\
{\rm // If the algorithm rejects, $\Phi$ is unsatisfiable} \\
%{\rm // If there is no such set $S$, return \false.} \\
for all $x\in V(\Phi)$ do begin \\
\hspace{.5cm}    $S$ := $\{x\}$ \\
\hspace{.5cm}    $recheck$ := \true;  $correct$ := \true \\
\hspace{.5cm}    while $recheck \wedge correct$ do begin \\
\hspace{1cm}       $recheck$ := \false  \\
\hspace{1cm}       for all constraints $\psi$ of $\Phi$ such that $(V(\psi)\cap S) \neq\emptyset$ do begin \\
%\hspace{1.5cm}          let $\{y_{p_1},\dots,y_{p_l}\}$ be $S \cap \{y_1,\dots,y_k\}$ \\
\hspace{1.5cm}         if there is no min-set of $\psi$ containing 
$S \cap V(\psi)$ then  $correct$ := \false \\
\hspace{1.5cm}        else begin \\
\hspace{2cm}          $S$ := $S\;\cup$ the minimal min-set of $\psi$ containing $S \cap V(\psi)$ \\
\hspace{2cm}          if $S$ \textrm{changed} then $recheck$ := \true \\
\hspace{1.5cm}        end \\
\hspace{1cm}      end \\
\hspace{.5cm}    end \\
\hspace{.5cm}    if $correct$ then return $S$ \\
end \\
reject 
\end{tabular}}

\caption{A polynomial time algorithm that computes free sets for min-intersection and shuffle closed constraint languages.}
\label{fig:alg-mi}
\end{center}
\end{figure}

\begin{lemma}\label{lem:lcsp-pp-ic-sub-correct}
The procedure {\tt FindFreeSetIC} in
Figure~\ref{fig:alg-mi} returns a free set $S$ of $\Phi$, or rejects.
If it rejects, $\Phi$ is unsatisfiable.
\end{lemma}
\begin{proof}
Suppose that 
the algorithm returns a set $S$.
The variable $\it correct$ must then be equal to \true.
When the while loop terminates,
$recheck$ equals \false, 
and so for all constraints $\psi\in\Phi$ such that $V(\psi)\cap
S\neq\emptyset$ the set $S$ did not change.
This implies that for all these constraints the minimal min-set of $\psi$ containing
$S\cap V(\psi)$ is equal to $S\cap V(\psi)$.
We conclude that $S$ is a
free set of $\Phi$.

We now have to argue that in case that $\Phi$ is satisfiable,
the algorithm does not reject.
If $\Phi$ has a solution, then there is some set $S'$ of variables that have the minimal value in this solution. Consider a run of the while loop in the procedure
{\tt FindFreeIC} for some variable $x\in S'$. 
In the beginning, it holds that
$S= \{x\} \subseteq S'$. 
For each constraint $\psi$ 
from $\Phi$ we have that $S' \cap V(\psi)$
is a min-set of $\psi$ if $S' \cap V(\psi)$ is non-empty.
Therefore, the program
variable $correct$ cannot be
set to \false\ while $S\subseteq S'$. Because we always add only
variables of the \emph{minimal} min-set of $\psi$ containing $S\cap V(\psi)$ to $S$, all these variables are always in $S'$. 
Therefore, $S$ remains a subset of $S'$ all the time,
and the algorithm does not reject.
\end{proof}

\begin{theorem}\label{thm:mi-alg}
If $\bB$ is preserved by $\mi$ there is an algorithm solving 
$\Csp(\bB)$ in time $O(n^3m)$.
\end{theorem}
\begin{proof}
We use the procedure 
{\tt FindFreeSetIC} in Figure~\ref{fig:alg-mi}
for the sub-routine {\tt FindFreeSet}
in Figure~\ref{fig:alg-pp}. 
Lemma~\ref{lem:lcsp-pp-compose} and Lemma~\ref{lem:lcsp-pp-ic-sub-correct} imply the correctness of these algorithms.
\end{proof}

\subsection*{An algorithm for languages preserved by \mx}
\label{ssect:alg-mx}
Finally, we consider languages $\bB$ preserved by $\mx$. 
Let $R$ be a relation from $\bB$.
%Let $\psi=R(x_1,...,x_k)$ be a constraint from an
%  instance $\Phi$ of $\Csp(\bB)$.
For a tuple $t \in R$, we define $\chi_{\min}(t)$ 
to be a vector from $\{0,1\}^k$ such that $\chi_{\min}(t)[i]=1$ 
if and only if $t[i]$ 
is minimal in $t$.  
We define $\chi_{\min}(R)$ to be $\{\chi_{\min}(t)\; | \; t\in R\}$.  
Since $R$ is preserved by $\mx$ and hence min-xor closed by Lemma~\ref{lem:mx-mx},  
the set $\chi_{\min}(R)$ is closed under addition 
of distinct vectors over $GF(2)$, and hence in  particular closed under the Boolean minority operation $\minority(x,y,z)=x \oplus y \oplus z$. By Theorem~\ref{thm:schaefer},
$\chi_{\min}(R)\cup\{0^k\}$ 
is exactly the set of solutions of a system of linear equations.

\begin{theorem}\label{thm:mx-alg}
If $\bB$ is preserved by $\mx$ there is an algorithm solving 
$\Csp(\bB)$ in time $O(n^4)$.
\end{theorem}
\begin{proof}
To find a free set of variables of an instance $\Phi$ of $\Csp(\bB)$ (if it exists), we first construct a system $S$ of linear equations over $GF(2)$ with variable set $\{x_v \; | \; v \in V\}$ and linear equations as described above for each constraint
in $\Phi$. It is well-known that a solution of $S$ that is distinct from $0^n$ can be computed in cubic time (by Gaussian elimination). 
If there is such a solution, then the set of variables mapped to $1$ 
is a free set of $\Phi$.
If the system has no such solution, then there is no free set of variables,
and there is no solution for $\Phi$. 
Now the claim follows from Lemma~\ref{lem:lcsp-pp-compose}
as in Theorem~\ref{thm:min-alg} and Theorem~\ref{thm:mi-alg}.
\end{proof}

\section{Classification}
\label{sect:tcsp-classification}
This section combines the previous results 
to show that every temporal constraint language has a polynomial-time
constraint satisfaction problem, or is NP-complete.  

\subsection{Classification in the presence of $<$}

\begin{lemma}\label{lem:lcsp-viol-btw-inj-tuples}
Let $f$ be a binary operation violating $\Betw$ and preserving $<$. Then
there are $t_1,t_2\in \Betw$ such that $f(t_1,t_2)$
has three distinct entries and $f(t_1,t_2)\not\in\Betw$.
\end{lemma}
\begin{proof}
Since $f$ violates $\Betw$, there are two triples $t_1,t_2\in\Betw$ such
that $t:=f(t_1,t_2)\not\in\Betw$.  Because $f$ preserves $<$, we 
can assume without loss of generality that
$t_1[1]<t_1[2]<t_1[3]$ and $t_2[1]>t_2[2]>
t_2[3]$. If $t$ has three distinct entries (in this case, 
we also say that $t$ is injective), we are done. Otherwise
we distinguish two cases:
\begin{enumerate}
\item $t[1]=t[2]=t[3]$: In that case, take a triple $s_1$
such that $s_1[1]<t_1[1]$, $s_1[2]=t_1[2]$, and
$s_1[3]=t_1[3]$. We also choose a triple $s_2$ such that
$t_2[2]<s_2[1]<t_2[1]$, $s_2[2]=t_2[2]$, and
$s_2[3]=t_2[3]$. It is straightforward to check that $s_1[1]<
s_1[2]<s_1[3]$ and $s_2[1]>s_2[2]>s_2[3]$ and thus
both triples belong to $\Betw$. Now, consider $s:=f(s_1,s_2)$.
We have that $s[2]=t[2]$,
$s[3]=t[3]$, and $s[1]<t[1]=s[2]=s[3]$ because $f$ preserves $<$.
Therefore $s\not\in \Betw$.  Take $s_1$ instead of $t_1$,
$s_2$ instead of $t_2$ and proceed with case 2.
\item If exactly two entries in $t$ have the same value, let $i,j$ be
their indices and let $k$ be the index of the entry with the unique value.  We
assume that $t[k]>t[i]$ (the other case is symmetric). It is
straightforward to verify that there is an entry in $t$ such that making
the value of this entry smaller would make $t$ injective and it would still
not be in $\Betw$. We can assume without loss of generality that $i$ is an index
of such an entry. We choose $s_1$ so that $s_1[i]<t_1[i]$,
$s_1[j]=t_1[j]$, $s_1[k]= t_1[k]$, and
$s_1[1]<s_1[2]<s_1[3]$. We choose $s_2$ such that
$s_2[i]<t_2[i]$, $s_2[j]=t_2[j]$,
$s_2[k]=t_2[k]$, and $s_2[1]>s_2[2]>s_2[3]$.  Note that $s_1,s_2\in \Betw$. The tuple
$s:=f(s_1,s_2)$ satisfies $s[i]<t[i]$,
$s[j]=t[j]$, and $s[k]=t[k]$. By the choice of $i$ we
conclude that $s$ is injective, $s\not\in \Betw$ and we are done.
\end{enumerate}
\end{proof}

We use Ramsey theory via Theorem~\ref{thm:multivariate-canonical-violation}
 to prove the following.

\begin{lemma}\label{lem:lcsp-lex-or-pp}
Let $f$ be a binary operation that preserves $<$ and violates $\Betw$. 
Then $f$ generates $\lele$, $\dll$, $\pp$, or $\dpp$.
\end{lemma}

\begin{proof}
If $f$ violates $\Betw$ and preserves $<$, then
Lemma~\ref{lem:lcsp-viol-btw-inj-tuples} asserts that there are
$t_1,t_2\in \Betw$ such that $t:=f(t_1,t_2) \not\in \Betw$
and $t$ is injective. As $f$ preserves $<$, we can assume without loss of
generality that $t_1[1]<t_1[2]<t_1[3]$ and
$t_2[1]>t_2[2]>t_2[3]$ (otherwise, we apply the argument to $f(y,x)$).

Either the triple $t$ satisfies
$t[1]>t[2]<t[3]$ or $t[1]<t[2]>t[3]$. In the
first case, let $S_1:=\{x\in\mathbb{Q}\; | \; t_1[1]<x<t_1[2]\}$,
$S_2:=\{x\in\mathbb{Q}\; | \; t_1[3]<x\}$, $T_1:=\{y\in\mathbb{Q}\; | \;
t_2[3]<y<t_2[2]\}$, and $T_2:=\{y\in\mathbb{Q}\; | \;
t_2[1]<y\}$. In the second case, let $S_1:=\{x\in\mathbb{Q}\; |
\; t_1[2]<x<t_1[3]\}$, $S_2:=\{x\in\mathbb{Q}\; | \;
x<t_1[1]\}$, $T_1:=\{y\in\mathbb{Q} \; | \; t_2[2]<y<t_2[1]\}$,
and $T_2:=\{y\in\mathbb{Q}\; | \; y<t_2[3]\}$.  See
Figure~\ref{fig:lcsp-ll-grids} for an illustration of these sets.

\begin{figure}
\begin{center}
\includegraphics[scale=.8]{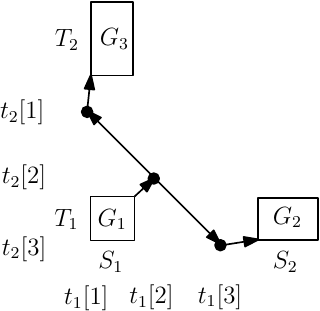}\hskip1cm\includegraphics[scale=.8]{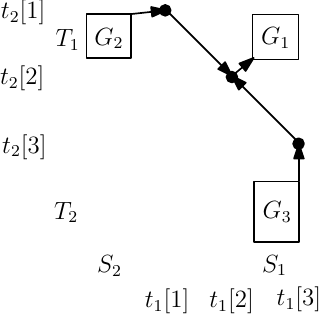}
\caption{Grids chosen for the application of the product Ramsey theorem. The depicted
ordering on the values of $f$ follows from the choice of $t_1,t_2$
and because $f$ preserves $<$.}
\label{fig:lcsp-ll-grids}
\end{center}
\end{figure}

For each $k \in \mathbb N$, we define sets $S^{(k)}_1, T^{(k)}_1, S^{(k)}_2, T^{(k)}_2$ as follows.
Apply Lemma~\ref{thm:product-ramsey} to the grid $S_1\times T_1$
 (both $S_1$ and $T_1$ are infinite and in particular larger than ${\bf R}({\bf R}(k))$),
and obtain subsets $U^{(k)}\subseteq S_1$ and
$V^{(k)}\subseteq T_1$ such that $|U^{(k)}|\geq {\bf R}(k)$, $|V^{(k)}|\geq
{\bf R}(k)$, and $f$ is canonical on $U^{(k)} \times V^{(k)}$.
Similarly, we apply Theorem~\ref{thm:product-ramsey} to the grid
$U^{(k)}\times T_2$ and obtain subsets
$S^{(k)}_1\subseteq U^{(k)}$ and $T^{(k)}_2\subseteq T_2$ of cardinality at least $k$ such that $f$ is homogenous on $S^{(k)}_1\times T^{(k)}_2$. 
We finally apply Theorem~\ref{thm:product-ramsey} 
to the grid $S_2\times V^{(k)}$ 
and obtain subsets $S^{(k)}_2\subseteq S_2$ and
$T^{(k)}_1\subseteq V^{(k)}$ of cardinality at least $k$
such that $f$ is canonical on $S^{(k)}_2\times T^{(k)}_1$.
Note that $f$ 
is in particular canonical on $S^{(k)}_1 \times T^{(k)}_1$. 

There are just $6^3$ possibilities for how $f$ behaves on those grids for given $k$. Hence, there is an infinite set $K\subseteq\mathbb{N}$ such that
$f$ behaves in the same way on $S^{(k)}_1\times T^{(k)}_1$
for all $k \in K$,
in the same way on $S^{(k)}_1\times T^{(k)}_2$ for all $k \in K$, 
and in the same way on $S^{(k)}_2\times T^{(k)}_1$ for all $k \in K$.

The following observations will be obvious by inspection
of Figure~\ref{fig:lcsp-ll-grids}, left side.
In case that $S^{(k)}_1$ is before 
%(i.e., in Figure~\ref{fig:lcsp-ll-grids} below
$S^{(k)}_2$ (that is, all elements in $S^{(k)}_1$ are smaller than all elements in $S^{(k)}_2$) and $T^{(k)}_1$
is before $T^{(k)}_2$, then by the choice of $S^{(k)}_1$, $S^{(k)}_2$, $T^{(k)}_1$, and
$T^{(k)}_2$, and because $f$ preserves $<$, we have 
$$f(x,y)<f(t_1[2],t_2[2])< f(t_1[1],t_2[1]) < f(x',y')$$ 
for all $(x,y)\in S^{(k)}_1\times T^{(k)}_1$ and 
$(x',y')\in(S^{(k)}_1\times T^{(k)}_2)$. 
Similarly, $$f(x,y)<f(t_1[2],t_2[2])< f(t_1[3],t_2[3]) < f(x'',y'')$$ 
for all $(x,y)\in S^{(k)}_1\times T^{(k)}_1$ and 
$(x'',y'') \in (S^{(k)}_2\times T^{(k)}_1)$. 
The other case is that $S_2^{(k)}$ is before $S^{(k)}_1$ and $T^{(k)}_2$ is before $T^{(k)}_1$ (see the right side of Figure~\ref{fig:lcsp-ll-grids} for an illustration). In this case $f(x,y)>f(t_1[2],t_2[2])>f(x',y')$ for all
$(x,y)\in S^{(k)}_1\times T^{(k)}_1$ and $(x',y')\in (S^{(k)}_1\times
T^{(k)}_2)\cup(S^{(k)}_2\times T^{(k)}_1)$. 

First suppose that $f$ is dominated by the same argument
on all the grids $S^{(k)}_1\times T^{(k)}_1$, $S^{(k)}_1\times T^{(k)}_2$, and $S^{(k)}_2\times T^{(k)}_1$ for all $k\in K$. We can assume that $f$ is dominated on these grids by 
the second argument; otherwise we swap the arguments of $f$.
Let $g,h \in \{\lex_{y,x},\lex_{y,-x},p_y\}$ be such that $f$ behaves like $g$ on $S^{(k)}_1\times T^{(k)}_1$ and like $h$ on $S^{(k)}_2\times T^{(k)}_1$. Then by the above observations and local interpolation $f$ generates 
$[g|h]$ if $S_1$ is before $S_2$, and $[h|g]$ if $S_2$ is before $S_1$. Moreover, we show that $f$ also generates $\lex$.
\begin{itemize}
\item If $g$ or $h$ is $\lex_{x,y}$ or $\lex_{y,x}$,
then $f$ clearly generates $\lex$. 
\item If $g$ or $h$ is
$\lex_{x,-y}$ or $\lex_{y,-x}$, then $f$ generates $\lex$ as well, as $\lex(x,-\lex(x,-y))$ 
behaves like $\lex(x,y)$.
\item If $g$ is $p_y$ and $h$ is $p_y$, then $f$ generates $\lex$ by Lemma~\ref{lem:gen}.
\end{itemize}
Note that the operation $[g|h]$ satisfies the conditions in Lemma~\ref{lem:genl-l}, and hence $\{\lex,[g|h]\}$ generates $[\lex_{y,x}|\lex_{y,x}]$. By Lemma~\ref{lem:gen}, $f$ generates $\lele$.

Now we consider the case that $f$ is dominated by different arguments
on the grids $S^{(k)}_1\times T^{(k)}_1$ and $S^{(k)}_1\times T^{(k)}_2$, or by different arguments on the grids $S^{(k)}_1\times T^{(k)}_1$ and $S^{(k)}_2\times T^{(k)}_1$, for all $k\in K$.
We only consider the first case; the second case is symmetric under swapping the arguments of $f$.
Let $g,h$ be from $\{\lex_{x,y},\lex_{x,-y},\lex_{y,x},\lex_{y,-x},p_x,p_y\}$ such that $f$ behaves like $h$ on the grids $S^{(k)}_1\times T^{(k)}_1$ and like $g$ on the grids $S^{(k)}_2\times T^{(k)}_1$.
Again, by local interpolation $f$ generates $[h|g]$ if $S_1$ is before $S_2$, and $[g|h]$ if $S_2$ is before $S_1$.
We assume without loss of generality that $f$ generates $[h|g]$
(in the other case we can exchange the names of $h$ and $g$ and proceed in the same way).

If $h$ is $p_y$ and $g$ is $p_x$, then 
$[h|g]$ behaves like $\pp$; hence $f$ generates $\pp$ and we are done. Dually, if $h$ is $p_x$ and $g$ is $p_y$,
then $f$ generates $\dpp$. 
In all other cases, either $h$ or $g$ is from $\lex_{x,y}, \lex_{y,x}, \lex_{x,-y}$, or $\lex_{y,-x}$, and
thus $f$ generates $\lex$ as we have already seen before.
But then Lemma~\ref{lem:genl-l} shows that $f$ generates $\lele$ or $\dll$.
\end{proof}

\begin{figure}[h]
\begin{center}
\includegraphics[scale=0.5]{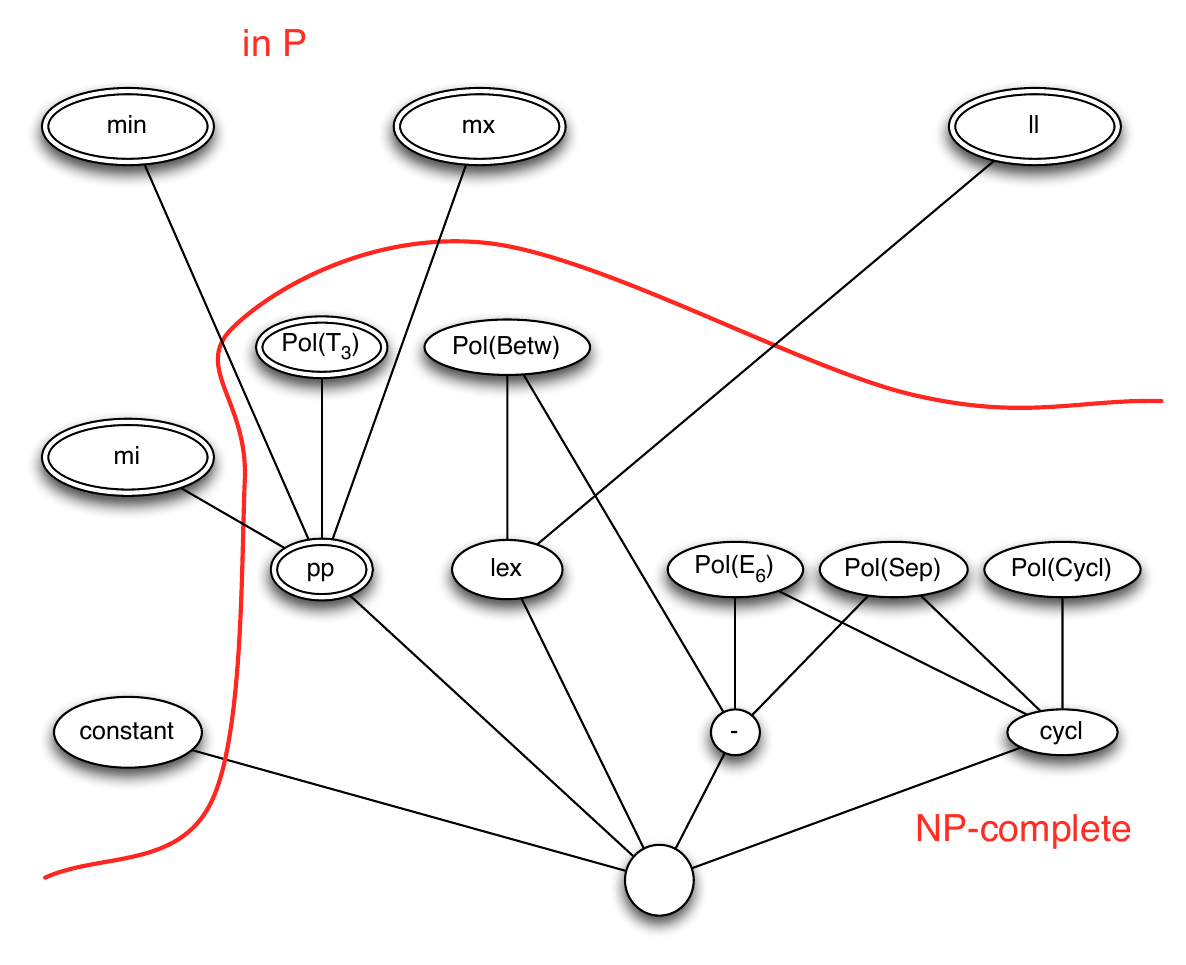}
\caption{An illustration of the classification result for temporal constraint languages. %that contain $<$. 
Double-circles mean that the corresponding operation has a dual generating a distinct clone which is not drawn in the figure.}
\label{fig:tcsp-main}
\end{center}
\end{figure}

\subsection{Summary}
We summarize our findings in the following classification statement; also see Figure~\ref{fig:tcsp-main}.

\begin{theorem}\label{thm:pre-main}
Let $\bB$ be a temporal constraint language. Then one of the following applies.
\begin{itemize}
\item $\bB$ is preserved by at least one of the following nine operations: $\lele, \min, \mi, \mx$, their duals, or a constant operation. 
%In this case, for every finite signature reduct $\bB'$ of $\bB$ the problem $\Csp(\bB)$ can be solved in polynomial time. 
\item $\Betw$, $\Cycl$, $\Sep$, $T_3$, $-T_3$, or 
$I_6$ is primitive positive definable in 
$\bB$.
%In this case, there is a finite signature reduct $\bB'$ of $\bB$ such that 
%$\Csp(\bB)$ is NP-complete.
\end{itemize}
\end{theorem}
\begin{proof}
%It is easy to check that each of the operations above violates each of the relations above. 
%If $\bB$ is preserved by a constant operation, then tractability follows from Proposition~\ref{prop:const-core}. 
%For the case that $\bB$ is preserved by $\lele$ or $\dll$ we have presented a polynomial-time algorithm in Theorem~\ref{thm:ll-alg}. 
%If $\bB$ is preserved by $\min,\mi,\mx$ or one of their duals,
%tractability is shown in Section~\ref{sect:shuffle-algs}.
Theorem~\ref{thm:cores} asserts that one of the following cases is true:
\begin{enumerate}
\item There is a primitive positive definition of $\Cycl$, $\Betw$, or $\Sep$ in $\bB$.
%,and the reduct $\bB'$ of the relations mentioned in this primitive positive definition 
%have an NP-hard CSP.
\item $\pol(\bB)$ contains a constant operation.
\item $\pol(\bB)$ contains all permutations of $\mathbb{Q}$. In this case,
Theorem~\ref{thm:ecsps} shows that $\bB$ either has a binary injective
polymorphism $g$, or the relation $I_6$ has a primitive positive definition in $\bB$. 
%Again, the relations mentioned in the primitive positive definition of $E_6$ 
In the first case, by composing $g$ with a permutation, we see that \emph{all} binary
injective operations preserve $\bB$, and hence in particular the operation $\lele$
is a polymorphism of $\bB$.
\item all $f \in\pol(\bB)$ preserve $<$.
\end{enumerate}
We are done in all cases except the fourth.
Also, we can assume that $\bB$ has a polymorphism $f$ that violates $\Betw$. 
By Lemma~\ref{lem:binary}, we can
assume that $f$ is binary. 
Then Lemma~\ref{lem:lcsp-lex-or-pp} implies that the operation $f$ 
generates \pp, \dpp, \lele, or \dll. If $f$ generates \lele\ or \dll\  there is nothing to show.
If $f$ generates $\pp$ then Lemma~\ref{lem:lcsp-pp-one-closure} shows that either $T_3$ has a primitive positive definition in $\bB$, 
or $\bB$ is preserved by $\min$, $\mi$, or $\mx$.
Dually, if $f$ generates $\dpp$ then either $-T_3$ has a primitive positive definition in $\bB$, or $\bB$ is preserved by one of the duals of $\min$, $\mi$, or $\mx$, 
which completes the proof.
\end{proof}

With the previous theorem it is easy to obtain the full complexity
classification for temporal constraint satisfcation problems, and 
finally show Theorem~\ref{thm:tcsp-main}.

% 8/11: Unfortunately, have for identities involving nesting no general argument
% that says that after expansions with constants we still have a Taylor modulo
% inner and outer automorphisms -- so we cannot use a general argument
% for the disjointness of the cases in the temporal case with this approach.
% However, relations vs polymorphism list is disjoint, and also purely mathematical,
% so we can go with this one

\begin{proof}[Proof of Theorem~\ref{thm:tcsp-main}]
When $\bB$ is preserved by
$\lele$, $\min$, $\mi$, $\mx$, one of their duals, or the constant operation,
then $\bB$ has an at most ternary weak near unanimity polymorphism modulo endomorphisms; this is immediate for
the commutative binary functions $\mx$, $\min$, their duals, and for the constant function. 
For $\lele$, this has been shown in Theorem~\ref{thm:ll-wnu}, and for $\mi$ in Theorem~\ref{thm:wnu-mi}. For dual $\mi$ and dual $\lele$ the dual argument works. 
 
%then every expansion $\bC$ of $\bB$ by finitely many constants
%is preserved by an operation $f$ satisfying 
%$f(x,y) = \alpha f(\beta y, \beta x)$ for $\alpha,\beta \in \Aut((\mQ;<))$. 
%The operations $\min$, $\mx$, their duals, and the constant binary operations are commutative, so in these cases Proposition~\ref{prop:eq-const-pres} implies that $\bC$ is even preserved
%by an operation $f$ satisfying $f(x,y) = %\alpha(f(y,x))$ for an $\alpha 
%\in \Aut(\bC)$. 
%For $\mi$ or by $\lele$ the statement has been shown in Proposition~\ref{prop:good-mi} and 
%Proposition~\ref{prop:good-ll}.

Now let $\bB'$ be a finite signature reduct of $\bB$.
If $\bB'$ is preserved by a constant operation, then tractability of $\Csp(\bB')$ follows from Proposition~\ref{prop:const-core}. 
For the case that $\bB'$ is preserved by $\lele$ or $\dll$ we have presented a polynomial-time algorithm for $\Csp(\bB')$ in Theorem~\ref{thm:ll-alg}. 
If $\bB'$ is preserved by $\min,\mi,\mx$, or one of their duals,
tractability of $\Csp(\bB')$ is shown in Section~\ref{sect:shuffle-algs}.

Now suppose that $\bB$ is not preserved by one of the listed operations.
Then by Theorem~\ref{thm:pre-main} we know that one of
the relations $\Betw$, $\Cycl$, $\Sep$, $T_3$, $-T_3$, or 
$I_6$ has a primitive positive definition in $\bB$. Each of those relations together
with finitely many constants primitively
positively interprets $(\{0,1\};\OIT)$:
\begin{itemize} 
\item For $(\mQ;\Betw,0)$ a primitive positive interpretation of $(\{0,1\};\NAE)$ has been shown in Proposition~\ref{prop:betw-hard},which also gives a primitive positive interpretation of $(\{0,1\};\OIT)$ in $(\mQ;\Betw,0)$
via Theorem~\ref{thm:simulates-tame}.
\item a primitive positive interpretation of $(\{0,1\}; \OIT)$ in $({\mathbb Q}; \Cycl)$ with parameters
has been given in Theorem~\ref{thm:cycl-hard}.
\item The structure $(\mQ; \Sep,0,1)$ primitively positively interprets $(\{0,1\};\OIT)$  
by Proposition~\ref{prop:lcsp-sep-hard}.
\item The structure $({\mathbb Q}; T_3,0)$ primitively positively interprets $(\{0,1\};\OIT)$  
by Proposition~\ref{prop:Shard}; the proof for $-T$ is dual.
\item $I_6$ primitively positively interprets $(\{0,1\};\OIT)$  by Proposition~\ref{prop:1in3hard}.
\end{itemize}
Finally, recall from Theorem~\ref{thm:t-endo}
 that if $\bB$ does not have 
a constant endomorphism, then it is a model-complete core, and hence 
Corollary~\ref{cor:ua-dich-disj} shows that 
the two cases in the statement of Theorem~\ref{thm:tcsp-main} are distinct. 
\end{proof}

%By inspection of all the temporal relations that were used to
%show hardness, we can also describe the main result
%relationally as follows.
%\begin{corollary}\label{cor:hardrels}
%If there is a primitive positive
%definition of $\Betw$, $\Cycl$, $\Sep$, $T_3$, $-T_3$, or 
%$E_6$ in $\bB$, then CSP$(\bB)$ is NP-complete. Otherwise, CSP$(\bB)$ is tractable.
%\end{corollary}
%\begin{proof}
%The result follows from the proof of the previous theorem and the theorems referenced therein, and Theorem~\ref{thm:ecsps}. 
%\end{proof}

See Figure~\ref{fig:summary} for an overview over the nine 
largest tractable temporal constraint languages; the entries also 
mention \emph{typical relations} for the respective
language, i.e., a set of relations that is contained in the language, but not contained in any other of the nine languages -- hence, these
relations show that all the languages are distinct.

\begin{figure}[h]
\begin{center}
\begin{tabular}{|llll|}
\hline
Polymorphism & Typical Relations & Complexity & Reference \\
\hline
min & $\{U,<\}$ & $O(n^2m)$ & Theorem~\ref{thm:min-alg} \\
mi & $\{I \}$ & $O(n^3m)$ & Theorem~\ref{thm:mi-alg} \\
mx & $\{X\}$ & $O(n^4)$ & Theorem~\ref{thm:mx-alg} \\
max = dual min & $\{-U,<\}$ & $O(n^2m)$ & \\
dual mi & $\{-I\}$ &  $O(n^3m)$ & \\
dual mx & $\{-X\}$ & $O(n^4)$ &  \\
ll & $\{(u \neq v) \vee (x > y) \vee (x > z)\}$ & $O(nm)$ & Theorem~\ref{thm:ll-alg} \\
dual ll & $\{(u \neq v) \vee (x < y) \vee (x < z)\}$ & $O(nm)$  & \\
constant & $\{(x \leq y \leq z) \vee (z \leq y \leq x)\}$ & $O(m)$ & \\
\hline
\end{tabular}
\caption{Summary of the various tractable languages. For the last three operations, the typical relations are given by their first-order definition; in all other cases, see Section~\ref{sect:shuffle}.}
\label{fig:summary}
\end{center}
\end{figure}

\subsection{Decidability of Tractability} 
We want to remark that the so-called \emph{meta-problem} for
tractability is decidable; this is formally stated in the following
corollary.

\begin{corollary}
There is an algorithm that, given quantifier-free first-order formulas
 $\phi_1,\dots,\phi_n$ that define over $(\mathbb Q;<)$
 the relations $R_1,\dots,R_n$, decides
whether $\Csp(\mQ;R_1,\dots,R_n)$ is tractable or NP-complete.
\end{corollary}
\begin{proof} 
Follows from Theorem~\ref{thm:pp-decidability} 
in combination with Theorem~\ref{thm:pre-main}. 
\end{proof}

\chapter{Non-Dichotomies}
\label{chap:nodich}
\begin{center}
\includegraphics[scale=0.7]{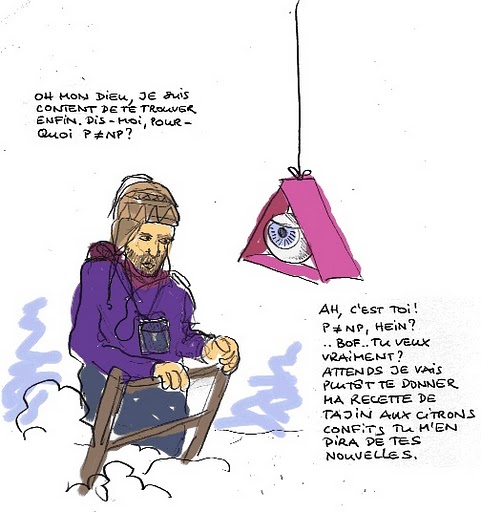}
\end{center}
% THERE IS A THIRD METHOD: in contrast to the first,
% we don't reduce a class of problems 
% that doesn't have a dichotomy to our class,
% but we rather do this with the machine descriptions of this class.
% Illustrate this with monotone SNP (make the obstructions connected by blowing up the input signature).

\vspace{.4cm}
There are basically two methods for proving that a subclass of NP
does not have a complexity dichotomy.
The first is to show that for every problem in NP there is a polynomial-time equivalent problem in the subclass. By polynomial-time equivalent we mean that there are polynomial-time Turing reductions between
the two problems.
The non-dichotomy result then follows from Ladner's theorem~\cite{Ladner}, which asserts that there are problems in NP that are
neither in P nor NP-complete, unless P=NP.
This method has been applied to show that, for example, the class
of monotone SNP does not exhibit a complexity dichotomy~\cite{FederVardi}.
We will apply this technique in Section~\ref{sect:arithmetic} and Section~\ref{sect:cspsnp} to give two different proofs of the
fact that the class of all constraint satisfaction problems with infinite domains does not have a complexity dichotomy. 
%In fact, in Section~\ref{sect:arithmetic} we show
%that there is a single infinite structure $\bC$ such that every recursively enumerable problem is polynomial-time equivalent to a CSP whose template is first-order definable in $\bC$.

The second technique to show a non-dichotomy is to directly use Ladner's proof technique,
which is sometimes called \emph{delayed diagonalization}. We will
use this method in Section~\ref{sect:coNP} to show that there
are $\omega$-categorical structures $\bB$ such that
$\Csp(\bB)$ is in coNP, but neither in P nor coNP-complete
(unless P=coNP). The question whether there are 
$\omega$-categorical structures $\bB$ such that $\Csp(\bB)$ is in NP $\setminus$ P but not NP-complete is still open.

This chapter contains results from~\cite{BodirskyGrohe} (in Section~\ref{sect:coNP})
as well as previously unpublished results.

\section{Arithmetical Templates}
\label{sect:arithmetic}
In this section we 
show that for every computational decision problem there exists
a polynomial-time equivalent constraint satisfaction problem with an infinite template $\bB$. This result was first shown in~\cite{BodirskyGrohe}. Here we present a new proof that uses Matiyasevich's theorem. In fact, we prove a stronger result, namely the existence of a single structure $\bC$ such that for every recursively enumerable problem $\mathcal P$ there is a structure $\bB$ with a first-order definition in  $\bC$ such that CSP$(\bB)$  is polynomial-time equivalent to $\mathcal P$. A second proof, based on the results
in Section~\ref{ssect:snp} of Chapter~\ref{chap:intro}, can be found 
in the next section.

Previously, Bauslaugh~\cite{BauslaughH} showed
that for every recursive function $f$ there exists 
an infinite structure $\bB$ such that CSP$(\bB)$ is decidable, but has time complexity at least $f$. More recently, Schwandtner gave
upper and lower bounds in the exponential time hierarchy for
some infinite domain CSPs~\cite{Schwandtner}; but these bounds leave an exponential gap. 

In this section we make essential use of the following theorem, which is due to %Martin
Davis, %Yuri 
Matiyasevich, %Hilary 
Putnam, and %Julia 
Robinson.

% Problem: overlooked some \infty solutions 
% for some components. More importantly:
% this value infty could be simulated by 0,
% by shifting all other values, and modifying
% the reduct relations appropriately!!

\begin{theorem}[See e.g.~\cite{Matiyasevich}]\label{thm:matiyasevitch}
A subset of $\mathbb Z$ is recursively enumerable
if and only if it has a primitive positive definition in $({\mathbb Z}; *,+,1)$, the integers with addition and multiplication.
\end{theorem}

\begin{theorem}
For every recursively enumerable problem $\mathcal P$ there exists a relational
structure $\bB$ with a first-order (in fact, a primitive positive) definition in $({\mathbb Z}; *,+,1)$ such that $\Csp(\bB)$ is polynomial-time Turing equivalent to $\mathcal P$.
\end{theorem}
\begin{proof}
Code $\mathcal P$ as a set $L$ of natural numbers,
viewing the binary encodings of natural numbers as bit strings. More precisely, 
$s \in \mathcal P$ if and only if the number 
represented in binary by the string $1s$ is in $L$.
That is, we append the symbol $1$ at the front so that for instance $00 \in \mathcal P$ and $01 \in \mathcal P$ correspond
to different numbers in $L$. 
%Let $N$ be this subset of $\mathbb N$ together with a new element $\infty$.
Now consider the structure $\bB := ({\mathbb Z}; S,D,L',N)$ 
where 
\begin{itemize}
\item $S$ is the binary relation defined by 
$$S(x,y) \Leftrightarrow \big((y=x+1 \wedge x \geq 0) \vee (x=y=-1) \big)$$
\item $D$ is the binary relation defined by 
$$D(x,y) \Leftrightarrow \big((y=2x \wedge x \geq 0) \vee (x=y=-1) \big) $$
\item $L' := L \cup \{-1\}$
\item $N := \{0\}$ 
\end{itemize}
Clearly, if $\mathcal P$ is recursively enumerable,
then $L$ and $L'$ are recursively enumerable, too. 

We have to verify that $\Csp(\bB)$ is polynomial time equivalent to $\mathcal P$. We first show that there is a polynomial-time reduction from $\mathcal P$ to CSP$(\mathfrak B)$.
View an instance of $\mathcal P$ as a number $n \geq 0$ as above,
and let $\eta(x)$ be a primitive positive definition for $x=n$ in $\mathfrak B$.
It is possible to find such a definition in polynomial time by repeatedly doubling ($y=x+x$) and incrementing ($y=x+1$) 
the value $0$ (this also follows from the more general 
Lemma~\ref{lem:pp-rational}).  
It is clear that $n$ codes a yes-instance of $\mathcal P$ if and only if
 $\exists x (\eta(x) \wedge L'(x))$ is true in $\mathfrak B$.

To reduce CSP$(\mathfrak B)$ to $\mathcal P$, 
we present a polynomial-time algorithm for CSP$(\bB)$
that uses an oracle for $\mathcal P$  (so our reduction will be a polynomial-time 
Turing reduction).
Let $\phi$ be an instance of
$\Csp(\bB)$, and 
let $H$ be the undirected graph
whose vertices are the variables $W$ of $\phi$, and which has an
edge between $x$ and $y$ if $\phi$ contains the constraint 
$S(x,y)$ or the constraint $D(x,y)$. 
Compute the connected components of $H$.
If a connected component does not contain
$x$ with a constraint $N(x)$ in $\phi$,
then we can set all variables of that component
to $-1$ and satisfy all constraints involving those
variables. 

Otherwise, suppose that we have a component
$C$ that does contain $x_0$ with a constraint $N(x_0)$. Observe that by connectivity, if there exists a solution, then all variables in $C$ must take non-negative value. Consider the following linear
system: for each constraint of the form $S(x,y)$ 
for $x,y \in C$ we add $y=x+1$ and $x \geq 0$
to the system, and for each constraint
of the form $D(x,y)$ for $x,y \in D$ we add
$z=2x$ and $x \geq 0$. Subject to $x_0=0$ this system has either one or no solution. We can check in polynomial time whether a linear system with 2 variables per constraint has no integer solution~\cite{BNO},
and if there is no solution, the algorithm rejects.
Otherwise, the algorithm assigns to each variable
$x \in C$ its unique integer value, and
if $\phi$ contains a constraint $L'(x)$, 
we call the oracle for $\mathcal P$ with the binary encoding of this value. If any of those oracle calls has a negative result, reject.
Otherwise, we have found an assignment that 
satisfies all constraints, and accept. 
\end{proof}

The universal-algebraic approach fails badly when
it comes to analysing the computational
complexity of $\Csp(\bB)$:  the semi-lattice operation $(x,y) \mapsto \max(x,y)$ preserves $\bB$ for all structures $\bB$ considered in the previous proof, and from that we cannot draw any consequences for the computational complexity of $\Csp(\bB)$. 

\section{CSPs in SNP}
\label{sect:cspsnp}
Another proof that shows that every problem in NP
is polynomial-time Turing equivalent to an infinite domain CSP
is based on a result by Feder and Vardi, and the results from Section~\ref{ssect:snp-csp}. % in Chapter~\ref{chap:intro}.

\begin{theorem}[Theorem 3 in~\cite{FederVardi}]\label{thm:feder-vardi-snp}
Every problem in NP is equivalent  to a problem in monotone SNP under polynomial-time reductions.
\end{theorem}

We show the following.

\begin{proposition}\label{prop:snp-csnp}
Every problem in monotone SNP is equivalent to a problem in monotone connected SNP under polynomial-time Turing reductions.
\end{proposition}

\begin{proof}
Let $\Phi$ be a monotone SNP sentence of the form $\exists R_1,\dots,R_k \; \forall x_1,\dots,x_l. \; \phi$ for $\phi$ quantifier-free and in conjunctive normal form. 
The sentence $\Psi$ that we are going to 
construct from $\Phi$ has an additional free relation symbol $E$, 
and an existentially quantified relation symbol $T$, and is defined by
 $$\exists R_1,\dots,R_k,T \; \forall x_1,\dots,x_l. \; \psi$$ where
$\psi$ is the quantifier-free first-order formula with the following clauses.
\begin{enumerate}
\item $\neg E(x_1,x_2) \vee T(x_1,x_2)$;
\item $\neg T(x_1,x_2) \vee \neg T(x_2,x_3) \vee T(x_1,x_3)$;
\item $\neg T(x_1,x_2) \vee T(x_2,x_1)$;
\item for each clause $\phi'$ of $\phi$ with variables $x_1,\dots,x_q$, the clause
$$\phi' \vee \bigvee_{i<j<q} \neg T(x_i,x_j) \; .$$
\end{enumerate}
The sentence $\Psi$ is clearly connected and monotone.
We are therefore left with the task to verify that $\Phi$ and $\Psi$ are equivalent under polynomial-time Turing reductions.

We start with the reduction from $\Phi$ to $\Psi$. When $\bA$ is a finite $\tau$-structure, we
expand $\bA$ to a $(\tau \cup \{E\})$-structure $\bA'$ by choosing for $E$ the full binary relation. Then also $T$ must denote
the full binary relation (so that the clauses from item (1), (2), and (3) above are satisfied), and the clauses introduced in (4) are equivalent to $\phi'$. Hence, $\Phi$ holds on $\bA$ if and only if $\Psi$ holds on $\bA'$.

For the reduction from $\Psi$ to $\Phi$, let $\bA$ be an instance of $\Psi$. We can compute the connected components $C_1,\dots,C_k$ of the $\{E\}$-reduct of $\bA$ in polynomial time in the size $\bA$. For each of those connected components $C$, we evaluate
$\Phi$ on the $\tau$-reduct $\bA_C$ of $\bA[C]$. If for one component this evaluation is negative, then $\bA[C]$ and consequently $\bA$ do not satisfy $\Psi$. 
Otherwise, for each $C$ there exists an $\tau \cup \{R_1,\dots,R_k\}$-expansion of $\bA_C$ that
satisfies $\phi$. Let $\bA'$ be the expansion of the disjoint union of all those $(\tau \cup \{R_1,\dots,R_k\})$-structures
by the relation $T$ denotes the equivalence relation with equivalence classes $C_1,\dots,C_k$.
Clearly, all clauses from items (1), (2), and (3) in the definition of $\Psi$ are satisfied by $\bA'$. 
Each $q$-tuple $(a_1,\dots,a_q)$ from elements of $\bA'$ either contains entries from different components,
and hence satisfies the disjunctions from item (4), or contains only entries from the same component $C$,
but in this case the tuple also satisfies the disjunctions from item (4) since $\bA_C$ satisfies $\Phi$.
\end{proof}

%The following is an immediate consequence of the preceding theorem and Theorem~\ref{cor:snp-csp}).

\begin{corollary}\label{cor:np-csp}
For every problem in NP 
there is a structure $\bB$
such that the problem is polynomial-time Turing equivalent to $\Csp(\bB)$. 
\end{corollary}
\begin{proof} 
By Theorem~\ref{thm:feder-vardi-snp}, every problem in NP is equivalent to a monotone SNP sentence $\Phi$ under
polynomial-time reductions. We have shown in Proposition~\ref{prop:snp-csnp} that $\Phi$ is equivalent
to a monotone connected SNP sentence $\Psi$, and by Theorem~\ref{cor:snp-csp} there exists an infinite structure $\bB$ such that $\Psi$ describes $\Csp(\bB)$. 
\end{proof}

\begin{figure}[h]
\begin{center}\includegraphics[scale=.5]{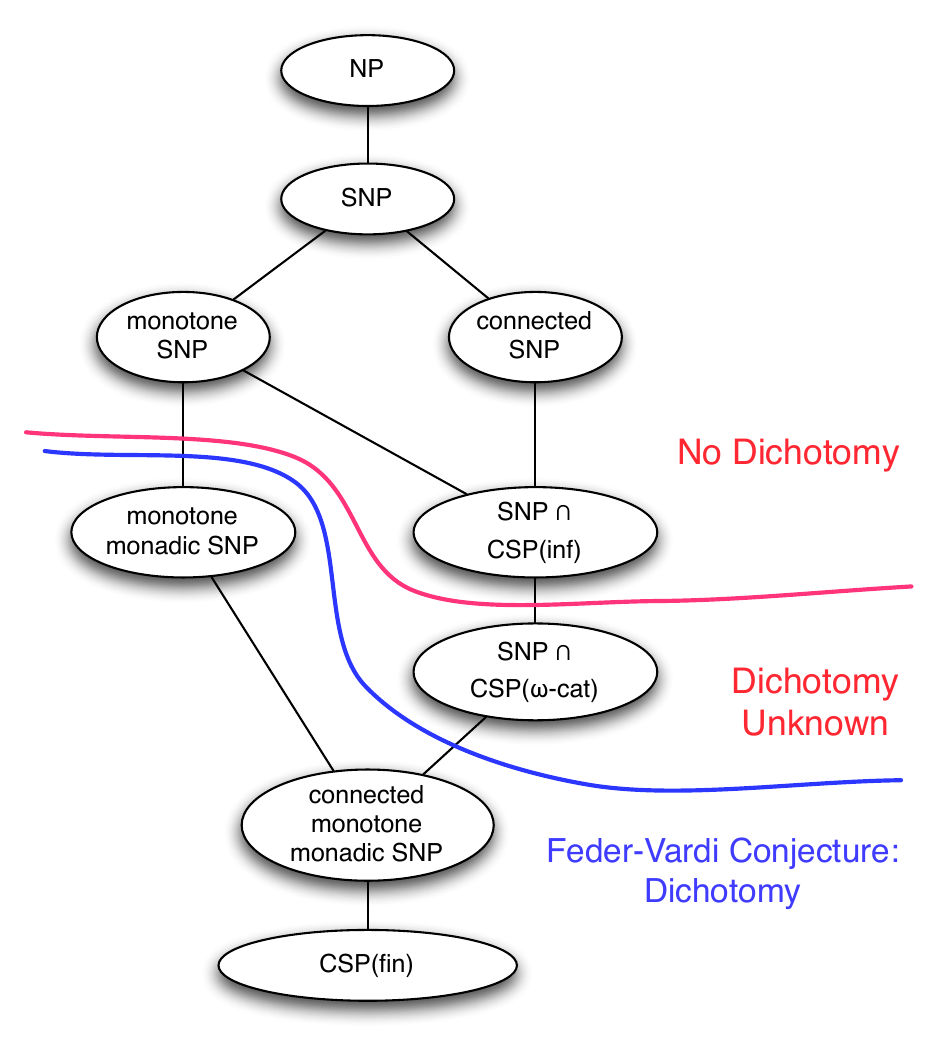} 
\end{center}
%\vspace{1cm}
\caption{Dichotomies and non-dichotomies for fragments of SNP.
CSP(inf) (CSP($\omega$-cat), CSP(fin)) refers to the class of all problems $\Csp(\bB)$ where $\bB$ is an infinite ($\omega$-categorical, finite, respectively) structure with finite relational signature.
}
\label{fig:snp}
\end{figure}

In Figure~\ref{fig:snp} the diagram about the fragments of SNP from Section~\ref{sect:snp} %in Chapter~\ref{chap:intro} 
has been decorated with information about the complexity classification status.

\section{coNP-intermediate $\omega$-categorical Templates}
\label{sect:coNP}
In this section we show that there exists an $\omega$-categorical directed graph $\bB$
such that $\Csp(\bB)$ is in coNP, but neither coNP-complete nor in P (unless
coNP=P).  All structures in this section will be \Fresse\ limits of classes of directed graphs.

Let $\cal N$ be a class of finite tournaments, and recall that $\text{Forb}(\cal N)$, the class
of all finite digraphs that does not embed a tournament from $\cal N$, is an
amalgamation class (Example~\ref{expl:Henson}).
We write $\bB_{\cal N}$ for
the \Fresse-limit of $\text{Forb}(\cal N)$. Observe that for finite $\cal N$
the problem $\Csp(\bB_{\cal N})$ can be solved in deterministic polynomial
time, because for a given instance $\bA$ of this problem an algorithm simply has
to check whether there is a homomorphism from one of the structures in $\cal
N$ to $\bA$, which is the case if and only if there is a homomorphism from $\bA$
to $\bB_{\cal N}$.

When proving that there are uncountably many homogeneous digraphs,
Henson specified an infinite set $\cal T$ of 
tournaments $T_3,T_4,\dots$ with the property that
$T_i$ does not embed into $T_j$ if $i \neq j$.
The tournament $T_n$, for $n \geq 3$, in Henson's set $\cal T$ 
has vertices
$0, \dots, n+1$, and the following edges:
\begin{itemize}
\item $(i,j)$ for $j=i+1$ and $0 \leq i \leq n$;
\item $(0,n+1)$;
\item $(j,i)$ for $j>i+1$ and $(i,j) \neq (0,n+1)$.
\end{itemize}
% the reduction should also not have too few clauses.

\begin{proposition}\label{prop:coNP}
The problem $\Csp(\bB_{\cal T})$ is coNP-complete.
\end{proposition}
\begin{proof}
The problem is contained in coNP, because 
we can efficiently test whether a sequence $v_1,\dots,v_k$ of distinct vertices
of a given directed graph $\bA$ induces $T_k$ in $\bA$,
i.e., whether $(v_i,v_j)$ is an arc in $\bA$ if and only if $(i,j)$ is an arc in $T_k$, for all $i,j \in \{1,\dots, k\}$.
If for all such sequences of vertices this test is negative,
we can be sure that $\bA$ is from $\text{Forb}(\cal T)$, and
hence homomorphically maps to $\bB_{\cal T}$. 
Otherwise, $\bA$ embeds a structure from $\cal T$,
and hence does not homomorphically map to $\bB_{\cal T}$.

The proof of coNP-hardness goes by reduction from the 
complement of the NP-complete 3SAT problem (see Example~\ref{expl:sat}),
and is inspired by a classical reduction from
3-SAT to Clique. For a given 3-SAT instance, we create an
instance $\bA$ of $\Csp(\bB_{\cal T})$ as follows:
%by first constructing an
%undirected graph $\bA$, and then orienting the edges of $\bA$.
If \[\{x_0^1,x_0^2,x_0^3\}, \dots, \{x_{k+1}^1,x_{k+1}^2,x_{k+1}^3\}\] are 
the clauses of the 3-SAT formula (we assume without loss of generality
that the 3-SAT instance has at least three clauses), then 
the vertex set of $\bA$ is 
$$\{(0,1),(0,2),(0,3),\dots,(k+1,1),(k+1,2),(k+1,3)\} \; ,$$
and the arc set of $\bA$ consists of all pairs $((i,j),(p,q))$ of vertices such
that $x_i^j \neq \neg x_p^q$ and such that $(i,p)$ is an arc in $T_k$.

We claim that a 3-SAT instance is unsatisfiable if and only
if the created instance $\bA$ 
homomorphically maps to $\bB_{\cal T}$.
The 3-SAT instance is satisfiable iff there is a mapping
from the variables to true and false such that in each clause
at least one literal, say $x_0^{j_0},\dots,x_{k+1}^{j_{k+1}}$, is true.
This is the case if and only if the vertices $(0,j_1),\dots,(k+1,j_{k+1})$
induce $T_k$ in $\bA$, i.e., $((i,j_i),(p,j_p))$ is an edge if and only
if $(i,p)$ is an edge in $T_k$. This is the case if and only if
$T_k$ embeds into $\bA$.
To conclude, it suffices to prove that $T_k$ embeds into $\bA$ if and only if $\bA$ does not 
homomorphically map to $\bB_{\cal T}$.
It is clear that if $T_k$ embeds into $\bA$, then
$\bA$ does not 
homomorphically map to $\bB_{\cal T}$.
Conversely, if $\bA$ does not homomorphically
embed to $\bB_{\cal T}$, then there exists a $j$ such
that there is an embedding $e$ of $T_j$ into $\bA$. 
Then for any $(i,j)$, $(p,q)$ in the image of $e$
we have that $(i,p)$ is an edge of $T_k$.
Therefore, the mapping that sends an element 
$u$ of $T_j$ to the first component of $e(u)$ is
an embedding of $T_j$ into $T_k$. 
Since $T_j$ and $T_k$ are homomorphically
inequivalent for all distinct $j,k \geq 3$ we obtain that 
$j = k$ and that $T_k$ embeds into $\bA$, which finishes the proof.
\end{proof}

We now modify the proof of Ladner's Theorem given in~\cite{Papa}
(which is basically Ladner's original proof)
to create a subset ${\cal T}_0$ of 
$\cal T$ such that $\Csp(\bB_{{\cal T}_0})$ is in coNP,
but neither in P nor coNP-complete (unless coNP=P).
One of the ideas in Ladner's proof is to  \emph{`blow holes into SAT'},
such that the resulting problem is too sparse to be NP-complete
and to dense to be in P. Our 
modification is that we do not blow holes into a computational problem itself, but that we \emph{`blow holes into the obstruction set $\cal T$ of $\Csp(\bB_{{\cal T}})$'}.  

In the following, we fix one of the standard encodings of graphs as strings over the alphabet $\{0,1\}$.
Let $M_1, M_2, \dots$ be an enumeration of all polynomial-time bounded Turing machines, and let $R_1,R_2,\dots$ be an enumeration
of all polynomial time bounded reductions. We assume that these enumerations are effective; it is well-known that such enumerations exist.

The definition of ${\cal T}_0$ uses a Turing machine $F$ that
computes a function $f \colon \mathbb N \rightarrow \mathbb N$, which is defined below. The set ${\cal T}_0$ is then defined as follows.
\begin{align*}
{\cal T}_0 = \{ T_n  \; | \; f(n) \text{ is even } \} 
\end{align*}
The input number $n$ is given to the machine $F$ in unary representation. The computation of $F$ 
proceeds in two phases. In the first phase, $F$ simulates itself\footnote{Note that by the fixpoint theorem of recursion theory we can assume
that $F$ has access to its own description.}
on input $1$, then on input $2$, $3$, and so on, until the number of computation steps of $F$ in this phase exceeds $n$ (we can always maintain a counter during the simulation to recognize when to stop). 
Let $k$ be the value $f(i)$ 
for the last input $i$ for which the simulation was
completely performed by $F$.

In the second phase, the machine stops if phase two takes more than $n$ computation steps, and $F$ returns $k$. We distinguish whether $k$ is even or odd. If $k$ is even, 
all directed graphs $\bA$ on $s=1,2,3,\dots$ vertices are enumerated.
%(all graphs with $s$ vertices 
%appear in this enumeration before all graphs with $s+1$ vertices, 
%for all $s$). 
For each directed graph $\bA$ in the enumeration
the machine $F$ simulates $M_{k/2}$ on the encoding of $\bA$.
Moreover, $F$ computes whether 
$\bA$ homomorphically maps to $\bB_{{\cal T}_0}$.
This is the case if for all structures $T_l \in \cal T$ that embed into $\bA$ the value of $f(l)$ is even.
So $F$ tests for $l=1,2,\dots,s$ whether $T_l$ embeds to $\bA$
($F$ uses any straightforward exponential time algorithm for this purpose), and if it does, simulates itself on input $l$ 
to find out whether $f(l)$ is even.
If 
\begin{itemize}
\item[(1)] $M_{k/2}$ rejects and 
$\bA$ homomorphically maps to $\bB_{{\cal T}_0}$,
or 
\item[(2)] $M_{k/2}$ accepts and 
$\bA$ does not homomorphically map to $\bB_{{\cal T}_0}$,
\end{itemize}
then $F$ returns $k+1$ (and $f(n)=k+1$).

The other case of the second phase is that $k$ is odd.
Again $F$ enumerates all directed graphs $\bA$ on $s=1,2,3,\dots$
vertices, and simulates the computation of $R_{\lfloor k/2 \rfloor}$
on the encoding of $\bA$. Then $F$ computes whether 
the output of $R_{\lfloor k/2 \rfloor}$ encodes a directed graph $\bA'$ that homomorphically maps to $\bB_{{\cal T}_0}$.
The graph $\bA'$ homomorphically maps to $\bB_{{\cal T}_0}$
iff for all tournaments $T_l$ that embed into $\bA'$ the value $f(l)$ is even.
Whether $T_l$ embeds into $\bA'$ is tested with a straightforward exponential-time algorithm. To test whether $f(l)$ is even, $F$ simulates itself on input $l$. 
Finally, $F$ tests with a straightforward exponential-time algorithm whether $\bA$ homomorphically maps to $\bB_{\cal T}$.
If 
\begin{itemize}
\item[(3)] $\bA$ homomorphically maps to $\bB_{\cal T}$ and
$\bA'$ does not homomorphically map to $\bB_{{\cal T}_0}$, or 
\item[(4)] $\bA$ does not homomorphically map to $\bB_{\cal T}$ and
$\bA'$ homomorphically maps to $\bB_{{\cal T}_0}$, 
\end{itemize}
then 
$F$ returns $k+1$.

\begin{lemma}\label{lem:mon}
The function $f$  is a non-decreasing function, that is, 
for all $n$ we have $f(n) \leq f(n+1)$.
\end{lemma}
\begin{proof}
We inductively assume that $f(s-1) \leq f(s)$
for all $s \leq n$, and have to show that $f(n) \leq f(n+1)$.
Since $F$ has more time to simulate itself when we run it on
$n+1$ instead of $n$, the value $i$ computed in the first phase of $F$
cannot become smaller.
By inductive assumption, $k=f(i)$ cannot become smaller as well.
In the second phase, we either return $k$ or $k+1$.
Hence, if $k$ becomes larger in the first phase, the output
of $F$ cannot become smaller. 
If $k$ does not become larger, then the only difference between
the second phase of $F$ for input $n+1$ compared to input $n$
is that there is more time for the computations.
Hence, if the machine $F$ on input $n$ verifies 
condition (1),(2),(3),(4) for some graph $\bA$ (and hence returns $k+1$), then $F$
also verifies this condition for $\bA$ on input $n+1$, and returns $k+1$
as well. Otherwise, $f(n)=k$, and also here $f(n+1) \geq f(n)$ holds.
\end{proof}

%It is impossible that $f(n)$ is constant for all but finitely many inputs.
\begin{lemma}\label{lem:inc}
For all $n_0$ there exists an $n > n_0$
such that $f(n) > f(n_0)$ (unless coNP $\neq$ P).
\end{lemma}

\begin{proof}%[of Lemma~\ref{lem:inc}]
Assume for contradiction that there exists an $n_0$ such that
$f(n)$ equals a constant $k_0$ for all $n \geq n_0$. 
Then there also exists an $n_1$ such that for all $n \geq n_1$
the value of $k$ computed by the first phase of $F$ on input $n$
is $k_0$. 

If $k_0$ is even, 
then on all inputs $n \geq n_1$ the second phase of $F$ 
simulates $M_{k_0/2}$ on encodings of an enumeration of graphs.
Since the output of $F$ must be $k_0$, for all graphs neither (1) nor (2)
can apply. Since this holds for all $n \geq n_1$,
the polynomial-time bounded machine $M_{k_0/2}$ correctly decides $\Csp(\bB_{{\cal T}_0})$, and hence $\Csp(\bB_{{\cal T}_0})$
is in P. But then there is the following polynomial-time algorithm 
that solves $\Csp(\bB_{\cal T})$, a contradiction to coNP-completeness of $\Csp(\bB_{\cal T})$ (Proposition~\ref{prop:coNP}) and our assumption that coNP $\neq$ P.

\begin{center}
\fbox{
\begin{tabular}{l}
Input: A directed graph $\bA$. \\ \\
If $\bA$ homomorphically maps to $\bB_{{\cal T}_0}$ then accept. \\
Test whether one of the finitely
many graphs in ${\cal T} \setminus {\cal T}_0$ embeds into $\bA$.\\
Accept if none of them embeds into $\bA$. \\
Reject otherwise.
\end{tabular}}
\end{center}

If $k_0$ is odd, then on all inputs $n \geq n_1$ the
second phase of $F$ does not find a graph $\bA$ for which (3)
or (4) applies, because the output of $F$ must be $k_0$. 
Hence, $R_{\lfloor k_0/2 \rfloor}$ is a polynomial-time reduction
from $\Csp(\bB_{\cal T})$ to $\Csp(\bB_{{\cal T}_0})$,
and by Proposition~\ref{prop:coNP} the problem $\Csp(\bB_{{\cal T}_0})$ is coNP-hard. But note that because
$f(n)$ equals the odd number $k_0$ for all but finitely many $n$, 
the set ${\cal T}_0$ is finite. Therefore,
$\Csp(\bB_{{\cal T}_0})$ can be solved in polynomial time,
contradicting our assumption that coNP $\neq$ P.
\end{proof}

\begin{theorem}
$\Csp(\bB_{{\cal T}_0})$ is in coNP, but neither in P nor coNP-complete (unless coNP=P).
\end{theorem}
\begin{proof}
It is easy to see that $\Csp(\bB_{{\cal T}_0})$ is in coNP.
On input $\bA$ the algorithm non-deterministically chooses
a sequence of $l$ vertices, and checks in polynomial time 
whether this sequence induces a copy of $T_l$.  If yes, the algorithm computes
$f(l)$, which can be done in linear time by executing $F$ on the unary representation of $l$. 
If $f(l)$ is even, the algorithm accepts. Recall that 
$\bA$ does not homomorphically map to 
$\bB_{{\cal T}_0}$ iff a tournament $T_l \in {\cal T}_0$ 
embeds into $\bA$, which is the case iff there is an accepting computation path for the above non-deterministic algorithm.

Suppose that $\Csp(\bB_{{\cal T}_0})$ is in $P$.
Then for some $i$ the machine $M_i$ decides 
$\Csp(\bB_{{\cal T}_0})$. By Lemma~\ref{lem:mon}
and Lemma~\ref{lem:inc} there exists an $n_0$ such that
$f(n_0)=2i$. Then there must also be an $n_1 > n_2$ such that
the value $k$ computed during
the first phase of $F$ on input $n_1$ equals $2i$.
Since $M_i$ correctly decides $\Csp(\bB_{{\cal T}_0})$, 
the machine $F$ returns $2i$ on input $n_1$. 
By Lemma~\ref{lem:mon}, the machine $F$ also returns $2i$
for all inputs from $n_1$ to $n_2$, and by induction
it follows that it $F$ returns $2i$ for \emph{all} inputs larger
than $n \geq n_0$, in contradiction to Lemma~\ref{lem:inc}.

Finally, suppose that $\Csp(\bB_{{\cal T}_0})$ is coNP-complete.
Then for some $i$ the machine $R_i$ is a valid reduction
from $\Csp(\bB_{\cal T})$ to $\Csp(\bB_{{\cal T}_0})$.
Again, by Lemma~\ref{lem:mon}
and Lemma~\ref{lem:inc} there exists an $n_1$ such that
the value $k$ computed during
the first phase of $F$ on input $n_1$ equals $2i$.
Since the reduction $R_i$ is correct, the machine $F$
returns $2i$ on input $n_1$, and in fact returns $2i$ on all inputs
greater than $n_1$.  This contradicts Lemma~\ref{lem:inc}. 
\end{proof}

\chapter{Future Work}
\label{chap:concl}
We conclude the thesis by mentioning four promising directions of future work.

\tocless\section{Phylogeny Constraints}
We have presented a classification of the complexity of $\Csp(\bB)$ for all structures $\bB$ with a first-order
definition over $({\mathbb Q};<)$, or over the random graph $({\mathbb V};E)$.
One might ask which other structures, besides $({\mathbb Q};<)$ and the random graph $({\mathbb V};E)$,
are interesting and promising candidates for such a classification. 
A very interesting candidate is the structure $({\mathbb L};|)$ (or equivalently, over a relatively 3-transitive $C$-set), introduced in Section~\ref{ssect:c-relation}. 
The class of CSPs that can be formulated with templates that can be defined over this structure is very large
and contains many problems that have been independently studied in the literature, capturing for instance the rooted triple satisfaction problem and the quartet satisfiability problem from phylogenetic analysis. 
All the tools we needed for complexity classification are available: $({\mathbb L};|)$ is homogeneous, 
and an appropriate order expansion of it is Ramsey (see Example~\ref{expl:ordered-c-relation}). 

\tocless\section{Datalog}
%\subsection{Datalog}\label{ssect:datalog}
Feder and Vardi~\cite{FederVardi} observed that 
all the known algorithms for solving $\Csp(\bB)$, 
for a finite structure $\bB$,  are either based on algebraic
algorithms that can be seen as generalizations of Gaussian elimination, or based on simple `constraint propagation',
or combinations of these two paradigms. 
This is still the case today. An elegant way to formalize
algorithms that perform constraint propagation is Datalog.
Datalog can be seen as conjunctive queries that have been
extended by a recursion mechanisms; alternatively, one can
view Datalog as Prolog (see e.g.~\cite{SchoeningLogic}) without function symbols. In the context of constraint satisfaction
Datalog has been introduced in~\cite{FederVardi} and further studied in~\cite{KolaitisVardi}. Some of the early contributions were equivalent
characterizations of the expressive power of Datalog
in terms of \emph{bounded treewidth duality} and \emph{existential pebble games}. 

Recently, Barto and Kozik~\cite{BoundedWidth}
presented an exact characterization of those
CSPs where $\Csp(\bB)$ can be solved by a Datalog program.
The characterization is universal-algebraic (see Chapter~\ref{chap:algebra}), and confirming 
a conjecture of Larose and Zadori~\cite{LaroseZadori}. It
was later shown to be equivalent to a conjecture
 made already by Feder and Vardi in~\cite{FederVardi}, see~\cite{AbilityToCount}.

Datalog programs are very useful to solve infinite-domain constraint satisfaction problems as well. It has been shown in~\cite{BodDalJournal} that
when $\bB$ is an $\omega$-categorical structure, then the characterizations
of the expressive power of Datalog in terms of bounded treewidth duality
and existential pebble games remain valid. This has been applied
to show that several fundamental infinite-domain CSPs in the 
literature cannot be solved by Datalog~\cite{ll,phylo-long}.
It would be very interesting to have an algebraic characterization of the expressive power of Datalog
for CSPs with $\omega$-categorical templates. 

\tocless\section{Topological Clones}
We have seen in Section~\ref{sect:bi-interpret} that the topological automorphism group
of an $\omega$-categorical structure $\Gamma$ describes $\Gamma$ up to 
bi-interpretability; that is, two $\omega$-categorical structures $\Gamma$ and $\Delta$ whose
automorphism groups are isomorphic as topological groups are first-order bi-interpretable. 

We have also seen that the right tool for the complexity study of CSPs is \emph{primitive positive} interpretability, 
and not first-order interpretability; see Section~\ref{sect:pseudo-var}. So it is natural to ask in this context whether primitive positive 
interpretability can be characterized in terms of the polymorphism clone viewed as a \emph{topological clone},
that is, viewed as an abstract clone equipped with the topology of point-wise convergence. We did not define
abstract clones; but in this context it suffices to know that they relate to clones in the same way as
permutation groups relate to abstract groups. 

A partial result in this direction has been obtained in joint work with Junker~\cite{BodJunker};
one of the results proven there is that two $\omega$-categorical structures without constant endomorphism are \emph{existential positive bi-interpretable}\footnote{A first-order interpretation is called \emph{existential positive} if all the involved formulas of the interpretation are existential positive; \emph{existential positive bi-interpretations} are defined by a similar modification of first-order bi-interpretations.} if and only if their transformation monoids, viewed as topological monoids, are isomorphic. 

The goal to lift this further to primitive positive interpretability  amounts to showing that the topological clone of $\bB$ characterizes the pseudo-variety generated by the polymorphism algebra of $\bB$, 
via Theorem~\ref{thm:pp-interpret}.

\vspace{1cm}
\tocless\section{A Logic for P?}
In Section~\ref{sect:snp} we have seen a logical characterization of the complexity class NP:
by Fagin's theorem, a problem is in NP if and only if it can be described in existential second-order logic. 
A similar logic for the complexity class P is not known. 
The question whether there exists a logic for P has been formalized by Gurevich~\cite{Gurevich}
and became one of the most influential questions in finite model theory~\cite{GroheLogicforP}.

One approach to shed some light on this question is to identify large fragments of existential second-order logic 
such that the set of sentences in this fragment
that describe problems in P has an effective enumeration.
For example, consider the logic of \emph{connected monotone SNP}. Theorem~\ref{thm:feder-vardi-snp} and Proposition~\ref{prop:snp-csnp} show that every problem in NP is polynomial-time equivalent to a problem in connected monotone SNP. The proof of Theorem~\ref{thm:feder-vardi-snp} and Proposition~\ref{prop:snp-csnp} is constructive in the sense
that from a non-deterministic Turing machine we can
effectively construct the corresponding connected monotone SNP sentence. So if there were an algorithm
that enumerates those connected monotone SNP sentences that describe a problem in P, then the question to Gurevich's question is positive (Gurevich conjectured that the answer is negative). 
Such an algorithm probably does not exist. But it might exist for fragments of connected monotone SNP. 
Recall that every sentence in connected monotone SNP 
describes a CSP. Hence, complexity classification for infinite domain constraint satisfaction can also be motivated by the quest for a logic for P.

\bibliographystyle{abbrv}
\bibliography{../../global}
\end{document}